%% file: PhysReport.tex
\documentclass[11pt,twoside,a4paper,fleqn,titlepage]{article}
\usepackage[english]{babel}
\usepackage[dvipdfm]{graphicx}
\usepackage{subfigure}
\usepackage{styles/cite}
\usepackage{hyperref}
\usepackage{amsmath,amsfonts,amssymb,mathrsfs}

\makeatletter 
\long\def\@makecaption#1#2{\footnotesize
\vskip\abovecaptionskip 
\sbox\@tempboxa{#1: #2}
\ifdim \wd\@tempboxa >\hsize
#1: #2\par \else
\global \@minipagefalse
\hb@xt@\hsize{\hfil\box\@tempboxa\hfil}
\fi
\vskip\belowcaptionskip} 
\makeatother
\input styles/ISS-Physics-Group-Report_def.tex
\input 02-Standard-Neutrino-Model/def_temp.tex

\input 03-New-physics-and-the-SnuM/def_temp.tex
\input 05-Performance/def_temp.tex
\input 06-Alternatives/def_temp.tex
\begin{document}
%
%
\pagestyle{plain}
\setcounter{secnumdepth}{5}
\setcounter{tocdepth}{3}
\selectlanguage{english}
\makeatletter
\renewcommand{\l@subsection}{\@dottedtocline{2}{1.0em}{2.3em}}
\renewcommand{\l@subsubsection}{\@dottedtocline{3}{2.0em}{3.2em}}
\renewcommand{\l@paragraph}{\@dottedtocline{4}{3.0em}{4.1em}}
\renewcommand{\l@subparagraph}{\@dottedtocline{5}{4.0em}{5em}}
\makeatother
%
\pagestyle{empty}
\begin{figure}
  \vspace{-1cm}
  \begin{center}
    \includegraphics[width=\textwidth]{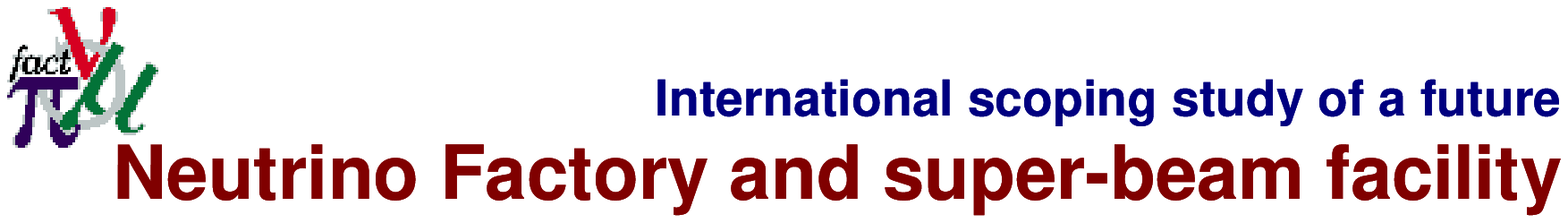}
  \end{center}
  \includegraphics[width=0.15\textwidth]{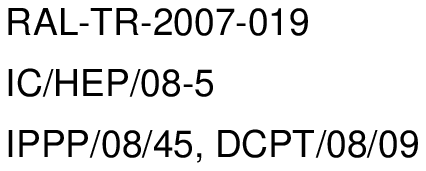}
\end{figure}
\begin{center}
  {\bf\LARGE 
    Physics at a future Neutrino Factory \\
    and super-beam facility
  }
\end{center}
\vspace{0.5cm}

\input 00-Abstract/00-Abstract

\vspace{0.5cm}
\begin{center}
  {\Large \bf \it The ISS Physics Working Group} \\
\vspace{1cm}
{\bf Editors:}  \\
  S.F.~King\footnote{
    University of Southampton, School of Physics and Astronomy,
    Highfield, Southampton S017 1BJ, United Kingdom.
  },
  K.~Long\footnote{
  Imperial College London, Blackett Laboratory, Department
  of Physics, Exhibition Road, London, SW7 2AZ, United
  Kingdom.
  },
  Y.~Nagashima\footnote{
  Department of Physics, Osaka University, Toyonaka,
  Osaka 560-0043, Japan. 
  },
  B.L.~Roberts\footnote{
    Boston University, Department of Physics, 590 Commonwealth Avenue,
    Boston, MA 02215, USA
  }, and
  O.~Yasuda\footnote{
  Department of Physics, Tokyo Metropolitan University,
  Hachioji, Tokyo 192-0397, Japan.
  }.
  \vspace{1cm}   \\
  $21^{\rm st}$ February 2009
\end{center}
\vfill
\cleardoublepage
%
%
\cleardoublepage
\input 00-AuthorList/00-AuthorList
\cleardoublepage
%
\vfill
\pagestyle{plain}
\pagenumbering{roman}                   
\setcounter{page}{1}
\input 00-ExecSumm/00-ExecSumm.tex
\vfill
\cleardoublepage
%
\pagestyle{plain}
\parindent 10pt
%
\tableofcontents
\cleardoublepage
\pagenumbering{arabic}                   
\setcounter{page}{1}
%
\input 01-Introduction/01-Introduction

\input 02-Standard-Neutrino-Model/02-Standard-Neutrino-Model

\input 03-New-physics-and-the-SnuM/03-New-physics-and-the-SnuM

\input 04-BSnuM/04-BSnuM

\input 05-Performance/05-Performance

\input 06-Alternatives/06-Alternatives

\input 07-Muon-Physics/07-Muon-Physics

%
\section*{Acknowledgements}

We would like to thank the management of Rutherford Appleton
Laboratory, CCLRC and PPARC, for their support during the study and to
Professor~J.~Wood in particular for initiating the study.
During the course of the year, we were welcomed at a number of
laboratories across the world and therefore thank the management of
the CERN, KEK, RAL laboratories and the University of Irvine for
hosting the ISS plenary meetings.
The work of the group benefitted greatly from the Physics Working
Group meetings at Imperial, Boston, and Valencia; we would like to
thank the groups involved for welcoming us to their laboratories.
Finally, our warm and sincere thanks go to all those who contributed
to the plenary and working group meetings, either through making
presentations or by contributing to the discussion, and to those who
contributed to the production of this report.

The authors acknowledge the support of CARE, contract number
RII3-CT-2003-506395. 
The work was supported by the Science and Technology Facilities
Council under grant numbers 
PPA/G/S2003/00512, PP/B500790/1, PP/B500815/1, PP/B500882/1, and
through SLAs with STFC supported laboratories. 
The authors also acknowledge the support of the World Premier
International Research Center Initiative (WPI Initiative), MEXT,
Japan.
This research was partially supported by the Director, Office of
Science, Office of High Energy Physics, of the U.S. Department of
Energy, under contract nos. DE-AC02-05CH11231, DE-AC02-98CH10886,
DE-AC02-07CH11359, and DE-AC03-76SF00098.
The work was also supported by the U.S. National Science Foundation 
under grants PHY-0355245 and PHY-04-57315. 

The authors acknowledge the support of Grants-in-Aid
for Scientific Research, Japan Society for the Promotion of
Science.
%
\cleardoublepage
\bibliographystyle{styles/utcaps}
\bibliography{PhysReport}
\cleardoublepage
\input 08-Appendix/08-Appendix
%
\end{document}

%% file: styles/ISS-Physics-Group-Report_def.tex
\def\half{{\scriptscriptstyle 1/2}}

\def\eV{\ensuremath{\text{e}\text{V\/}}}

\def\GeV{\ensuremath{\text{Ge}\text{V\/}}}

\def\cm{\,\text{cm}}

\def\SU2U1{{\rm SU}(2)\times{\rm U}(1)}

\def\det{{\rm det}}
\def\dof{{\rm dof}}

\def\exp{{\rm exp}}

\def\min{{\rm min}}

\mathchardef\qsm=63
\mathchardef\pls=43
\mathchardef\mns=512
\mathchardef\plm=518
\mathchardef\eql=61
\mathchardef\smallleft=300
\mathchardef\smallright=301
\mathchardef\perslsh=47
\mathchardef\les=316
\mathchardef\gre=318
\mathchardef\leq=532
\mathchardef\grq=533
\chardef\usc=95
\chardef\til=126

\def\sqr#1#2#3{{\vcenter{\hrule height.#3ex\hbox{\vrule width.#2ex height#1ex
    \kern#1ex\vrule width.#3ex}\hrule height.#2ex}}}

\def\angleto{\vrule width.035em height2.1ex depth-.56ex\unskip\kern-.6ex\to}
\def\perchc#1{{\raise.4ex\hbox{$\mkern4mu#1{\it\perslsh}_
             {\mkern-5mu\scriptscriptstyle{{\rm o}\!{\rm o}}}^
             {\mkern-12.8mu\scriptscriptstyle{\rm o}}$}}}

\catcode`\@=11 
\def\parenbar{\mathpalette\p@renb@r}
\def\p@renb@r#1#2{\vbox{%
  \ifx#1\scriptscriptstyle \dimen@.7em\dimen@ii.2em\else
  \ifx#1\scriptstyle \dimen@.8em\dimen@ii.25em\else
  \dimen@1em\dimen@ii.4em\fi\fi \offinterlineskip
  \ialign{\hfill##\hfill\cr
    \vbox{\hrule width\dimen@ii}\cr
    \noalign{\vskip-.3ex}%
    \hbox to\dimen@{$\mathchar300\hfil\mathchar301$}\cr
    \noalign{\vskip-.3ex}%
    $#1#2$\cr}}}
\catcode`\@=12 

\newbox\struttbox
\setbox\struttbox=\hbox{\vrule height1.65ex depth.485ex width0pt}
\def\strutt{\relax\ifmmode\copy\struttbox\else\unhcopy\struttbox\fi}
\def\stru#1#2{\relax\ifmmode\hbox{\vrule height#1 depth#2 width0pt}
\else\vrule height#1 depth#2 width0pt\fi}

\def\ronum#1{\uppercase\expandafter{\romannumeral#1}}
\def\ronuml#1{\expandafter{\romannumeral#1}}

\DeclareMathAlphabet{\mathbf}{OT1}{cmr}{bx}{sl}

\newcommand{\ra}        {$\rightarrow$}

\newcommand{\gsim}      {\mbox{\raisebox{-0.4ex}{$\;\stackrel{>}{\scriptstyle \sim}\;$}}}
\newcommand{\lsim}      {\mbox{\raisebox{-0.4ex}{$\;\stackrel{<}{\scriptstyle \sim}\;$}}}

 {\end{list}}
 {\end{list}}
 {\end{list}}

\catcode`\@=11 
\newlength{\@fninsert}
\newlength{\@fnwidth}
\renewcommand{\@makefntext}[1]%
  {\noindent\makebox[\@fninsert][r]{\@makefnmark}\hfil%
  \parbox[t]{\@fnwidth}{#1}}
\catcode`\@=12 
\catcode`\@=11 
\renewcommand\section{\@startsection{section}{1}{\z@}%
                                   {-3.5ex \@plus -1ex \@minus -.2ex}%
                                   {2.3ex \@plus.2ex}%
                                   {\normalfont\Large\bfseries}}
\renewcommand\subsection{\@startsection{subsection}{2}{\z@}%
                                   {-3.25ex\@plus -1ex \@minus -.2ex}%
                                   {1.5ex \@plus .2ex}%
                                   {\normalfont\large\bfseries}}
\renewcommand\subsubsection{\@startsection{subsubsection}{3}{\z@}%
                                   {-3.25ex\@plus -1ex \@minus -.2ex}%
                                   {1.5ex \@plus .2ex}%
                                   {\normalfont\large\bfseries}}
\renewcommand\paragraph{\@startsection{paragraph}{4}{\z@}%
                                   {3.25ex \@plus1ex \@minus.2ex}%
                                   {1.2ex \@plus .2ex}%
                                   {\normalfont\normalsize\bfseries}}
\catcode`\@=12 

%% file: 02-Standard-Neutrino-Model/def_temp.tex
\newcommand{\raw}{\rightarrow}
\newcommand{\<}{\langle}
\renewcommand{\>}{\rangle}
\newcommand{\tc}{ \theta_{13} }
\newcommand{\tatm}{ \theta_{23} }

\def\nue{\ensuremath{\nu_{e}}}
\def\nubare{\ensuremath{\overline{\nu}_{e}}}

\def\numu{\ensuremath{\nu_{\mu}}}
\def\nubarmu{\ensuremath{\overline{\nu}_{\mu}}}

\def\simge{\mathrel{%
   \rlap{\raise 0.511ex \hbox{$>$}}{\lower 0.511ex \hbox{$\sim$}}}}
\def\simle{\mathrel{
   \rlap{\raise 0.511ex \hbox{$<$}}{\lower 0.511ex \hbox{$\sim$}}}}

\newcommand{\kl}{\mbox{KamLAND~}}
\newcommand{\obb}{0\mbox{$\nu\beta\beta$}}
\newcommand{\meff}{\mbox{$\langle m \rangle$}}
\newcommand{\onbb}{neutrinoless double beta decay }
\newcommand{\nme}{\mbox{nuclear matrix elements}}
\newcommand{\meffnh}{\mbox{$\langle m \rangle$}^{\rm NH}}
\newcommand{\ms}{\Delta m^2_{21}}
\newcommand{\sss}{\sin^2 \theta_{12}}
\newcommand{\csh}{\sin^2 \theta_{13}}
\newcommand{\ma}{\Delta m^2_{31}}
\newcommand{\sch}{\sin^2 \theta_{13}}
\newcommand{\meffih}{\mbox{$\langle m \rangle$}^{\rm IH}}

\newcommand{\meffihmin}{\mbox{$\langle m \rangle$}^{\rm IH}_{\rm min}}

\newcommand{\meffnhmax}{\mbox{$\langle m \rangle$}^{\rm NH}_{\rm max}}

\newcommand{\SPbea}{\begin{equation} \begin{array}{c}}
\newcommand{\SPeea}{ \end{array} \end{equation}}
\newcommand{\bad}{\begin{array}{ccc}}
\newcommand{\SPea}{\end{array}} 
\newcommand{\pmns}{\mbox{$ U_{\rm PMNS}$}}
\newcommand{\dmsol}{\mbox{$\Delta m^2_{\odot}$~}}
\newcommand{\dma}{\mbox{$\Delta m^2_{\rm A}$ }}
\newcommand{\deltaatm}{\mbox{$\Delta m^2_{31}$}}
\newcommand{\mmin}{\mbox{$m_{\mbox{}_{\rm MIN}}$}}
\newcommand{\deltasol}{\mbox{$ \Delta m^2_{\odot}$}}
\newcommand{\mefff}{\mbox{$ < \! m \! > $}}
\newcommand{\hbeta}{$\mbox{}^3 {\rm H}$ $\beta$-decay }

%% file: 03-New-physics-and-the-SnuM/def_temp.tex
\let\vev\VEV

\def\e6{$E(6)$}
\def\10{$SO(10)$}
\def\21{$SU(2) \otimes U(1) $}

\def\422{$SU(4) \otimes SU(2) \otimes SU(2)$ }
\def\321{$SU(3) \otimes SU(2) \otimes U(1)$ }

\newcommand{\eVq}  {\rm{eV}^2}
\newcommand{\Sol}  {\textsc{sol}} 
\newcommand{\Atm}  {\textsc{atm}} 

\newcommand{\Dms}  {\Delta m^2_\Sol}
\newcommand{\Dma}  {\Delta m^2_\Atm}
\newcommand{\Dml}{\Delta m^2_\Lsnd}

\newcommand{\be}{\begin{equation}}
\newcommand{\ee}{\end{equation}}
\newcommand{\beba}{\begin{equation}\begin{array}{lcl}}
\newcommand{\eaee}{\end{array}\end{equation}}

\def\simlt{\mathrel{\lower2.5pt\vbox{\lineskip=0pt\baselineskip=0pt
           \hbox{$<$}\hbox{$\sim$}}}}
\def\simgt{\mathrel{\lower2.5pt\vbox{\lineskip=0pt\baselineskip=0pt
           \hbox{$>$}\hbox{$\sim$}}}}

\newcommand{\ba}{\begin{eqnarray}}
\newcommand{\ea}{\end{eqnarray}}
\usepackage{epsfig}
\def\lsim{\mathrel{\rlap{\lower4pt\hbox{\hskip1pt$\sim$}}
    \raise1pt\hbox{$<$}}}                
\def\gsim{\mathrel{\rlap{\lower4pt\hbox{\hskip1pt$\sim$}}
    \raise1pt\hbox{$>$}}}                

\def\half{\textstyle{\frac{1}{2}}}

\def\ltap{\ \raisebox{-.4ex}{\rlap{$\sim$}} \raisebox{.4ex}{$<$}\ }
\def\gtap{\ \raisebox{-.4ex}{\rlap{$\sim$}} \raisebox{.4ex}{$>$}\ }
\def\simlt{\stackrel{<}{{}_\sim}}
\def\simgt{\stackrel{>}{{}_\sim}}
\def\be{\begin{equation}}
\def\ee{\end{equation}}
\def\beq{\begin{equation}}
\def\eeq{\end{equation}}

\def\SU{\text{SU}}
\def\SO{\text{SO}}

\def\<{\left\langle}
\def\>{\right\rangle}

\def\D{\mathrm{d}} 

\def\SU{\text{SU}}
\def\SO{\text{SO}}

\def\beq{\begin{equation}}
\def\eeq{\end{equation}}

\def\<{\left\langle}
\def\>{\right\rangle}

\DeclareMathOperator{\diag}{diag}

\def\gtap{\ \raisebox{-.4ex}{\rlap{$\sim$}} \raisebox{.4ex}{$>$}\ }
\def\ltap{\mathrel{
   \rlap{\raise 0.511ex \hbox{$<$}}{\lower 0.511ex \hbox{$\sim$}}}}
\def\ra{\rightarrow}
\newcommand{\baz}{\begin{array}{cc}}

\def\gtap{\mathrel{ \rlap{\raise 0.511ex \hbox{$>$}}{\lower 0.511ex
   \hbox{$\sim$}}}} 
\def\ltap{\mathrel{ \rlap{\raise 0.511ex
   \hbox{$<$}}{\lower 0.511ex \hbox{$\sim$}}}}
     \newcommand{\betabeta}{\mbox{$(\beta \beta)_{0 \nu}$}}

\newcommand{\LNP}{\Lambda_{\text{NP}}}
\newcommand{\LEW}{\Lambda_{\text{EW}}}
\newcommand{\SUW}{\text{SU(2)}_L}

\newcommand{\Meff}{M_{\text{eff}}}
\newcommand{\dmm}[1]{(\Delta m^2_{#1})_m}
\newcommand{\Eres}{E_{\text{res}}}
\newcommand{\ENP}{E_{\text{NP}}}
\newcommand{\ete}{\epsilon_{e\tau}}
\newcommand{\etu}{\epsilon_{\mu\tau}}
\newcommand{\eue}{\epsilon_{e\mu}}
\newcommand{\ve}{\varepsilon}

\newlength{\myem}
\newcommand{\sep}[1]{#1}
\newcounter{mysubequation}[equation]
\renewcommand{\themysubequation}{\alph{mysubequation}}
\newcommand{\mytag}{\stepcounter{mysubequation}%
\tag{\theequation\protect\sep{\themysubequation}}}
\newcommand{\globallabel}[1]{\refstepcounter{equation}\label{#1}}

\newcommand{\fracwithdelims}[4]{\left#1 \frac{#3}{#4} \right#2}

\newcommand{\as}{^{*)}}

\newcommand{\CL}   {C.L.}
\newcommand{\SlKm} {\textsc{sol+kam}}
\newcommand{\KtK}  {\textsc{k2k}}

\newcommand{\Lsnd} {\textsc{lsnd}}
\newcommand{\Sbl}  {\textsc{sbl}}
\newcommand{\Nev}  {\textsc{nev}}

%% file: 05-Performance/def_temp.tex
\newcommand{\bi}{\begin{itemize}}
\newcommand{\ei}{\end{itemize}}
\newcommand{\bea}{\begin{eqnarray}}
\newcommand{\eea}{\end{eqnarray}}

\newcommand{\ldm}{\Delta m_{31}^2}
\newcommand{\sdm}{\Delta m_{21}^2}

\newcommand{\stheta}{\sin^2  2 \theta_{13} }

\newcommand{\JHFHK}{{\sc T2HK}}

\def\nue{\ensuremath{\nu_{e}}}
\def\nubare{\ensuremath{\overline{\nu}_{e}}}

\def\numu{\ensuremath{\nu_{\mu}}}
\def\nubarmu{\ensuremath{\overline{\nu}_{\mu}}}

\def\simge{\mathrel{%
   \rlap{\raise 0.511ex \hbox{$>$}}{\lower 0.511ex \hbox{$\sim$}}}}
\def\simle{\mathrel{
   \rlap{\raise 0.511ex \hbox{$<$}}{\lower 0.511ex \hbox{$\sim$}}}}

\newcommand{\eg}{{\it e.g.}}

\newcommand{\cf}{{\it cf.}}

\def\nue{\ensuremath{\nu_{e}}}
\def\nubare{\ensuremath{\overline{\nu}_{e}}}

\def\numu{\ensuremath{\nu_{\mu}\ }}
\def\nubarmu{\ensuremath{\overline{\nu}_{\mu}}}

\newcommand{\BB}{$\beta$B}

\def\anue{\overline{{\mathrm\nu}}_{\mathrm e}}
\def\anumu{\overline{{\mathrm\nu}}_{\mathrm \mu}}

\newcommand{\bbph}{\mathop{\beta\beta}}

%% file: 06-Alternatives/def_temp.tex
\newcommand{\sig}{$3\sigma$}
\newcommand{\sa}{\sin^2 \theta_{23}}
\newcommand{\mat}{\Delta m^2_{31}{\mbox {(true)}}}
\newcommand{\sat}{\sin^2 \theta_{23}{\mbox {(true)}}}
\newcommand{\scht}{\sin^2 \theta_{13}{\mbox {(true)}}}

%% file: 00-Abstract/00-Abstract.tex
\centerline{\bf Abstract}

The conclusions of the Physics Working Group of the International
Scoping Study of a future Neutrino Factory and super-beam facility
(the ISS) are presented. 
The ISS was carried out by the international community between NuFact05,
(the 7th International Workshop on Neutrino Factories and Superbeams,
Laboratori Nazionali di Frascati, Rome, June 21-26, 2005) and NuFact06
(Irvine, California, 24{30 August 2006). 
The physics case for an extensive experimental programme to understand
the properties of the neutrino is presented and the role of
high-precision measurements of neutrino oscillations within this
programme is discussed in detail. 
The performance of second generation super-beam experiments, beta-beam
facilities, and the Neutrino Factory are evaluated and a quantitative
comparison of the discovery potential of the three classes of facility
is presented. 
High-precision studies of the properties of the muon are complementary
to the study of neutrino oscillations. 
The Neutrino Factory has the potential to provide extremely intense
muon beams and the physics potential of such beams is discussed in the
final section of the report.

%% file: 00-AuthorList/00-AuthorList.tex
{\noindent
  A.~Bandyopadhyay, S.~Choubey, R.~Gandhi, S.~Goswami
  \\{\it
    Harish-Chandra Research Institute, Chhatnag Road, Jhunsi,
    Allahabad, 211 019, India
  } 
  \par \filbreak
  B.L.~Roberts
  \\{\it
    Boston University, Department of Physics, 590 Commonwealth Avenue,
    Boston, MA 02215, USA
  } 
  \par \filbreak
  J.~Bouchez
  \\{\it
    Service de Physique des Particules, CEA -- Saclay, 
    F-91191 Gif sur Yvette Cedex, France
  } 
  \par \filbreak
  I.~Antoniadis, J.~Ellis, G.F.~Giudice, T.~Schwetz\footnote{
Present address:~Max-Planck-Institut f\"ur Kernphysik,
Postfach 103980, 69029 Heidelberg, Germany}
  \\{\it 
    Department of Physics, CERN Theory Division, 1211 Geneva 23,
    Switzerland
  }
  \par \filbreak
  S.~Umasankar
  \\{\it
    Institute of Mathematical Sciences, Taramani, C.I.T. Campus,
    Chennai 600113, India
  } 
  \par \filbreak
  G.~Karagiorgi\footnote{
Present address:~Massachusetts Institute of Technology, Cambridge, MA 02139, USA},
A.~Aguilar-Arevalo, J.~M.~Conrad
\footnote{
Present address:~Massachusetts Institute of Technology, Cambridge, MA 02139, USA},
M.~H.~Shaevitz
  \\{\it
  Department of Physics, Columbia University, New York, NY 10027, USA
  }
  \par \filbreak
  S.~Pascoli
  \\{\it
    University of Durham, Department of Physics, 
    Ogen Center for Fundamental Physics, South Road,
    Durham, DH1 3LE, United Kingdom
  } 
  \par \filbreak
  S.~Geer
  \\{\it 
  Fermilab, Batavia, IL 60510-0500, USA 
  }
  \par \filbreak
  J.E.~Campagne
  \\{\it 
  LAL, Universit\'{e} Paris-Sud 11, B\^{a}timent 200, F-91898 Orsay
  cedex, France
  }
  \par \filbreak
  M.~Rolinec
  \\{\it
    Physik-Department T30d, Technische Universitaet Muenchen, 
    James-Franck-Strasse, 85748 Garching, Germany
  } 
  \par \filbreak
  A.~Blondel, M. Campanelli
  \\{\it
    Departement de Physique Nucleaire et Corpusculaire (DPNC),
    Universite de Geneve, Geneve, Switzerland
  } 
  \par \filbreak
  J.~Kopp, M.~Lindner
  \\{\it
    Max-Planck-Institut f\"ur Kernphysik,
    Postfach 103980, 69029 Heidelberg, Germany
  } 
  \par \filbreak
  J.~Peltoniemi
  \\{\it
    Centre for Underground Physics in Pyh\"asalmi,
    Universty of Oulu, Oulu, Finland
  } 
  \par \filbreak
  P.J.~Dornan, K.~Long, T.~Matsushita, C.~Rogers, Y.~Uchida
  \\{\it
    Imperial College London, Blackett Laboratory, Department of Physics,
    Exhibition Road, London, SW7 2AZ, United Kingdon
  } 
  \par \filbreak
  M.~Dracos
  \\{\it
    Centre de Recherches Nucleaire, Hautes Energies,
    Institut National de Physique Nucleaire et des Particules,
    Universite Louis Pasteur, Paris, France
  } 
  \par \filbreak
  K.~Whisnant
  \\{\it 
  Department of Physics and Astronomy, Iowa State University,
  Ames, IA 50011, USA
  }
  \par \filbreak
  D.~Casper, Mu-Chun~Chen\footnote{And 
    Theoretical Physics Department, Fermilab, Batavia, IL 60510-0500, USA 
                       }
  \\{\it
    Department of Physics and Astronomy, University of California, 
    Irvine, CA 92697, USA
  }
  \par \filbreak
  B.~Popov
  \\{\it
    Joint Institute for Nuclear Research, Joliot-Curie 6, 141980, Dubna,
    Moscow Region, Russia
  } 
  \par \filbreak
  J.~\"Ayst\"o
  \\{\it
    Department of Physics, University of Jyv\"askyl\"a, PB 35 (YFL) FIN-40351 Jyv\"askyl\"a, Finland
  } 
  \par \filbreak
  D.~Marfatia
  \\{\it
    Department of Physics and Astronomy, University of Kansas,
    Lawrence, KS 66045, USA
  } 
  \par \filbreak
  Y.~Okada, H.~Sugiyama\footnote{
Present address:~Department of Physics, Ritsumeikan University, Kusatsu, Shiga 525-8577, Japan}
  \\{\it 
    Theory Group, KEK, 1-1 Oho, Tsukuba, Ibaraki 305-0801, Japan
  }
  \par \filbreak
  K.~Jungmann
  \\{\it
    Kernfysisch Versneller Instituut, University of Groningen, Zernikelaan 25, 9747AA
    Groningen, The Netherlands}
  } 
  \par \filbreak
  J.~Lesgourgues
  \\{\it 
    Laboratoire de Physique Theorique, LAPTH, F-74941, Annecy-le-Vieux
    Cedex, France 
  }
  \par \filbreak
  M.~Zisman
  \\{\it
    Lawrence Berkeley National Laboratory, Physics Division,
    1 Cyclotron Road, 50R4049, Berkeley, CA 94720-8153 USA
  } 
  \par \filbreak
  M.A.~T\'ortola\footnote{
Present address:~II. Institut f\"ur theoretische Physik,
Universit\"at Hamburg, Luruper Chausee 149, 22761 Hamburg, Germany}
  \\{\it
  CFTP and Departamento de Fisica, Instituto Superior Tecnico. Av. Rovisco
  Pais, 1049-001 Lisboa, Portugal
  }
  \par \filbreak
  A.~Friedland
  \\{\it
    Los Alamos National Lab, P.O. Box 1663, Los Alamos, NM 87545, USA
  } 
  \par \filbreak
  S.~Davidson
  \\{\it
    IPN de Lyon, Universit\'e  Lyon 1, CNRS,
    4 rue Enrico Fermi, Villeurbanne, F-69622 cedex France
  } 
  \par \filbreak
  S.~Antusch\footnote{
Present address:~Max-Planck-Institut f\"ur Physik (Werner-Heisenberg-Institut),
    F\"ohringer Ring 6, 80805 M\"unchen, Germany
                     },
  C.~Biggio\footnote{
Present address:~Max-Planck-Institut f\"ur Physik (Werner-Heisenberg-Institut),
F\"ohringer Ring 6, 80805 M\"unchen, Germany},
A.~Donini, E.~Fernandez-Martinez\footnote{
Present address:~Max-Planck-Institut f\"ur Physik (Werner-Heisenberg-Institut),
F\"ohringer Ring 6, 80805 M\"unchen, Germany},
B.~Gavela, M.~Maltoni,
  J.~Lopez-Pavon, S.~Rigolin\footnote{
Present address:~Dipartimento d Fisica ~Galileo Galilei~,
Universit~a di Padova, via Marzolo 8, I-35131 Padova, Italia}
  \\{\it
    Universidad Autonoma de Madrid, Facultad de Ciencias C-XVI,
    Departamento de Fisica Teorica (IFT), UAM-CSIC,
    Cantoblanco, Madrid 28049, Spain
  } 
  \par \filbreak
  N.~Mondal
  \\{\it
    Tata Institute of Fundamental Research, School of Natural Sciences,
    Homi Bhabha Road, Mumbai 400005, India
  } 
  \par \filbreak
  V.~Palladino
  \\{\it
    Universit\`a Federico II and INFN  Napoli, Complesso Universitario di Monte S. Angelo, via Cintia, I-80126 Napoli, Italy
  } 
  \par \filbreak
  F.~Filthaut
  \\{\it
    Katholieke Universiteit Nijmegen, HEFIN - High Energy Physics Inst.,
    P.O. Box 9010, NL-6500 GL Nijmegen, Netherlands
  } 
  \par \filbreak
  C.~Albright\footnote{And 
    Theoretical Physics Department, Fermilab, Batavia, IL 60510-0500, 
    USA 
                      }
  \\{\it 
    Department of Physics, Northern Illinois University, DeKalb,
  IL 60115, USA 
  }
  \par \filbreak
  A.~de~Gouvea
  \\{\it
    Northwestern University  633 Clark Street  Evanston, IL 60208, USA
  } 
  \par \filbreak
  Y.~Kuno, Y.~Nagashima
  \\{\it
    Department of Physics, Osaka University, 
    Toyonaka, Osaka 560-0043, Japan
  } 
  \par \filbreak
  M.~Mezzetto
  \\{\it
    Univ. di Padova, Dept. di Fisica, via Marzolo 8, I-35100 Padua, Italy
  } 
  \par \filbreak
  S.~Lola
  \\{\it 
    Department of Physics, University of Patras, GR-26100 Patras, Greece
  }
  \par \filbreak
  P.~Langacker
  \\{\it 
    Department of Physics and Astronomy, University of Pennsylvania,
    Philadelphia, PA 19104 
  }
  \par \filbreak
  A.~Baldini
  \\{\it
    INFN Pisa, Edificio C - Polo Fibonacci Largo B. Pontecorvo, 3, I-56127 Pisa, Italy
  } 
  \par \filbreak
  H.~Nunokawa
  \\{\it
    Departamento de F\'{\i}sica, Pontif{\'\i}cia Universidade Cat{\'o}lica
    Ddo Rio de Janeiro, C. P. 38071, 22452-970, Rio de Janeiro, Brazil
  } 
  \par \filbreak
  D.~Meloni
  \\{\it
  INFN and Dipto.~di Fisica, Universit\`a degli Studi di Roma
  ``La Sapienza'', P.le A. Moro 2, I-00185, Rome, Italy
  } 
  \par \filbreak
  M.~Diaz
  \\{\it 
    Departamento de Fisica, Universidad Catolica de Chile, Avenida
    Vicuna Mackenna 4860, Santiago, Chile
  }
  \par \filbreak
  S.F.~King
  \\{\it
    University of Southampton, School of Physics and Astronomy, Highfield,
    Southampton S017 1BJ, United Kingdom
  } 
  \par \filbreak
  K.~Zuber
  \\{\it
    University of Sussex, Department of Physics and Astronomy, Falmer,
    Brighton, Sussex BN1 9QH, United Kingdom
  } 
  \par \filbreak
  A.~G.~Akeroyd\footnote{
Present address:~Department of Physics, National Central University, Jhongli 320, Taiwan}
  \\{\it 
    Department of Physics, National Cheng Kung University, Tainan 701, Taiwan
  }
  \par \filbreak
  Y.~Grossman\footnote{
Present address:~Newman Laboratory of Elementary Particle Physics
Cornell University, Ithaca, NY 14853, USA}
  \\{\it
    Department of Physics, Technion, Haifa 32000, Israel
  } 
  \par \filbreak
  Y.~Farzan
  \\{\it
    Institute for Studies in Theoretical Physics and Mathematics,
    P.O. Box 19395-1795, Tehran, Iran
  } 
  \par \filbreak
  K.~Tobe\footnote{
Present address:~Department of Physics, Nagoya University, Nagoya 464-8602, Japan}
  \\{\it 
    Department of Physics, Tohoku University,
    Aoba-ku, Sendai, Miyagi 980-8578, Japan
  }
  \par \filbreak
  Mayumi~Aoki\footnote{
Present address:~Department of Physics, Tohoku University, Sendai 980-8578, Japan}
  \\{\it 
    Theory Group, Institute of Cosmic Ray Research, University of Tokyo,
    5-1-5 Kashiwanoha, Kashiwa, Chiba 277-8582, Japan
  }
  \par \filbreak
  H.~Murayama
  \\{\it
    Institute for the Physics and Mathematics of the Universe, 
    University of Tokyo, Kashiwa, Japan 277-8568 \\
    Berkeley Center of Theoretical Physics, University of California, 
    Berkeley, CA 94720 \\
    Theoretical Physics Group, Lawrence Berkeley National Laboratory, 
    MS 50A-5104, Berkeley, CA 94720
  }
  \par \filbreak
  N.~Kitazawa, O.~Yasuda
  \\{\it
    Department of Physics, Tokyo Metropolitan University,
    Hachioji, Tokyo 192-0397, Japan
  } 
  \par \filbreak
  S.~Petcov\footnote{
Also at:~Institute of Nuclear Research and Nuclear Energy, Bulgarian Academy of Sciences, 1784, Sofia, Bulgaria},
A.~Romanino
  \\{\it
  International School for Advanced Studies (SISSA/ISAS), via Beirut 2-4,
    I-34100 Trieste, Italy
  } 
  \par \filbreak
  P.~Chimenti, A.~Vacchi
  \\{\it
    INFN presso il Dipartimento di Fisica dell'Universit\`{a} di Trieste,
    Via Valerio, 2, I - 34127 Trieste, Italia
  } 
  \par \filbreak
  A.Yu.~Smirnov\footnote{
Also at:~Institute for Nuclear Research, Russian Academy of Sciences, Moscow, Russia}
  \\{\it
    Abdus Salam International Centre for Theoretical Physics
    Strada Costiera 11, I-34014 Trieste, Italy
  } 
  \par \filbreak
  E.~Couce, J.J.~Gomez-Cadenas, P.~Hernandez, M.~Sorel, J.~W.~F.~Valle
  \\{\it
    Instituto de F\'{i}sica Corpuscular, IFIC, CSIC and Universidad de 
    Valencia, Spain
  } 
  \par \filbreak
  P.F.~Harrison
  \\{\it Department of Physics, University of Warwick, Coventry, 
         Warwickshire, CV4 7AL, England
  }
  \par \filbreak
  C.~Lunardini\footnote{
Present address:~Arizona State University, Tempe, AZ 85287-1504, USA and
RIKEN BNL Research Center, Brookhaven National Laboratory, Upton, NY 11973, USA}
  \\{\it
  Institute for Nuclear Theory and Department of Physics, University 
  of Washington, Seattle, WA 98195, USA
  } 
  \par \filbreak
  J.K.~Nelson
  \\{\it
    Department of Physics, College of William and Mary, P.O. Box 8795,
    Williamsburg, VA 23187-8795, USA
  } 
  \par \filbreak
  V.~Barger, L.~Everett, P.~Huber\footnote{
Present address:~Department of Physics, IPNAS, Virginia Tech, Blacksburg, VA 24061, USA}
  \\{\it
    Department of Physics, University of Wisconsin,
    2506 Sterling Hall, 1150 University Avenue, Madison, WI 53706, USA
  } 
  \par \filbreak
  W.~Winter
  \\{\it
    Universit\"at W\"urzburg, Lehrstuhl fur Theoretische Physik II, Am Hubland,
    D-97074 W\"urzburg, Germany
  } 
  \par \filbreak
  W.~Fetscher
  \\{\it
    Institute for Particle Physics (IPP),
    ETH Z\"urich, HPK G 30, Schafmattstr. 20, CH-8093 Z\"urich, Switzerland
  } 
  \par \filbreak
  A.~van~der~Schaaf
  \\{\it
    Physik-Institut der Universit\"at Z\"urich, CH-8057 Z\"urich, Switzerland
  } 
  \par \filbreak
}

%% file: 00-ExecSumm/00-ExecSumm.tex
\section*{Executive summary}

The International Scoping Study of a future Neutrino Factory and
super-beam facility (the ISS) was carried by the international
community between NuFact05, (the $7^{\rm th}$ International Workshop
on Neutrino Factories and Superbeams, Laboratori Nazionali di
Frascati, Rome, June 21--26, 2005) and NuFact06 (Irvine, California,
24--30 August 2006).  
The physics case for the facility was evaluated and options for the
accelerator complex and the neutrino detection systems were studied.
The principal objective of the study was to lay the foundations for a
full conceptual-design study of the facility. 
The plan for the scoping study was prepared in collaboration by the
international community that wished to carry it out; the ECFA/BENE
network in Europe, the Japanese NuFact-J collaboration, the US
Neutrino Factory and Muon Collider collaboration and the UK Neutrino
Factory collaboration. 
STFC's Rutherford Appleton Laboratory was the host laboratory for the
study.
The study was directed by a Programme Committee advised by a
Stakeholders Board. 
The work of the study was carried out by three working groups: the
Physics Group; the Accelerator Group; and the Detector Group.  
Four plenary meetings at CERN, KEK, RAL, and Irvine were held during
the study period; workshops on specific topics were organised by the
individual working groups in between the plenary meetings. 
The conclusions of the study was presented at NuFact06.
This document, which presents the Physics Group's conclusions, was
prepared as the physics section of the ISS study group.
More details of the ISS activities can be found at
http://www.hep.ph.ic.ac.uk/iss/.

Neutrino oscillations are the sole body of experimental evidence for
physics beyond the Standard Model of particle physics. 
The observed properties of the neutrino--the large flavour mixing and
the tiny mass--are believed to be consequences of phenomena which
occur at energies never seen since the Big Bang.
Neutrino facilities to pursue the study of oscillation phenomena are
therefore complementary to high-energy colliders and competitive
candidates for the next world-class facilities for particle physics.
Neutrino oscillations also provide a window on important issues in
astrophysics and cosmology. 
Ongoing and approved experiments utilise intense pion
beams (super-beams) to generate neutrinos. 
They are designed to seek and measure the third mixing
angle, $\theta_{13}$, of the neutrino mixing matrix (the `PMNS' matrix), but
will have little or no sensitivity to matter-antimatter symmetry
violation. 
Several neutrino sources have been conceived to reach high sensitivity
and to allow the range of measurements necessary to remove all
ambiguities in the determination of oscillation parameters.
The sensitivity of these facilities is well beyond that of the
presently approved neutrino oscillation programme. 
Studies so far have shown that the Neutrino Factory, an intense
high-energy neutrino source based on a stored muon beam, gives the
best performance over virtually all of the parameter space; its time
scale and cost, however, remain important question marks. 
Second-generation super-beam experiments using megawatt proton drivers
may be an attractive option in certain scenarios. 
Super-beams have many components in common with the Neutrino Factory. 
A beta-beam, in which electron neutrinos (or anti-neutrinos) are
produced from the decay of stored radioactive ion beams, in
combination with a second-generation super-beam, may be a competitive
option. 

The role of the ISS Physics Group was to establish the strong
physics case for the various proposed facilities and to find optimum
parameters for the accelerator and detector systems from the physics
point of view.
The first objective of this report, therefore, is to try to answer the
big questions of neutrino physics; questions such as the origin of
neutrino mass, the role that neutrinos played in the birth of the
universe, and what the properties of the neutrino can tell us about
the unification of matter and force.
These questions form the basis for the clarification of the physics
cases for various neutrino facilities. 
Since it is not (yet) possible to answer these questions in general,
studies have concentrated on more specific issues that may lead to
answers to the big questions.
In particular, studies have addressed such issues as:
\begin{enumerate}
  \item
    The relevance of neutrino physics to the understanding of dark
    matter and dark energy, the connection between neutrino mass and
    leptogenesis and galaxy-cluster formation;
  \item
    The connection of predictions at the grand-unification scale with
    low energy phenomena in the framework of the see-saw mechanism and
    supersymmetric extensions of the Standard Model; and
  \item
    The understanding of flavour and the connection between quarks and
    leptons, the possible existence of hidden flavour quantum numbers
    that may be connected with the small mixings among the quarks, the
    mass hierarchy of the quarks and charged leptons, and the relationship
    of these phenomena with the neutrino mass matrix. 
\end{enumerate}
  
The second objective of this report is to review the predictions of
the various models for the physics that gives rise to neutrino
oscillations and to review the associated phenomenology that is
relevant for precision measurements of neutrino oscillations.
For this purpose, we have evaluated the degree to which the various
facilities, alone or in combination, can distinguish between the
various models of neutrino mixing and determine optimum parameter sets
for these investigations.
A class of directly-testable predictions is afforded by the fact that
the GUT and family symmetries result in relationships between the
quark- and lepton-mixing parameters. 
These relationships can be cast in the form of sum rules. 
One example that can be used to discriminate amongst various models is: 
\begin{equation}
  \label{Eq:T12Sigma}
  \theta^\Sigma_{12} \equiv \theta_{12} - \theta_{13}\cos(\delta) \, ,
  \nonumber
\end{equation}
where $\theta^{\Sigma}_{12}$ can be predicted in classes of flavour
models while $\theta_{12}$ (the solar mixing angle) and
$\theta_{13}\cos (\delta )$ (the product of the small mixing angle and
the cosine of the CP violating phase) are measured experimentally. 
Another class of test is afforded by the investigation of the
unitarity of the PMNS matrix.
While the quark-mixing (CKM) matrix is constrained to be unitary in
the Standard Model, the PMNS matrix, which originates from physics
beyond the Standard Model, may not be exactly unitary; this is the
case, for example, in see-saw models. 
The third class of the test is the existence of flavour-changing
interactions that might appear at the production point, in the 
oscillation that occur during propagation, or at the point of
detection.
The possible strong correlations between lepton-flavour violation and
neutrino oscillations are also discussed.

The potential of non-accelerator, long-baseline neutrino oscillation
measurements were also considered.
Significant improvements in the precision with which the
solar parameters are known could be made using a new long-baseline
reactor experiment or by using gadolinium loading of the water in the 
Super-Kamiokande detector to increase its sensitivity to solar
neutrinos.
A large, underground, magnetised-iron detector could be used to
improve the precision of the atmospheric mixing parameters, to
determine the octant degeneracy, and to search for deviations from
maximal atmospheric mixing.
   
The third, and key, objective of this report is to present the first
detailed comparison of the performance of the various facilities. 
Using realistic specifications, we have estimated the likely
performance, tried to find optimum combinations of facilities,
baselines, and neutrino energies, and attempted to identify some
staging scenarios.
The cases considered are described in detail in the main report, only
a brief summary is given here.
Although the Neutrino Factory can achieve very large data samples with
small backgrounds, it operates at energies considerably higher than
the first oscillation peak $(E_{max}/\GeV=L/564 ~{\rm km})$.
Because of this, at intermediate values of 
$\theta_{13}$ ($10^{-3} \lsim \sin^2 2 \theta_{13} \lsim 10^{-2}$)
the Neutrino Factory with only one golden-channel 
($\nu_e\to \nu_{\mu}$ and $\bar{\nu}_e\to \bar{\nu}_{\mu}$) detector (at, say, 4000~km) can not 
resolve all parameter degeneracies and the precision of the
measurement of a particular parameter is reduced by correlations 
among the parameters. 
These problems can be resolved in one of three ways:
\begin{enumerate}
  \item
    Placing a second detector at a different baseline (i.e. varying 
    the ratio $L/E$ );
  \item
    Adding a detector sensitive to the silver channel 
    ($\nu_e\to \nu_{\tau}$); or  
  \item
   Using an improved detector with lower neutrino-energy threshold and
   better energy resolution.
\end{enumerate}
Possible configurations for each alternative, alone and in
combination, were investigated to find an optimum performance of the
Neutrino Factory.
It was shown that a considerable reduction of parent muon-energy down
to $\sim 25$~GeV is feasible without a significant loss of
oscillation-physics output, provided a detector performance improved
with respect to the one assumed in earlier studies can be achieved.
      
To make direct, quantitative comparisons of the various facilities,
the GLoBES package was used. 
Three representative super-beam configurations were considered:
the SPL, a super-beam directed from CERN to the Modane laboratory;
T2HK, an upgrade of the J-PARC neutrino beam illuminating a detector
close to Kamioka, and the WBB, a wide-band, on-axis beam from BNL or
FNAL to a deep underground laboratory in the US. 
Each super-beam was assumed to illuminate a megaton-class water
Cherenkov detector. 
The beta-beam options considered were the CERN baseline scheme in
which helium and neon ions are stored with a relativistic $\gamma$ of
100 and an optimised beta-beam for which $\gamma=350$.
Two Neutrino Factory options were considered: a conservative option
with a single 50~kton detector sited at a baseline of 4000~km from a
50~GeV Neutrino Factory; and the optimised Neutrino Factory (see the
full report) with two detectors, one at a baseline of 4000~km and
the second at the magic baseline~($\sim 7500~km$).
The result of the comparisons may be summarised as follows: 
for the options considered, the Neutrino Factory has the best
discovery reach for $\sin^2 2 \theta_{13}$ followed by the beta-beam and
the super-beam, while the $\sin^2 2 \theta_{13}$ reach for resolving
the sign of the atmospheric mass difference is mainly controlled by
the length of the baseline.  
For large values of $\theta_{13}$ 
($\sin^2 2 \theta_{13} \gsim 10^{-2}$), the three classes of facility
have comparable sensitivity for the discovery of CP violation; the best
precision on individual parameters being achieved at the Neutrino
Factory using optimised detectors. 
The reduction of systematic uncertainties is the key issue at large
$\theta_{13}$; by reducing systematic uncertainties, the super-beam
may be favourably compared with the conservative Neutrino Factory.
For intermediate values of $\theta_{13}$ 
($10^{-3} \lsim \sin^2 2 \theta_{13} \lsim 10^{-2}$), the super-beams
are outperformed by the beta-beam and the Neutrino Factory and the
best CP coverage is achieved by the beta-beam. 
For small values of $\theta_{13}$
($\sin^2 2 \theta_{13} \lsim 10^{-3}$), the Neutrino Factory
out-performs the other options.
Note, the comparisons are made using three performance indicators
only ($\sin^22\theta_{13}$, the sign of mass hierarchy and the CP
violating phase $\delta$). 
If other physics topics, such as the search for $e,\mu,\tau$ flavour
anomalies, were to be emphasised, the relative performance may be
different. 

The final contribution to this report reviews the muon physics that
can be performed with the intense muon beams that will be available at
the Neutrino Factory.
The study of rare, lepton-flavour violating processes in muon decay,
and the search for a permanent electric-dipole moment of the muon, are
complementary to precision studies of neutrino physics; often
sensitive to the same underlying physics. 
The complementarity and the potential of a muon-physics programme at
the Neutrino Factory is investigated.
It will be important in the coming years to establish quantitatively
the  synergy between muon physics and the study of neutrino
oscillations and to develop a plan for the co-existence of muon and
neutrino programmes at the Neutrino Factory facility. 

A significant amount of conceptual design work and hardware R\&D is
required before the performance assumed for each of the facilities can
be realised.
Therefore, an energetic, programme of R\&D into the accelerator
facilities and the neutrino detectors must be established with a view
to the timely delivery of conceptual design reports for the various
facilities.

%% file: 01-Introduction/01-Introduction.tex
%
\section{Introduction}
\label{Sect:Intro}
%
\input 01-Introduction/01-01-Elevator/01-01-Elevator
%
\input 01-Introduction/01-02-Context/01-02-Context
%
\input 01-Introduction/01-03-Implications/01-03-Implications
%
\input 01-Introduction/01-04-Experiment/01-04-Experiment
%
\input 01-Introduction/01-05-Study/01-05-Study

%% file: 01-Introduction/01-01-Elevator/01-01-Elevator.tex
\subsection{Neutrino In a Nutshell}
\label{Sect:Intro:Elevator}
Elusive, mysterious, yet abundant.  
Neutrinos are elementary particles, just like the electrons in our
bodies.
Neutrinos are so elusive that we don't feel ten trillions of them
going through our body every second.  
They were discovered fifty years ago, but still pose many  
mysteries, defying our efforts to understand them due to their
elusiveness.  
Yet neutrinos are the most numerous matter particles in the universe;
there are about a billion neutrinos for every single atom.

Slowly we began to appreciate the important roles that neutrinos have
played in shaping the universe as we see today.  
We already know that stars would not burn without neutrinos.  
Neutrinos played an important role in producing the various chemical
elements that we need for daily life.  
Given that we (atoms) are completely outnumbered by neutrinos, it is
quite certain that they played even more important roles.

There was a major surprise eight years ago when we discovered that  
neutrinos have a tiny, but non-zero,  mass quite against the
expectations of our best theory. 
This discovery opened up new important roles for neutrinos.
We are all of a sudden grappling with new exciting questions about
neutrinos that may lead to revolutionary understandings on how the
universe came to be. 

Because they have mass, neutrinos may have played an important role  
in shaping the galaxies, and eventually stars and planets.  
Neutrinos  may actually be their own anti-particles; this may be the
reason why the Universe did not end up empty but has atoms in it.
Neutrinos seem to be telling us profound facts about the way matter
and forces are unified, and how the three types of neutrino are related
to each other; we have yet to decipher their message.  In addition, neutrinos may actually be the reason why the universe
exists at all.

We are only beginning to understand neutrinos and their roles in how  
the universe works.  
It will take many experimental approaches to get the full picture.  
A neutrino factory discussed here will most likely be an essential
component of this programme. 

%% file: 01-Introduction/01-02-Context/01-02-Context.tex
\subsection{Neutrino physics as part of the High Energy Physics Programme}
\label{Sect:Intro:context}
 
The present is a very interesting time in the field of
fundamental physics: over the past four decades, an impressive
theoretical framework, the Standard Model, has been established.
The Standard Model is capable of explaining how nature works at the
smallest, experimentally-accessible distance scales; yet a handful
of phenomena seem decisively to elude an explanation within the
Standard Model and are therefore clues to a more fundamental
understanding.
These observations provide the only clues we have that our
understanding of fundamental physics is incomplete. 
The experimental and theoretical pursuit of these clues drives the
high energy physics programme and is likely to guide the bulk of the
research in this area over the coming decades. 
In this brief sub-section, the forces currently driving research in
fundamental physics are discussed and the possible interplay between the
component parts of the research programme are investigated; the
objective being to establish the context within which the future
experimental neutrino-physics programme must be developed.
 
Non-zero neutrino masses cannot be explained by the Standard Model.
To allow for massive neutrinos, the Standard Model must be modified
qualitatively. 
There are several distinct ways in which the Standard Model can be
modified to accommodate neutrino mass, some of which will be discussed
in detail in the next section. 
Our current experimental knowledge of the properties of the neutrino
property does not allow us to choose a particular ``new Standard
Model'' over all the others. 
We do not know, for example, whether neutrino masses are to be
interpreted as evidence of new, very light, fermionic degrees of 
freedom (as is the case if the neutrinos are Dirac fermions), new,
very heavy, degrees of freedom (as is the case if the canonical see-saw
mechanism is responsible for tiny Majorana neutrino masses), or
whether a more complicated electroweak-symmetry-breaking sector is
required.
To make progress, it is imperative that new probes of neutrino
properties be vigorously developed. 
This is the main driving force of all experimental endeavours discussed
in this study.   
 
According to the Standard Model, the Lagrangian of nature is invariant
under an $SU(3)_c\times SU(2)_L\times U(1)_Y$ local symmetry, but the
quantum numbers of the vacuum are such that this gauge symmetry is
spontaneously broken to $SU(3)_c\times U(1)_{\rm EM}$. 
The physics responsible for this electroweak-symmetry breaking is not
known.
The Standard Model states that electroweak symmetry breaking arises
due to the dynamics of a scalar field -- the Higgs field. 
While the Standard Model explanation for this phenomenon is in
(reasonable) agreement with precision electroweak measurements, the
definitive prediction -- the existence of a new, fundamental scalar
boson, the Higgs boson -- has yet to be confirmed experimentally.

Even if the standard mechanism of electroweak symmetry breaking is
realised in nature, our theoretical understanding of particle physics
strongly hints at the possibility that there are more degrees of
freedom at or slightly above the electroweak-symmetry-breaking scale
(around 250~GeV). 
Furthermore, it is widely anticipated that these new degrees of
freedom will serve as evidence of new organising principles; examples
of such principles include supersymmetry and the existence of new
dimensions of space.

The pursuit of the mechanism responsible for electroweak symmetry
breaking is the driving force behind the current and the future 
high-energy-collider physics programme, which aims at exploring the high
energy frontier. 
In the near future, the Large Hadron Collider (LHC) is expected to
supplant the ongoing Tevatron collider as the highest energy particle
accelerator in the world. 
It is widely expected that the LHC will reveal the mechanism of
electroweak symmetry breaking and provide evidence of new heavy
degrees of freedom. 
Anticipating the potential findings of the LHC-based experiments, the
collider physics community is currently planning a high intensity,
high precision, high energy  electron collider -- the International
Linear Collider (ILC).
The ILC should be able to study in detail the electroweak-symmetry
breaking sector and reveal the properties of the new physics at the
electroweak scale.

Finally, several very different but equally impressive measurements of
the mass-energy budget of the Universe have revealed beyond reasonable
doubt the existence of what is referred to as `dark matter'. 
After several years of observation, it is now clear that the Standard
Model does not contain the degrees of freedom necessary to explain the
dark matter. 
Other than its gravitational properties, very little is known about
dark matter. 
It could consist of weakly interacting fundamental particles, but it
may also consist of very heavy, very weakly interacting states, or
more exotic objects.   
Experimental searches for dark matter are currently among the highest
priorities of the fundamental physics research programme. 
Experiments that are sensitive to dark matter vary from 
direct-detection experiments, to neutrino telescopes, to gamma-ray
observatories. 
The hope is that the pursuit of the dark-matter clue will not only
reveal the existence of a new form of matter but will also provide
other clues that will allow a more satisfying understanding of the
composition of the universe and its behaviour in the first instants
after the Big Bang to be developed.

On top of the dark-matter problem, we are now faced with a seemingly
deeper puzzle -- the existence of dark energy. 
We are still far from properly decoding what this puzzle means, and it
is not clear how progress will be made towards resolving this most
mysterious issue. 
New experiments are being devised to study in more detail the
properties of dark energy. 
The results of these experiments may play a large role in modifying
our picture of how nature works at its most fundamental level.

The different probes discussed above not only address different clues
regarding new fundamental physics, but also complete and complement
one-another. 
The amount of synergy among the different experiments cannot be over
emphasised.
Consider the following examples of such synergies:
\begin{enumerate}
  \item
    While new physics at the electroweak scale is usually best studied
    using a high energy collider, there are several new-physics
    phenomena that will only manifest themselves in neutrino
    experiments, including new light, very-weakly-coupled degrees of
    freedom that could be related to dark matter or dark energy;
  \item
    The knowledge of neutrino properties is essential for the
    understanding of certain dark-matter searches (for example those
    performed using neutrino telescopes);
  \item
    A high-energy collider may provide the only means of studying in
    any detail the property of dark matter particles; and
  \item
    A proper understanding of the origin of neutrino mass can only
    be obtained after the mechanism of electroweak-symmetry breaking
    is properly understood.
\end{enumerate}

It is important to bear in mind that we do not know what the next set
of clues will be, or where they will come from. 
It may turn out, for example, that neutrino experiments provide our
only handle on grand unification and other types of very high energy
physics, or that astrophysical searches for the properties of dark
energy will reveal a direct window on quantum gravity. 
Or it may turn out that collider experiments will be able to study
directly string-theoretical effects (this may be the case if there
really are large extra dimensions). 
Only a comprehensive pursuit of the questions that we can formulate
today will allow us to reach the next stage in understanding
fundamental physics -- and ask a new set of more fundamental, deeper
questions tomorrow.  
In this sense, we perceive the physics discussed here to be on equal
footing with other studies of fundamental importance to our field,
including the direct searches for dark matter, satellite missions that
will measure the acceleration of the universe, or collider experiments
at the energy frontier. 
These are all different, complementary ways of addressing the
different questions that we cannot answer given our current 
understanding of fundamental physics.

%% file: 01-Introduction/01-03-Implications/01-03-Implications.tex
\subsection{Implications and opportunities}
\label{Sect:Intro:Implications}

Fundamental fermions, are classified in three generations, each
generation containing six quarks (two flavours, three colours) and two
leptons.
The measured properties of these particles exhibit a clear
`horizontal' hierarchy in which the mass of fermions carrying the same
Standard Model quantum numbers increases with generation number. 
Within a generation, the fermion properties also exhibit `vertical'
patterns, for example, the sum of the electric charge of the members
of a particular generation is zero. 
The quarks come in three colour varieties, the source of the strong
force.  
Under the weak force, both the quarks and the leptons within a
particular generation transform as a doublet.
In contrast to the general expectation, the mixing angles among lepton
flavours have turned out to be different to the quark-mixing angles.
Many of the properties of the neutrino are unique, not shared by the
other fundamental fermions.
Firstly, it has neither colour nor electric charge,
hence is the only fundamental fermion that feels solely the weak force
in addition to the gravitational force.
This fact becomes important when cosmological impact of the neutrino is discussed. 
Secondly, neutrino masses are tiny compared to the masses of all other
fundamental fermions.  
Thirdly, the neutrino could be a Majorana particle; a fermion which
cannot be distinguished from its own antiparticle\footnote{
The observation of double beta-decay processes in which no neutrino is 
produced would imply that the neutrino is its own antiparticle
\cite{Schechter:1981bd}.}. 

The physics of flavour seeks to provide an explanation of these
observed patterns. 
The vertical patterns noted above can be explained in `Grand Unified
Theories' (GUTs) in which the fermions are assigned to representations
of a large symmetry group such as $SO(10)$. 
The horizontal, or family patterns, can be explained by assuming a
family symmetry such as $SU(3)_{\rm family}$. 
Some models that incorporate GUT and family symmetries with
super-symmetric extensions come within the realm of string theories
that incorporate extra dimensions.  
Understanding the symmetry structure seems to be a promising
strategy to arrive at a description of the physics of flavour.

\begin{figure}
  \begin{center}
    \begin{tabular}{cc}
      \mbox{
        \includegraphics[width=0.7\textwidth]%
          {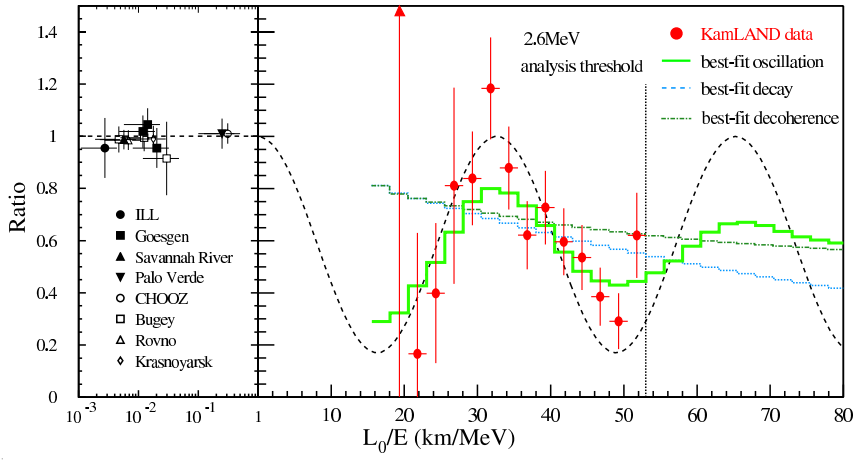}
      } \\
    \end{tabular}
  \end{center}
  \caption{
    The ratio of the measured to the predicted neutrino flux is
    plotted as a function of $L/E$.
    The anti-electron-neutrino contribution to the reactor neutrino
    flux measured by the KamLAND collaboration
    \protect\cite{Araki:2004mb} is shown.  (Figure courtesy of K. Inoue)
  }
  \label{Fig:OscilationSignature}
\end{figure}

Neutrino oscillations are a phenomenon in which the neutrino changes
flavour as it propagates. 
It was predicted by Pontecorvo
\cite{Pontecorvo:1957cp,Pontecorvo:1957qd} and Maki, Nakagawa and
Sakata \cite{Maki:1962mu} and the first clear evidence for neutrino
oscillations was presented by Super-Kamiokande in 1998 in observations
of the zenith angle distributions of atmospheric neutrinos
\cite{Fukuda:1998mi}. 
The first indication, however, dates back as early as 1970, when the
Homestake group detected a deficit in the solar-neutrino flux 
compared to that predicted by the Standard Solar Model
\cite{Cleveland:1998nv}.
The long-standing "solar neutrino puzzle" was finally proved to be a
result of oscillations by SNO in 2001 \cite{Ahmad:2001an}. 
The first observation of neutrino oscillations from terrestrial
neutrino sources was obtained by KamLAND by measuring the energy
spectrum of neutrinos produced in nuclear reactors
\cite{Araki:2004mb}.
The result of the KamLAND measurement, shown in figure
\ref{Fig:OscilationSignature}, exhibits the expected oscillatory
behaviour and constitutes compelling evidence for neutrino
oscillations \cite{Ashie:2004mr}.

Neutrino oscillations occur because of flavour mixing and the tiny,
but different, masses of the neutrinos. 
The see-saw mechanism, the most attractive and promising scheme to
explain the tiny mass, requires the presence of very heavy Majorana
neutrinos. 
In such models, neutrino oscillations are a consequence of the physics
which pertains at an extremely large energy scale.  
See-saw models are able to explain the striking difference between the
quark- and lepton-mixing angles in a natural way. 
If the heavy Majorana neutrino is abundant in the early Universe and
decays preferentially into matter leptons in a CP violating process,
then the lepton asymmetry would be converted into a baryon asymmetry a
split second later during the electroweak era. 
This process is referred to as ``leptogenesis'' and is a primary theory  
to explain our matter universe.  
The neutrino is the most abundant of the matter fermions in the
Universe; with a billion neutrinos for each of the other known
matter particles, only the ubiquitous photon is more abundant. 
Hence, the tiny neutrino mass could contribute a non-negligible
fraction of the dark matter and is known to play an important role in
the formation of large-scale structure in the Universe. 

Because of its direct connection with phenomena at energies never seen
since the Big Bang, the precise determination of the masses and mixing
angles of the 3 families of neutrino is a unique window onto these
early times and provides a path to the possible unification of all
forces.
Measurements of neutrino oscillations can be used to determine the three mixing angles and the CP violating phase of the lepton-mixing
matrix (the PMNS, Pontecorvo-Maki-Nakagawa-Sakata matrix) and two
mass-squared differences. 
Examining neutrino oscillations is a most direct way to distinguish
between the various possible theories of the physics of flavour
and to understand the origin of neutrino mass.
It is also a logical place to seek for the origin of the CP
violation.

Taking a different perspective, ever since Pauli's 1930 prediction,
the unveiling of the properties of the neutrino has always heralded a
new epoch in the history of elementary particle physics.
Including the neutrino as a player of beta decay, Fermi formulated the
first successful theory of weak interactions. 
The absence of right-handed neutrinos has manifested itself as the
`V-A' structure of the weak interaction and the pursuit of its
origin lead to the discovery of the chiral gauge theory
which forms the foundation of modern particle theories.
The realisation of intense neutrino beams immediately resulted in the
discovery of the neutral current, establishing the unification of the
electroweak interactions.
The ability of neutrino interactions to distinguish flavour and
handedness has been extensively utilised in deep inelastic interactions
to clarify the structure of the nucleon and to establish the
asymptotic freedom of QCD.
The recent discovery of neutrino oscillations could be regarded as
another epoch-making observation. 
So far it is the only experimental evidence for, and a vital clue to,
the physics beyond the Standard Model.
Both the mysteries, and the brilliant record, of the neutrino can be
attributed to its unique and characteristic insensitivity to both the strong and the electromagnetic forces. 
There are ample reasons to believe that this asset remains valid in
uncovering the veils that surround the neutrino.  
Neutrino facilities to pursue oscillation phenomena are complementary
to high-energy colliders and are competitive candidates for the next
world-class facilities for particle physics.

%% file: 01-Introduction/01-04-Experiment/01-04-Experiment.tex
\subsection{Precision measurements and sensitive searches}
\label{Sect:Intro:Experiment}

Experimentally, there are several different approaches to elucidate
the properties of the neutrino. 
In this report we concentrate mainly on accelerator-based facilities
illuminating massive, underground detectors. 
Other complementary means are also taken into account and
are described in section \ref{Sect:Alternatives}. 
Among them, reactor-based oscillation experiments may play a crucial
role in untangling the degeneracies inherent in oscillation
measurements.
Other techniques include double-beta decay, a unique tool to test the
Majorana nature of the neutrino \cite{Schechter:1981bd}. 
Massive, underground detectors, while serving as a far detector for the
oscillation experiments, also have an important role in their own
right as telescopes for neutrino astronomy and as a possible window on
grand unified theories by way of searching for proton decay.

Theories that purport to explain neutrino oscillations have
consequences for the properties of the charged leptons, such as
flavour changing process in lepton decay or lepton-induced reactions.
Considering that very intense muon beams will be available 
 as a by-product of the Neutrino Factory, it is natural to include muon
physics as an indispensable ingredient of the study. 
Section \ref{Sect:MuonPhysics} discusses in detail the opportunities
that high-statistics studies of the properties of the muon have to
offer.

The present generation of neutrino-oscillation experiments
\cite{Ahn:2006zz,Michael:2006rx,Kodama:1999hg}, reviewed in section
\ref{SubSect:SnuM} below, are designed to measure the smallest
neutrino mixing parameter if it is not `too small'.  
They utilise intense pion beams (super-beams) to generate neutrinos. 
They are designed to seek and measure the third mixing
angle~$\theta_{13}$ of the PMNS matrix, but will have little or no
sensitivity to matter-antimatter symmetry violation. 
Several neutrino sources, including second generation super-beams,
beta-beams, and the Neutrino Factory have been envisaged to reach high
sensitivity and redundancy well beyond that which can be achieved in
the presently-approved neutrino-oscillation programme.  
Section \ref{Sect:Performance} reviews the detailed performance of
each of these classes of facility and presents quantitative
comparisons of the physics potential. 
Their essential features are briefly introduced below.

The super-beam is a natural extension of the conventional neutrino
beam and the current and approved experiments are mostly of this type
\cite{Itow:2001ee,Mezzetto:2003mm,Campagne:2004wt,Ayres:2004js,MenaRequejo:2005hn,Mena:2005ri,Ishitsuka:2005qi,Hagiwara:2005pe,Hagiwara:2006vn,Kajita:2006bt,Diwan:2006qf}.
The neutrino beam is produced through pion and kaon decay and hence
these facilities provide beams in which $\nu_{\mu}$ and
$\bar{\nu}_{\mu}$ dominate the neutrino flux.
However, these beams also contain $\nu_e(\bar{\nu}_e)$ from kaon and
muon decay which constitute an irreducible background for the
oscillation signal $\nu_{\mu}(\bar{\nu}_{\mu})\to \nu_e(\bar{\nu}_e)$.
In addition, the selection of samples of $\nu_e$ ($\bar{\nu}_e$) is prone
to neutral current contamination. 
The principal source of systematic uncertainty arises from the fact
that the spectral shape of the pions and kaons is not well known.
In the second generation super-beam experiments, the emphasis is on large
detector mass, i.e. the collection of large data samples, and on muon
and electron particle identification.
These detector solution most often adopted for second-generation
super-beam experiments is the megaton-scale water Cherenkov counter.
Liquid-argon detectors or large volume scintillator detectors have
also been considered.

The beta-beam \cite{Zucchelli:2002sa}, in which electron neutrinos
or (anti-neutrinos) are produced from the decay of stored radio-active
ion beams, provides essentially background-free pure ``golden-channel''
($\nu_e\to\nu_{\mu}$), i.e. ``the appearance of wrong-sign muons''.
Unlike the Neutrino Factory, the beta-beam does not need a magnetised
detector, because there is no contamination from anti-neutrinos.
This allows the beta-beam to use a very massive detector just as the
second generation super-beam does
\cite{Burguet-Castell:2003vv,Mezzetto:2005ae}.
Simultaneous operation of the beta-beam and the second generation 
super-beam has also been considered \cite{Campagne:2006yx}.
The disadvantage of the beta-beam is lack of the ``silver channel'',
the $\nu_e \to \nu_{\tau}$, transition for most of the case studied.

The Neutrino Factory \cite{Geer:1997iz,DeRujula:1998hd}, 
an intense high-energy neutrino source derived
from the decay of a stored muon beam, has access to all channels of
neutrino-flavour transition including the golden channel.
However, to reject beam-induced muon-neutrino events requires that the
detector be magnetised.
This leads most naturally to the magnetised iron calorimeter design.
Another unique feature of the Neutrino Factory is the
possibility to observe the silver channel. 
This can be achieved using either emulsion based detectors or a
magnetised liquid-argon time-projection chamber.

Studies
\cite{MuColl,Finley:2000cn,Ozaki:2001bb,Gruber:2002tn,Alsharoa:2002wu,Zisman:2003bh}
so far have shown that the Neutrino Factory gives the best performance
over virtually all of the parameter space; its time scale and cost
remain, however, important question marks. 
Super-Beams have many components in common with the Neutrino
Factory. 
A beta-beam may be competitive with the Neutrino Factory in some
parameter space, but, being relatively new in this field, needs
further study to fully explore its capability.

There is an important issue common to all the facilities that must be borne in mind. 
A typical oscillation experiment, trying to determine the small mixing
angle $\theta_{13}$ and the CP violating phase $\delta$, generally
suffers from correlation/degeneracy problem, described in detail in
section \ref{SubSect:SnuMDegenCorrel}.
These correlations and degeneracies reduce the sensitivity typically
by one order of magnitude over that given by the statistical and
systematic uncertainties.
This happens because a canonical oscillation experiment measures only
two transition rates, $P(\nu_{\alpha}\to \nu_{\beta})$ and
$P(\bar{\nu}_{\alpha}\to \bar{\nu}_{\beta})$ and the expression for
these probabilities is a quadratic function of two unknown variables,
$\sin2\theta_{13}$ and $\sin \delta$. 
Given two measured values at fixed energy, $E$, and baseline, $L$, the
solution of the equations has an extra, fake, $(\theta_{13}, \delta)$
solution which is referred as the intrinsic degeneracy. 
Our ignorance of the sign of $\Delta m^2_{23}$ (the sign degeneracy),
and the indistinguishability of $\theta_{23}$ from $\pi/2-\theta_{23}$
(the octant degeneracy) results in a total eight-fold degeneracy. 
The problem can be resolved in one of three ways:
\begin{enumerate}
  \item
    To place a second detector at different value of $L/E$;
  \item
    To add a different channel (or to combine data from a
    complementary source, for example from a reactor experiments); and
  \item
    To use an improved detector with lower threshold and better energy
    resolution.
\end{enumerate}
Method (3) may be regarded as a variant of (1) from the physics point
of view because it is essentially equivalent to widening the energy
spectrum.  
This is the reason why the consideration of synergy among the
proposed, as well as the current experiments, is particularly
important in oscillation physics. 
More often than not, the combination of experiments of different
design can achieve a sensitivity that far exceeds what a mere
statistical gain would suggest. 
For instance, NOvA \cite{Ayres:2004js} expects to enhance its
sensitivity by combining with a proposed reactor experiment and
an upgraded version of NOvA \cite{MenaRequejo:2005hn,Mena:2005ri}
proposes to put a second detector at a different off axis angle
(i.e. energy).
T2KK  \cite{Ishitsuka:2005qi,Hagiwara:2005pe}, a variation of T2HK
\cite{Itow:2001ee}, proposes to split their megaton water detector in
two and place one in Korea; a remedy using two different
baselines~(L=295~km and $\simeq$ 1050~km).  
The on-axis WBB
\cite{Diwan:2006qf,Diwan:2003bp,Diwan:2004bt}, a very long-baseline
wide-band beam from FNAL or BNL to Henderson or Homestake mine in the
US, on the other hand, takes advantage of its wide spectrum to 
resolve the problem. 
It has also been shown that the combination of atmospheric-neutrino
data with T2HK \cite{Huber:2005ep} or a low energy beta-beam
\cite{Campagne:2006yx} is extremely helpful in resolving the
degeneracies related to the mass hierarchy and the octant degeneracy. 
These examples illustrate the importance of working towards the
identification of an optimum combination of the various facilities.

For all detector concepts, there are important questions concerning
cost, feasibility and time scales.
In addition, there are design optimisations to be made, e.g. between
energy and angle resolution, optimum baseline length and detector
mass. 
The study of detector concepts for the near detector stations will be
an important aspect of the neutrino physics because of its access to
many reactions complementary to the oscillation process.

For the neutrino-physics community to arrive at a consensus on the
best possible neutrino-oscillation programme to follow the present
generation of experiments requires a detailed evaluation of the
performance and cost of the various options and of the timescale on
which each can be implemented. Further R\&D programmes into the
accelerator systems and the neutrino detectors must be carried out.

%% file: 01-Introduction/01-05-Study/01-05-Study.tex
\subsection{What the study tried to achieve}
\label{Sect:Intro:Study}

The role of the Physics Group of the ISS study was to establish the
strong physics case for the various proposed facilities and to find
the optimum parameters of the accelerator facility and detector
systems from physics point of view. 
The first objective of this report, therefore, is to try to identify
the big questions of neutrino physics such as the origin of neutrino
mass, the role of the neutrino in the birth of the universe, what the
properties of the neutrino can tell us about the unification of matter
and force.
These questions lay down the basis for making the physics case for the
various neutrino facilities.
Since it is not (yet) possible to answer these questions in general,
studies have concentrated on more specific issues that may lead to
answers to the big questions. 
A class of directly-testable predictions is afforded by the fact that
GUT and family symmetries result in relationships between the quark-
and lepton-mixing parameters; such relationships can be cast in the
form of sum rules. 
  
The second objective was to look for possible clues of new physics in a
`bottom-up' approach. 
For this purpose, we have evaluated the degree to which the various
facilities, alone or in combination, can distinguish between the
various models of neutrino mixing and determined optimum parameter
sets for these investigations.
One example is to search for the existence of a sterile neutrino. 
Although the anomaly presented by LSND
\cite{Athanassopoulos:1997pv} was not confirmed by MiniBooNE
\cite{Aguilar-Arevalo:2007it}, the question is important enough to be
pursued further.
The second example is the unitary triangle: while the CKM matrix in
quark sector is constrained to be unitary in the Standard Model, the
PMNS matrix originates from physics beyond the Standard Model and, in
see-saw models, may not be exactly unitary.
The third example is the existence of flavour-changing interactions
that might appear at the production point, in the oscillation stage, or
at the detection point.  
Possible strong correlations between lepton-flavour violation and
neutrino oscillations were also discussed. 
Other approaches to the determination of the three-flavour parameters
(i.e. non-accelerator based measurements) were also considered.
For example, the possibility of a new long-baseline reactor
experiment and the loading of the water in the Super-Kamiokande
detector with gadolinium to improve the solar-neutrino parameters, or
a large, underground, magnetised-iron detector to improve the
atmospheric-neutrino parameters and to test for deviation from 
maximal-mixing and determine the octant degeneracy were also
discussed.
   
The third, and the key, objective of this report is to present the
first detailed comparison of the performance of the various
facilities. 
Utilising realistic specifications, we have estimated likely
performances, tried to find an optimum combination of facilities,
baselines and neutrino energies, and to come up with some staging
scenarios.

Although past studies have shown that the Neutrino Factory can be
considered as an excellent, and perhaps as an ultimate, facility, 
many questions remain open. 
For instance, the performance of the Neutrino Factory at large
$\theta_{13}~(\sin^22\theta_{13} \gtrsim 10^{-2})$ where most
super-beam experiments work is only now being studied in detail
\cite{Geer:2007kn}.
A question that must therefore be asked is: ``Can the Neutrino
Factory remain competitive if $\theta_{13}$ turns out to be large?''.
Another concern is the cost of the accelerator facility and the
detector systems.
One estimate \cite{Ozaki:2001bb} in previous studies gives a total
cost of $1500 {\rm~M\$} + 400{\rm~M\$}\times E/20(\GeV)$. 
Therefore, the second question is: "What is the minimum energy that
will deliver the physics?". 
The Neutrino Factory operates at energies considerably higher than the
first oscillation peak $(E_{max}/\GeV=L/564 ~{\rm km})$. 
The canonical operating condition in past studies has been to use a
parent muon beam of 50~GeV and a 50~kton magnetised-iron detector at a
distance of 3000~--~4000~km \cite{Cervera:2000vy}.
Because of its operation at high energy with a single detector, it
suffers from the degeneracy problem at intermediate values of
$\theta_{13}$ ($10^{-3} \lsim \sin^2 2 \theta_{13} \lsim 10^{-2}$). 
It has been shown that remedies exist through the addition of either
a second detector at the `magic' baseline ($L \simeq 7500$~km)
\cite{Huber:2003ak} or the silver channel \cite{Donini:2002rm}. 
Both of these solutions require the second detector. 
So, the third question is "Can a single-detector configuration with
improved performance do any better, and if two detectors are
unavoidable, which combination is the best?".
In order to answer those questions, an extensive investigation in
the parameter space $(E_{\mu} - L)$ was carried out. 
Then, various combinations have been compared with the intention to
identify both a conservative option and an improved set of detector
configurations with possible staging. 
        
Direct, quantitative comparison of the various facilities is a
highlight of the study. 
The GLoBES package~\cite{Huber:2004ka,Huber:2007ji} was used. 
Other codes, Valencia and Madrid, showed good agreement with GLoBES in 
a test using a single reference input.
A realistic set of detector specifications and a precise normalisation 
of neutrino flux and cross sections were prepared. 
The comparisons are made for three performance indicators
only ($\sin^22\theta_{13}$, the sign of mass hierarchy, and the CP
violation phase $\delta$).  
If other physics topics, such as $e,\mu-\tau$ flavour anomaly
searches, are emphasised, the relative importance may be different.

The final contribution to this report reviews the muon physics that
can be performed with the intense muon beams that will be available at
the Neutrino Factory.
The study of rare processes in muon decay and muon-electron and
muon-nucleon scattering is complementary to precision studies of
neutrino oscillations; often sensitive to the same underlying physics.
The complementarity and the potential of a muon-physics programme at
the Neutrino Factory is investigated.
It will be important in the coming years to establish quantitatively
the qualitative synergy between muon physics and the study of neutrino
oscillations.

This report is organised as follows. 
First, in section \ref{SubSect:SnuM}, we give a review of the present
generation of experiments, state what is needed to complete the
picture and explain the degeneracy problem. 
Next we expand upon the physics motivation for the
neutrino-oscillation programme in sections \ref{Sect:NPandtheSNuM} and 
\ref{Sect:NewPhys}; section \ref{Sect:NPandtheSNuM} contains a
`big-picture' description of neutrino physics addressing such
questions as the origin of neutrino mass, extra dimensions, flavour
symmetry, and the role of the neutrino in unification and in
cosmology, while section \ref{Sect:NewPhys} takes a phenomenological
approach to consider how measurements of neutrino properties may
provide clues to new physics through studies such as the search for
sterile neutrinos, the investigation of the leptonic unitary triangle,
and the search for non-standard interactions in the oscillation
experiments.
Section \ref{Sect:Performance} deals with the physics potential of
the proposed facilities: the super-beam; the beta-beam; and the
Neutrino Factory.  
Direct comparison of various facilities is given here. 
Alternative experiments which can complement the oscillation
experiments are described in section \ref{Sect:Alternatives}. 
The final section \ref{Sect:MuonPhysics} is devoted to muon physics.

%% file: 02-Standard-Neutrino-Model/02-Standard-Neutrino-Model.tex
\section{The Standard Neutrino Model}
\label{SubSect:SnuM}

\input 02-Standard-Neutrino-Model/02-01-Introduction/02-01-Introduction

\input 02-Standard-Neutrino-Model/02-02-Present-status/02-02-00-Present-status

\input 02-Standard-Neutrino-Model/02-03-Completing-the-picture/02-03-Completing-the-picture

\input 02-Standard-Neutrino-Model/02-04-Degeneracies-and-correlations/02-04-Degeneracies-and-correlations

%% file: 02-Standard-Neutrino-Model/02-01-Introduction/02-01-Introduction.tex
\subsection{Introduction}
\label{SubSect:SnuMPhenom}

The ``standard neutrino-mixing model'' emerged as a result of the
remarkable progress made in the past decade in the studies of neutrino
oscillations.
The experiments with solar, atmospheric, and reactor neutrinos
\cite{Cleveland:1998nv,Fukuda:1996sz,Abdurashitov:2002nt,Kirsten:2003ev,Cattadori:2005gq,Fukuda:2002pe,Ahmad:2001an,Ahmad:2002jz,Ahmad:2002ka,Ahmed:2003kj,Fukuda:1998mi,Eguchi:2002dm}  
have provided compelling evidences for the existence of neutrino
oscillations driven by non-zero neutrino masses and neutrino mixing.
Evidence for neutrino oscillations were also obtained in the long-baseline
accelerator neutrino experiments K2K~\cite{Ahn:2002up,Aliu:2004sq} and 
MINOS~\cite{Michael:2006rx}.

We recall that the idea of neutrino mixing and neutrino oscillations
was formulated in
\cite{Pontecorvo:1957cp,Pontecorvo:1957qd,Maki:1962mu}. 
It was predicted in 1967~\cite{Pontecorvo:1967fh} that the existence of 
$\nu_e$ oscillations would cause a ``disappearance'' of solar $\nu_e$
on the way to the Earth. 
The hypothesis of solar-$\nu_e$ oscillations, which (in one variety or
another) were considered from $\sim$1970 on as the most natural
explanation of the
observed~\cite{Cleveland:1998nv,Fukuda:1996sz,Abdurashitov:2002nt,Kirsten:2003ev,Cattadori:2005gq} 
solar-neutrino, $\nu_e$, deficit (see, e.g.,
references~\cite{Bilenky:1978nj,Wolfenstein:1977ue,Mikheev:1986gs,Bilenky:1987ty,Petcov:1997qv,Gonzalez-Garcia:2002dz}), 
has been convincingly confirmed in the measurement of the
solar-neutrino flux through the neutral-current (NC) reaction on
deuterium by the SNO
experiment~\cite{Ahmad:2001an,Ahmad:2002jz,Ahmad:2002ka,Ahmed:2003kj},  
and by the first results of the KamLAND experiment~\cite{Eguchi:2002dm}.
The combined analysis of the solar-neutrino data obtained by the
Homestake, SAGE, GALLEX/GNO, Super-Kamiokande, and the SNO
experiments, and of the KamLAND reactor $\bar{\nu}_e$ data
\cite{Eguchi:2002dm}, 
established large mixing-angle (LMA), MSW oscillations
\cite{Wolfenstein:1977ue,Mikheev:1986gs} as the dominant mechanism
giving rise to the observed solar-$\nu_e$ deficit (see, e.g.,
\cite{Bandyopadhyay:2003pk}).
The Kamiokande experiment \cite{Fukuda:1996sz} provided the first
evidence for oscillations of atmospheric neutrinos, $\nu_{\mu}$ and
$\bar{\nu}_{\mu}$, while the data from the Super-Kamiokande experiment
made the case for atmospheric-neutrino oscillations convincing  
\cite{Fukuda:1998mi}.
Indications for $\nu$-oscillations were also reported by the LSND 
collaboration \cite{Athanassopoulos:1997pv} but are dis-favoured by
the recent MiniBooNE measurement \cite{Aguilar-Arevalo:2007it}.

Compelling confirmation of oscillations in ($\nu_{\mu}$,
$\bar{\nu}_{\mu}$), and reactor, $\bar{\nu}_e$ was provided by
$L/E$-dependence observed by Super-Kamiokande~\cite{Ashie:2004mr} and
by the spectral distortion observed by the KamLAND and K2K
experiments~\cite{Araki:2004mb,Aliu:2004sq}.
For the first time the data exhibit directly the effects of the 
oscillatory dependence on $L/E$ and $E$ characteristic of
neutrino-oscillations in vacuum \cite{Gribov:1968kq}.
As a result of these developments, the oscillations of solar $\nu_e$,
atmospheric $\nu_{\mu}$ and $\bar{\nu}_{\mu}$, accelerator
$\nu_{\mu}$ (at $L\sim 250$ km and $L\sim 730$~km) and reactor
$\bar{\nu}_e$ (at $L\sim 180$ km), driven by non-zero $\nu$-masses and
$\nu$-mixing, can be considered as practically established.

All existing $\nu$-oscillation data, except the data of LSND
experiment \cite{Athanassopoulos:1997pv}, can be described assuming 
three-neutrino mixing in vacuum.
Let us recall that in the LSND experiment indications for $\bar
\nu_{\mu}\to\bar \nu_{e}$ oscillations with $(\Delta
m^{2})_{\rm{LSND}}\simeq 1~\rm{eV}^{2}$ were obtained.
The minimal four-neutrino-mixing scheme which could incorporate the
LSND indications for $\bar \nu_{\mu}$ oscillations is disfavoured by
the existing, long-baseline data \cite{Maltoni:2004ei} and by the
recent MiniBooNE data \cite{Aguilar-Arevalo:2007it}.
The $\nu$-oscillation explanation of the LSND results is possible
assuming five-neutrino mixing \cite{Sorel:2003hf}.

The three-neutrino mixing scheme will be referred to in what follows
as the ``Standard Neutrino Model'' (S$\nu$M). 
It is the minimal neutrino mixing 
model which can account for the oscillations 
of solar ($\nu_e$), atmospheric ($\nu_{\mu}$ and 
$\bar{\nu}_{\mu}$), reactor ($\bar{\nu}_e$)
and accelerator ($\nu_{\mu}$) neutrinos.
In the S$\nu$M, the (left-handed) fields of 
the flavour neutrinos $\nu_e$, $\nu_\mu$ and $\nu_\tau$ 
in the expression for the weak charged 
lepton current are linear combinations 
of fields of three neutrinos 
$\nu_j$, $j=1,2,3$, having definite mass $m_j$:
\SPbea \label{3numix}
\left( \bad 
\nu_{e \mathrm{L}} \\[0.2cm]
\nu_{\mu \mathrm{L}} \\[0.2cm]
\nu_{\tau \mathrm{L}} \\
\SPea   \right)
= \pmns \, \left( \bad 
\nu_{1 \mathrm{L}} \\[0.2cm]
\nu_{2 \mathrm{L}} \\[0.2cm]
\nu_{3 \mathrm{L}} \\
\SPea  \right)
= \left( \bad 
U_{e 1} & U_{e 2} & U_{e 3} \\[0.2cm]
U_{\mu 1} & U_{\mu 2} & U_{\mu 3} \\[0.2cm]
U_{\tau 1} & U_{\tau 2} & U_{\tau 3} \\
\SPea   \right)\, 
\left( \bad 
\nu_{1 \mathrm{L}} \\[0.2cm]
\nu_{2 \mathrm{L}} \\[0.2cm]
\nu_{3 \mathrm{L}} \\
\SPea   \right) 
\SPeea
\noindent where 
$\pmns$ is the Pontecorvo-Maki-Nakagawa-Sakata 
(PMNS) 
neutrino mixing matrix \cite{Pontecorvo:1957cp,Pontecorvo:1957qd,Maki:1962mu}, 
$\pmns \equiv U $.
The PMNS mixing matrix 
can be parametrised by 3 angles, 
and, depending on whether the massive 
neutrinos $\nu_j$ are Dirac or Majorana 
particles, by 1 or 3 CP-violation ($CPV$) phases 
\cite{Bilenky:1980cx,Schechter:1980gr,Doi:1980yb,Schechter:1980gk}.
In the standard parameterisation 
(see, e.g., \cite{Bilenky:2001rz}),
$\pmns$ has the form:
\SPbea 
\label{eq:Upara}
\pmns = \left( \bad 
c_{12} c_{13} & s_{12} c_{13} & s_{13} e^{-i \delta}  \\[0.2cm] 
-s_{12} c_{23} - c_{12} s_{23} s_{13} 
& c_{12} c_{23} - s_{12} s_{23} s_{13} e^{i \delta} 
& s_{23} c_{13} 
\\[0.2cm] 
s_{12} s_{23} - c_{12} c_{23} s_{13} e^{i \delta} & 
- c_{12} s_{23} - s_{12} c_{23} s_{13} e^{i \delta} 
& c_{23} c_{13} 
\\ 
\SPea   \right) 
   {\rm diag}(1, e^{i \alpha/2}, e^{i \beta/2}) \, , 
\SPeea
\noindent where 
$c_{ij} = \cos\theta_{ij}$, 
$s_{ij} = \sin\theta_{ij}$,
the angles $\theta_{ij} = [0,\pi/2]$,
$\delta = [0,2\pi]$ is the 
Dirac $CPV$ phase and
$\alpha,\beta$ 
are two Majorana 
CP-violation phases \cite{Bilenky:1980cx,Schechter:1980gr,Doi:1980yb,Schechter:1980gk}. 
One can identify 
$\dmsol = \Delta m^2_{21} > 0$ with the neutrino mass squared 
difference responsible for the solar-neutrino oscillations.
In this case $|\dma|=|\Delta m^2_{31}|\cong |\Delta m^2_{32}|
\gg \Delta m^2_{21}$ is the 
neutrino mass-squared 
difference driving the dominant atmospheric-neutrino oscillations,
while $\theta_{12} = \theta_{\odot}$ and 
$\theta_{23} = \theta_{\rm A}$ are the solar 
and atmospheric neutrino mixing angles, respectively.
The angle $\theta_{13}$ is 
the so-called ``CHOOZ mixing angle'' -- it is
constrained by the data from the CHOOZ 
and Palo Verde experiments~\cite{Apollonio:1999ae,Boehm:1999gk}.

Let us recall that  the properties of 
Majorana particles  are very different from 
those of Dirac particles.
A massive Majorana neutrino $\chi_k$ 
with mass $m_k > 0$ 
can be described (in local quantum field theory)
by a 4-component, complex spin-$1/2$ field, $\chi_k(x)$,
which satisfies the Majorana condition:
\beq
C~(\bar{\chi}_k(x))^{{\rm T}} = \xi_k \, \chi_k(x), ~~|\xi_k|^2 = 1\,,
\label{MajCon}
\eeq
\noindent where $C$ is the charge conjugation matrix.
The Majorana condition is invariant under 
proper Lorentz transformations.
It reduces by two the number of independent 
components in $\chi_k(x)$.

The condition (\ref{MajCon}) is invariant with
respect to $U(1)$ global gauge transformations
of the field $\chi_k(x)$ carrying a $U(1)$ charge $Q$,
$\chi_k(x) \rightarrow e^{i\alpha Q}\chi_k(x)$, only if $Q = 0$.
As a result and in contrast to the Dirac fermions: 
i) the Majorana particles $\chi_k$ 
cannot carry non-zero additive quantum 
numbers (lepton charge, etc.);  
and ii) the Majorana fields $\chi_k(x)$ 
cannot ``absorb'' phases.  This is the reason 
why the PMNS matrix  contains two additional 
CP-violating phases in the case 
when the massive neutrinos $\nu_k$ are 
Majorana fermions \cite{Bilenky:1980cx}, $\nu_k \equiv \chi_k$.
It follows from the above that the Majorana-neutrino field,
$\chi_k(x)$, describes the two spin states of a spin 1/2, 
absolutely-neutral particle, which is identical with
its antiparticle, $\chi_k \equiv \bar{\chi}_k$.
If CP-invariance holds, Majorana neutrinos 
have definite CP-parity, $\eta_{CP}(\chi_k) = \pm i$:
\beq
U_{CP}~\chi_k(x)~U^{-1}_{CP} = \eta_{CP}(\chi_k)~\gamma_{0}~\chi_k(x'),~~
\eta_{CP}(\chi_k) = \pm i~.
\label{CPMaj} 
\eeq
It follows from the Majorana condition that 
the currents $:\bar{\chi}_{k}(x)O^{i}\chi_{k}(x):~ \equiv 0$,
${\rm for}~O^{i}=\gamma_{\alpha}$;
$\sigma_{\alpha\beta}$; $\sigma_{\alpha\beta}\gamma_5$.
This means that Majorana neutrinos
cannot have non-zero $U(1)$ charges and 
intrinsic magnetic- and electric-dipole moments.
Dirac fermions can possess non-zero
lepton charge and intrinsic 
magnetic- and electric dipole-moments
\footnote{Let us add, finally, that
Majorana neutrinos have in addition to the standard 
propagator (formed by the neutrino field and 
its Dirac conjugate field), 
{\it two non-trivial non-standard (Majorana)} propagators.
If $\nu_j(x)$ in equation (\ref{3numix}) 
are massive Majorana neutrinos,
the process of $\betabeta$-decay,
$(A,Z) \rightarrow (A,Z+2) + e^- + e^-$, 
for example, can proceed by exchange of virtual 
neutrinos $\nu_j$ due to the one of these 
Majorana propagators.
For Dirac fermions, the two analogous non-standard 
propagators  are identically equal to zero.
For further detailed discussion of the properties of 
Majorana neutrinos (fermions) see, e.g., \cite{Bilenky:1987ty,Valle:1990pk}.
}.

The existing data allow a determination of 
$\dmsol$, $\sin^2\theta_{12}$, and of
$|\dma|$, $\sin^22\theta_{23}$
with a relatively good precision (see, e.g.  
\cite{Bandyopadhyay:2004da,Bandyopadhyay:2005zz,Schwetz:2006dh} and
subsections \ref{SubSubSect:SolarandReactor} and \ref{SubSubSect:Atmospheric}).
For the best fit values we have:
$\dmsol = 8.0\times 10^{-5}~{\rm eV^2}$, 
$\sin^2\theta_{12} = 0.30$, 
$|\dma| = 2.5\times 10^{-3}~{\rm eV^2}$, 
$\sin^22\theta_{23} = 1$. 
Thus, $\dmsol \ll |\dma|$.
It should be noted, however, that 
the sign of $\dma$ is not 
fixed by current data.
The present atmospheric-neutrino data is essentially insensitive to
$\theta_{13}$, satisfying the upper limit on $\sin^2\theta_{13}$
obtained in the CHOOZ experiment \cite{Hosaka:2006zd}.
The probabilities of survival of 
solar $\nu_e$ and reactor $\bar{\nu}_e$,
relevant for the interpretation of
the solar neutrino, KamLAND and CHOOZ 
neutrino oscillation data, depend 
on $\theta_{13}$ in the case of interest, 
$|\Delta m^2_{31}|\gg \Delta m^2_{21}$:\\

$P^{3\nu}_{\rm KL} \cong \sin^4\theta_{13} + 
\cos^4\theta_{13}\left [1 - \sin^2 2\theta_{12}
\sin^2\frac{\Delta m^2_{21}{\rm L}}{4E} \right ]$,~ \\

$P^{3\nu}_{\rm CHOOZ} \cong 1 -  
\sin^2 2\theta_{13}
\sin^2\frac{\Delta m^2_{\rm 31}{\rm L}}{\rm 4E}$,\\

$P^{3\nu}_{\odot} \cong \sin^4\theta_{13} + 
\cos^4\theta_{13}~P^{2\nu}_{\odot}
(\Delta m^2_{21},\theta_{12};N_e\cos^2\theta_{13})$,\\

\noindent where $P_{\odot}^{2\nu}$ 
is the solar $\nu_e$ survival probability \cite{Petcov:1987zj,Petcov:1988wv,Petcov:1997am} 
corresponding to 2-$\nu$ oscillations
driven by $\Delta m^2_{21}$ and $\theta_{12}$,
in which the solar $e^-$ number density $N_e$
is replaced by $N_e \cos^2\theta_{13}$ \cite{Petcov:1987cd},
$P^{2\nu}_{\odot} = \bar{P}^{2\nu}_{\odot} + P^{2\nu}_{\odot~{\rm osc}}$,
$P^{2\nu}_{\odot~{\rm osc}}$ being 
an oscillating term 
\cite{Petcov:1987zj,Petcov:1988wv,Petcov:1997am} and\\
\begin{equation}
\bar{P}^{2\nu}_{\odot} = \frac {1}{2} + 
(\frac{1}{2} - P')\cos2\theta_{12}^m\, \cos 2\theta_{12}, 
\label{Psol}
\end{equation}
\begin{equation}
P'= \frac{e^{-2\pi r_{0}\frac{\Delta m^2_{21}}{2E}\sin^2\theta_{12}}
- e^{-2\pi r_{0}\frac{\Delta m^2_{21}}{2E}}}
{1 - e^{-2\pi r_{0}\frac{\Delta m^2_{21}}{2E}}}. 
\label{Psolexp}
\end{equation}
\noindent  Here $\bar{P}^{2\nu}_{\odot}$
is the average probability \cite{Haxton:1986dm,Parke:1986jy,Petcov:1987zj,Petcov:1988wv,Petcov:1997am},
$P'$ is the ``double exponential'' 
jump probability \cite{Petcov:1987zj,Petcov:1988wv,Petcov:1997am}, 
$r_0$ is the 
scale-height of the change of $N_e$ along the 
$\nu$-trajectory in the Sun 
\cite{Petcov:1987zj,Petcov:1988wv,Petcov:1997am,Krastev:1988ci,Bruggen:1995hr,Petcov:1989du}, and 
$\theta_{12}^m$ is the mixing angle 
in matter, 
which in the vacuum limit 
coincides with $\theta_{12}$. 
In the LMA solution region of interest,
$P^{2\nu}_{\odot~{\rm osc}} \cong 0$ 
\cite{Petcov:1989du}.
Performing a combined analysis of
the solar-neutrino, CHOOZ, and KamLAND data, 
one finds \cite{Bandyopadhyay:2004da,Bandyopadhyay:2005zz,Schwetz:2006dh}:
$\sin^2\theta_{13} < 0.040$ at 99.73\% C.L.

It follows from the results described above 
that the atmospheric-neutrino mixing  
is close to maximal, $\theta_{23} \cong \pi/4$,
the solar-neutrino mixing angle 
$\theta_{12} \cong \pi/3$, and the 
CHOOZ angle $\theta_{13} < \pi/15$.
Correspondingly, the pattern of 
neutrino mixing is drastically different 
from that of the quark mixing. 
A comprehensive theory of 
flavour and of neutrino mixing 
must be capable of explaining 
this remarkable difference.
The current theoretical ideas 
about the possible origin of the 
pattern of neutrino mixing 
are reviewed in Section \ref{Sect:NPandtheSNuM}.

As we have seen, the fundamental 
parameters characterising 
the S$\nu$M are:
i) the 3 angles $\theta_{12}$, $\theta_{23}$, $\theta_{13}$;
ii) depending on the nature of massive neutrinos 
$\nu_j$ - 1 Dirac ($\delta$), 
or 1 Dirac + 2 Majorana 
($\delta,\alpha,\beta$),
CP-violation phases; and 
iii) the 3 neutrino masses, $m_1,~m_2,~m_3$.
This makes 9 additional parameters 
in the Standard Model of particle interactions.

It is convenient to express the two 
larger neutrino masses in terms of the 
third mass and the measured 
$\dmsol = \Delta m^2_{21}> 0$ and $\dma$.
We have remarked earlier that the 
atmospheric-neutrino, K2K, and MINOS data
do not allow one to determine the sign of 
$\dma$. This implies that, if we identify
$\dma$ with $\Delta m^2_{31(2)}$
in the case of 3-neutrino mixing, 
one can have $\Delta m^2_{31(2)} > 0$
or $\Delta m^2_{31(2)} < 0$. 
The two possible signs of
$\dma$ correspond to two 
types of $\nu$-mass spectrum:
\begin{itemize}
  \item
    {\it Normal ordering}: $m_1 < m_2 < m_3$, 
    $\dma = \Delta m^2_{31} >0$, 
    $m_{2(3)} = (m_1^2 + \Delta m^2_{21(31)})^{1\over{2}}$; and
  \item
    {\it Inverted ordering}
    \footnote{
      In the convention we use (called A), the neutrino masses are not
      ordered in magnitude according to their index number: 
      $\Delta m^2_{31} < 0$ corresponds to $m_3 < m_1 < m_2$.
      We can also always number the neutrinos with definite mass in
      such a way that \cite{Bilenky:1996cb} $m_1 < m_2 < m_3$. 
      In this convention (called B), we have in the case of
      inverted-hierarchy spectrum: $\dmsol = \Delta m^2_{32}$,  
      $\dma = \Delta m^2_{31}$. 
      Convention B is used, e.g., in
      \cite{Bilenky:2001rz,Pascoli:2002xq}. 
    }: $m_3 < m_1 < m_2$, $\dma = \Delta m^2_{32}< 0$, 
    $m_{2} = (m_3^2 + \Delta m^2_{23})^{1\over{2}}$, 
    $m_{1} = (m_3^2 + \Delta m^2_{23} - \Delta m^2_{21})^{1\over{2}}$.
\end{itemize}
Depending on the values of the lightest neutrino mass, 
${\rm min}(m_j)$, the neutrino mass spectrum can also be:
\begin{itemize}
  \item
    {\it Normal Hierarchy (NH)}: $m_1 {\small \ll m_2 \ll }m_3$,
    $m_2 \cong (\dmsol)^ {1\over{2}}\sim 0.009$~eV, 
    $m_3 \cong |\dma|^{1\over{2}} \sim 0.05$ eV; 
  \item
    {\it Inverted Hierarchy (IH)}: $m_3 \ll m_1 < m_2$, 
    with $m_{1,2} \cong |\dma|^{1\over{2}}\sim 0.05$ eV; or 
  \item
    {\it Quasi-Degenerate (QD)}: $m_1 \cong m_2 \cong m_3 \cong m_0$,
    $m_j^2 \gg |\dma|$, $m_0 \gtap 0.10$ eV.\\ 
\end{itemize}
One of the principal goals of the future 
studies of neutrino mixing is to determine 
the basic parameters of the S$\nu$M and to test 
its validity.

The possibilities of measuring with high
precision the basic parameters of S$\nu$M 
$\dmsol$, $\sin^2\theta_{\odot}$
$|\dma|$, $\sin^22\theta_{23}$, 
$\sin^2\theta_{13}$, of determining
${\rm sgn}(\deltaatm)$
and of searching for the effects of 
CP-violation due to the Dirac phase $\delta$, 
in neutrino oscillation experiments, 
will be discussed in detail below.
It is well-known that the neutrino-oscillation experiments
are not sensitive to the 
absolute scale of neutrino
masses. Information on the absolute 
neutrino-mass scale, or on ${\rm min}(m_j)$,
can be derived in \hbeta experiments 
and from cosmological and astrophysical data 
(see sections \ref{Sect:0nu2betaExp} and \ref{Para:Currentbounds}).
The most stringent upper bounds on the $\bar{\nu}_e$ mass  
and on the sum of neutrino masses will be discussed briefly 
in sections \ref{Sect:0nu2betaExp} and \ref{subsubsect:leptogenesis}.
These bounds lead to the conclusion that 
neutrino masses satisfy 
$m_j \ltap 1$ eV and thus 
are much smaller than the masses 
of the charged leptons and quarks.
A comprehensive theory of neutrino 
mixing should be able to explain this 
enormous difference between the neutrino 
and charged-fermion masses.
The theoretical aspects of the problem 
of neutrino mass generation 
and of the smallness of neutrino masses 
are reviewed in section \ref{susect:origin-nu-mass}.

Neutrino-oscillation experiments 
are also insensitive
to the nature -- Dirac or Majorana, 
of massive neutrinos 
and, correspondingly, 
to the two CP-violating, Majorana phases in the PMNS matrix 
\cite{Bilenky:1980cx,Langacker:1986jv} since the latter do not enter
into the expressions for the probabilities for neutrino oscillations. 
The only realistic experiments 
which could verify that the massive neutrinos 
$\nu_j$ are Majorana particles are, at present,
the neutrinoless
double-beta ($\betabeta$-) decay experiments.
The physics potential of these experiments is 
discussed in section \ref{Sect:0nu2betaExp}.
Even if massive neutrinos are 
proven to be Majorana fermions,
measuring the Majorana CP phases would be extremely challenging.
It is quite remarkable, however, 
that the Majorana CP-violating 
phase(s) in the PMNS matrix, through leptogenesis (see section 
\ref{subsubsect:leptogenesis}, may result in the baryon asymmetry of the
Universe \cite{Pascoli:2006ie,Pascoli:2006ci,Petcov:2006bn}.

The existing data on neutrino oscillation, 
as we will see, allow a determination of
$\dmsol$, $\sin^2\theta_{\odot}$,
$|\dma|$, and $\sin^22\theta_{23}$,
at 3$\sigma$ with an uncertainty of approximately 
$\sim$12\%, $\sim$24\%, 
$\sim$28\% and $\sim$15\%, respectively. 
These parameters can, and very likely will, 
be measured with much higher accuracy 
in the future: the indicated 3$\sigma$ 
errors in the determination, for instance, of 
$\dmsol$ and $\sin^2\theta_{\odot}$,
can be reduced to
\cite{Choubey:2004bf,Bandyopadhyay:2004cp,Bandyopadhyay:2003du} 
4\% and 10\%,
as will be reviewed below. 
``Near'' future experiments with reactor $\bar{\nu_e}$ 
can improve the current sensitivity to the value 
of $\sin^2\theta_{13}$ by a factor of between 5 and 10.
The type of neutrino-mass spectrum, i.e. ${\rm sgn}(\deltaatm)$, can
be determined by studying the oscillations of neutrinos and
antineutrinos, say, 
$\nu_{\mu} \leftrightarrow \nu_e$
and $\bar{\nu}_{\mu} \leftrightarrow \bar{\nu}_e$,
in which matter effects are sufficiently large.
If $\sin^22\theta_{13}\gtap 0.05$
and $\sin^2\theta_{23}\gtap 0.50$,
information on ${\rm sgn}(\Delta m^2_{31})$
might be obtained in 
atmospheric neutrino experiments
by investigating the effects 
of the sub-dominant transitions
$\nu_{\mu(e)} \rightarrow \nu_{e(\mu)}$
and $\bar{\nu}_{\mu(e)} \rightarrow \bar{\nu}_{e(\mu)}$ 
of atmospheric neutrinos which traverse 
the Earth \cite{Chizhov:1998ug,Bernabeu:2003yp,Palomares-Ruiz:2004tk}. 
For $\nu_{\mu(e)}$ (or $\bar{\nu}_{\mu(e)}$) 
crossing the Earth's core, new types of resonance-like
enhancement of the oscillation probabilities may
take place due to the mantle-core constructive-interference effect
(neutrino oscillation length resonance (NOLR)) 
\cite{Petcov:1998su,Chizhov:1999az,Chizhov:2000tn,Chizhov:1999he}.
As a consequence of this effect,
the corresponding 
$\nu_{\mu(e)}$ (or $\bar{\nu}_{\mu(e)}$)
transition probabilities can be maximal
\cite{Chizhov:1999az,Chizhov:2000tn,Chizhov:1999he}.
For $\Delta m^2_{31}> 0$, the neutrino transitions
$\nu_{\mu(e)} \rightarrow \nu_{e(\mu)}$
are enhanced, while for $\Delta m^2_{31}< 0$
the enhancement of antineutrino transitions
$\bar{\nu}_{\mu(e)} \rightarrow \bar{\nu}_{e(\mu)}$
takes place, which might allow 
to determine ${\rm sgn}(\Delta m^2_{31})$.

It should be emphasised that the CP-violation
in the lepton sector is one of the most 
challenging frontiers in future studies 
of neutrino mixing. The experimental searches 
for $CP$-violation in neutrino oscillations
can help answer fundamental questions about 
the status of CP-symmetry in the 
lepton sector at low energy.
The observation of
leptonic $CP$-violation at low energies
will have far reaching consequences. 
It can shed light, in particular, 
on the possible origin of the baryon asymmetry 
of the Universe. As was realised recently
\cite{Pascoli:2006ie,Pascoli:2006ci}, 
the CP-violation necessary 
for the generation of the baryon asymmetry 
can be due exclusively to the Dirac 
(and/or Majorana) CP-violating phase 
in the PMNS matrix.
Thus, there can be a direct relation
between  low energy CP-violation 
in the lepton sector, observable, e.g., in neutrino 
oscillations, and the matter-antimatter 
asymmetry of the Universe.    
These results underline the importance of understanding the 
status of the CP-symmetry in the lepton sector 
and, correspondingly, of the experiments aiming 
to measure the CHOOZ angle
$\theta_{13}$ and of the experimental 
searches for CP-violation in neutrino oscillations.

%% file: 02-Standard-Neutrino-Model/02-02-Present-status/02-02-00-Present-status.tex
\subsection{Review of the present generation of experiments}
\label{SubSect:SnuMReview}

\input 02-Standard-Neutrino-Model/02-02-Present-status/02-02-01-Solar-and-reactor

\input 02-Standard-Neutrino-Model/02-02-Present-status/02-02-02-Atmospheric

\input 02-Standard-Neutrino-Model/02-02-Present-status/02-02-03-LBL

\input 02-Standard-Neutrino-Model/02-02-Present-status/02-02-04-Neutrinoless-Double-Beta.tex

\input 02-Standard-Neutrino-Model/02-02-Present-status/02-02-05-Update

%% file: 02-Standard-Neutrino-Model/02-02-Present-status/02-02-01-Solar-and-reactor.tex
\subsubsection {Solar and reactor neutrino experiments}
\label{SubSubSect:SolarandReactor}

Measurements of the solar-neutrino flux were the first to indicate
that neutrinos undergo flavour oscillations.
The first indications that the solar-neutrino flux was smaller than
that predicted by the Standard Solar Models came from Davis'
experiment at Homestake (USA) \cite{Cleveland:1998nv}.
The results of this experiment, have been confirmed by a series of
solar neutrino experiments, the SAGE experiment in Russia
\cite{Abdurashitov:2002nt}, the Gallex and GNO experiments in Italy  
\cite{Hampel:1998xg,Altmann:2005ix}, the Kamiokande and
Super-Kamiokande (SK) in Japan  \cite{Fukuda:2002pe,Smy:2003jf}
and finally by the Sudbury Neutrino Observatory (SNO) in Canada 
\cite{Ahmad:2001an,Ahmad:2002jz,Ahmad:2002ka,Ahmed:2003kj,Aharmim:2005gt}.
In particular, the neutral current (NC) to charged current (CC)
ratio from the SNO data in 2002 \cite{Ahmad:2002jz}
established the presence of an active 
neutrino flavour other than $\nue$ in the observed solar neutrino 
flux at the 5.3$\sigma$ level, putting to rest any doubt about the 
existence of flavour oscillations of solar neutrinos. 
Further evidence was provided by the statistically powerful 
NC data from the salt phase of the SNO experiment
\cite{Ahmed:2003kj,Aharmim:2005gt}.
The cumulative result of solar neutrino data, collected from different 
experiments over a period of more than four decades, culminated 
in the emergence of the `Large Mixing Angle' (LMA) solution as the
most favoured explanation of the solar neutrino problem.

\begin{figure}
  \begin{center}
    \includegraphics[width=13.0cm]
      {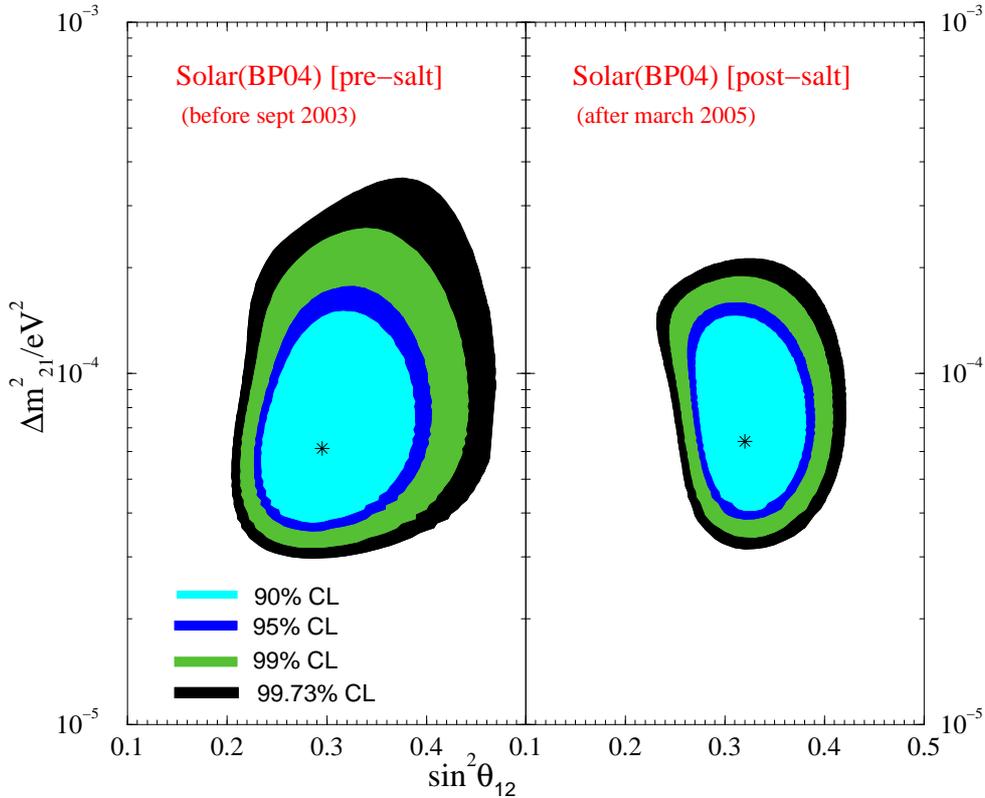}
    \caption{
      \label{fig:solar}
      The 90\%, 95\%, 99\% and 99.73\% 
      C.L. contours (for two degrees of freedom (dof)) show the 
      allowed areas from the global analysis of the solar neutrino 
      data with (right-hand panel) and without (left-hand panel) the
      SNO salt phase data.  Taken with kind permission of
	International Journal of Modern Physics
	from figure 1 in reference \cite{Goswami:2006hc}.
	Copyrighted by World Scientific Publishing Company.
    }
  \end{center}
\end{figure}

Figure \ref{fig:solar} shows the confidence level contours in the
$\Delta m_{21}^2-\sin^2\theta_{12}$ plane, allowed from the
global analysis of all solar neutrino data combined
\cite{Bandyopadhyay:2004da,Choubey:2005es,Goswami:2006hc}. 
To illustrate the effect of the results from the 
salt phase data from SNO, the figure shows in the right-hand 
and left-hand panels, the allowed areas obtained with and without the
salt phase SNO results respectively. 
The high statistics NC to CC ratio in SNO salt data, causes the
shrinking of the allowed regions. 
In particular, the upper bound on both $\Delta m_{21}^2$ and
$\sin^2\theta_{12}$ is seen to improve remarkably.

The \kl reactor anti-neutrino experiment in Japan
\cite{Araki:2004mb,Eguchi:2002dm}, specifically designed to test the
LMA region of the solar neutrino parameter space,
presented its first results in 2002, confirming the LMA 
solution \cite{Eguchi:2002dm}. 
The higher statistics data from this experiment released in 2004
\cite{Araki:2004mb} not only confirmed the observed depletion 
of the reactor antineutrinos from 
the first results \cite{Eguchi:2002dm}, 
but for the first time 
unambiguously showed the existence of an $L/E$ dependence in its 
positron spectrum, confirming that the 
observed $\anue$ flavour oscillations 
were indeed due to neutrino mass and mixing. 

\begin{figure}
  \begin{center}
    \includegraphics[width=13.0cm]
      {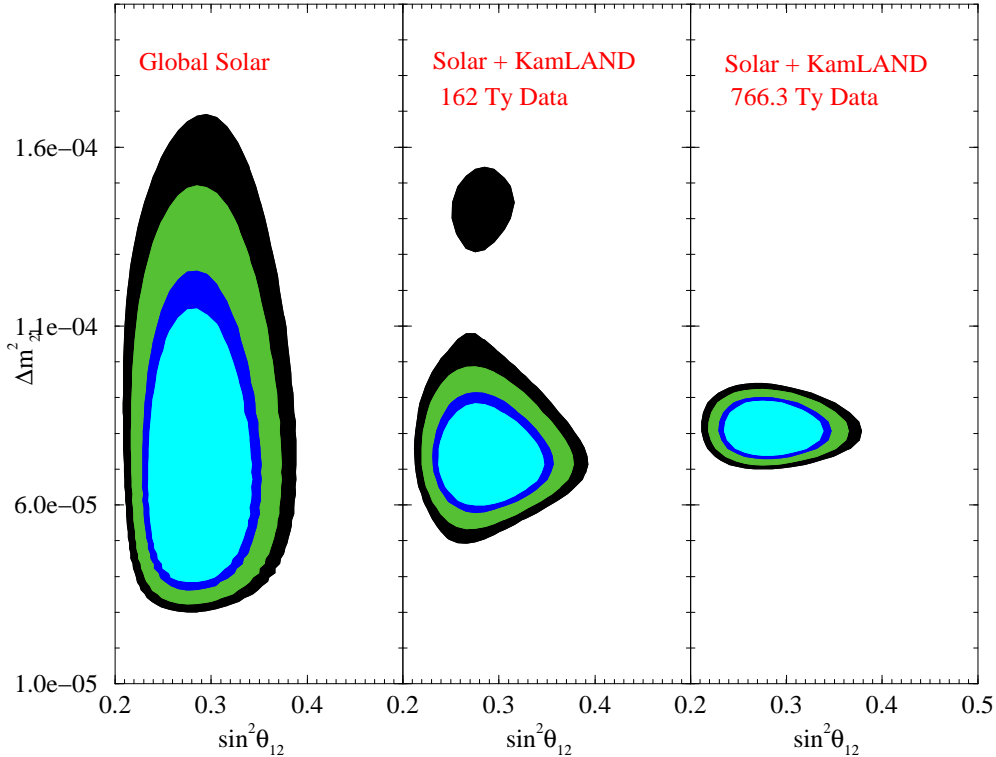}
    \caption{
      \label{fig:solarpluskl}
      The 90\%, 95\%, 99\% and 99.73\% C.L. contours (2 dof) show the
      allowed areas from the global analysis of the 
      solar neutrino data (left-hand panel) and
      solar neutrino data combined with the first KamLAND results
      \protect\cite{Eguchi:2002dm}
      (middle panel) and second KamLAND results \protect\cite{Araki:2004mb}
      (right-hand panel).  Taken with kind permission of
	International Journal of Modern Physics
	from figure 2 in reference \cite{Goswami:2006hc}.
	Copyrighted by World Scientific Publishing Company.
    }
  \end{center}
\end{figure}

Figure \ref{fig:solarpluskl} \cite{Choubey:2005es,Goswami:2006hc}
shows the impact of the first and second set of data from the 
KamLAND experiment on the solar neutrino oscillation 
parameter space. 
The current $3\sigma$ allowed range of $\Delta m_{21}^2$ and
$\sin^2\theta_{12}$ obtained in the analysis of Bandyopadhyay 
{\it et al.} \cite{Choubey:2005es,Goswami:2006hc,Bandyopadhyay:2004da}
is given in Table \ref{tab1} along with their corresponding ``spread''
defined as:
\be
  {\rm spread} = \frac{ \mathscr{P}_{max} - \mathscr{P}_{min}}
  {\mathscr{P}_{max} + \mathscr{P}_{min}}\times 100~,
  \label{error}
\ee
where $\mathscr{P}_{min}$ and $\mathscr{P}_{max}$ are the minimum and
maximum allowed values of the parameter $\mathscr{P}$ at $3\sigma$.
The allowed regions were derived on the assumption of CPT invariance
and that $\theta_{13}$ is negligible.
Also given in Table \ref {tab1} are
the bounds on $\Delta m_{21}^2$ and $\sin^2\theta_{12}$ that are
expected to be obtained when additional data from the running SNO and 
KamLAND experiments becomes available. 
For SNO, the analysis assumes 
that the third and final phase of the experiment will
measure the same NC and CC rates as the salt phase, but with 
reduced errors of 6\% and 
5\% respectively \cite{Poon:2005qu}. For KamLAND,
the prospective 3 kTy data is simulated
at $\Delta m_{21}^2=8.0\times 10^{-5}$ eV$^2$
and $\sin^2\theta_{12}=0.3$ and a systematic error of 5\%
is assumed.
Better measurement of charged-current (CC) and neutral-current (NC)
rates in SNO is expected to improve the limits on
$\sin^2\theta_{12}$.
The sensitivity of the KamLAND experiment to the shape  
of the reactor-induced $\anue$ positron spectrum, gives the experiment
a tremendous ability to constrain $\Delta m_{21}^2$. 
However, we can see from table \ref{tab1}, KamLAND is not as
sensitive to the mixing angle $\theta_{12}$
\cite{Bandyopadhyay:2003du,Bandyopadhyay:2003ks}. 
The uncertainty in $\Delta m_{21}^2$ is expected to reduce 
to 6\% at $3\sigma$ with 3~kTy of data from KamLAND. The uncertainty in 
$\sin^2\theta_{12}$ is expected to improve after the phase-III results 
from SNO to 18\% at $3\sigma$. This would improve to about 16\% if the 
SNO phase-III projected results are combined with the 3 kTy 
simulated data from KamLAND. However, we 
note that even with the combined data from 
phase-III of SNO and 3 kTy statistics from KamLAND, the uncertainty 
on $\sin^2\theta_{12}$ would stay 
well above the 10-15\% level at $3\sigma$. 
\begin{table}
  \begin{tabular}{ccccc}
    \hline
    {Data set} 
    & Range of 
    & Spread in  
    & Range of
    & Spread in \cr
    used 
    & $\Delta m^2_{21}$ 
    & $\Delta m^2_{21}$
    & $\sin^2\theta_{12}$
    & $\sin^2\theta_{12}$ \cr
    \hline\hline
    {only solar} & $(3.3 - 18.4)$$\times 10^{-5}$ eV$^2$ &{69\%} 
    & $0.24-0.41$ &26\%\cr
    {solar + 766.3 Ty KL}& $(7.2 - 9.2)$$\times 10^{-5}$ eV$^2$ 
    & 12\% & $0.25-0.39$ & 22\% \cr
    \hline
    {solar(SNO3) + 766.3 Ty KL} & $(7.2 - 9.2)$$\times 10^{-5}$ eV$^2$ 
    & 12\%  & $0.26 - 0.37$ & 18\% \cr
    {solar(SNO3) + 3KTy KL}  & $(7.6 - 8.6)$$\times 10^{-5}$ eV$^2$ 
    & 6\%& $0.26 - 0.36$ & 16\% \cr \hline
    \end{tabular} \\
    \caption{
    The 3$\sigma$ allowed ranges (1 dof) 
    and \% spread  of $\Delta m^2_{21}$ and
    $\sin^2\theta_{12}$ obtained using current and expected future 
    data from the current generation of solar neutrino and 
    KamLAND experiments.
  }
  \label{tab1}
\end{table}

\begin{figure}
  \begin{center}
    \includegraphics[width=11.0cm]
    {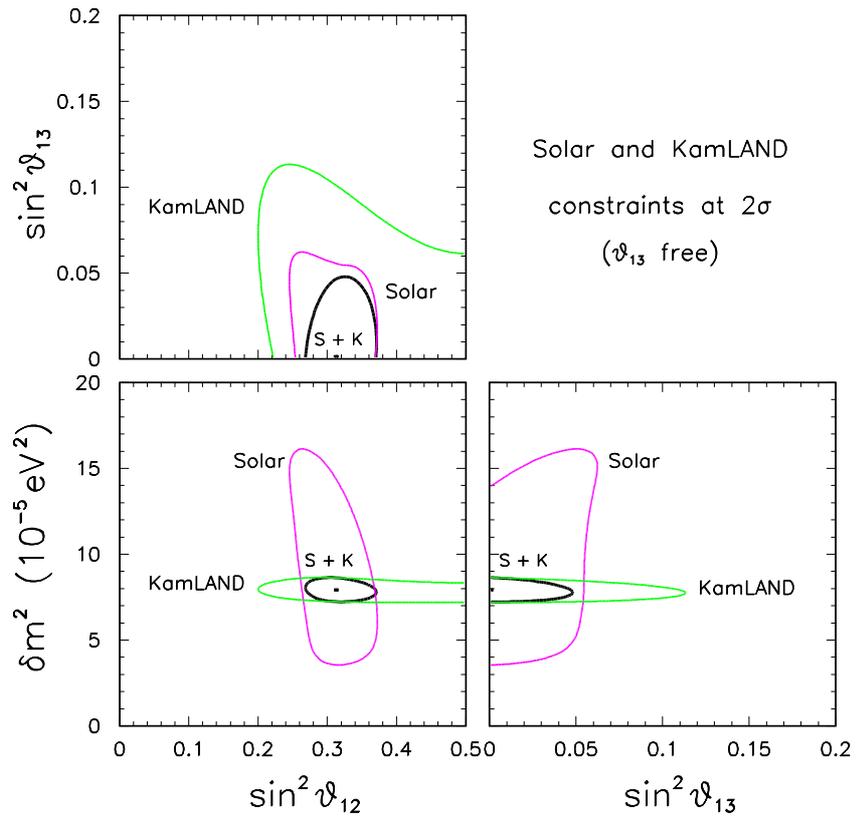}
    \caption{
      \label{fig:solarplusKLth13free}
      Three flavour analysis of solar and KamLAND data (both separately
      and in combination) in the 
      ($\Delta m_{21}^2 (\equiv \delta m^2)$, $\sin^2\theta_{12}$, 
      $\sin^2\theta_{13}$). The contours show the $2\sigma$ 
      allowed regions corresponding to $\Delta \chi^2=4$.
	Taken with kind permission of Progress in Particle and Nuclear Physics
	from figure 14 in reference \cite{Fogli:2005cq}.
	Copyrighted by Elsevier Science B.V.
    }
  \end{center}
\end{figure}

In our discussion so far, we have assumed the mixing angle 
$\theta_{13}$ to be zero. 
If $\theta_{13}$ is allowed to vary freely,
then the allowed regions obtained are those shown in 
figure \ref{fig:solarplusKLth13free} \cite{Fogli:2005cq}. 
Note that this figure shows 
only the $2\sigma$ contours and uses the confidence-level definition
appropriate for one degree of freedom. 
The data do not exclude the possibility that $\theta_{13} = 0$.
The KamLAND experiment places an upper bound on the value of 
$\theta_{13}$ by taking into account the neutrino energy spectrum as
well as the absolute rate. 
By lowering the value of $\theta_{12}$, the anti-correlation between
$\theta_{12}$ and $\theta_{13}$ can be used to explain the KamLAND
rate data for a wide range of values of $\theta_{13}$.
In contrast, the KamLAND data on the positron energy spectrum can be 
explained only for a certain range of $\theta_{12}$. 
This imposes an upper limit on the allowed value of 
$\theta_{13}$. 
For the solar neutrinos, the upper limit on
$\theta_{13}$ comes mainly from the difference in the 
$\theta_{12}$--$\theta_{13}$ anti-correlation between the 
low- and high-energy end of the solar-neutrino spectrum. 
The tension between the low energy solar neutrino data from 
SAGE, GALLEX, and GNO and the high energy $^8B$ data 
from SK and SNO, results in a reasonably tight upper bound 
on $\theta_{13}$ \cite{Goswami:2004cn,Bandyopadhyay:2004cp}. 
Together, the data from solar-neutrino experiments and KamLAND put a  
rather stringent limit of $\sin^2\theta_{13} > 0.05$ at $2\sigma$
\cite{Fogli:2005cq}. 

%% file: 02-Standard-Neutrino-Model/02-02-Present-status/02-02-02-Atmospheric.tex
\subsubsection{Atmospheric neutrino experiments}
\label{SubSubSect:Atmospheric}

\begin{figure}
  \begin{center}
    \includegraphics[width=8.0cm]
      {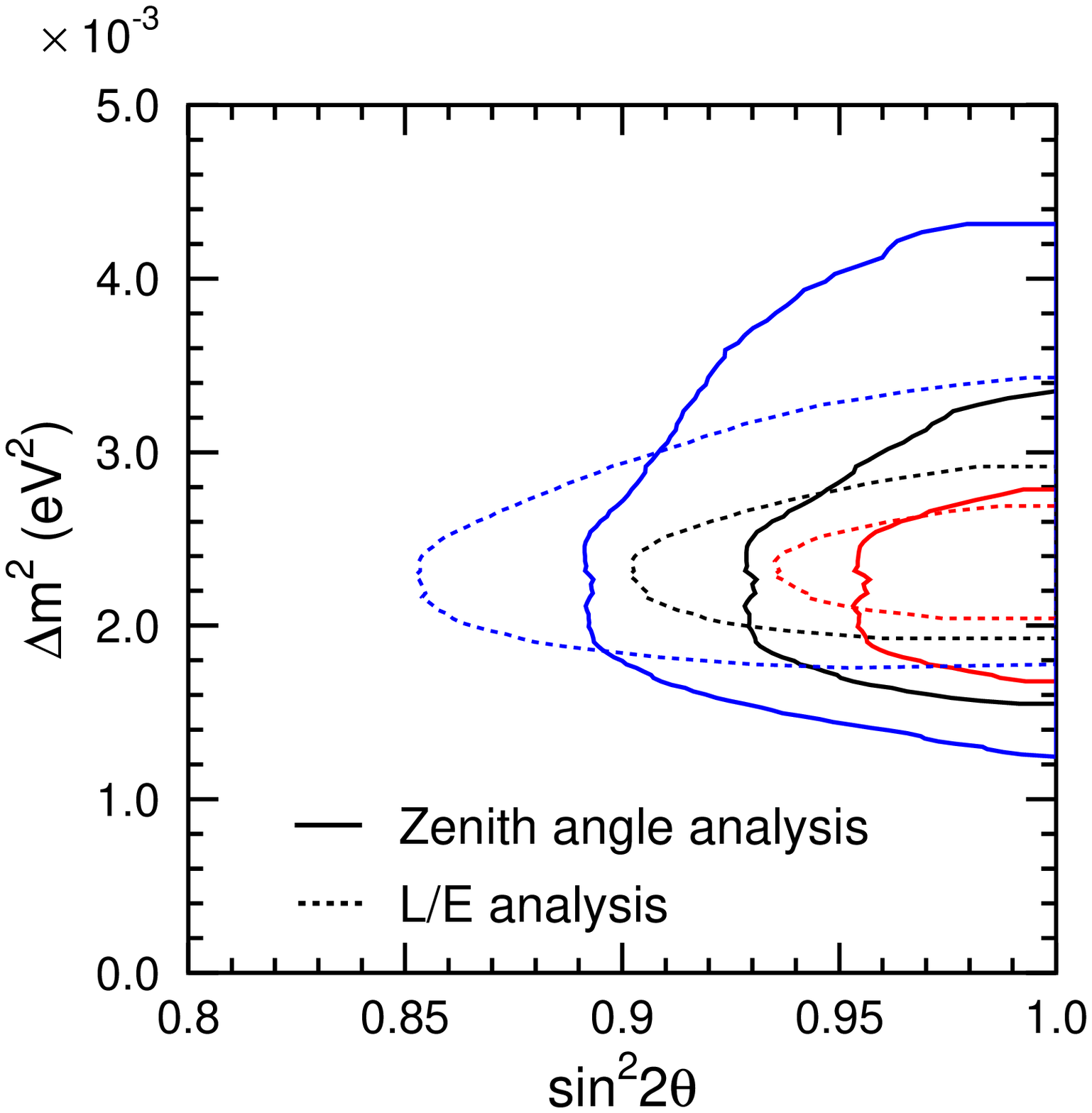}
    \caption{
      \label{fig:SKatmallowed}
      The 68\% (red lines), 90\% (black lines) and 99\% (blue lines)
      C.L. (2 dof) allowed oscillation parameter regions obtained in
      two-generation framework by the SK collaboration. The solid
      lines are with the analysis of the zenith angle binned data,
      while the dashed lines are obtained using the $L/E$ binned
      analysis.  Taken with kind permission of the Physical Review
	 from figure 42 in reference \cite{Ashie:2005ik}.
    Copyrighted by the American Physical Society.
    } 
  \end{center}
\end{figure}

The parameters $\Delta m_{32}^2$ ($\approx \Delta m_{31}^2$) and 
$\sin^2\theta_{23}$ are constrained by the zenith-angle dependence
of the atmospheric-neutrino data obtained by the Super-Kamiokande
experiment (SK) \cite{Fukuda:1998mi,Ashie:2005ik}.
The results from the earlier Kamiokande 
\cite{Fukuda:1994mc,Hatakeyama:1998ea}, 
MACRO \cite{Ambrosio:2003yz,Ambrosio:2004ig},
and Soudan-2 \cite{Sanchez:2003rb}
experiments are in agreement with the SK data. 
Figure \ref{fig:SKatmallowed}
\cite{Ashie:2005ik}
shows the allowed areas in the 
$\Delta m_{31}^2$--$\sin^22\theta_{23}$ parameter space, 
from a two-generation analysis. 
The allowed regions are obtained by fitting both the zenith-angle data
\cite{Ashie:2005ik} and the $L/E$ dependent data \cite{Ashie:2004mr}
from SK.

The values of $\Delta m_{31}^2$ and $\sin^22\theta_{23}$ are also
constrained by the results from the K2K \cite{Aliu:2004sq}
and MINOS \cite{Michael:2006rx}
long-baseline experiments. While K2K has finished its 
run, MINOS has declared its first results in the summer of 2006.
Both K2K and MINOS results are consistent with the 
SK atmospheric neutrino data, and while the allowed 
range of values for 
$\sin^22\theta_{23}$ is still controlled mainly 
by the SK atmospheric data, the results from the long-baseline 
experiments have an impact on 
the allowed range of values for $\Delta m_{31}^2$. 

\begin{figure}[htbp]
  \begin{center}
    \includegraphics[width=17.0cm]
      {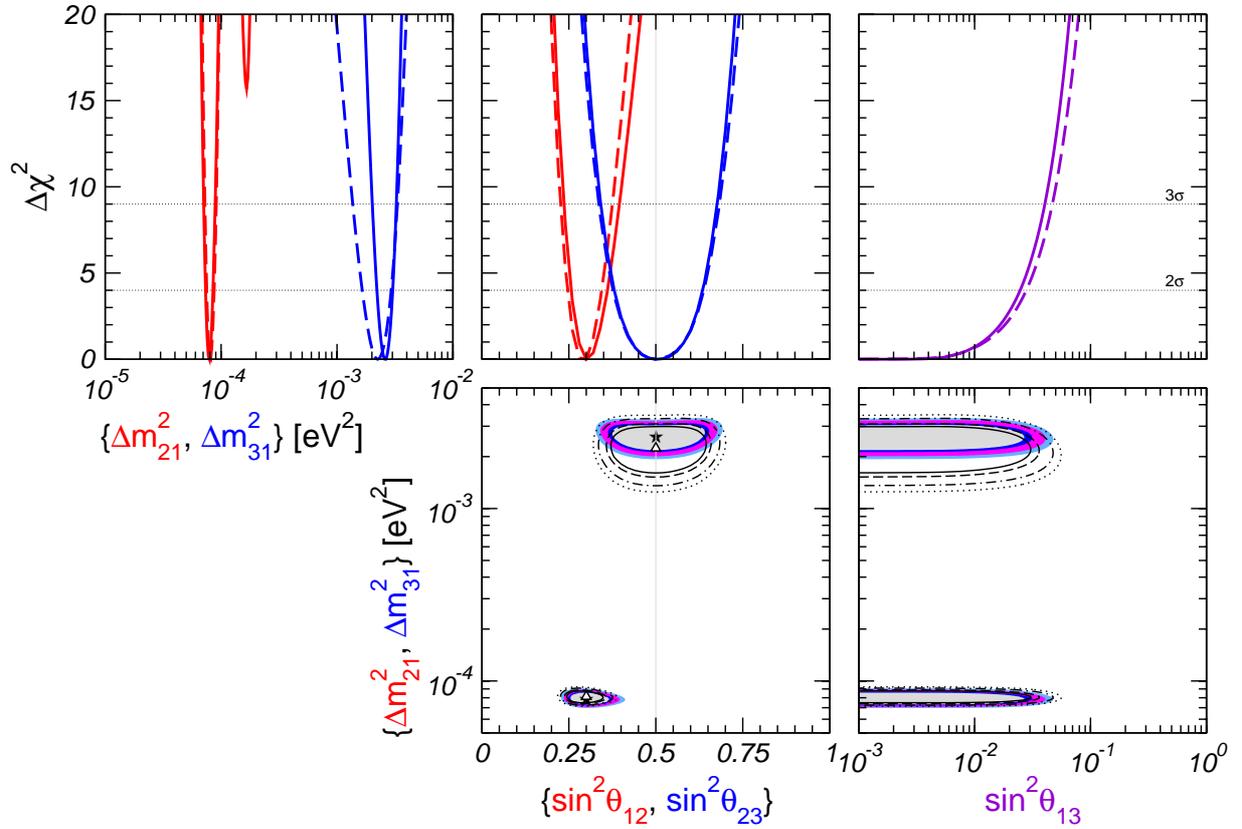}
    \caption{
      \label{fig:allowedareassolplusatm}
      Projections of the allowed regions from the global oscillation
      data at 90\%, 95\%, 99\%, and 3$\sigma$ C.L. for 2 dof for
      various parameter combinations. Also shown is $\Delta \chi^2$ as
      a function of the oscillation parameters $\sin^2\theta_{12},
      \sin^2\theta_{23}, \sin^2\theta_{13}, \Delta m^2_{21}, \Delta
      m^2_{31}$, minimized with respect to all undisplayed
      parameters. 
      Dashed lines and empty regions correspond to the global analysis
      before this update, while solid lines and colored regions show
      our most recent results.
    Taken with kind permission of New Journal of Physics
    from figure 12 in reference \cite{Maltoni:2004ei}.
    Copyrighted by Deutsche Physikalische Gesellschaft \& Institute of Physics
    }
  \end{center}
\end{figure}

Figure \ref{fig:allowedareassolplusatm} \cite{Maltoni:2004ei}
shows the projected allowed 
areas obtained from a full three-generation analysis of 
the global data from all solar, atmospheric, 
long-baseline, and reactor-neutrino experiments. Filled 
regions correspond to allowed areas with the latest MINOS 
\cite{Michael:2006rx}
and SNO \cite{Aharmim:2005gt} results, 
while the hollow regions correspond to 
the allowed areas obtained without these updates. In the 
$\Delta \chi^2$ versus parameter curves in the figure, the solid
lines are for the full data set, while the dashed lines are without
the new SNO \cite{Aharmim:2005gt} and MINOS \cite{Michael:2006rx}
results.   
The impact 
of the MINOS data on the allowed values of $\Delta m_{31}^2$
is clearly visible. The best-fit for $\Delta m_{31}^2$ 
shifts to a larger value compared to that obtained from
the SK atmospheric-neutrino data alone. The range
of allowed values for $\Delta m_{31}^2$ is also 
significantly changed.   
While the upper bound on
$\Delta m_{31}^2$ is hardly affected, the lower limit on 
this parameter is considerably improved. The current 
limits on all the oscillation parameters can be directly
read from this figure.
\begin{figure}[htbp]
  \begin{center}
    \includegraphics[width=8.0cm]
      {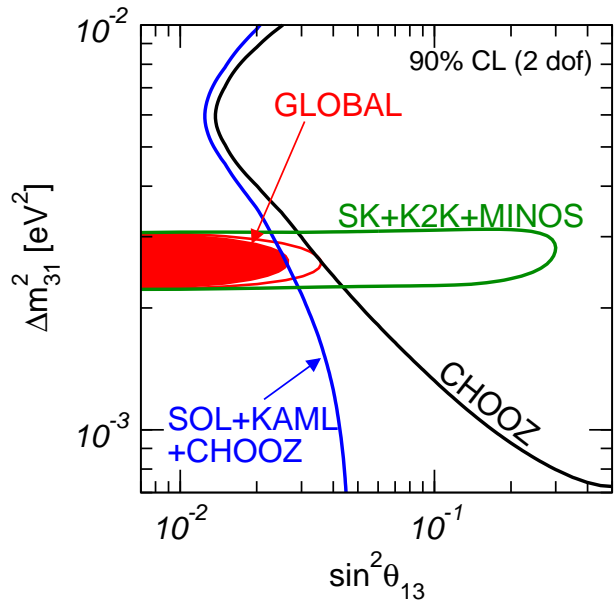}
    \caption{
      \label{fig:allowedth13solplusatm}
      90\% C.L. upper bound on $\sin^2\theta_{13}$ (2
      dof) from the combination of all neutrino oscillation data as
      a function of $\Delta m_{31}^2$.
    Taken from figure C2 in \cite{Maltoni:2004ei} (v6).
    }
  \end{center}
\end{figure}

Figure 
\ref{fig:allowedth13solplusatm} shows the 
90\% C.L. upper limit on 
$\theta_{13}$ and how it depends on the different data sets.
One can note from this figure that the bound from the 
solar+KamLAND combined analysis is comparable to the 
one obtained using the atmospheric+K2K+MINOS results. 
The 90\%($3\sigma$) bounds (1 dof) on $\sin^2\theta_{13}$ from 
an analysis of different sets of data read as \cite{Maltoni:2004ei}
\begin{equation}\label{eq:th13-06}:
    \sin^2\theta_{13} \le \left\lbrace \begin{array}{l@{\qquad}l}
    0.033~(0.071) & {\rm (solar+KamLAND)} \\
    0.026~(0.054) & {\rm (CHOOZ+atmospheric+K2K+MINOS)} \\
    0.020~(0.040) & {\rm (global ~data)}
\end{array} \right.
\end{equation}
The best-fit values and 
allowed range of values of the oscillation parameters at 
different C.L. obtained by Maltoni {\it et al.} in 
\cite{Maltoni:2004ei}
are shown in Table \ref{tab:globalallowedrange}.
\begin{table}[htbp]
  \begin{center}
    \begin{tabular}{|c|c|c|c|c|}
    \hline
        parameter & best fit & 2$\sigma$ & 3$\sigma$ & 4$\sigma$
        \\
        \hline
        $\Delta m^2_{21}\: [10^{-5}$ eV$^2]$
        & 7.9 & 7.3--8.5 & 7.1--8.9 & 6.8--9.3\\
        $\Delta m^2_{31}\: [10^{-3}$ eV$^2]$
        & 2.6 & 2.2--3.0 & 2.0--3.2 & 1.8--3.5\\
        $\sin^2\theta_{12}$
        & 0.30 & 0.26--0.36 & 0.24--0.40 & 0.22--0.44\\
        $\sin^2\theta_{23}$
        & 0.50 & 0.38--0.63 & 0.34--0.68 & 0.31--0.71 \\
        $\sin^2\theta_{13}$
        & 0.000 &  $\leq$ 0.025 & $\leq$ 0.040  & $\leq$ 0.058 \\
        \hline
    \end{tabular}
    \caption{ 
      \label{tab:globalallowedrange} 
      Best-fit values, 2$\sigma$,
      3$\sigma$, and 4$\sigma$ intervals (1 dof) for the
      three--flavour neutrino oscillation parameters from global data
      including solar, atmospheric, reactor (KamLAND and CHOOZ) and
      accelerator (K2K and MINOS) experiments.
    }
  \end{center} 
\end{table}

%% file: 02-Standard-Neutrino-Model/02-02-Present-status/02-02-03-LBL.tex
\subsubsection{Long-baseline neutrino-oscillation experiments}

In 1962, just a few years after neutrinos were observed directly for
the first time using the intense flux generated in a nuclear
reactor\cite{Cowan:1956xc}, the AGS proton accelerator at Brookhaven
was used to show that a second generation of neutrinos exists
\cite{Danby:1962nd}.
In this experiment, a 15~GeV proton beam impinged on a beryllium
target, producing pions, which decayed into muons and neutrinos.
13.5m of steel separated the volume where the pions decayed and the spark
chambers detected the muons created by the neutrinos penetrating the
steel. 

Today, the same fundamental principles are used to study the phenomenon of
neutrino oscillations. The energies of the neutrinos are fixed at a GeV or more
due to the production mechanism, therefore, to probe the oscillations first
seen in atmospheric neutrinos, the distances between neutrino source and target
have stretched to hundreds of kilometres, giving rise to their collective name
of long-baseline (LBL) neutrino-oscillation experiments.

At the time of writing, two such experiments, K2K and MINOS, have
demonstrated that neutrinos disappear from their muon neutrino beams
in a way that is consistent with neutrino oscillations. 
A third LBL beam, providing neutrinos with energies running up to of
tens of GeV, has just started operating from CERN to Gran Sasso.
This facility will test whether the $\nu_\mu$-disappearance signals
are actually accompanied by conversions of $\nu_\mu$ into $\nu_\tau$,
by looking for tau production in a beam that is originally free of tau
neutrinos. 

The K2K (KEK-to-Kamioka) experiment was formally proposed in 
1995 \cite{k2kproposal}, after the first indications of oscillations
were seen in Kamiokande, IMB, and Soudan-II atmospheric neutrino data,
but before the confirmation by Super-Kamiokande, and indeed before the
completion of the 50~kt water Cherenkov detector.

K2K had a baseline of 250km, and the muon-neutrino energy was a GeV or
so. 
The beam was created from a 12~GeV proton beam, the hadrons from which
were focussed in a horn-shaped electromagnetic volume to increase the
beam intensity. 
A dedicated detector complex, with a 1~kt water Cherenkov tank,
fine-grained detectors, and a muon ranger, was  located 100~m from the
end of the pion-decay volume, and measured the beam before it started
oscillating on its way to Kamioka. 
Super-Kamiokande was used as the far detector, and the first
beam-induced neutrino event was observed in the summer of 1999.

\begin{figure}
  \begin{center}
    \includegraphics[width=7.0cm]%
      {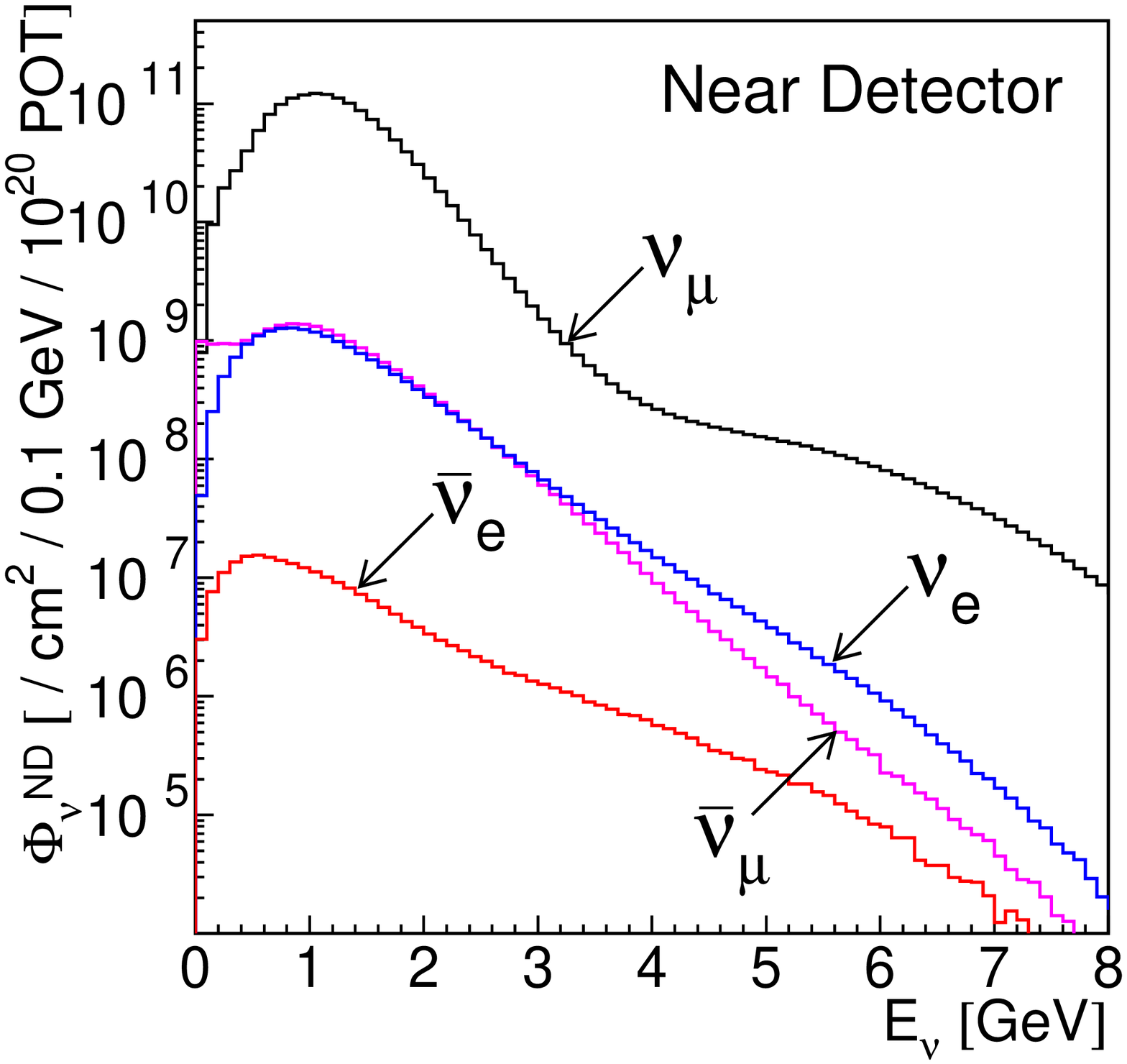}
    \hspace{0.5cm}
    \includegraphics[width=8.0cm]%
      {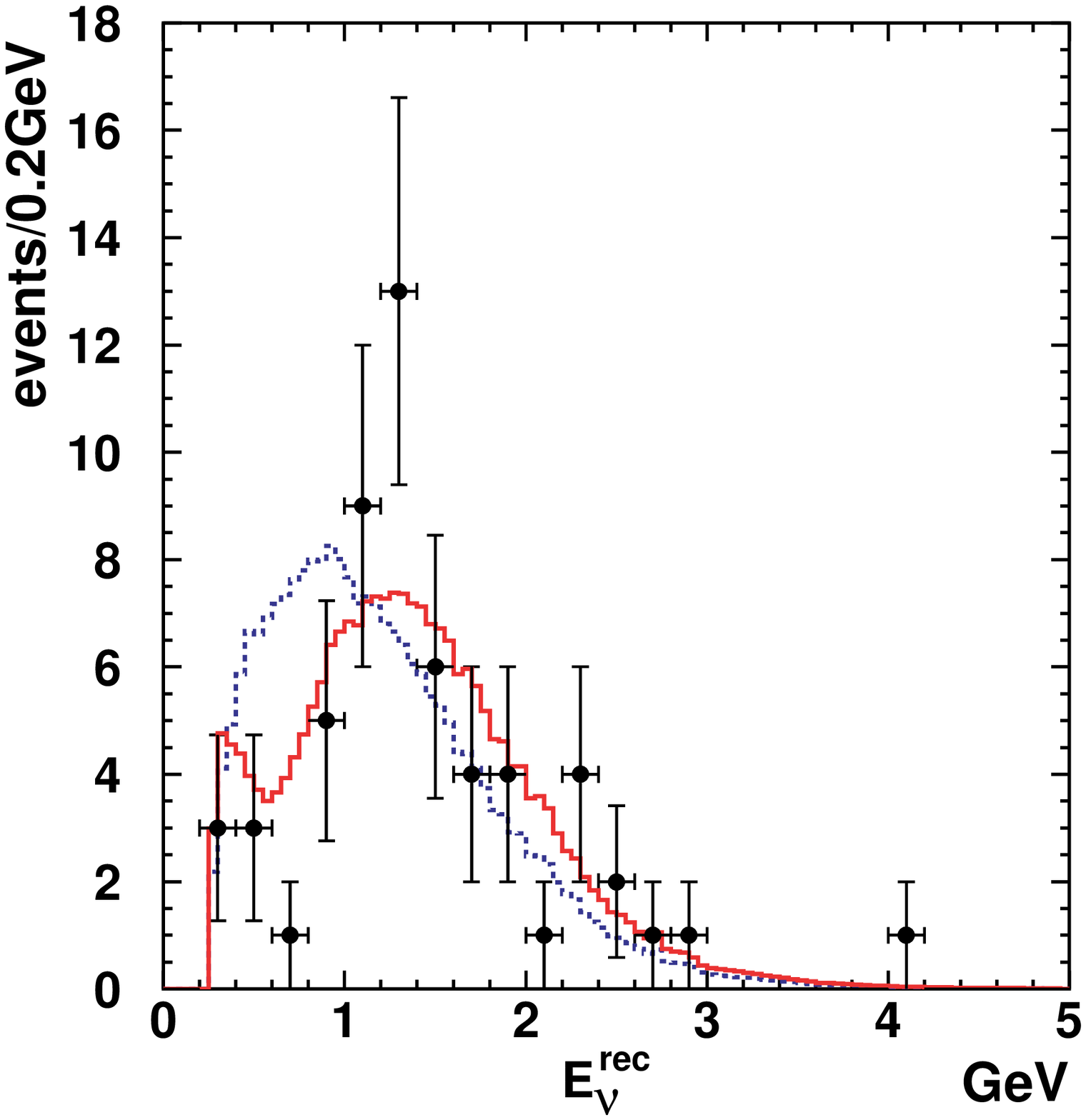}
  \end{center}
  \caption{
    Left: The energy spectrum for each type of neutrino at the K2K
    Near Detector, estimated by MC simulations. The neutrino beam
    consists of 97.3\% muon neutrinos. 
    Right: The 58 fully-contained muon-like single-ring events, out of
    the 112 beam-originated neutrino events in K2K. 
    The muon energies and directions can be reconstructed for these
    events, allowing their parent neutrino energies to be estimated
    under the assumption that they are from quasi-elastic
    interactions. 
    The solid line is the best fit spectrum with neutrino oscillation
    and the dashed line is the expectation without oscillation, both
    normalised to the number of events seen \protect\cite{Ahn:2006zz}.
	Both figures taken with kind permission of Physical Review
	from figures 6 and 43 in referencefrom \cite{Ahn:2006zz}.
        Copyrighted by the American Physical Society.
  }
  \label{fig:K2K-Beam-and-Results}
\end{figure}
Five and a half years after commissioning, K2K running ended late in
2004.
The final oscillation analysis \cite{Ahn:2006zz} was performed using a
data set corresponding to 0.922$\times 10^{20}$ protons on target.
The estimated beam spectra for different neutrino types are shown in
figure \ref{fig:K2K-Beam-and-Results}. 
112 beam-originated neutrino events were observed, where the expected number in
the absence of oscillations was $158.1^{+9.2}_{-8.6}$. 
Of these events, 58 were
single-ring muon-like events fully-contained within the Super-Kamiokande
detector. 
The energies and directions of the muons in fully-contained events
can be reconstructed, and because of the simple kinematics of the charged-current quasi-elastic (CCQE) events that make up much of the cross
section around 1~GeV, it is possible to estimate the energy of the incoming
neutrinos. 
Such a spectrum is shown in figure \ref{fig:K2K-Beam-and-Results}, for
the 58 events, with unoscillated and best-fit oscillated curves, normalised to
the number of events seen.
These results support maximal mixing, with best-fit two-neutrino
oscillation parameters of $\sin^2 2\theta = 1$ and 
$\Delta m^2 = 2.8 \times 10^{-3} \mathrm{eV}^2$. 
The 90\% C.L. range for $\Delta m^2$ at $\sin^2 2\theta = 1$ is
between 1.9 and 3.5 $\times 10^{-3} \mathrm{eV}^2$. 

\begin{figure}[htbp]
  \begin{center}
    \includegraphics[width=10.0cm]%
      {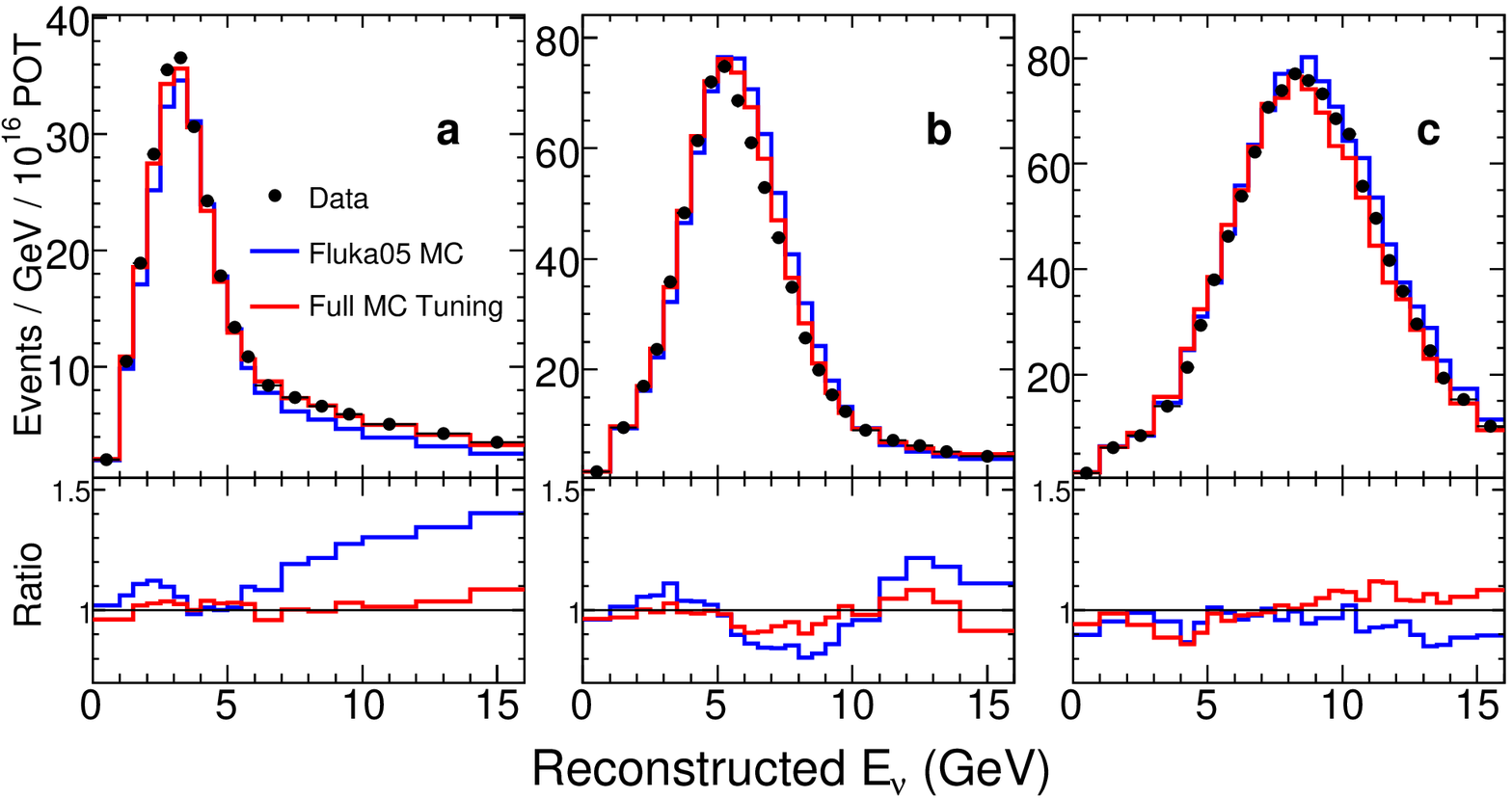}\\
    \includegraphics[width=11.0cm]%
      {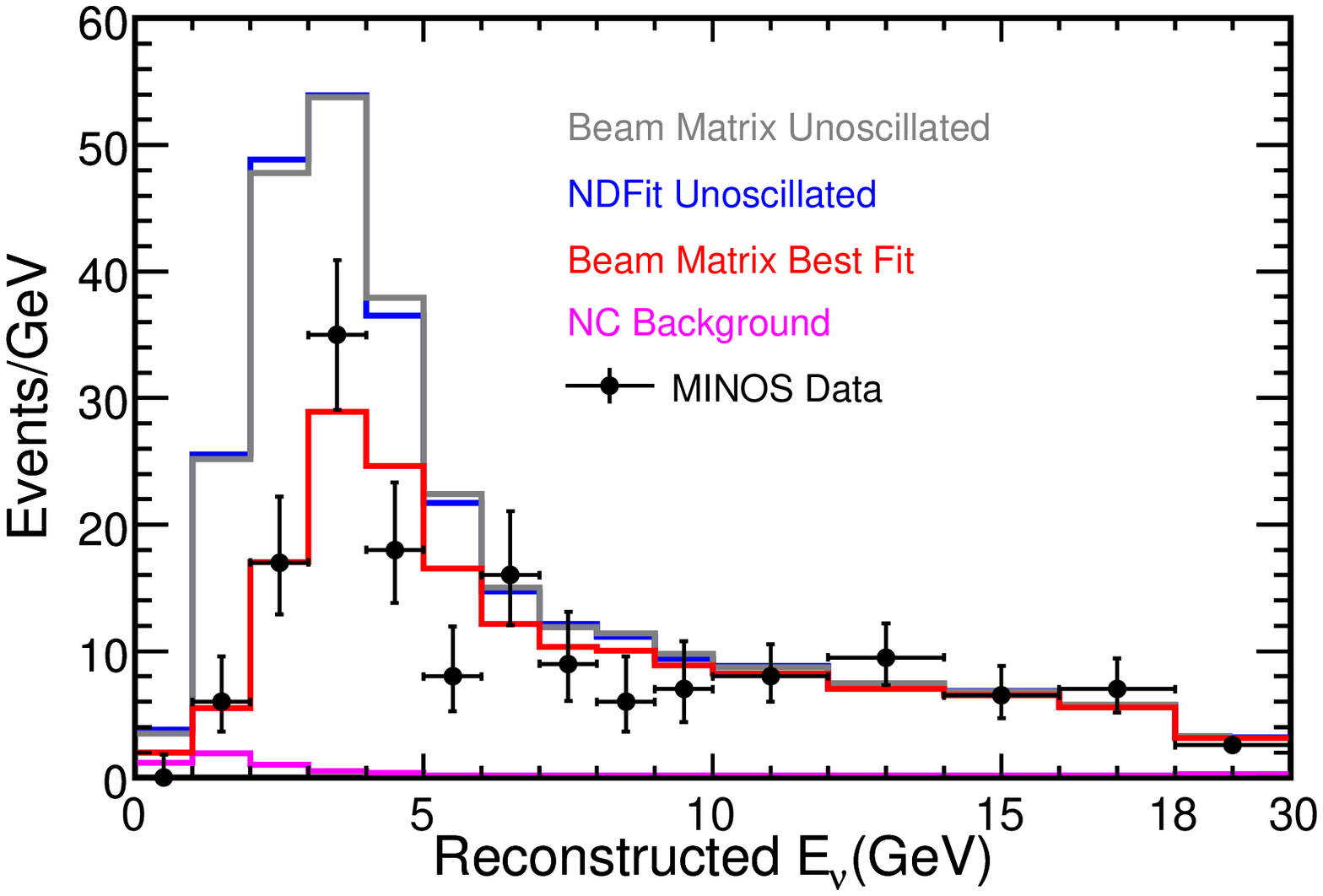}
  \end{center}
  \caption{
    Top: MINOS neutrino beam spectra at the Near Detector, for three
    beam configurations. 
    Bottom: The final far detector spectrum and predicted
    distributions, after the first full year of MINOS running 
    ($1.27 \times 10^{20}$ protons on target)
    \protect\cite{Michael:2006rx}. 
    Two different methods of near-to-far extrapolation are shown for
    the unoscillated spectrum.
	Both figures taken from with kind permission of Physical Review Letters
	from figures 2 and 3 in \cite{Michael:2006rx}.
        Copyrighted by the American Physical Society.
  }
  \label{fig:MINOS-Beam-and-Results}
\end{figure}

The MINOS (Main Injector Neutrino Oscillation Search) experiment was
also proposed in 1995, with a neutrino beam pointed from Fermilab to
the Soudan mine in Minnesota, with a baseline of 735~km. 
The beam has a system of movable focussing horns to allow the beam
energy spectrum to be altered.  
Three different spectra are shown in the upper plot in figure
\ref{fig:MINOS-Beam-and-Results}.
Both near and far detectors consist of a steel and
plastic-scintillator sandwich structure, the performance of which was
studied in detail in test beam  work at CERN \cite{Adamson:2006xv}. 

\begin{figure}[hbtp]
  \begin{center}
    \includegraphics[width=10.0cm]%
      {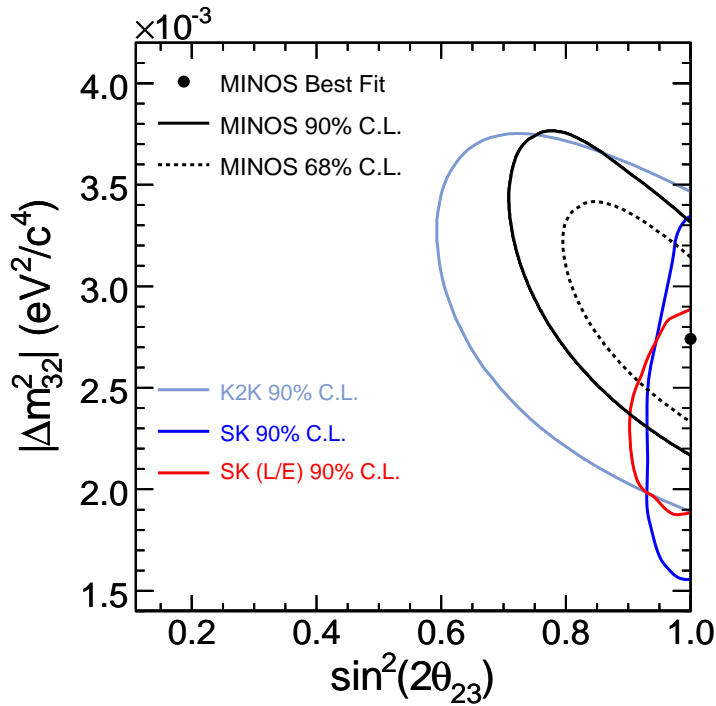}
  \end{center}
  \caption{
    Confidence intervals from the MINOS experiment
    \protect\cite{Michael:2006rx}. 
    Results from K2K \protect\cite{Aliu:2004sq} and Super-K
    \protect\cite{Ashie:2004mr,Ashie:2005ik} are also shown.
	Taken with kind permission of Physical Review Letters
	from figure 4 in \cite{Michael:2006rx}.
        Copyrighted by the American Physical Society.
  }
  \label{fig:MINOS-Oscillation-Contours-With-K2K-Super-K}
\end{figure}

The experiment started running in the spring of 2005, and within a
year had gathered data corresponding to $1.27 \times 10^{20}$ protons
on target.  
The data are shown in the lower plot in figure
\ref{fig:MINOS-Beam-and-Results}.
The MINOS results support maximal mixing, with best fit parameters of
$|\Delta m^2_{32}| = 2.74^{+0.44}_{-0.26} \times 10^{-3}
\mathrm{eV}^2$ and $\sin^2 2\theta_{23} > 0.87$ at 68\% C.L.
The oscillation parameters from the K2K and MINOS experiments,
together with results from Super-Kamiokande are shown in
figure \ref{fig:MINOS-Oscillation-Contours-With-K2K-Super-K}. 
MINOS will run for five years, with the goal of accumulating 
$16 \times 10^{20}$ protons on target.
This data set should improve our knowledge of the oscillation
parameters substantially.
Both the experiments described here are linked, if only indirectly, to
future projects to make precision measurements of the oscillation
parameters and to probe the third mixing angle. 
These projects, T2K and NO$\nu$A, are discussed below.

%% file: 02-Standard-Neutrino-Model/02-02-Present-status/02-02-04-Neutrinoless-Double-Beta.tex
\subsubsection{$0\nu\beta\beta$ Experiments}
\label{Sect:0nu2betaExp}

Establishing whether the neutrino is a Dirac or a Majorana fermion
is of fundamental importance for understanding the origin of
neutrino masses and mixing (see, e.g., \cite{Petcov:2004dz}).
Let us recall that the neutrinos, $\nu_j$,
with definite mass, $m_j$, will be Dirac fermions if 
particle interactions conserve 
some additive lepton number, e.g., the total
lepton number $L = L_e + L_{\mu} + L_{\tau}$. 
If no lepton number is conserved, 
the neutrinos will be Majorana fermions 
(see, e.g., \cite{Bilenky:1987ty}). 
The heavy neutrinos are
predicted to be Majorana in nature
by the see-saw mechanism 
\cite{Minkowski:1977sc}, which also provides an  
attractive explanation of the
smallness of neutrino masses 
and, through the leptogenesis theory 
\cite{Fukugita:1986hr}, of the observed baryon 
asymmetry of the Universe.
The observed patterns of 
neutrino mixing 
and of neutrino mass-squared differences
driving the solar and the dominant 
atmospheric-neutrino oscillations,
can be related to massive Majorana neutrinos 
and the existence of an approximate symmetry in the lepton sector
corresponding to the conservation of the
non-standard lepton number
$L' = L_e - L_{\mu} - L_{\tau}$
(see, e.g., \cite{Frampton:2004ud}).

The only experiments which have the 
potential of establishing the
Majorana nature of massive neutrinos are the
$\betabeta$-decay experiments searching 
for the process $(A,Z) \rightarrow (A,Z+2) + e^- + e^-$
(for reviews see, e.g., 
\cite{Bilenky:1987ty,Morales:2002zf,Aalseth:2004hb,Petcov:2004wz,Petcov:2005yq,Pascoli:2003xb}).
The observation of \betabeta-decay
and the measurement of the corresponding 
half-life with sufficient accuracy,
would not only be a proof that total 
lepton number is not conserved, but might also provide
unique information on: i) the type of neutrino-mass spectrum; ii) the
absolute scale of neutrino masses; and iii) the Majorana
CP-violating phases in the neutrino mixing matrix 
\cite{Bilenky:1980cx,Schechter:1980gr,Doi:1980yb,Bilenky:2001rz,Bilenky:2001xq,Bilenky:1999wz,Barger:1999na,Vissani:1999tu,Czakon:2000vz,Klapdor-Kleingrothaus:2000gr,Pascoli:2001by,Pascoli:2002qm,Pascoli:2002xq,Pascoli:2002ae,Pascoli:2003ke,Murayama:2003ci,Bilenky:1996cb,Rodejohann:2000ne,Matsuda:2000by,Matsuda:2000iw,Feruglio:2002af,Nunokawa:2002iv,Pascoli:2005zb,Choubey:2005rq,Lindner:2005kr}.

If the $\nu_j$ are Majorana fermions,
obtaining information about the Majorana CP phases in $\pmns$
will be remarkably difficult
\cite{Bilenky:2001rz,Bilenky:2001xq,Pascoli:2005zb,Barger:2002vy,deGouvea:2002gf}. 
In a large class of supersymmetric theories
which include the see-saw neutrino-mass-generation mechanism,
the phases $\alpha$ and $\beta$ 
can affect significantly the predictions for the 
rates of lepton-flavour violating (LFV) decays such as
$\mu \rightarrow e + \gamma$, $\tau \rightarrow \mu + \gamma$, etc.
(see, e.g., \cite{Pascoli:2003rq,Petcov:2005yh,Petcov:2006pc}).

Under the assumptions of massive, Majorana neutrinos, 
three-neutrino mixing,
and \betabeta-decay being generated solely through the (V-A)
charged-current weak interaction mediated by the exchange of the three
Majorana neutrinos, the $\betabeta$-decay amplitude  
has the form (see, e.g., \cite{Bilenky:2001rz,Bilenky:2001xq}):
$A\betabeta \cong \mefff~M$, where 
$M$ is the corresponding 
nuclear matrix element (NME) which does not 
depend on the neutrino mixing parameters, and:
\begin{equation}
\meff=\left| m_1 |U_{\mathrm{e} 1}|^2 
+ m_2 |U_{\mathrm{e} 2}|^2e^{i\alpha}
+ m_3 |U_{\mathrm{e} 3}|^2e^{i\beta} \right|\,,
\label{effmass2}
\end{equation}
is the effective Majorana 
mass in \betabeta-decay,
$|U_{\mathrm{e}1}|$=$c_{12}c_{13}$,
$|U_{\mathrm{e}2}|$=$s_{12}c_{13}$, 
$|U_{\mathrm{e}3}|$=$s_{13}$.
In the case of CP-invariance 
one has \cite{Wolfenstein:1981rk,Valle:1982yw,Bilenky:1984fg,Kayser:1984ge},
$\eta_{21} \equiv e^{i\alpha}$=$\pm 1$, 
$\eta_{31}\equiv e^{i\beta}$=$\pm 1$; $\eta_{21(31)}$ being the  
relative CP-parity of Majorana neutrinos 
$\nu_{2(3)}$ and $\nu_1$. 

Information on the absolute 
scale of neutrino masses
can be derived in \hbeta experiments 
\cite{Lobashev:2003kt,Eitel:2005hg}
and from cosmological and astrophysical data. 
The most stringent upper 
bounds on the $\bar{\nu}_e$ mass 
were obtained in the Troitzk~\cite{Lobashev:2003kt} 
and Mainz~\cite{Eitel:2005hg} experiments: 
\beq
m_{\bar{\nu}_e} < 2.3 \mathrm{eV}~~~\mbox{at}~95\%~\mathrm{C.L.} 
\label{H3beta}
\eeq
\noindent We have $m_{\bar{\nu}_e} \cong m_{1,2,3}$
in the case of the QD $\nu$-mass spectrum.
The KATRIN experiment~\cite{Eitel:2005hg}
is planned to reach a sensitivity  
of  $m_{\bar{\nu}_e} \sim 0.20$~eV,
i.e.\ it will probe the region of the QD 
spectrum. The 
CMB data of the 
WMAP experiment, combined with data from large-scale structure surveys
(2dFGRS, SDSS), lead to a limit on the sum 
of $\nu_j$ masses (see, e.g., \cite{Tegmark:2005cy,Elgaroy:2004rc}):
\beq
\sum_{j} m_{j} \equiv \Sigma < (0.4 \mbox{--} 1.7)~
{\rm eV~~~\mbox{at}~95\%~C.L.} 
\label{WMAP}
\eeq
Data on weak lensing of galaxies, 
combined with data from the WMAP and PLANCK
experiments, may allow $\Sigma$ to be determined 
with an uncertainty of 
$\sim 0.04$~eV \cite{Hannestad:2006as,Lesgourgues:2004ps}.
It proves convenient to express   
\cite{Petcov:1993rk,Pascoli:2007qh} the three neutrino masses
in terms of $\dmsol$ and $\dma$, measured 
in neutrino-oscillation experiments,
and the absolute neutrino-mass scale
determined by ${\rm min}(m_j)$  
\cite{Bilenky:2001rz,Bilenky:2001xq,Petcov:2004wz,Petcov:2005yq,Pascoli:2003xb}.
In both the normal- and the inverted-hierarchy, one has:
$\dmsol$=$\Delta m_{21}^2 > 0$, 
$m_2$=$(m_1^2 + \dmsol)^{\frac{1}{2}}$.
For normal ordering, 
$\dma$=$\Delta m_{31}^2 > 0$
and $m_3$=$(m_1^2 + \dma)^{\frac{1}{2}}$,
while if the spectrum is with inverted ordering,
$\mmin$=$m_3$, $\dma$=$\Delta m_{23}^2 > 0$ and 
$m_1$=$(m_3^2 + \dma - \dmsol)^{\frac{1}{2}}$. 
Thus, given $\dma$, $\dmsol$, $\theta_{\odot}$ 
and $\theta_{13}$, $\meff$ depends 
on $\min(m_j)$, Majorana phases $\alpha$, $\beta$ and 
the type of $\nu$-mass spectrum. 
\begin{figure}
  \includegraphics[width=7.5cm,clip=]%
    {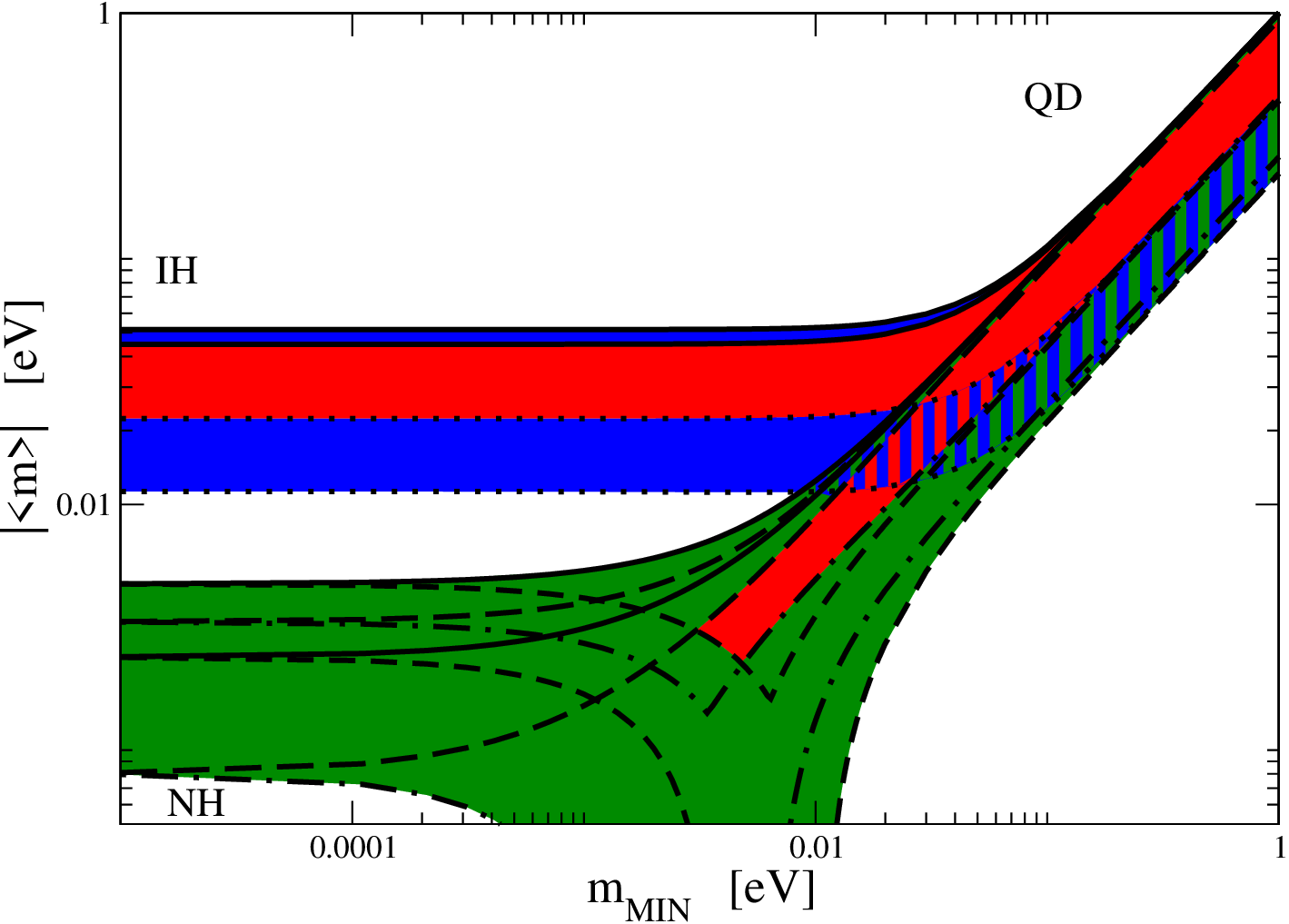}
\hspace{0.2cm}
  \includegraphics[width=7.5cm,clip=]%
    {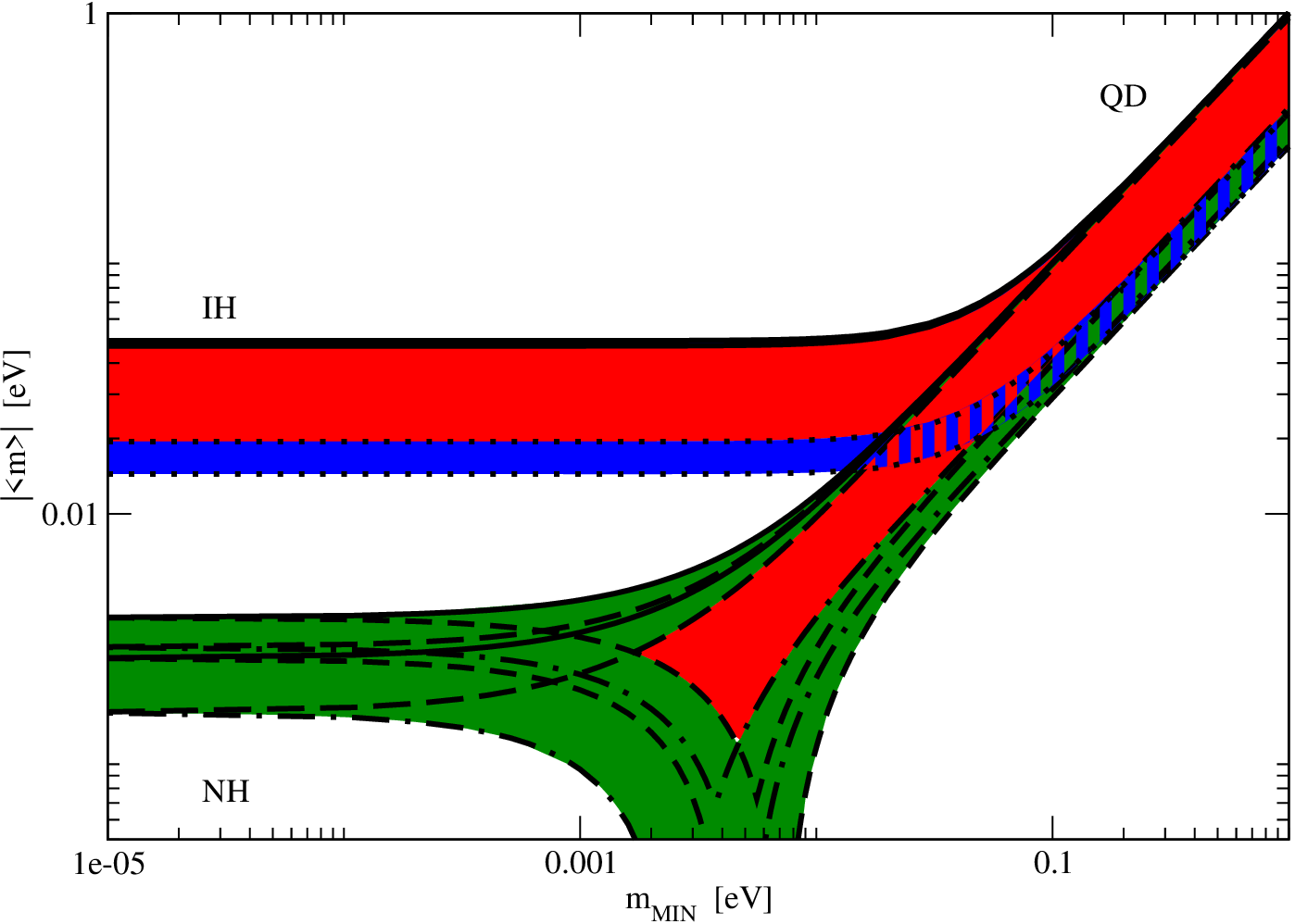}
  \caption{
    The value of $\meff$ as a function of $min(m_j)$, obtained using
    i) the 95\% C.L. allowed ranges of $\dmsol$, $|\dma|$,  $\sin^2
    \theta_\odot$ and $\sin^2 \theta_{13}$ (left panel), and ii)
    prospective 2$\sigma$ uncertainty in $\meff$, corresponding to
    input 1-$\sigma$ experimental errors in $\deltasol \!\!$,
    $\deltaatm \!\!$ and $\sin^2 \theta_\odot$ of 2\%, 2\% and 4\% and
    $\sin^2\theta_{13} = 0.010 \pm 0.006$ (right panel).
    The best fit values and the $2\sigma$ ranges used in the analysis
    are given in equations (2.1) - (2.4) in \cite{Pascoli:2007qh}.
    The regions
    shown in red/grey correspond to violation of CP-symmetry.
    Taken with kind permission of Physical Review from figures 1 and 2 in
    reference \cite{Pascoli:2007qh}.
    Copyrighted by the American Physical Society.
  }
  \label{Fig1}
\end{figure}

The problem of obtaining 
the allowed values of 
$\meff$ given the constraints on the 
parameters following from neutrino-oscillation data,
and, more generally, of the physics potential of 
\betabeta-decay experiments,
was first studied in \cite{Petcov:1993rk,Pascoli:2007qh} 
and subsequently in
\cite{Petcov:2004wz,Petcov:2005yq,Pascoli:2003xb}. 
Detailed analyses were performed
more recently in \cite{Pascoli:2005zb,Choubey:2005rq,Lindner:2005kr,Pascoli:2007qh}.
The results  are illustrated in Fig.\ref{Fig1}. 
The main features of the 
predictions for $\meff$   
are
\cite{Bilenky:2001rz,Bilenky:2001xq,Pascoli:2001by,Pascoli:2002qm,Pascoli:2002xq}
(figure \ref{Fig1}, left panel):
\begin{enumerate}
  \item
    For the NH spectrum, $\meff{\small \cong}|\sqrt{\dmsol}
    s^2_{12}$+$\sqrt{\dma} s^2_{13}e^{i(\alpha-\beta)}|{\small \ltap}
    0.005$~eV; 
  \item
    For the IH spectrum, $\meff{\small \cong}\sqrt{|\dma|}{\small
    (1-}\sin^22\theta_{\odot}\sin^2\frac{\alpha}{2})^{\frac{1}{2}}$,
    thus $\meff{\small \ltap \sqrt{|\dma|}\ltap}$0.055 eV and
    $\meff{\small\gtap \sqrt{|\dma|}\cos2\theta_{\odot} \gtap}
    0.013$~eV, the bounds corresponding to the values
    $\alpha$=$0;~\pi$; and 
  \item
    For the QD spectrum, $\meff{\small
    \cong}m_0{\small(1-}\sin^22\theta_{\odot}
    \sin^2\frac{\alpha}{2})^{\frac{1}{2}}$, $m_0 {\small
    \gtap}\meff{\small \gtap m_0\cos2\theta_{\odot}\gtap 0.03}$~eV,
    with $m_0{\small\gtap 0.1}$ eV, $m_0~{\small <2.3}$ eV
    \cite{Eitel:2005hg} or $m_0~{\small \ltap0.5}$ eV
    \cite{Tegmark:2005cy,Elgaroy:2004rc}.
\end{enumerate}
For the IH (QD) spectrum we have:
$\sin^2(\alpha/2){\small\cong
(1-}\meff^2/\tilde{m}^2)/\sin^{2}2\theta_\odot$, 
$\tilde{m}^2$=$|\dma|~(m_0^2)$. Thus, 
a measurement of $\meff$ (and $m_0$ for QD spectrum) 
can allow to determine $\alpha$.\\
\indent  Many experiments have searched for
$\betabeta$-decay \cite{Morales:2002zf}. 
The best sensitivity was achieved in
Heidelberg-Moscow $^{76}$Ge experiment 
\cite{Klapdor-Kleingrothaus:2000dg}: $\meff<$(0.35 - 1.05) eV (90\% C.L.),
where a factor of 3 uncertainty in the relevant NME 
(see, e.g., \cite{Rodin:2003eb,Rodin:2005dp,Poves})
is taken into account.
The IGEX collaboration has obtained
\cite{Aalseth:2000ud}: $\meff <$ (0.33 - 1.35) eV (90\% C.L.).
A positive signal at $>$3$\sigma$,
corresponding to
$\meff = (0.1 - 0.9)~{\rm eV}$,
is claimed to be observed \cite{Klapdor-Kleingrothaus:2004wj}. 
Two experiments, NEMO3 (with $^{100}$Mo and 
$^{82}$Se) \cite{Arnold:2004cd}
and CUORICINO (with $^{130}$Te) \cite{Capelli:2005jf},
designed to reach a sensitivity to $\meff\sim$ of
$\meff\sim (0.2--0.3)$~eV, published 
first results: $\meff<(0.7--1.2)$~eV \cite{Arnold:2004cd} 
and $\meff< (0.2 -- 0.9)$~eV
\cite{Capelli:2005jf} (90\% C.L.),
where estimated uncertainties in the NME
are accounted for. Most importantly,
a number of projects aim at sensitivity of
$\meff\sim$(0.01--0.05) eV 
\cite{Avignone:2005hj}:
CUORE ($^{130}$Te), GERDA ($^{76}$Ge),
SuperNEMO ($^{100}$Mo),
EXO ($^{136}$Xe), MAJORANA ($^{76}$Ge),
MOON ($^{100}$Mo), 
XMASS ($^{136}$Xe), CANDLES ($^{48}$Ca), etc. 
These experiments will probe the region 
corresponding to IH and QD spectra
and test the positive result 
claimed in \cite{Klapdor-Kleingrothaus:2004wj}.\\ 
\indent  The existence of significant 
lower bounds on $\meff$
in the cases of IH and QD spectra 
\cite{Pascoli:2002xq},
which lie either partially (IH spectrum) or completely
(QD spectrum) within the range of sensitivity of 
the next generation of \betabeta-decay experiments,
is one of the most important features of
the predictions of $\meff$. These minimal values are given, 
up to small corrections, by $\dma \cos2\theta_{\odot}$ and
$m_0 \cos2\theta_{\odot}$. According to the
combined analysis of the solar- and reactor-
neutrino data \cite{Bandyopadhyay:2004da,Bandyopadhyay2005,Schwetz:2006dh}
including the latest SNO and KL results: 
i) the possibility of  $\cos2\theta_{\odot}$ = 0
is excluded at $\sim$6$\sigma$;
ii) the best fit value of 
$\cos2\theta_{\odot}$ is 
$\cos2\theta_{\odot}$= 0.38; and
iii) at 95\% C.L. one has
for $\sin^2\theta_{13}$= 0~(0.02),
$\cos{2\theta_{\odot}} \gtap$0.28~(0.28).
The quoted results on $\cos{2\theta_{\odot}}$
together with the range of possible values of 
$|\dma|$ and $m_0$, 
lead to the 
significant and robust lower 
bounds on $\meff$ in the cases of the
IH and the QD spectrum
\cite{Pascoli:2002xq,Pascoli:2002ae,Pascoli:2003ke,Murayama:2003ci}. 
At the same time 
one can always have $\meff$ = 0 
in the case of spectrum with 
(partial) normal hierarchy \cite{Pascoli:2001by,Pascoli:2002qm}. 
As figure \ref{Fig1} indicates, 
$\meff$ cannot exceed $\sim 6$ meV
for the NH neutrino mass spectrum. 
This implies that 
$max(\meff)$ in the case of the NH 
spectrum is considerably smaller than
$\min(\meff)$ for the IH and the QD spectra.
This makes it possible that information about the type of 
neutrino-mass spectrum may be obtained from a measurement of 
$\meff \neq 0$ \cite{Pascoli:2002xq}.
In particular, a positive result in the future generation
of \betabeta-decay experiments with $\meff > 0.01$ eV
would imply that the 
NH spectrum is strongly disfavored (if not excluded).
Prospective experimental errors 
in  the values of the oscillation parameters
(figure \ref{Fig1}, right panel),
in $\meff$ and the sum of neutrino masses,
and the uncertainty in the relevant NME
\cite{Rodin:2003eb,Rodin:2005dp,Poves}, 
can weaken but do not invalidate these results 
\cite{Pascoli:2002ae,Pascoli:2003ke,Murayama:2003ci,Pascoli:2001by,Pascoli:2002qm,Pascoli:2005zb}.

As figure \ref{Fig1} indicates, a measurement
of $\meff \gtap 0.01$ eV would either:
i) determine a relatively narrow 
interval of possible values 
of the lightest $\nu$-mass \mmin; or 
ii) would establish an upper limit on $\mmin$. 
If an upper limit on $\meff$ is experimentally 
obtained below 0.01 eV,
this would lead to a significant upper limit
on $\mmin$.

The possibility of establishing  CP-
violation in the lepton sector 
due to Majorana CPV  
phases has been studied in
\cite{Bilenky:2001rz,Bilenky:2001xq,Barger:2002vy,deGouvea:2002gf} and
in much greater detail in
\cite{Pascoli:2001by,Pascoli:2002qm,Pascoli:2005zb}. 
It was found that it is very challenging:
it requires quite accurate measurements 
of $\meff$ (and of $m_0$ for QD spectrum), 
and holds only for a limited range of 
values of the relevant parameters.
More specifically \cite{Pascoli:2001by,Pascoli:2002qm,Pascoli:2005zb}, 
establishing at 2$\sigma$
CP-violation associated with
Majorana neutrinos 
in the case of QD spectrum requires, 
for $\sin^2\theta_{\odot}$=0.31 in particular, 
a relative experimental 
error on the measured value of 
$\meff$ and $m_0$ smaller than 15\%,
a ``theoretical uncertainty'' 
$F{\small \ltap}$1.5 in the value of
$\meff$ due to an imprecise 
knowledge of the corresponding NME, 
and value of the relevant Majorana
CPV phase $\alpha$ typically within the ranges 
of ${\small\sim (\pi/4 - 3\pi/4)}$ and
${\small\sim (5\pi/4 - 7\pi/4)}$ (figure \ref{Fig1}, right-hand panel). 

The knowledge of the NMEs with 
sufficiently small uncertainty
is crucial for obtaining quantitative information on
the neutrino-mixing parameters from a measurement of
$\betabeta$-decay half-life.
Possible tests of the NME calculations 
are discussed in \cite{Bilenky:2004um}.

%% file: 02-Standard-Neutrino-Model/02-02-Present-status/02-02-05-Update.tex
\subsubsection{Recent progress in measurements of neutrino oscillations}

Several results confirming the hypothesis of neutrino oscillations and
improving the precision with which the various parameters are known
have been reported since the ISS concluded.
Comprehensive reviews of these results can be found in, for
example, \cite{Dore:2008dp,Amsler:2008zzb}.
While it is not appropriate to attempt a complete review here, it is
of interest to note, briefly, the most important developments.

KamLAND has reported results based on a four-fold increased in
exposure and with an improved analysis leading to a significant
reduction in the systematic error \cite{Abe:2008ee}. 
The new data and analysis yield 
$\Delta m^2_{21} = 7.58^{+0.14}_{-0.13}{\rm (stat)}^{+0.15}_{-0.15}{\rm(syst)} \times 10^{-5}$~eV$^2$
and 
$\tan^2 \theta_{12} = 0.56^{+0.10}_{-0.07}{\rm(stat)}^{+0.10}_{-0.06}{\rm (syst)}$.

MINOS has provided an improved measurement of $\Delta m^2_{32}$ based
on two-years of running and $3.36 \times 10^{20}$ protons on
target \cite{Adamson:2007gu,Adamson:2008zt,Adamson:2008jh}.
The results confirm the neutrino-mixing hypothesis and yield
$\Delta m^2_{32} = 2.43 \pm 0.13 \times 10^{-3}$~eV$^2$.
The OPERA experiment, which took its first data in August 2006, has
been commissioned, recording $\sim 640$ neutrino events from an
exposure corresponding to $\sim 16 \times 10^{17}$ protons on target.
The experiment is now poised to begin the search for direct
evidence of $\nu_\mu \rightarrow \nu_\tau$.

MiniBOONE has carried out a detailed evaluation of backgrounds to the
$\nu_e$ appearance signal and published improved
results \cite{AguilarArevalo:2007it,Adamson:2008qj,AguilarArevalo:2008rc}
that indicate that a two-neutrino oscillation explanation of the data
from Bugey, KARMEN-2, LSND, and MiniBOONE is only possible at the
3.94\% level.
MiniBOONE is now taking data with an anti-neutrino beam.
The results of this phase of the experiment will be of interest since
the LSND experiment was also carried out in an $\bar{\nu}_\mu$ beam.

Progress has also been made in the development of the reactor-neutrino
programme and the preparation of the T2K and NO$\nu$A experiments.
The interested reader is referred to references 
\cite{AguilarArevalo:2007it,Adamson:2008qj,AguilarArevalo:2008rc}
for further information.

%% file: 02-Standard-Neutrino-Model/02-03-Completing-the-picture/02-03-Completing-the-picture.tex
\subsection{Completing the picture}
\label{sec:pic}

The measurements of the neutrino-oscillation parameters reviewed above
hint at new interactions present at an extremely large mass scale,
$\Lambda$.
In scattering experiments, for example at hadron or lepton colliders,
these new interactions are suppressed by powers of $\Lambda$.
In contrast, neutrino oscillations are widely believed to be a direct
consequence of the physics at the large mass scale; hence,
measurements of neutrino oscillations probe physics at a uniquely high
mass scale.
The measurements reviewed above have established the presence of
neutrino oscillations and have determined a number of relevant
parameters. 
To complete the picture, a dedicated experimental programme is
required; the elements of this experimental programme are
\cite{Blondel:2006su}):
\begin{itemize}
  \item The search for neutrinoless double-beta decay, to establish
        whether neutrinos are Majorana particles
        \cite{Vogel:2000vc,Elliott:2002xe};
  \item The determination of the neutrino-mass scale by direct
        measurement (see for example \cite{Osipowicz:2001sq}) or
        through cosmology (see for example
        \cite{Bilenky:2002aw,Dolgov:2002wy});
  \item The determination of the neutrino-mass hierarchy by combining
        neutrino-oscillation measurements with the results of direct
        neutrino-mass measurements and searches for $0\nu\beta\beta$
        decay;
  \item The determination of the small mixing angle $\theta_{13}$
        through measurements of the sub-dominant neutrino
        oscillations; 
  \item The precise determination of the mixing angle $\theta_{23}$ to
        seek to establish whether $\theta_{23}$ is maximal;
  \item The search for leptonic CP violation in neutrino oscillations; and
  \item The search for sterile light neutrinos through the observation
        of a third mass-squared difference in neutrino oscillations.
        The resent measurements from MiniBooNE \cite{Aguilar-Arevalo:2007it}
        dis-favour a sterile-neutrino interpretation of the LSND
        results \cite{Aguilar:2001ty}.
\end{itemize}

\subsubsection{Bounds on $\theta_{13}$ from approved experiments}

The present generation of long-baseline oscillation experiments
(K2K \cite{Ahn:2006zz} at KEK, MINOS \cite{Michael:2006rx} at the NuMI beam
and ICARUS \cite{Arneodo:2001sg} and OPERA \cite{Kodama:1999hg}  
at the CNGS beam, see table \ref{tab:beams1}),
are expected to measure $\sin^2 2 \theta_{23}$ and 
$|\Delta m^2_{31}|$ with a precision of $\sim 10\%$, if 
$|\Delta m^2_{31}| >  10^{-3}$ eV$^2$. 
These experiments could, in principle, measure $\theta_{13}$ through 
$\nu_{\mu} \rightarrow \nu_e$ oscillations even though they are not
optimized for such a measurement. 
MINOS is expected to reach a sensitivity of $\sin^2
\theta_{13} \le 0.02$ at a confidence level (CL) of 90\% in 5 years
\cite{Michael:2006rx}.
The main limitation of the MINOS experiment is the poor
electron-identification efficiency of the detector.  
Thanks to the high density and high granularity of the emulsion
cloud chamber (ECC) structure, the OPERA detector
is better suited for electron detection and can reach 
$\sin^2 \theta_{13} \le 0.015$ at 90\% CL 
(for $\Delta m^2_{31} = 2.5 \times 10^{-3}$ eV$^2$), 
after five years exposure to the CNGS beam at nominal intensity
\cite{Komatsu:2002sz,Migliozzi:2003pw}. 
\begin{table*}
 \begin{tabular}{lccccc}
 \hline
  Neutrino facility  &Proton momentum (GeV/$c$)&L (km)& $E_{\nu}$
 (GeV) & pot/yr ($10^{19}$)\\
 \hline
   KEK PS      \cite{Ahn:2006zz}          &   12    & 250    &   1.5      &  2             \\
   FNAL NuMI   \cite{Ables:1995wq}        &  120    & 735    &   3        &  20$\div$ 34   \\
   CERN CNGS   \cite{Acquistapace:1998rv} &  400    & 732    &   17.4     &  4.5$\div$ 7.6 \\
 \hline
 \end{tabular}
 \caption{
   Main parameters for present long-baseline neutrino beams
 }
 \label{tab:beams1}
\end{table*}

\begin{figure}
  \vskip 1.0cm
  \begin{center}
  \mbox{\epsfig{file=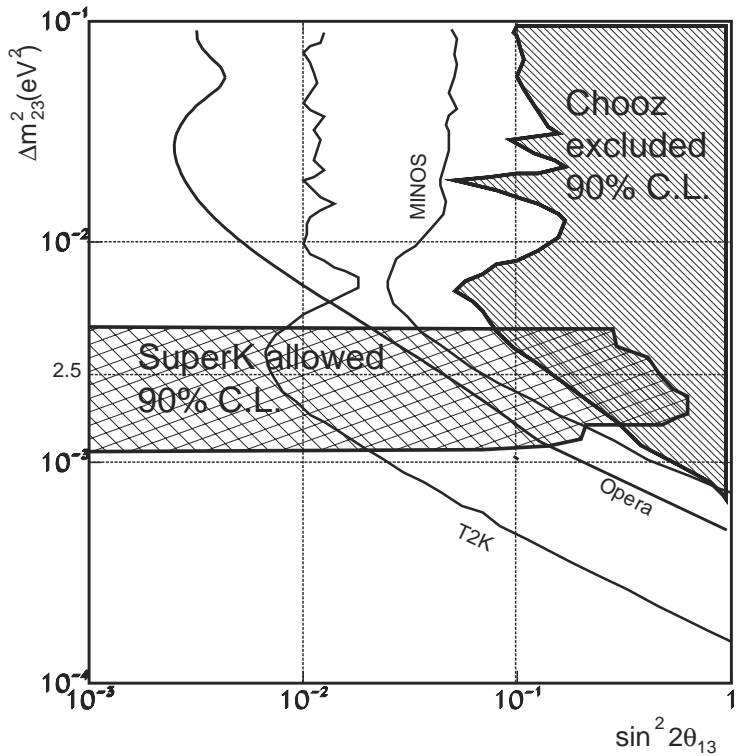, width=0.6\textwidth}}
  \end{center}
  \vskip -0.5cm
  \caption{ Expected $\theta_{13}$-sensitivity (in vacuum and for $\delta_{CP} = 0$)
                  for MINOS, OPERA and for the next T2K experiment, compared to the CHOOZ exclusion plot. 
                  Taken from reference~\protect\cite{Guglielmi:2005hi}.}
  \label{fig:exclusion}
 \end{figure}

The $\theta_{13}$-sensitivity of the present LBL experiments
(including the T2K, that will be discussed in more detail below) is
shown in figure~\ref{fig:exclusion}.  
The sensitivity of such experiments to $\theta_{13}$ is limited by the
power of the proton driver and by the $\nu_e$ contamination of the
beam.
In particular, the CNGS beam, which has been optimised for $\tau$
production, has a mean energy about ten times larger than the first
$\nu_\mu \to \nu_e$ oscillation peak at a baseline of 732~Km.

Another approach to search for non-vanishing $\theta_{13}$ (and the
$\theta_{23}$-octant \cite{Yasuda:2003qg}) is to look at \nubare\
disappearance using reactor neutrinos. 
The relevant oscillation probability is:
\begin{equation}
  \label{eq:pee}
   P(\bar\nu_e\rightarrow\bar\nu_e)\simeq 1 -
   \sin^22\theta_{13}\sin^2\left(\frac{\Delta m^2_{31}L}{4E}\right)+\ldots ,
\end{equation}
which does not depend on $\theta_{23}$ or $\delta_{CP}$.
At the baselines relevant for reactor-neutrino experiments, the
dependence of the oscillation probability on $\Delta m^2_{21}$ and
$\theta_{12}$ is negligible.
Therefore, this approach allows an unambiguous measurement of
$\theta_{13}$ free of correlations and degeneracies (see section
\ref{SubSect:SnuMDegenCorrel}), though it requires a very precise
knowledge of the absolute flux.  
The Double-Chooz experiment \cite{Ardellier:2004ui,Ardellier:2006mn}
will employ a near and far detector, located at baselines of 0.2~Km
and 1.05~Km respectively. 
Both detectors will be based on gadolinium-loaded liquid scintillator
with a fiducial mass of 10.16~tonne.
Antineutrinos will be detected using the delayed coincidence of the
positron from the inverse $\beta$-decay and the photons from
neutron capture.
The direct comparison of the event rates in the two detectors will
allow the cancellation of many of the systematic errors. 
After 5 years of data taking, this experiment will reach a
$\theta_{13}$-sensitivity of $\sin^2 \theta_{13}  \le 0.0025$ at
$90$\% CL. 
Another reactor experiment has been recently proposed in Japan
\cite{Aoki:2006bk}. 
This experiment has an expected sensitivity of 
$\sin^2 \theta_{13} \le 0.0038$ at $90$\% CL. 

Present LBL and reactor-neutrino experiments can not address the 
other issues raised above; the baselines are too short to take
advantage of matter effects required to identify the mass hierarchy,
and they are not designed to look for CP-violation.

%% file: 02-Standard-Neutrino-Model/02-04-Degeneracies-and-correlations/02-04-Degeneracies-and-correlations.tex
\subsection{Degeneracies and correlations}
\label{SubSect:SnuMDegenCorrel}

We will follow reference~\cite{Donini:2003vz} to introduce the degeneracy
problem. Other approaches have been proposed in
references
\cite{Minakata:2001qm,Minakata:2002qe,Minakata:2002qi,Minakata:2002jv,Leung:2003cb,Yasuda:2003qg}. 

\subsubsection{%
  Appearance channels: %
  $\boldsymbol{\nu_e} \boldsymbol{\to} \boldsymbol{\nu_\mu}, \boldsymbol{\nu_\tau}$ and %
  $\boldsymbol{\nu_\mu} \boldsymbol{\to} \boldsymbol{\nu_e}$ %
}
\label{sec:emu}

\begin{figure}[t!]
  \begin{center}
    \begin{tabular}{ll}
      \hskip -0.5cm
        \epsfig{file=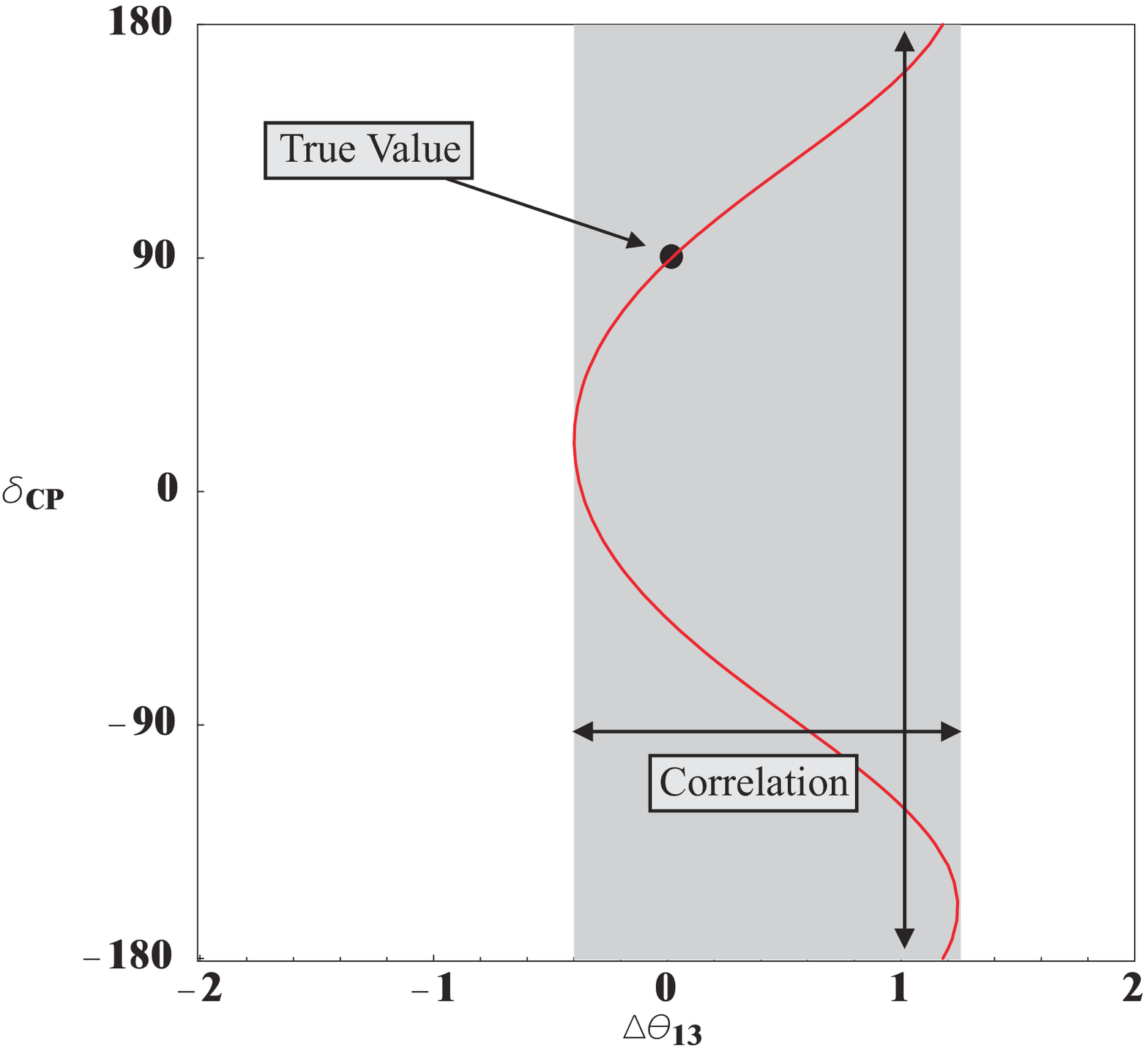, width=7cm} &
        \hskip -0.5cm
        \epsfig{file=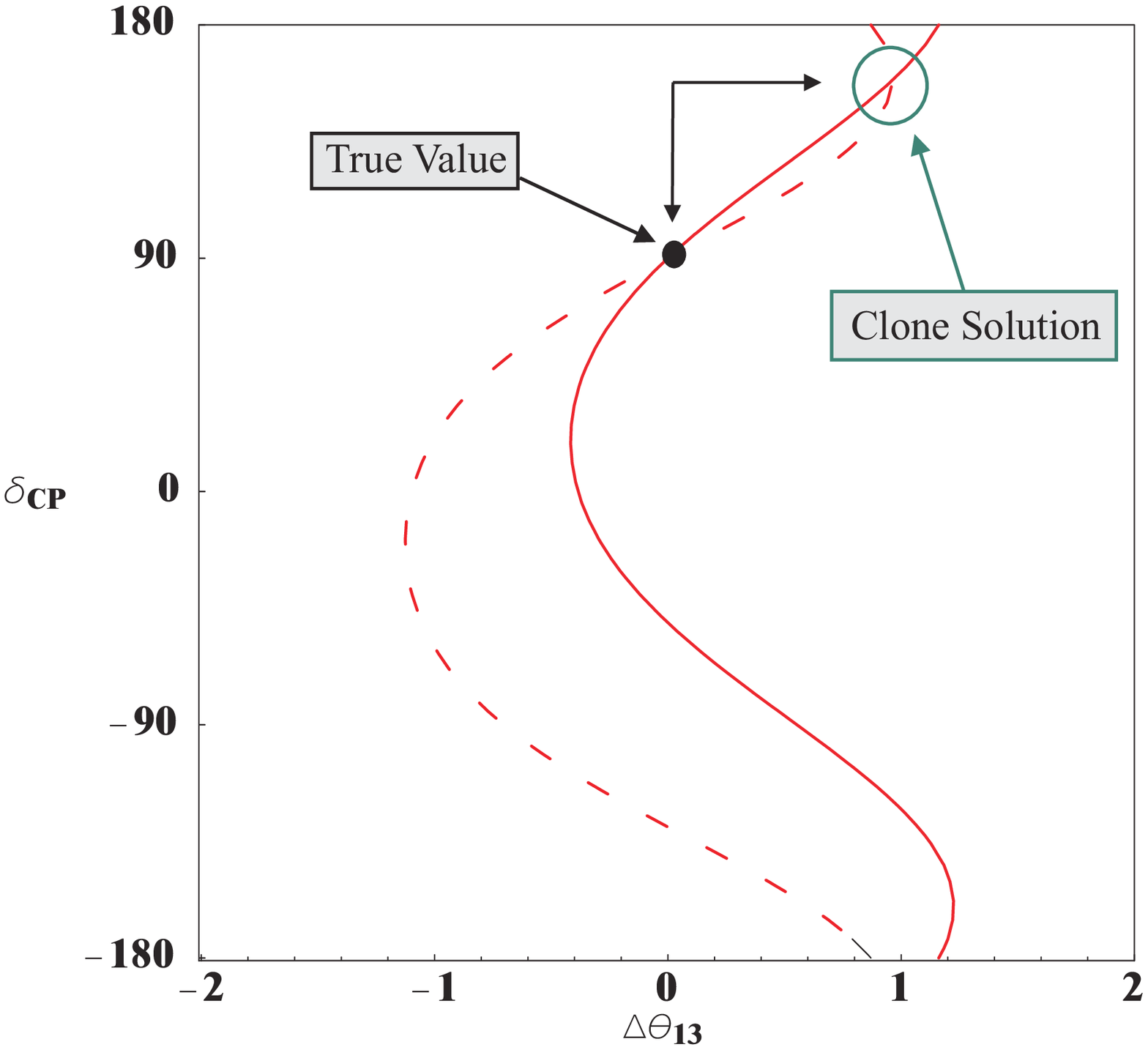, width=7cm} \\
    \end{tabular}
  \end{center}
  \caption{ Correlation of $\theta_{13}$ and $\delta$: 
    Left: if only neutrinos (or antineutrinos) are measured: continuum degeneracy;
    Right: if both neutrinos (full line) and antineutrinos (dashed line) are measured: twofold degeneracy.
    Taken with kind permission of Nuclear Physics B Proceedings Supplements
    from figures 1 and 2 in reference \cite{Rigolin:2005mr}.
    Copyrighted by Elsevier Science B.V.
  }
  \label{fig:deg0}
\end{figure}

It was originally pointed out in
reference~\cite{Burguet-Castell:2001ez} that a measurement of the
appearance probability 
$P( \nu_\alpha \to \nu_\beta ) = P_{\alpha \beta}$ for a
neutrino-oscillation experiment with a fixed baseline ($L$) and energy
($E$) can not be used to determine uniquely the oscillation parameters.
Indeed, taking ($\bar\theta_{13},\bar\delta$) as the `true' values, 
the equation 
\be
  \label{eq:equi0}
  P_{\alpha\beta} (\bar\theta_{13},\bar\delta) = P_{\alpha\beta}
  (\theta_{13},\delta)
\ee
has a continuous number of solutions. 
The locus of points in the ($\theta_{13},\delta$) plane satisfying
this equation is called an `equiprobability' curve. 
As can be seen from figure \ref{fig:deg0}(left), the strong
correlation between $\theta_{13}$ and $\delta$ \cite{Cervera:2000kp}
defines a strip in the ($\theta_{13},\delta$) plane compatible with
$P_{\alpha \beta} (\bar\theta_{13},\bar \delta)$.

Consider now an experiment that can measure both neutrino ($+$) and
antineutrino ($-$) appearance oscillation probabilities, 
at the same $L/E$.
The system of equations:
\be
  \label{eq:equi1}
  P^\pm_{\alpha \beta} (\bar \theta_{13},\bar \delta) = P^\pm_{\alpha \beta}
  (\theta_{13}, \delta)
\ee
describes two equiprobability curves, see figure \ref{fig:deg0}(right).
The system has two solutions: the input pair 
($\bar \theta_{13},\bar \delta$) and a second, ($L/E$)-dependent,
point.
The `continuum degeneracy' has been solved, but a discrete ambiguity in the measurement of the physical values of $\theta_{13}$ and $\delta$
is still present; the `intrinsic degeneracy' or `intrinsic clone'
\cite{Burguet-Castell:2001ez}. 

\begin{figure}
  \begin{center}
    \begin{tabular}{ll}
      \hskip -0.5cm
      \epsfig{file=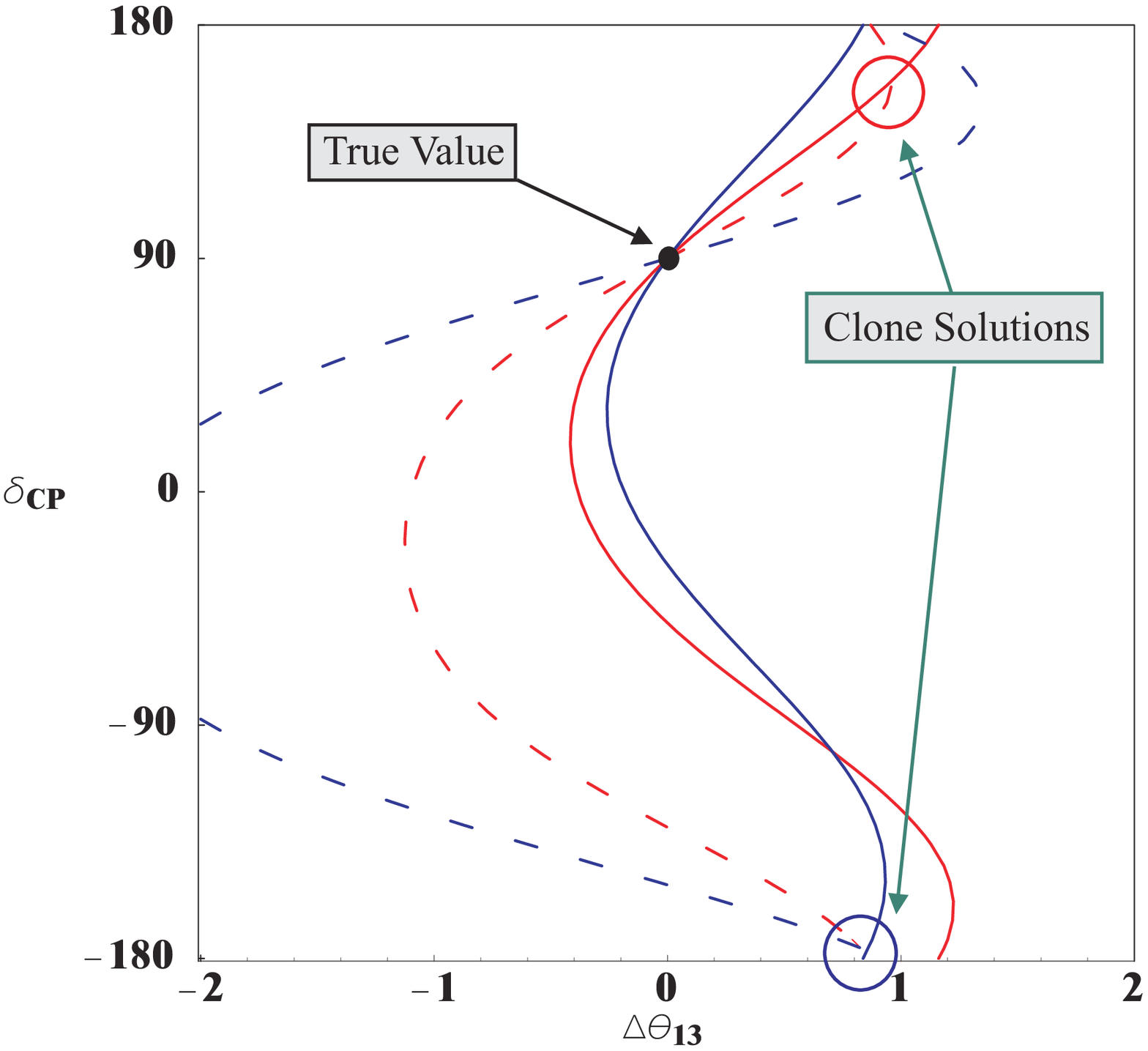, width=7cm} &
      \hskip -0.5cm
      \epsfig{file=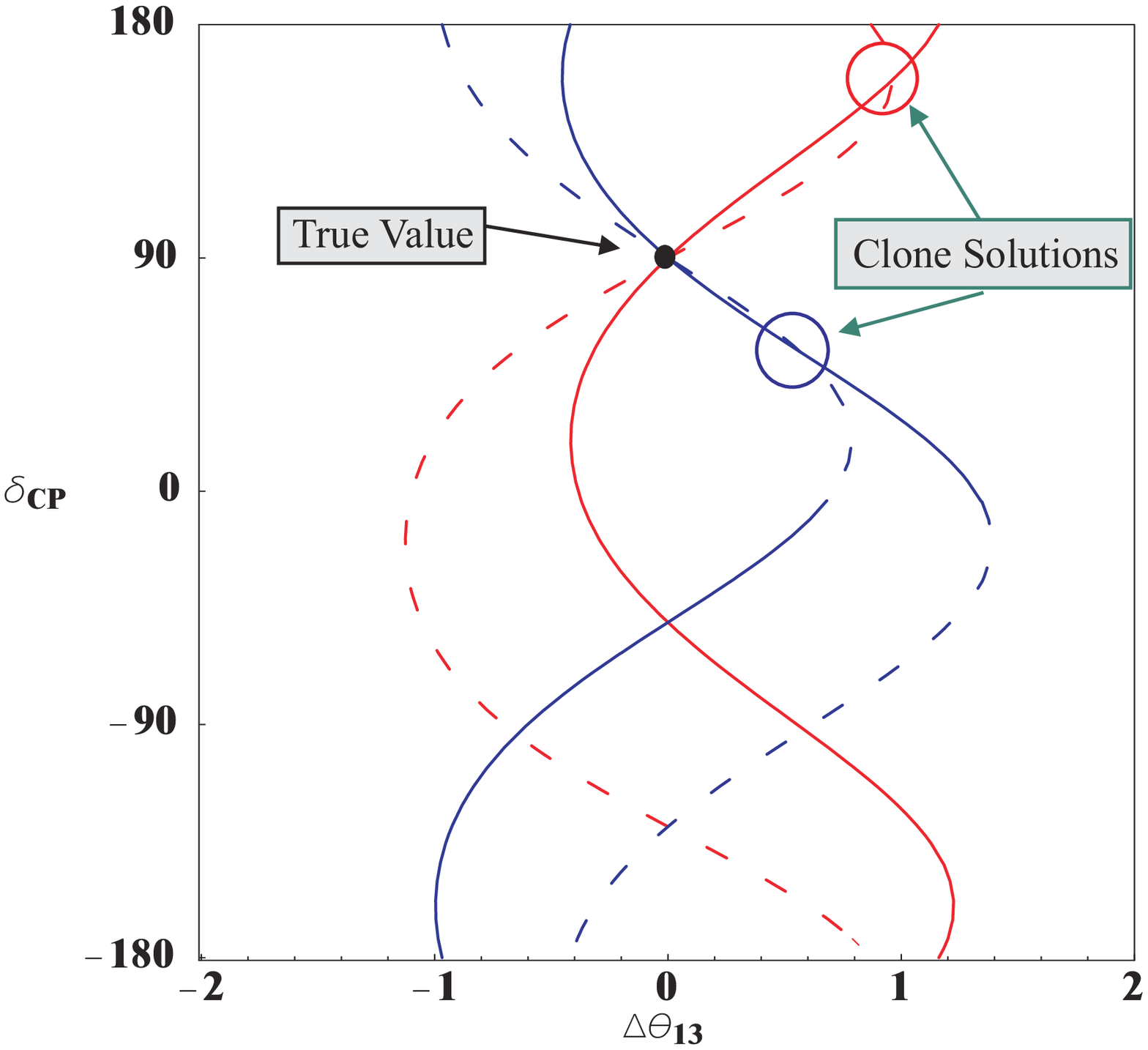, width=7cm} \\
    \end{tabular}
  \end{center}
  \caption{
    Solving the intrinsic degeneracy using: 
    Left: same oscillation channel, but two different baselines; 
    Right: same $L/E$, but two different oscillation channels
    (i.e. golden and silver).
    Taken with kind permission of Nuclear Physics B Proceedings Supplements
    from figures 3 and 4 in reference \cite{Rigolin:2005mr}.
    Copyrighted by Elsevier Science B.V.
  }
  \label{fig:deg2}
\end{figure}

More information is needed to solve the intrinsic degeneracy.
This information can be obtained either by making independent
measurements at different values of $L/E$ or by making use of
independent oscillation channels. 
The value of $L/E$ may be varied, for example, by measuring the
Neutrino Factory beam at a number of baselines
\cite{Burguet-Castell:2001ez} and \cite{Huber:2006wb}, by varying the
neutrino-beam energy at a beta-beam facility \cite{Donini:2006dx}, or
by measuring precisely the neutrino-energy 
spectrum in a liquid-argon detector \cite{Rubbia:2001pk}.
In figure \ref{fig:deg2}(left) it can be seen that experiments with
different baselines have intrinsic clones in different regions of the
($\theta_{13},\delta$) plane. 
If the clones are well separated, the degeneracy can be solved.  
The equiprobability curves for the two oscillation channels 
$\nu_e\raw \nu_\mu$ and $\nu_e \raw \nu_\tau$ measured at a particular
$L/E$ are shown in figure \ref{fig:deg2}(right).
The figure shows that the intrinsic clones for the two channels appear
in different regions of the parameter space, making it possible to
resolve the intrinsic degeneracy.

Two other sources of ambiguities are also present
\cite{Minakata:2001qm,Fogli:1996pv,Barger:2001yr}: 
\begin{itemize}
  \item Atmospheric-neutrino experiments measure $\nu_\mu$
        disappearance or $\nu_\mu \to \nu_\tau$ appearance for which 
        the leading terms in the expressions for the oscillation
        probabilities depend quadratically on $\Delta m^2_{13}$,
        therefore the sign of $\Delta m^2_{13}$ is not known
        \cite{Akhmedov:2004ny}; and
  \item At leading order, the oscillation probabilities for
        $\nu_\mu,\nu_e$ disappearance and $\nu_\mu \to \nu_\tau$
        appearance depend upon $\sin^2 2 \theta_{23}$.
        Therefore only the difference of $\theta_{23}$ from
        $45^\circ$ (maximal mixing) is known, i.e. it is not known
        whether $\theta_{23}$ is smaller or greater than $45^\circ$. 
\end{itemize}
As a consequence, future experiments must measure the two continuous  
variables $\theta_{13}$ and $\delta$ as well as the two discrete
variables: 
\be
  s_{atm} = {\rm sign} [ \Delta m^2_{23} ] \, , \qquad 
  s_{oct} = {\rm sign} [ \tan (2 \theta_{23}) ] \, . 
\ee
These two variables assume the values $\pm 1$ depending on the
sign of $\Delta m^2_{23}$ ($s_{atm}=1$ for $m_3^2>m_2^2$ and
$s_{atm}=-1$ for $m_3^2<m_2^2$) and $\theta_{23}$ ($s_{oct}=1$ for
$\theta_{23}<\pi/4$ and $s_{oct}=-1$ for $\theta_{23}>\pi/4$). 
Therefore, taking into account the present ignorance on the neutrino
masses and mixing matrix, equation (\ref{eq:equi1}) must be rewritten, more
precisely, as: 
\bea
  \label{eq:equi0int} 
  P^\pm_{\alpha \beta} (\bar \theta_{13},\bar \delta; \bar
  s_{atm},\bar s_{oct}) &=& 
  P^\pm_{\alpha \beta} (\theta_{13},\delta; s_{atm}=\bar s_{atm};
  s_{oct}=\bar s_{oct})\, , 
\eea
where $\bar s_{atm}$ and $\bar s_{oct}$ have been included as input
parameters in addition to $\bar \theta_{13}$ and $\bar \delta$. 
In equation (\ref{eq:equi0int}) we have implicitly assumed that 
the sign of $\Delta m^2_{23}$ and the octant for $\theta_{23}$
are unknown.
The following systems of equations should be considered:
\bea
  \label{eq:equi0sign} 
  P^\pm_{\alpha \beta} (\bar \theta_{13}, \bar \delta; \bar s_{atm},
    \bar s_{oct}) &=& 
  P^\pm_{\alpha \beta} (\theta_{13}, \delta; s_{atm} = - \bar s_{atm};
    s_{oct} = \bar s_{oct}) \\  
  \label{eq:equi0t23} 
  &=&
  P^\pm_{\alpha \beta} (\theta_{13}, \delta; s_{atm} =  \bar s_{atm};
    s_{oct} = -\bar s_{oct} ) \\ 
  \label{eq:equi0t23sign} 
  &=&
  P^\pm_{\alpha \beta} (\theta_{13}, \delta; s_{atm} = -  \bar
  s_{atm}; s_{oct} = - \bar s_{oct} ) \, . 
\eea
These new sets of equiprobability systems arise when we equate the
measured probability (l.h.s.) with the theoretical probabilities
obtained including one of the three possible wrong guesses of
$s_{atm}$ and $s_{oct}$ (r.h.s.).

Solving the four systems of equations
(\ref{eq:equi0int})--(\ref{eq:equi0t23sign}) will yield the true solution
plus additional `clones', forming an  eightfold-degeneracy
\cite{Barger:2001yr}. 
These eight solutions are respectively:  
\begin{itemize}
  \item The true solution and its intrinsic clone, obtained
        solving the system  in equation (\ref{eq:equi0int});
  \item The $\Delta m^2_{23}$-sign clones (hereafter called `sign'
        clones) of the true and intrinsic solution, obtained solving
        the system in equation (\ref{eq:equi0sign}); 
  \item The $\theta_{23}$-octant clones (hereafter called `octant'
        clones) of the true and intrinsic solution, obtained solving
        the system in equation (\ref{eq:equi0t23}); and
  \item The $\Delta m^2_{atm}$-sign $\theta_{23}$-octant clones
        (hereafter called `mixed' clones) of the true and intrinsic
        solution, obtained solving the system in equation
        (\ref{eq:equi0t23sign}). 
\end{itemize}
Notice, however, that transition probabilities are not the
experimentally measured quantities.  
Experimental results are given in terms of the number of charged
leptons observed in a specific detector. 
For the Neutrino Factory `golden channel' ($\nu_e \to \nu_\mu$), for
example, one counts the number of muons with charge opposite to the
charge of the muons circulating in the storage ring. 
If the detector can measure the final state lepton and hadron energies
with enough precision, events can be grouped in energy bins of width
$\Delta E$.  
The number of muons in the $i^{\rm th}$ energy bin for the input pair ($\bar
\theta_{13},\bar\delta$), for a parent muon energy $\bar E_\mu$, is
given by: 
\bea
  N^i_{\mu^\mp} (\bar \theta_{13}, \bar \delta)
  &=& \left \{ \frac{d \sigma_{\nu_\mu (\bar \nu_\mu)} (E_\mu, E_\nu)
  }{d E_\mu} 
                           \, \otimes \,
                 P^\pm_{e\mu} (E_\nu, \bar \theta_{13}, \bar \delta)           
                           \, \otimes \,
                 \frac{d \Phi_{\nu_e (\bar \nu_e) } (E_\nu, \bar E_\mu)}{d E_\nu} 
                      \right \}_{E_i}^{E_i + \Delta E_\mu} 
  \label{eq:convogolden}
\eea
where $\otimes$ stands for a convolution integral, $N^i$ is the number
of events in bin $i$, $E_\nu$ is the neutrino energy, $E_\mu$ is the
scattered muon energy, $\sigma_{\nu_mu (\bar{\nu}_\mu)}$, is the
neutrino charged-current scattering cross section, and $\Phi$ is the
neutrino flux.
Solving the following systems of equations, for a given energy bin and
fixed input parameters ($\bar\theta_{13},\bar\delta$): 
\bea
\label{eq:ene0}
N^i_{\mu^\pm}(\bar \theta_{13}, \bar \delta; \bar s_{atm}, \bar s_{oct}) &=& 
N^i_{\mu^\pm} ( \theta_{13},  \delta; s_{atm} = \bar s_{atm}, s_{oct} = 
\bar s_{oct})\\ 
\label{eq:ene0t23}
&=& 
N^i_{\mu^\pm} ( \theta_{13},  \delta; s_{atm} = \bar s_{atm}, s_{oct} = -\bar
s_{oct})\\ 
\label{eq:ene0sign}
&=&
N^i_{\mu^\pm} ( \theta_{13}, \delta; s_{atm} = -\bar s_{atm}, s_{oct} = \bar
s_{oct}) \\  
\label{eq:ene0t23sign}
&=& 
N^i_{\mu^\pm} ( \theta_{13},  \delta; s_{atm} = -\bar s_{atm}, s_{oct}
= -\bar s_{oct}) \, ,
\eea
yields the eight solutions corresponding to the $i$-th bin. 

The existence of unsolved degeneracies results in a loss of
sensitivity to the unknowns $\theta_{13},\delta,s_{atm}$ (see below). 
The best way to solve the degeneracies is to perform a set of
complementary measurements; experiments must have different baselines,
good energy resolution, and access to different channels. 
There is no `synergy' in experiments at the same $L/E$ measuring
the same channel \cite{Donini:2004hu}. 
A method to look for optimal combinations of measurements based on
solving the set of systems of equations
(\ref{eq:ene0})--(\ref{eq:ene0t23sign}) has been presented in
reference~\cite{Donini:2003vz}. 
Most of the previous considerations also apply to the T-conjugated
transition $\nu_\mu\to\nu_e$ and to $\nu_e\to\nu_\tau$ (the Neutrino
Factory `silver channel').  

\subsubsection{Disappearance channels: %
  $\boldsymbol{\nu_\mu} \boldsymbol{\to} \boldsymbol{\nu_\mu}$ %
}
\label{sec:mumu}

An independent measurement of the atmospheric parameters
$\theta_{23}$ and $\Delta m^2_{23}$ can be made via the 
$\nu_\mu$-disappearance channel using a conventional neutrino beam or
the Neutrino Factory. 
It is expected that this kind of measurement will reduce the error on
the atmospheric-mass difference to less than 10\% with a few years of
data if $\Delta m^2_{23} \geq 2.2 \times 10^{-3}$ eV$^2$
\cite{Petyt:1998zd}. 
The expected error on the atmospheric angle depends on the value of
$\theta_{23}$ itself, the smallest error being achieved for large, but
non-maximal, mixing \cite{Minakata:2004pg}.
It is interesting to study in detail the parameter correlations and
degeneracies that affect this measurement and that can induce large
uncertainties. 
The vacuum-oscillation probability expanded to the second order  
in the small parameters $\tc$ and $(\Delta_{12}L/E)$
\cite{Akhmedov:2004ny} is: 
\bea
  P(\nu_\mu \to \nu_\mu) & = & 1-  \left [ \sin^2 2 \theta_{23}
    -s^2_{23} \sin^2 2 \theta_{13} \cos 
    2\theta_{23} \right ]\, \sin^2\left(\frac{\Delta_{23} L}{2}\right)
  \cr 
  &   & - \left(\frac{\Delta_{12} L}{2}\right) [s^2_{12} \sin^2 2
  \theta_{23} + \tilde{J}  
  s^2_{23} \cos \delta] \, \sin(\Delta_{23} L) \cr
  &   & - \left(\frac{\Delta_{12} L}{2}\right)^2 [c^4_{23} \sin^2
  2\theta_{12}+ 
  s^2_{12} \sin^2 2\theta_{23} \cos(\Delta_{23} L)] \, ,
  \label{eq:probdismu}
\eea
where $\tilde{J}=\cos \theta_{13} \sin 2\theta_{12}\sin
2\theta_{13}\sin 2\theta_{23}$ and $\Delta_{23}=\Delta m^2_{23}/2 E$,
$\Delta_{12}=\Delta m^2_{12}/2 E$. 
The dominant contribution comes from first term in the first
parenthesis which is symmetric under $\tatm \to \pi/2-\tatm$.
This symmetry is lifted by the other terms which introduce a mild
CP-conserving $\delta$-dependence, albeit through sub-leading
effects which are very difficult to isolate. 

Since the $s_{atm} = {\rm sign}(\Delta m^2_{23})$ is unknown,
two systems of equations must be solved: 
\be
  \label{eq:enedismu}
  N^{\pm}_{\mu\mu}( \bar \theta_{23}, \Delta m^2_{atm}; \bar s_{atm})= 
  N^{\pm}_{\mu\mu} ( \theta_{23}, |\Delta m^2_{23}|; \bar s_{atm}) \, ;
\ee
and
\be
\label{eq:enedismusign}
N^{\pm}_{\mu\mu}( \bar \theta_{23}, \Delta m^2_{atm}; \bar s_{atm})=
N^{\pm}_{\mu\mu} ( \theta_{23}, |\Delta m^2_{23}|; -\bar s_{atm}) \, , 
\ee
where $\bar s_{atm}$ is the physical mass hierarchy.
For non-maximal $\bar \theta_{23}$, four different solutions are
obtained.
For $| \Delta m^2_{23}| \sim \Delta m^2_{atm}$ equation
(\ref{eq:enedismu}) yields two solutions, the input value
$\theta_{23} = \bar \theta_{23}$ and 
$\theta_{23} \simeq \pi/2 - \bar \theta_{23}$.
The second solution is not exactly 
$\theta_{23} = \pi/2 - \bar \theta_{23}$ due to the small
$\theta_{23}$-octant asymmetry;  
Two more solutions from equation (\ref{eq:enedismusign}) at a
different value of $|\Delta m^2_{23}|$  are also present
\cite{Donini:2004iv}. 
In equation (\ref{eq:probdismu}) we can see that changing the sign of
$\Delta m^2_{23}$ makes the second term positive; a change that must
be compensated with an increase in $|\Delta m^2_{23}|$ to give  
$P^\pm_{\mu\mu}(\Delta m^2_{atm}; \bar s_{atm}) =
P^\pm_{\mu\mu}(|\Delta m^2_{23}|; - \bar s_{atm})$.

\begin{figure}[t!]
  \begin{center}
    \begin{tabular}{cc}
      \hspace{-1.0cm} \epsfxsize8.25cm\epsffile{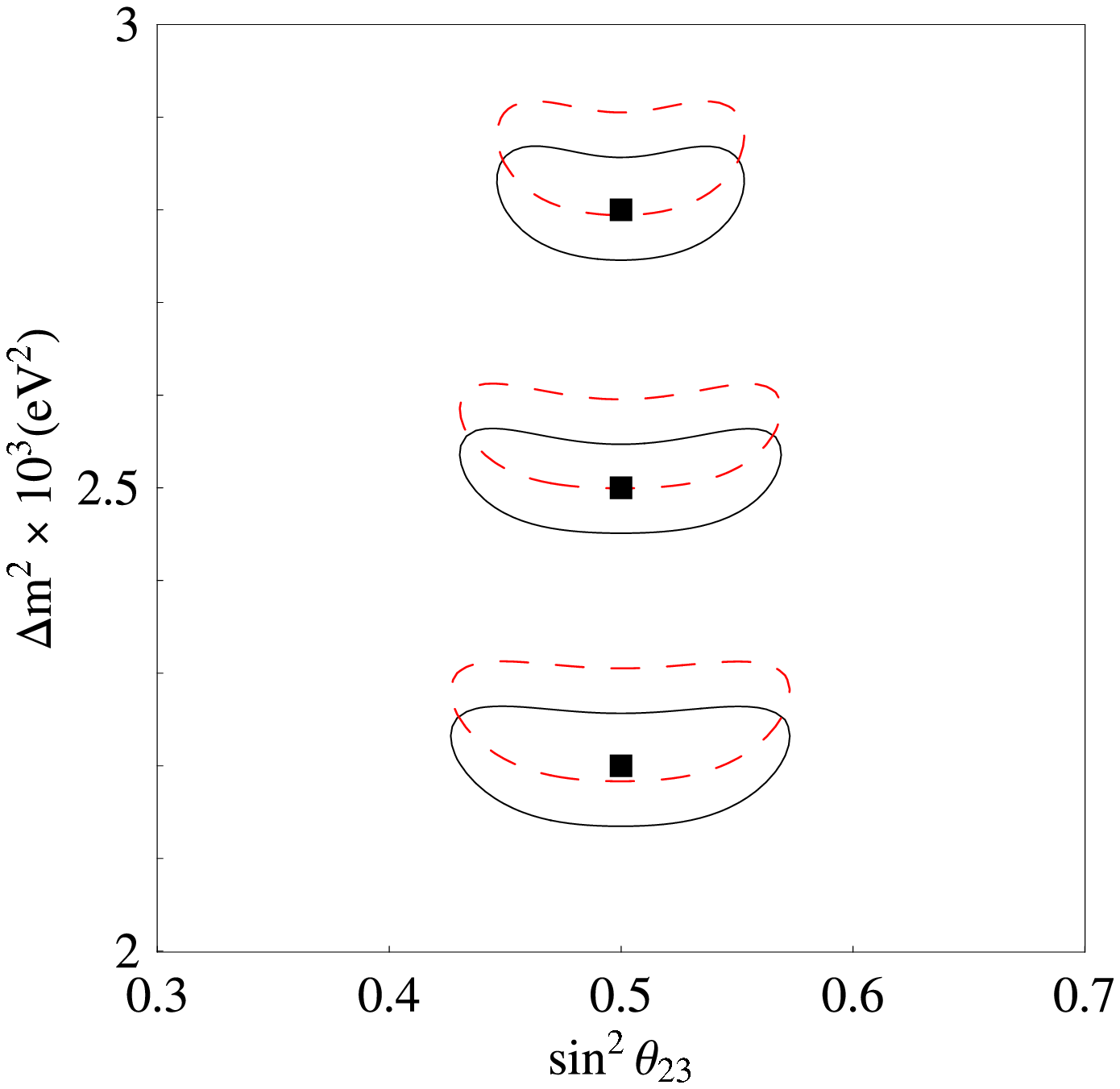} &
      \hspace{-0.5cm} \epsfxsize8.25cm\epsffile{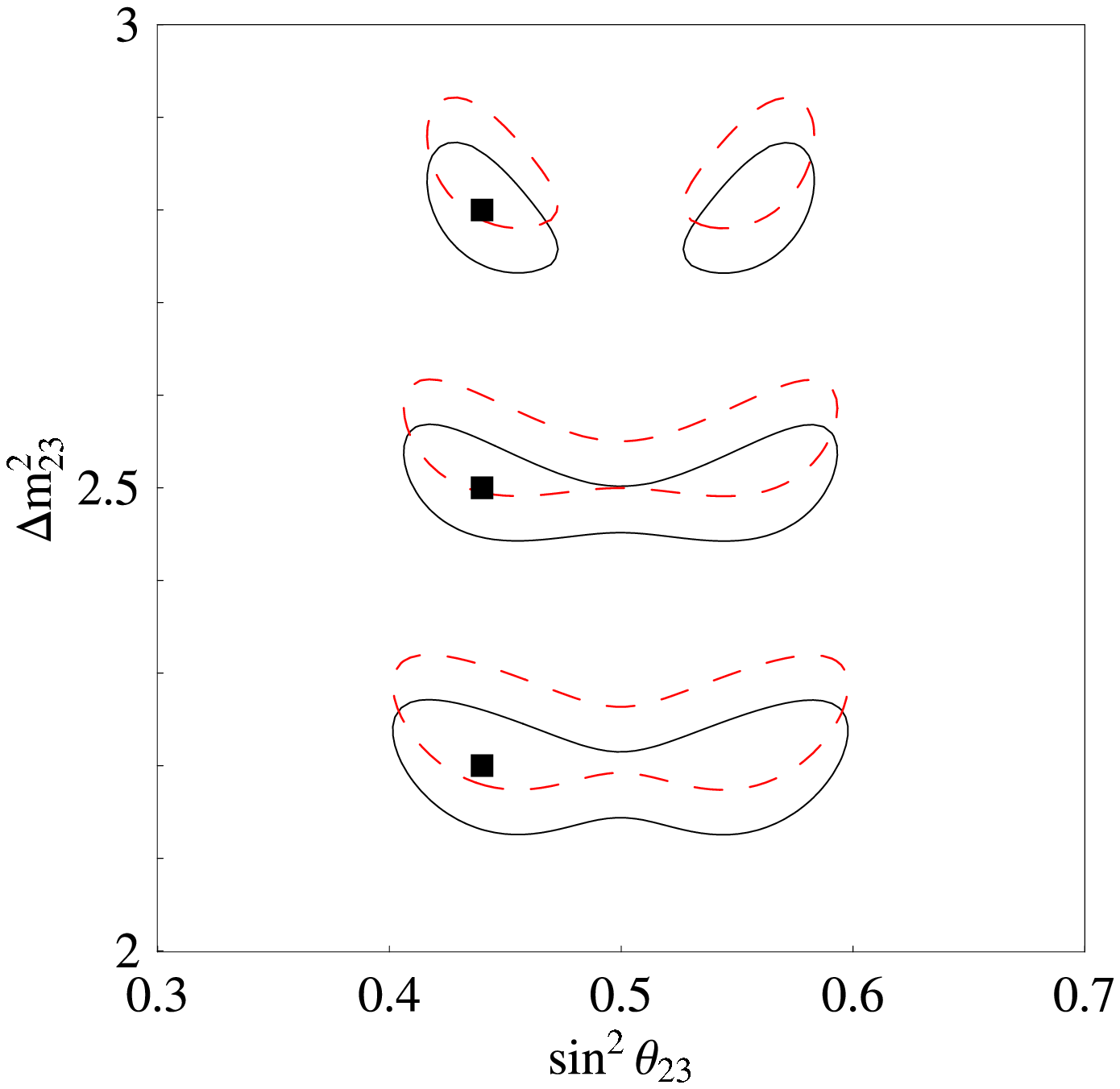} \\
    \end{tabular}
  \end{center}
  \caption{
    The sign degeneracy at T2K-I; left: $\theta_{23} = 45^\circ$;
    right: $\theta_{23} = 41.5^\circ$.
    Taken with kind permission of Nuclear Physics B from figure 4
    in reference \cite{Donini:2005db}.
    Copyrighted by Elsevier Science B.V.
  }  
  \label{fig:degen}
\end{figure}

The result of a fit to the disappearance-channel data at the T2K phase
I experiment is shown in figure \ref{fig:degen} for three different
values of the atmospheric mass difference
$\Delta m^2_{23} = (2.2, 2.5, 2.8) \times 10^{-3}$~eV$^2$.
Fixed values of the solar parameters have been used, $\Delta m^2_{12}
= 8.2 \times 10^{-5}$ eV$^2$; $\theta_{12} = 33^\circ$. 
For maximal mixing, $\theta_{23}=45^\circ$, figure~\ref{fig:degen}
(left), two solutions are found at 90 \% CL when both choices of
$s_{atm}$ are considered. 
On the other hand, using a non-maximal atmospheric angle 
$\theta_{23} = 41.5^\circ$ ($\sin^2 \theta_{23} = 0.44$) four
degenerate solutions are found, figure~\ref{fig:degen}(right). 
In general, a two-fold or four-fold degeneracy must be discussed 
in the disappearance channel. 

Notice how the disappearance sign clones appear at a value of
$|\Delta m^2_{23}|$ higher than the input value.
This is expected from equation (\ref{eq:probdismu}); the shift in the
vertical axis is a function of $\theta_{13}$ and $\delta$ which, in this
case, has been kept fixed at $\theta_{13} = 0^\circ = \delta$. 
The degeneracy can be softened or solved by using detectors at
baselines long enough that matter effects can be exploited
\cite{Donini:2005db}.

\subsubsection{A matter of conventions}
\label{sec:sb:conve}

It is useful to open here a short parenthesis to address a problem
that arose recently concerning the `physical' meaning of the variables
used to fit the `atmospheric' mass difference, $\Delta m^2_{atm}$.
Notice, first of all, that the experimentally measured solar-mass
difference $\Delta m^2_{sol}$ can be unambiguously identified with the
three-family parameter $\Delta m^2_{12} = m^2_2 - m^2_1$.
This is not true for the experimentally measured atmospheric mass
difference $\Delta m^2_{atm}$.  
Since the sub-leading solar effects are, at present, barely seen in
atmospheric neutrino experiments we can define the three-family
parameter to be used in the fits in a number of ways: by using 
$\Delta m^2_{23} = m^2_3 - m^2_2$; $\Delta m^2_{13} = m^2_3 -m^2_1$;   
or $\Delta m^2 = \left ( \Delta m^2_{23} + \Delta m^2_{13} \right )/2$ 
\cite{Fogli:2001wi}.
A good description of the data will be obtained with either choice.
When measurements of the atmospheric mass-squared difference with a
precision at the level of $10^{-4}$ eV$^2$ are available,
however, the different choices of the fitting parameter will give 
different results.  

\begin{figure}[t!]
  \begin{center}
    \begin{tabular}{ccc}
      \hspace{-1.0cm} \epsfxsize5cm\epsffile{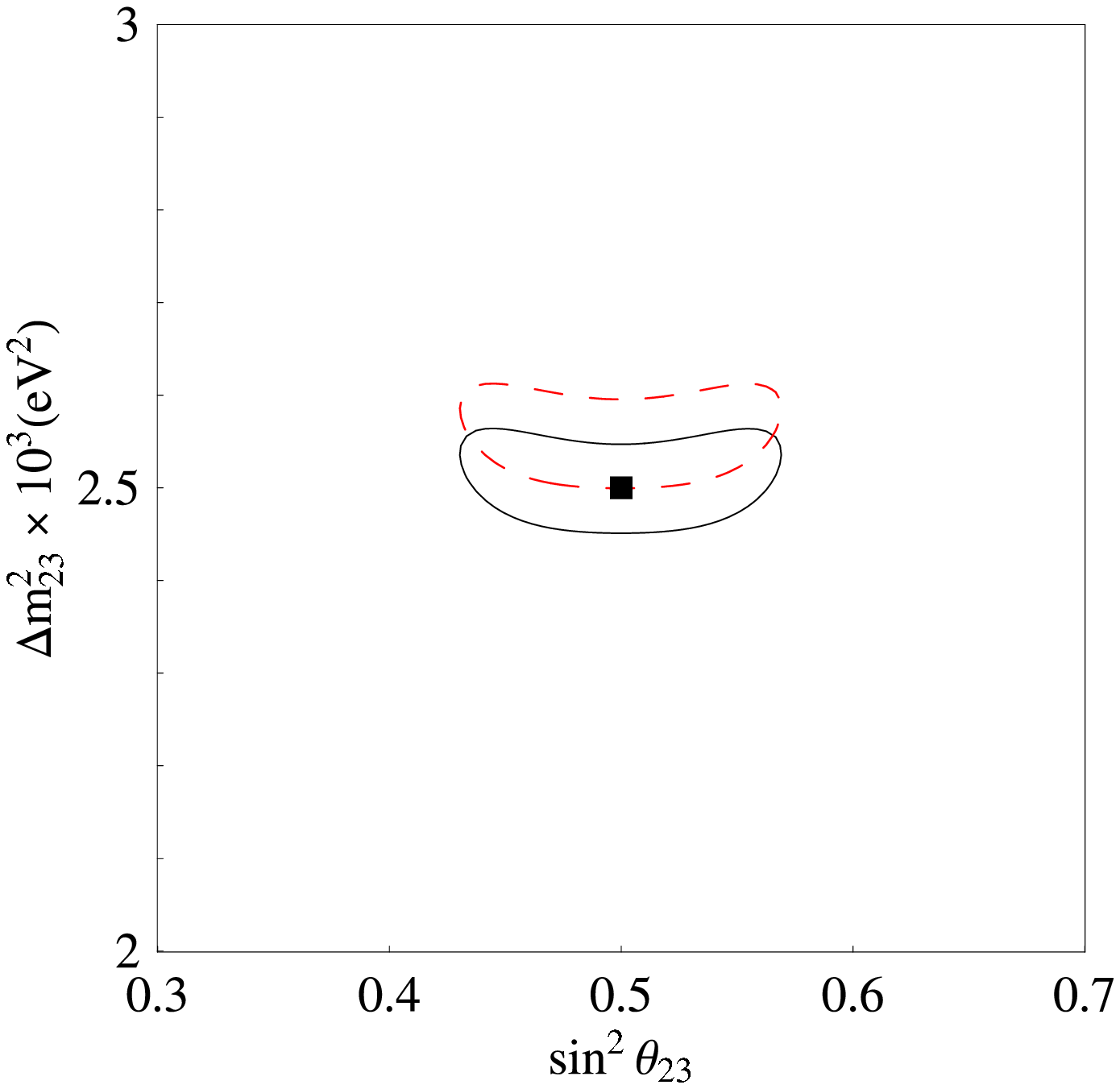} &
      \hspace{-0.5cm} \epsfxsize5cm\epsffile{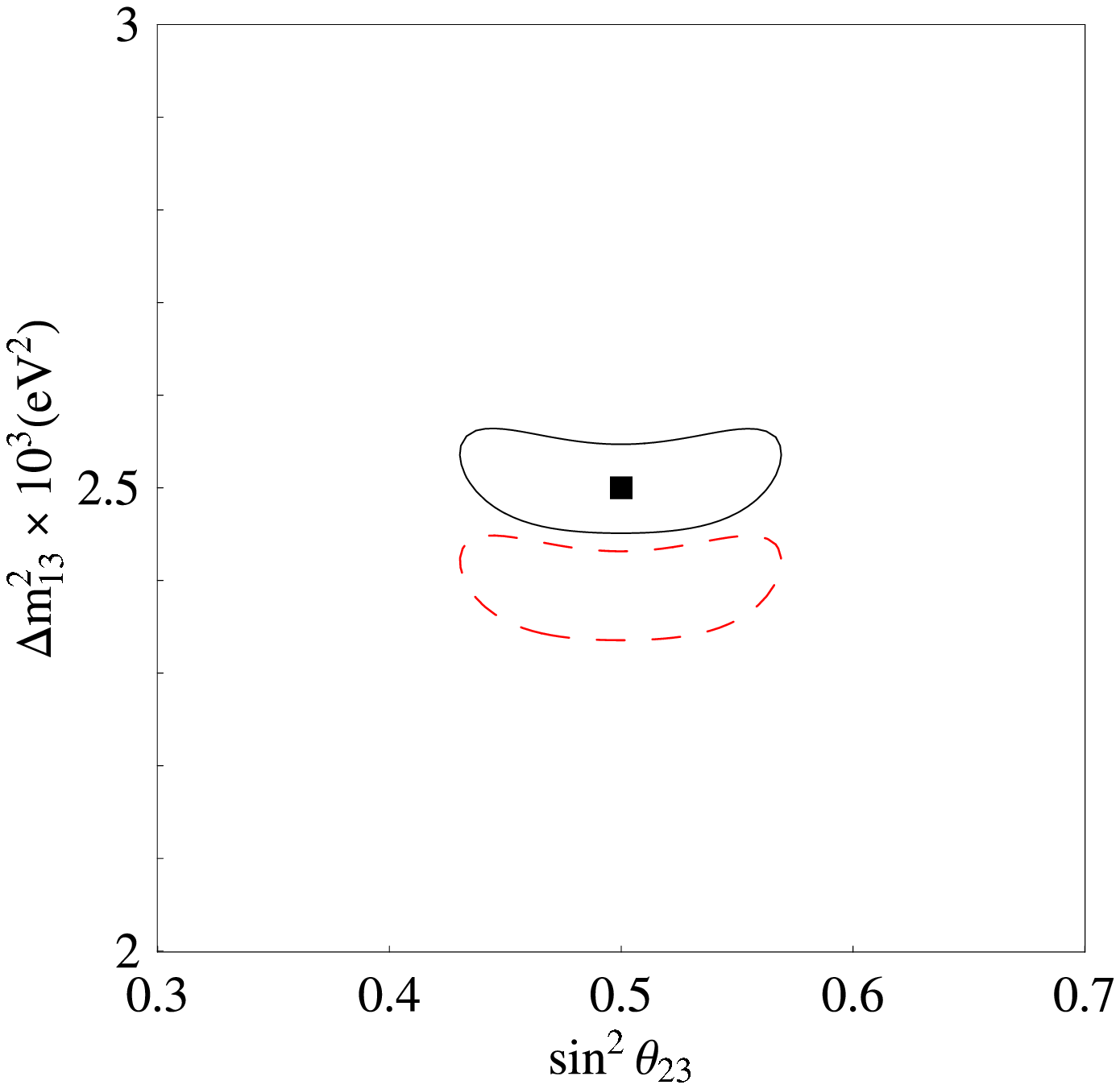} &
      \hspace{-0.5cm} \epsfxsize5cm\epsffile{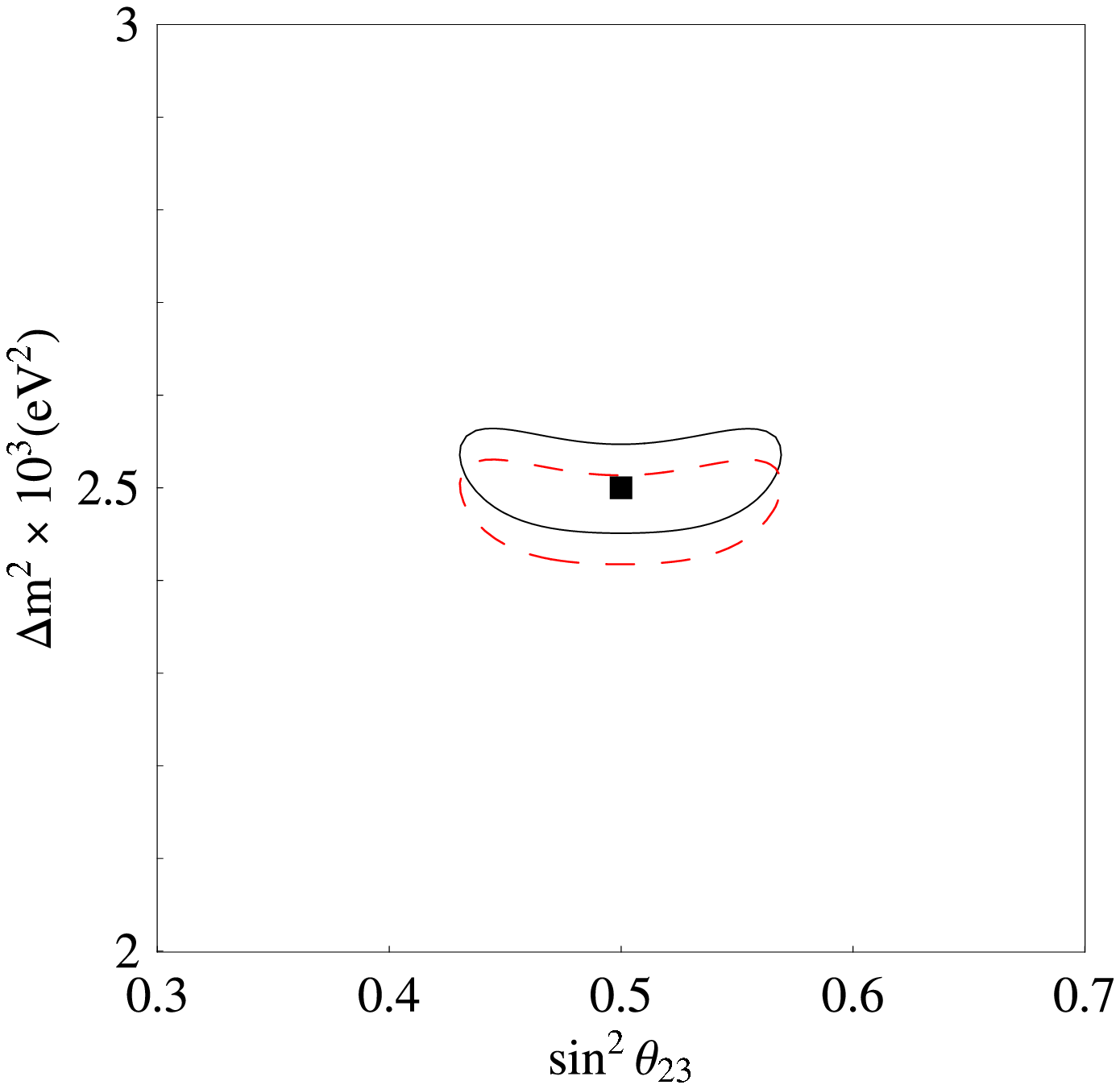} \\
    \end{tabular}
  \end{center}
  \caption{
    Different choices of the three-family ``atmospheric'' mass
    difference; left: $\Delta m^2_{23}$; middle: $\Delta m^2_{13}$;
    right: $\Delta m^2$.
    Taken with kind permission of Nuclear Physics B from figure 5
    in reference \cite{Donini:2005db}.
    Copyrighted by Elsevier Science B.V.
  }  
  \label{fig:conve}
\end{figure}

This effect can be observed in figure~\ref{fig:conve}, where the three
choices introduced above are compared. 
The three panels show the 90\% CL contours resulting from a fit
to the experimental data corresponding to the input value, $\Delta
m^2_{atm} = 2.5 \times 10^{-3}$, in normal hierarchy, but fitted using
in turn $\Delta m^2_{23}$ (left panel), $\Delta m^2_{13}$ (middle
panel) and $\Delta m^2$ (right panel). 
It can be seen that the contour corresponding to the normal hierarchy,
$s_{atm} = \bar s_{atm}$, is always located around the input value.
On the other hand, the contour obtained for the inverted hierarchy is
located above, below, or on top of the input value depending on the
choice of fitting variable. 
This is a consequence of the fact that the difference between 
each of the possible choices is $O (\Delta m^2_{12})$.

For three-family mixing, three `frequencies' can be defined, the
shortest being the solar-oscillation frequency (unambiguously related
to the mass difference $\Delta m^2_{12}$). 
In the case of the  normal hierarchy, the middle frequency is related
to $\Delta m^2_{23}$ and the longest one to $\Delta m^2_{13}$. 
In the case of the inverted hierarchy these two frequencies are
interchanged and the middle frequency will be related to $\Delta
m^2_{31}$ and not to $\Delta m^2_{32}$. 
For this reason, it has been suggested that the analysis of the normal
and inverted hierarchies should be presented using variables which 
maintain the ordering of the oscillation frequencies. 

\subsubsection{Disappearance channels: %
  $\boldsymbol{\nu_e} \boldsymbol{\to} \boldsymbol{\nu_e}$ %
}
\label{sec:ee}

A beta-beam or Neutrino Factory can exploit the $\nu_e$-disappearance
channel to measure the solar parameters $\Delta m^2_{12},\theta_{12}$
or $\theta_{13}$ in a degeneracy-free environment.
The $\nu_e$-disappearance probability does not depend on $\delta$ or
on $\theta_{23}$. 
The $\theta_{13}$ measurement is, therefore, not affected by
$(\theta_{13}-\delta)$ correlations or the $s_{oct}$ ambiguity. 
The $\nu_e \to \nu_e$ matter-oscillation probability, expanded at
second order in the small parameters $\theta_{13}$ and 
$(\Delta m^2_{12}L/E)$, is \cite{Donini:2005rn}: 
\be
  P^\pm_{ee} =1-\left(\frac{\Delta_{23}}{B_\mp}\right)^2
    \sin^2(2\theta_{13}) \, \sin^2\left(\frac{B_\mp\,L}{2}\right) 
    - \left(\frac{\Delta_{12}}{A}\right)^2 \sin^2(2\theta_{12})
    \,\sin^2\left(\frac{A\,L}{2}\right) \, ,
  \label{eq:disnue}
\ee
where $\Delta_{23}=\Delta m^2_{23}/2 E$, 
$\Delta_{12}=\Delta m^2_{12}/2 E$, $A = \sqrt{2} G_F N_e$, and
$B_\mp=|A\mp\Delta_{23}|$.
This equation describes reasonably well the behaviour of the transition
probability in the energy range covered by the beta-beam facilities
presently considered.
Two sources of ambiguities are still present in $\nu_e$-disappearance
measurements, $s_{atm}$ (for large values of $\theta_{13}$, i.e. in
the `atmospheric' region) and the $\theta_{13}-\theta_{12}$
correlation (for small values of $\theta_{13}$, i.e. in the `solar'
region). 
A beta-beam could in principle improve the precision with which the
solar parameters are known through $\nu_e$ disappearance
measurements.
This is not the case for a beta-beam facility in which the neutrino
energy of $\sim 100$~MeV is matched to a baseline of $\sim 100$~km.
For such a facility, at large $\theta_{13}$, the second term in 
equation (\ref{eq:disnue}) dominates over the last term.
On the other hand, for small $\theta_{13}$ the statistics is too low
to improve upon the present uncertainties on $\theta_{12}$ and 
$\Delta m^2_{12}$ (note that the energy and baseline of the
low-$\gamma$ beta-beam has not been chosen to perform this task).
It has been shown that if systematic errors cannot be controlled to
better than at 5\%, the beta-beam disappearance channel does not
improve the CHOOZ bound on $\theta_{13}$ \cite{Donini:2004iv}.

Equation (\ref{eq:disnue}) can also be applied to reactor-neutrino
experiments which aim at a precise measurement of $\theta_{13}$ in a 
`degeneracy-free' regime. 
For the typical baseline and energy of a reactor experiment (e.g., 
$L = 1.05$ km and $\langle E_\nu \rangle = 4$ MeV for the Double-Chooz
proposal \cite{Ardellier:2004ui,Ardellier:2006mn}) we can safely
consider antineutrino propagation in vacuum. 
As a consequence, no sensitivity to $s_{atm}$ is expected at these
experiments, since $B_\mp \to \Delta_{23}$ for $\Delta_{23} \gg A$. 
It is very difficult for reactor experiments to test small values of
$\theta_{13}$, and thus the $\theta_{13}-\theta_{12}$ correlation
(significant only in the ``solar'' region) can also be neglected.

%% file: 03-New-physics-and-the-SnuM/03-New-physics-and-the-SnuM.tex
\section{Implications for new physics and cosmology}
\label{Sect:NPandtheSNuM}

\input 03-New-physics-and-the-SnuM/03-00-Introduction/03-00-Introduction

\input 03-New-physics-and-the-SnuM/03-01-Origin-of-small-mass/03-01-Origin-of-small-mass

\input 03-New-physics-and-the-SnuM/03-02-Uni-and-flav/03-02-Uni-and-flav

\input 03-New-physics-and-the-SnuM/03-03-Cosmology/03-03-Cosmology

%% file: 03-New-physics-and-the-SnuM/03-00-Introduction/03-00-Introduction.tex
Neutrino mass is the first example of physics beyond the Standard
Model. 
The extreme smallness of neutrino masses, compared to
charged fermion masses, and the large mixing angles, are both
mysteries that make more acute the flavour problem in the Standard
Model: why are there three families of quarks and leptons with the
masses and mixings that are observed? 
Although there are many ideas concerning the underlying mechanism by
which neutrino mass is generated, at present none of the proposed
mechanisms have any experimental foundation; to make
real progress more data is required. 
The neutrino masses and mixings are as fundamental as those
of the quarks, yet the precision with which the neutrino-mixing
parameters are known is very poor when compared to the precision of
the quark parameters. 
Some of the neutrino parameters, such as the reactor angle and the 
CP-violating phase, have yet to be measured, and the sign of 
the atmospheric mass-squared difference is undetermined. 
If neutrino are Dirac fermions, then neutrino masses may arise in a
manner similar to that which generates the masses of the other charged
fundamental fermions.
However, if neutrinos are Majorana particles, then the mass-generation
mechanism may be quite different.
These issues, which have profound implications for particle physics
and cosmology, will be discussed in detail in this section.

%% file: 03-New-physics-and-the-SnuM/03-01-Origin-of-small-mass/03-01-Origin-of-small-mass.tex
\subsection{The origin of small neutrino mass}
\label{susect:origin-nu-mass}
This section will address the implications of see-saw mechanisms,
supersymmetry and R-parity violation, extra dimensions, string
theory, and TeV scale mechanisms for small neutrino masses
on the properties of the neutrinos.

\subsubsection{See-Saw mechanisms}

The charged-fermion spectrum already contains quite strong hierarchies
with the electron mass being a few million times smaller than the 
top-quark mass. 
Neutrino masses are also very small compared to charged-fermion masses,
with the atmospheric neutrino mass being a few million times smaller
than the electron mass. 
Such severe fermion mass hierarchies demand some
explanation. 
One simple approach is based on the see-saw mechanism and its
generalisation to include the charged-fermion masses by Froggatt and
Nielsen \cite{Froggatt:1978nt}. 
The idea is that all Yukawa couplings are of order unity, but lowest
order Yukawa couplings to Higgs fields are forbidden by some symmetry;
neutrino masses are further suppressed by the fact that
right-handed neutrinos are very heavy. 
Small effective Yukawa couplings and small Majorana masses are then
generated at higher order, suppressed by ratios of vacuum expectation
values (vevs) to heavy field masses.
The see-saw mechanism thus provides a convincing explanation for the
smallness of neutrino masses. 
Here we review its simplest form, the type I see-saw mechanism and its 
generalisation to the type II see-saw mechanism.

Before discussing the see-saw mechanism, the different types of
neutrino mass that are possible will be reviewed.  
So far we have been assuming that neutrino masses are Majorana masses
of the form: 
\begin{equation}
  m^\nu_{LL}\overline{\nu_\mathrm{L}} \nu_\mathrm{L}^\mathrm{C}
  \label{mLL}
\end{equation}
where $\nu_\mathrm{L}$ is a left-handed neutrino field and
$\nu_\mathrm{L}^\mathrm{C}$ is the CP conjugate of a left-handed
neutrino field, in other words a right-handed anti-neutrino field. 
Majorana masses imply lepton-number violation.
Note that lepton-number violation is forbidden by gauge invariance at
the renormalisation level in extensions of the Standard Model in which
the Higgs sector only contains doublets.
The simplest version of the see-saw mechanism assumes that
Majorana-mass terms are generated 
through the interactions of the right-handed neutrinos
\cite{Minkowski:1977sc,Gell-Mann:1980vs,Yanagida:1979as}. 

If we introduce right-handed neutrino fields then there are two sorts
of additional neutrino mass terms that are possible:
additional Majorana masses of the form:
\begin{equation}
  M_\mathrm{RR}\overline{\nu_\mathrm{R}}\nu_\mathrm{R}^\mathrm{C}\; 
  + \mbox{\rm hermitian~conjugate}\; ,
  \label{MRR}
\end{equation}
where $\nu_\mathrm{R}$ is a right-handed neutrino field,
$\nu_\mathrm{R}^\mathrm{C}$ is the CP conjugate of a right-handed
neutrino field, in other words a left-handed antineutrino field; and
Dirac masses of the form: 
\begin{equation}
  m^\nu_\mathrm{LR}\overline{\nu_\mathrm{L}}\nu_\mathrm{R}\;
  + \mbox{\rm hermitian~conjugate}\; .
  \label{mLR}
\end{equation}
Such Dirac mass terms conserve lepton number, and are not forbidden
by electric-charge conservation.

Once this is done, the types of neutrino mass described
in equations~(\ref{MRR}), (\ref{mLR}) (but not equation~(\ref{mLL})
since we do not assume direct mass terms, e.g.\ from Higgs triplets,
at this stage) are permitted, and we have the mass matrix:
\begin{equation}
  \left(
    \begin{array}{cc} 
      \overline{\nu_\mathrm{L}} & \overline{\nu^\mathrm{C}_\mathrm{R}}
    \end{array} \\ 
  \right)
  \left(
    \begin{array}{cc}
      0 & m^\nu_\mathrm{LR}\\
      m^{\nu T}_\mathrm{LR} & M_\mathrm{RR} \\
    \end{array}
  \right)
  \left(
    \begin{array}{c} 
      \nu_\mathrm{L}^\mathrm{C} \\ \nu_\mathrm{R} 
    \end{array} \\
  \right)\; 
  + \mbox{\rm hermitian~conjugate}\; .
  \label{matrix}
\end{equation}
Since the right-handed neutrinos are electroweak singlets,
the Majorana masses of the right-handed neutrinos, $M_\mathrm{RR}$,
may be orders of magnitude larger than the electroweak
scale. In the approximation that $M_\mathrm{RR}\gg m^\nu_\mathrm{LR}$
the matrix in equation~(\ref{matrix}) may be diagonalised to
yield effective Majorana masses of the type in equation~(\ref{mLL}):
\begin{equation}
  m^\nu_\mathrm{LL}=-m^\nu_\mathrm{LR}M_\mathrm{RR}^{-1}m^{\nu T}_\mathrm{LR}\; .
  \label{see-saw}
\end{equation}
The effective left-handed Majorana masses, $m^\nu_\mathrm{LL}$, are
naturally suppressed by the heavy scale, $M_\mathrm{RR}$.
In a one-family example, if we take $m^\nu_\mathrm{LR}=M_W$ and 
$M_\mathrm{RR}=M_{\mathrm{GUT}}$, 
then we find $m^\nu_\mathrm{LL}\sim 10^{-3}$ eV which looks good for solar
neutrinos.
Atmospheric neutrino masses would require
a right-handed neutrino with a mass below the GUT scale.

With three left-handed neutrinos and three right-handed neutrinos the
Dirac masses, $m^\nu_\mathrm{LR}$, are a $3\times 3$ (complex) matrix
and the heavy Majorana masses, $M_\mathrm{RR}$, form a separate 
$3\times 3$ (complex, symmetric) matrix.  
The light effective Majorana masses
$m^\nu_\mathrm{LL}$ are also a $3\times 3$ (complex symmetric) matrix
and continue to be given by equation~(\ref{see-saw}) which is now interpreted
as a matrix product. 
From a model-building perspective the fundamental
parameters which must be input into the see-saw mechanism are the
Dirac mass matrix $m^\nu_\mathrm{LR}$ and the heavy right-handed
neutrino Majorana mass matrix $M_\mathrm{RR}$.  The light effective
left-handed Majorana mass matrix $m^\nu_\mathrm{LL}$ arises as an
output according to the see-saw formula in equation~(\ref{see-saw}).

\begin{figure}
  \begin{center} 
    \ensuremath{
      \vcenter{
        \hbox{
          \includegraphics[scale=1]{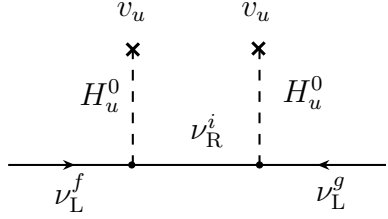}
        }
      }
    }
    \caption{
      \label{fig:TypeIDiagrams} 
      Diagram illustrating the type I see-saw mechanism.   
    } 
  \end{center}
\end{figure}

The version of the see-saw mechanism discussed so far is sometimes
called the type I see-saw mechanism. 
It is the simplest version of the see-saw mechanism, and can be
thought of as resulting from integrating out heavy right-handed
neutrinos to produce the effective dimension-5 neutrino mass operator:
\beq
  -\frac{1}{4} (H_\mathrm{u} \cdot L^T) \,\kappa \, (H_\mathrm{u}\cdot L)\; ,
  \label{dim5}
\eeq
where the dot indicates the $\SU (2)_\mathrm{L}$-invariant product,
and:
\beq
  \kappa = 2 \,Y_{\nu} M_\mathrm{RR}^{-1} Y_{\nu}^T   \, ,
  \label{kappa}
\eeq
with $Y_{\nu}$ being the neutrino Yukawa couplings and 
$m^\nu_\mathrm{LR}=Y_\nu v_\mathrm{u}$ with $v_\mathrm{u} = \< H_\mathrm{u}\>$.
The type I see-saw mechanism is illustrated diagrammatically in
figure~\ref{fig:TypeIDiagrams}. 

In models with a left-right symmetric particle content such as minimal
left-right symmetric models, Pati-Salam models, or Grand Unified
Theories (GUTs) based on $\SO (10)$, the type I see-saw mechanism is
often generalised to a type II see-saw (see e.g.
~\cite{Magg:1980ut,Lazarides:1980nt,Schechter:1980gr,Mohapatra:1980yp,Wetterich:1981bx}), 
where an additional direct mass term,
$m_{\mathrm{LL}}^{\mathrm{II}}$, for the light neutrinos is present. 

With such an additional direct mass term, the general neutrino mass
matrix is given by:
\begin{eqnarray}
  \,
  \left( \begin{array}{cc} \overline{\nu_{\mathrm{L}}}  &
  \overline{\nu^{\mathrm{C} }_\mathrm{R}}   \end{array}   \right) 
  \left( \begin{array}{cc} 
  m^{\mathrm{II}}_{\mathrm{LL}} \vphantom{\nu_{\mathrm{L}}^{\mathrm{C} }}&
  m^\nu_{\mathrm{LR}}\\[1mm]
  m^{\nu T}_{\mathrm{LR}} \vphantom{\nu_{\mathrm{L}}^{\mathrm{C} }}&
  M_{\mathrm{RR}}
  \end{array}   \right)
  \,
  \left( \begin{array}{c} 
  \nu_{\mathrm{L}}^{\mathrm{C} }  \\[1mm]
  {\nu_\mathrm{R}}   \end{array}   \right) .
\end{eqnarray}
Under the assumption that the mass eigenvalues $M_{\mathrm{R}i}$ of 
$M_{\mathrm{RR}}$ are very large compared to the components of  
$m^{\mathrm{II}}_{\mathrm{LL}}$ and $m_{\mathrm{LR}}$, the mass matrix
can approximately be diagonalised yielding effective Majorana masses:
\begin{eqnarray}
  \label{eq:TypIIMassMatrix}
  m^\nu_{\mathrm{LL}} \approx 
  m^{\mathrm{II}}_{\mathrm{LL}} + m^{\mathrm{I}}_{\mathrm{LL}} \, ,
\end{eqnarray} 
with :
\begin{eqnarray}
  m^{\mathrm{I}}_{\mathrm{LL}} \approx - m^\nu_{\mathrm{LR}}
  \,M^{-1}_{\mathrm{RR}}\,m^{\nu T}_{\mathrm{LR}} \, ,
\end{eqnarray}
for the light neutrinos. 

\begin{figure}
  \begin{center}
    \ensuremath{
      \vcenter{
        \hbox{
          \includegraphics[scale=1]{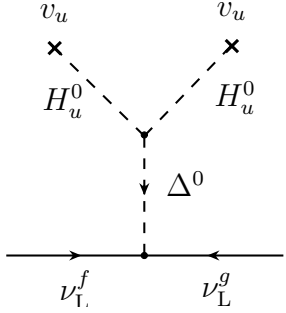}
        }
      }
    }
    \caption{
      \label{fig:TypeIIDiagrams}
      Diagram leading to a type II contribution
      $m^{\mathrm{II}}_{\mathrm{LL}}$ to the neutrino mass matrix via an
      induced vev of the neutral component of a triplet Higgs $\Delta$. 
    } 
  \end{center}
\end{figure}

The direct mass term, $m^{\mathrm{II}}_{\mathrm{LL}}$, can also
provide a naturally small contribution to the light-neutrino masses if
it stems, e.g., from a see-saw suppressed induced vacuum-expectation
value. 
We will refer to the general case, where both possibilities are
allowed, as the type II see-saw mechanism. 
Realising the type II contribution by generating the dimension-5
operator in equation~(\ref{dim5}) via the exchange of heavy Higgs triplets
of SU(2)$_\mathrm{L}$ is illustrated diagrammatically in
figure~\ref{fig:TypeIIDiagrams}. 

\subsubsection{Supersymmetry and R-parity Violation}

Another example of the origin of small neutrino masses is R-parity
violating supersymmetry (SUSY) (for a review see \cite{Dreiner:1997uz}). 
Here, the left-handed neutrinos mix with neutralinos after SUSY
breaking, leading to small, loop suppressed, Majorana masses.
The masses depend on the SUSY mass spectrum.
Should SUSY be discovered, and the mass spectrum determined, at
high-energy colliders, the theory could be used to predict the
Majorana masses.

In any supersymmetric extension of the Standard Model it is possible to 
introduce interactions that break R-parity, defined as $R=(-1)^{3B+L+2S}$ 
\cite{Hall:1983id}, where $L$, $B$, and $S$ are the lepton number, baryon 
number, and spin, respectively. The interactions that can contribute to 
the neutrino masses must also violate lepton number, and are 
given by \cite{Barbier:2004ez}:
\begin{equation}  
W_{RpV}=\varepsilon_{ab}\left[ 
\half\lambda_{ijk}\widehat L_i^a\widehat L_j^b\widehat R_k 
+\lambda'_{ijk}\widehat L_i^a\widehat Q_j^b\widehat D_k 
+\epsilon_i\widehat L_i^a\widehat H_u^b\right] 
\label{eq:Wsuppot} 
\end{equation} 
The trilinear R-Parity violating (TRpV) parameters $\lambda_{ijk}$ and
$\lambda'_{ijk}$ are dimensionless Yukawa couplings that violate lepton 
number keeping baryon number conserved. The baryon number violating 
interactions (of the form $\half\lambda''UDD$) can also be included,
leading to proton decay.
The present limit on the lifetime of the proton \cite{Yao:2006px} leads to
stringent constraints on products of $\lambda$ couplings, although
such constraints can be relaxed in the case of Split Supersymmetry  
\cite{Nath:2006ut}.

The bilinear R-Parity violating (BRpV) parameters, $\epsilon_i$,
induce sneutrino vacuum expectation values $v_i$, as well as mixing
between particles and sparticles. 
In particular, neutrinos mix with neutralinos forming a set 
of seven neutral fermions $F^0_i$.
A low energy see-saw mechanism induces the tree-level neutrino-mass
matrix \cite{Nowakowski:1995dx}:
\begin{equation}
  {\bf M}_{\nu}^{(0)}=-m\cdot{\cal M}_{\chi^0}^{-1}\cdot m^T=
  \frac{M_1 g^2 \!+\! M_2 {g'}^2}{4\,\det({\cal M}_{\chi^0})}
  \left[
    \begin{matrix}
      \Lambda_1^2 & \Lambda_1 \Lambda_2 & \Lambda_1 \Lambda_3 \cr
      \Lambda_1 \Lambda_2 & \Lambda_2^2 & \Lambda_2 \Lambda_3 \cr
      \Lambda_1 \Lambda_3 & \Lambda_2 \Lambda_3 & \Lambda_3^2
    \end{matrix}
  \right] 
  \; ,
  \label{treemassm}
\end{equation}
where ${\cal M}_{\chi^0}$ is the Minimal Supersymmetric Standard
Model (MSSM) (for a review see \cite{Chung:2003fi}) 
neutralino mass matrix and the
parameters $\Lambda_i\equiv\mu v_i+\epsilon_i v_d$ are proportional to 
the sneutrino vevs in the basis where the $\epsilon_i$ terms are rotated
away from the superpotential. Note that if this is done BRpV reappears in
the soft terms \cite{Diaz:1997vq}. 

The tree-level neutrino mass matrix has only one non-zero eigenvalue, equal to 
the trace of the matrix in equation~(\ref{treemassm}), and therefore proportional 
to $|\vec\Lambda|^2$. If the above tree-level contribution dominates over 
one-loop graphs, the square of this eigenvalue would be equal to the 
atmospheric mass-squared difference, $\Delta m^2_{31}\approx m_3^{(0)2}$, and 
the atmospheric and reactor angles would be given by 
$\tan^2\theta_{23}^{(0)}\approx\Lambda_2^2/\Lambda_3^2$ and 
$\tan^2\theta_{13}^{(0)}\approx\Lambda_1^2/(\Lambda_2^2+\Lambda_3^2)$
respectively. 
Without one-loop corrections, the solar mass-squared difference 
and the solar angle remain undetermined. 

\begin{figure}
\begin{center} 
\ensuremath{
\vcenter{
\hbox{
\includegraphics[scale=1]{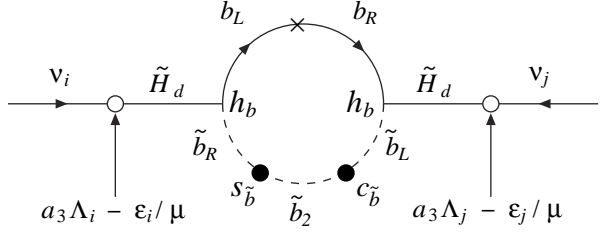}}}}
\caption{Pictorial representation of the bottom-sbottom loops contributing
to the neutrino mass matrix, with Rp violated bilinearly in the open
circles.}
\label{Fig:BRpVPic1}
\end{center}
\end{figure}

Once the one-loop corrections are included \cite{Hirsch:2000ef} the symmetry 
of the neutrino mass matrix in equation~(\ref{treemassm}) is broken and thus
the solar mass squared difference is generated radiatively. 
The one-loop corrected neutrino-mass matrix has the general form:
\begin{equation}
  M^{\nu}_{ij}=A\Lambda_i\Lambda_j+
  B(\epsilon_i\Lambda_j+\epsilon_j\Lambda_i)+C\epsilon_i\epsilon_j \; ,
  \label{deltapi}
\end{equation}
where $A^{(0)}=(g^2M_1+g'^2M_2)/4\det({\cal M}_{\chi^0})$ is the only non-zero
coefficient at tree-level. In BRpV most particles contribute in loops to the
neutrino mass matrix. An important loop is the one involving bottom quarks
and squarks, which is shown in figure \ref{Fig:BRpVPic1}.
The external arrows represent the flow of lepton number, while the internal 
ones show the flow of the bottom-quark electric charge, and the cross signals 
a mass insertion. 
The complete dashed line represents a single scalar 
propagator corresponding to the heavy bottom squark $\tilde b_2$, with the
full circles pictorially showing the component of this mass eigenstate in 
left and right sbottoms. The external lines are the neutrino states which
define the basis used to write the neutrino mass matrix in 
equation~(\ref{treemassm}). 
The open circles pictorially represent the component 
of these neutrinos in higgsinos and indicates the place where R-Parity is
violated. A similar graph with the light sbottom, $\tilde b_1$, is
obtained replacing $c_{\tilde b}\rightarrow-s_{\tilde b}$ and 
$s_{\tilde b}\rightarrow c_{\tilde b}$. The sum of these two graphs
contributes to the coefficient $C$ in the following way:
\begin{equation}
  C^{(\tilde b)}=-{\frac{N_c
  m_b}{16\pi^2\mu^2}}h_b^2\sin(2\theta_{\tilde b}) \;\Delta
  B_0^{\tilde b_1\tilde b_2} \, ,
  \label{C:bsb}
\end{equation}
where $N_c=3$ is the number of colours, and we have defined:
\begin{equation}
  \Delta B_0^{\tilde b_1\tilde b_2}\equiv
  B_0(0;m^2_b,m^2_{\tilde b_2})-B_0(0;m^2_b,m^2_{\tilde b_1})
  \approx-\ln(m^2_{\tilde b_2}/m^2_{\tilde b_1}) \; .
\end{equation}
The result in equation~(\ref{C:bsb}) can be understood with the help
of the graph presented above. 
It is proportional to the bottom-quark mass due to the mass insertion,
and to the square of the bottom Yukawa coupling due to the vertices. 
The sbottom mixing contributes with the factor 
$\sin(2\theta_{\tilde b})$, and the higgsino-neutrino mixing accounts for the
factor $\epsilon_i\epsilon_j/\mu^2$, where the $\epsilon$ parameters have been
factored out from $C$. The contribution is finite because Veltman
functions \cite{Passarino:1978jh}
from $\tilde b_2$ and $\tilde b_1$ are subtracted from each other. The 
contribution to the $B$ parameter can be obtained using
$B^{(\tilde b)}=-a_3\mu C^{(\tilde b)}$, with
$a_3=v_u(g^2M_1+g'^2M_2)/4\det({\cal M}_{\chi^0})$, as can be inferred
from the neutralino-neutrino mixing shown in the graph. There is also
a contribution $A^{\tilde b}$, but it is in general a small correction
to $A^{(0)}$. 

There are similar loops with charged scalars $S_i^+$ (charged Higgs 
bosons mixing with charged sleptons \cite{Diaz:2003as}) together with 
charged fermions $F_j^+$ (charginos mixing with charged leptons 
\cite{Faessler:1997db}). 
Among these are the charged Higgs and stau contributions which have
the same form as that given in equation~(\ref{C:bsb}) with the replacements
$b\rightarrow\tau$, $\tilde b\rightarrow\tilde\tau$, and taking $N_c=1$ .
There are also loops with neutral scalars $S^0_i$ (neutral Higgs bosons mixing
with sneutrinos \cite{Grossman:1997is}) together with the neutral fermions 
$F^0_j$ mentioned above.

\begin{figure}
  \centerline{
    \protect\hbox{
      \epsfig{file=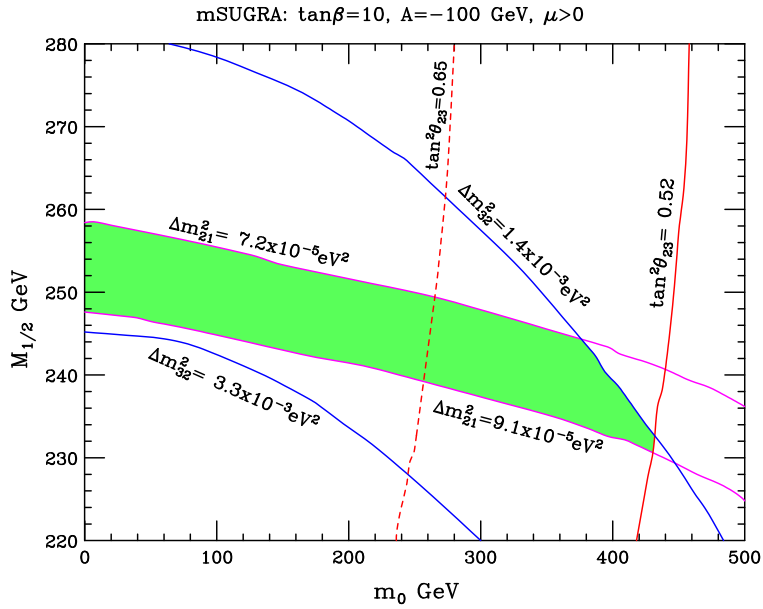,width=0.5\textwidth,angle=90}
    }
  }
  \caption{
    Region of parameter space where solutions satisfy all 
    experimental constraints.
    Adapted with kind permission of the European Physical Journal from 
    figure 6 in reference \cite{Diaz:2004fu}.
    Copyrighted by Springer Berlin/Heidelberg. 
  }
  \label{m0mhalf}
\end{figure}

BRpV can successfully be embedded in supergravity \cite{Diaz:1997xc}, 
although with non-universal $\epsilon_i$ terms at the GUT scale (as well as 
bilinear soft terms $B_i$, associated to $\epsilon_i$). By definition, 
the coefficients $A$, $B$, and $C$ in equation~(\ref{deltapi}) depend
exclusively on the universal scalar mass $m_0$, gaugino mass
$M_{1/2}$, and trilinear parameter $A_0$ at the GUT scale, and the
values of $\tan\beta$ and $\mu$ at the weak scale. In
figure~\ref{m0mhalf} we see the region of the $m_0-M_{1/2}$ plane
consistent with neutrino experimental data, for fixed values of the
BRpV parameters $\epsilon_1=-0.0004$, $\epsilon_2=0.052$, 
$\epsilon_3=0.051$ GeV, and $\Lambda_1=0.022$, $\Lambda_2=0.0003$, 
$\Lambda_3=0.039$ GeV${}^2$ \cite{Diaz:2004fu}. In this scenario, the solar 
mass-squared difference strongly limits the universal gaugino mass from 
above and below. Large values of the universal scalar mass are limited 
mainly by the atmospheric mass-squared difference. 

This model can be tested at colliders, and the main signal that differentiates
it from the MSSM is the decay of the lightest neutralino which decays
only in RpV modes. 
In the scenario of figure~\ref{m0mhalf} the neutralino mass is 99 GeV
and decays to an on-shell $W$, satisfying:
\begin{equation}
  \frac{B(\chi^0_1\rightarrow We)}{B(\chi^0_1\rightarrow W\mu)}=
  \frac{\Lambda_1^2}{\Lambda_2^2} \; .
\end{equation}
Such ratios can be directly related to neutrino mixing 
angles \cite{Hirsch:2003fe}.
\begin{figure}
  \centerline{\protect\hbox{\epsfig{file=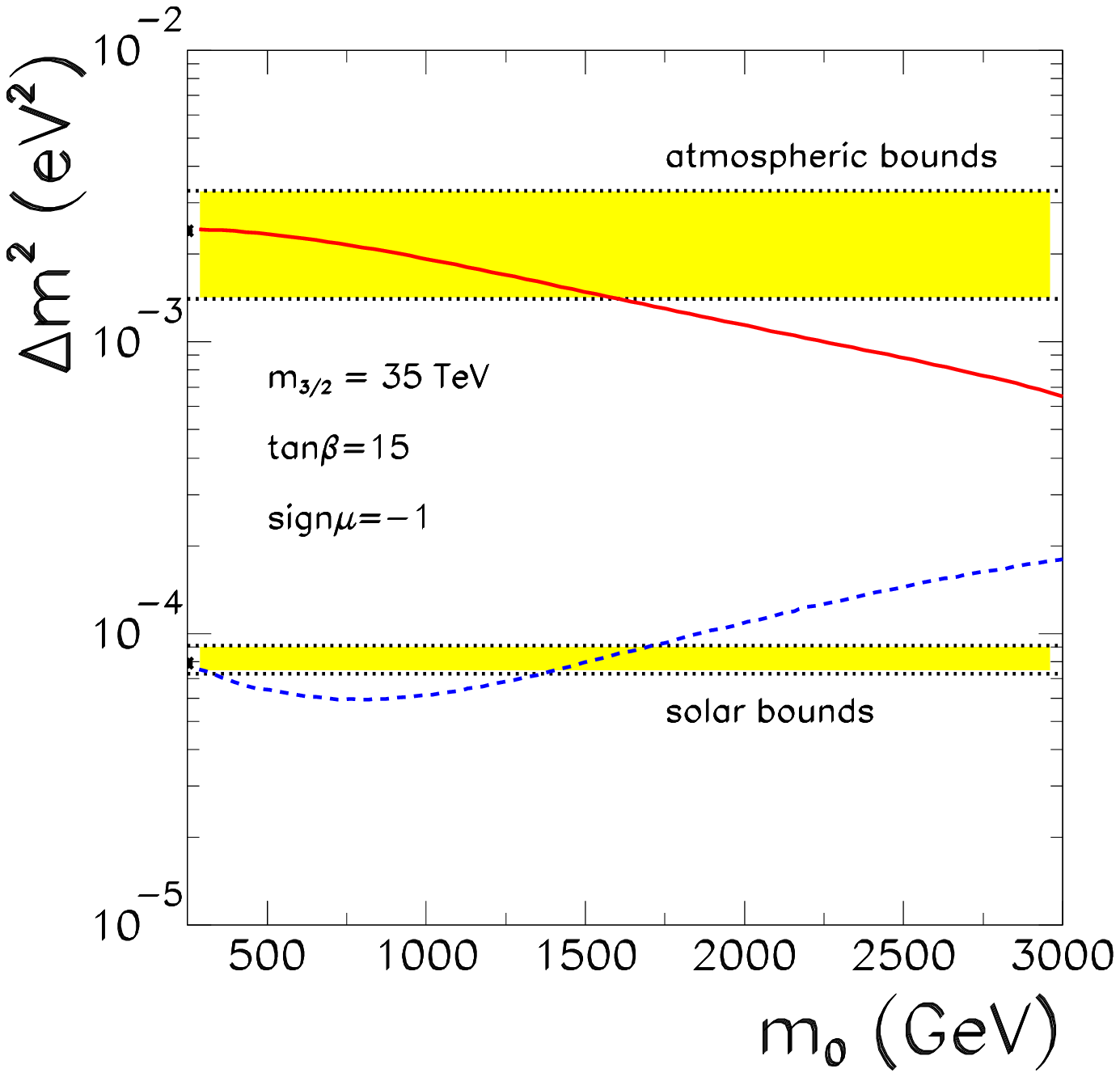,width=0.5\textwidth}}
  \protect\hbox{\epsfig{file=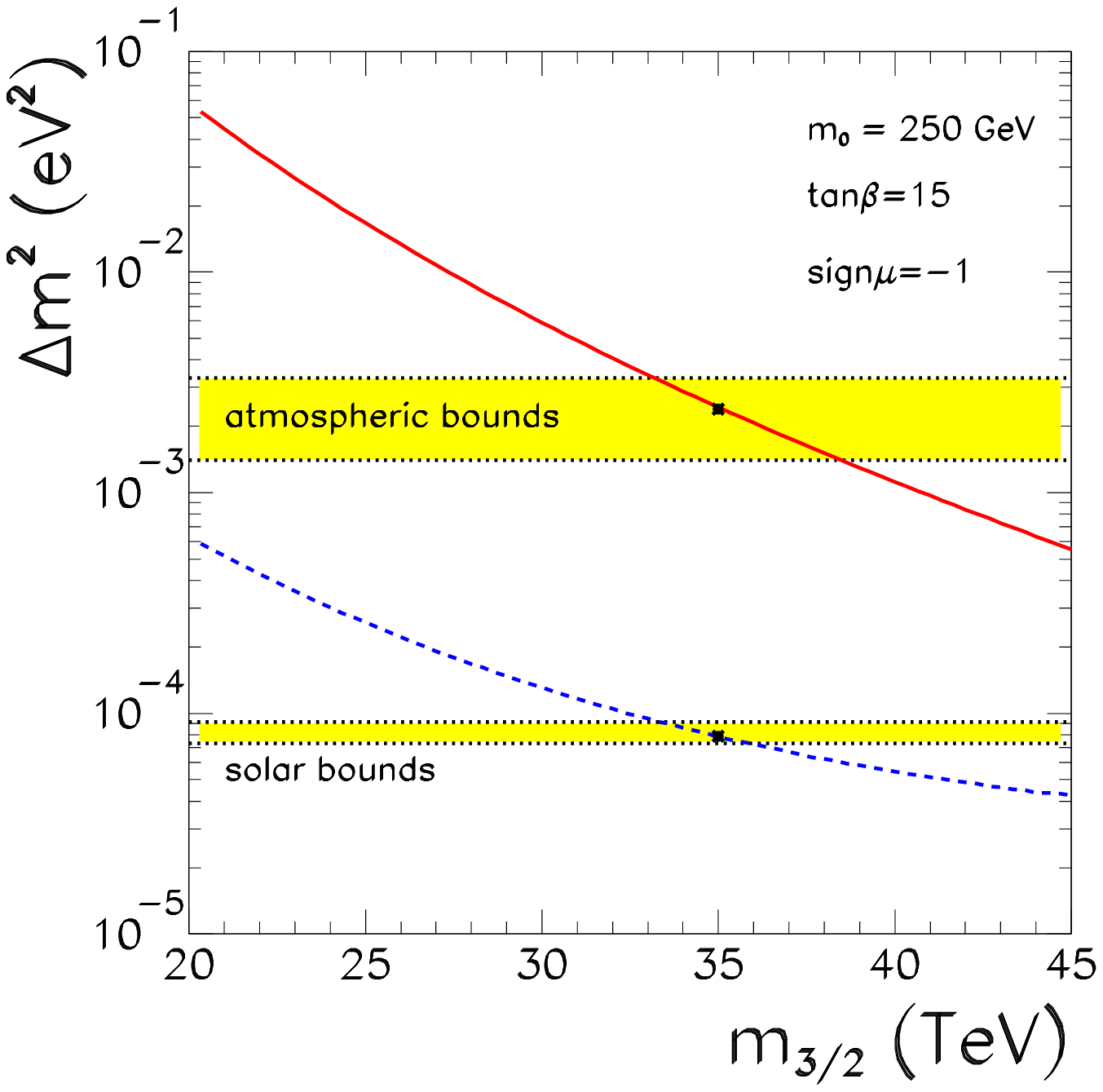,width=0.5\textwidth}}}
  \caption{
    Atmospheric and solar mass-squared differences as a function of
    susy masses in AMSB.
    Taken with kind permission of Physical Review from figures 1 and 3 in
    reference \cite{deCampos:2004iy}.
    Copyrighted by the American Physical Society.
  }
  \label{fig3}
\end{figure} 

Other scenarios have been studied, for example Anomaly Mediated
Super-Symmetry Breaking (AMSB) \cite{DeCampos:2001wq}, and Gauge
Mediated Supersymmetry Breaking \cite{Chun:1999kd}, and Split
Supersymmetry \cite{Chun:2004mu}. 
In the case of AMSB we see in figure~\ref{fig3} how the solar and
atmospheric mass-squared differences depend on the universal scalar
and gaugino masses, for fixed values of the BRpV parameters
$\epsilon_1=-0.015$, $\epsilon_2=-0.018$, $\epsilon_3=0.011$ GeV, and
$\Lambda_1=-0.03$, $\Lambda_2=-0.09$, $\Lambda_3=-0.09$ GeV${}^2$.

\begin{figure}
\begin{center}
\ensuremath{
\vcenter{
\hbox{
\includegraphics[scale=1]{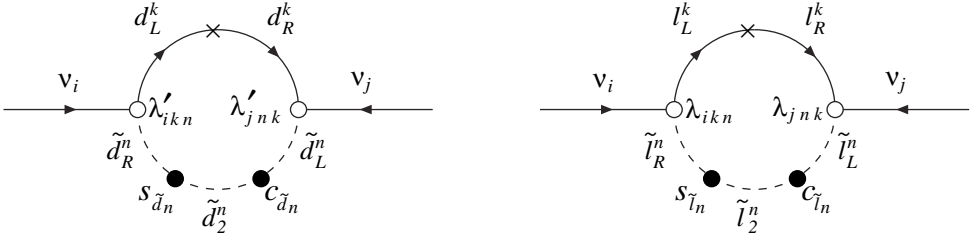}}}}
\end{center}
\vspace{0pt}
\caption{Pictorial representation of the fermion-sfermion loops contributing
to the neutrino mass matrix, with Rp violated trilinearly in the open
circles.}
\label{RPVFig2}
\end{figure}

TRpV interactions do not contribute to neutrino masses at tree-level
\cite{Abada:1999ai}. The one-loop contributions to these diagrams are
given by the diagrams shown in figure \ref{RPVFig2}.
The convention for the graphs is the same as before. Analogous graphs are 
obtained for the light scalars $\tilde d^n_1$ and $\tilde l^n_1$ with the 
replacement $c_{\tilde d_n}\rightarrow-s_{\tilde d_n}$ and 
$s_{\tilde d_n}\rightarrow c_{\tilde d_n}$. The mixing angles are:
\begin{equation}
  \sin(2\theta_{\tilde d_n})=\frac{2(M^{\tilde d\,2}_{LR})_n}
  {M^{\tilde d\,2}_{Ln}-M^{\tilde d\,2}_{Rn}}
  \,,\qquad
  \sin(2\theta_{\tilde l_n})=\frac{2(M^{\tilde l\,2}_{LR})_n}
  {M^{\tilde l\,2}_{Ln}-M^{\tilde l\,2}_{Rn}}\; .
\end{equation}
The contribution to the neutrino mass matrix due to TRpV is:
\begin{equation}
  (\Delta M^\nu_{ij})^{\lambda'}=\frac{N_c}{16\pi^2}\sum_{kn}
  \left(\lambda'_{ikn}\lambda'_{jnk}+\lambda'_{ink}\lambda'_{jkn}\right)
  \sin(2\theta_{\tilde d_n})\,m_{d_k}\Delta
  B_0(0;m^2_{d_k};m^2_{\tilde d_n}) \; .
  \label{DMnulambda}
\end{equation}
Note that the contribution to the neutrino mass matrix is symmetric in 
the indices $i$ and $j$ \cite{AristizabalSierra:2004cy}. A similar
contribution holds for leptons and sleptons inside the loop, replacing
$\lambda'$ by $\lambda$ couplings.
In the approximation where only particles of the third generation 
contribute inside the loops, the shift to the neutrino mass matrix
from TRpV is:
\begin{equation}
  (\Delta M^\nu_{ij})^{TRpV}=
  D\lambda'_{i33}\lambda'_{j33}+E\lambda_{i33}\lambda_{j33} \; ,
\end{equation}
which can be added to equation~(\ref{deltapi}).
In this way, BRpV and TRpV, together or separated, can explain the neutrino
masses and oscillations observed in experiments.

\subsubsection{Extra Dimensions}
\label{sec:neutr-extra-dimens}

\begin{figure}
  \begin{center}
    \includegraphics[width=.4\linewidth]{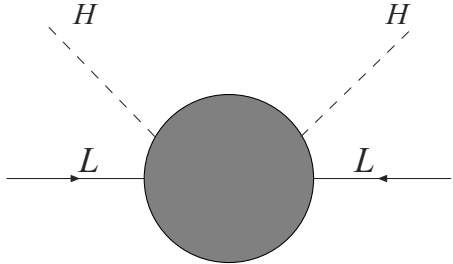}
  \end{center}
  \caption{
    Dimension five operator responsible for neutrino mass.
    $L$ denotes any of the three lepton doublets and $H$ is the SM
    scalar doublet.  
  }
  \label{fig:d-5}
\end{figure}

The basic gauge-theoretic way to account for small neutrino masses is
to ascribe them to the violation of lepton number by adding to the SM
an effective dimension-five operator ${\cal O} = \lambda L H L H$
\cite{Weinberg:1980bf} (see figure \ref{fig:d-5}).
The favourite scenario realising this idea is the
``see-saw'' mechanism, which requires the presence of singlet
``right''-handed neutrinos, which mix with the ordinary SU(2) doublet
``left''-handed neutrinos~\cite{Valle:2006vb}. The suppression of the
neutrino masses results from the structure of the full mass
matrix~\cite{Schechter:1980gr,Schechter:1981cv}. In the simplest
versions of this mechanism the mass of the extra states should be
about ten orders of magnitude larger than the electroweak scale.

Recently, there have been a number of attempts to explain neutrino
oscillations in theories with large internal dimensions and a low
fundamental scale
\cite{Arkani-Hamed:1998rs,Antoniadis:1998ig,Antoniadis:1990ew,Antoniadis:1993jp,Lykken:1996fj}. 
A convenient, perturbatively-calculable framework is type I string
theory with D-branes. The SM is then localised on a stack of D-branes,
transverse to some large extra dimensions, where gravity propagates.
D-brane models offer a novel scenario to account for neutrino
masses~\cite{Dienes:1998sb,Arkani-Hamed:1998vp,Faraggi:1999bm,Dvali:1999cn,Mohapatra:1999zd,Barbieri:2000mg};
right-handed neutrinos are assumed to propagate in the bulk while
left-handed neutrinos, being a part of the lepton doublet, live on the
SM branes. As a result, the Dirac neutrino mass is naturally
suppressed by the bulk volume.  Adjusting this volume, so that the
string scale lies in the TeV range, leads to tiny neutrino masses
compatible with current experimental data.

Indeed, the relation between the string scale, $M_s$, and
the four-dimensional Planck mass, $M_P$, is:
\ba
 M_P^2=\frac{8}{g^4} V_b M_s^2\; ,
 \label{mp2}
\ea 
where $g$ is the SM gauge coupling and $V_b$ is the volume of the bulk
in string units. The simplest way to introduce a right-handed neutrino
is to identify it with an open string excitation on some (stack of)
brane(s) extended in the bulk. Moreover the SM Higgs and lepton
doublets must come from open strings stretched between the SM and bulk
branes and thus, living at their intersection, they couple to
the bulk neutrino state. More precisely, its kinetic term is: 
\ba
  S_{kin}=V_b\int d^4x\sum_m\left\{
  {\bar\nu}_{Rm}{\slash\!\!\!\partial}\nu_{Rm}
  +{\bar\nu}_{Rm}^c{\slash\!\!\!\partial}\nu_{Rm}^c+ {m\over
    R}\nu_{Rm}\nu_{Rm}^c+c.c.\right\} \; ,
  \label{nuR}
\ea 
where the sum is extended over all Kaluza-Klein (KK) excitations,
denoted collectively by $m$. For simplicity we assumed a toroidal
compactification for $n$ extra dimensions of common radius $R$, with
$V_b=(2\pi R)^n$ in string units. The two states $\nu_R$ and $\nu^c_R$
correspond to the left and right four-dimensional (4d) components of
the higher-dimensional spinor. The zero-mode $\nu_{R0}$ will be
identified with the right-handed-neutrino state, while $\nu^c_{R0}$
may be projected out from the spectrum by an orbifold projection and
is not relevant for our purposes.

The interaction of the bulk neutrino with the localised Higgs and lepton 
doublets reads:
\ba
  S_{int}=\lambda\int d^4x H(x)L(x)\nu_R(x,y=0)\, ,
  \label{inter}
\ea 
where it has been assumed that the SM brane stack is localised at the
origin of the bulk and the coupling, $\lambda$, is in general of order
$g^2$ ($\lambda$ is equal to $g^2$ in the simplest 3-brane realisation
of the SM).  
By expanding $\nu_R$ in KK modes, one gets the mass terms: \ba
S_{mass}={\lambda v\over\sqrt{V_b}}\sum_m\nu_L\nu_{Rm}\, ,
\label{mass}
\ea where $v$ is the Higgs expectation value, $\langle H\rangle=v$.
Note that the apparent mixing of $\nu_L$ with all KK excitations can
be neglected since its strength (\ref{mass}) is much smaller than the
KK mass $gv/R^{n/2}<<1/R$, or equivalently $gv<<R^{n/2-1}$, which is
valid for any $n\ge 2$. As a result, the right-handed neutrino is
essentially the zero mode $\nu_{R0}$ and taking into account the
normalisation of its kinetic term (\ref{nuR}), one obtains a
Dirac neutrino mass, $m_\nu\nu_L\nu_{Rm}$, with:
\ba
  m_\nu\simeq{\lambda v\over\sqrt{V_b}}\simeq
  {\sqrt{8}\lambda\over g^2}\, v{M_s\over M_p}\, ,
  \label{mnu}
\ea
which is of the order of $10^{-3}$ to $10^{-2}$ eV for $M_s\sim 1-10$
TeV. 

The extra dimensional neutrino-mass suppression mechanism described
above can be destabilized by the presence of a large Majorana 
neutrino-mass term. Indeed, in the absence of any protecting symmetry, the
lepton-number-violating dimension-5 effective operator in
figure~\ref{fig:d-5} will be present.  This would lead, in the case of
TeV-string-scale models, to an unacceptable Majorana mass term of the
order of a few GeV. Even if we manage to eliminate this operator in
some particular model, higher-order operators would also give
unacceptably large contributions, since in low-scale gravity models
the ratio between the Higgs vacuum expectation value and the
string scale is of order ${\cal O}(1/10-1/100)$.

An elegant way to avoid this problem was suggested in
reference \cite{Ioannisian:1999sw}. 
It consists of assuming that the bulk
sector, where the SM singlet states live, is eight-dimensional.
There is, however, a general theorem that states that in eight
dimensions there can be no massive Majorana
spinor~\cite{Weinberg:1984vb,Shimizu:1984ik,Wetterich:1983bg,Finkelstein:1984bt}.
Moreover, further unwanted large L-violating contributions to neutrino
masses could be prevented by imposing lepton-number conservation
leaving only the Dirac mass (\ref{mnu}).
Indeed, lepton number often arises as an anomalous abelian gauge
symmetry associated to the $U(1)_b$ of the bulk (stack of) brane(s),
possibly in a linear combination with other $U(1)$'s
\cite{Ibanez:2001nd,Antoniadis:2002qm}. The anomaly is canceled by
shifting an axion field from the closed string (Ramond-Ramond) sector
\cite{Sagnotti:1992qw,Ibanez:1998qp}.  As a result, the gauge boson
becomes massive, while lepton number remains unbroken as an effective
global symmetry in perturbation theory \cite{Poppitz:1998dj}.  
The gauge coupling, $g_b$, of the bulk $U(1)_b$ gauge boson is
extremely small since it is suppressed by the volume of the bulk
$V_b$:  
\ba
  \frac{1}{g_b^2}=\frac{1}{g^2}V_b=\frac{g^2}{8}\frac{M_P^2}{M_s^2}\, ,
  \label{gb}
\ea
where in the second equality we used equation~(\ref{mp2}). 
It follows that $g_b\simeq 10^{-16} - 10^{-14}$ for $M_s\sim 1-10$
TeV. 
Such a theory would lead to light Dirac neutrino masses, in
contrast with general four-dimensional gauge-theoretic expectations
which lead to Majorana neutrinos~\cite{Valle:2006vb}.

If the $U(1)_b$ gauge boson is light, it would be copiously produced
in stellar processes, leading to supernova cooling through energy loss
in the bulk of extra dimensions.  There are strong constraints coming
from supernova observations. Note that the corresponding process is
much stronger than the production of gravitons because of the
non-derivative coupling of the gauge-boson interaction
\cite{Arkani-Hamed:1998vp}. In fact, for the case of $n$ large
transverse dimensions of common radius $R$, satisfying $m_A, R^{-1}<<T$
with $m_A$ the gauge boson mass and $T$ the supernova temperature, the
production rate, $P_A$, is proportional to:
\ba 
  P_A\sim g_b^2\times
  [R(T-m_A)]^n\times\frac{1}{T^2}\simeq\frac{T^{n-2}}{M_s^n}\, , 
\ea
where the factor $[R(T-m_A)]^n$ counts the number of KK excitations of
the $U(1)_b$ gauge boson with mass less than $T$.  This rate can be
compared with the corresponding graviton production:
\ba
  P_G\sim\frac{1}{M_P^2}\times (RT)^n\simeq\frac{T^n}{M_s^{n+2}}\, , 
\ea
showing that for $n=2$ (sub)millimeter extra dimensions, it is
unacceptably large, unless the bulk gauge boson acquires a mass
$m_A\simgt 10$ MeV.  For $n\ge 3$, the supernova bound becomes much
weaker and $m_A$ may be much smaller \cite{Antoniadis:2002cs}. Such a
light gauge boson can mediate short distance forces within the range
of table-top experiments that test Newton's law at very short distances
\cite{Hoyle:2004cw,Long:2003ta,Decca:2005qz,Smullin:2005iv}.

These theories also lead to novel ways to
generate neutrino oscillations. The interaction term in equation
(\ref{inter}) involves in general all left-handed neutrinos and
additional Higgs doublets: 
\ba 
  \sum_{i=1}^3\lambda_i L_i H_i\nu_R\ \rightarrow\
  \sum_{i=1}^3\lambda_i\,v_i\, \nu_{iL}\, \nu_R\, ,
  \label{threenu}
\ea 
where $i$ is a generation index and for each generation $i$, $H_i$ is
one of the possible available Higgs doublets $H_d$ or $H_u$, providing
also masses to down or to up quarks, with $v_i=\langle H_i\rangle$ the
corresponding vev.  The above couplings give mass to one linear
combination of the weak eigenstates $\nu_{iL}$, while the other two
remain massless. 
The mass is given by equation (\ref{mnu}) with $\lambda v$
replaced by $\sqrt{\sum_{i=1}^3\lambda_i^2 v_i^2}$.  The right-handed
neutrino, being a bulk state, has a tower of KK excitations. 
The mixing of these states with the ordinary neutrinos may have an
impact upon neutrino oscillations.  

\paragraph{The effect of extra dimensions}
\label{sec:extra-dimens-effects}

The most important features of the data on neutrino oscillations that
are relevant for the present discussion are:
\begin{enumerate}
  \item The existence of spectral distortions indicative of neutrino
        oscillations; 
  \item The solar mixing angle is large but significantly non-maximal;
  \item The atmospheric best-fit mixing angle is maximal;
  \item Both solar and atmospheric oscillation data strongly as well
        as the recent MiniBoone data \cite{Aguilar-Arevalo:2007it}
        disfavour the presence of sterile neutrino states in the
        channel to which the relevant neutrino is oscillating.
\end{enumerate}

There are several discussions in the literature
\cite{Dienes:1998sb,Arkani-Hamed:1998vp,Faraggi:1999bm,Dvali:1999cn,Mohapatra:1999zd,Barbieri:2000mg}
regarding neutrino masses and oscillations in the context of extra
dimensions. 
Most of these discussions are restricted to the
case of an effectively one-dimensional bulk.  
This simple one-dimensional bulk picture is not
realistic~\cite{Altarelli:2002hx}, as it is at odds with the current
global status of neutrino-oscillation data given in
reference \cite{Maltoni:2004ei} and described above. 
Indeed, such a picture violates at
least one of the four points mentioned above.
In addition there is also a serious theoretical problem, since
one-dimensional propagation of massless bulk states gives rise to
linearly growing fluctuations which, in general, yield large
corrections to all couplings of the effective field theory,
destabilizing the hierarchy~\cite{Antoniadis:1998ax}.

In the case of a two-dimensional bulk the situation is significantly
improved \cite{Antoniadis:2002qm}. 
Indeed, there is enough structure
to describe both solar and atmospheric oscillations by
introducing a single bulk-neutrino pair, using essentially the two
lowest frequencies of the neutrino-mass matrix: the mass of the zero
mode (equation (\ref{mnu})), arising via the electroweak Higgs
phenomenon, which is suppressed by the volume of the bulk, and the
mass of the first KK excitation.  
The former is used to reproduce the solar-neutrino data.
The later is used to explain atmospheric-neutrino oscillations, which
have a higher oscillation frequency, with an amplitude which is
enhanced due to logarithmic corrections of the two-dimensional bulk
\cite{Antoniadis:1998ax}.  
One can see, however, that at least condition (4) above is violated,
as there is a significant sterile component at least in one of the
channels of neutrino conversion, corresponding to the KK excitations
of the bulk right-handed neutrino, and this is highly disfavoured by
the global fits of neutrino oscillations~\cite{Maltoni:2004ei}.

One way out is to introduce three bulk neutrinos and explain
the observed neutrino oscillations in the traditional way
\cite{Davoudiasl:2002fq}. In this case, $\nu_R$ in equation
(\ref{threenu}) would carry a generation index $i$ and all left-handed
neutrinos would acquire Dirac-type masses with the zero modes of the
bulk states. 
Moreover, the effect of KK mixing can be suppressed by
appropriately decreasing the size of the extra dimensions and thus
increasing the value of the string scale.  
Thus, in this limit one
would obtain the generic case of three Dirac neutrinos, and the 
lepton-mixing matrix depends on precisely three angles and one CP phase, as
the quark-mixing matrix. Correspondingly, the oscillation pattern is
``generic'' without special predictions. Having Dirac instead of
Majorana neutrinos can be experimentally tested by searching for the
existence of processes like $0\nu\beta\beta$.

On the other hand, ``extra-dimensional'' signatures may be present in
oscillations at a sub-leading level, as non-standard interactions
(see \cite{Valle:2006vb} for a short discussion). 
The Neutrino Factory will provide an interesting laboratory to probe for the
possible presence of such effects.

\subsubsection{String Theory}

There has been relatively little work on the implications of
superstring theories for neutrino masses. However, it is known that
some of the ingredients employed in Grand Unified Theories and other
four-dimensional models may be difficult to implement in known types of
constructions. For example, the chiral supermultiplets that survive in
the effective four-dimensional field theory are generally bi-fundamental
in two of the  gauge-group factors (including the case
of fundamental under one factor and charged under a $U(1)$) for 
lowest-level heterotic constructions;  or either bi-fundamental,
adjoint, antisymmetric, or symmetric
for intersecting brane constructions.  This makes it difficult to
break the GUT symmetry, and even more so to find the high-dimensional
Higgs representations (such as the {\bf 126} of $SO(10)$) usually
employed in GUT models for neutrino and other fermion masses. Thus,
it may be difficult to embed directly many of the models, especially
GUT models involving high-dimensional representations rather than
higher-dimensional operators, in a string framework. Perhaps more likely
is that the underlying string theory breaks directly to an effective
four-dimensional theory including the Standard Model and perhaps other
group factors~\cite{Langacker:2003xa}. Some of the aspects of grand
unification, especially in the gauge sector, may be maintained in such
constructions. However, the GUT relations for Yukawa couplings are often
not retained~\cite{Dine:1985vv,Breit:1985ud,Witten:1985bz} because the
matter multiplets of the effective theory may have a complicated origin in
terms of the underlying string states.  Another difference is that Yukawa
couplings in string-derived models may be absent due to symmetries in
the underlying string construction, even though they are not forbidden
by any obvious symmetries of the four-dimensional theory, contrary to
the assumptions in many non-string models. Finally, higher-dimensional
operators, suppressed by inverse powers of the Planck scale, are common.

Much activity on neutrino masses in string theory occurred following
the first superstring revolution. In particular, a number of authors
considered the implications of an $E_6$ subgroup of the heterotic $E_8
\times E_8$ construction 
\cite{Dine:1985vv,Font:1989aj,Mochinaga:1993td,Nandi:1985uh}.  
Assuming
that the matter content of the effective theory involves three {\bf
27}s, one can avoid neutrino masses altogether by fine-tuned
assumptions concerning the Yukawa couplings~\cite{Dine:1985vv}. 
However, it is
difficult to implement a canonical type I see-saw.  Each {\bf 27} contains
two Standard Model singlets, which are candidates for right-handed
neutrinos, and for a field which could generate a large Majorana mass for
the right-handed neutrinos if it acquires a large vacuum expectation value
and has an appropriate trilinear coupling to the neutrinos. However, there
are no such allowed trilinear couplings involving three {\bf 27}s (this
is a reflection of the fact that the {\bf 27} does not contain a {\bf 126}
of the $SO(10)$ subgroup). $E_6$ string-inspired models were constructed
to get around this problem by invoking additional fields not in the 
{\bf 27} \cite{Witten:1985bz,Mohapatra:1986bd} or higher-dimensional
operators~\cite{Nandi:1985uh}, typically leading to extended versions
of the see-saw model involving fields with masses or vevs at the TeV
scale.

Similarly, more recent heterotic and intersecting brane constructions,
e.g., involving orbifolds and twisted sectors, may well have the
necessary fields for a type I see-saw, but it is again required that
the necessary Dirac Yukawa couplings and Majorana masses for the
right-handed neutrinos be present simultaneously.  
Dirac couplings need not emerge at the renormalisable level, but can
be of the form:
\begin{equation}
  \langle S'_1 \cdots S'_{d-3}\rangle N LH_u/M_{\rm PL}^{d-3} \; ,
\end{equation}
where the $S'_i$ are Standard Model singlets which acquire large
expectation values ($d=3$ corresponds to a renormalisable
operator). 
Similarly, Majorana masses can be generated by the operators:
\begin{equation}
  \langle S_1 \cdots S_{n-2}\rangle N N/M_{\rm  PL}^{n-3}\; .
\end{equation}
Whether such couplings are present at the appropriate orders depends on the
underlying string symmetries and selection rules, which are often very restrictive.
It is also necessary for the relevant $S$ and $S'$ fields to acquire the 
large expectation values that are needed, presumably without breaking
supersymmetry at a large scale.   
Possible mechanisms involve
approximately flat directions of the potential, e.g.,\ associated with
an additional $U(1)'$ gauge symmetry~\cite{Cleaver:1997nj,Langacker:1998ut}, string
threshold corrections~\cite{Cvetic:1992ct,Haba:1993yj,Haba:1994ym}, or
 hidden sector condensates~\cite{Faraggi:1993zh}.

There have been surprisingly few investigations of neutrino masses in
explicit semi-realistic string constructions. 
It is difficult to obtain canonical Majorana masses in intersecting
brane constructions \cite{Blumenhagen:2005mu} because there are no
interactions involving the same intersection twice. 
Two detailed studies \cite{Ibanez:2001nd,Antoniadis:2002qm}
of non-supersymmetric models with a low string scale concluded that
lepton number was conserved, though a small Dirac mass might emerge
from a large internal dimension. 
Large enough internal dimensions for the supersymmetric case may be
difficult to achieve, at least for simple toroidal orbifolds. 

There are
also difficulties for heterotic models.
An early study of $Z_3$ orbifolds
yielded no canonical Dirac neutrino Yukawa couplings
\cite{Font:1989aj} at low order. 
Detailed analyses of free-fermionic models and their flat directions
were carried out in \cite{Faraggi:1993zh,Coriano:2003ui} 
and \cite{Ellis:1997ni,Ellis:1998nk}. Both studies concluded that small Majorana
masses could be generated if one made some assumptions about dynamics
in the hidden sector. In  \cite{Faraggi:1993zh,Coriano:2003ui} the masses
were associated with an extended see-saw involving a low mass scale.
The see-saw found in  \cite{Ellis:1997ni,Ellis:1998nk} was of the
canonical type I type, but in detail it was rather different to
GUT-type models.  
A see-saw was also claimed in a heterotic $Z_3$ orbifold model with
$E_6$ breaking to $SU(3)\times SU(3)\times SU(3)$ \cite{Kim:2004pe}. 
A recent study of $Z_6$ orbifold constructions found Majorana-type
operators \cite{Kobayashi:2004ya}, but (to the order studied) the
$S_i$ fields did not have the required expectation values when
$R$-parity is conserved. 

In \cite{Giedt:2005vx} a large class of vacua of the bosonic $Z_3$ orbifold were
analysed with emphasis on the neutrino sector to determine whether the minimal type I
see-saw is common, or if not to find possible
guidance to model building, and possibly to get clues concerning textures and mixing
if examples were found.
 Several examples from each of 20 patterns of vacua were studied, and the non-zero
superpotential terms through degree 9 determined.
There were a huge number of $D$-flat directions, with the number reduced
greatly by the $F$-flatness condition.
Only two of the patterns had Majorana mass operators, while none
had simultaneous Dirac operators of low enough degree to allow neutrino
masses larger than $10^{-5}$ eV. (One apparently successful model was
ruined by off-diagonal Majorana mass terms.) It is not clear whether this failure
to obtain a minimal see-saw is a feature of the particular class of construction,
or whether it is suggesting that string constraints and selection rules might
make string vacua with minimal see-saws rare. Systematic analyses of
the neutrino sector of other classes of constructions would be very useful.

There are other possibilities for obtaining small neutrino masses in string 
constructions, such as
extended see-saws \cite{Faraggi:1993zh,Coriano:2003ui} and
small Dirac masses from higher dimension operators \cite{Langacker:1998ut}. 
Small Dirac neutrino masses in models with anisotropic compactifications 
motivated by type I strings \cite{Ibanez:1998rf} have been 
discussed recently in \cite{Antusch:2005kf}.  
The possibility of embedding type II
see-saw ideas (involving Higgs triplets) in heterotic string constructions
was considered in \cite{Langacker:2005pf}. It is
possible to obtain a Higgs triplet of $SU(2)$ with non-zero hypercharge
in a higher level construction (in which $SU(2) \times SU(2)$ is broken
to a diagonal subgroup). In this case, because of the underlying $SU(2)
\times SU(2)$ symmetry the Majorana mass matrix for the light neutrinos
should involve only off-diagonal elements (often with one of the three
off-diagonal elements small or vanishing). This leads to phenomenological
consequences very different from those of triplet models that have been motivated by
grand unification or bottom-up considerations, including an inverted hierarchy,
two large mixings, a value of $U_{e3}$ induced from the charged-lepton
mixings that is close to the current experimental lower limit, and an
observable neutrinoless double-beta decay rate.  
This string version of the triplet model is a top-down motivation for
the $L_e-L_\mu-L_\tau$-conserving models that have previously been
considered from a bottom-up point of view \cite{Mohapatra:2005wg}, but
has the advantage of allowing small mixings from the charged-lepton
sector. 
A recent study indicates that it may also be
possible to generate a type II see-saw in intersecting D6-brane models
involving $SU(5)$ grand unification, although the examples constructed are
not very realistic ~\cite{Cvetic:2006by}.

These comments indicate that string constructions may be very different from
traditional grand unification or bottom-up constructions, mainly because of the
additional string constraints and symmetries encountered. Versions of the minimal see-saw
(though perhaps with non-canonical family structure) are undoubtedly present 
amongst the large landscape of string vacua, though perhaps they are rare. 
One point of view is to simply focus on the search for such string vacua. However,
another is  to keep an open mind about other possibilities that may appear
less elegant from the bottom-up point of view but which may occur
more frequently in the landscape.

\subsubsection{TeV scale mechanisms for small neutrino masses}

Neutrino mass may arise in a class of non-SUSY models 
via $L=2$ scalar-lepton-lepton Yukawa interactions.
The Lagrangian can be written generically as follows:
\begin{equation}
  -{\cal L}^{yuk}= f_{ij}H^{++}l_il_j +  g_{ij}H^+l_i\nu_j  + 
  h_{ij}H^0\nu_i\nu_j + \mbox{\rm hermitian~conjugate} \; .
\label{gen_lagran}
\end{equation}
Here $H^{\pm\pm},H^\pm$ and $H^0$ are doubly-charged, singly-charged
and neutral scalars respectively which originate from an $SU(2)_{L,R}$
isospin singlets ($I=0$) or triplets ($I=1$). Each scalar is assigned
$L=2$. 
The charged leptons ($l^\pm$) and neutrinos ($\nu$) may be of either
chirality. 
Four examples of models which utilise various terms in ${\cal
L}^{yuk}$ to generate neutrino mass are listed below: 
\begin{itemize}
  \item {\it The left-right symmetric model}:
        TeV scale breaking of $SU(2)_R$ via the right-handed scalar
        triplet vacuum expectation value which gives rise to 
        a TeV scale see-saw mechanism \cite{Mohapatra:1979ia};
  \item {\it Higgs Triplet Model}:
        Tree-level neutrino mass for the observed neutrinos
        proportional to $SU(2)_L$ triplet scalar vev (no right-handed
        neutrino) \cite{Gelmini:1980re}; 
  \item {\it Zee model}:
        Radiative neutrino mass at 1-loop via $SU(2)_L$ singlet scalar
        $H^\pm$ \cite{Zee:1980ai}; and
  \item {\it Babu model}:
        Radiative neutrino mass at 2-loop via $SU(2)_L$ singlet 
        scalars $H^{\pm\pm}$ and $H^\pm$ \cite{Babu:1988ki}.
\end{itemize}

All the above models can provide TeV-scale mechanisms
of neutrino mass generation consistent with current
neutrino-oscillation experiments. 
New particle discovery
(e.g. $Z',W',H^{\pm\pm})$ at the Large Hadron Collider (LHC)
is also a possibility if 
$M_{Z'},M_{W'}< 3-4$ TeV, $M_{H^{\pm\pm}}<1$ TeV.
Precision measurements of the neutrino-mass matrix
at a Neutrino Factory would
provide valuable information on the Yukawa couplings
$f,g,h$. Such couplings also induce lepton-flavour
violating (LFV) decays (e.g. $\mu \to eee, \mu \to e \gamma$)
~\cite{Bilenky:1987ty,Petcov:1982en},
which might also form part of the research programme at a Neutrino
Factory. Importantly, any signal for 
$\mu \to e \gamma$ from the MEG experiment can be interpreted
in the above models. 
The first pair of models above can accommodate any value of $\sin\theta_{13}$
and any of the currently allowed mass hierarchies, 
normal (NH), inverted (IH) and degenerate (DG).
The second pair of models above are more predictive for $\sin\theta_{13}$
and accommodate specific neutrino mass hierarchies.
A distinctive feature of all the models is the synergy between
precision measurements of oscillation parameters (at a Neutrino Factory),
LFV decays of $\mu$ and $\tau$, and direct searches for the $L=2$
scalars, all of which involve the couplings $f,g,h$.

{\bf Left-Right Symmetric Model}

The left-right (LR) symmetric model \cite{Pati:1974yy}
is an extension of the Standard Model
based on the gauge group $SU(2)_R \otimes SU(2)_L \otimes U(1)_{B-L}$.
The LR-symmetric model has many virtues, e.g.:
\begin{itemize}
  \item The restoration of parity as an original symmetry of the
        Lagrangian which is broken spontaneously by a Higgs vev; and
  \item The replacement of the arbitrary SM hypercharge $Y$ by the
        theoretically more attractive $B-L$. 
\end{itemize}
Although the Higgs sector is
arbitrary, a theoretically and phenomenologically 
appealing way to break the $SU(2)_R$ gauge symmetry 
is by invoking Higgs isospin-triplet representations. 
Such a choice conveniently 
allows the implementation of a low energy see-saw mechanism
for neutrino masses. 
A right-handed neutrino is required by the $SU(2)_R$ gauge group
and leptons are assigned to multiplets with quantum numbers 
($T_L, T_R, B-L$):
\begin{eqnarray}
\label{eq:matter}
  L_{iL}&=&\left( \begin{array}{c}
\nu^\prime_i\\l^\prime_i\end{array}\right)_{L}: \left(1/2:0:-1\right),~
L_{iR}=\left( \begin{array}{c}
\nu^\prime_i\\l^\prime_i\end{array}\right)_{R}: \left(0:1/2:-1\right) \,.
\end{eqnarray}
Here $i=1,2,3$ denotes generation number.
The Higgs sector consists of a bidoublet Higgs field, $\Phi$,
and two triplet Higgs fields, $\Delta_L$ and $\Delta_R$:
\begin{eqnarray} 
  \Phi & = &
    \left( \begin{array}{cc} \phi_1^0  \phi_2^+\\\phi_1^- 
    \phi_2^0\end{array}\right): \left(1/2:1/2:0\right) \ , \nonumber \\
  \Delta_{L} & = &
    \left( \begin{array}{cc}
    \delta_{L}^+/\sqrt{2}  \delta_{L}^{++} \\ \delta_{L}^0 
    -\delta_{L}{^+}/\sqrt{2}
    \end{array}\right): \left(1:0:2\right)  \ ,      \\
  \Delta_{R} & = & 
    \left( \begin{array}{cc}
    \delta_{R}^+/\sqrt{2} \delta_{R}^{++} \\ \delta_{R}^0 
    -\delta_{R}{^+}/\sqrt{2}
    \end{array}\right): \left(0:1:2\right) \ . \nonumber
  \label{trip_rep}
\end{eqnarray}
The vevs for these fields are as follows:
\begin{eqnarray}
\langle\Phi\rangle&=&\left( \begin{array}{cc} \kappa_1 & 0\\ 0 &
\kappa_2\end{array}\right)\frac{1}{\sqrt{2}},~~ 
\langle\Delta_{L}\rangle=\left( \begin{array}{cc}0& 0\\ v_{L} & 0\end{array}\right)\frac{1}{\sqrt{2}} \ ,~~ 
\langle\Delta_{R}\rangle=\left( \begin{array}{cc}0& 0\\ v_{R} & 0\end{array}\right)\frac{1}{\sqrt{2}} \ .
\end{eqnarray}
The gauge groups 
$SU(2)_R$ and $U(1)_{B-L}$ are spontaneously broken at the scale $v_R$.
Phenomenological considerations require $v_R \gg \kappa 
=\sqrt{\kappa_1^2+\kappa_2^2}\sim \frac{2M_{W_1}}{g}$.
The vev $v_L$ does not play a role in the breaking of the 
gauge symmetries and is constrained to be small ($v_L < 8$ GeV)
in order to comply with the measurement of 
$\rho=M^2_Z\cos^2\theta_W/M^2_W\sim 1$. 
The Lagrangian responsible for generating neutrino mass is as follows:
\begin{eqnarray}
-{\cal L}=\bar L_L(y_D\Phi+\tilde y_D\tilde \Phi)L_R 
+iy_M(L_L^TC\tau_2\Delta_LL_L+L_R^TC\tau_2\Delta_RL_R) 
+ \mbox{\rm hermitian~conjugate} \; ,
\end{eqnarray}
where $y_M$ is a $3\times 3$ Majorana-type Yukawa coupling matrix.
Expanding the terms proportional to $y_M$ results in a
Lagrangian of the form of equation (\ref{gen_lagran}) with 
$y_M=f=\sqrt{2} g=h$.
The $6 \times 6$ mass matrix for the neutrinos can be written 
in the block form:
\begin{eqnarray}
M^{LR}_\nu=\left( \begin{array}{cc}
M_L & m_D
\\
m_D^T & M_R
\end{array}\right) \; .
\label{eq:Mnu}
\end{eqnarray}
Each entry is given by
\begin{eqnarray}
m_D
={1\over \sqrt{2}}\left(y_D\kappa_1+{\tilde y_D \kappa_2}\right);\;\; 
M_R=\sqrt{2}h v_R; \;\; 
M_L=\sqrt{2}h v_L . 
\end{eqnarray}
The neutrino mass matrix is diagonalised by a $6\times 6$ unitary matrix 
$V$
as $V^TM_\nu V=M_\nu^{diag}={\rm diag}(m_1,m_2,m_3,M_1,M_2,M_3)$, 
where $m_i$ and $M_i$
are the masses for neutrino mass eigenstates.
The small neutrino masses $m_i$ are generated by the 
Type II see-saw mechanism. Obtaining eV scale neutrino masses 
with $h={\cal O} (0.1-1)$
requires $M_L$ (and consequently $v_L$)
to be at the eV scale. 
In LR-model phenomenology, with $v_R\sim$ TeV,
it is customary to arrange the Higgs potential such that $v_L=0$
\cite{Deshpande:1990ip}.
In this case the masses of the light neutrinos arise from  
the Type I see-saw mechanism and are approximately
$m_i\sim m_D^2/M_R$. 
In order to realise the low-energy 
($\sim {\cal O}(1-10)$ TeV) scale for the right-handed
Majorana neutrinos, 
the Dirac mass term, $m_D$, should be {$\cal O$} (MeV), 
which for $\kappa_2\sim 0$
corresponds to $y_D\sim 10^{-6}$ 
(i.e. comparable in magnitude to
the electron Yukawa coupling). 
The LR model with $v_R$ of order a TeV predicts
lepton-flavour violating (LFV) decays of the muon and tau mediated by
$H^{\pm\pm}$ with a rate $\sim |hh|^2/M_{H^{\pm\pm}}^4$
\cite{Cirigliano:2004mv}, and a rich phenomenology in direct searches
at the LHC. 

{\bf Higgs Triplet Model}

In the Higgs Triplet Model (HTM) \cite{Schechter:1980gr},
\cite{Cheng:1980qt}
\footnote{
The model of \cite{Gelmini:1980re} contains a triplet majoron
and was excluded by LEP data. A viable extension of the HTM which
contains a singlet majoron (referred to as the "123" model) was
introduced in ref. \cite{Schechter:1981cv}.}
a single $I=1$, $Y=2$ complex $SU(2)_L$ triplet $\Delta_L$
(see equation (\ref{trip_rep}))
is added to the SM with the Yukawa coupling:
\begin{equation}
iy_M(L_L^TC\tau_2\Delta_LL_L) + {\rm hermitian~conjugate} \; .
\label{HTM_lagran}
\end{equation}
Expanding equation (\ref{HTM_lagran}) results in 
equation (\ref{gen_lagran}) with $y_M=f=\sqrt{2} g=h$.
No right-handed neutrino is introduced, and 
the light neutrinos receive a Majorana mass proportional
to the left-handed triplet vev $(v_L)$
leading to the following neutrino mass matrix:
\begin{equation}
{\cal M}^{HTM}_{\nu}=\sqrt 2 v_{L}h_{ij} \; .
\label{HTMneutmass}
\end{equation}
The presence of a trilinear coupling $\mu\Phi^Ti\tau_2\Delta_L^\dagger\Phi$ 
(where $\Phi$ is the SM Higgs doublet with vev $v$) in the
Higgs potential ensures a non-zero $v_L\sim \mu v^2/M^2$,
where $M$ is the mass of the triplet scalars.
Taking $M$ to be at the TeV scale results in $v_L\sim \mu$. 
From equation (\ref{HTMneutmass}) it is apparent that
the HTM does not provide predictions for the
elements of ${\cal M}_{\nu}$ but instead 
accommodates the observed values (as does the LR model).
However, combining accurate measurements
of the neutrino oscillation parameters with 
any signals in LFV processes involving the muon or the tau
\cite{Chun:2003ej} and/or direct observation of 
$H^{\pm\pm}$ \cite{Akeroyd:2005gt} would
enable this mechanism of neutrino mass generation to be
tested. From equation (\ref{HTMneutmass})
$h_{ij}$ is directly related to the neutrino
masses and mixing angles as follows:
\begin{equation}
h_{ij}=\frac{1}{\sqrt{2}v_L}V_{_{\rm PMNS}}diag(m_1,m_2,m_3)
V_{_{\rm PMNS}}^T \; .
\label{hij}
\end{equation}
Observation of LFV decays of the muon for example at MEG and/or
of the tau (at a Super B Factory) together with discovery of 
$H^{\pm\pm}$ (at LHC) would permit measurements of $h_{ij}$.
A Neutrino Factory would greatly reduce the experimental error
in the right-hand side of equation(\ref{hij}) and allow
the above identity in the HTM to be checked precisely.

{\bf One loop radiative mechanism via a singly-charged, singlet scalar 
     (Zee model)} 

\begin{figure}[ht]
\begin{center}
\includegraphics[width= 10 cm, angle= 0]{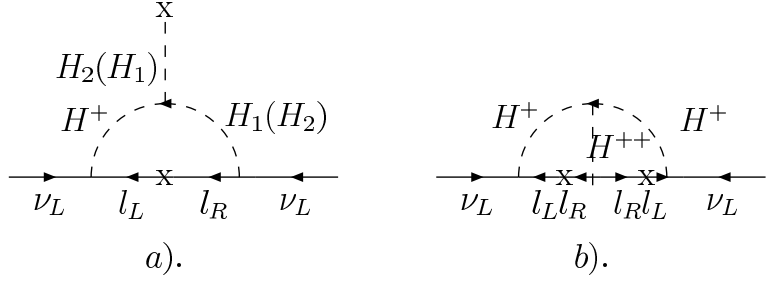}
\end{center}
\caption{Diagram for neutrino mass generation in
a) Zee model, and b) Babu model}
\label{zee_babu}
\end{figure}

A singly-charged, singlet scalar is added to the Two Higgs Doublet
Model (2HDM) extension of the SM.
Neutrino mass is generated radiatively via a 1-loop diagram
figure \ref{zee_babu}a
in which the mixing between the charged singlet and doublet scalars
(proportional to a trilinear coupling $\mu$) is crucial \cite{Zee:1980ai}.
The relevant Lagrangian is:
\begin{equation}
{\cal L}^{Zee}=   g_{ab}\left( L^{T i}_{aL} C L^j_{bL} \right)
\epsilon_{ij} H^+  +\sum_{i=1,2}y_k\overline L_L H_i l_R + 
\mbox{\rm hermitian conjugate} \; ,
\end{equation}
where $y_k$ is the Yukawa coupling of the doublet $H_k$ to the leptons.
If only one of the Higgs doublets couples to leptons 
(referred to as the ``minimal Zee model'') 
the resulting neutrino mass matrix
is symmetric with vanishing diagonal elements:
\begin{equation}
{\cal M}^{Zee}_\nu=\begin{pmatrix}
0 & m_{e \mu} & m_{e \tau} \cr
                 m_{e \mu} & 0 & m_{\mu \tau} \cr
                 m_{e \tau} & m_{\mu \tau} & 0
\end{pmatrix}
\end{equation}
where~\cite{Petcov:1982en}
\begin{equation}
m_{ij}=g_{ij}(m^2_{l_j}-m^2_{l_i})\mu F\frac{1}{16\pi^2}\frac{1}
{m^2_{S_1}-m^2_{S_2}}{\rm ln}\frac{m^2_{S_1}}{m^2_{S_2}}
\end{equation}
and $m_{l_i}$ are the
charged lepton masses, $M_{S_i}$ are the charged scalar masses
and $F=\cot\beta$ ($\tan\beta$) for Type I (II) couplings of the
doublets to the leptons.
The above mass matrix predicts the solar angle to be almost maximal,
which is now ruled out at the $6\sigma$ level (see section \ref{SubSubSect:SolarandReactor}).
However, allowing both Higgs doublets to couple
to the leptons (the ``general Zee model'') leads 
to non-zero diagonal elements in ${\cal M}^{Zee}_\nu$
\cite{Balaji:2001ex}. The non-maximal solar angle can
then be accommodated, $\sin\theta_{13}\ne 0$ is
expected, and an inverted hierarchical neutrino mass pattern
is predicted.

{\bf Two loop radiative mechanism via singly and doubly-charged,
     singlet scalars (Babu model)}  

$SU(2)_L$ singlet charged scalars $H^{\pm\pm}$ and $H^\pm$ are added
to the SM Lagrangian \cite{Babu:1988ki} with the following Yukawa 
couplings:
\begin{equation}
{\cal L} = \ f_{ab} \left( l^T_{aR} C l_{bR} \right) H^{++}
+ g_{ab}\left( L^{T i}_{aL} C L^j_{bL} \right)
\epsilon_{ij} H^+ \ + \mbox{\rm hermitian~conjugate} \; .
\end{equation}
No right-handed neutrino is
introduced. A Majorana mass for the light neutrinos 
arises at the two loop level (figure \ref{zee_babu}b)
in which the lepton number violating trilinear coupling 
$\mu H^\pm H^\pm H^{\pm\pm}$ plays a crucial role. 
The explicit form for ${\cal M}_{\nu}$ is as follows:
\begin{eqnarray}
  {\cal M}^{Babu}_{\nu} = \zeta
  \times \left(
\begin{array}{ccc}
\epsilon^2 \omega_{\tau \tau} + 2 \epsilon \epsilon' \omega_{\mu \tau} + 
\epsilon'^2 \omega_{\mu \mu}\ , &
\epsilon \omega_{\tau \tau}  + \epsilon' \omega_{\mu \tau} - 
\epsilon \epsilon' \omega_{e \tau}
  &
-\epsilon \omega_{\tau\tau} -\epsilon' \omega_{\mu\mu} - \epsilon^2 
\omega_{e\tau}  \\
 &  \hbox{~~~~~~~~~} - \epsilon'^2 \omega_{e \mu} \ , &
 \hbox{~~~~~~~~~} - \epsilon \epsilon' \omega_{e\mu} \\
. & \omega_{\tau\tau} -2 \epsilon' \omega_{e\tau} + \epsilon'^2 
\omega_{ee} \ , &
-\omega_{\mu\tau} -\epsilon \omega_{e\tau} + \epsilon' \omega_{e\mu}  \\
 & & \hbox{~~~~~~~~~} + \epsilon \epsilon' \omega_{ee} \\
. & . & \omega_{\mu\mu} + 2 \epsilon \omega_{e\mu} 
+ \epsilon^2 \omega_{ee}
\end{array}
\right) \; ,
\end{eqnarray}
where $\epsilon=g_{e\tau}/g_{\mu\tau}$, $\epsilon'=g_{e\mu}/g_{\mu\tau}$,
$\omega_{ab}=f_{ab}m_a m_b$ ($m_a,m_b$ are charged lepton masses)
and $\zeta$ is given by:
\begin{equation}
\zeta=\frac{8\mu g^2_{\mu\tau}\tilde I}{(16\pi^2)^2m_{H^\pm}^2} \; .
\end{equation}
Here $\tilde I$ is a dimensionless quantity of ${\cal O}(1)$
originating from the loop integration.
The expression for ${\cal M}_{\nu}$ involves 9 arbitrary couplings.
Since the model predicts one massless neutrino (at the two-loop level),
quasi-degenerate neutrinos are not permitted
and only normal-hierarchy (NH) and inverted-hierarchy (IH) mass
patterns can be accommodated. 
The $g$ couplings (contained in 
$\epsilon$ and $\epsilon'$) are directly related to the elements
of ${\cal M}_{\nu}$, and thus would be obtained precisely at a 
Neutrino Factory. In the scenario of NH, $\epsilon\approx \epsilon'\approx 
\tan\theta_{12}/\sqrt 2$
and $\sin\theta_{13}$ is close to zero. Since $\epsilon,\epsilon'<1$
one may neglect those terms in ${\cal M}_{\nu}$ which 
are proportional to the electron mass (i.e. $\omega_{ee},\omega_{e\mu},
\omega_{e\tau}$).
This simplification leads to the following prediction: $f_{\mu\mu}:f_{\mu\tau}:
f_{\tau\tau}\approx 1:m_\mu/m_\tau:(m_\mu/m_\tau)^2$.
In the case of IH, large values  
are required for $\epsilon,\epsilon'(>5)$, and thus neglecting
  $\omega_{ee},\omega_{e\mu},
\omega_{e\tau}$ in ${\cal M}_{\nu}$
may not be entirely justified. However, if such terms are neglected
then the above prediction for the ratio
of $f_{\mu\mu}:f_{\mu\tau}:f_{\tau\tau}$ also holds approximately 
for the case of IH. A lower bound on $s_{13}>0.05$ can also be derived.
If the 2-loop diagram is solely responsible for the generation
of the neutrino mass matrix the Babu model requires 
$g,f_{\mu\mu}\sim 10^{-2}$.
Such relatively large couplings may lead to observable rates
for LFV decays of muons and taus.

%% file: 03-New-physics-and-the-SnuM/03-02-Uni-and-flav/03-02-Uni-and-flav.tex
\subsection{Unification and Flavour}

A survey of the theoretical models that have been developed to
explain the physics of flavour is presented in this section.
Measurables that can be used to distinguish between the various models
is also presented. 
These measurables include the mixing angles themselves and
combinations of mixing angles, the latter are referred to as `sum
rules'.
This section also contains a discussion of lepton-flavour violation.

\subsubsection{Model survey}

To understand the origin of the postulated forms
of the Yukawa matrices, one must appeal to some sort of
Family symmetry, $G_{{\rm Family}}$.
In the framework of the see-saw mechanism, new physics beyond the
Standard Model is required to cause lepton-number conservation to be
violated and to generate right-handed neutrino masses at
around the GUT scale. 
This is exciting since it implies that
the origin of neutrino masses is related to a
GUT symmetry group $G_{{\rm GUT}}$, which unifies the
fermions within a family.
Putting these ideas together leads to the development of a framework
for physics beyond the SM which is based on $N = 1$ super-symmetry
with commuting GUT and Family symmetry groups,
$G_{{\rm GUT}}\times G_{{\rm FAM}}$.
There are many possible candidate GUT and Family symmetry
groups. 
Unfortunately the model dependence does not end there; the
details of the symmetry-breaking vacuum plays a crucial role in
specifying the model and in determining the masses and mixing angles.
These models may be classified according to the particular
GUT and Family symmetry that is assumed. 

It may be possible to use precise measurements of the oscillation
parameters to distinguish between different models.
A survey of over sixty neutrino-mass models has been performed. 
The survey included:
\begin{itemize}
  \item Models with assumptions about the structure of the mixing 
        matrix ('texture' assumptions);
  \item Models based on lepton symmetries such as $A_{4}$, $S_{3}$, or
        $L_{e}-L_{\mu}-L_{\tau}$; and
  \item Models based on GUT symmetries such as $SU(5)$, flipped
        $SU(5)$, $SO(10)$, $E_{6}$, or $E_{8} \times E_{8}$. 
\end{itemize}
These models are reviewed briefly below with emphasis on how the
different predictions arise from different symmetry-breaking
patterns. 
A detailed, tabulated summary of the predictions for all three angles
with references to models that have been included in our survey can be
found in reference \cite{Albright:2006cw}.   

{\noindent \bf Models with Lepton Symmetries based on {\boldmath $\mu-\tau$} Symmetry}

The maximal (or near maximal) mixing observed in atmospheric neutrinos
strongly suggests a $\mu-\tau$ symmetry in the neutrino-mass
matrix.
There are two ways to realise the $\mu-\tau$ symmetry which give rise
to maximal mixing in the atmospheric-neutrino sector, 
$\theta_{23} = \frac{\pi}{4}$ \cite{Mohapatra:2004mf}. 
The first possibility is of the following form:
\begin{equation}
  M_{\nu} \simeq 
  \left(
    \begin{array}{ccc}
      0 & 0 & 0\\
      0 & 1 & 1\\
      0 & 1 & 1
    \end{array}
  \right) 
  \; ,
\end{equation}
which gives rise to the normal mass hierarchy. 
In this case, when the
$\mu-\tau$ symmetry is exact, the 1-3 mixing angle vanishes,
$\sin\theta_{13}=0$. 
In addition, the mass splitting in the solar
neutrino sector vanishes, $\Delta m_{12}^{2} = 0$. 
Non-vanishing $\Delta m_{12}^{2}$ can be generated in a $\mu-\tau$
symmetric way by adding small parameters of the order of
$\mathcal{O}(\epsilon \ll 1)$,  
\begin{equation}
  M_{\nu} \simeq \frac{\sqrt{\Delta m_{13}^{2}}}{2} \left(
  \begin{array}{ccc}
    c\epsilon & d \epsilon & d\epsilon\\
    d\epsilon & 1 + \epsilon & -1\\
    d\epsilon & -1 & 1 + \epsilon
  \end{array}\right) 
  \; ,
\end{equation}
where the coefficients $c$ and $d$ are of order $1$. This leads to,
\begin{equation}
\theta_{13} = 0, \; \theta_{23} = \frac{\pi}{4}, \; \tan 2\theta_{12}
\simeq \frac{2\sqrt{2}d}{(1-c)} \; , 
\end{equation} 
and the parameter $\epsilon$ is fixed by the ratio of $\Delta
m_{13}^{2}$ and $\Delta m_{12}^{2}$ as: 
\begin{equation}
  \epsilon = \frac{4}{1+c+\sqrt{(c-1)^{2} + 8d^{2}}} 
             \sqrt{\frac{\Delta m_{12}^{2}}{\Delta m_{13}^{2}}} \; . 
\end{equation}
In order to generate non-zero $\theta_{13}$, the $\mu-\tau$ symmetry
has to be broken.  
How the symmetry breaking occurs dictates the size of the
$\theta_{13}$ angle. 
The $\mu-\tau$ symmetry breaking also causes
$\theta_{23}$ to differ from $\frac{\pi}{4}$, i.e. the mixing is no
longer maximal.    
The breaking of the $\mu-\tau$ symmetry can generally be parametrised
as:
\begin{equation}
M_{\nu} \simeq \frac{\sqrt{\Delta m_{13}^{2}}}{2} \left(
\begin{array}{ccc}
c\epsilon & d \epsilon & b\epsilon\\
d\epsilon & 1 + a\epsilon & -1\\
b\epsilon & -1 & 1 + \epsilon
\end{array}\right) \; ,
\end{equation}
where the parameter $a$ is of order unity. If the breaking is introduced in the $e$-sector, that is, 
$a = 1$, $b \ne d$, one then has:
\begin{equation}
\epsilon = \frac{4}{\sqrt{1+8d^{2}}} \sqrt{\frac{\Delta m_{12}^{2}}{\Delta m_{13}^{2}}}, \quad 
\tan 2 \theta_{12} \simeq \frac{2(b+d)}{(1-c)} \; ,
\end{equation}
and a non-vanishing $\theta_{13}$ angle:
\begin{equation}
\theta_{13} = (b-d) \sqrt{\frac{\Delta m_{12}^{2}}{\Delta m_{13}^{2}}} \; .
\end{equation}
A non-vanishing deviation of the atmospheric mixing angle from
$\frac{\pi}{4}$ can exist with magnitude
$\frac{\pi}{4} - \theta_{23} \sim \mathcal{O}(\epsilon^{2})$. 
The breaking of the
$\mu-\tau$ symmetry can also be introduced in the $\mu-\tau$
sector. This is characterised by $a \ne 1$ and $b = d$.  
In this case, the parameter $\epsilon$ is related to 
$\Delta m_{12}^{2}$ and $\Delta m_{13}^{2}$ by:
\begin{equation}
\epsilon = \frac{4}{c + \frac{1}{2}(1+a) + \sqrt{(c -
\frac{1}{2}(1-c))^{2} + 8 d^{2}}} \sqrt{\frac{\Delta
m_{12}^{2}}{\Delta m_{13}^{2}}} \; .  
\end{equation}
Thus, the predictions for $\sin\theta_{13}$ and $\pi / 4 -
\theta_{23}$ strongly depend on the symmetry-breaking pattern.  
Table~\ref{table:lep.sym.nor} summarises the predictions for
$\theta_{13}$ and for $\frac{\pi}{4} - \theta_{23}$ for various
symmetry-breaking scenarios.  
\begin{table}[t!]
\caption{Predictions for $\theta_{13}$ and for the deviation
$(\theta_{23}-\pi/4)$ in models with softly broken $\mu-\tau$ symmetry
for different symmetry breaking directions. This table is taken from
reference \cite{Mohapatra:2004mf}.} 
\begin{center}
\begin{tabular}{|c|c|c|}
\hline
symmetry breaking & $\theta_{13}$ & $\theta_{23}-\frac{\pi}{4}$\\
\hline
none & 0 & 0\\
$\mu-\tau$ sector only & $\sim \Delta m_{12}^{2}/\Delta m_{13}^{2}$ &
$\lesssim 8^{o}$\\ 
$e$-sector only & $\sim \sqrt{  \Delta m_{12}^{2}/\Delta m_{13}^{2}
}$  & $\lesssim 4^{o}$ \\ 
dynamical & $\sim \sqrt{  \Delta m_{12}^{2}/\Delta m_{13}^{2}  }$ &
large\\ 
\hline
\end{tabular}
\end{center}
\label{table:lep.sym.nor}
\end{table}%

The inverted mass hierarchy can be obtained when the neutrino mass
matrix is of the form: 
\begin{equation}
  M_{\nu} \simeq 
  \left(
    \begin{array}{ccc}
      0 & 1 & 1\\
      1 & 0 & 0\\
      1 & 0 & 0
    \end{array}
  \right) 
  \; .
\end{equation}
This mass matrix has an enhanced $L_{e}-L_{\mu}-L_{\tau}$ symmetry
~\cite{Petcov:1982ya,Schechter:1981hw} and is a special case of the following mass matrix:
\begin{equation}
M_{\nu} \simeq \sqrt{\Delta m_{13}^{2}} \left(
\begin{array}{ccc}
0 & \sin\theta & \cos\theta\\
\sin\theta & 0 & 0\\
\cos\theta & 0 & 0
\end{array}\right) \; .
\end{equation}
In the exact $L_{e}-L_{\mu}-L_{\tau}$ symmetric limit, this leads to
the following predictions \cite{Petcov:1982ya}:
\begin{equation}
\Delta m_{12}^{2} = 0, \quad \theta_{13}=0, \quad \theta_{12} = \frac{\pi}{4}, \quad \sin^{2}2\theta_{23} = \sin^{2} 2\theta \quad .
\end{equation}
Since $\theta_{12} \ne \frac{\pi}{4}$, 
the $L_{e}-L_{\mu}-L_{\tau}$ symmetry has to be softly broken. The
soft breaking of the $L_{e}-L_{\mu}-L_{\tau}$ symmetry can be
introduced by adding small $e-e$, $\mu-\mu$, $\mu-\tau$ and
$\tau-\tau$ couplings:
\begin{equation}
M_{\nu} \simeq \sqrt{\Delta m_{13}^{2}} \left(
\begin{array}{ccc}
z & \sin\theta & \cos\theta\\
\sin\theta & y & d\\
\cos\theta & d & x
\end{array}\right), \qquad 
x,\; y, \; d \ll 1 \qquad .
\end{equation}
For non-zero $x, \; y$ and $d$, one has:
\begin{equation}
\sin^{2} 2 \theta_{12} \simeq 1 - \biggl( \frac{\Delta
m_{12}^{2}}{4\Delta m_{13}^{2}} - z \biggr)^{2} \; . 
\end{equation} 
The breaking of the $\mu-\tau$ symmetry can arise in the $\mu-\tau$
sector, {\it i.e.},  
$\cos\theta = \sin\theta = 1/\sqrt{2}$ and $x \ne y$, 
which leads to:
\begin{equation}
\theta_{13} = \frac{1}{2}(x-y), \quad \frac{\Delta m_{12}^{2}}{\Delta
m_{13}^{2}} = 2 (x+y+z+d) \quad . 
\end{equation}
The breaking of the $\mu-\tau$ symmetry can also be introduced in the
$e$-sector by having $\cos\theta \ne \sin\theta$ and $x = y$.
This leads to $\theta_{13} \simeq -d \cos 2\theta_{23}$. 
In the inverted hierarchy case, the correlations among the 
neutrino-mixing angles is not as strong as in the normal-hierarchy
case. 

{\noindent \bf Single-RH neutrino dominance}

Single-RH neutrino dominance (SRND), proposed in
\cite{King:1999cm}, can be implemented in many classes of
model; it is therefore a mechanism rather than a model. 
SRND provides a natural way to generate large mixing angles. 
In the simplified case, with only the second and
third families, the Dirac neutrino-mass matrix and RH Majorana
neutrino-mass matrix are generally of the form:
\begin{equation}
 M_{D} = \left(\begin{array}{ccc}
 \cdot & \cdot & \cdot\\
 \cdot & a & b\\
 \cdot & c & d
 \end{array}\right), \quad 
 M_{R} = \left(\begin{array}{ccc}
 \cdot & \cdot & \cdot\\
 \cdot & x & 0\\
 \cdot & 0 & y
 \end{array}\right) \quad , 
\end{equation}
in the basis where the RH Majorana neutrino-mass matrix is diagonal. 
The effective light neutrino-mass matrix is then given by:
\begin{equation}
m_{\nu} = -M_{D} \cdot M_{R}^{-1} \cdot M_{D}^{T} = \left(
\begin{array}{ccc}
\cdot & \cdot & \cdot\\
\cdot & \frac{a^{2}}{x} + \frac{b^{2}}{y} & \frac{ac}{x} + \frac{bd}{y}\\
\cdot & \frac{ac}{x}+\frac{bd}{y} & \frac{c^{2}}{x} + \frac{d^{2}}{y}
\end{array}\right) \; .
\end{equation}
If one RH neutrino dominates, that is, if $y \gg x$, then the sub-determinant in the $\mu-\tau$ block is roughly 
of the order $\sim m_{2}\cdot m_{3}$.   The normal hierarchy is
obtained for $m_{2} \ll m_{3}$. The atmospheric mixing angle is
roughly given by $\tan\theta_{23} \sim (a/c)$. For $a \sim c$, large
mixing angles can arise naturally. The two-family case can be
generalised to the three-family case when sequential dominance
with three RH neutrinos is implemented~\cite{Antusch:2004gf}. 

{\noindent \bf Models with GUT Symmetries}

Grand Unified Theories based on $SO(10)$ accommodate all 16 fermions
(including the right-handed neutrinos) in a single spinor
representation. Furthermore, $SO(10)$ provides a framework in which
the see-saw mechanism arises naturally. Models based on $SO(10)$
combined with a continuous, or discrete, flavour symmetry group have
been constructed to understand the flavour problem, especially the
small neutrino masses and the large leptonic mixing angles. These
models can be classified according to the family symmetry that is
implemented as well as the Higgs representations introduced in the
model. 
For reviews, see, for example, reference \cite{Chen:2003zv}. 
Phenomenologically, the resulting mass matrices can be either
symmetric, lop-sided, or asymmetric.

Due to the product rule, $16 \otimes 16 = 10 \oplus 120_{a}  \oplus
126_{s}$, the only Higgs particles that can couple to the matter
fields at tree level are in the 10, $\overline{120}$, and
$\overline{126}$ representations of SO(10). The Yukawa matrices
involving the 10 and $\overline{126}$ are symmetric under
interchange of family indices, while the matrix involving the 
$\overline{120}$ is anti-symmetric. The Majorana mass term for the RH
neutrinos can arise either from a renormalisable operator involving
the $\overline{126}$, or from a non-renormalisable operator that
involves the $16$s. The case of $\overline{126}$ has the advantage
that R-parity is preserved automatically.  

Two large mixing angles in the leptonic sector may arise in two ways:

\begin{enumerate}
  \item {\it Symmetric mass textures:} 
  This scenario is realised if $SO(10)$ is broken through the left-right 
  symmetry-breaking route. In this case, both the large solar mixing 
  angle and the maximal atmospheric mixing angle come from  
  the effective neutrino-mass matrix. A characteristic of this class
  of models is that  
  the predicted value for the $|U_{e\nu_{3}}|$ element tends to be
  larger than the value predicted by models in class (ii) below. This
  GUT-symmetry-breaking pattern gives rise to the following relations
  among various mass matrices:
  \begin{equation}
  M_{u} = M_{\nu_{D}}, \quad M_{d} = M_{e} \; ,
  \end{equation}
  up to some calculable, group-theoretical factors which are useful in
  obtaining the Jarlskog relations among masses for the charged
  leptons and down-type quarks when combined with family symmetries.  
  The value of $U_{e3}$ is predicted to be large, 
  close to the sensitivity of current experiments. The prediction for
  the rate of $\mu \rightarrow e\gamma$ is about two orders  
  of magnitude below the current experimental bound.

  In a particular model constructed by Chen and
  Mahanthappa~\cite{Chen:2000fp}, the Higgs sector contains fields in
  10, 45, 54, 126 representations, with the 10 and 126 breaking the
  electro-weak symmetry and generating fermions masses, and the 45,
  54, 126 breaking the SO(10) GUT symmetry. The mass hierarchy can
  arise if there is an $SU(2)_{H}$ symmetry acting non-trivially on
  the first two generations such that the first two generations
  transform as a doublet and the third generation transforms as a
  singlet under $SU(2)_{H}$, which breaks down in two steps,  
  $SU(2) {\epsilon M \atop \rightarrow} U(1) {\epsilon^{\prime} M
  \atop \rightarrow} \mbox{`nothing'}$, $\epsilon^{\prime} \ll
  \epsilon \ll 1$. The mass hierarchy is generated by the
  Froggatt-Nielsen mechanism \cite{Froggatt:1978nt}. 
  The resulting mass matrices at the GUT scale are given by:
  \begin{equation}  
    {M_{u,\nu_{LR}}=
    \left( \begin{array}{ccc}
    {0} & 
    {0} & 
    {\left<10_{2}^{+} \right> \epsilon'}\\
    {0} & 
    {\left<10_{4}^{+} \right> \epsilon} & 
    {\left<10_{3}^{+} \right> \epsilon} \\
    {\left<10_{2}^{+} \right> \epsilon'} & \
    {\left<10_{3}^{+} \right> \epsilon} &
    {\left<10_{1}^{+} \right>}
    \end{array} \right)
    = 
    \left( \begin{array}{ccc}
    {0} & 
    {0} & 
    {r_{2} \epsilon'}\\
    {0} & 
    {r_{4} \epsilon} & 
    {\epsilon} \\
    {r_{2} \epsilon'} & 
    {\epsilon} & 
    {1}
    \end{array} \right) M_{U}} \; ,
  \end{equation}
  \begin{equation}  
    {M_{d,e}=
    \left(\begin{array}{ccc}
    {0} & 
    {\left<10_{5}^{-} \right> \epsilon'} & 
    {0} \\
    {\left<10_{5}^{-} \right> \epsilon'} &  
    {(1,-3)\left<\overline{126}^{-} \right> \epsilon} & 
    {0}\\ 
    {0} & 
    {0} & 
    {\left<10_{1}^{-} \right>}
    \end{array} \right)  
    =
    \left(\begin{array}{ccc}
    {0} & 
    {\epsilon'} & 
    {0} \\
    {\epsilon'} &  
    {(1,-3) p \epsilon} & 
    {0}\\
    {0} & 
    {0} & 
    {1}
    \end{array} \right) M_{D}} \; .
  \end{equation}
  The right-handed neutrino mass matrix is of the same form as 
  $M_{\nu_{LR}}$:
  \begin{equation} 
    {M_{\nu_{RR}}=  
    \left( \begin{array}{ccc}
    {0} & 
    {0} & 
    {\left<\overline{126}_{2}^{'0} \right> \delta_{1}}\\
    {0} & 
    {\left<\overline{126}_{2}^{'0} \right> \delta_{2}} & 
    {\left<\overline{126}_{2}^{'0} \right> \delta_{3}} \\ 
    {\left<\overline{126}_{2}^{'0} \right> \delta_{1}} & 
    {\left<\overline{126}_{2}^{'0} \right> \delta_{3}} &
    {\left<\overline{126}_{1}^{'0} \right> }
    \end{array} \right)
    = 
    \left( \begin{array}{ccc}
    {0} & 
    {0} & 
    {\delta_{1}}\\
    {0} & 
    {\delta_{2}} & 
    {\delta_{3}} \\ 
    {\delta_{1}} & 
    {\delta_{3}} & 
    {1}
    \end{array} \right) M_{R} \; .
    \label{Mrr}}
  \end{equation}
  Note that, since the $\overline{126}$-dimensional Higgs
  representation is used to generate the heavy 
  Majorana neutrino-mass terms, R-parity is preserved at all energies. 
  The effective neutrino mass matrix is:
  \begin{equation}  
    {\label{eq:Mll}
    M_{\nu_{LL}}=M_{\nu_{LR}}^{T} M_{\nu_{RR}}^{-1} M_{\nu_{LR}}
    = \left( 
    \begin{array}{ccc}  
    {0} &   {0} &   {t} \\
    {0} &   {1} &   {1+t^{\prime}} \\  
    {t} &   {1+t^{\prime}} &   {1} 
    \end{array} \right) \frac{d^{2}v_{u}^{2}}{M_{R}}} \, ,
  \end{equation}
  and causes the atmospheric mixing angle to be maximal and the solar
  mixing angle to be large.
  The form of the neutrino mass matrix in
  this model is invariant under the see-saw mechanism. The value of
  $U_{e3}$ is related to the ratio $\sim \sqrt{\Delta
  m_{12}^{2}/\Delta m_{13}^{2}}$, which is predicted to be close to
  the sensitivity of current experiments. The prediction for the rate
  of $\mu \rightarrow e\gamma$ is about two orders of magnitude below
  the current experimental bound. 

  \item {\it Lopsided mass textures for charged fermions:} 
  In this scenario, the large atmospheric-mixing angle comes from 
  the unitary matrix that diagonalises the charged-lepton mass
  matrix. This scenario is realised in models with $SU(5)$ as the
  intermediate symmetry which gives rise to the so-called ``lopsided''
  mass textures, due to the $SU(5)$ relation:
  \begin{equation}
    M_{e} = M_{d}^{T}.
  \end{equation}
  Due to the lopsided nature of $M_{e}$ and $M_{d}$, 
  the large atmospheric neutrino mixing is 
  related to the large mixing in the $(23)$ sector of 
  the RH charged-lepton diagonalisation matrix, 
  instead of $V_{cb}$.   
  Thus it explains why $V_{cb}$ is small while $U_{\mu\nu_{3}}$ is
  large.
  The large solar mixing angle comes from the diagonalisation matrix 
  for the neutrino mass matrix. Because the two large mixing angles
  come from different sources, the constraint on $U_{e\nu_{3}}$ is not
  as strong as in class (1). In fact, the prediction for
  $U_{e\nu_{3}}$ in this class of models tends to be quite small. On
  the other hand, this mechanism also predicts an enhanced decay rate
  for the flavour-violating process $\mu \rightarrow e \; \gamma$
  which is close to current experimental limit. As R-parity is broken
  by the vev of the $16$ dimensional Higgs, a separate `matter parity'
  must be imposed to distinguish the particles from their SUSY
  partners. 

  In a particular model constructed by Albright and
  Barr~\cite{Albright:2001uh}, the Higgs sector of the model contains
  Higgs particles in the $10, 16, 45$, with $\left< 16_{H_{1}}
  \right>$ breaking $SO(10)$ down to $SU(5)$ and $\left<16_{H_{2}}
  \right>$ breaking the  EW symmetry. The lopsided textures arise due
  to the operator $\lambda (16_{i}16_{H_{1}})(16_{j}16_{H_{2}})$ which
  gives rise to mass terms for the charged leptons and down quarks which
  satisfy the SU(5) relation $M_{d} = M_{e}^{T}$. When other
  operators are included, the lopsided structure of $M_{e}$ results,
  provided the coupling $\sigma$ is of order 1:
  \begin{eqnarray}
    M_{u,\nu_{LR}} = 
    \left(\begin{array}{ccc}
    \eta &    0 &    0\\
    0 &    0 &    (1/3,1) \epsilon\\
    0 &    -(1/3,1)\epsilon &    1
    \end{array}\right)    \cdot m_{u}
    \\
    M_{d} =     \left(\begin{array}{ccc}
    \eta &    \delta &    \delta^{'}e^{i\phi}\\
    \delta &    0 &    \sigma+\epsilon/3\\
    \delta^{'}e^{i\phi} &    -\epsilon/3 &    1
    \end{array}\right)   \cdot m_{d}
    , \quad
    M_{e} =     \left(\begin{array}{ccc}
    \eta &    \delta &    \delta^{'}e^{i\phi}\\
    \delta &    0 &    -\epsilon\\
    \delta^{'}e^{i\phi} &    \sigma+\epsilon &    1
    \end{array}    \right)\cdot m_{d}.
  \end{eqnarray}
  The large mixing in $   U_{e,L}$ leads to the large atmospheric
  mixing angle. Meanwhile, because large mixing in $   U_{e,L}$
  corresponds to large mixing in $U_{d,R}$, the CKM mixing angles
  remain small. A unique prediction of the lopsided models is the
  relatively large branching ratio for LFV processes, e.g. $\mu
  \rightarrow e \gamma$. By considering a RH Majorana neutrino-mass
  term of the following form, a large solar mixing angle can arise for
  some choice of the parameters in $M_{\nu_{RR}}$, leading to a large
  value for the solar mixing angle:
  \begin{equation}
    M_{\nu_{RR}} = \left(\begin{array}{ccc}
    c^{2}\eta^{2} &    -b\epsilon\eta &    a \eta\\
    -b\epsilon\eta &    \epsilon^{2} &    -\epsilon\\
    a \eta &    -\epsilon &    1
    \end{array}\right)   \cdot \Lambda_{R}, \quad
    M_{\nu}^{eff} =    \left(\begin{array}{ccc}
    0 &    -\epsilon &    0\\
    -\epsilon &    0 &    2\epsilon\\
    0 &    2\epsilon &    1
    \end{array}   \right)
    \frac{m_{u}^{2}}{\Lambda_{R}} \; .
  \end{equation}
\end{enumerate}

{\noindent \bf Models with renormalisation-group enhancements}

It is possible to obtain large neutrino mixing
angles through renormalisation-group evolution. 
Assuming that the CKM matrix and the leptonic mixing matrix are
identical at the GUT scale, which is a natural consequence of
quark-lepton unification, two large neutrino mixing angles can be
generated by renormalisation-group evolution 
\cite{Mohapatra:2003tw}.
The only requirement for this mechanism to work is that the masses of
the three neutrinos are nearly degenerate 
($m_{3} \gtrsim m_{2} \gtrsim m_{1}$) and have the same CP parity.
The one-loop renormalisation-group equation (RGE) of the effective
left-handed Majorana neutrino mass operator is given by:
\begin{equation}
  \label{rgem}
  \frac{d m_{\nu}}{dt}=-\{\kappa_{u}m_{\nu}+m_{\nu}P+P^{T}m_{\nu}\},
\end{equation}
where $t \equiv \ln \mu$ and $\mu$ is the energy scale. 
In the MSSM, $P$ and $\kappa_{u}$ are given by:
\begin{eqnarray}
  P & = & -\frac{1}{32\pi^{2}} \frac{Y_{e}^{\dagger}Y_{e}}{\cos^{2} \beta} 
  \simeq -\frac{1}{32\pi^{2}} \frac{h_{\tau}^{2}}{\cos^{2}\beta} diag(0,0,1)
  \equiv diag(0,0,P_{\tau}) \; ; \\ 
  \kappa_{u} & = &
  \frac{1}{16\pi^{2}}\biggl[ \frac{6}{5}g_{1}^{2} + 6g_{2}^{2} 
  - 6 \frac{Tr(Y_{u}^{\dagger}Y_{u})}{\sin^{2}\beta} \biggr] 
  \simeq  
  \frac{1}{16\pi^{2}} \biggl[ \frac{6}{5}g_{1}^{2} + 6g_{2}^{2} 
  - 6 \frac{h_{t}^{2}}{\sin^{2}\beta} \biggr] \; ;  
\end{eqnarray}
respectively, where $g_{1}^{2}=\frac{5}{3}g_{Y}^{2}$ is the $U(1)$
gauge coupling constant, $Y_{u}$ and $Y_{e}$ are the $3 \times 3$
Yukawa coupling matrices for the up quarks and charged leptons
respectively, and $h_{t}$ and $h_{\tau}$ are the $t$- and
$\tau$-Yukawa couplings. 
One can then follow the ``diagonalise-and-run'' procedure and obtain
the RGEs at scales between $M_{R} \ge \mu \ge M_{SUSY}$  for the mass
eigenvalues and the three mixing angles, assuming CP violating phases
vanish: 
\begin{eqnarray}
  \frac{d \; m_{i}}{d t} & = & 
    -4 P_{\tau} m_{i} U_{\tau \nu_{i}}^{2} - m_{i} \kappa_{u}, \quad (i=1,2,3)
    \; ; \label{nurge1}  \\
  \frac{d \; s_{23}}{d t} & = & -2P_{\tau} c_{23}^{2} 
    (-s_{12}U_{\tau\nu_{1}}\nabla_{31} + c_{12}U_{\tau\nu_{2}}\nabla_{32})
    \; ; \\
  \frac{d \; s_{13}}{dt} & = & -2P_{\tau}c_{23}c_{13}^{2}
    (c_{12}U_{\tau\nu_{1}}\nabla_{31}+s_{12}U_{\tau\nu_{2}}\nabla_{32})
    \; ; \\
  \frac{d \; s_{12}}{dt} & = & -2P_{\tau}c_{12}
    (c_{23}s_{13}s_{12}U_{\tau\nu_{1}}\nabla_{31}
    -c_{23}s_{13}c_{12}U_{\tau\nu_{2}}\nabla_{32}
    +U_{\tau\nu_{1}}U_{\tau\nu_{2}}\nabla_{21}) \; ;
  \label{nurge4}
\end{eqnarray}
where $\nabla_{ij} \equiv (m_{i}+m_{j})/(m_{i}-m_{j})$. 
Because the leptonic-mixing matrix is identical to the CKM matrix, we
have, at the GUT scale, the following initial conditions, 
$s_{12}^{0} \simeq \lambda$, 
$s_{23}^{0} \simeq \mathcal{O}(\lambda^{2})$ and  
$s_{13}^{0} \simeq \mathcal{O}(\lambda^{3})$, 
where $\lambda$ is the Wolfenstein parameter.  
When the masses $m_{i}$ and $m_{j}$ are nearly degenerate, 
$\nabla_{ij}$ approaches infinity. 
Thus it drives the mixing angles to become large. 
Starting with the values of 
$(m_{1}^{0},m_{2}^{0},m_{3}^{0}) = (0.2983,0.2997,0.3383)$~eV 
at the GUT scale, the solutions at the weak scale for the masses are
$(m_{1},m_{2},m_{3}) = (0.2410,0.2411,0.2435)$~eV, which correspond to
$\Delta m_{13}^{2} = 1.1 \times 10^{-3} \mbox{eV}^{2}$ and 
$\Delta m_{\odot}^{2} = 4.8 \times 10^{-5} \mbox{eV}^{2}$.  
The mixing angles predicted at the weak scale are
$\sin^{2}2\theta_{23} = 0.99$, $\sin^{2} 2\theta_{12} = 0.87$ and
$\sin\theta_{13} = 0.08$. 
Because the masses are larger than $0.1$ eV, they are testable at the
present searches for the neutrinoless double beta decay. 

{\noindent \bf Predictions for the Oscillation Parameters}

\begin{figure}
\begin{tabular}{cc}
\resizebox{78mm}{!}
 {\includegraphics{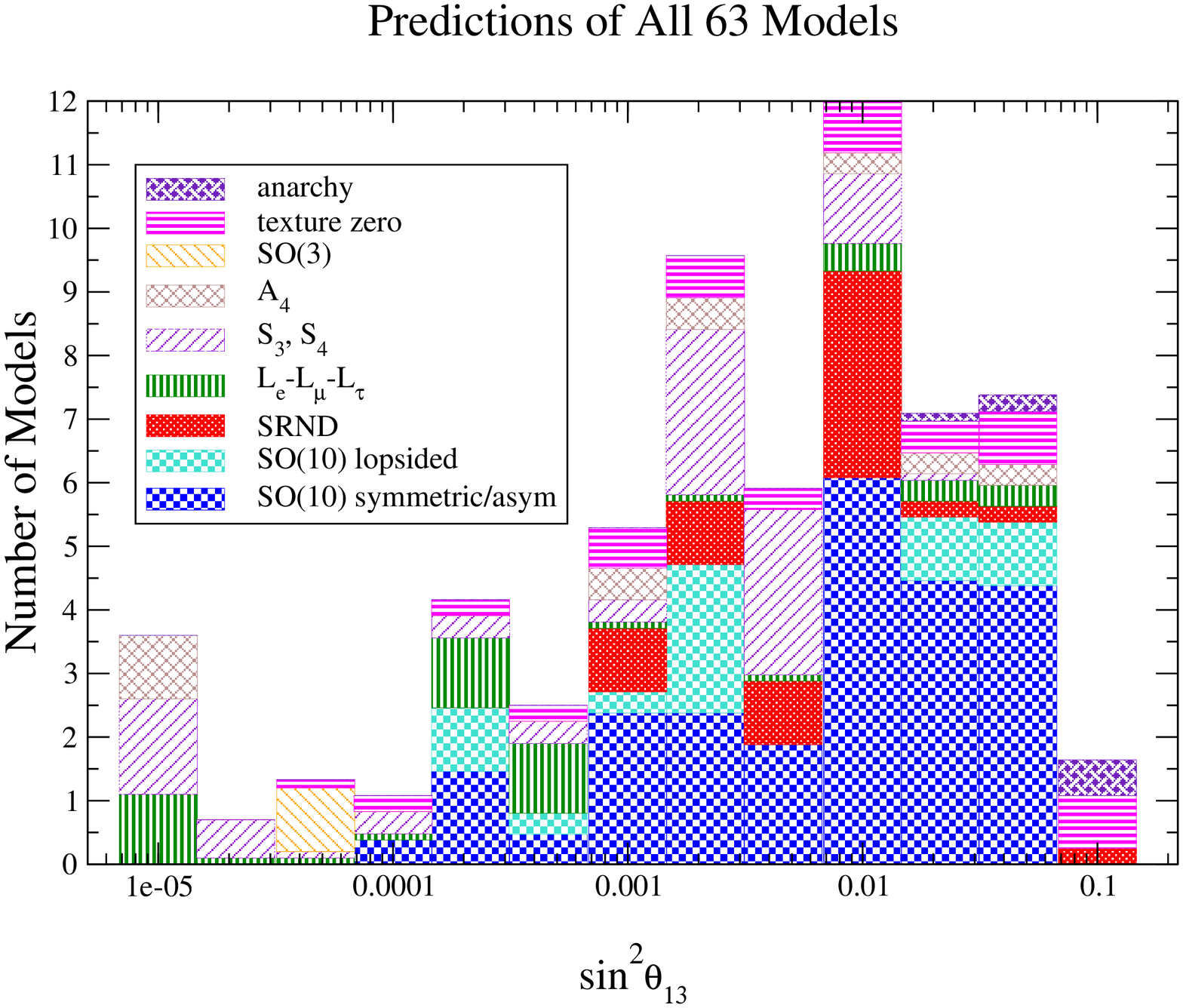}}&
\resizebox{78mm}{!}
 {\includegraphics{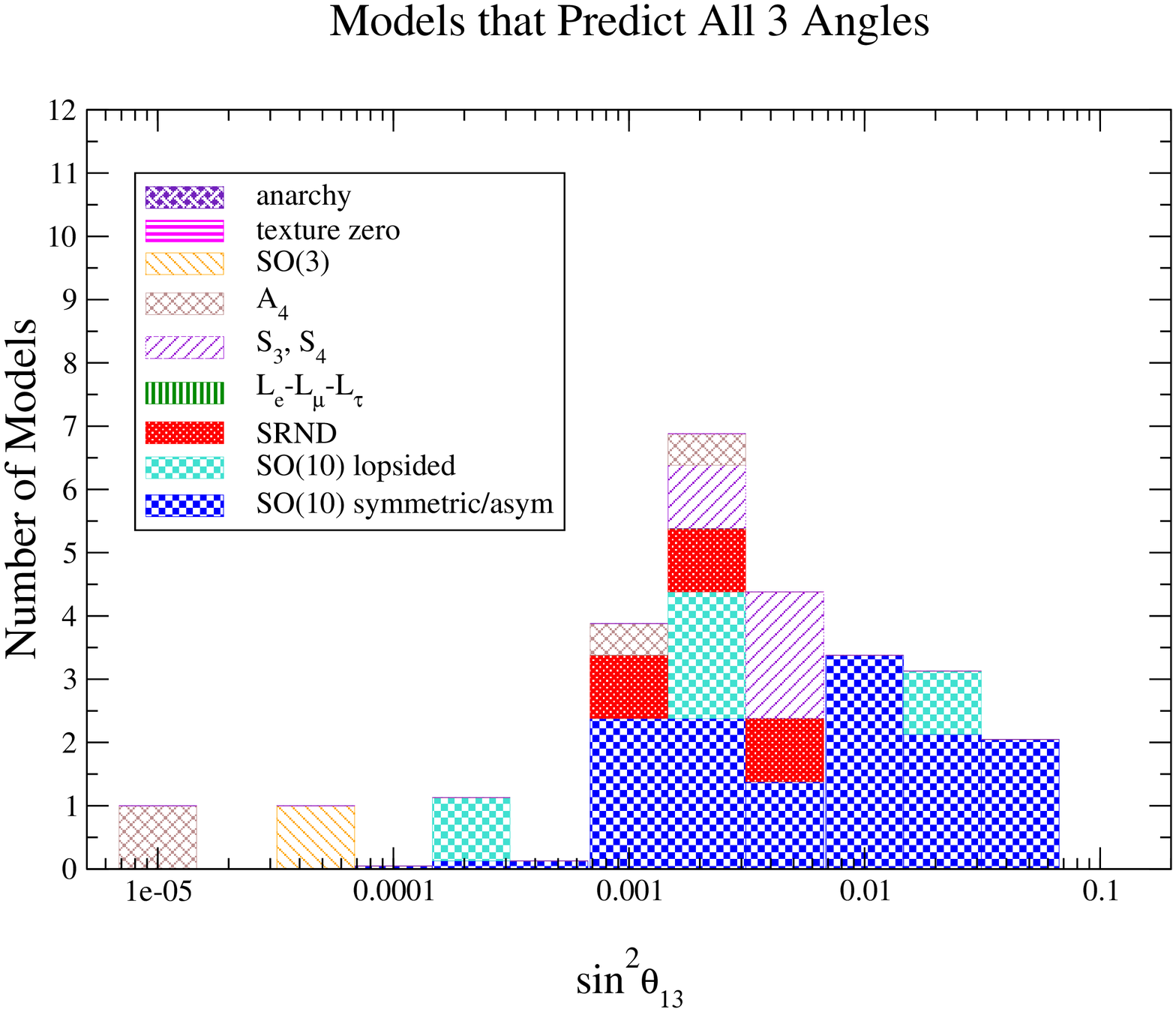}}
\end{tabular}
  \caption{
    \label{fig:all} 
    Histogram of the number of models for each
    $\sin^{2}\theta_{13}$. The diagram on the left includes all sixty
    models, while the diagram on the right includes only those that
    give predictions for all three leptonic mixing angles.
    Taken with kind permission of Physical Review from figures 1 and 2 in
    reference \cite{Albright:2006cw}.
    Copyrighted by the American Physical Society.
  }
\end{figure}

\begin{figure}
\begin{tabular}{cc}
\resizebox{78mm}{!}
 {\includegraphics[scale=0.35]{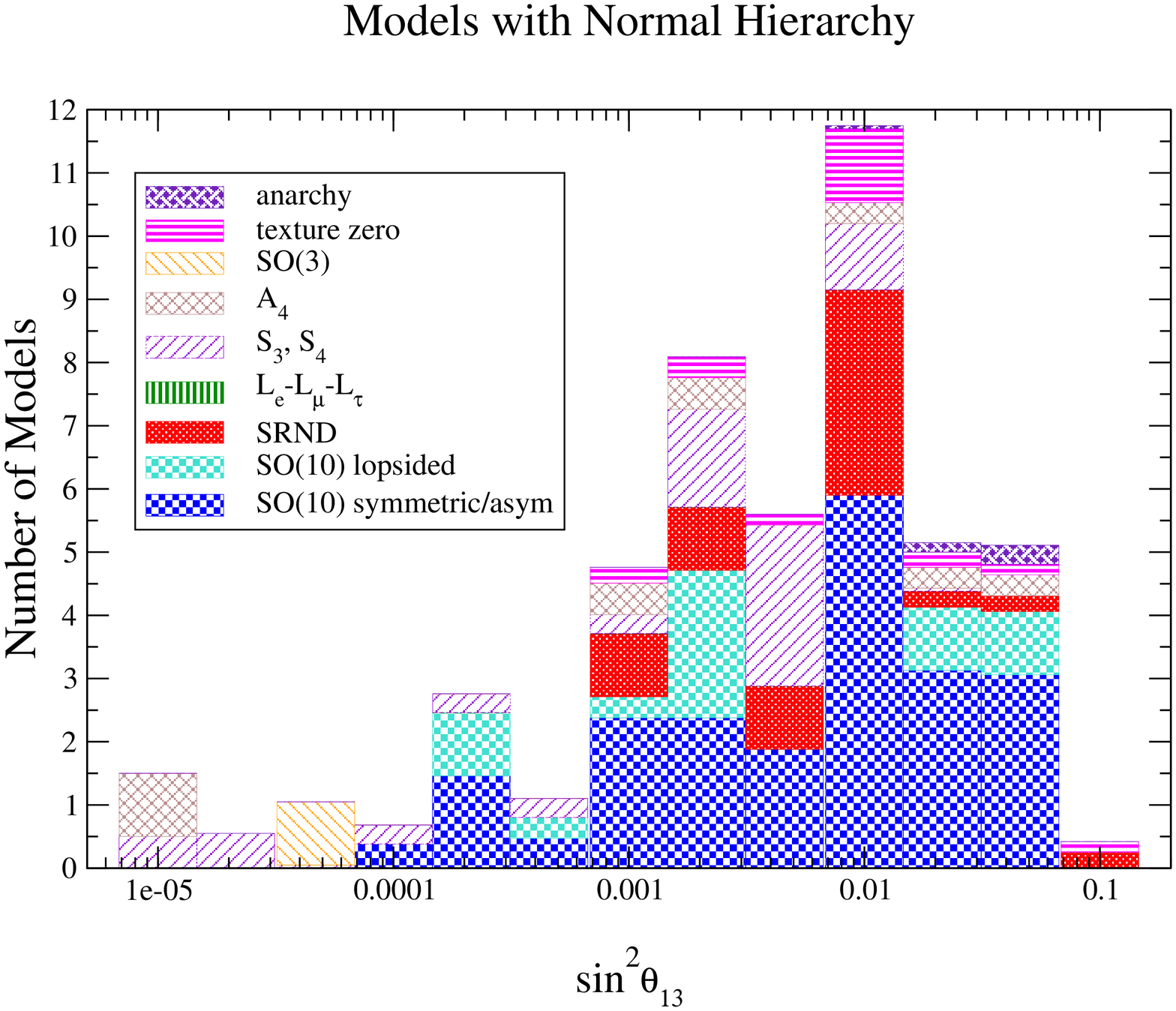}}&
\resizebox{78mm}{!}
 {\includegraphics[scale=0.35]{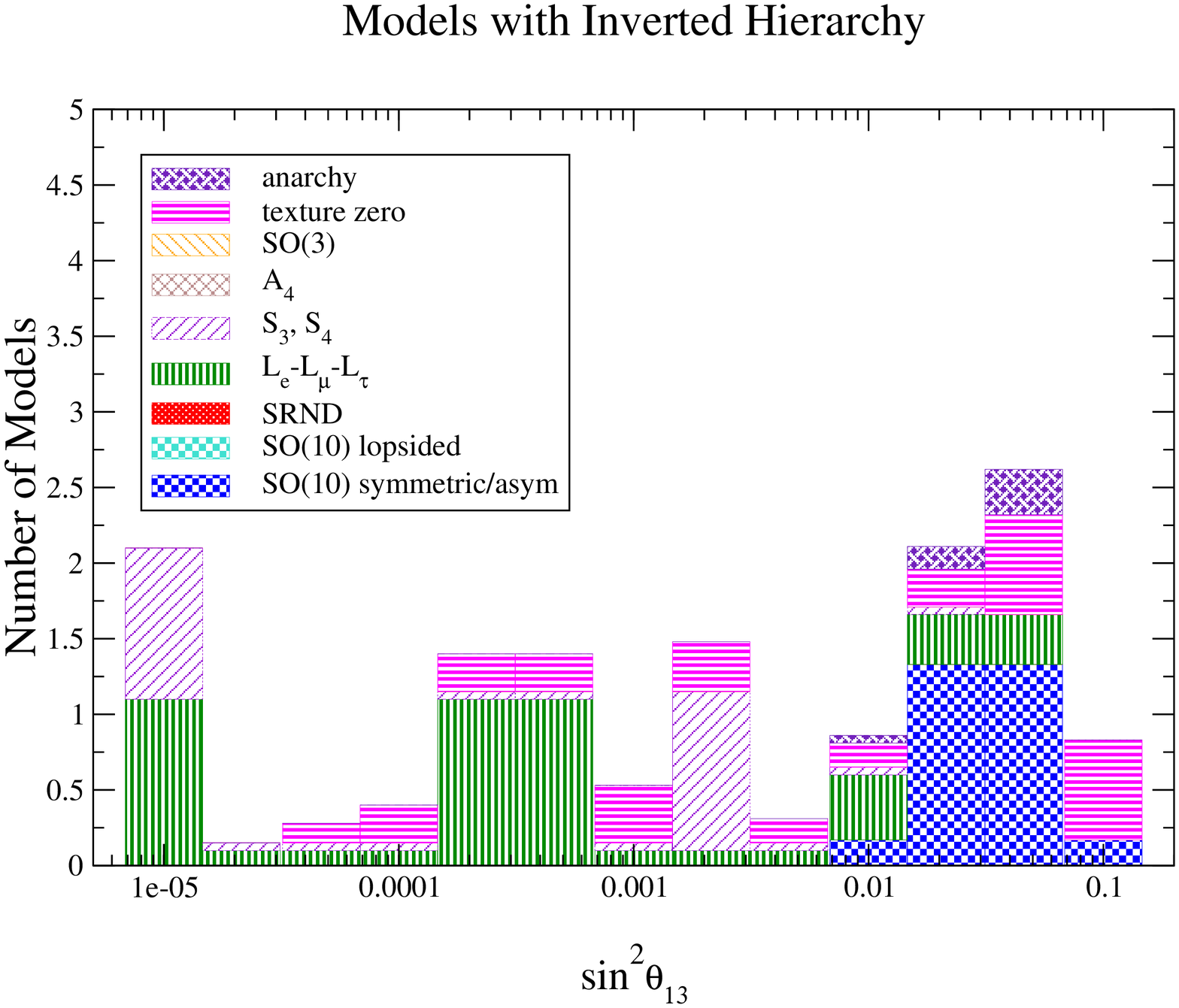}}
\end{tabular}
  \caption{
    \label{fig:hierarchyMot}
    Histogram of the number of models for each
    $\sin^{2}\theta_{13}$. The diagram on the left includes models
    that predict normal mass hierarchy, while the diagram on the right
    includes models that predict inverted mass hierarchy.
    Taken with kind permission of Physical Review from figure 3 in
    reference \cite{Albright:2006cw}.
    Copyrighted by the American Physical Society.
  }
\end{figure}

In the literature, there are thirty models based on $SO(10)$, six
models that utilise single-RH-neutrino dominance mechanism, five based
on $L_{e} - L_{\mu} - L_{\tau}$ symmetry, ten based on $S_{3}$
symmetry, three on $A_{4}$ symmetry, one on $SO(3)$ symmetry, and
three based on texture-zero assumptions.  
The predictions of these models for $\sin^{2}\theta_{13}$ are
summarised in figures \ref{fig:all} and \ref{fig:hierarchyMot}. 
In some models, a range of values (rather than a single value) is
given for $\theta_{13}$. 
If these values range over $N$ bins for $\sin^{2}\theta_{13}$ in a
particular model, a weight of $1/N$ is assigned for each bin.
As a result, non-integer values for the number of models for some
values of $\sin^{2}\theta_{13}$ can arise.

Figure~\ref{fig:all} shows the histogram of the number of models for
each $\sin^{2}\theta_{13}$ including all sixty models and one
including only models that predict all three mixing angles. 
An observation one can draw immediately is that the predictions of 
$SO(10)$ models are larger than $10^{-4}$, and the median value is
roughly $\sim 10^{-2}$.  
Furthermore, $\sin^{2}\theta_{13} < 10^{-4}$ can only
arise in models based on leptonic symmetries. 
However, these models are not as predictive as the GUT models, due to
the uncertainty in the charged-lepton mixing matrix. 
In this case, to measure $\theta_{13}$ 
will require a neutrino superbeam or the Neutrino Factory. 
In table \ref{tbl:sensitivity} the reach of future experiments is
summarised.

In figure \ref{fig:hierarchyMot}, histograms of the number of models
for each $\sin^{2}\theta_{13}$ value are shown for both normal and
inverted neutrino-mass hierarchies.   
From these two diagrams, one finds that there are more models that
predict the normal hierarchy than the inverted hierarchy. 
This could merely be a result of the theorists' prejudice
for the  model. What is more important is the correlation between the
type of the hierarchy and the predicted values for $\theta_{13}$. In
the normal-hierarchy case, the predicted values tend to be larger,
while in the inverted case, the distribution is quite uniform. The
normal hierarchy arises in $SO(10)$ models with type-I see-saw, models
with single-RH-neutrino dominance, and models based on $SO(3)$ and
$A_{4}$ lepton symmetries, while the inverted hierarchy arises in
models based on $L_{e}-L_{\mu}-L_{\tau}$, $S_{3}$, and $S_{4}$ lepton
symmetries. 
\begin{table}
  \caption{
    \label{tbl:sensitivity}
    A summary of the current
    experimental limit on $\theta_{13}$ and the reach of future
    experiments.
  } 
  \begin{center}
  \begin{tabular}{|c|c|c|}
  \hline
   & $\sin^{2} 2\theta_{13}$ & $\sin\theta_{13}$ \\
   \hline
   current limit & $10^{-1}$ & 0.16\\
   \hline
   reactor & $10^{-2}$ & 0.05\\
   conventional beam & $10^{-2}$ & 0.05\\
   superbeam & $3 \times 10^{-3}$ & $2.7 \times 10^{-2}$\\
  neutrino factory & $(5-50) \times 10^{-5}$ & $(3.5-11) \times 10^{-3}$\\
  \hline
  \end{tabular}
  \end{center}
  \label{tbl:limits}
\end{table}
\vspace{0.5cm}

In conclusion, predictions for $\theta_{13}$ range from zero to the
current experimental limit.   
For models based on GUT symmetries, the normal mass hierarchy can be 
generated naturally. 
The inverted hierarchy may also be obtained in these models with a
type-II see-saw, even though some fine-tuning is needed. 
Predictions for $\theta_{13}$ in these models tend to be large, with a
median value $\sin^{2}\theta_{13} \sim 0.01$. 
On the other hand, models based on leptonic symmetries can give
rise to inverted hierarchies and the predictions for $\theta_{13}$ 
can be quite small. 
Therefore, models based on lepton symmetries will be favoured if
$\theta_{13}$ turns out to be tiny and the inverted hierarchy is
observed.
However, if $\theta_{13}$ turns out to be large, the two different
classes would not be distinguishable. 
A precise measurement for the deviation of $\theta_{23}$ from $\pi/4$
can also be crucial for distinguishing different models. 
This is especially true for models based on lepton symmetries in which
the deviation strongly depends on how the symmetry breaking is
introduced into the models. 
Precision measurements are thus indispensable in order to
distinguish different classes of models. 

\subsubsection{Sum Rules}

In the previous section, the predictions of various models of neutrino
masses have been reviewed. Many particularly attractive classes of
models lead to interesting predictions for the neutrino-mass matrix
$m_\nu$, such as for instance tri-bimaximal or bimaximal
mixing. 
Measurements of neutrino oscillation determine matrix elements of the
neutrino-mixing matrix, $U_{\rm PMNS}$, which may be written as the
product of $V_{\nu_\mathrm{L}}$, that diagonalises the neutrino-mass
matrix and $V_{e_\mathrm{L}}$, which diagonalises the charged-lepton
mass matrix, i.e. 
$U_{\mathrm{PMNS}} = V_{e_\mathrm{L}} V^\dagger_{\nu_\mathrm{L}}$.
Often, the essential predictions of flavour models are hidden
due to the presence of the charged lepton corrections.  
In many cases it can be shown that a combination of the
measurable parameters $\theta_{12}$, $\theta_{13}$, and $\delta$
can be combined to yield a prediction for the 1-2 mixing of the
neutrino-mass matrix \cite{King:2005bj,Antusch:2005kw}, i.e.\ to
$\arcsin(\tfrac{1}{\sqrt{3}})$ for tri-bimaximal and $\tfrac{\pi}{4}$
for bimaximal mixing, for example. 
In an SO(3) family-symmetry model based on the see-saw mechanism with
sequential dominance that predicts tri-bimaximal mixing via vacuum
alignment, such a `sum rule' has been obtained in reference
\cite{King:2005bj}. 
In reference \cite{Antusch:2005kw}, it has been shown that neutrino
sum rules are not limited to one particular model, but apply to large
classes of models under very general assumptions, to be specified
below. 
Examples for sum rules with theory predictions of tri-bimaximal and
bimaximal neutrino mixing, respectively, are
\cite{King:2005bj,Masina:2005hf,Antusch:2005kw}:
\begin{eqnarray}
  \label{sumrules1}
  \theta_{12} - \theta_{13}\cos (\delta) &\approx&  \arcsin
  \tfrac{1}{\sqrt{3}} \; ; \\
  \label{sumrules2}\theta_{12} - \theta_{13}\cos (\delta) &\approx&
  \tfrac{\pi}{4} \; . 
\end{eqnarray}
Neutrino sum rules \cite{King:2005bj,Antusch:2005kw} are thus a means
of exploring the structure of the neutrino mass matrix in the presence
of charged-lepton corrections and of testing whole classes of models. 
The sum rules, such as those of equations (\ref{sumrules1}) and
(\ref{sumrules2}), can only be tested to high-enough precision in the
most accurate experimental facilities such as the Neutrino Factory. 

{\noindent \bf Charged-lepton corrections and sum rules}

To illustrate the use of sum rules in testing theories of
the neutrino-mass matrix in the presence of charged-lepton
corrections, consider two examples, bimaximal
\cite{Barger:1998ta} and tri-bimaximal \cite{Harrison:2002er}
neutrino mixing, where the predicted neutrino-mixing angles are:
\begin{eqnarray}
  \left.
  \begin{array}{llll}
    \theta^\nu_{12} = \pi/4, &\theta^\nu_{12} = \pi/4,& 
    \theta^\nu_{13} = 0 & \mbox{for bimaximal neutrino mixing; and} \\
    \theta^\nu_{12} =\arcsin (\tfrac{1}{\sqrt{3}}) ,& \theta^\nu_{12} = \pi/4,& 
    \theta^\nu_{13} = 0 & \mbox{for tri-bimaximal neutrino mixing.}
  \end{array}\right. 
  \label{Eq:BiTriPred}
\end{eqnarray}
A similar, but physically different form, was proposed earlier
\cite{Wolfenstein:1978uw}.
The leptonic-mixing matrix is the product of
$V_{\nu_\mathrm{L}}$ and $V_{e_\mathrm{L}}$, and therefore corrections
to the predictions for the neutrino-mixing angles given in equations 
(\ref{Eq:BiTriPred}) arising from the charged-lepton mixing matrix must
be evaluated to obtain estimates of the mixing angles that are
accessible experimentally.
 
The charged-lepton corrections can be evaluated if it is assumed 
that the charged-lepton mixing matrix has a CKM-like structure, 
i.e. the charged-lepton mixing angles $\theta^{e}_{ij}$ are
small and dominated by a 1-2 mixing $\theta^{e}_{12}$. 
This is the case in many generic classes of flavour model in the
context of GUTs in which quarks and leptons are assigned to
representations of the unified gauge symmetries  
\cite{King:2005bj,King:2003rf,deMedeirosVarzielas:2005ax}. 
For $\theta^\nu_{13} = 0$, which is the case in the examples mentioned
above, such charged-lepton corrections lead to the following PMNS
mixing angles \cite{Antusch:2005kw}:  
\begin{subequations}
  \begin{eqnarray}
    \theta_{23} &\approx& \theta^\nu_{23} \; ,\\
    \label{Eq:t13}\theta_{13} &\approx& \sin(\theta^\nu_{23})\,\theta^{e}_{12}\; ,\\
    \label{Eq:t12}\theta_{12} &\approx& \theta^\nu_{12} +  \cos(\theta^\nu_{23})\,\theta^{e}_{12} 
    \cos(\delta)\; .
  \end{eqnarray}
\end{subequations} 
The quantity $\delta$ which appears on the right-hand side of
equation~(\ref{Eq:t12}) is the Dirac CP phase observable in neutrino
oscillations. 
For bimaximal and tri-bimaximal mixing, this implies that
$\theta_{23} \approx \pi/4$ and leads to the prediction 
$\theta_{13} \approx \tfrac{1}{\sqrt{2}}\theta^{e}_{12}$. 
Substituting the expressions for $\theta_{13}$ and $\theta_{23}$ into
equation (\ref{Eq:t12}) results in the following sum rules
\cite{King:2005bj,Masina:2005hf,Antusch:2005kw}:
\begin{eqnarray}
  \label{Eq:ModelSumrules}\theta_{12} - \theta_{13}\cos(\delta) &\approx& \theta^\nu_{12} \;=\;
  \left\{
    \begin{array}{ll}
      \tfrac{\pi}{4} & \mbox{for bimaximal neutrino mixing,} \\
      \arcsin (\tfrac{1}{\sqrt{3}}) & \mbox{for tri-bimaximal neutrino mixing.} \\
    \end{array}
  \right.
\end{eqnarray}
Therefore, in the case of bimaximal or tri-bimaximal neutrino
mixing, precise measurements of the leptonic mixing parameters
$\theta_{13}$, $\theta_{12}$, and $\delta$ allow the prediction for
$\theta^\nu_{12}$ in equation (\ref{Eq:ModelSumrules}) to be tested
without assuming any particular value for $\theta^{e}_{12}$. 

More generally, if it is assumed that $\theta^\nu_{13} \approx 0$,
$\theta^{e}_{13} \approx 0$ and $\theta^{e}_{23} \approx 0$, and 
assuming $\theta_{23} \approx \pi/4$, then
\cite{Antusch:2005kw}:
\begin{subequations}
  \label{Eq:relations}
  \begin{eqnarray}
    \label{Eq:sumrule}
    \theta_{12} - \theta_{13}\cos(\delta)  &\approx& \theta^{\nu}_{12}
  \quad \quad \mbox{($\theta^{\nu}_{12}$ from ``$m_\nu$-theory black box'')} \; ; {\rm and } \\
    \label{Eq:t13prediction} 
  \theta_{13} &\approx& \tfrac{1}{\sqrt{2}} \theta^{e}_{12} \quad \qquad  
    \mbox{($\theta^{\mathrm{e}}_{12}$ from ``GUT black box'')}.
  \label{sumrule}
  \end{eqnarray}
\end{subequations}
A measurement of the combination of PMNS parameters:
\beq 
  \label{Eq:T12Sigma}
  \theta^\Sigma_{12}\equiv \theta_{12} - \theta_{13}\cos(\delta)
\eeq
can be used to constrain the neutrino mixing $\theta^{\nu}_{12}$ by
means of the sum rule in equation~(\ref{Eq:sumrule}).
In many unified flavour models, the Cabibbo angle, $\theta_C$ is
related to $\theta^{\mathrm{e}}_{12}$; equation
(\ref{Eq:t13prediction}), therefore, can be used to relate $\theta_{13}$
to $\theta_C$.
Hence, a precise measurement of $\theta_{13}$ may be used to test such
GUT predictions.

{\noindent \bf Sum Rules and Sensitivities of Future Experiments}

\begin{figure}
  \begin{center}
    \ensuremath{
      \vcenter{
        \hbox{
          \includegraphics[scale=0.5]{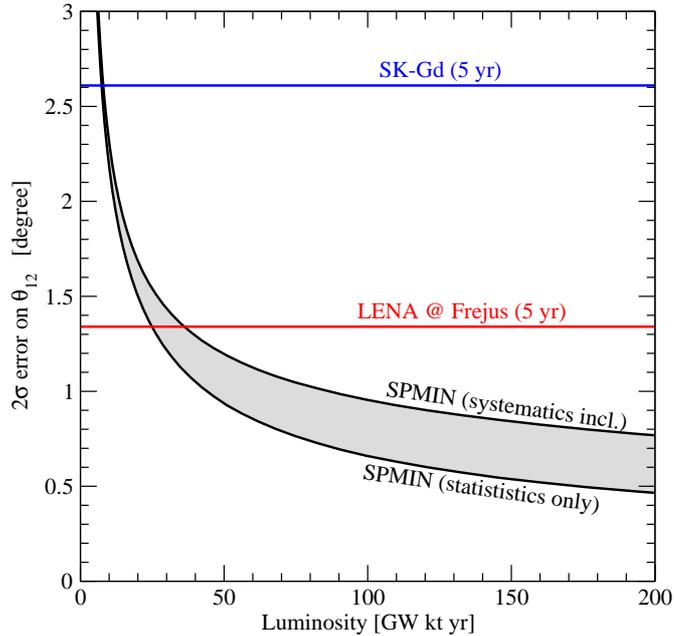}
        }
      }
    }
    \vspace{0.5cm}
    \caption{
      The $2\,\sigma$ error on $\theta_{12}$ as a function of the
      exposure for a so called SPMIN experiment.  }
    \label{fig:th12}
  \end{center}
\end{figure}

For $\theta_{12}^\Sigma$ to be used to discriminate between the
various models, precise, independent measurements of $\theta_{12}$
and on $\theta_{13}\cos(\delta)$ are required, 
(for more details see \cite{Antusch:2007rk}). 
$\theta_{12}$ can be measured using solar neutrinos or using
the neutrinos generated in nuclear reactors;
a comparison of these options indicates that the best precision on
is obtained using the latter \cite{Bandyopadhyay:2004cp}. 
An experiment optimised for the measurement of $\theta_{12}$, the
`Survival Probability MINimum' (SPMIN) experiment, has been proposed
\cite{Bandyopadhyay:2004cp}.   
In this experiment a single detector is placed at a baseline of
$\sim 60\,\mathrm{km}$ so that the first oscillation 
minimum is right in the middle of the neutrino energy spectrum.
The dependence of the $2\sigma$ error on $\theta_{12}$ on the exposure
in units of $\mathrm{GW}\,\mathrm{kt}\,\mathrm{y}$ is shown 
in figure \ref{fig:th12}. 
The following systematic uncertainties were considered: normalisation,
$5\%$; beam tilt, $2\%$; energy scale, $0.5\%$, reactor power, $2\%$;
and burn-up, $2\%$. 
At large exposures these systematic uncertainties are as large as the
statistical uncertainty. 
The figure also shows the performance that would be obtained if the
water in the Super-Kamiokande detector were doped with gadolinium
to make the detector sensitive to neutrinos from the nuclear
reactors in Japan \cite{Choubey:2004bf}.
Another alternative, LENA, a $40\,\mathrm{kt}$ liquid scintillator
detector that has been proposed for the Frejus laboratory in France,
would be sensitive to neutrinos produced in the French nuclear
reactors \cite{Petcov:2006gy}.
These experiments would yield $2\,\sigma$ errors on $\theta_{12}$ of
$2.6^\circ$ and $1.35^\circ$ respectively. 
The SPMIN experiment has a greater sensitivity than either of these
proposals since the baseline has been chosen to be optimal.

Long-baseline experiments, which are sensitive to $\delta$ and
$\theta_{13}$ but have little sensitivity to $\theta_{12}$, must be
used to determine $\theta_{12}^\Sigma$. 
The precision with which $\theta_{12}^\Sigma$ can be determined, has
been estimated under the assumption that $\theta_{12}$ has been
measured in a reactor experiment.
Three cases have been considered corresponding to $2\,\sigma$
errors on $\theta_{12}$ of $=0.75^\circ$, $1.35^\circ$, and
$2.6^\circ$ respectively.  
For comparison, note that the current error on $\theta_{12}$ is
$5.6^\circ$~\cite{Maltoni:2004ei}.  
To estimate the precision on the quantity $\theta_{12}^\Sigma$ the
general procedure described in \cite{Huber:2002mx} has been followed.
The analysis therefore includes the uncertainties on $\theta_{13}$ and
$\delta$, including correlations, as well as the uncertainties on
$\theta_{12}$, $\Delta m^2_{21}$, $\theta_{23}$, $\Delta m^2_{31}$ and
the matter density. 
The inclusion of the correlation between $\theta_{13}$ and $\delta$ is
crucial since the relevant oscillation probability contains terms
which go as $\theta_{13} \sin\delta$ and $\theta_{13} \cos\delta$.
However, the $L/E$ dependence of these two terms is different and
therefore experiments covering different $L/E$ ranges may have very
different sensitivities to $\theta_{12}^\Sigma$. 
For these reasons the accuracy on the combination
$\theta_{13}\cos\delta$ may be very different from the precision with
which either $\theta_{13}$ or $\cos\delta$ can be determined
individually. 

Numerical estimates of the precision with which $\theta_{12}^\Sigma$
can be determined were made using the assumptions for the various
oscillation parameters defined in section \ref{Sect:Performance}.
The calculations are performed with GLoBES~\cite{Huber:2004ka,Huber:2007ji}. 
The cases considered are (see section \ref{SubSect:Perf:SB}): T2HK --
an upgrade of the Japanese superbeam programme; SPL to Frejus -- a
European, CERN based superbeam facility; WBB -- a US experiment
employing a wide band neutrino beam; a conservative
Neutrino Factory (NFC) and an optimistic Neutrino Factory NFO (as
defined in section \ref{SubSect:Perf:NF}); and a $\gamma=350$
$\beta$-beam (BB350) as described in \cite{Burguet-Castell:2005pa}
(see section \ref{SubSect:Perf:BetaBeam}).

\begin{figure}
  \begin{center}
    \ensuremath{
      \vcenter{
        \hbox{
          \includegraphics[width=\textwidth]{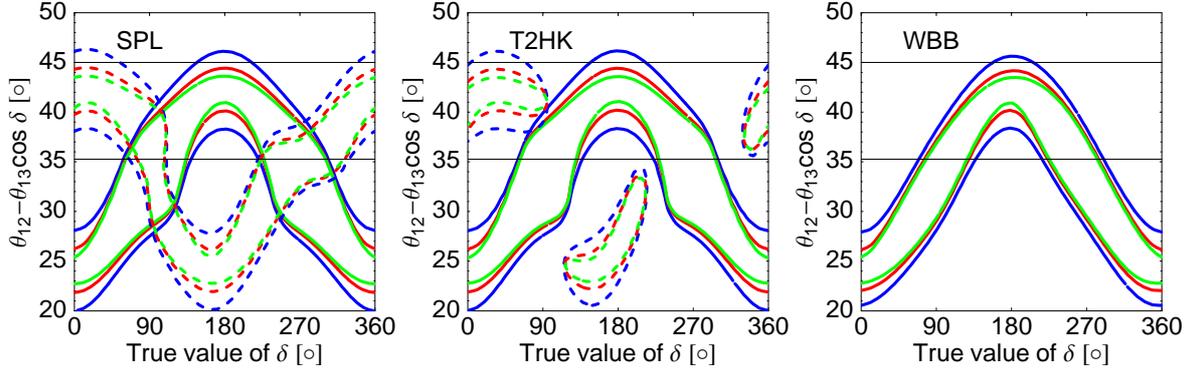}
        }
      }
    }
   \vspace{0.5cm}
   \caption{
      The $3\,\sigma$ allowed interval for the combination of physical 
parameters $\theta_{12}^\Sigma = \theta_{12} - \theta_{13}\cos(\delta)$ 
      (defined in equation~(\ref{Eq:T12Sigma}))
      as a function of the
      true value of $\delta$ for $\sin^22\theta_{13}=10^{-1}$. The left
      hand panel is for SPL, whereas the middle one is for T2HK and the
      right hand one for WBB. The
      dashed lines are for the $\mathrm{sgn}\Delta m^2_{31}$ degenerate
      solution. The colours indicate different errors on $\theta_{12}$:
      blue -- $2.8^\circ$, red -- $1.35^\circ$ and green -- $0.75^\circ$.
      For the true value of $\theta_{12}$, $\sin^2 \theta_{12}=0.3$
      ($\theta_{12}=33.12^\circ$) has been used.
      The horizontal lines show the case of bimaximal and tri-bimaximal
      neutrino mixing.
    Taken with kind permission of the Journal of High Energy Physics 
    from figure 2 in reference \cite{Antusch:2007rk}.
    Copyrighted by SISSA.
    }
    \label{fig:lth13}
  \end{center}
\end{figure}

Figure \ref{fig:lth13} shows the $3\,\sigma$ allowed interval in
$\theta_{12}^\Sigma$ as a function of the true value of $\delta$ for
$\sin^22\theta_{13}=10^{-1}$. 
The plot shows three different experiments from left to right: SPL,
T2HK, and WBB. 
All three have good sensitivity to $\theta_{12}^\Sigma$. 
The presence of the mass-hierarchy-degenerate solutions (dashed lines)
limits the usefulness of SPL and T2HK severely. 
These experiments are not able to distinguish between bimaximal and
tri-bimaximal mixing (horizontal lines). 
This problem is absent for WBB for which the accuracy on
$\theta_{12}^\Sigma$ is also somewhat better.  
 
\begin{figure}
  \begin{center}
    \ensuremath{\vcenter{\hbox{\includegraphics[width=\textwidth]{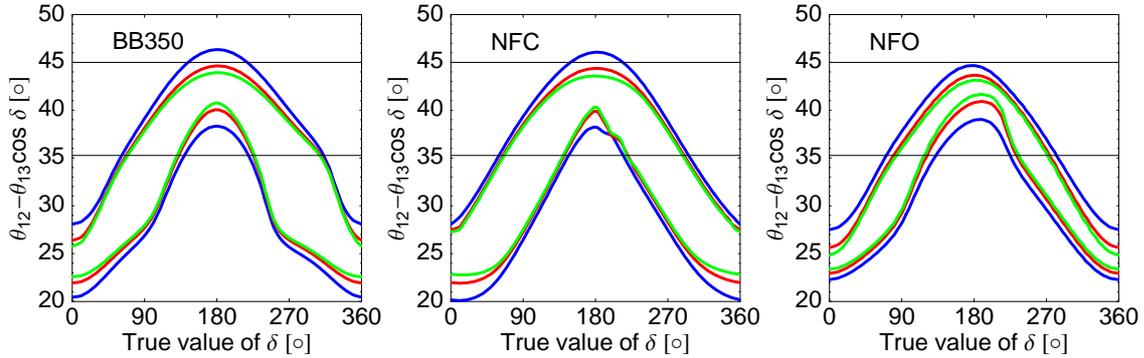}}}}
    \vspace{0.5cm}
    \caption{
      The $3\,\sigma$ allowed interval for the combination of physical 
parameters $\theta_{12}^\Sigma = \theta_{12} - \theta_{13}\cos(\delta)$ 
      (defined in equation~(\ref{Eq:T12Sigma})) as a function of the
      true value of $\delta$ for $\sin^22\theta_{13}=10^{-1}$. The left
      hand panel is for BB350, whereas the middle one is for NFC and the
      right hand one for NFO. The
      dashed lines are for the $\mathrm{sgn}\Delta m^2_{31}$ degenerate
      solution. The colours indicate different errors on $\theta_{12}$:
      blue -- $2.8^\circ$, red -- $1.35^\circ$ and green -- $0.75^\circ$.
      For the true value of $\theta_{12}$, $\sin^2 \theta_{12}=0.3$
      ($\theta_{12}=33.12^\circ$) has been used.
      The horizontal lines show the case of bimaximal and tri-bimaximal
      neutrino mixing.
    Taken with kind permission of the Journal of High Energy Physics 
    from figure 2 in reference \cite{Antusch:2007rk}.
    Copyrighted by SISSA.
    }
    \label{fig:lth132}
  \end{center}
\end{figure}

Figure \ref{fig:lth132} shows the results for: BB350, NFC, and
NFO. 
Each of these experiments is unaffected by the mass-hierarchy
degeneracy problem mentioned above for the large value of
$\theta_{13}$ considered. 
NFO offers the best sensitivity.
The conservative Neutrino Factory option compares well to BB350,
whereas the performance on $\delta$ and $\theta_{13}$ individually is
much worse than for BB350 (see also section \ref{SubSect:Perf:NF}).
The reason for this is that an experiment for which events are centred
around the first oscillation maximum, such as a $\beta$-beam or a
superbeam, is sensitive mainly to the $\theta_{13}\sin\delta$ term. 
The Neutrino Factory, however, produces the bulk of the events
above the first oscillation maximum and thus is much more sensitive to
the $\theta_{13}\cos\delta$ term.

\begin{figure}
  \begin{center}
    \ensuremath{\vcenter{\hbox{\includegraphics[width=0.6\textwidth]{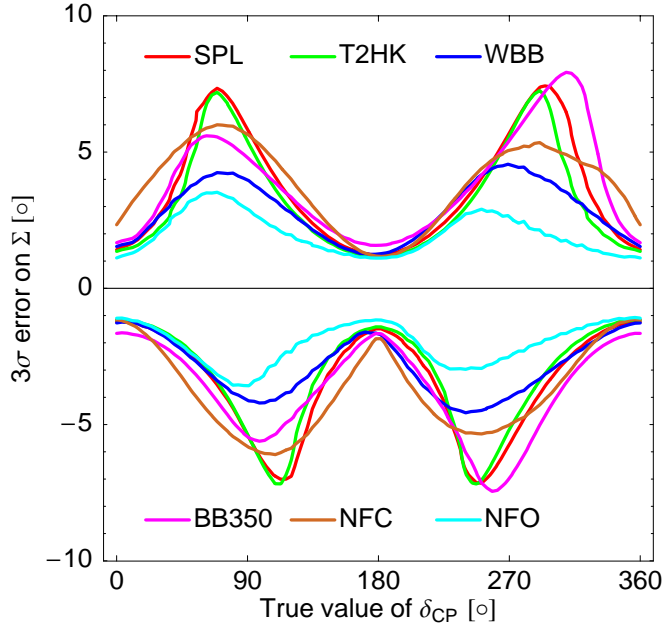}}}}
    \vspace{0.5cm}
    \caption{
      The $3\,\sigma$ error in degrees for $\theta_{12}^\Sigma$ as a
      function of the 
      true value of $\delta$ for $\sin^22\theta_{13}=10^{-1}$. The
      different coloured lines are for different experiments as given in
      the legend. The $\mathrm{sgn}\Delta m^2_{31}$ degenerate solution
      has been omitted. The error on $\theta_{12}$ is $0.75^\circ$.
    }
    \label{fig:sum}
  \end{center}
\end{figure}

So far, results for large $\theta_{13}$ only have been shown.
However, the relative performance of the various options does not
change very much with $\theta_{13}$.
In contrast, each of the options considered except the Neutrino
Factory suffers from the mass-hierarchy degeneracy problem if
$\theta_{13}$ is too small. 
For intermediate values of $\sin^22\theta_{13}\simeq10^{-2}$ the
accuracy of the measurement of $\theta_{12}$ is the dominating factor
and the performance of the various experiments is similar if the
mass-hierarchy problem is ignored.
The true value of $\theta_{12}$ used in the plots is 
$\theta_{12}=33.12^\circ$ ($\sin^2 \theta_{12}=0.3$). 
For larger (smaller) values of true $\theta_{12}$, the bands and
islands in figures \ref{fig:lth13} and \ref{fig:lth132} are shifted up
(down) accordingly. 
The performance of all experiments at large
$\sin^22\theta_{13}=10^{-1}$ is summarised in figure \ref{fig:sum}. 
An interesting observation from this figure is that the WBB performs
second only to NFO. 
The NF is particularly well suited to the determination of the
combination $\theta_{13}\cos\delta$, making this the machine of choice
for testing the sum rule, even for large $\theta_{13}$. 

\subsubsection{Cabibbo Haze in Lepton Mixing}

As a step toward an explanation of the physics of flavour, a
phenomenological approach was advocated recently in which
parametrisations of the lepton-mixing matrix were developed as an
expansion in $\lambda\equiv\sin\theta_c\simeq0.22$ in analogy with
Wolfenstein's parametrisation of quark mixing
\cite{Datta:2005ci,Everett:2005ku,Wolfenstein:1983yz}. 
In addition to its practical advantages for phenomenology,  
the Wolfenstein parametrisation hints at a guiding principle for
flavour theory by providing a framework for examining quark mixing in
the $\lambda \rightarrow 0$ limit.  
Quark-lepton unification implies that if Cabibbo-sized perturbations
are present in the quark sector, such perturbations will also be
manifest in the lepton sector.   
Due to the presence of large angles, however, the lepton-mixing matrix
is unknown in the $\lambda\rightarrow 0$ limit (unlike the quark
mixings, which vanish). 
Hence, if the limit of zero Cabibbo-angle is meaningful for theory,
there is a `Cabibbo haze' in lepton mixing, in which the initial
or `bare' values of the mixings are screened by Cabibbo-sized effects.

Cabibbo effects therefore represent deviations from bare mixings. 
They can be deviations from zero mixing (as in the quark sector); in
this approach such effects are likely to represent the dominant source
of $\theta_{13}$. 
For $\theta_{23}$ and $\theta_{12}$ (and possibly $\theta_{13}$),
Cabibbo-sized perturbations represent deviations from (presumably
large) non-zero initial values.  
Parametrisations are categorised according to the bare mixings and the
structure of the allowed perturbations.  
Perturbations which are linear in $\lambda$ yield shifts of 
$\simlt \theta_c\simeq 13^\circ$, while ${\cal O}(\lambda^2)$ 
shifts are $\sim 3^\circ$.  
CP-violating phases can enter the ${\cal O}(\lambda)$ shifts but may
only occur at sub-leading order, in which case the effective phase is
suppressed and the size of $\theta_{13}$ does not dictate the size of
CP-violating observables. 

One aim of this approach is to obtain an efficient parametrisation of
the lepton-mixing matrix in analogy to the Wolfenstein parametrisation
for the quark-mixing matrix. 
However, current data is clearly consistent with many possible
Wolfenstein-like parametrisations.  
One reason is that there is a wide range of possible bare mixing
parameters and Cabibbo shifts, though some particular values may be
singled out by well-motivated flavour theories.  
Another reason is the current precision of the data.  
Recast in terms of the Cabibbo angle, the error bar on $\theta_{12}$
is of $O(\lambda^2)$, while the uncertainties in $\theta_{23}$ and
$\theta_{13}$ are of $O(\lambda)$. 
Although it is not possible to single out a particular
parametrisation, the approach provides an organising principle for
categorising the many top-down flavour models based on a $\lambda$ 
expansion.  
The approach also provides a useful framework in which to interpret
the results of future experiments, such as the programme to measure
$\theta_{13}$.  
Future facilities are expected to reach the ${\cal O}(\lambda^2)$
range, which will yield important insight into the nature of lepton
mixing in the $\lambda\rightarrow 0$ limit.  

The classification scheme proceeds as follows.  
Recall that the Wolfenstein parametrisation is based on the idea that
the hierarchical quark mixing angles can be understood as a $\lambda$
expansion, with:
\be 
  \mathcal{U}_{\rm CKM}=1+\mathcal{O}(\lambda). 
\ee 
In the lepton sector, a similar parametrisation requires a $\lambda$
expansion of the form :
\be 
  \mathcal{U}_{\rm PMNS}=\mathcal{W}+\mathcal{O}(\lambda). 
\ee 
The starting matrix $\mathcal{W}$, which is dictated by the (unknown)
underlying flavour theory, is then perturbed multiplicatively by a
unitary matrix $\mathcal{V}(\lambda)$, which in turn is assumed to
have a $\lambda$ expansion:
\be
  \mathcal{V}(\lambda)=1+\mathcal{O}(\lambda).
\ee
For the quarks, the starting matrix is the identity matrix and the
perturbation matrix takes the Wolfenstein form. 
For the leptons, the structure of the allowed perturbations depend on
the details of $\mathcal{W}$.  
Due to  Cabibbo haze, $\mathcal{W}$ can take different forms which are
characterised by the number of large angles.  
For simplicity attention will be restricted here to the best-motivated
scenario, in which the bare solar and atmospheric mixings, $\eta_{12}$
and $\eta_{23}$, are non-zero and the bare $\theta_{13}$ vanishes (see
\cite{Everett:2005ku} for a more general analysis). 
In this case  $\mathcal{W}$ is of the form:
\be 
  {\cal W}={\cal R}_1(\eta_{23}){\cal R}_{3}(\eta_{12})\equiv 
  \begin{pmatrix}
    1&0&0\\ 
    0&\cos \eta_{23}&\sin\eta_{23} \\ 0&-\sin\eta_{23} 
    &\cos\eta_{23}
  \end{pmatrix} \,
  \begin{pmatrix}
    \cos\eta_{12} &\sin\eta_{12}&0\\ 
    -\sin\eta_{12}&\cos \eta_{12}&0\\ 0&0&1
  \end{pmatrix}
  \mathcal{P}, 
\ee
where $\mathcal{P}$ is a diagonal phase matrix of the form:
\be 
  \mathcal{P}= \begin{pmatrix}e^{i\alpha_1}&0&0\\ 0&e^{i\alpha_2} &0\\ 
  0&0&e^{i\alpha_3}
  \end{pmatrix}, 
\ee 
which encodes the two physical Majorana CP-violating phases 
$\alpha_{12}\equiv\alpha_1-\alpha_2$ and 
$\alpha_{23}\equiv\alpha_2-\alpha_3$. 

Unlike the quark sector, generically the perturbations do not commute
with the starting matrix:
\be 
  [{\mathcal W}\,,\,{{\mathcal V}(\lambda)}\,]~\neq~0\ .
\ee 
Hence, there are several possible implementations of Cabibbo shifts:  
\begin{itemize} 
  \item {\it Right Cabibbo shifts}. 
        The perturbations can be introduced as a multiplication of
        $\mathcal{V}(\lambda)$ on the right:
        \be 
          {\cal U}_{\rm PMNS}={\cal W}\,{\cal V}(\lambda). 
          \label{eqright};
        \ee
  \item {\it Left Cabibbo shifts}.  
        The perturbations can be implemented as a multiplication of
        $\mathcal{V}(\lambda)$ on the left:  
        \be 
          {\cal U}_{\rm PMNS}={\cal V}(\lambda)\,{\cal W};
          \label{eqleft}
        \ee 
  \item {\it Middle Cabibbo shifts}.
        The perturbations can be sandwiched between the rotation
        matrices of $\mathcal{W}$: 
        \be 
          {\cal U}_{\rm PMNS} = 
              {\cal R}_{1}\,{\cal V(\lambda)}\,{\cal R}_3\mathcal{P},
          \label{eqmidleft} 
        \ee 
        or
        \be 
          {\cal U}_{\rm PMNS} =
              {\cal R}_{1}\,{\cal V}(\lambda)\,{\cal R}_3\mathcal{P}.
          \label{eqmidright}
          \ee
\end{itemize} 
To see that this encompasses all possibilities, recall that the
assumption of Cabibbo haze is that the lepton-mixing matrix has a
$\lambda$ expansion:
\be
  \mathcal{U}_{\rm PMNS}(\lambda)=\sum^{\infty}_{n=0}\lambda^nW_n,
\ee
in which $W_0\equiv\mathcal{W}$. 
This can be expressed as a right Cabibbo shift:
\be
  \mathcal{U}_{\rm PMNS}(\lambda)=\mathcal{W} 
  \sum^{\infty}_{n=0}\lambda^n(\mathcal{W}^{-1}W_n) 
  \equiv\mathcal{W}\mathcal{V}(\lambda), 
\ee
with $\mathcal{V}=\sum^{\infty}_{n=0}\lambda^n(\mathcal{W}^{-1}W_n)$.
It can also be expressed as a middle Cabibbo shift (dropping
$\mathcal{P}$ for simplicity): 
\bea
  \mathcal{U}_{\rm PMNS}(\lambda)&=&\mathcal{R}_1 \mathcal{R}_3 
  \mathcal{V}(\lambda) ; \nonumber \\  
  &=&\mathcal{R}_1\left (\mathcal{R}_3
  \mathcal{V}(\lambda)\mathcal{R}^{-1}_3 \right)
  \mathcal{R}_3\equiv \mathcal{R}_1
  \mathcal{V}^{\prime}(\lambda)\mathcal{R}_3. 
\eea
The generalisation to left shifts is straightforward.  
Note that since $\mathcal{V}$, by assumption, is given by:
\be
  \mathcal{V}(\lambda)=1+\sum^{\infty}_{i=1}\lambda^nV_n,
\ee
$\mathcal{V}^\prime$ can also be written in an analogous form:
\be
  \mathcal{V}^\prime(\lambda)=\mathcal{R}_3\mathcal{V}\mathcal{R}^{-1}_3 
=1+\sum^{\infty}_{i=1}\lambda^n(\mathcal{R}_3V_n\mathcal{R}^{-1}_3).
\ee
Hence, the decomposition into right, left, or middle shifts is
meaningful for a specific choice of $\mathcal{V}$.  
To leading order in $\lambda$, $\mathcal{V}$ is assumed to be:
\be
  \mathcal{V}=
  \begin{pmatrix}
    1 & 
    a_1\lambda&  
    c_1\lambda \\
    -a^*_1\lambda &
    1 & 
    b_1\lambda \\ 
    -c^*_1\lambda 
    & -b^*_1\lambda & 1
  \end{pmatrix}
  +\mathcal{O}(\lambda^2),
\ee
which encompasses the Wolfenstein form ($a_1=1$, $b_1=c_1=0$, and
higher order terms  $b_2=A$ and $c_3=A (\rho-\frac{1}{2}-i\eta)$, in
self-evident notation), and allows for more general perturbations.
Finally, as the shifts in the mixing angles are clearly dominated by
perturbations linear in $\lambda$, it is useful to categorise models
further as single, double, or triple shifts according to the number of
such ${\cal O}(\lambda)$ perturbations in $\mathcal{V}$. 

Given these ingredients, a systematic classification of possible
models was presented in \cite{Datta:2005ci,Everett:2005ku}, to which
the reader is referred for further details.  
Here attention will be focused on one subset of examples.  
It is straightforward to obtain the following general results for the
$O(\lambda)$ shifts in the mixing angles (including phases):
\begin{itemize}
  \item {\it Right shifts}: 
  \bea
    \theta_{12}&=&\eta_{12}+\lambda |a_1|\cos(\alpha_{12}+\phi_{a_1})\\
    \theta_{23}&=&\eta_{23}+\lambda 
    (\cos\eta_{12}|b_1|\cos(\alpha_{23}+\phi_{b_1})
    -\sin\eta_{12}|c_1|\cos(\alpha_{12}-\alpha_{23}+\phi_{c_1})) \\
    \theta_{13}&=&\lambda|b_1e^{i\alpha_{23}}\sin\eta_{12}
    +c_1e^{i(\alpha_{12}-\alpha_{23})}\cos\eta_{12}| ;
  \eea
  \item {\it Left shifts}: 
  \bea
    \theta_{12}&=&\eta_{12}+\lambda(\cos\eta_{23} 
    |a_1|\cos\phi_{a_1}-\sin\eta_{23}|c_1|\cos\phi_{c_1})\\
    \theta_{23}&=&\eta_{23}+\lambda|b_1|\cos\phi_{b_1}\\
    \theta_{13}&=&\lambda|\sin\eta_{23}a_1 + 
    \cos\eta_{23} c_1|;
  \eea
  \item {\it Middle shifts}: 
  \bea
    \theta_{12}&=&\eta_{12}+\lambda|a_1|\cos\phi_{a_1}\\
    \theta_{23}&=&\eta_{23}+\lambda|b_1|\cos\phi_{b_1}\\
    \theta_{13}&=&\lambda|c_1|.
  \eea
\end{itemize}
Each scenario displays distinct correlations between the Cabibbo
shifts of the mixing angles.  
Note that certain shifts are sized by factors dependent on the 
bare-mixing parameters.  
In addition, the shifts in the mixing angles depend on the Majorana
phases $\alpha_{12},\,\alpha_{23}$ only in the right Cabibbo shift
scenario. 
The reason is that generically:
\be
  [{\mathcal P}\,,\,{{\mathcal V}(\lambda)}\,]\neq 0, 
\ee
and hence the right shifts can be rewritten as follows:
\bea
  \mathcal{U}_{\rm PMNS}&=&\mathcal{W}\mathcal{P}\mathcal{V}\nonumber \\
  &=&\mathcal{W}(\mathcal{P}\mathcal{V}\mathcal{P}^{-1})\mathcal{P} \equiv 
  \mathcal{W}\mathcal{V}_{\mathcal{M}}\mathcal{P}.
\eea
$\mathcal{V}_{\mathcal{M}}$ can be obtained from $\mathcal{V}$ through
the replacements $a_i\rightarrow a_ie^{i\alpha_{12}}$, 
$b_i\rightarrow b_ie^{i\alpha_{23}}$, and 
$c_i\rightarrow c_ie^{i(\alpha_{12}-\alpha_{23})}$. 
 
How might certain examples emerge from the viewpoint of flavour
theory?  
One class of examples occur within grand unified models in which the
fermion Dirac-mass matrices obey $SU(5)$ and $SO(10)$. GUT relations
based on the simplest Higgs structures and the down-quark mass matrix
is further assumed to be symmetric, such that 
$\mathcal{M}_d= \mathcal{M}_d^T\sim \mathcal{M}_e$ and
$\mathcal{M}_u\sim \mathcal{M}_\nu$.  
In such models, the quark and lepton mixing matrices are related
\cite{Ramond:2003kk,Ramond:2004en,Ramond:2004xe,Ramond:2004xf}: 
\be 
  {\mathcal U}^{}_{\rm PMNS} =
    {\mathcal U}^{\dagger}_{\rm CKM}\,\,{\mathcal F},
  \label{udagf}
\ee 
where $\mathcal{F}$ is a matrix which encodes the effects of the
neutrino see-saw; in these models, $\mathcal{F}$ must contain two
large angles.  
In the language of this classification scheme, this scenario is an
example of a left Cabibbo single-shift model, in which $\mathcal{F}$
plays the role of $\mathcal{W}$ and $\mathcal{V}$ takes the form of
${\mathcal U}^{\dagger}_{\rm CKM}$.  
Other possible examples include models based on quark-lepton
complementarity, in which case $\mathcal{W}$ is a bimaximal-mixing
matrix and $\mathcal{V}$ has $a_1\neq 0$, $b_1=0$, and $c_1$ may or
may not vanish depending on the details of the model.  
Different predictions for $\theta_{13}$ are implied in these cases
depending on whether the model is a right, left, or middle Cabibbo
shift model.
Tri-bimaximal mixing scenarios are models in which $\mathcal{W}$
takes on the standard tri-bimaximal form, and $\mathcal{V}$ has
$a_1=b_1=0$ and $c_1$ may or may not be zero, with a range of
predictions for $\theta_{13}$ depending on the shift scenario. 

Turning now to the issue of CP violation, the parametrisations also
display different predictions for the leptonic Dirac and Majorana
phases, depending on the details of how and whether phases enter
$\mathcal{W}$ and $\mathcal{V}$.   
Here, only Dirac-type CP violation is considered (as CP-violating
observables sourced by Majorana phases are helicity suppressed and
thus difficult to observe).  
For models in which $\mathcal{W}$ has two large angles (the reader is
once again referred to \cite{Everett:2005ku} for a more general
discussion), the invariant measure of  Dirac CP violation: 
\be 
  \mathcal{J}_{\rm CP} =
  {\rm Im}(\mathcal{U}_{\alpha i}\mathcal{U}_{\beta j} 
  \mathcal{U}^*_{\beta i} \mathcal{U}^*_{\alpha j}) 
  \simeq  \sin 2\theta_{12}\sin 2\theta_{23}\sin 2\theta_{13} \sin\delta,
  \label{jarlskogdef} 
\ee
vanishes in the $\lambda \rightarrow 0$ limit, and a non-zero value can
be generated in two ways: 
\begin{itemize}
  \item {\it Complex $\mathcal{V}(\lambda)$:}
        $\mathcal{V}(\lambda)$ can be the source of CP-violating
        phases, which can be ${\cal O}(1)$ (as in the quark
        sector). 
        Models can be categorised in terms of whether CP
        violation enters at leading or higher order in $\lambda$, and
        whether the effective leptonic phase is predicted to be ${\cal
        O}(1)$ or further suppressed;
  \item {\it  Bare Majorana phases:}
        Majorana phases can also provide a source for Dirac CP
        violation once the Cabibbo-sized perturbations are switched
        on.  
        For left and middle Cabibbo shifts, this does not
        occur. 
        However, for right Cabibbo shifts it does, as such
        shifts encode $\mathcal{P}$ through the modification
        $\mathcal{V} \rightarrow \mathcal{V}_{\mathcal{M}}$. 
\end{itemize}
Consider equation (\ref{udagf}) as an illustrative example.  
If $\mathcal{V}$ is of the Wolfenstein form (complex $O(\lambda^3)$
terms), $\mathcal{J}_{\rm CP}$ is:
\be
  \mathcal{J}_{\rm CP}=\frac{1}{4}A\lambda^3\eta \cos\eta_{23}
  \sin 2\eta_{23} \sin 2\eta_{12}.
  \label{udagfj}
\ee
Note that in this model, the shifts in the angles are given to
$O(\lambda^2)$ by: 
\bea 
  \theta_{12}&=&\eta_{12}-\lambda\cos\eta_{23} \, , \\ 
  \theta_{23}&=&\eta_{23}-\lambda^2(A+\frac{1}{4}\sin 2\eta_{23}) 
           \, {\rm~and} \\
  \theta_{13}&=&-\lambda\sin\eta_{23} \, . 
\eea
The effective leptonic phase is $\delta \sim O(\lambda^2)$, in
contrast to the $O(1)$ CKM phase.  
This suppression occurs because the phases in $\mathcal{V}$ arise in
subdominant contributions to the mixing angles. 
Models with this feature demonstrate that while the magnitude of
$\theta_{13}$ is clearly correlated with the prospects for the
observability of lepton-sector CP violation, it is not the whole
story because the CP-violating phase itself may be suppressed. 
  
In summary, we are beginning to read the new lepton data, but there is
much work to do before a satisfactory and credible theory of flavour
is proposed. 
In the meantime, it is illustrative to examine the lepton sector
through the lens of quark-lepton unification, and investigate
parametrisations of the lepton-mixing matrix which include
Cabibbo-sized effects. 
The approach emphasises the need for precision measurements, as
present data are insufficient for singling out a particular
parametrisation.  
Should the limit of zero Cabibbo mixing prove to be
meaningful for theory, with improved data we may be able to see
flavour patterns through the Cabibbo haze. 

\subsection{Lepton-flavour violation}

Searching for lepton-flavour violation in charged-lepton decays
is an important way to look for new physics beyond the Standard 
Model \cite{Kuno:1999jp}. 
Since the early days of muon experiments, processes such as 
$\mu \to e \gamma$ have been searched for, and the absence
of such processes has lead us to consider the separate conservation of
electron and muon numbers. 
The discovery of two flavours of neutrino in 1962 at BNL indicated
that lepton-flavour conservation is indeed realised in nature to a
good degree of accuracy.  

The situation has changed since the discovery of the neutrino
oscillations. 
The separate conservation of each lepton number individually is likely
violated. 
However, lepton-flavour violation can be observed in 
charged-lepton processes depends on how neutrino mass is
generated. 
In the simple Dirac-neutrino, or the see-saw, framework, 
lepton-flavour violating processes in muon decays
are suppressed by more than twenty orders of magnitude below 
the present experimental upper bounds. 
On the other hand, lepton-flavour violation becomes large, if some 
new particles or interactions exists at the TeV scale. 
Therefore, searching for lepton-flavour violation in muon and tau
decay processes provides important information on the origin 
of the neutrino mass. 
   
{\noindent \bf Lepton-flavour violation in three-muon processes}

Among the various lepton-flavour violating processes, three-muon 
processes, $\mu \to e \gamma$, $\mu \to 3e$, and $\mu - e$ conversion
in muonic atoms, are particularly important. 
The current experimental upper bound for $\mu \to e \gamma$
\cite{Brooks:1999pu} is at the $10^{-11}$ level and about one order of
magnitude smaller for the other two processes
\cite{Bellgardt:1987du,Bertl:2001fu}.  
Although the $\mu -e$-conversion process has the smallest upper bound,
the process which imposes the strongest constraints on the theoretical
parameters depends on the model under consideration. 
Muonium-anti-muonium conversion is another process which violates
the conservation of electron and muon numbers but conserves the total
lepton number. 
This process is sensitive to new physics which changes the
muon and electron numbers by two units. 
Upper bounds on the branching ratios of tau-lepton-flavour violating
processes have been improved recently at KEK and the SLAC-B-factory
experiments, and have reached the level of $10^{-7}$ and below
depending on the decay mode in question
\cite{Hayasaka:2007vc,Miyazaki:2007jp,Miyazaki:2006sx,Yusa:2004gm,Aubert:2006cz,Aubert:2005wa,Aubert:2005tp,Aubert:2005ye,Aubert:2003pc}.
Generally speaking, three-muon processes put stringent 
constraints on models that yield lepton-flavour violation and the
study of correlations among the varous processes is useful to identify
the correct model. 

In near future, the MEG experiment is expected to improve the
search-limit on the $\mu \to e \gamma$ process by more than  
two orders of magnitude. 
If lepton-flavour violation is discovered, the next steps will be to 
discover the nature of lepton-flavour violation and to distinguish
between the different models.
The following techniques can be used to do this:
\begin{itemize}
  \item 
    The ratio of the branching ratios of $\mu \to 3 e$ ($\mu-e$
    conversion) and $\mu \to e \gamma$ depends on what kinds of
    operator are responsible for lepton-flavour violation. 
    In particular, if all three processes are generated by
    the same photonic-dipole-type operator, the following relations
    are hold:
    \begin{eqnarray}
      B(\mu^+ \to e^+e^+e-) &\sim& 6\times10^{-3}B(\mu\to e\gamma),
        \label{3e-rel}\\
      \frac{\sigma(\mu^- Ti \to e^- Ti)}{\sigma(\mu^- Ti \to capture)}&\sim& 
        4 \times 10^{-3}
        B(\mu\to e\gamma).
        \label{mue-rel}
    \end{eqnarray}
    This is a good approximation, for example, for most 
    supersymmetric models. 
    On the other hand, if lepton-flavour violation is generated by
    tree-level processes, $\mu \to 3 e$ and/or $\mu-e$ conversion, the
    branching fractions could be much larger than that of 
    $\mu \to e \gamma$;
  \item 
    Angular distributions in polarised muon decays provide information
    on the chiral and CP structures of lepton-flavour violating
    operators \cite{Okada:1999zk}.
    For the $\mu \to e \gamma$ search with a polarised $\mu^+$, the 
    $\mu^+ \to e^+_L \gamma$ and $\mu^+ \to e_R \gamma$ operators are
    distinguished by the angular distribution of the positron-momentum
    direction with respect to the initial muon-polarisation direction.
    The chiral structure carrys information on the origin of the
    lepton-flavour violating interaction. 
    In supersymmetric models, for example, the chirality depends on
    whether the flavour mixing exists in the right- or left-handed  
    slepton sector, and this distinction could provide very important
    clues to the interaction at the GUT scale; and
  \item 
    In the $\mu-e$ conversion search, branching-ratio measurements
    of different atoms provides one means of discriminating between
    the different operators \cite{Kitano:2002mt}.
    The atomic-number dependence of the $\mu-e$-conversion rate
    differs for different types of quark-level operators. 
    For example, we can distinguish scalar, vector, and photon-dipole
    type operators by comparing branching fractions measured using
    different nuclei, for example a low-atomic-number nucleus such as
    aluminium and a heavy nucleus such as lead.  
\end{itemize}
These techniques would provide information on different aspects of
lepton-flavour violating interactions, and are the basic steps to
required to clarify the nature of new interactions.  
          
{\noindent \bf Supersymmetry and muon lepton-flavour violating processes}
 
Among the new physics models explored by searches for lepton-flavour
violation, supersymmetry is the most important.
Since supersymmetry requires the introduction of a supersymmetric
partner for each particle in the Standard Model, sleptons should
exist. 
Mass terms for the slepton depend on supersymmetry-breaking terms,
which do not have an a-priori relation with lepton mass terms. 
In fact, the flavour mixing in the  
slepton-mass matrix is strongly constrained by the lepton
flavour-violating processes. 
This is a part of the flavour problem in supersymmetric models,
some mechanism is needed to suppress flavour-changing neutral-current
processes in the quark and the lepton sectors. 
A solution to this problem is one of necessary conditions for a
realistic supersymmetric model, and a variety of
supersymmetry-breaking mechanisms are proposed.  
In principle, we will be able to identify the correct scenario by
looking at the super-particle mass spectrum in energy frontier
experiments at the LHC and the International Linear Collider.

\begin{figure}
  \begin{center}
    \includegraphics[width=10cm]{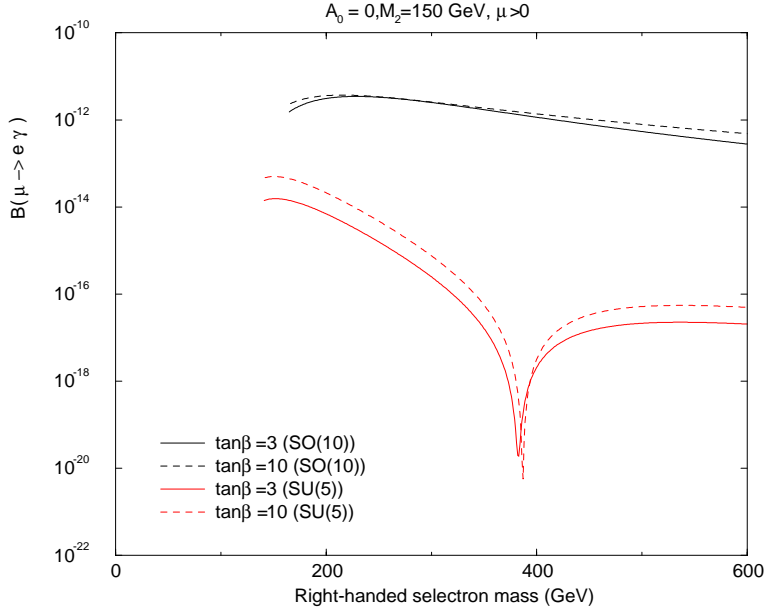}
  \end{center}
  \caption{
    $\mu \to e \gamma$ branching ratios
    for SU(5) and SO(10) SUSY GUT.
  Adapted with permission of Reviews of Modern Physics from
  figures 8 and 13 in reference \cite{Kuno:1999jp}.
  Copyrighted by the American Physical Society.
  }
  \label{fig:br_eg_sgut}
\end{figure}

Searches for lepton-flavour violating processes have a role to play 
in the determination of the off-diagonal elements of the slepton-mass
matrix. 
The determination of these elements is particularly important because
these elements carry information at very high energy scales such as
the GUT scale and the see-saw neutrino scales
\cite{Hall:1985dx,Borzumati:1986qx}.
Even if we take a scenario where off-diagonal slepton terms are absent
at the Planck scale, renormalisation effects due to large Yukawa
coupling constants can induce sizable off-diagonal terms.
In SUSY-GUT models, the large top Yukawa coupling constant a source of
lepton-flavour violation because quarks and leptons are connected to
each other above the GUT scale
\cite{Barbieri:1994pv,Barbieri:1995tw}. 
A typical example is shown in figure \ref{fig:br_eg_sgut} for SU(5)
and S0(10) SUSY GUTs. 
The branching ratio is expected to be close to the current experimental 
upper limit for the SO(10) case.

In the supersymmetric see-saw model, a potentially large Yukawa coupling 
is provided by the neutrino Yukawa coupling constants. 
The off-diagonal term in the left-handed slepton-mass matrix is give
by:
\begin{eqnarray}
  (m_{\tilde{l}_{L}}^{2})_{ij} \simeq -\frac{1}{8\pi^{2}}
  (y_{\nu})_{ki}^* (y_{\nu})_{kj} m_{0}^{2}
  (3 + |A_{0}|^{2})\ln({M_{P}\over M_{R}}),
  \label{eq:mssmrn}
\end{eqnarray}
where $M_P$ and $M_R$ are the Planck mass and the right-handed 
neutrino mass respectively, $m_0$ is the universal scalar mass,
$A_{0}$ is the universal triple-scalar-coupling constant for
supersymmetry-breaking terms, and $y_{\nu}$ is the neutrino Yukawa
coupling constant. 
Since the see-saw relation suggests that the Yukawa coupling is
proportional to the square-root of $M_R$, the lepton-flavour violating  
branching ratio is proportional to $M_R^2$. 
Although the flavour structure of $y_{\nu}$ is not directly related
to the flavour mixing in the PMNS matrix, it is natural to expect
sizable off-diagonal elements from the large neutrino mixing. 
In fact, the $\mu \to e \gamma$
branching ratio can reach the experimental bound for 
$M_R = 0(10^{13})-0(10^{14})$ GeV 
\cite{Hisano:1995nq,Hisano:1995cp,Hisano:1998fj}.

\begin{figure}
  \begin{center}
    \includegraphics[width=14cm]{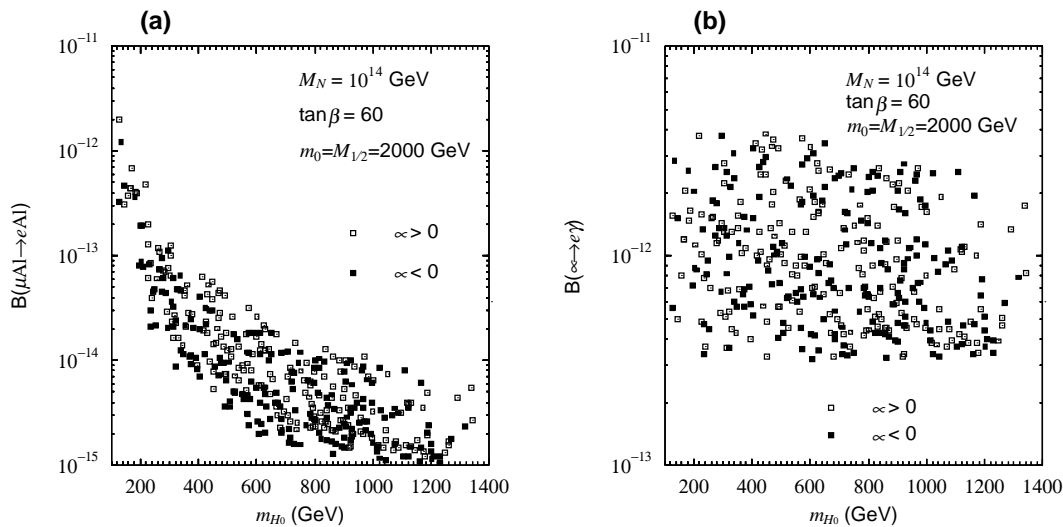}
  \end{center}
  \caption{
    $\mu-e$ conversion branching ratio
    in aluminium nucleus and $\mu \to e \gamma$ branching ratio
    as a function heavy CP even Higgs boson mass 
    in the supersymmetric see-saw model.
    Taken with kind permission of Physical Letters from figure 1 in
    reference \cite{Kitano:2003wn}.
    Copyrighted by Elsevier B.V.
  }
  \label{fig:br_higgs}
\end{figure}

There is an interesting special case which can be realised 
for a larger value of the ratio of the two Higgs vacuum expectation 
values $(\tan{\beta})$ \cite{Babu:2002et,Sher:2002ew,Dedes:2002rh}. 
In this case, supersymmetric loop corrections
to the Higgs-lepton vertex can generate a large lepton-flavour
violating coupling. 
As a result, heavy Higgs-boson exchange diagrams
can be dominant, and the $\mu -e$ conversion process is enhanced 
relative to the $\mu \to e \gamma$ process \cite{Kitano:2003wn}. 
An example is shown in figure \ref{fig:br_higgs}. 
For a smaller heavy-Higgs-boson mass, the two branching ratios can be 
more similar. 
For the same parameter space, we can confirm that the dominant
operator is of the scalar type from the atomic-number dependence of 
the $\mu-e$ conversion rate.

{\noindent \bf Other theoretical models}

There are many new-physics models that predict sizable rates for
muon lepton-flavour violating processes \cite{Kuno:1999jp}. 
In many cases, the lepton-flavour violation is related to the physics
of neutrino-mass generation, namely the interaction responsible for
the neutrino mixings also induces lepton-flavour violation. 
This is the case for the supersymmetric see-saw model discussed
above. 
Other examples are the Zee model \cite{Hasegawa:2003by}, Dirac-type
bulk neutrinos in the warped extra dimension \cite{Kitano:2000wr},
the triplet-Higgs model \cite{Chun:2003ej,Kakizaki:2003jk}, and 
the non-supersymmetric left-right symmetric model
\cite{Cirigliano:2004mv,Boyarkin:2004zd,Akeroyd:2006bb}.
Supersymmetric, with R-parity violation, can be considered 
to be in this category, since neutrino masses can be generated
from R-parity violating couplings \cite{deGouvea:2000cf}. 
Since each model introduces lepton-flavour violation in 
a different way, the phenomenological features can be quite
different and measurements will provide important clues to identify
the correct model of neutrino-mass generation.   

The triplet-Higgs model provides a simple way to generate neutrino
masses from a small triplet vacuum-expectation value. 
In this model, the triplet Higgs and lepton coupling generating
neutrino mass also induces a doubly-charged Higgs boson and lepton
coupling.     
The neutrino-mixing matrix has a direct relation to the
doubly-charged-Higgs-boson coupling. 
Since the doubly-charged Higgs boson gives a   
tree-level contribution to the $\mu \to 3e$ process, this can 
dominate over the other two processes. 
On the other hand, the $\mu \to e \gamma$ and the $\mu-e$ conversion
branching ratios become similar. 

The left-right symmetric model also has the triplet-Higgs field. 
In this case, however, neutrino masses can be generated by the see-saw
mechanism. 
The right-handed neutrino-mass term arises in  
association with $SU(2)_L\times SU(2)_R \times U(1)_{B-L}$ symmetry
breaking to the Standard Model gauge groups. 
If this scale is close to the TeV scale, observable lepton-flavour
violating effects are generated through the doubly-charged Higgs boson 
and lepton couplings. 
Unlike the triplet-Higgs model, the relationship
between neutrino mixing and lepton-flavour violation is not
straightforward.
A generic feature is that the $\mu \to 3e$ branching ratio is larger
by two orders of magnitude compared to the $\mu \to e \gamma$ and the
$\mu-e$ conversion branching ratios. 

In this way, muon lepton-flavour violating processes provide
one way to explore physics beyond the Standard Model.
This is particularly important because neutrino oscillations
are clear evidence of new physics, and the origin of neutrino
masses is still unknown. 
There are various scenarios for neutrino-mass generation, each
with different features that may give rise to observable signals for 
lepton-flavour violation in charged-lepton processes.
The experimental pursuit of $\mu \to e \gamma$, $\mu \to 3e$,
and $\mu-e$ conversion is important if the origin of flavour mixing in
the lepton sector is to be determined.

%% file: 03-New-physics-and-the-SnuM/03-03-Cosmology/03-03-Cosmology.tex
\subsection{Cosmology}

\subsubsection{Neutrinos and Large Scale Structure}

The observation of cosmological perturbations -- such as Cosmic
Microwave Background (CMB) anisotropies, or the large-scale density
perturbations reconstructed, e.g., from the galaxy distribution in the
Universe -- are known to provide good measurements of many
cosmological parameters. 
For instance, the spectrum of cosmological perturbations
is very sensitive to the abundance of ultra-relativistic particles in
the early Universe. 
This can be used to make a good estimate of the number of neutrinos
which were in thermal equilibrium at that time, parametrised by an
effective number, $N_{\rm eff}$.  
The standard scenario with three neutrino flavours and no
other relativistic relics in the Universe (apart from photons)
corresponds to $N_{\rm eff}=3$, while scenarios with one light sterile
neutrino originally in thermal equilibrium corresponds to 
$N_{\rm eff}=4$; relaxing the thermal equilibrium assumption, the last 
scenario would give $3 < N_{\rm eff}~\leq~4$. 
Current cosmological bounds give $N_{\rm eff}=3.8^{+2.0}_{-1.6}$ at
2$\sigma$
\cite{Hannestad:2005jj,Hannestad:2006mi,Crotty:2003th,Pierpaoli:2003kw,Seljak:2006bg,Mangano:2006ur,Hamann:2007pi},  
which is compatible with the standard scenario, but also with the
presence of extra relativistic relics. Future experiments are expected
to reach a 1$\sigma$ sensitivity of 0.3 in approximately five years
from now, and should be able to confirm the standard $N_{\rm eff}=3$
cosmological scenario with better accuracy than Big Bang
nucleosynthesis bounds. 
In the rest of this section, it will be assumed, for simplicity, that
$N_{\rm eff}=3$.

Neutrino masses are more difficult to measure than $N_{\rm eff}$
because they are too small to contribute more than $\sim 1\%$ of
the current energy density of the Universe. 
Fortunately, the formation of structures (galaxies and clusters)
during the matter-dominated epoch is quite sensitive even to small
neutrino masses.

\paragraph{Impact of neutrinos on structure formation: theoretical
predictions} 

The process of galaxy formation depends very much on the velocity
dispersion of the components contributing to the matter of the
Universe (for a review, see \cite{Lesgourgues:2006nd}).
If all non-relativistic components (such as baryons and Cold Dark
Matter, CDM) have a very small velocity dispersion the process of
gravitational collapse reaches its maximal efficiency. 
The matter (or energy) density contrast starts from very small values
in the early Universe, with Fourier modes 
$\delta_k= [\delta \rho_k /\bar{\rho}]$ of order $10^{-5}$. 
On wavelengths corresponding today to the Large Scale Structure (LSS)
of the Universe, the density contrast starts to be amplified during
the radiation dominated epoch, but at a slow (logarithmic)
rate. 
Efficient structure formation begins after the time of
radiation-matter equality, when the photon pressure cannot resist the
gravitational in-fall. 
At this point, the rate of linear structure formation is given by
$\delta_k \propto a$, where $a$ is the scale factor. 
This simple law is the result of a balance between gravitational
collapse and the expansion of the Universe (which tends to increase
all distances, and therefore to damp gravitational forces and to slow
down structure formation). 
A crucial observation is that $\delta_k \propto a$ is a solution of
the Einstein equation only under the condition that the same species
contributes to both gravitational collapse and to the expansion
(through the Friedmann law). 
This process brings $\delta_k$ from order $10^{-5}$ to order one, i.e. to
the non-linear regime, starting with the smallest wavelengths.
The non-linear evolution is very difficult to simulate numerically,
but many current and future observations are based on large enough
wavelengths or redshifts for probing the linear (or mildly non-linear)
regime, for which theoretical predictions are well under control.

If neutrinos have a small mass (it will be assumed first, for
simplicity, that only one species is massive), there will be a
constant fraction of non-relativistic matter in the form of neutrinos
between the time at which the neutrino became non-relativistic and
today. 
Non-relativistic neutrinos have a much larger velocity dispersion than
CDM particles, only two or three orders of magnitude smaller than the
speed of light:
\begin{equation}
  v \equiv\frac{\langle p \rangle}{m}
  \simeq\frac{3 T_{\nu}}{m}
  \simeq 150 (1+z) \left( \frac{1 \, \mathrm{eV}}{m} \right)
  {\rm km}\,{\rm s}^{-1}~,
\end{equation}
where $z=(a_0/a-1)$ is the redshift.  
The neutrinos cannot cluster on scales smaller than the total distance
over which they travel on average between the early Universe and today
(this distance, called the free-streaming length, is insensitive to
the precise choice of `time zero').
Indeed, on such scales, the neutrinos experience free diffusion
instead of being trapped inside gravitational potential wells. 
Therefore, we could expect naively that on scales smaller than the
free-streaming scale, the density contrast $\delta_k$ of the total
non-relativistic matter should be reduced by a fraction $f_{\nu}$,
where $f_{\nu}$ is the relative contribution of neutrinos to the total
non-relativistic matter density. 
Fortunately, the effect is stronger than this since neutrinos not only
do not participate in the gravitational collapse on small scales but
neutrinos also slow down the growth of the density contrast of other
matter components, CDM, and baryons.

The balance between gravity and expansion described above is broken in
presence of neutrinos. 
On scales smaller than the free-streaming length, neutrinos do not
participate in the gravitational collapse, but contribute to the
expansion because their homogeneous background density appears in the
Friedmann equation.
So, massive neutrinos give rise to more expansion for the same amount
of matter subject to gravitational clustering than would be the case
if neutrinos were massless. 
As a consequence, the growth rate of the density contrast $\delta_k$
is reduced on those scales to $\delta_k \propto a^{1- (3/5)f_\nu}$. 
If the neutrino mass is very small, this reduction is tiny, but it
accumulates over an extended period of time, so that today $\delta_k$
can be significantly smaller than in the massless case; typically the 
relative reduction is given by a factor $-4 f_{\nu}$.

\begin{figure}
  \begin{center}
    \includegraphics[width=0.73\textwidth]{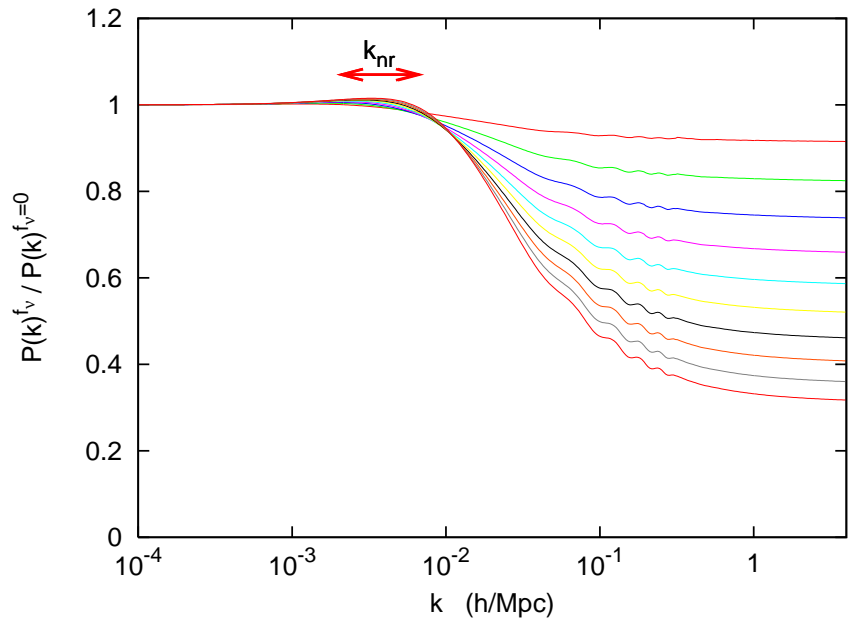}\\
  \end{center}
  \caption{
    Ratio of the matter power spectrum including three degenerate
    massive neutrinos with density fraction $f_{\nu}$ to that with
    three massless neutrinos.  
    The parameters 
    $(\omega_{\rm m}, \, \Omega_{\Lambda})=(0.147,0.70)$ are kept
    fixed, and from top to bottom the curves correspond to 
    $f_{\nu}=0.01, 0.02, 0.03,\ldots,0.10$.  
    The individual masses $m_{\nu}$ range from $0.046$~eV to 
    $0.46$~eV, and the scale $k_{\rm nr}=2 \pi/ \lambda_{\rm nr}$
    from $2.1\times10^{-3}h\,$Mpc$^{-1}$ to
    $6.7\times10^{-3}h\,$Mpc$^{-1}$ as shown on the top of the 
    figure.
    Taken with kind permission of Physics Reports from figure 13
    in reference \cite{Lesgourgues:2006nd}.
    Copyrighted by Elsevier Science B.V.
  }
  \label{fig_tk} 
\end{figure}

In total, the signature of massive neutrinos on the total matter power
spectrum at redshift $z$, denoted 
$P(k,z) \equiv \langle | \delta_k (z)|^2 \rangle$, is the sum of two
effects: 
\begin{enumerate}
  \item As a function of $k$, the matter power spectrum $P(k,z)$ is 
        step-like suppressed for wavelengths smaller than the
        free-streaming length (see figure \ref{fig_tk}).
        More precisely, for any observable redshift, what matters
        is the free-streaming length at the time of the
        non-relativistic transition, since $P(k,z)$ goes through a
        maximum when:
        \begin{equation}
          \lambda_{\rm nr}
          \simeq 350 \,\, \Omega_{\rm m}^{-1/2} 
          \left( \frac{1 \, \mathrm{eV}}{m} \right)^{1/2}
          h^{-1}\,\mathrm{Mpc}~,
        \end{equation}
        where $\Omega_m \simeq 0.3$ is the matter-density fraction
        today. 
        The relative amplitude of the small-scale suppression today is
        well approximated by $-8 f_{\nu}$. Note that no other
        cosmological parameters have such a step-like effect on
        $P(k,z)$; and
  \item As a function of $z$ or $a$, the matter power spectrum 
        undergoes a different evolution on large scales (with 
        $P(k,z) \propto a^2$) and small scales (with 
        $P(k,z) \propto a^{2-(6/5)f_{\nu}}$). 
        This is an absolutely unique effect of dark-matter particles
        with a large velocity dispersion, no other known ingredient
        can justify such a scale-dependent growth factor. 
\end{enumerate}
Note that both effects depend primarily on the total neutrino mass
$M_{\nu}=\sum_i m_{{\nu}i}$ (summed over the three mass eigenstates):
\begin{equation}
  f_{\nu} = \frac{\Omega_{\nu}}{\Omega_m} 
  \simeq \frac{M_{\nu}}{14 {\rm eV}}~.
\end{equation}
Since massive neutrinos have such distinct signatures on LSS, the
total neutrino mass may, in principle, be extracted with a precision 
which depends upon:
\begin{itemize}
  \item The statistical and systematic uncertainties of the
        experimental data; large uncertainties will result in a
        confusion between the effect of massive neutrinos and that of
        other cosmological parameters; and
  \item The priors that are considered acceptable for the underlying
        cosmological model. 
\end{itemize}

\paragraph{Current bounds} 
\label{Para:Currentbounds}
Currently, the combination of up-to-date CMB and LSS data is
compatible with the simplest version of the $\Lambda$CDM scenario,
containing three species of massless neutrinos. Still, cosmological
observations provide a stringent upper bound on the total neutrino
mass.  
This bound is not unique since it depends on the exact data set
considered and on the theoretical priors.
The data sets used to determine the bound include CMB anisotropy
measurements (from WMAP \cite{Spergel:2006hy} and other experiments
probing smaller angular scales). 
CMB anisotropies have a weak dependence on neutrino masses,
but by accurately measuring other cosmological parameters CMB data
plays a crucial role in reducing parameter degeneracies.

Current large-scale structure data consists of several types of
complementary observations. One of them is the galaxy-galaxy two-point
correlation function, best measured by the 2dF \cite{Cole:2005sx} 
and SDSS \cite{Tegmark:2003uf,Tegmark:2006az} groups. 
This observable can be used over a range of scales and directly
reflects the shape of the linear power spectrum predicted by the
theory, modulo an unknown normalisation factor called the
light-to-mass bias, the galaxy-galaxy correlation function probes the 
shape, but not the amplitude of the primordial spectrum. 
Therefore, an accurate determination of the shape of the spectrum is
sufficient for detecting the characteristic step-like suppression
caused by massive neutrinos.
However, the galaxy-galaxy correlation function does not probe an
extended range of scales; it is limited on small scales by the
fact that it is difficult to compare the theory with the data
for strongly non-linear scales $k > 0.2 h^{-1}$Mpc; and it is limited
on large scales by selection effects (i.e., if galaxies are too far
from us, they are also too faint to be accurately sampled).

To complement this type of observation, the light-to-mass bias can
be measured (for example by using higher-order correlations beyond the
two-point correlation function). 
Determination of the light-to-mass bias is important for the
determination of the neutrino mass because it fixes the amplitude of
the matter power spectrum on small scales, while on large scales CMB
experiments provide an accurate normalisation. 
The comparison of the two measurements provides some constraints on the
step-like suppression caused by massive neutrinos.

Instead of being computed in three-dimensional space, the
galaxy-galaxy two-point correlation function can be measured in
angular space. 
This method offers greater sensitivity to the acoustic oscillations
imprinted on the baryon density before photon decoupling. 
This type of data is usually called Baryonic Acoustic Oscillation
(BAO) data. 
The latest BAO data, obtained by the SDSS collaboration
\cite{Eisenstein:2005su}, provides more precise constraints on
cosmological parameters (including the neutrino mass) than can be
obtained using the three-dimensional galaxy-galaxy power spectra at
present.

Measurements of the matter power spectrum over a wide range of 
scales on both sides of the characteristic scale $\lambda_{\rm nr}$
are required for the determination of neutrino mass.
Since the limitation on small scales is given by the transition
to the non-linear regime, it would be very useful to measure the
matter power spectrum at large redshift, i.e. far back in time, when
the non-linear scales were confined to smaller wavelengths than
today. 
This can be done using the Lyman-$\alpha$ forest data coming from a 
detailed analysis of quasar spectra, obtained for instance by the SDSS
collaboration \cite{McDonald:2004eu}. 
For each spectrum, one can identify a waveband corresponding to
Lyman-$\alpha$ absorption along the line of sight; the wavelength at
which the Lyman-$\alpha$ absorption band appears depends upon the
redshift of the galaxy in question.
The Lyman-$\alpha$ forest data is a tracer of matter fluctuations at
redshifts in the range $2 < z < 3$, this is to be compared with
current date on the galaxy-galaxy correlation function which spans
redshifts in the range $0<z<0.2$.
Therefore, Lyman-$\alpha$ forest data can probe very small scales
which are strongly non-linear today, but were mildly non-linear
at the time of the transition. 
The data can be related to the theoretical linear power spectrum.
However, there is still some controversy about various aspects which
might lead to an underestimation of systematic uncertainties.

\begin{figure}
  \begin{center}
    \includegraphics[width=.80\textwidth]{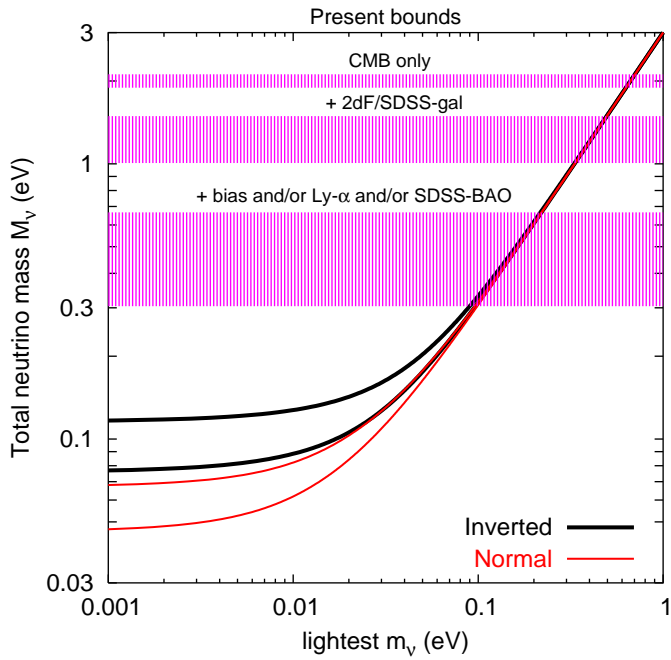}
  \end{center}
  \caption{
    Various current upper bounds (2$\sigma$ level)
    from cosmological data on the total neutrino mass, compared to the
    values in agreement with neutrino oscillation data (at the 3 $\sigma$
    level): for normal hierarchy, the total neutrino mass as a function of
    the lightest eigenstate mass is to be found between the two red lines;
    for inverted hierarchy, between the two black lines.
    Taken with kind permission of Physics Reports from figure 19
    in reference \cite{Lesgourgues:2006nd}.
    Copyrighted by Elsevier Science B.V.
  }
  \label{fig_current}
\end{figure}

A graphical summary is presented in figure \ref{fig_current}, where the
cosmological bounds found in the literature correspond to the
horizontal bands. 
The three bands correspond to different types of data
and the thickness of each band roughly describes the spread
of values obtained by different authors \cite{Lesgourgues:2006nd}
(see reference \cite{Hannestad:2007tu} for an update).
The upper band corresponds to the constraints obtained from CMB data
only.
These bounds are very robust because the CMB probes the density
contrast deep into the linear regime.  
The 2$\sigma$ limits on $M_{\nu}$ derived from current CMB data range
from 2~eV to 3~eV.
The middle band includes three-dimensional measurements of the
galaxy-galaxy correlation function in addition to CMB data. 
Here, the light-to-mass bias is left as a free parameter,
so the data only measures the shape of the matter power spectrum.
The corresponding robust and conservative bounds on the neutrino mass
are in the range $0.9-1.7$~eV.
Finally, the lower band includes data with more controversial
systematic uncertainties, or for which the comparison between theory
and observations is non-trivial and subject to caution, the
light-to-mass bias determination and/or Lyman-$\alpha$ forest data
and/or BAO angular spectrum. 
In this case, the 2$\sigma$ upper limit on $M_{\nu}$ ranges typically
from 0.2 to 0.9 eV.
One can see from figure\ \ref{fig_current} that current cosmological
data probe the region where the 3 neutrino states are degenerate, with
a mass $M_\nu/3$. If one trusts the most aggressive combination of data
sets (in particular, from Lyman-$\alpha$ forests), this region can be
considered as entirely excluded by cosmological observations.

\paragraph{Future prospects} 

In order to improve the bounds on $M_{\nu}$ significantly, or to
detect a non-zero value, it is necessary to observe large-scale
structures both:
\begin{itemize}
  \item On larger scales than today, in order to increase the
        lever-arm on the matter power spectrum towards large
        wavelengths, and also to reduce the statistical (sampling)
        error on all scales; and
  \item At higher redshift to probe smaller scales in the linear
        or mildly non-linear regime and thereby to increase the lever
        arm towards small wavelengths. 
        In addition, measurements at high redshift are sensitive to
        the modified growth rate of density contrasts which are
        imprinted by neutrinos on small scales. 
\end{itemize}
More precise CMB data would also be useful to constrain more strongly 
other cosmological parameters so further reducing parameter
degeneracies. 

\begin{figure}
  \begin{center}
    \vspace{-2cm}
    \hspace{-1.5cm}
    \includegraphics[width=.90\textwidth]{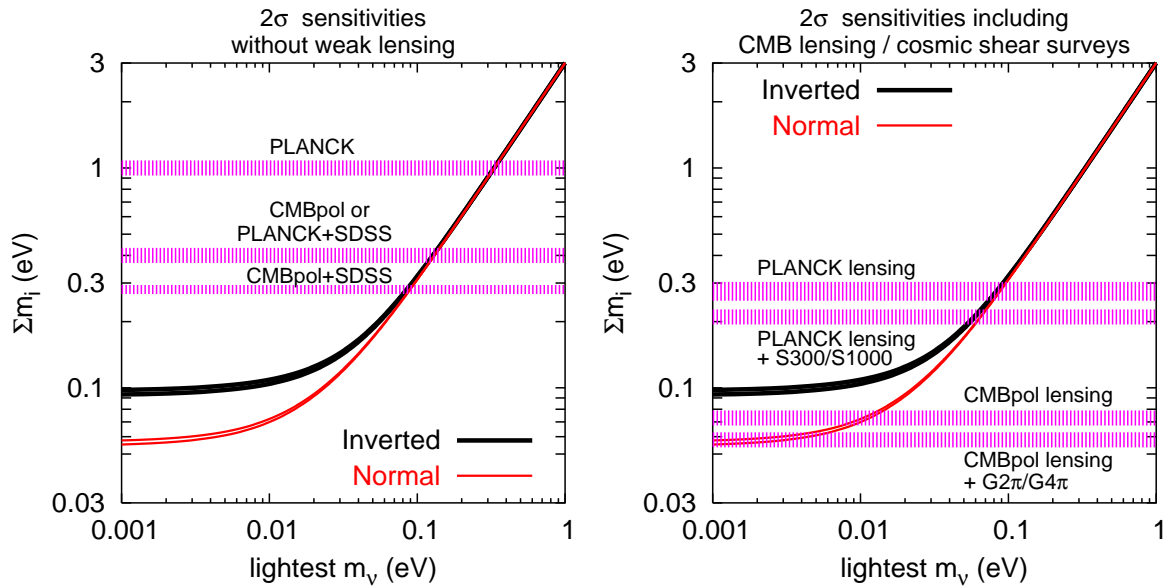}
    \caption{
      Forecast 2$\sigma$ sensitivities to the total neutrino mass from
      future cosmological experiments, compared to the values in
      agreement with present neutrino oscillation data (assuming a
      future determination at the $5\%$ level). 
      Left: sensitivities expected for future CMB experiments (without
      lensing extraction), alone and combined with the completed SDSS
      galaxy redshift survey. 
      Right: sensitivities expected for future CMB experiments
      including lensing information, alone and combined with future
      cosmic shear surveys.  Here CMBpol refers to a hypothetical CMB
      experiment roughly corresponding to the Inflation Probe mission.
    Taken with kind permission of Physics Reports from figure 23
    in reference \cite{Lesgourgues:2006nd}.
    Copyrighted by Elsevier Science B.V.
    }
  \label{fig_future}
  \end{center}
\end{figure}

The expected sensitivity of different cosmological data to $M_{\nu}$
is shown in figure \ref{fig_future}.
The figure also shows the values of $M_{\nu}$ which are allowed in two
of the possible three-neutrino schemes (see reference
\cite{Lesgourgues:2006nd} for details). 
The left-hand panel shows the expected sensitivity of future CMB and
galaxy-redshift surveys.  
The Planck satellite will provide a measurement with a $2\sigma$
sensitivity of the order of 1~eV; the same data combined with the
completed results of the SDSS galaxy redshift survey should reach
0.4~eV \cite{cosmo:planck}. 
NASA is studying a number of projects with even better sensitivity and 
resolution, under the generic name of the `Inflation Probe'
\cite{cosmo:ip}.
Taking the sensitivity of one of these projects, CMBpol, as a
benchmark, yields an expected sensitivity of 0.4~eV for CMBpol alone,
and a sensitivity slightly better than 0.3~eV when the CMBpol data is
combined with SDSS.

More spectacular improvements can be expected from weak lensing
experiments, the goal of which is to deduce the surrounding
gravitational potential and matter distribution from the distortion of
the images of galaxies or from the anisotropy patterns in the CMB
radiation itself. 
It would be impossible to estimate lensing effects by observing a
single galaxy or a single CMB pixel.
However, lensing distortions can be accurately deduced from a
statistical analysis of many groups of galaxies or extended regions in
CMB maps.
CMB lensing measurements offer a unique opportunity to probe density
contrasts at very high redshift (up to $z \sim 3$).
However, galaxy weak lensing observations can reach a higher
signal-to-noise ratio and can be used for tomography.
By classifying the source galaxies in redshift bins, one can
reconstruct the gravitational potential distribution at different
redshifts, and follow the growth of perturbations as a function of
redshift. 
This has been shown to be particularly useful for probing the
neutrino-mass effect on small scales.  

The right-hand pane of figure \ref{fig_future} shows the expected
sensitivity of future CMB experiments such as Planck and CMBpol,
including the lensing data extracted from the same experiment. 
The 2$\sigma$ sensitivity to $M_{\nu}$ is as good as 0.3~eV and
0.07~eV respectively. 
These forecasts should be interpreted with care because it has been
assumed that astrophysical foregrounds can be removed accurately from
the CMB map.
The improvement in sensitivity that can be obtained by adding data
from galaxy weak-lensing surveys is also shown. 
S300/S1000 refers to experiments involving a spatial telescope
scanning galaxies in a small region of the sky (300 to 1000 squared
degrees). 
G2$\pi$/G4$\pi$ refers to plausible ground-based experiments
probing half of the sky or all of it. 
Such experiments are planned for the near future
(see \cite{Lesgourgues:2006nd,Hannestad:2006as} and references
therein).  
The sensitivity of Planck plus S300/S1000 is of the order of 0.2~eV,
while CMBpol combined with a full-sky galaxy scan would reach a
2$\sigma$ sensitivity equal to the minimum value of $M_{\nu}$ in the
case of normal hierarchy (of order $0.05$ eV). 
The combination of measurements of the CMB with future galaxy-cluster
surveys (derived from the same weak lensing observations as well as
X-ray and Sunyaev--Zel'dovich surveys) should yield a similar
sensitivity \cite{Wang:2005vr,Hannestad:2007cp}. 

\subsubsection{Leptogenesis}
\label{subsubsect:leptogenesis}
The origin of the matter-antimatter asymmetry is one of the most
important questions in cosmology. 
The  presently observed baryon asymmetry is \cite{Spergel:2006hy}:
\begin{equation}
  Y_B=\frac{n_B-n_{\bar B}}{s}\simeq 6.1 \times 10^{-10}\,.
\end{equation}
where $Y_B$ is the baryon to photon ratio at recombination.
In 1967 A. Sakharov suggested that the baryon density can be explained
in terms of microphysical laws \cite{Sakharov:1967dj}. 
Three conditions need to be fulfilled: there must exist a mechanism
by which baryon number conservation is violated; the conservation of C
and CP must be violated; and there must be a period in which the
Universe is out of thermal equilibrium.
Several mechanism have been proposed to explain the baryon
asymmetry, many of which are dis-favoured by cosmological or
theoretical considerations.

Leptogenesis has emerged as a successful mechanism for explaining the
origin of the baryon asymmetry of the Universe \cite{Fukugita:1986hr}.
Assuming that $B-L$ is conserved both at the perturbative and the
non-perturbative level, then if a net $B-L$ (for example a net lepton
number) could be created, then the `sphaleron' process would convert
the net $B-L$ into a net baryon and lepton number of comparable
magnitude. 
Leptogenesis is particularly appealing because it takes place in the
context of see-saw models
\cite{Minkowski:1977sc,Gell-Mann:1980vs,Yanagida:1979as}, 
which, naturally  explain the smallness of neutrino masses.
As discussed above, the see-saw mechanism requires the existence of
heavy right-handed (RH) Majorana neutrinos, singlet with respect to
the Standard Model gauge symmetry group.
Introducing a Dirac neutrino mass term and a Majorana mass term for
the right-handed neutrinos via the Lagrangian:
\begin{equation} 
  \label{eq:L}
  -{\cal{L}} = \overline{\nu_{Li}} \, (m_D)_{ij} \, N_{Rj} \, + 
  \frac{1}{2} \, \overline{(N_{Ri})^c} \, (M_R)_{ij} \, N_{Rj} ~, 
\end{equation}
leads, for sufficiently large $M_R$, to the well known see-saw
formula  for the low-energy neutrino mass matrix, $m_\nu$
\cite{Minkowski:1977sc,Gell-Mann:1980vs,Yanagida:1979as}:
\begin{eqnarray} \label{eq:seesaw}
  m_\nu  &   \simeq - m_D \ M_R^{-1} \ m_D^T~, \\
         & = U \   D_m  \ U^T \; ,
\end{eqnarray}
where terms of order ${\cal O}(M_R^{-2})$ have bee neglected, $D_m$ is
a diagonal matrix containing the masses $m_{1,2,3}$ of the three light
massive Majorana neutrinos, and $U$ is the unitary PMNS matrix.

The CP-violating and out-of-equilibrium decays of RH neutrinos produce
a lepton asymmetry \cite{Fukugita:1986hr} that can be  converted into
a baryon asymmetry through anomalous electroweak processes
\cite{Kuzmin:1985mm,Harvey:1990qw}. 
The requisite CP-violating decay asymmetry is caused by the
interference of the tree-level contribution and the one-loop
corrections in the decay rate of  the heavy Majorana neutrinos,
$N_i \ra \Phi^- \, \ell^+$ and $N_i \ra \Phi^+ \, \ell^-$:
\begin{equation} 
  \label{eq:eps}
  \begin{array}{ll} 
    \varepsilon_i & = \frac{\D \Gamma (N_i \ra \Phi^- \, \ell^+) - 
    \Gamma (N_i \ra \Phi^+ \, \ell^-)}{\D \Gamma (N_i \ra \Phi^- \, \ell^+) +  
    \Gamma (N_i \ra \Phi^+ \, \ell^-)} \\
                    &  \simeq  \D \frac{\D 1}{\D 8 \, \pi \, v^2} 
    \sum_{j\neq i} \frac{{\rm Im} (m_D^\dagger m_D)^2_{ij}}{(m_D^\dagger m_D)_{ii}} \, \left( f(x_j) + 
    g(x_j) \right)~,
  \end{array} 
\end{equation}
where $\Phi$ and $\ell$ indicate the Higgs field and the charged
leptons, respectively.
Here $v \simeq 174$ GeV is the electroweak-symmetry-breaking scale and
$x_j \equiv M_j^2 / M_i^2$. The functions $f$ and $g$ stem from vertex 
\cite{Fukugita:1986hr,Luty:1992un,Flanz:1994yx,Plumacher:1996kc} and
from self-energy
\cite{Covi:1996wh,Flanz:1996fb,Covi:1996fm,Pilaftsis:1997jf,Buchmuller:1997yu}
contributions:
\begin{equation} 
  \label{eq:fapprox}
  \begin{array}{ll} 
    f(x) = \sqrt{x} \left(1 - 
      (1 + x) \, \ln \left(\frac{\D 1 + x}{\D x}\right) \right) 
      \; {\rm ; and} \\
    g(x) = \frac{\D \sqrt{x}}{\D 1 - x} \; . 
  \end{array}
\end{equation}
For $x \gg 1$, i.e. for hierarchical heavy Majorana neutrinos, 
$f(x) + g(x) \simeq -\frac{3}{2\sqrt{x}}$.  
Under these assumptions, the baryon asymmetry is obtained via:
\begin{equation} \label{eq:YB}
  Y_B = a \, \frac{\kappa}{g^\ast} \, \varepsilon_1~,
\end{equation}
where $a \simeq -1/2$ is the fraction of the lepton asymmetry
converted into a baryon asymmetry \cite{Kuzmin:1985mm,Harvey:1990qw}, 
$g^\ast \simeq 100$ is the number of massless degrees of freedom at
the time of the decay, and $\kappa$ is a efficiency factor that is
obtained by solving the Boltzmann equations.
Typically, one gets $Y_B \sim 6 \times 10^{-10}$ when 
$\varepsilon_1 \sim (10^{-6} - 10^{-7})$ and 
$\kappa \sim (10^{-3} - 10^{-2})$.  
Note that this estimate of $Y_B$ is valid in the supersymmetric
theories too
\cite{Covi:1996wh,Flanz:1996fb,Covi:1996fm,Pilaftsis:1997jf,Buchmuller:1997yu}.

The results reported so far are valid if the individual lepton 
flavours (which indicate the lepton-mass eigenstates at the 
temperature of leptogenesis) are effectively indistinguishable.
Recently, it has been pointed out that if this assumption
does not hold, the evolution of the
lepton asymmetry in each flavour $\alpha$, $Y_{L, \alpha}$,
needs to be considered separately and the resulting final baryon
asymmetry can be different from the one obtained from equation
(\ref{eq:YB}) \cite{Abada:2006ea}. 
Following reference \cite{Abada:2006ea}, consider the case of
hierarchical heavy neutrinos for which the generation of the lepton
asymmetry is dominated by the decay of the lightest with mass $M_1$.
The lepton-flavour asymmetry is proportional to the flavour
CP-asymmetry, $\epsilon_{\alpha \alpha}$, of the decay of $N_1$ into
the leptons of flavour $\alpha$: 
\begin{equation}
  \label{eq:asymmflavour}
  \epsilon_{\alpha \alpha} = \frac{1}{8 \pi v^2} \frac{1}{(m_D^\dagger m_D)_{11}}
  \sum_j \mathrm{Im} \Big( (m_D)_{\alpha 1} (m_D^\dagger m_D)_{1 j} (m_D)_{\alpha j }^\ast 
  \, \left( f(x_j) + 
  g(x_j) \right) \Big) ~.
\end{equation}
By solving the coupled Boltzmann equations for the asymmetries
corresponding
to the indistinguishable flavours, one obtains an efficiency factor for each
flavour $\alpha$. When computing the final baryon asymmetry, this flavour
efficiency factor weights the decay asymmetries as
$\varepsilon_{\alpha\alpha}$.
If $M_1 \ltap 10^9$~GeV, the $\mu$ and $\tau$ Yukawa couplings are in
equilibrium and the three flavours need to be considered separately.  
For $10^9 \ {\rm GeV} \ltap M_1 \ltap 10^{12}$~GeV, only the
interactions mediated by the $\tau$ Yukawa coupling are in equilibrium 
and the problem reduces to an effective two-flavour case.
Finally, if $M_1 \gtap 10^{12}$~GeV, the Yukawa interactions
are all out of equilibrium and all flavours are indistinguishable. 
In this case, the results of one flavour are recovered.

In the MSSM, flavour effects are relevant for even larger temperature
ranges \cite{Antusch:2006cw}. 
Here, the one-flavour formul\ae can only be applied for
temperatures larger than $(1+\tan^2 \beta)\times 10^{12}$~GeV, since
the squared charged-lepton Yukawa couplings in the MSSM are multiplied by
this factor. Consequently, charged $\mu$- and $\tau$-lepton Yukawa
couplings are in thermal equilibrium for 
$(1+\tan^2 \beta)\times 10^5 \: \mbox{~GeV} \ll M_1
\ll (1+\tan^2 \beta)\times 10^{9} \: \mbox{GeV}$ and all flavours in the
Boltzmann equations are to be treated separately. 
For $(1+\tan^2 \beta)\times
10^9 \: \mbox{GeV} \ll M_1 \ll (1+\tan^2 \beta)\times 10^{12} \:
\mbox{GeV}$,
only the $\tau$ Yukawa coupling is in equilibrium and only the $\tau$
flavour
is treated separately in the Boltzmann equations, while the $e$ and $\mu$
flavours are indistinguishable.

Establishing a connection between the parameters
at low energy (neutrino masses, mixing angles, and CP-violating
phases), measurable in principle in present and future experiments,
and at high energy (relevant in leptogenesis) has been 
intensively investigated.
The number of parameters in the full Lagrangian
of models which implement the see-saw mechanism
is larger than the ones in the low-energy sector: 
in the case of three light neutrinos and three heavy ones,
at high energy the theory contains, in the neutrino sector, 18
parameters of which 12 are real.
At low energy only 9 are accessible - 3 angles, 3 masses and 3
phases. 
The decoupling of the heavy right-handed neutrinos 
implies the loss of information on 9 of the parameters required to
specify the theory at high energy.
This implies that reconstructing the high-energy
parameters entering in the see-saw models from the measurement
of the masses, angles, and CP-violating phases of $m_\nu$
depends on the specific model considered. 

Using the weak basis in which both $M_R$ and the charged-lepton mass 
matrix are real and diagonal, it is useful to parametrise the Dirac
mass by the bi-unitary or the orthogonal parametrisations:
\begin{itemize}
  \item {\it Bi-Unitary parametrisation:}
        The complex $3 \times 3$ Dirac mass matrix can be written in
        the form \cite{Pascoli:2003uh}:
        \begin{equation} 
          \label{eq:mdULUR}
          m_D = U_L^\dagger \, m_D^{\rm diag} \, U_R~,  
        \end{equation}
        where $U_L$ and $U_R$ are unitary $3 \times 3$ matrices and
        $m_D^{\rm diag}$ is a real diagonal matrix. 
        All the CP-violating phases are contained in $U_L$ and $U_R$;
        and 
  \item {\it Orthogonal parametrisation:}
        By using the see-saw formula, equation (\ref{eq:seesaw}), we
        can express $m_D$ as \cite{Casas:2001sr,Pascoli:2003rq}: 
        \begin{equation}
          \label{eq:mdO}
          m_D =  i \, U \, D_m^{1/2}  \, R \, M_R^{1/2}~,
        \end{equation}
        where $D_m$ is the diagonal real matrix which contains the
        low-energy light neutrino masses, and  $R$ is a complex
        orthogonal matrix. 
        $R$ contains 3 real parameters and 3 phases.
\end{itemize}
The use of these parametrisations clarifies the dependence of
leptogenesis and LFV charged-lepton decays, on the different
parameters entering in $m_D$.

For leptogenesis, in the case of one effective flavour,
the decay asymmetry $\varepsilon_1$ depends  on the hermitian matrix
$m_D^\dagger m_D$: 
\begin{equation} 
  \label{eq:mddmd}
  m_D^\dagger m_D = 
  \left\{ 
    \baz 
    U_R^\dagger \, (m_D^{\rm diag})^2 \, U_R~, & \mbox{bi--unitary;} \\[0.3cm]
    M_R^{1/2} \, R^\dagger \, D_m \, R \, M_R^{1/2}~,
    & \rm orthogonal. \\
    \end{array}  
  \right. 
\end{equation}
Notice that the PMNS unitary mixing matrix does not enter
explicitly into the expression for the lepton asymmetry.
However, it has been pointed out that if this approximation does not
hold, single-flavour asymmetries need to be considered.
In this case, the flavour CP asymmetry may be written:
\begin{equation}
  \label{eq:asymmetryR}
  \epsilon_{\alpha \alpha} = - \frac{3 M_1}{16 \pi v^2} 
  \frac{\mathrm{Im} \Big( \sum_{\beta \rho} m_\beta^{1/2} m_\rho^{3/2}
  U^\ast_{\alpha \beta} U_{\alpha \rho} R_{ \beta 1} R_{ \rho 1}\Big)}
  {\sum_\beta m_\beta |R_{ \beta 1}|^2} \; .
\end{equation}
It is important to notice that in this case the lepton asymmetry
depends also on the CP-violating phases in $U$.
In the interesting case of $R$ real, the asymmetry does not cancel out
and will be determined by the values of the low-energy Dirac and
Majorana CP-violating phases, which are measurable in principle in
future experiments. 

In the bi-unitary parametrisation, the neutrino mass matrix $m_\nu$ can
be written as:
\begin{equation} \label{eq:mnupara}
  m_\nu = 
  - U_L^\dagger \, m_D^{\rm diag} \, U_R \, M_R^{-1} \,  U_R^T \, 
  m_D^{\rm diag} \, U_L^\ast~,
\end{equation}
showing that the phases in $U$ receive contributions from CP-violation
both in the right-handed sector, responsible for leptogenesis, and in
the left-handed one, which enters in lepton-flavour-violating processes. 
Due to the complicated way in which the high-energy phases and real
parameters enter in $m_\nu$, equation (\ref{eq:mnupara}), if there is
CP-violation at high energy, as required by the leptogenesis
mechanism, we can expect in general to have CP-violation at
low-energy, as a complete cancellation would require some fine-tuning
or special forms of $m_D$ and $M_R$.

More specifically, from equation (\ref{eq:mnupara}), it can be seen that,
in general, there is no one-to-one link between low energy
CP-violation in the lepton sector and the baryon asymmetry;
a measurement of the low-energy CP-violating phases does not allow
the leptogenesis phase to be reconstructed.
However, if the number of parameters in $m_D$ is reduced, a
one-to-one correspondence between high-energy and low-energy
parameters might be established.
For example, in certain classes of neutrino-mass models with sequential
right-handed neutrino dominance, a strong link between the leptonic-CP
violating phase, $\delta$, and the CP-violation required for
leptogenesis can be established, and flavour-dependent effects have a
significant effect \cite{Antusch:2006cw}. 
In other classes of models such strong links were not
found. Links can also be achieved in models which allow for CP-violation.
For example, this can be achieved in models which allow for
CP-violation only in the right-handed sector, that is in $U_R$.
It has been shown recently that, to the extent that the different
flavours can be distinguished, leptogenesis depends only on the phases
in the PMNS mixing matrix, if $R$ is real. 
Each model of neutrino mass generation should be studied separately in
detail to establish the feasibility of the leptogenesis mechanism 
\cite{Davidson:2001zk,Ellis:2001xt,Branco:2002kt,Ellis:2002xg,Frampton:2002qc,Davidson:2002em,Branco:2002xf,King:2002qh}. 

In conclusion, the observation of \betabeta-decay, implying the
violation of the global lepton number (one of the main conditions for
leptogenesis), and of leptonic CP-violation in neutrino oscillations 
and/or neutrinoless double-beta decay is crucial in understanding the
origin of the baryon asymmetry.
The observation of leptonic-CP violation itself would be a strong
indication, though not a proof, that leptogenesis is the explanation
for the observed baryon asymmetry of the Universe.

\subsubsection{Neutrinos and Inflation}

In the previous sections, we have seen that neutrinos may have played
crucial roles in shaping the Universe of today (the large-scale
structure) and in the removal of the anti-matter from the early
Universe (leptogenesis). 
The neutrino may also may also be the key to the process of inflation
by which the Universe went through a period of exponential growth.
The `stretching' of the Universe during inflation is held to explain
the uniformity of the today's Universe.

In nearly all discussions about the birth of the Universe, it is
assumed that the Universe was originally microscopically small.  
There is a good reason for this; communication between different parts
of the Universe has a `speed limit', the speed of light, $c$.
No regions of space `know' about other regions if they are separated
by more than the distance $c \tau$ where $\tau$ is the age of the
Universe at a particular moment.  
It is very difficult to conceive of a process that can create the
Universe that is larger than $c\tau$.  
It is much more natural to think that the Universe was born small, but
that there was a mechanism to stretch it to a macroscopic size later,
much larger than that allowed by the assumed speed limit $c$.

Support for this view can be found in the CMB.
The temperature of cosmic microwave background is the same, 
to better than one part in a hundred thousand or so, in every
direction.  
The microwave photons from the different directions come from opposite
ends of the Universe that could never have been in communication with
each other. 
The CMB distribution reflects the temperature distribution of the
early Universe. 
Therefore, the uniformity of the CMB implies that different regions of
the early Universe, which are not necessarily causally connected, are
nonetheless at the same temperature.
This is the `horizon problem': what is the mechanism which gave rise
to such a uniform temperature distribution.

Another well-known problem is the `flatness problem'.  
When the Universe was born, no known microphysics can determine what
kind of space, namely the topology and the local curvature, should be
chosen.  
At the time of big-bang nucleosynthesis (the best-tested aspect of
the description of the early Universe when it was about a second to a
minute old) the Universe must have been extremely flat at the level of
$10^{-20}$. 
This requirement becomes much stronger if we contemplate even earlier
times.
What mechanism squashed the Universe so flat?

Finally, the large-scale structure of the Universe discussed above
suggests that the Universe originated from a small fluctuation
in the energy density at the level of $10^{-5}$, which somehow appears
to be correlated in different parts of the space --
i.e., the initial density perturbation appears acausal.  
In addition, the spectrum of the fluctuation is nearly independent of
the distance scales, suggesting that it was generated by some kind of
self-replicating mechanism.  
The fluctuations themselves are Gaussian in nature.

Cosmological inflation is currently the only way to answer these
profound questions and explain the empirical observations of the
large-scale structure of the Universe \cite{Guth:1980zm,Sato:1980yn}.  
Inflation stretches the Universe exponentially from the microscopic
size at its birth to a macroscopic size which leads to the vast
Universe as observed today.  
At the same time, even a bumpy space gets flattened once it is
exponentially stretched because what we see today is only a tiny
portion of the entire space.
Also, because the entire Universe originated from a small patch which
was in communication, the sky in all directions must look the same.  
In a surprising way, quantum fluctuations in an exponentially
expanding Universe soon become classical because the natural wave
length exceeds the causally-connected region of space, and as the
Universe keeps expanding it generates itself many times leading to a
scale-invariant Gaussian spectrum of density fluctuations
\cite{Starobinsky:1982ee}. 

For concreteness, consider a simple model of inflation based on a
scalar field ($\phi$) with just a mass term, namely a quadratic
potential \cite{Linde:1983gd}:
\begin{equation}
  {\cal L} = \frac{1}{2} (\partial \phi)^2 - \frac{1}{2} m^2 \phi^2.
\end{equation}
Such a scalar field can drive inflation. 
The equation of motion of the scalar field is:
\begin{equation}
  \ddot{\phi} + 3 H \dot{\phi} + m^2 \phi = 0,
  \label{Eq:Infl1}
\end{equation}
while the expansion rate of the Universe $H = \dot{a}/a$, if dominated
by this scalar field, is given by:
\begin{equation}
  H^2 = \frac{8\pi}{3} G_N \frac{1}{2}(\dot{\phi}^2 + m^2 \phi^2).
  \label{Eq:Infl2}
\end{equation}
This coupled equation has a very simple solution if $\phi \gtrsim
M_{Pl} = G_N^{-1/2}$.  It can be shown that the $\ddot{\phi}$ and
$\dot{\phi}^2$ in the above equations can be safely neglected
(the `slow-roll' condition).  
In this case equations (\ref{Eq:Infl1}) and (\ref{Eq:Infl2}) can be
combined into a single equation:
\begin{equation}
  3 \sqrt{\frac{8\pi}{3}}\ \frac{1}{M_{Pl}} m \phi \dot{\phi} + m^2
  \phi = 0 \; ,
\end{equation}
and hence:
\begin{equation}
  \phi(t) = \phi(0) - \frac{m M_{Pl}}{\sqrt{24\pi}} t \; .
\end{equation}
At the same time, the Universe expands as:
\begin{equation}
  a(t) = a(0) \exp \left[ \sqrt{\frac{4\pi}{3}} \frac{1}{M_{Pl}}
    \left(\phi(0)t - \frac{m M_{Pl}}{2\sqrt{24\pi}}\ t^2\right)\right]
  \; .
\end{equation}
For $t \ll \phi(0)/(m M_{Pl})$ the second term in the parentheses can
be ignored, and the expansion of the Universe is exponential.  
This way, the initial microscopic size of the Universe can be made
macroscopically large.  
The curvature is squashed exponentially as $(a(0)/a(t))^2$ solving the
flatness problem, and also the horizon problem by assuming that the
$e$-folding is large enough ($\gtrsim 60$) so that the initial horizon
contains the entire visible Universe of today.

\begin{figure}
  \centering
  \includegraphics[width=0.8\textwidth]{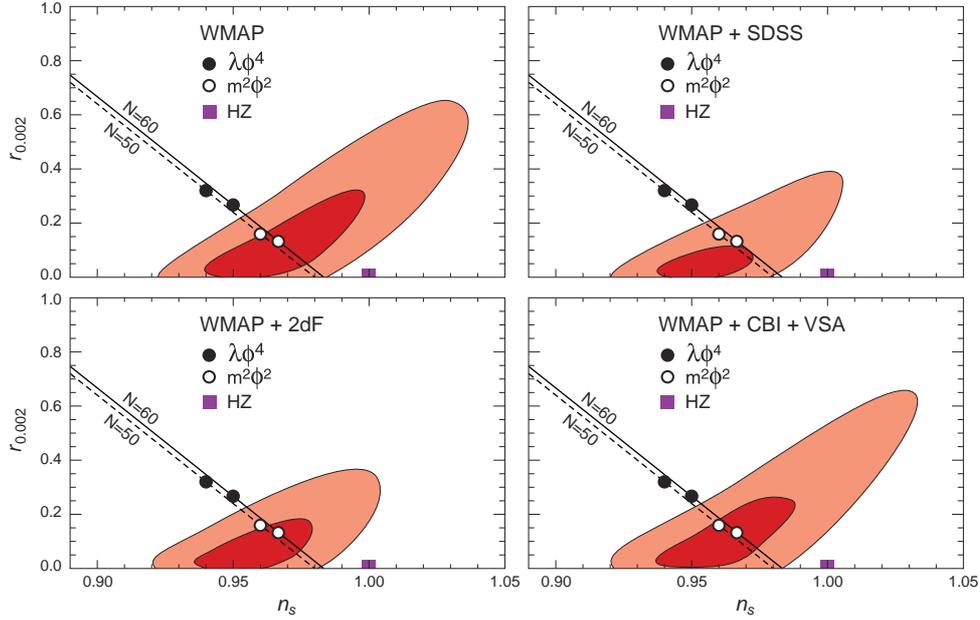}
  \caption{Global fits to the cosmological data versus predictions of
    simple inflation models \cite{Spergel:2006hy}. 
    Adapted with kind permission of the Astrophysical Journal Supplement Series
    from figure 14 in reference \cite{Spergel:2006hy}.
    Copyrighted by the American Astronomical Society.
} 
  \label{fig:WMAP}
\end{figure}

To obtain the correct size of the density fluctuations, we need 
$m \simeq 2\times 10^{13}$ GeV \cite{Ellis:2003sq}.  
It is remarkable that the simple quadratic potential is consistent
with available cosmological data, including the upper limit on the
tensor component (see figure \ref{fig:WMAP} \cite{Spergel:2006hy}).  
A natural question from the particle physics point of view is what
is this scalar field?
The most likely candidate is a gauge singlet, to maintain the form of 
the potential against radiative corrections.
There are no such fields within the Standard Model or its minimal
supersymmetric extension.

However, the mass of the scalar required to generate the quadratic
potential is similar to the mass of right-handed neutrinos required by
the see-saw mechanism.
The RH neutrinos are also naturally gauge singlets.
It is therefore tempting to consider that neutrinos have something to
do with inflation.
If nature is supersymmetric, the right-handed neutrinos needed in the
see-saw mechanism have superpartners (sneutrinos) which are scalar
fields.
The sneutrino potential is quadratic, making the right-handed
sneutrinos candidates for the inflaton field \cite{Murayama:1992ua}.  
Moreover, once the inflation is over, the right-handed sneutrino
oscillates around the origin and decays, reheating the Universe to an
ordinary thermal bath.  
This process is the same as that assumed in the discussion of
leptogenesis above, and hence can generate the baryon asymmetry. 
Because the Universe is dominated by the right-handed neutrino at this
point, leptogenesis is more efficient than conventional thermal
leptogenesis.  
Using the available neutrino data and assuming hierarchical spectrum
of right-handed neutrinos, the resulting lepton asymmetry is given by:
\begin{equation}
  \frac{n_L}{s} \simeq 1.5 \times 10^{-10}
  \frac{T_{RH}}{10^6~\mbox{GeV}}\ \delta,
\end{equation}
where $\delta$ is the CP violating phase in the neutrino mixing
\cite{Hamaguchi:2001gw}.  
Interestingly, leptogenesis is possible with a relatively low
reheating temperature $T_{RH} \sim 10^6$~GeV, low enough to avoid the
cosmological problem of gravitinos even for the case of hadronic decay
\cite{Kawasaki:2004qu} 
which imposes strong constraints on thermal leptogenesis
with hierarchical right-handed neutrinos 
\cite{Buchmuller:2004nz,Antusch:2006gy}. 

There is, however, an important issue to be addressed for the simple
quadratic form of the potential to extend beyond the Planck-scale
amplitude.
In fact, supergravity tends to modify this form and requires a
somewhat special K\"ahler potential to maintain the quadratic form
\cite{Murayama:1993xu} (see also \cite{Kadota:2005mt,Ellis:2006ar} for
more recent discussions). 
The most important test of the quadratic potential is its prediction
of the tensor component $r \simeq 0.15$.  
This relatively large tensor component will be probed in the near
future by $B$-mode polarisation measurements of the cosmic microwave
background \cite{PolarBear}.
An alternative scenario of sneutrino inflation, where a small $r$ is
predicted and which is therefore easily distinguishable from the above
model,
is hybrid inflation with a sneutrino inflaton field \cite{Antusch:2004hd}.
There is therefore an intriguing coincidence between the properties of scalar
fields required to drive inflation, for the see-saw mechanism, and for
leptogenesis.  
It points to a remarkable possibility that the neutrino is the
mother of the Universe.

%% file: 04-BSnuM/04-BSnuM.tex
\section{Effects of New Physics beyond the Standard Neutrino Model}
\label{Sect:NewPhys}

\input 04-BSnuM/04-00-Introduction/04-00-Introduction

\input 04-BSnuM/04-01-Sterile-neutrino/04-01-01-Sterile-neutrino

\input 04-BSnuM/04-02-Mass-varying-neutrinos/04-02-Mass-varying-neutrinos

\input 04-BSnuM/04-03-CPTLIV/04-03-CPTLIV

\input 04-BSnuM/04-04-Unitarity/04-04-Unitarity

\input 04-BSnuM/04-05-Nonstandard-Interaction/def_temp.tex
\input 04-BSnuM/04-05-Nonstandard-Interaction/04-05-Nonstandard-Interaction

%% file: 04-BSnuM/04-00-Introduction/04-00-Introduction.tex
Almost all experimental results to date are consistent with the
Standard Neutrino Model (see section \ref{SubSect:SnuM}). 
These results are most often used to determine the parameters of the
S$\nu$M.
To go beyond the S$\nu$M, there are two complementary approaches.
The first is the `theoretical approach' in which models are
constructed which solve one or more of the problems of the S$\nu$M,
this was the approach taken in section \ref{Sect:NPandtheSNuM}.
These models may predict new phenomena or predict small deviations
from the results of the S$\nu$M.
Present and future experiments may support, constrain, or contradict
these models. 
The Standard Model itself was established in this way, the minimal
super-symmetric standard model (MSSM) and other extension of the SM
are expected to be tested in future experiments especially at the LHC
\cite{Weiglein:2004hn}.
The second approach is the `phenomenological approach' in which
possible effects of unknown physics are described in a
model-independent way. 
Experiments may give constraints on these parameters, giving very
important information for the development of a complete theoretical
description. 
An example of this approach is the model-independent parameterisation
of new physics effects in the vacuum polarisation of electroweak gauge
bosons  \cite{Peskin:1990zt,Peskin:1991sw}.
The strong experimental constraints on the relevant parameters have
allowed technicolor models to be rejected.

Neutrino experiments which are being carried out or are in preparation
are optimised to measure precisely the parameters of the S$\nu$M.
The second-generation neutrino facility, on the other hand, should be
designed not only to determine these parameters but also to have
sensitivity to signatures of physics beyond the S$\nu$M.
In the following, the possibility of detecting various new-physics
effects in neutrino oscillations will be discussed.

%% file: 04-BSnuM/04-01-Sterile-neutrino/04-01-01-Sterile-neutrino.tex
\subsection{Sterile neutrinos}

Neutrinos which have no Standard Model couplings are referred to as
`sterile'. 
They arise in many extensions of the SM which include singlet-fermion
states and the corresponding mass eigenstates. 
The new sterile-neutrino states can mix with ordinary neutrinos and
generate effects that may be observed in terrestrial, cosmological, and
astrophysical experiments. 

\subsubsection{Theoretical issues}

If sterile neutrinos are present, it is necessary to explain the
origin of their masses and of the mixing with active neutrinos.
Small masses and large mixings can arise, for example, via
higher-dimensional operators in the superpotential
\cite{Langacker:1998ut} which induce an intermediate-scale expectation
value, $v_S$, for a singlet field.
The magnitude of $v_S$ is between the electroweak scale and a large
energy scale, $M$.
The masses of the sterile neutrinos are found to be suppressed by
powers of $v_S/M$.
Sterile neutrinos with masses from 100~MeV to few~GeV are required to
generate the observed light-neutrino masses in theories with dynamic
electroweak-symmetry breaking
\cite{Appelquist:2002me,Appelquist:2003uu,Appelquist:2003hn,Appelquist:2004ai}.
Models with `mirror matter' contain mirror neutrinos which would be
light for reasons similar to the reasons for which their ordinary
partners are light
\cite{Foot:1991bp,Foot:1991py,Foot:1993yp,Foot:1995pa,Berezhiani:1995yi,Berezhiani:1995am,Mohapatra:1996yy,Berezhiani:2000gw}.
The interactions between active and sterile neutrinos would be
mediated by operators of the type 
$\nu \phi \nu^\prime \phi^\prime/M_P$, where the prime refers to the
mirror world and $M_P$ is the Planck mass.
Singlet neutrinos could be the supersymmetric partners of the moduli
field \cite{Benakli:1997iu} or the singlets contained in
representations of $E_6$
\cite{Chacko:1999aj,Frank:2004vg,Frank:2005rb}. 
In these cases it can be argued that the singlet mass would be of
order TeV$^2/M_P$, the TeV mass scale arising from supersymmetry
breaking. 
Sterile neutrinos can easily be embedded in models based on extra
dimensions, the sterile neutrinos can be new singlet fermions
propagating in the bulk of a higher-dimensional theory with naturally
small masses \cite{Mohapatra:1999af}.
In addition, such theories predict a tower of Kaluza-Klein modes 
which can generate interesting observational signatures, neutrino
oscillations in particular \cite{Cacciapaglia:2003dx}.

\subsubsection{Phenomenology of light sterile neutrinos}

Here we consider sterile neutrinos with masses up to a few eV that mix
with ordinary neutrinos. 
The main signals for such sterile neutrinos arise in neutrino
oscillations.
The implications of the LSND measurements is postponed to section
\ref{Para:LSNDChallenge}.
A detailed discussion of the bounds summarised below is given in
\cite{Cirelli:2004cz,Maltoni:2004ei}.

{\noindent \em Reactor- and accelerator-neutrino experiments}

\noindent In these experiments active--sterile neutrino oscillations
would take place, implying a reduction of the observed flux at the far
detector. 
Reactor-neutrino experiments are sensitive to the mixing with $\nu_e$,
$U_{es}$. 
The CHOOZ and Bugey experiments put bounds as strong as 
$|U_{es}|^2 \ltap 0.01$ for particular neutrino-mass ranges.
Mixing of sterile neutrinos with $\nu_\mu$ may be tested in
accelerator experiments for which muon neutrinos are the dominant
beam contribution at the source.
Data from the CDHS and CCFR disappearance experiments allow limits on
$U_{ \mu s}$ to be derived.
In addition, appearance experiments sensitive to the transition 
$\nu_\mu \rightarrow \nu_e$ (KARMEN) and 
$\nu_\mu \rightarrow \nu_\tau$ (NOMAD and CHORUS) probe a combination
of the mixing angles, $U_{\mu s}$, $U_{ e s}$ and $U_{\mu s}$, $
U_{\tau s}$, respectively.
A combined analysis of recent data from Super-Kamiokande (SK), K2K,
MACRO  have yielded the constraint $|U_{ \mu s}|^2\ltap 0.065$ at
99\%~C.L.~\cite{Maltoni:2004ei}.

{\noindent \em Solar neutrino experiments and KamLAND}

The data from these experiments can be used to constrain the mixing
of sterile neutrinos with $\nu_e$. 
The MSW enhancement of electron-neutrino oscillations, with its
characteristic energy dependence, makes it possible to search for a
sterile-neutrino component down to mass-squared differences as small
as $\Delta m^2 \sim 10^{-8}$~eV$^2$.
For large mixing angles, the search can be extended to masses as small
as $\Delta m^2 \sim 10^{-12}$~eV$^2$. 
In some models sterile neutrinos produce effects at
energies below an MeV; data from the SNO and SK experiments already
dis-favour models which modify the energy distribution for neutrino
energies greater than a few MeV.
The low-energy ($E_\nu < 1$~MeV) region was accessible only to the
Gallium experiments. 
In the future, Borexino will be able to test part of this interesting 
region through the analysis of, for example, diurnal or seasonal
variations in the neutrino spectra.

{\noindent \em Neutrinoless double-beta decay}

\noindent Sterile neutrinos that are Majorana particles and mix with  
electron neutrinos would contribute to the effective Majorana mass on
which the half-life of the double-beta-decay process depends. 
In particular, the effective mass would be:
\begin{equation}
  |< m>| = \left| \sum_{i=1,2,3} m_i  \ U_{e i}^2 + m_s  \ U_{e s}^2 \right|,
\end{equation}
where $m_i$ are the masses of the light, ordinary neutrinos and $m_s$
indicates the mass of the sterile neutrino.  
Notice that $U_{e s}^2 = |U_{e s}|^2 \, e^{i \beta_s}$, where
$\beta_s$ is a Majorana CP--violating phase. 
Due to the presence of the Majorana phases the contributions in 
$|<m>|$ can be constructive or partially cancel \cite{Bilenky:2001xq}.
A future measurement of $|<m>|$ with values outside the range
predicted in the case of three light neutrinos might be a signal for
the presence of sterile neutrinos.

\subsubsection{Signatures of heavy sterile neutrinos}

The signatures of sterile neutrinos with masses $m_s \gg 100$~eV 
depend strongly on the flavour with which the sterile neutrino mixes
and on the sterile-neutrino mass \cite{Pascoliinprep}.
For masses $30 \ {\rm eV} \lesssim m_N \lesssim 1$~MeV, the most sensitive
probe is the search for kinks close to the end-point of $\beta$-decay
spectra \cite{Shrock:1980vy,Shrock:1980ct}.
The bounds are typically in the $|U_{es}|^2 \sim 10^{-2}$--$10^{-3}$
range.
For heavier masses, a very powerful probe of the mixing of a heavy
neutrino with both $\nu_e$ and $\nu_\mu$ are peak searches in leptonic
decays of pions and kaons \cite{Shrock:1980vy,Shrock:1981wq}. 
A heavy neutrino can be produced in such decays and the lepton
spectrum would show a monochromatic line at:
\begin{equation}
  E_l = \frac{m_M^2 + m_l^2 - m_s^2}{2 m_M},
\end{equation}
where $E_l$ and $m_l$ are, respectively, the lepton energy and mass,
and $m_M$ is the meson mass.
The mixing angle controls the branching ratio of this process and can
be constrained by the height of the peak. 
Notice that these bounds are very robust because they rely only on the
assumption that a heavy neutrino exists and mixes with $\nu_e$ and/or
$\nu_\mu$.
The limits for $|U_{es}|^2$ are as strong as 
$10^{-8}$--few~$\times 10^{-7}$ for masses around 100 MeV.
For masses up to 34~MeV, the most stringent constraints on the mixing
with muon neutrinos come from pion decays with 
$|U_{\mu s}|^2 \ltap \mathrm{few} \ 10^{-5}$, while for higher masses
kaon decays are used and lead to limits as strong as 
$|U_{\mu s}|^2 \ltap 10^{-6}$ \cite{Kusenko:2004qc}.

Another strategy to search for a heavy sterile neutrino is to look for
the products of its decay.
A sterile neutrino, $\nu_s$, would be produced in every process in which
active neutrinos are emitted, with a branching ratio depending on the
mixing-matrix element $|U_{ls}|^2$.
It would subsequently decay into neutrinos and other visible particles
such as electrons, muons, and pions.
Searches for the visible products were performed and were used to
constrain the mixing parameters.
These bounds are less robust than the ones previously discussed. 
In fact, if the dominant decay modes of the heavy neutrinos are into 
invisible particles, these bounds would be weakened, if not completely
evaded. 
In reactors and in the Sun only low mass, $m_N <$~few~MeV, heavy
sterile neutrinos mixed with $\nu_e$ can be produced.
The bounds, obtained by looking for decays into electron-positron
pairs are, typically,  $|U_{es}|^2\ltap 10^{-4}$.
For higher masses, heavy sterile neutrinos mixed with 
$\nu_{e, \mu, \tau}$ can be produced in meson and vector bosons
decays. 
There are two different types of experiments. 
In beam-dump experiments, $\nu_s$ are usually produced by the decay of
mesons, $\pi$, K and D, and the detector is located far away from the
production site. 
Alternatively, the production can happen in the detector itself.
The limits depend strongly on the mass range. 
Typical values for the limits are: 
$|U_{es}|^2\ltap 10^{-9}$--$10^{-4}$, if $m_s \sim 0.02$~GeV -- $0.4$~GeV;
$|U_{es}|^2\ltap 10^{-7}$--$10^{-6}$, if $m_s \sim 0.4$~GeV -- 2~GeV; and
$|U_{es}|^2\ltap \mathrm{few} \ 10^{-5}$, if $m_s \sim 2$~GeV -- 80~GeV.
Similar bounds hold for the mixing with $\nu_\mu$ while $|U_{\tau
s}|^2$ is constrained to be smaller than at most $10^{-5}$. 
For a detailed review see references
\cite{Kusenko:2004qc,Pascoliinprep}.

If heavy, sterile neutrinos are Majorana particles, they would mediate
$\Delta L=2$ processes such as neutrinoless double beta-decay.
New processes would be allowed and could also be resonantly enhanced
for some mass ranges. 
A very sensitive probe of the mixing with muon neutrinos is given by
the rare kaon decay 
$K^+ \rightarrow \pi^- \mu^+ \mu^+$
\cite{Littenberg:1991ek,Appel:2000tc}, as well as the nuclear
transition 
$\mu^- + (A,Z) \rightarrow \mu^+ + (A,Z-2)$ \cite{Missimer:1994xd}. 
Heavy-quark meson decays, e.g. $D^+ \rightarrow K^- (\pi^-) \mu^+ \mu^+$,
were also studied \cite{Weir:1989sq}.
Recently, bounds were obtained from the process
$\Xi^- \rightarrow p \mu^- \mu^-$
\cite{Littenberg:1991rd,Rajaram:2005bs}.

\subsubsection{Sterile neutrins and cosmology and astrophysics}

Sterile neutrinos, if mixed with the active neutrinos, would be
copiously produced in the early Universe and in astrophysical objects
such as supernovae. 
Since the presence of sterile neutrinos would significantly affect
the evolution of such events, it is possible to constrain
sterile-neutrino models using astrophysical and cosmological
observations (\cite{Dolgov:2002wy}).

\paragraph{Light, sterile neutrinos}

If light, sterile neutrinos, with masses $m_s < 10 $~eV, were produced
in the early Universe, they would generate various effects.
At Big Bang Nucleosynthesis (BBN), they would contribute to the energy
density in relativistic particles, modifying the expansion rate of the
Universe and consequently the $n/p$ ratio.
Different analyses have been performed and provide bounds on the
number of neutrinos, typically $N_\nu \ltap 3.24\pm 1.2$ at
95\%~C.L. \cite{Cyburt:2004yc} (see also reference
\cite{Cirelli:2004cz}). 
The presence of a neutrino asymmetry affects the reactions in which
neutrinos are involved and could weaken the bounds quoted above.
For a detailed recent analysis see, e.g., reference \cite{Chu:2006ua}.
The number of relativistic degrees of freedom at photon decoupling 
can be probed by CMB observations and is constrained to be  
$N_\nu = 3 \pm 2$
\cite{Crotty:2003th,Pierpaoli:2003kw,Barger:2003zg}.

Finally, light sterile neutrinos affect large-scale structure
formation, making structures less clustered due to the free-streaming
of these particles. 
Two parameters are relevant for these studies: the temperature
at which these particles become non-relativistic, $T\sim m_s/3$; and
the energy density, $\Omega_s h^2$. 
As the total energy-density in light degrees of freedom is constrained
to be less than 1\%, it is possible to put strong bounds on the mass
of light-sterile neutrinos. 
Supernov\ae are also sensitive probes of the existence of sterile
neutrinos \cite{Kainulainen:1990bn,Abazajian:2001nj}. 
The data from SN1987A strongly constrain the mixing angles,
future experiments might allow these bounds to be strengthened. 
Sterile neutrinos would be produced in the core of supernovae and
would escape carrying away a sizable fraction of the energy.
The limit $|U_{l s}|^2 \ltap 10^{-10}$ can be derived from such an
analysis, while for large values of the mixing, $|U_{l s}|^2 \gtap
10^{-2}$, the sterile neutrinos would be effectively trapped and no
bound applies. 
In addition, MSW oscillation in sterile neutrinos can take place for
specific ranges of parameters and can modify the flux of electron
anti-neutrinos. 
These bounds should be used with care as there is not yet a full
understanding of the initiation and evolution of supernovae.

\paragraph{KeV sterile neutrinos}

Sterile neutrinos with masses in the few-KeV range have been advocated 
as a source of dark matter
\cite{Dodelson:1993je,Dolgov:2000ew,Abazajian:2001nj,Asaka:2005an}. 
Sterile neutrinos could have been produced via scattering-induced 
conversion of active neutrinos \cite{Dodelson:1993je,Abazajian:2001nj}.
In this case they would constitute a warm dark-matter candidate 
with interesting features for structure formation.
A bound of $m_s> 10$~KeV applies in this case
\cite{Viel:2005qj,Seljak:2006qw,Viel:2006kd} from Lyman-$\alpha$
observations.  
In the presence of a large lepton asymmetry, the conversion can be
resonantly enhanced and the resulting spectrum would be non-thermal,
allowing for cool and cold dark matter as well \cite{Shi:1998km}.
Other mechanisms of production in which sterile neutrinos are colder
than in the case of a thermal spectrum at structure formation allow
the 10~KeV limit reported above to be relaxed down to masses as small
as a few KeV \cite{Asaka:2006ek,Kusenko:2006rh}. 
These massive neutrinos would decay into a neutrino and a photon, 
contributing to the diffuse extragalactic background
radiation
\cite{Abazajian:2001vt,Boyarsky:2005us,Boyarsky:2006jm,Boyarsky:2006zi}.
The observations typically exclude a large fraction of the parameter
space required for dark matter. 
Future observations and in particular the Chandra X-ray observatory 
have the potential of strengthening these bounds or of detecting X-ray
fluxes from clusters of galaxies. 
Weaker bounds on mixing angles and masses can also be obtained from
the contribution of sterile neutrinos to BBN and to the CMB. 
These bounds are not competitive with the ones from X-ray observations
and structure formation. 
The decays of sterile neutrinos into photons could have affected star
formation, as they can catalyse the production of molecular hydrogen
and favour star formation \cite{Biermann:2006bu}.
Sterile neutrinos in the same mass and mixing ranges can explain the
very high velocities of pulsars. 
In the presence of the strong magnetic fields of newly born neutron stars,
they can be emitted asymmetrically generating a strong kick which
boosts the star. 
The required values of the mixing angle are in the range
$10^{-5}$--$10^{-4}$, depending on the mass and on the type of
conversion (resonant or non-resonant) of active-sterile neutrinos in
the star core \cite{Kusenko:1998bk,Fuller:2003gy,Barkovich:2004jp}. 
Larger values of the mixing angles are excluded by considerations
similar to those which apply in the case of light neutrinos in
supernovae. 

\paragraph{MeV-GeV mass sterile neutrinos}

Heavy sterile neutrinos, once produced in the early Universe, would
decay rapidly into light particles; mainly neutrinos, electrons, and 
pions.
They would affect the predictions of BBN for the abundance of light
elements and in particular of $\mbox{}^4$He \cite{Dolgov:2000jw}.
The main effect would be to increase the energy density, leading to a
faster expansion of the Universe and to an earlier freeze out of the
$n/p$-ratio. 
In addition, the decay of $\nu_s$ into light neutrinos, in particular,
$\nu_e$, would modify the neutrino-energy spectrum and the equilibrium
of the $n-p$ reactions.
In principle, SN1987A data could also be used to exclude sterile
neutrinos with mixing angles $10^{-7} \ltap |U_{l s}|^2 \ltap 10^{-2}$
and masses $m_s \ltap T_{\rm core}$, where $T_{\rm core}=30-80$~MeV is
the temperature of the neutron star core. 
For masses larger than $ T_{\rm core}$, the production of sterile
neutrinos is suppressed by the Boltzmann factor.  
The emission of sterile neutrinos from the core depends on the mixing
with active neutrinos and the emission history might be very
complicated \cite{Abazajian:2001nj}. 
More detailed analyses should be performed for reliable bounds to be
derived.
  
Notice that all the cosmological bounds quoted above depend on the
density of sterile neutrinos in the early Universe. 
If they were not efficiently produced, these limits would be weakened
or not apply at all.
This is the case in the presence of mirror neutrinos with very small
mass splittings or if there is a very late phase transition such that
sterile and active neutrinos are unmixed at higher temperatures, or if
the reheating temperature is as low as few MeV \cite{Gelmini:2004ah}. 

\subsubsection[The LSND challenge]{The LSND challenge{\protect\footnote{
In April 2007 the Miniboone group announced the
data \cite{Aguilar-Arevalo:2007it} which disfavours the
simplest sterile neutrino schemes (the two flavour scheme
as well as the (3+1)-scheme described in \ref{sec:3+1}).
However, the (3+2)-scheme with two sterile neutrinos
(section \ref{sec:3+2}) can make a difference between
$\nu$ and $\bar{\nu}$ (section \ref{sec:5nu-cp}) and
as was shown by \cite{Maltoni:2007zf},
the (3+2)-scheme is not dead at the time of writing
since the Miniboone group hasn't published the
$\bar{\nu}$ data yet.  Until Miniboone announces the negative result
for $\bar{\nu}$, therefore, the scenario which is described
in sections \ref{sec:3+2} and \ref{sec:5nu-cp} is still acceptable.
Furthermore, even if Miniboone announces the negative result for $\bar{\nu}$
in the future, and even if the schemes like (3+1) and (3+2) are dead,
there still remains a possibility for sterile neutrino scenarios whose
mixing angles are small enough to satisfy the Miniboone constraint,
and the effect of these scenarios could reveal as violation of three
flavour unitarity in the future neutrino experiments.  Such scenarios are
as probable as all other possibilities described in the rest of the
section \ref{Sect:NewPhys} because none of the latter
has ever been supported by any experiment so far.
So also from that point of view, it is still useful to have the descriptions
in \ref{Para:LSNDChallenge}.}}}
\label{Para:LSNDChallenge}

The LSND experiment \cite{Aguilar:2001ty} at LANSCE in Los Alamos took
data from 1993--1998 and observed an excess of $87.9\pm22.4\pm6.0$
events in the $\bar\nu_\mu\to\bar\nu_e$ appearance channel,
corresponding to a transition probability of
$P=(0.264\pm0.067\pm0.045)\%$, $\sim 3.3\sigma$ away from zero. 
To explain this signal with neutrino oscillations requires a
mass-squared difference $\Delta m^2 \sim 1~\eVq$. 
Such a value is inconsistent with the mass-squared differences
required by the solar and reactor experiments and that required by the
atmospheric and long-baseline experiments within the S$\nu$M.
Moreover, the KARMEN experiment at the neutron spallation source,
ISIS, at the Rutherford Appleton Laboratory studied the same
appearance channel ($\bar\nu_\mu\to\bar\nu_e$) between 1997 and 2001 at
a slightly different baseline than LSND, but did not observe a
positive signal \cite{Armbruster:2002mp}. 
A combined analysis of LSND and KARMEN data has been performed in
reference \cite{Church:2002tc}. 

The MiniBooNE experiment \cite{Monroe:2004su} at Fermilab has been
designed to test the indication for oscillations reported by LSND.
In April 2007 the Miniboone group announced the first oscillation
analysis \cite{Aguilar-Arevalo:2007it}.
The results dis-favour the simplest sterile-neutrino schemes (the two
flavour scheme as well as the (3+1)-scheme described in
\ref{sec:3+1}). 
However, the (3+2)-scheme, with two sterile neutrinos
(section \ref{sec:3+2}), can accommodate different oscillation
patterns for $\nu$ and $\bar{\nu}$ (see section \ref{sec:5nu-cp}) and,
as was shown in \cite{Maltoni:2007zf}, the (3+2)-scheme is not dead at
the time of writing since the Miniboone is yet to present $\bar{\nu}$
data.
Until Miniboone announces a negative result for $\bar{\nu}$,
therefore, the scenario which is described in sections \ref{sec:3+2}
and \ref{sec:5nu-cp} is still acceptable. 
Furthermore, even if Miniboone announces a negative result for
$\bar{\nu}$ in the future, and even if schemes like (3+1) and
(3+2) are dead, there still remains a possibility for sterile-neutrino
scenarios in which the mixing angles are small enough to satisfy the
Miniboone constraint, and the effect of these scenarios could be revealed
as a violation of three-flavour unitarity in future neutrino
experiments.  
Such scenarios are as probable as all other possibilities described in
the rest of section \ref{Sect:NewPhys} since there is no evidence as
yet for any of them. 
So, from this point of view, it is useful to consider scenarios
that seek to reconcile the evidence for $\bar\nu_\mu\to\bar\nu_e$
appearance from LSND with the other evidence for neutrino oscillations.
In the following we discuss the difficulties that must be overcome if
the LSND signal is to be explained by oscillations involving light
sterile neutrinos.

\paragraph{Four-neutrino oscillations}
\label{sec:4nu}

Three mass-squared differences are required to accommodate all
evidence for neutrino oscillations including that provided by LSND,
the third mass-squared difference being significantly larger than the
other two.
A sterile neutrino with mass in the eV range must be introduced
\cite{Peltoniemi:1992ss,Caldwell:1993kn,Peltoniemi:1993ec}.  
However, it turns out that in such four-neutrino models, it is not
possible to arrange the mixing so as to accommodate all the data
\cite{Maltoni:2002xd,Strumia:2002fw,Maltoni:2004ei}.

\begin{figure} 
  \centering
  \includegraphics[width=0.6\linewidth]{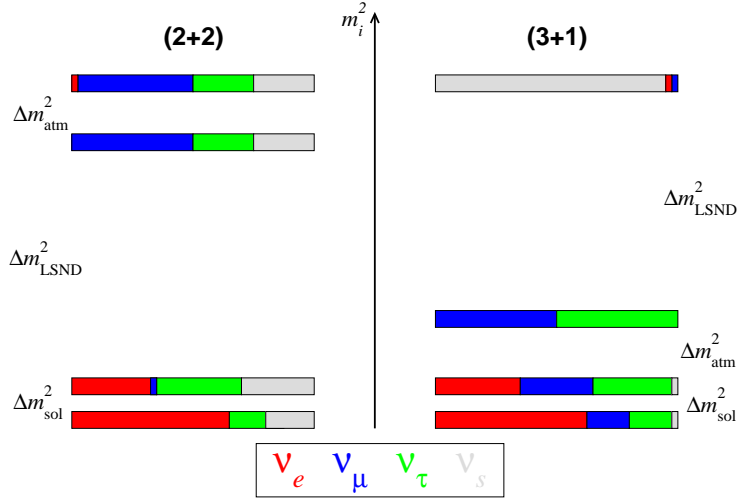}
  \caption{
    The two classes of four-neutrino mass spectra, (2+2) and (3+1).
  }
  \label{fig:4spectra}%
\end{figure}

Four-neutrino schemes are usually divided into the two classes (3+1)
and (2+2), as illustrated in figure \ref{fig:4spectra}.  
The (3+1) mass spectra can be considered as a small perturbation of
the standard three-active-neutrino scenario. 
In this case, solar- and atmospheric-neutrino oscillations are explained
mainly by active-neutrino oscillations, with mass-squared differences
$\Dms$ and $\Dma$, and the fourth neutrino state separated by $\Dml$
contains just a small component of the electron- and muon-neutrino
flavours to account for the LSND signal. 
In contrast, the (2+2) spectrum is intrinsically different from the
standard three-active-neutrino scenario as the sterile neutrino must
take part dominantly either in solar- or in atmospheric-neutrino
oscillations, or in both.

Neglecting CP violation, neutrino oscillations in four-neutrino
schemes are generally described by 9 parameters: 3 mass-squared
differences and 6 mixing angles.
A convenient parameterisation has been introduced in reference
\cite{Maltoni:2001bc}, in terms of $\Dms$, $\theta_\Sol$, $\Dma$,
$\theta_\Atm$, $\Dml$, and $\theta_\Lsnd$. 
These 6 parameters are similar to the two-neutrino mass-squared
differences and mixing angles and are directly related to the
oscillations in the solar, atmospheric, and LSND experiments.  
For the remaining 3 parameters one can use $\eta_s,\eta_e$ and
$d_\mu$. 
These quantities are defined by: 
\begin{eqnarray}
    \label{eq:def-eta}
    \eta_\alpha &=& \sum_i |U_{\alpha i}|^2 \quad
    \mbox{with $i\in$ solar mass states; and}\\%
    \label{eq:def-d}
    d_\alpha  &=& 1 - \sum_i |U_{\alpha i}|^2 \quad
    \mbox{with $i\in$ atmospheric mass states;}
\end{eqnarray}
where $\alpha = e,\mu,\tau,s$. Note that in (2+2) schemes the relation
$ \eta_\alpha = d_\alpha$ holds, whereas in (3+1) $\eta_\alpha$ and
$d_\alpha$ are independent. 
The physical meaning of these parameters is the following:
$\eta_\alpha$ is the fraction of $\nu_\alpha$ participating in solar
oscillations, and ($1-d_\alpha$) is the fraction of $\nu_\alpha$
participating in oscillations with frequency $\Dma$ (for further
discussions and details of the approximations adopted see reference
\cite{Maltoni:2001bc}).

\paragraph{(2+2): ruled out by solar and atmospheric data}
\label{sec:2+2}

\begin{figure}
   \centering
   \includegraphics[width=0.95\linewidth]{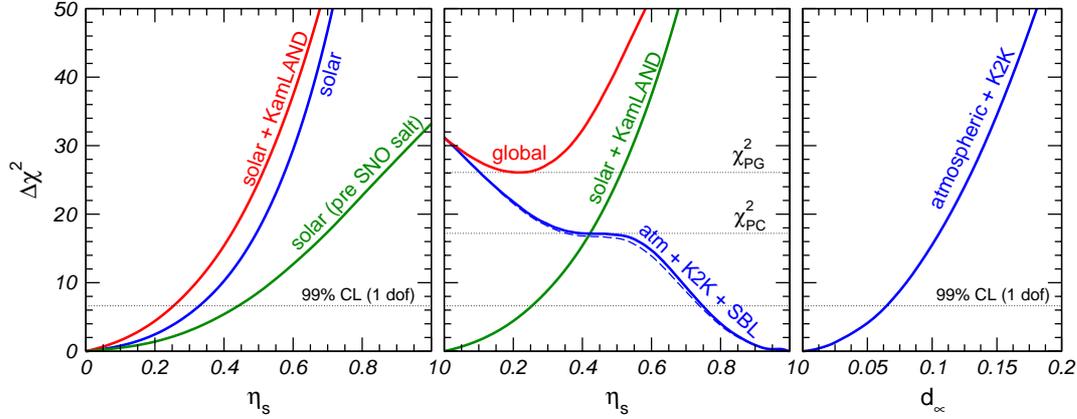}
   \caption{ 
     Left: $\Delta\chi^2$ as a function of $\eta_s$ from solar data
     before the SNO salt-phase results, from current solar data, and
     from solar+KamLAND data. Middle: $\Delta\chi^2_\Sol$,
     $\Delta\chi^2_{\Atm+\KtK+\Sbl}$ and $\bar\chi^2_\mathrm{global}$
     as a function of $\eta_s$ in (2+2) oscillation schemes. The
     dashed line corresponds to atmospheric and K2K data only
     (without SBL data). Right: $\Delta\chi^2_{\Atm+\KtK}$ as a
     function of $d_\mu$.
    Taken with kind permission of New Journal of Physics
    from figure 19 in reference \cite{Maltoni:2004ei}.
    Copyrighted by Deutsche Physikalische Gesellschaft \& Institute of Physics.
   }
   \label{fig:etas}
\end{figure}

The strong preference for oscillations into active neutrinos in solar
and atmospheric oscillations leads to a direct conflict in (2+2)
oscillation schemes \cite{Giunti:2000wt,Gonzalez-Garcia:2001uy}.
Thanks to recent solar neutrino data (in particular from the SNO-salt
phase \cite{Ahmed:2003kj}) in combination with the KamLAND experiment,
and Super-Kamiokande data on atmospheric neutrinos the tension in the
data has become so strong that (2+2) oscillation schemes are
essentially ruled out.
The left panel of figure \ref{fig:etas} shows the $\Delta \chi^2$ from
solar-neutrino data as a function of $\eta_s$, the parameter
describing the fraction of the sterile-neutrino participating in 
solar-neutrino oscillations. 
It is clear from the figure that the improved determination of the
neutral-current-event rate from the solar $^8$B flux implied by the
salt enhanced measurement in SNO \cite{Ahmed:2003kj} substantially
tightened the constraint on a sterile contribution; the 99\%~\CL\
bound improves from $\eta_s \le 0.44$ for pre-SNO-salt to $\eta_s
\le 0.33$ at the 99\%~\CL\  
Although KamLAND on its own is insensitive to a sterile neutrino
contamination, it contributes indirectly to the bound because of the
better determination of $\Dms$. 
The combined analysis leads to the 99\% \CL\ bound:
\begin{equation}
    \eta_s \le 0.25 \qquad\mbox{(solar + KamLAND).}
\end{equation}
In contrast, in (2+2) schemes atmospheric data prefer values of
$\eta_s$ close to 1. 
From the combined analysis of Super-Kamiokande atmospheric data, K2K
and short-baseline (SBL)
\cite{Armbruster:2002mp,Dydak:1983zq,Declais:1994su} neutrino data one
obtains the bound $\eta_s \ge 0.75$ at 99\%~\CL, in clear disagreement
with the bound from solar data.  
The middle panel of figure \ref{fig:etas} shows the $\Delta\chi^2$ for
solar data and for atmospheric+K2K combined with SBL data as a
function of $\eta_s$. 
Note that the main effect comes from atmospheric+K2K data; SBL
experiments contribute only marginally, as may be seen from the
dashed line. 
From this figure we also see that the `solar+KamLAND' and the
`atm+K2K+SBL' allowed domains overlap only at 
$\chi^2_\mathrm{PC} = 17.2$, i.e.\ at the $4.1\sigma$ level.
In the middle panel of figure \ref{fig:etas}, the `global'
$\bar{\chi}^2$ function defined as \cite{Maltoni:2003cu}: 
\begin{equation}\label{eq:chi2solatm}
    \bar\chi^2(\eta_s) \equiv
    \Delta\chi^2_{\SlKm}(\eta_s) +
    \Delta\chi^2_{\Atm+\KtK+\Sbl}(\eta_s)
\end{equation}
is shown.
In references \cite{Maltoni:2002xd,Maltoni:2003cu} a statistical
method to evaluate the disagreement of different data sets used in a
global analysis has been proposed. 
The `parameter goodness of fit' (PG) makes use of the $\bar\chi^2$
defined in equation \eqref{eq:chi2solatm}.
The result, $\chi^2_\mathrm{PG} \equiv \bar\chi^2_\mathrm{min} = 26.1$, 
corresponds to exclusion of (2+2) mass schemes at the 5.1$\sigma$
level.
Sub-leading effects beyond the approximations adopted in
\cite{Maltoni:2004ei} should not affect this result significantly
\cite{Pas:2002ah}. 

\paragraph{(3+1): strongly dis-favoured by SBL data}
\label{sec:3+1}

It has been known for a long time that (3+1) mass schemes are
dis-favoured by the comparison of SBL disappearance
data~\cite{Dydak:1983zq,Declais:1994su} with the LSND result
 \cite{Bilenky:1996rw,Okada:1996kw,Barger:1998bn,Bilenky:1999ny,Peres:2000ic,Giunti:2000ur,Grimus:2001mn}.
The reason is that in (3+1) schemes the relation 
$\sin^22\theta_\Lsnd = 4\, d_e\,d_\mu$ holds, and the parameters $d_e$ 
and $d_\mu$ (see equation (\ref{eq:def-d})) are strongly constrained by
$\nu_e$ and $\nu_\mu$ disappearance experiments, leading to a double
suppression of the LSND amplitude. 
In reference \cite{Bilenky:1999ny} it was shown that the up-down
asymmetry observed in atmospheric $\mu$ events leads to an additional
constraint on $d_\mu$ (see also reference \cite{Maltoni:2001mt}). 
The $\Delta\chi^2(d_\mu)$ from the fit to atmospheric+K2K data is
shown in the right panel of figure \ref{fig:etas}, and one obtains the
bound:
\begin{equation}\label{eq:dmu}
  d_\mu \le 0.065\quad\mbox{at 99\%~\CL}
\end{equation}

\begin{figure}
  \centering
  \includegraphics[height=8cm]{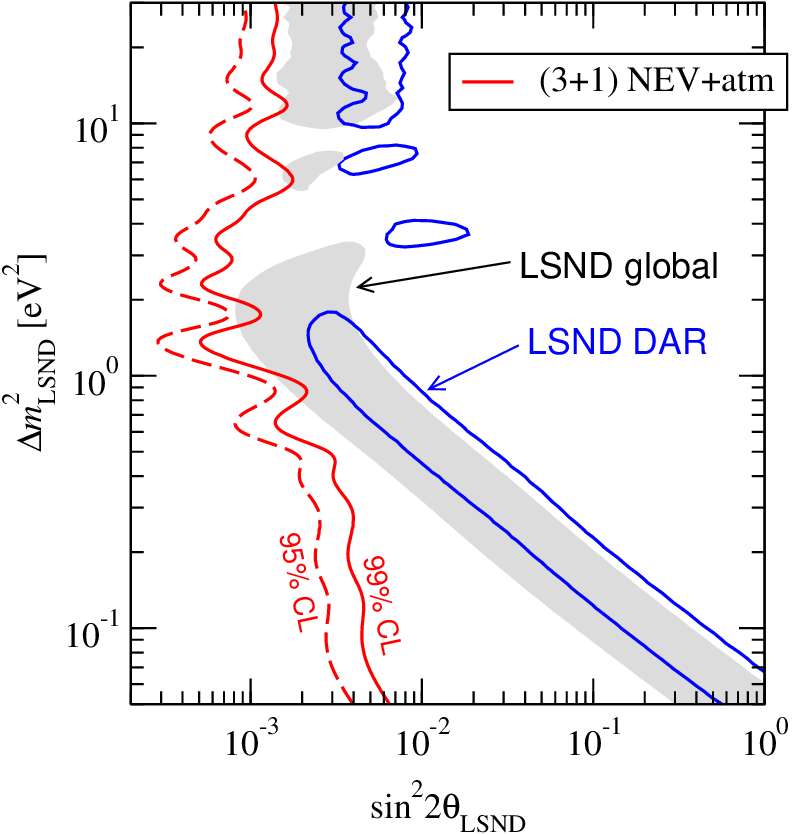}
  \caption{
    Upper bound on $\sin^22\theta_\Lsnd$
    from NEV, atmospheric and K2K neutrino data in (3+1)
    schemes. The bound is calculated for each $\Dml$ using the
    $\Delta \chi^2$ for 1~\dof\ Also shown are the regions allowed
    at 99\% \CL\ (2~\dof) from global LSND
    data~\protect\cite{Aguilar:2001ty} and decay-at-rest (DAR) LSND
    data~\protect\cite{Church:2002tc}.
    Taken with kind permission of New Journal of Physics
    from figure 20 in reference \cite{Maltoni:2004ei}.
    Copyrighted by Deutsche Physikalische Gesellschaft \& Institute of Physics.
  }
  \label{fig:3+1} 
\end{figure}

Figure \ref{fig:3+1} shows the upper bound on the LSND oscillation
amplitude, $\sin^22\theta_\Lsnd$, from the combined analysis of
no-evidence (NEV) and atmospheric neutrino data \cite{Grimus:2001mn}.  
Data from the Bugey \cite{Declais:1994su}, CDHS \cite{Dydak:1983zq},
KARMEN \cite{Armbruster:2002mp}, and CHOOZ \cite{Apollonio:2002gd}
experiments are included in the NEV data set;
the NOMAD experiment \cite{Astier:2003gs} gives additional constraints
in the region of high $\Dml$.
From this figure one can see that the bound is incompatible with the
signal observed in LSND at the 95\%~\CL\ 
Only marginal overlap regions exist between the bound and global LSND
data if both are taken at 99\%~\CL\ 
Using only the decay-at-rest LSND data sample \cite{Church:2002tc}, the
disagreement is even more severe. 
These results show that (3+1) schemes are strongly dis-favoured by SBL
disappearance data. 

\paragraph{Global fit in four-neutrino schemes}
\label{sec:global-fit-4nu}

The methods developed in \cite{Maltoni:2001bc} allow the oscillation
data to be fit using the four-neutrino model.
The result of such fits can be used to evaluate a goodness-of-fit
statistic using the PG method \cite{Maltoni:2003cu}, allowing the
different hypotheses to be compared. 
The global oscillation data were divided into the four data sets SOL,
ATM, LSND, and NEV and the PG method used to evaluate $\bar\chi^2$
\cite{Maltoni:2002xd}:
\begin{equation}\label{eq:chi2bar}
    \begin{array}{ccl}
        \bar\chi^2 &=&
        \Delta\chi^2_\Sol(\theta_\Sol,\Dms,\eta_s)
        + \Delta\chi^2_\Atm(\theta_\Atm,\Dma,\eta_s,d_\mu) \\
        &+& \Delta\chi^2_\Nev(\theta_\Lsnd,\Dml,d_\mu,\eta_e)
        + \Delta\chi^2_\Lsnd(\theta_\Lsnd,\Dml) \,,
    \end{array}
\end{equation}
where $\Delta\chi^2_X = \chi^2_X - (\chi^2_X)_\mathrm{min}$ 
($X$ = SOL, ATM, NEV, LSND), and $\bar\chi^2_\mathrm{min}$ is the
minimum value of the $\bar\chi^2$.
Table \ref{tab:pg} shows the contributions of the four data sets to
$\chi^2_\mathrm{PG} \equiv \bar\chi^2_\mathrm{min}$ for (3+1) and
(2+2) oscillation schemes. 
As expected, the main contribution to $\chi^2_\mathrm{PG}$ in (3+1)
schemes comes from SBL data due to the tension between LSND and NEV
data in these schemes.  
For (2+2) oscillation schemes a large contribution to
$\chi^2_\mathrm{PG}$ comes from solar and atmospheric data
as discussed in Sec.~\ref{sec:2+2}. 
The contribution from NEV data in (2+2) comes mainly from the tension
between LSND and KARMEN~\cite{Church:2002tc}, which does not depend on
the mass scheme. 
The parameter goodness of fit (PG) shown in the last column of
table \ref{tab:pg} is obtained by evaluating $\chi^2_\mathrm{PG}$ for 
four degrees of freedom \cite{Maltoni:2003cu}. 
This number of degrees of freedom corresponds to the four parameters 
$\eta_s, d_\mu, \theta_\Lsnd, \Dml$ describing the coupling of the
different data sets. 
\begin{table}
  \centering
  \catcode`?=\active \def?{\hphantom{0}}
  \begin{tabular}{|c|cccc|c|c|}
    \hline
    & SOL & ATM & LSND & NEV &   $\chi^2_\mathrm{PG}$ & PG \\
    \hline
    (3+1) & 0.0 & ?0.4 & 5.7 & 10.9 & 17.0 & $1.9 \times 10^{-3} \: (3.1\sigma)$ \\
    (2+2) & 5.3 & 20.8 & 0.6 & ?7.3 & 33.9 & $7.8 \times 10^{-7} \: (4.9\sigma)$ \\
    \hline
  \end{tabular}
  \caption{
    Parameter goodness-of-fit (PG) and the contributions of
    different data sets to $\chi^2_\mathrm{PG}$ in (3+1) and (2+2)
    neutrino-mass schemes~\cite{Maltoni:2004ei}.
  }
  \label{tab:pg}
\end{table}

The status of four-neutrino explanations of the LSND signal can be
summarised as follows:
\begin{itemize}
  \item Schemes of the (2+2) structure are ruled out at the 
        $\sim 5 \sigma$ level (PG of $7.8 \times 10^{-7}$) by the
        disagreement between the individual data sets.
        This result is very robust, independent of whether LSND is
        confirmed or disproved, and applies to all four-neutrino-mass
        models where two pairs of neutrino-mass states providing
        $\Dms$ and $\Dma$ are separated by a big mass gap;
  \item The explanation of the LSND effect within (3+1) schemes is
        dis-favoured at the $\sim 3\sigma$ level (PG of 0.19\%). 
        This result relies heavily on the null-result obtained from
        the SBL disappearance experiments Bugey and CDHS.
        Therefore, if the LSND appearance signal were to be confirmed
        by MiniBooNE, a (3+1) mass scheme should lead also to an
        observable signal for the $\nu_\mu$ disappearance signal in
        MiniBooNE. 
\end{itemize}

\paragraph{Five-neutrino oscillations}
\label{sec:3+2}

As a possible way out of the problems in four--neutrino schemes, 
a second sterile neutrino has been introduced in the
analysis, and a five-neutrino mass scheme of the type (3+2) considered
\cite{Sorel:2003hf}. 
In a manner similar to the (3+1) scheme, the active neutrinos are
contained mainly in the three lightest-mass states responsible for
solar and atmospheric oscillations.
the two states with masses in the eV range are available to explain
the LSND effect. 
The disagreement between the data sets measured by the parameter
goodness-of-fit is improved from 0.032\% for the (3+1) scheme to 2.1\%
for the (3+2) scheme. 
The best fit point for the (3+2) scheme gives the mass-squared
differences $\Delta m^2_{41} \simeq 0.9$~eV$^2$ and 
$\Delta m^2_{51} \simeq 22$~eV$^2$, but also solutions with only
sub-eV masses are found \cite{Sorel:2003hf}. 

Note that the possible conflicts of eV-scale sterile neutrinos with
cosmology (see e.g.\ \cite{Dodelson:2005tp}), which already appear for
four neutrinos, become more severe in the five-neutrino case and a
non-standard cosmological model must be constructed. 
Moreover, the (3+2)-best-fit point found in \cite{Sorel:2003hf} seems
to be disfavoured also by atmospheric-neutrino data;
as pointed out in \cite{Bilenky:1999ny}, atmospheric neutrinos provide
a constraint on the parameter $d_\mu$ (see equation
(\ref{eq:def-d})). 
In the (3+2) scheme this parameter is given by 
$d_\mu = |U_{\mu 4}|^2 + |U_{\mu 5}|^2$; with the best-fit values
$U_{\mu 4} = 0.204$, $U_{\mu 5} = 0.224$ one obtains 
$d_\mu \approx 0.09$, in conflict with the bound given in equation
(\ref{eq:dmu}) \cite{Sorel:2003hf}.  
Figure \ref{fig:etas} shows that this value of $d_\mu$ leads to a
$\Delta\chi^2 \approx 12.5$ from atmospheric and K2K data, and hence
is disfavoured at the $3.5\sigma$ level. 
Therefore, a re-analysis of the (3+2) scenario, including the
constraint from atmospheric data, is required for this model to be
considered viable.

\paragraph{Unconventional manifestations of leptonic-CP violation?}
\label{sec:5nu-cp}

Neutrino models involving active/sterile neutrino mixing at the LSND
\cite{Aguilar:2001ty} neutrino-mass-splitting scale via at least two 
sterile-neutrino states would open the possibility for further
manifestations of leptonic-CP violation, including ones that could be
measurable with neutrino-appearance experiments at short baselines
\cite{Karagiorgi:2006jf}.

For $N$ neutrino species, there are, in general,$(N-1)$ independent
mass splittings, $N(N-1)/2$ moduli of parameters in the unitary mixing
matrix, and $(N-1)(N-2)/2$ Dirac CP-violating phases that may be
observed in oscillations. 
In short-baseline (SBL) experiments that are sensitive only to
$\nu_{\mu}\to\nu_{\not{\mu}}$,  
$\nu_e\to\nu_{\not{e}}$, and $\nu_{\mu}\to\nu_e$
transitions, the set of observable parameters is reduced
considerably. 
First, oscillations due to atmospheric- and solar-mass splittings can
be neglected or, equivalently, one can set $m_1=m_2=m_3$. 
Second, mixing-matrix elements that measure the $\tau$-neutrino
flavour fraction of the various neutrino mass eigenstates do not enter
in the oscillation probability. 
In this case, the number of observable parameters is restricted to
$(N-3)$ independent mass splittings, $2(N-3)$ moduli of mixing matrix
parameters, and $(N-3)(N-4)/2$ CP-violating phases. 
Therefore, for the (3+2) sterile-neutrino models \cite{Sorel:2003hf}
that we wish to discuss here, that is for the $N=5$ case, there are
two independent mass splittings $\Delta m^2_{41}$ and $\Delta
m^2_{51}$, four moduli of mixing matrix parameters 
$|U_{e4}|,\ |U_{\mu 4}|,\ |U_{e5}|,\ |U_{\mu 5}|$, 
and one CP-violating phase. 
The convention used in the following for this CP-phase is:
\begin{equation}
  \label{eq:cpv_at_sbl_eq1}
  \phi_{45}=arg(U_{\mu 5}^*U_{e5}U_{\mu 4}U_{e4}^* )
\end{equation}
\noindent Under these assumptions, the relevant oscillation probabilities
can be rewritten as:
\begin{equation}
  \label{eq:cpv_at_sbl_eq2}
  P(\nu_{\alpha}\to\nu_{\alpha}) = 
  1-4[(1-|U_{\alpha 4}|^2-|U_{\alpha 5}|^2)\cdot
  (|U_{\alpha 4}|^2\sin^2 x_{41}+|U_{\alpha 5}|^2\sin^2 x_{51})+
  |U_{\alpha 4}|^2|U_{\alpha 5}|^2\sin^2 x_{54}] \, {\rm and}
\end{equation}
\begin{eqnarray}
\label{eq:cpv_at_sbl_eq3}
P(\nu_{\alpha}\to\nu_{\beta}) = 
4|U_{\alpha 4}|^2|U_{\beta 4}|^2\sin^2 x_{41}+ 
      4|U_{\alpha 5}|^2|U_{\beta 5}|^2\sin^2 x_{51}+ \nonumber \\
      8|U_{\alpha 5}||U_{\beta 5}||U_{\alpha 4}||U_{\beta 4}|
\sin x_{41}\sin x_{51}\cos (x_{54}+\phi_{45}) 
\end{eqnarray}
\noindent for $\alpha =\beta$ and $\alpha \neq \beta$, respectively.
The formulas for antineutrino oscillations are obtained by substituting
$\phi_{45}\to -\phi_{45}$.

We perform a combined analysis of SBL and atmospheric-neutrino data. 
The analysis uses the same seven SBL datasets as in 
reference~\cite{Sorel:2003hf}, including results on $\nu_{\mu}$ disappearance 
(from the CCFR84 \cite{Stockdale:1984cg} and CDHS \cite{Dydak:1983zq} 
experiments), $\nu_e$ disappearance (from the Bugey \cite{Declais:1994su} 
and CHOOZ \cite{Apollonio:2002gd} experiments), and
$\nu_{\mu}\to\nu_e$ oscillations (from the LSND \cite{Aguilar:2001ty},
KARMEN2 \cite{Armbruster:2002mp}, and NOMAD \cite{Astier:2003gs}
experiments). The assumptions used to describe SBL data 
are described  in reference~\cite{Sorel:2003hf}.
The atmospheric-neutrino constraints used in the analysis include 1489 days 
of Super-Kamiokande charged-current data \cite{Ashie:2005ik},
including the $e$-like and $\mu$-like data samples of sub- and
multi-GeV contained events, stopping events, and through-going
upward-going muon events. The assumptions used to describe
atmospheric data are described in 
Refs.~\cite{Maltoni:2004ei,Gonzalez-Garcia:2004wg}.
The atmospheric constraint also includes data on $\nu_{\mu}$ disappearance
from the long-baseline, accelerator-based experiment K2K
\cite{Ahn:2006zz}.

The purpose of this study is not only to determine what the allowed
values of the SBL CP-violation phase $\phi_{45}$ are from existing
SBL+atmospheric data, but also what the oscillation appearance
probabilities in neutrino and anti-neutrino running modes are to be
expected in the MiniBooNE experiment at Fermilab
\cite{MiniBooNE-RunPlan}, in the context
of (3+2) sterile neutrino models, and allowing for the possibility
of CP violation. The MiniBooNE experiment took
data in neutrino running mode between September 2002 and
January 2006, at which point the experiment started its
ongoing anti-neutrino run. Realistic estimates of the
oscillation probabilities to be expected at MiniBooNE are used
in the analysis, based on neutrino flux and cross-section 
expectations provided by the MiniBooNE 
Collaboration. In particular, the effect of ``wrong sign'' neutrinos in
computing the expected oscillation probabilities, which have
the effect of washing out CP-violating observables, is taken into
account. This effect is non-negligible since as much as one third
of the total interaction rate in anti-neutrino running mode
is expected to be due to neutrinos rather than anti-neutrinos; on
the other hand, the anti-neutrino contribution in neutrino running
mode is expected to be much smaller.

From the upcoming MiniBooNE appearance measurements in neutrino
and anti-neutrino running modes, the following CP-asymmetry
observable, $A_{CP}$, could be extracted:
\begin{equation}
\label{eq:cpv_at_sbl_eq4}
A_{CP} = \frac{p_{\hbox{\footnotesize BooNE}}-
                \bar{p}_{\hbox{\footnotesize BooNE}}}
              {p_{\hbox{\footnotesize BooNE}}+
                \bar{p}_{\hbox{\footnotesize BooNE}}} \; ,
\end{equation} 
\noindent where we have defined the oscillation probability in neutrino
(anti-neutrino) running mode as:
\begin{equation}
\label{eq:cpv_at_sbl_eq5}
\stackrel{\hbox{\small{(-)}}}{~p}_{BooNE}=
\frac{
\int dE\ [p(\nu_{\mu}\to\nu_e)\stackrel{\hbox{\small{(-)}}}{N_0}(\nu )+
 p(\bar{\nu}_{\mu}\to\bar{\nu}_e)\stackrel{\hbox{\small{(-)}}}{N_0}(\bar{\nu})]
}
{
\int dE\ [\stackrel{\hbox{\small{(-)}}}{N_0}(\nu
)+\stackrel{\hbox{\small{(-)}}}{N_0}(\bar{\nu})]
} \; ,
\end{equation}
\noindent where $E$ is the neutrino energy; $p(\nu_{\mu}\to\nu_e)$ and 
$p(\nu_{\bar{\mu}}\to\nu_{\bar{e}})$ are the oscillation probabilities 
given by equation (\ref{eq:cpv_at_sbl_eq3}), with $\phi_{45}=0$ or $\pi$ for 
the CP-conserving case, and $0<\phi_{45}<2\pi$ 
for the CP-violating case; $N_0(\nu )$ and $N_0(\bar{\nu})$ are the 
MiniBooNE neutrino and
anti-neutrino full-transmutation rate distributions in 
neutrino-running mode (that is, muon neutrino and anti-neutrino
fluxes multiplied by electron-neutrino and anti-neutrino
cross-sections), and $\bar{N}_0(\nu )$ and $\bar{N}_0(\bar{\nu})$ are 
the neutrino and anti-neutrino full-transmutation rate distributions in 
anti-neutrino-running mode.

\begin{table}
\centerline{
\begin{tabular}{ccccllllr} \hline 
Model & $\chi^2\ (d.o.f.)$ & $\Delta m^2_{41}\ (\hbox{eV}^2)$ &
$\Delta m^2_{51}\ (\hbox{eV}^2)$ & $|U_{e4}|$ & $|U_{\mu4}|$ & 
$|U_{e5}|$ & $|U_{\mu5}|$ & $\phi_{45}$ \\ \hline 
CPC & 141.4 (145) & 0.92 & 24 & 0.132 & 0.158 & 0.066 & 0.159 & 0 \\
CPV & 140.8 (144) & 0.91 & 24 & 0.127 & 0.147 & 0.068 & 0.164 & 1.8$\pi$
 \\ \hline
\end{tabular}
}  
\caption{\label{tab:cpv_at_sbl_tab1} Comparison of best-fit values 
for mass-splittings and mixing parameters for   
(3+2) CP-conserving and CP-violating models.} 
\end{table}

\begin{figure}
\centerline{
\includegraphics*[width=0.60\columnwidth, trim=30 30 0 0, angle=-90]
{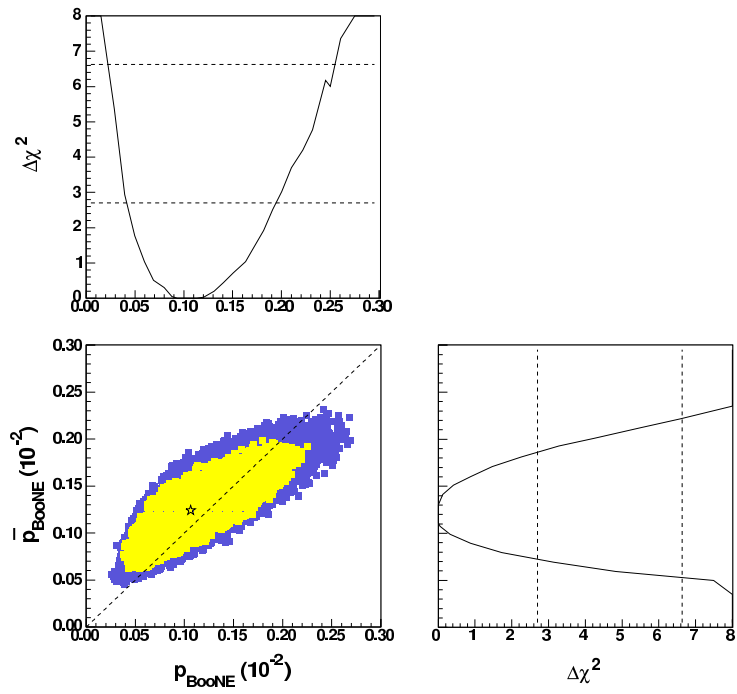}
}
\caption{\label{fig:cpv_at_sbl_fig4}Expected oscillation probabilities at 
MiniBooNE in neutrino and
anti-neutrino running modes, for CP-violating (3+2) models. The yellow 
(light grey) region corresponds to the 90\% CL allowed region; 
the blue (dark grey) region corresponds to the 99\% CL allowed 
region.
    Taken with kind permission of Physical Review from figure 4 in
    reference \cite{Karagiorgi:2006jf}.
    Copyrighted by the American Physical Society.
}
\end{figure}

\begin{figure}
\centerline{
\includegraphics*[width=0.60\columnwidth, trim=30 30 0 0, angle=-90]
{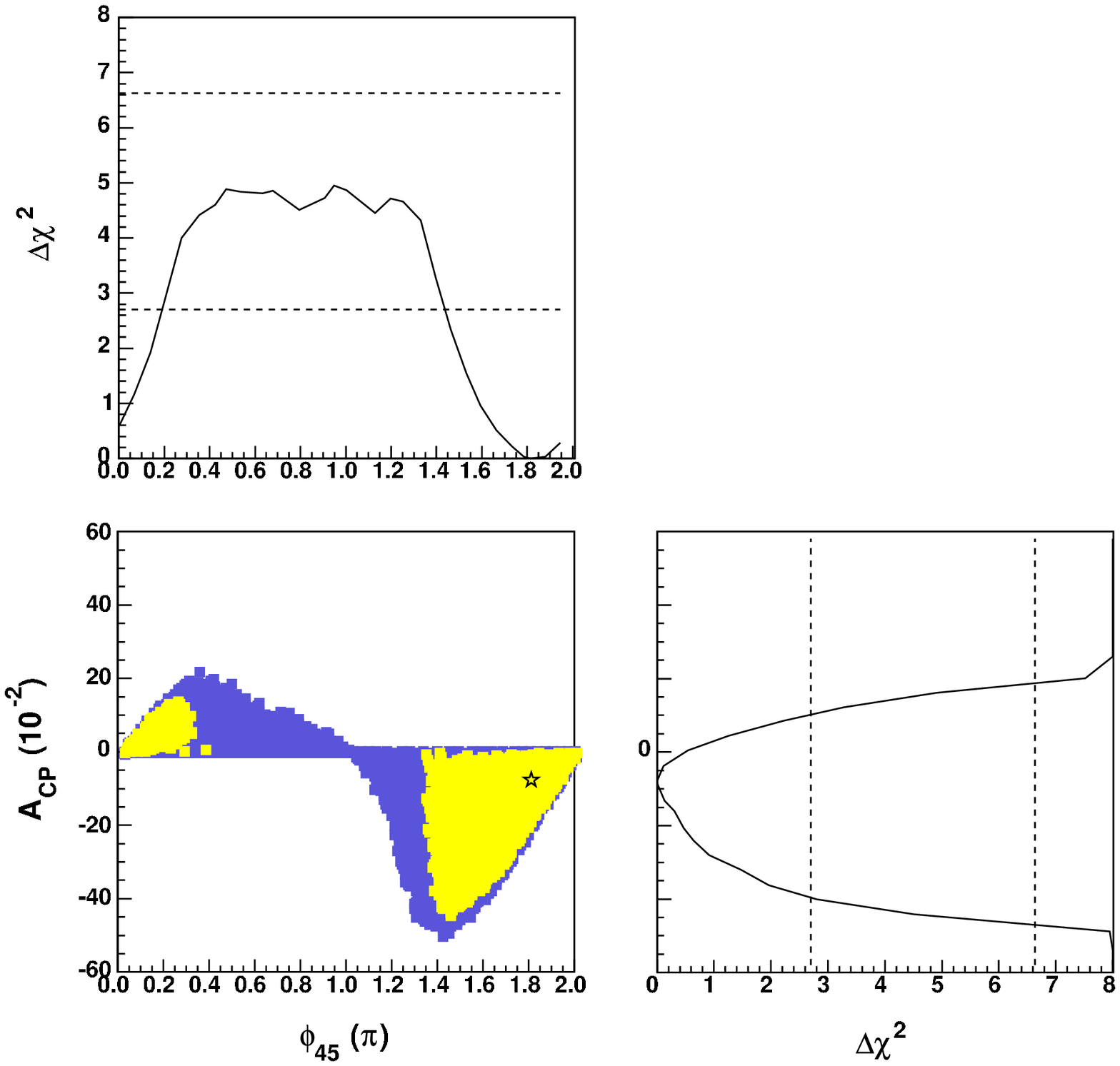}
}
\caption{\label{fig:cpv_at_sbl_fig5}Current limits on the CP-violating phase 
$\phi_{45}$ from current short-baseline results, and CP asymmetry 
measurement expected at MiniBooNE, $A_{CP}$, as a function of 
$\phi_{45}$. The yellow (light grey) region corresponds to the 90\% 
CL allowed region; the blue (dark grey) region corresponds to the 99\% 
CL allowed region.
    Taken with kind permission of Physical Review from figure 5 in
    reference \cite{Karagiorgi:2006jf}.
    Copyrighted by the American Physical Society.}
\end{figure} 
 
The main results of this study are given in 
table \ref{tab:cpv_at_sbl_tab1}, and in
figures \ref{fig:cpv_at_sbl_fig4} and \ref{fig:cpv_at_sbl_fig5}. 
Table \ref{tab:cpv_at_sbl_tab1} shows the best-fit model
parameters for CP-conserving and CP-violating (3+2) sterile
neutrino models; figure \ref{fig:cpv_at_sbl_fig4} shows the oscillation
probabilities to be expected at MiniBooNE in neutrino and
anti-neutrino running modes (equation (\ref{eq:cpv_at_sbl_eq5})), 
in a CP-violating, (3+2) scenario;
and figure \ref{fig:cpv_at_sbl_fig5} shows the $\phi_{45}$ values
(equation (\ref{eq:cpv_at_sbl_eq1})) allowed by current SBL+atmospheric 
constraints, and the
corresponding CP asymmetries expected at MiniBooNE
(equation (\ref{eq:cpv_at_sbl_eq4})), in this same scenario. 

The results shown can be summarised as follows. First, we find that
CP-violating, (3+2) models do not provide a significantly better
description of short-baseline and atmospheric data, compared to
CP-conserving, (3+2) models. On the other hand, even if only a small degree
of CP violation is marginally preferred, we also find that existing
data allow for all possible values for the single CP-violating phase that
could be observed at short baselines in (3+2) models, at 99\% C.L..
Finally, if leptonic-CP violation occurs and (3+2) sterile-neutrino
models are a good description of the data, we find that differences as
large as a factor of three between the electron (anti-)neutrino
appearance probabilities in neutrino and anti-neutrino running modes
at MiniBooNE, corresponding to $A_{CP}=-0.5$ in equation
(\ref{eq:cpv_at_sbl_eq4}), are possible.

\paragraph{More exotic explanations of LSND}

In view of these difficulties in finding an explanation of the LSND
result, several alternative mechanisms have been proposed.
Some of the proposed mechanisms involve speculative process such as:
non-standard neutrino interactions; violation of CPT invariance;
violation of Lorentz invariance; quantum decoherence; or mass-varying
neutrinos \cite{giunti-web}.
Many of the proposed mechanisms are unable to accommodate all of the
evidence for neutrino oscillations as well as the constraints from the
NEV experiments. 
Mechanisms which seem to be in agreement with all present data are:
the four-neutrino mass scheme plus CPT violation \cite{Barger:2003xm};
a model based on sterile neutrinos and large-extra dimensions
\cite{Pas:2005rb}; and a model with a decaying sterile
neutrino \cite{Palomares-Ruiz:2005vf}.  

In reference \cite{Barger:2003xm}, a four-neutrino scheme similar to
(3+1) is considered.
However, this model allows for different mixing matrices for neutrinos
and anti-neutrinos, violating CPT invariance and therefore avoiding
the constraints imposed by the NEV experiments.
In reference \cite{Pas:2005rb}, a new resonance effect is introduced by 
assuming `short-cuts' for sterile neutrinos through rather particular
extra dimensions.  
In contrast to such relatively exotic ideas, the scenario proposed in
reference \cite{Palomares-Ruiz:2005vf} involves a comparably `modest
amount' of non-standard physics. 
In this model a heavy neutrino, $n_4$, is introduced, with a small
mixing with muon neutrinos of $|U_{\mu 4}|^2 \sim 0.01$, such that a
small $n_4$ component is contained in the initial $\bar\nu_\mu$ beam
produced in the LSND experiment. 
The $n_4$ decays into a scalar particle and a light neutrino,
predominantly of the electron type, accounting in this way for the
$\bar\nu_e$ appearance in LSND. 
Values of $g m_4\sim$~few~eV are required, where $g$ is the
neutrino-scalar coupling constant and $m_4$ is the heavy-neutrino
mass. 
For example, one can take $m_4$ in the range from 1~keV to 1~MeV and 
$g \sim 10^{-6}$--$10^{-3}$, consistent with various bounds on such
couplings. 
Unlike the case of (3+1) four-neutrino oscillation schemes, the decay 
model is in complete agreement with the constraints from SBL
disappearance experiments. 
Testing the compatibility of LSND and all the null-result experiments,
one finds a parameter goodness-of-fit PG~$= 4.6\%$ for decay, which is
slightly better than the PG~$=2.1\%$ obtained in reference
\cite{Sorel:2003hf} for the (3+2) five-neutrino oscillation scenario.

%% file: 04-BSnuM/04-02-Mass-varying-neutrinos/04-02-Mass-varying-neutrinos.tex
\subsection{Mass Varying Neutrinos}

Abundant cosmological data indicate that the expansion of our Universe
is in an accelerating phase caused by a negative-pressure component
called dark energy. 
Dark energy is troubling because the acceleration of the Universe is a
very recent phenomenon in its expansion history. 
This `cosmic coincidence' problem can be expressed as follows: 
why are the dark-matter and dark-energy densities comparable today,
even though their ratio scales as $\sim 1/a^3$ (where $a$ is the scale
factor)?
The coincidence that the scale of dark energy $(2\times 10^{-3}$~eV)$^4$ 
is similar to the scale of the neutrino mass-squared difference squared 
($[0.01{\rm~eV}^2]^2$) was exploited recently to solve the coincidence
problem \cite{Fardon:2003eh}.
The assumption was made that neutrinos couple to dark energy by
in such a way as to make the dark-energy density a function of
neutrino mass.
The total energy density of neutrinos and dark energy was assumed to
be constant, i.e. to be independent of neutrino mass. 
Under these assumptions, changes in the neutrino-energy density and
the dark-energy density are correlated.
Over a wide range of values of $a$, neutrino masses must vary so as to
allow the total energy density remains constant.

A simple way to make the dark-energy density neutrino-mass dependent
is to introduce a Yukawa coupling between a sterile neutrino, $s$, and
a light, scalar field, $\phi$, called the acceleron. 
At scales below the sterile-neutrino mass, a Lagrangian of the form:
\begin{equation}
  \mathcal{L}= m_D\nu s + \lambda\phi s s + V_0(\phi)\,,
\end{equation}
where $\nu$ is a Standard Model left-handed neutrino, leads to an
effective potential for the acceleron (if neutrinos are
non-relativistic) given by:
\begin{equation}
  V = {m_D^2 \over {\lambda \phi}} n_\nu + V_0(\phi)\,.
\end{equation}
Thus, the effective potential of the acceleron at late times receives
a contribution equal to $m_\nu n_\nu$, where 
$m_\nu = m_D^2/(\lambda \phi)$ and $n_\nu$ are the active-neutrino
mass and number density, respectively. 
More elaborate supersymmetric models of neutrino dark energy have been
constructed in  \cite{Takahashi:2005kw,Fardon:2005wc}.

Model-independent tests of neutrino dark energy are cosmological
\cite{Fardon:2003eh,Peccei:2004sz}. 
A strict relationship between the equation of state of the combined
dark-energy neutrino fluid $w=p_{\rm nde}/\rho_{\rm nde}$ (where nde
denotes neutrino dark energy) and neutrino mass is predicted to be
\cite{Fardon:2003eh}:
\begin{equation}
  w=-1+ {m_\nu n_\nu \over V}\,.
\end{equation}
Further, since neutrino masses are predicted to scale with redshift
approximately as $a^{3}$ in the non-relativistic regime, cosmological
and terrestrial probes of neutrino mass could give conflicting
results.  

It has been argued that it is natural to expect couplings of the
acceleron to quarks and charged leptons to be generated radiatively
\cite{Kaplan:2004dq}.  
Moreover, Yukawa couplings of the acceleron to visible matter could be
low energy manifestations of non-renormalisable operators arising from
quantum gravity.
If the acceleron couples both to neutrinos and matter, it may be
possible to investigate this scenario through neutrino oscillations
\cite{Kaplan:2004dq,Zurek:2004vd}.
However, the coupling to matter is model-dependent.
The effective neutrino mass in matter is altered by the interactions
of the scalar which in turn modifies neutrino oscillations.

At low redshifts, the contribution to neutrino mass caused by the
interactions of the acceleron with electrons and neutrinos is of the
form \cite{Barger:2005mn}:
\begin{equation}
  M ={\lambda_{\nu}\over m^2_{\phi}}
  (\lambda_e n_e+ \lambda_{\nu} (n_{\nu}^{C\nu B} + 
  {\frac{m_{\nu}}{E_{\nu}}} n_{\nu}^{rel}))\,,
  \label{eq:m} 
\end{equation}
where $\lambda_{\nu}$ ($\lambda_e$) is the Yukawa coupling of the acceleron
to the neutrino (the electron).
In principle, $\phi$ has a mass, $m_{\phi}$, that depends on $n_e$ and
the $n_{\nu}$. 
This dependence is weak since the underlying assumption that has been
made in obtaining equation (\ref{eq:m}) is that $\phi$ evolves
adiabatically and remains at the minimum of its potential.
The number density of the cosmic neutrino
background in one generation of neutrinos and anti-neutrinos is 
$n_{\nu}^{C\nu B} \sim 112$ cm$^{-3} \sim 10^{-12}$ eV$^3$, 
the number density of relativistic neutrinos in the background frame is
$n_{\nu}^{rel}$, and the electron number density is $n_e$. 
We emphasise that  $m_{\nu}$ is the neutrino mass in a
background-dominated environment.  

In terrestrial environments, and even for applications to solar
neutrinos, the dominant contribution to the mass shift arises from the
$\lambda_e n_e$ term.  
In this case, one can  adopt a matter dependence of the
form \cite{Barger:2005mn}:
\begin{equation}
  M(n_e) = M^0 \left(n_e\over n_e^0\right)^k \,,
  \label{eq:M}
\end{equation}
where $M^0$ is the value at some reference density, $n_e^0$, and $k$ 
parametrises a power-law dependence of the neutrino mass on density. 
In principle, $M$ is expected to depend linearly on $n_e$, but,
phenomenologically, one may allow $k$ to deviate from unity.
The choice of reference density is arbitrary. 
If the environment that neutrinos traverse has a constant density (e.g
for passage through the earth's crust), then that density could be
taken to be the reference density. 
If neutrino propagation is adiabatic (as in the sun), the reference
density could be taken to be the density at which the neutrinos are
produced.
Implicit in the form of equation (\ref{eq:M}) is the assumption that the
neutrino number density has a negligible effect on neutrino
masses. 
Thus, it applies only in the current epoch when the cosmic neutrino
background number density (${\cal{O}}(10^{-12})$ eV$^3$) is tiny. 
At earlier epochs, the neutrino number density is orders of magnitude
larger and must be taken into account. For example, in the era of Big
Bang Nucleosynthesis (BBN), the neutrino number density is
${\cal{O}}(10^{30})$ eV$^3$. 
For the compatibility of mass-varying neutrinos (MaVaNs) with BBN see
\cite{Weiner:2005ac}. 

A simplifying assumption is that the heaviest neutrino has a mass of
${\cal{O}}(0.05)$~eV in the present epoch. 
As a result of their non-negligible velocities, the neutrino
`over-density' in the Milky Way from gravitational clustering can be
neglected \cite{Ringwald:2004np}.  
Then $m_\nu$ represents the masses of terrestrial neutrinos in
laboratory experiments such as those measuring tritium beta decay. 
Note that cosmological bounds on the sum of neutrino masses of
${\cal{O}}(1)$~eV are inapplicable to MaVaNs. 
Consequently, the usual relationship between neutrino dark matter and 
neutrinoless double beta decay is also rendered
inapplicable \cite{Barger:2002xm}. 
Moreover, it was pointed out that if the acceleron couples to highly
non-relativistic neutrino eigenstates, neutrino dark energy is unstable
\cite{Afshordi:2005ym,Takahashi:2006jt}. 
The assumption that the background neutrino masses are small
circumvents this instability problem.

For such light neutrinos, only model-dependent (neutrino oscillation)
tests of the MaVaN scenario are viable because the model-independent
(cosmological) tests become inoperable.  
There are two reasons for this: 
\begin{itemize}
  \item The effects of dark energy and a cosmological constant are
        almost the same in today's Universe; and
  \item If the light neutrinos do not cluster sufficiently, the local
        neutrino mass is the same as the background value, below the
        sensitivity of tritium beta-decay experiments.
        In this case, the high-redshift cosmological data (which
        should show no evidence for neutrino mass) and the data from 
        tritium beta-decay experiments will be consistent.
\end{itemize}

\begin{figure}
  \centering
  \leavevmode
  \mbox{
    \includegraphics[width=5in]{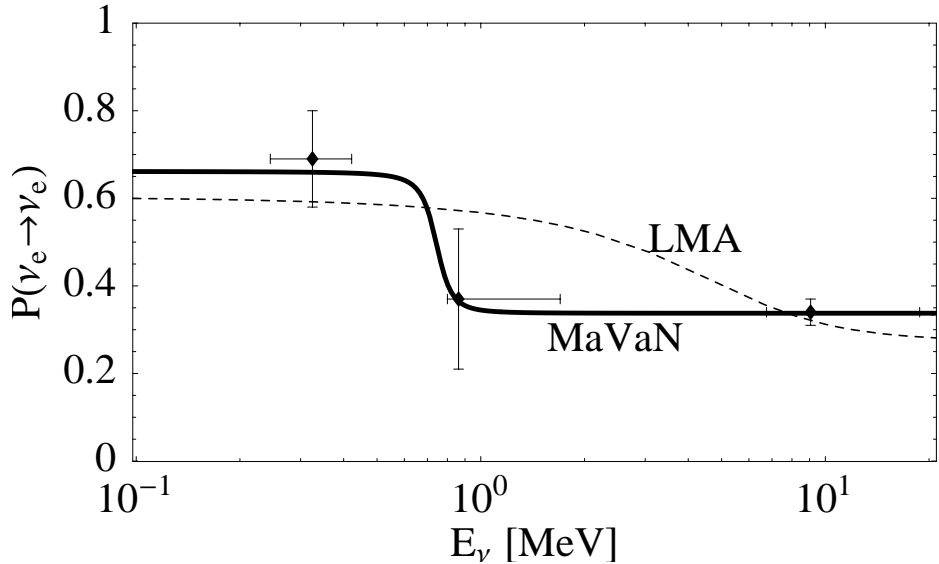}
  }
  \caption{
    $P(\nu_e \to \nu_e)$ vs. $E_\nu$ for MaVaN oscillations (solid
    curve).
    The dashed curve corresponds to conventional oscillations with the
    best-fit solution to KamLAND data.
    The data points and the procedure to extract them can be found in
    \cite{Barger:2005si,Barger:2001pf}. 
    Taken with kind permission of Physical Review Letters from figure 2 in
    reference \cite{Barger:2005mn}. 
    Copyrighted by the American Physical Society.
    Taken from reference \cite{Barger:2005mn}. 
  }
  \label{fig:Pee}
\end{figure}

It has been shown that oscillations of mass-varying neutrinos (that
result in exotic matter effects of the same size as standard matter
effects) lead to an improved agreement (relative to conventional
oscillations) with solar-neutrino data while remaining compatible with
KamLAND, CHOOZ, K2K, and atmospheric data \cite{Barger:2005mn}.
MaVaN oscillations are perfectly compatible with solar data because
the survival probability can change from a higher-than-vacuum value
(at low energies) to $\sin^2 \theta$ (at high energies) over a very
narrow range of energies as shown in figure \ref{fig:Pee}. 
An analysis of solar and KamLAND data concludes that the fit in the
LMA-II region is improved; while the region is excluded at more than 
the 4$\sigma$ C.L. in the standard oscillation analysis, it is allowed
at the 98.9\% C.L. for MaVaN oscillations
\cite{Gonzalez-Garcia:2005xu}.

Whether or not an explanation of solar-neutrino data requires MaVaN
oscillations will be answered by experiments that will measure 
the survival probability of MeV and lower energy neutrinos.
As shown in reference \cite{Barger:2005mh}, other tests in reactor and
long-baseline experiments emerge when the above scheme is embedded in
a comprehensive model that can explain all the available 
neutrino-oscillation data, including the LSND anomaly, and a null
MiniBooNE result.
This model requires large values of $\sin^22\theta_{13}$ in the range
$0.10 \lsim \sin^22\theta_{13} \lsim 0.30$.
Such values are not inconsistent with the CHOOZ reactor constraint on
$\bar\nu_e \to \bar\nu_e$ oscillations at the atmospheric scale
($L/E_\nu \simeq 250$~m/MeV) since the neutrino path in CHOOZ was
primarily in air. 
The relevant limit is that provided by Palo Verde for which the
neutrino path was through the ground; the Palo Verde limit is
significantly weaker than that provided by CHOOZ.
Such large values of $\sin^22\theta_{13}$, are likely to be measured
in experiments such as Angra, Daya Bay, or KASKA for which most of the
neutrino path is underground
\cite{Anderson:2004pk,Suekane:2003nh}.
These experiments will be sensitive to $\theta_{13}$ for 
$\sin^22\theta_{13} \ge 0.01$.
However, Double-CHOOZ \cite{Ardellier:2004ui,Berridge:2004gq}, which
should be sensitive to $\sin^22\theta_{13} \ge 0.03$, would see a null 
result since most of the neutrino path is in air.  
The MINOS experiment which is sensitive to 
$\sin^22\theta_{13} \ge 0.05$ at the 90\%~C.L. \cite{Barger:2001yx}, 
should also see a positive signal in the $\nu_\mu \to \nu_e$
appearance channel.

The idea of using reactor experiments with different fractions of air
and earth matter along the neutrino path to study MaVaN oscillations
has been further explored in \cite{Schwetz:2005fy}. 
For $\sin^22\theta_{13} \gsim 0.04$, two reactor experiments with
baselines of at least 1.5~km, one of which passes predominantly
through air, the other through the earth, can constrain an oscillation
effect which is different in air and matter at the level of a few
percent.
Neutrino super-beam experiments may probe mass-varying neutrinos in a  
controlled environment if the effects are large enough. 
It is worth investigating the sensitivity of long-baseline experiments
to non-standard matter effects in MaVaN  oscillations. 
A preliminary analysis can be found in reference \cite{Gu:2005pq}.

%% file: 04-BSnuM/04-03-CPTLIV/04-03-CPTLIV.tex
\subsection{CPT and Lorentz invariance violation}

CPT, the product of charge conjugation, parity, and time reversal is
one of the most fundamental symmetries in nature.
CPT invariance has a number of profound implications; for example, it
guarantees that the mass of particle and anti-particle are equal. 
Though there is no experimental evidence of CPT-invariance violation
(CPTV), the presence of a small violation is compatible with all
current data.  
On the other hand, CPTV and/or Lorentz-invariance violation (LIV) can
arise in string theories
\cite{Kostelecky:1988zi,Colladay:1998fq,Colladay:1996iz,Coleman:1998ti}. 
CPTV is closely related to LIV and it has been shown that CPTV
necessarily implies LIV \cite{Greenberg:2002uu}.  

The phenomenological consequences of CPTV and LIV in neutrino
oscillations have been widely discussed 
\cite{Barger:2000iv,Murayama:2000hm,Barenboim:2001ac,Barenboim:2002ah,Strumia:2002fw,Gonzalez-Garcia:2003jq,Hooper:2005jp,Bahcall:2002ia,Bilenky:2001ka,Kostelecky:2003cr,Kostelecky:2003xn,Datta:2003dg,Barger:2003xm,Murayama:2003zw,deGouvea:2004va,Coleman:1997xq,Foot:1998vr,Kostelecky:2003xn,Kostelecky:2004hg,Battistoni:2005gy,Hooper:2005jp,Datta:2003dg,Kostelecky:2003cr}.
In some cases, the phenomenological implications of the violation of
the Equivalence Principle (EPV) in neutrino oscillations 
\cite{Gasperini:1988zf,Halprin:1991gs,Pantaleone:1992ha,Bahcall:1994zw,Gago:1999hi} 
is identical to that of LIV and therefore the two can be treated in a
similar fashion \cite{Glashow:1997gx}.

One of the possible implications of CPTV is that the masses and/or the
mixings of neutrinos can be different from those of anti-neutrinos.
In this case, oscillation probabilities for neutrinos would differ
from those for anti-neutrinos even in the absence of CP violation or
matter effects.
CPTV has been suggested to reconcile LSND results with solar- as well as
atmospheric-neutrino observations in the framework of three neutrino
flavours \cite{Murayama:2000hm,Barenboim:2001ac,Strumia:2002fw}.
Atmospheric neutrinos probe oscillations for both neutrino and
anti-neutrino channels (though the contribution from neutrinos is
dominant) whereas solar neutrinos probe only the neutrino channel.  
As a result, once CPTV is allowed solar and atmospheric oscillations
can be described in terms of two mass squared differences $\Delta
m^2_{\odot}$ and $\Delta m^2_{\text{atm}}$ whereas atmospheric and
LSND anti-neutrino oscillations were driven by 
$\Delta m^2_{\text{atm}}$ and $\Delta m^2_{\text{LSND}}$.  
However, in view of the KamLAND results
\cite{Eguchi:2002dm,Araki:2004mb}, this scenario is strongly
dis-favoured, since KamLAND data are compatible with anti-neutrino
oscillations also characterised by $\Delta m^2_{\odot}$
\cite{Gonzalez-Garcia:2003jq,Schwetz:2003pv}.  

In the presence of LIV, the maximum velocity which a particle can
attain may differ from particle to particle.
For neutrinos the implication could be that velocity is dependent on
flavour.
Mixing between flavour  and velocity eigenstates will then lead to
neutrino oscillations even if neutrinos are massless
\cite{Coleman:1997xq}. 
The possibility of resonant conversions in the massless-neutrino
limit was first noted in the context of theories where neutrinos 
have non-standard interactions \cite{Valle:1987gv}. 
In such cases both the physics and the signatures are different from
those expected when Lorentz invariance is violated
\cite{Nunokawa:1996tg}.

\subsubsection{Direct bounds on CPTV} 

Without assuming any underlying model, CPTV can be constrained by
comparing the mixing parameters for neutrinos and anti-neutrinos
\cite{Minakata:2004jt}.
The current bound on the difference between $\sin^2{\theta_{12}}$ for 
neutrino and $\sin^2{\bar{\theta}_{12}}$ for antineutrinos is rather
weak \cite{Eguchi:2002dm,Araki:2004mb}.  
Even if it is assumed that $\bar{\theta}_{12}$ is in the first
octant, 
$|\sin^2{\theta_{12}} - \sin^2{\bar{\theta}_{12}}| \lsim 0.3$
at the 99.73\% C.L. 
For mass-squared differences, the current bound \cite{deGouvea:2004va}
is 
$|\Delta m^2_{21} - \bar{\Delta m^2_{21}} |\ \leq\  1.1 \times 10^{-4} \text{eV}^2$
at 99.73\% C.L., where $\bar{\Delta m^2_{21}}$ is the mass squared
difference of antineutrinos.
For comparison, the bound on the difference between the neutrino and
antineutrino $\theta_{23}$ mixing angle is
$-0.41 \ \leq \ \sin^2{\theta_{23}} - \sin^2{\bar{\theta}_{23}}\ \leq \ 0.45$
\cite{Gonzalez-Garcia:2004wg} at the 99.73\% C.L. level. 
For $\Delta m^2_{32}$ there are two results: one from the
Super-Kamiokande collaboration
$-1.9 \times 10^{-2} \mbox{ eV}^2 \ \leq \ |\Delta m^2_{32}| -
|\bar{\Delta m^2_{32}}| \  \leq \ 4.8 \times 10^{-3} \mbox{ eV}^2$
\cite{Saji:2004mw}; and the other from Gonzalez-Garcia et al.
$-10^{-2}  \mbox{ eV}^2\ \leq \ \Delta m^2_{32} - \bar{\Delta
m^2_{32}}\  \leq\ 3.4 \times 10^{-3} \mbox{ eV}^2$
\cite{Gonzalez-Garcia:2004wg};
both at the 99.73\% C.L. level. 

\subsubsection{CPTV/LIV Effect on conversion probability} 

Oscillations between two flavours, for example, between $\nu_\mu$
and $\nu_\tau$ in the presence of CPTV or LIV, can be described by the
Hamiltonian \cite{Coleman:1998ti,Gonzalez-Garcia:2005xw}:
\begin{equation} 
    {\rm H} \equiv
    \dfrac{\Delta m^2}{4 E}
    \mathbf{U}_\theta
    \begin{pmatrix}
	-1 & ~0 \\
	\hphantom{-}0 & ~1
    \end{pmatrix}
    \mathbf{U}_\theta^\dagger
    + \dfrac{\eta\, E^n}{2}
    \mathbf{U}_{\xi,\varphi}
    \begin{pmatrix}
	-1 & ~0 \\
	\hphantom{-}0 & ~1
    \end{pmatrix}
    \mathbf{U}_{\xi,\varphi}^\dagger \;,
\end{equation}
where $\Delta m^2$ is the mass-squared difference between the two
neutrino mass eigenstates, $\eta$ parametrises the size of the CPTV or
LIV effect.  
Here, $n$ is an integer where $n=0$ corresponds to CPTV and LIV, $n=1$
to LIV or EPV.   
The matrices $\mathbf{U}_\theta$ and $\mathbf{U}_{\xi,\varphi}$ are
given by: 
\begin{equation} \label{eq:rotat}
    \mathbf{U}_\theta =
    \begin{pmatrix}
	\hphantom{-}\cos\theta & ~\sin\theta \\
	-\sin\theta & ~\cos\theta
    \end{pmatrix}\,,
    \qquad
    \mathbf{U}_{\xi,\varphi} =
    \begin{pmatrix}
	\hphantom{-}\cos\xi \hphantom{e^{-i\varphi}} 
	& ~\sin\xi e^{\pm i\varphi} 
	\\
	-\sin\xi e^{\mp i\varphi} 
	& ~\cos\xi\hphantom{e^{-i\varphi}}
    \end{pmatrix}\,, 
\end{equation}
where $\varphi$ is the non-vanishing relative phase.  
Note that $n=0$ may also corresponds to the non-standard interaction
case described in \cite{Valle:1987gv,Nunokawa:1996tg}. 

If the CPTV or LIV strength is constant along the neutrino trajectory
the survival probability takes the form
\cite{Coleman:1998ti,Gonzalez-Garcia:2005xw}: 
\begin{equation} \label{eq:prob}
    P_{\nu_\mu \to \nu_\mu} = 1 - P_{\nu_\mu \to \nu_\tau} =
    1 - \sin^2 2\Theta \, \sin^2 \left( 
    \frac{\Delta m^2 L}{4E} \, \mathcal{R} \right) \,,
\end{equation}
with
\begin{align}
    \label{eq:Theta}
    \sin^2 2\Theta &= \frac{1}{\mathcal{R}^2} \left(
    \sin^2 2\theta + R_n^2 \sin^2 2\xi
    + 2 R_n \sin 2\theta \sin 2\xi \cos\varphi \right) \,,
    \\[2mm]
    \label{eq:Xi}
    \mathcal{R} &= \sqrt{1 + R_n^2 + 2 R_n \left( \cos 2\theta \cos 2\xi
      + \sin 2\theta \sin 2\xi \cos\varphi \right)}\; , \\
    R_n &= \frac{\eta E^n}{2} \, \frac{4E}{\Delta m^2} \,,
\end{align}
where, for simplicity, CPTV or LIV scenarios which can be
characterised by a unique parameter $\eta$ have been assumed.

The $n=0$ case can lead to CPTV and LIV by identifying 
$\eta = b_1-b_2$ where the $b_i$ are the eigenvalues of the
Lorentz-violating CPT-odd operator
\cite{Colladay:1996iz,Coleman:1998ti,Barger:2000iv}. 
The $n=1$ case can lead to LIV by identifying $\eta = c_1-c_2$ where
the $c_i$ are the maximal attainable velocities of $\nu_i$
\cite{Coleman:1997xq}.  
This case is phenomenologically equivalent to EPV
\cite{Glashow:1997gx} for the constant gravitational potential through
the identification $\eta = 2|\phi| (\gamma_1-\gamma_2)$ where $\phi$
is the gravitational potential, assumed to be constant, and $\gamma_i$
is the coupling of neutrinos to gravity
\cite{Gasperini:1988zf,Halprin:1991gs}. 

Atmospheric neutrino data can be used to constrain the possible CPTV,
LIV, or EPV effects.
For example, the following limits were derived in
\cite{Gonzalez-Garcia:2005xw,Gonzalez-Garcia:2004wg}:
\begin{eqnarray}
&&  \eta = b_1-b_2 \equiv \delta b\, \leq\, 5.0\times 10^{-23}~\text{GeV}\, \,,
\qquad ~~~~~~~
{\rm for\,\,\, CPTV\,,\, LIV}\ (n=0) \label{eq:cpt}     \\    
&& \eta = (c_1- c_2)\equiv\delta c/c 
\,\leq\, 1.6\times 10^{-24}\,, \qquad \qquad ~~~
{\rm for\, VLI}\ (n=1)\label{eq:vli} \\    
&&    \eta = 2 |\phi|(\gamma_1- \gamma_2) \equiv 2 |\phi| \Delta\gamma 
\,\leq\, 1.6\times 10^{-24}\,, \qquad
{\rm for\, EPV}\ (n=1)  \label{eq:vep} 
\end{eqnarray}

\subsubsection{Future Prospects} 

The bounds on the CPTV differences between neutrino and anti-neutrino
mixing parameters at a future Neutrino Factory have been studied for a
muon-beam energy of 50~GeV and a baseline of 7000~km
\cite{Bilenky:2001ka}.
Assuming a 10~kT detector and $10^{20}$ muon decays per year leads to
the following bounds:
\begin{eqnarray}
  \frac{ | | \Delta m^2_{32}| - |\bar{\Delta m^2_{32}} | | }
  {\frac{1}{2} (| \Delta m^2_{32}| + |\bar{\Delta m^2_{32}}|) }
  \ & \lsim\ &  8 \times 10^{-3} \; {\rm ; and}   \\
  \frac{ | \theta_{23} - \bar{\theta}_{23} | } 
  {\frac{1}{2}(\theta_{23} + \bar{\theta}_{23})}\  & \lsim\ &  4.2 \times
  10^{-2} \; .
\end{eqnarray}

The bound on the CPTV effect is determined by the parameter $\delta b$
through oscillation. 
For oscillation experiments with baselines shorter than $\sim 1000$~km
and energy $\sim 1-2$ GeV, such as T2K \cite{Itow:2001ee} and
NOvA \cite{Ayres:2004js}, the existing bounds described in the
previous section will not be improved.
On the other hand, future neutrino-oscillation experiments at the
Neutrino Factory with longer baseline, $L \gsim 3000$ km, and higher
energy, $\langle E \rangle > 10 $ GeV, are expected to improve the
present bounds.
Reference \cite{Barger:2000iv} estimates that the sensitivity on
$\delta b$ can be as small as $\sim 10^{-23}$~GeV for $L=2900$ km 
for a 10 kt detector and $10^{19}$ muon decays.

%% file: 04-BSnuM/04-04-Unitarity/04-04-Unitarity.tex
\subsection{Leptonic unitarity triangle and CP-violation}
\label{Subsubsect:triangleCP}

In the quark sector, the unitarity triangle has proved to be a very
useful representation of mixing and CP-violation. Similarly, in the
lepton sector, the unitarity triangle provides a convenient framework
for a variety of analyses including: analysing the experimental
results on lepton mixing; testing the unitarity of the mixing matrix
and searching for new neutrino states; establishing the violation of
the CP invariance and measuring the Dirac CP-violating phase $\delta$;
and searching for effects of new interactions of neutrinos.

In the following it will be assumed that there are only three
neutrinos so that mixing may be described using a $3 \times 3$
unitarity matrix. In the standard parametrisation, $U = U_{23}
I_{{\delta}} U_{13} I_{{\delta}} ^\dagger U_{12}$, where $U_{ij}$
are rotation matrices in the $ij$-plane, and $I_{\delta} \equiv
\diag (1, 1, e^{i\delta})$. The unitarity condition $U U^{\dagger} =
I$ leads to three `row equalities', $U_{\alpha i} U_{\beta i}^* =
0$, $\alpha \neq \beta$, or explicitly:
\begin{eqnarray}
  \begin{matrix}
    U_{e1}U_{\mu 1}^*
    +U_{e2} U_{\mu 2}^*+U_{e3}U_{\mu 3}^*=0, &\ \ \ \ & (a)   \cr
    U_{e1}U_{\tau 1}^* +U_{e2} U_{\tau 2}^*+U_{e3}U_{\tau 3}^*=0, &\ \
    \ \ & (b) \cr U_{\tau 1}U_{\mu 1}^* +U_{\tau 2} U_{\mu 2}^*+U_{\tau
    3}U_{\mu 3}^*=0. &\ \ \ \ & (c)
  \end{matrix}
  \label{unit}
\end{eqnarray}
In the complex plane, each term from the sums in equation (\ref{unit})
represents a vector. Equations (\ref{unit}) imply that the three terms
appearing in each equation form  a triangle, known as a `unitarity
triangle'. The expressions (\ref{unit}) also reflect the orthogonality
of the flavour states; the corresponding triangles are called the
flavour triangles, for example, equation (\ref{unit})a describes the
$e\mu$-triangle. The sides of triangle can be then enumerated by the
mass eigenstates:
\begin{equation}
  z_i \equiv U_{\alpha i}U_{\beta i}^*.
\end{equation}
The unitarity condition $U^{\dagger} U = I$ leads to the `column
equalities' $\sum_\alpha U_{\alpha i} U_{\alpha j}^* = 0$, (for $i
\neq j$). These equations also define triangles known as the mass
state triangles.

The shape and area of the  triangles are closely related to
CP-violation in leptonic mixing. Indeed, the Dirac CP-violating
phase vanishes if and only if the phases of all elements of the mixing 
matrix are factorisable: $U_{\alpha i}=e^{i(\sigma_\alpha
+\gamma_i)}|U_{\alpha i}|$. In this case, $U_{\alpha i}U_{\beta i}^*
=
 e^{i( \sigma_\alpha-\sigma_\beta)}|U_{\alpha i}||U_{\beta i}|$ and
therefore the unitarity triangles shrink to segments.
Recall that the CP-violating effects are determined by the Jarlskog
invariant:
\begin{equation}
  J_{CP} = {\rm Im} [U_{\alpha i} U_{\beta i}^* U_{\alpha j}^*
  U_{\beta j}] \; , \label{jarlskog}
\end{equation}
which in the standard parametrisation is given by:
\begin{equation}
  J_{CP} = s_{12}c_{12}s_{23}c_{23}s_{13}c_{13}^2\sin\delta,
\end{equation}
where $s_{12} \equiv \sin \theta_{12}$, etc. In particular, the
invariant determines the CP-asymmetries in neutrino
oscillations, $P(\bar{\nu}_\alpha \to \bar{\nu}_\beta)-P(\nu_\alpha
 \to \nu_\beta) \propto J_{CP}$.
The area of the triangle, $S$, is related to the Jarlskog invariant.
For the flavour triangle:
\be
  S = \frac{1}{2} |U_{\alpha i} U_{\beta i}^*|
      | U_{\alpha j} U_{\beta j}^*| \sin \phi_{ij} \; ,
  \label{area}
\ee where $\phi_{ij}$ is the angle between the sides $i$ and $j$.
Equations (\ref{jarlskog}) and (\ref{area}) can be combined to give: \be
  S = \frac{1}{2} J_{CP}.
  \label{relation1}
\ee The relation \ref{relation1} is the basis of the unitarity
triangle method for measuring the CP-violating phase
\cite{Aguilar-Saavedra:2000vr,Farzan:2002ct}. Reconstructing the
unitarity triangle is an alternative to the direct measurement of
the CP-asymmetries in transition probabilities $P(\bar{\nu}_\alpha
\to \bar{\nu}_\beta)-P(\nu_\alpha \to \nu_\beta)$
\cite{Freund:2000ti,Barger:2000ax,Kuo:1987km,Fritzsch:1999ee,Sato:2000wv,Dick:1999ed}.

\subsubsection{Properties of the leptonic triangles}

For very small 13 mixing, $\sin \theta_{13} \ll 0.15$, the unitarity
triangles are of two forms:
\begin{itemize}
  \item Triangles that include the element $U_{e3}$ and therefore have one small
        side and two nearly equal sides: e.g., the $e\mu$-triangle in which $|z_1| \approx |z_2|$ and
        $|z_3| \approx s_{13}/\sqrt{2}$; and
  \item Triangles that do not include the $U_{e3}$ element: e.g.,
        the $\mu\tau$-triangle for which $|z_1| \approx 1/6 + O(s_{13})$,
        $|z_2| \approx 1/3$ and $|z_3| \approx -1/2 +O(s_{13})$.
\end{itemize}
For $s_{13}$ saturating its upper bound, $s_{13}\sim 0.15$, the
sides of the triangle can be of the same size.

\begin{figure}
  \centering
  \includegraphics[bb=56 49 552 487,clip=true,width=0.6\linewidth]{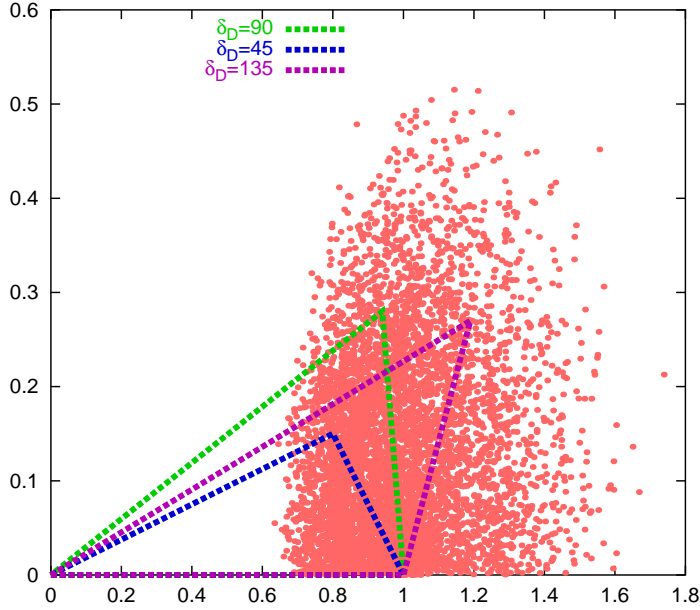}
  \caption{The
    unitarity $e\mu$-triangles. The horizontal side, $|U_{e1}U_{\mu 1}^*|$  is
  normalised to one. The triangles correspond to $s_{13}=0.15$ and
  different values of $\delta$. Each scatter point represents a
  possible position of vertex as the  mixing parameters
  pick up random values within the present uncertainty ranges: $\sin^2 \theta_{23}\in
[0.36,0.61]$, $\sin^2 \theta_{12} \in [0.27,0.37]$ and
$\sin^2\theta_{13}\in [0,0.031]$ and $\delta\in [0,2\pi]$ }
  \label{trianglenow}
\end{figure}

Figure \ref{trianglenow} shows  examples of the $e\mu$-triangle
constructed for $s_{12}=0.56$, $s_{23}=0.67$ (the best-fit values),
$s_{13}=0.15$, and for three different values of the CP-violating
phase. The horizontal side is normalised to 1. Each scatter point
represents the possible position of vertex if the values of mixing
parameters pick up different values within the present uncertainty
ranges: $\sin^2 \theta_{23}\in [0.36,0.61]$, $\sin^2 \theta_{12} \in
[0.23,0.37]$ and $\sin^2\theta_{13}\in [0,0.031]$ and $\delta$
varies between 0 and $2\pi$. Some  scatter points lie on the
horizontal axis. This reflects the fact that with the present data
it is not possible to establish CP-violation. Notice that
despite the fact that $s_{13}$ is relatively small, for a
considerable portion of the scatter points, the sizes of the all three
sides of the triangle are comparable. 
The triangles can take a particular
form if the mass matrix, and consequently the mixing matrix, have a
certain symmetry. The $\mu-\tau$ reflection symmetry defined as
$\nu_{\mu} \to \nu_{\tau}^c$, $\nu_{\tau} \to \nu_{\mu}^c$, where
the superscript $c$ denotes the C-conjugation
\cite{Harrison:2002et,Grimus:2003yn}, leads to isosceles
column-based triangles and congruent row-based $e \tau$- and $e
\mu$-triangles.

\subsubsection{Leptonic triangles and coherence of neutrino states
\label{section-coherence}} The charged-current (CC) coupling of
neutrino mass eigenstate, $\nu_i$, and the charged lepton, $\alpha$,
is given by $U_{\alpha i}$. As a result, by studying the CC
interactions of a neutrino beam of pure mass eigenstates, we can
derive the moduli of the mixing-matrix elements which give the sides
of the unitarity triangle. Unfortunately, terrestrial neutrino
beams are composed of flavour, rather than mass, eigenstates which
are coherent combinations of the mass eigenstates.  To create beams
of neutrino-mass eigenstates it is necessary to destroy the
coherence of the neutrino-flavour state produced through the weak
interaction. There are several circumstances in which the coherence
can be destroyed \cite{Farzan:2002ct}:
\begin{itemize}
  \item {\it Adiabatic conversion of the flavour neutrino state:}
        Suppose the neutrino flavour state, $\nu_{\alpha}$, is
        produced at densities much higher than the MSW-resonance
        density.
        Then, at the production point, the mixing in matter is
        strongly suppressed and $\nu_{\alpha}$  practically
        coincides with one of the energy (or effective-mass)
        eigenstates in matter: $\nu_{\alpha} \approx \nu_{im}$.
        Suppose this state propagates to a region of small (zero)
        density.
        If the propagation is adiabatic, then $\nu_{im} \to \nu_i$
        and at the exit from the matter layer, the beam will be pure
        $\nu_i$.
        Such a situation is approximately realised for the
        high-energy (with $E_\nu>10$~MeV) solar neutrinos; {\it i.e.,} $\nu_e$ produced in the center of the Sun
        is transformed into
        $\nu_2$ at the surface;
  \item {\it Neutrino decay:}
        If the heavier neutrinos on their way to the detectors decay into
        the lightest neutrino (plus another light or massless particle), regardless of the original flavour composition,
        we will
        obtain a
        flux which is purely composed of the lightest
        mass eigenstate; and
  \item {\it Decoherence:}
        One can also use a beam of several mass eigenstates provided
        that they are {\it incoherent}.
        The rates of processes induced by such a beam will be
        determined by the moduli of matrix elements.
        The effective loss of coherence can occur due to divergence of
        the neutrino wave-packets over long distances or the averaging
        of oscillations.
\end{itemize}
The decay and the loss of coherence both require astronomical
distances; moreover, the adiabatic conversion cannot be realised on
distances smaller than the solar radius. Obtaining pure neutrino
mass eigenstates therefore requires astrophysical sources of
neutrinos and astrophysical methods. In section \ref{icecube}, we
will discuss such methods.

To reconstruct the  unitarity triangle, the absolute values of the
elements of two rows (or equivalently two columns) in the mixing
matrix must be measured. To reconstruct the $e\mu$-triangle, three
quantities should be determined independently:
\begin{equation}
  |U_{e1} U_{\mu 1}^*|, ~~~~~ |U_{e2} U_{\mu 2}^*|,  ~~~~~|U_{e3}
  U_{\mu 3}^*|.
  \label{sides}
\end{equation}
The form of the triangle depends on the, as yet unknown, value of
$|U_{e3}|$. Assuming that only three neutrino species take part in
the mixing and that there is no other source of CP-violation apart
from the phases of the  mass-matrix elements, one can use the two
independent normalisation conditions:
\begin{equation}
  \sum_{i = 1,2,3} |U_{ei}|^2 = 1 ~, ~~~~ \sum_{i = 1,2,3}
  |U_{\mu i}|^2 = 1 \; ,
  \label{normalisation}
\end{equation}
to determine the length of the sides of the $e\mu$-triangle. Thus,
to find the sides of the $e\mu$-triangle it is enough to measure the
moduli of the four mixing matrix elements:
\begin{equation}
  |U_{e 1}|, ~~~~ |U_{\mu 1}|, ~~~~ |U_{e3}|, ~~~~ |U_{\mu 3}|.
  \label{elements}
\end{equation}

To prove the CP violation, the following inequalities
must be established:
\begin{eqnarray}
  \label{ineq-tri}
  |U_{e1} U_{\mu 1}^*| &<&  |U_{e2} U_{\mu 2}^*| + |U_{e3} U_{\mu 3}^*
  |;\cr |U_{e2} U_{\mu 2}^*| &<&  |U_{e1} U_{\mu 1}^*| +
  |U_{e3} U_{\mu 3}^*|.
\end{eqnarray}
Using the present information on the absolute value of the matrix
elements one can estimate the accuracy required.
According to equation (\ref{ineq-tri}) the quantities:
\begin{equation}
  A_1 \equiv - |U_{e1}||U_{\mu 1}| + |U_{e3}||U_{\mu3}| +\sqrt{(1 -
  |U_{e1}|^2 - |U_{e3}|^2)(1 - |U_{\mu 1}|^2 - |U_{\mu 3}|^2)}
  \label{measure1}
\end{equation}
and:
\begin{equation}
  A_2 \equiv  |U_{e1}||U_{\mu 1}| + |U_{e3}||U_{\mu3}| -\sqrt{(1 -
  |U_{e1}|^2 - |U_{e3}|^2)(1 - |U_{\mu 1}|^2 - |U_{\mu 3}|^2)}
  \label{measure2}
\end{equation}
are measures of CP violation; i.e., CP is conserved if either
$A_1$ or $A_2$ is zero. Setting $\theta_{12}$ and $\theta_{23}$ to
their best fit values  ($\sin^2 \theta_{12}=0.315$ and $\sin^2
\theta_{23}=0.45$), $\theta_{13}$  close to the present upper
bound ($U_{e3}=0.15$), and taking a maximal Dirac phase,
$\delta = 90^{\circ}$, we find $A_1= 0.09$ and $A_2=0.10$.

Notice that $A_{1,2}$ can  be negative if there is an extra neutrino
and as a result the 3$\times$3 active sub-matrix is not unitary.
Equations (\ref{measure1}) and (\ref{measure2}) imply that, in order to
establish CP-violation (non-zero values of $A_{1,2}$) the  absolute
errors, $\Delta |U_{\mu 3}|$, $\Delta |U_{e1}|$, $\Delta|U_{\mu 1}|$
and $\Delta |U_{\mu 3}U_{e 3}|$ (regardless of the measurement
method) should be smaller than a few percent even in the most
optimistic case -- namely, $U_{e3}$ as large as possible and 
$\delta = 90^{\circ}$. 
This seems quite challenging specially in the case of $|U_{\mu 1}|$.

If $|U_{e3}|$ turns out to be very small, the sides proportional to
$|U_{e3}|$ will be also  tiny and  it will be more
difficult to reconstruct the $e \mu $-triangle and to check the
inequalities (\ref{ineq-tri}). A similar situation occurs for the
$uc$-triangle in the quark sector. In this connection, it was
proposed in \cite{Zhang:2004hf} to reconstruct the 12-triangle which
is made up of $|U_{e1}U_{e2}^*|$, $|U_{\mu 1} U_{\mu 2}^*|$, and
$|U_{\tau 1}U_{\tau 2}^*|$. Notice that although all the sides of
the 12-triangle are comparable, in the limit of small $s_{13}$, the
height of this triangle will be small. This creates another problem
for measuring the area. 
Within the tri-bimaximal scenario, it may be simpler to reconstruct
the 23-triangle (the triangle made up of $|U_{e3}U_{e2}^*|$, 
$|U_{\mu 3} U_{\mu 2}^*|$, and $|U_{\tau 3}U_{\tau 2}^*|$)
\cite{Bjorken:2005rm,Ahuja:2007cu}.
Up to now, there is no direct information
about the values of $|U_{\tau 1}|$ and $|U_{\tau 2}|$. Moreover,
both the creation of intense $\nu_{\tau}$-beams and the detection of
$\nu_{\tau}$ seem to be difficult. Reconstructing the 23- and
12-triangles  does not therefore seem very promising from a
practical point of view.  If we do not want to make any theoretical
pre-assumptions about the mass texture, the $e\mu$-triangle seems to
be a more promising option to reconstruct, especially if, for
$s_{13}$ close to the present upper bound, as demonstrated in figure
\ref{trianglenow}, all the sides of the $e \mu$-triangle are
comparable. Throughout the present analysis we therefore focus on
the $e\mu$-triangle.
\subsubsection{The unitarity triangle and oscillation experiments}
\label{terrestrial}

In this section, we describe  the set of oscillation measurements
that have been suggested in \cite{Farzan:2002ct} to determine the $e
\mu$-triangle. Since the oscillation probabilities depend not only
on the moduli of the mixing-matrix elements, but also on the unknown
relative phases ($\delta_x$), the strategy is to select
configurations of oscillation measurements for which the dominant
effect is determined by moduli:
\begin{equation}
  P_{\alpha \beta} = P_{\alpha \beta}(|U_{e i}|,|U_{\mu i}|) +
  \Delta P_{\alpha \beta} (\delta_x), ~~~~ \alpha, \beta = e, \mu~,
  \label{reduced}
\end{equation}
where $\Delta P \ll P$.
The hierarchy of mass splittings:
\begin{equation}
  \epsilon \equiv \frac{\Delta m_{12}^2}{\Delta m_{23}^2} \simeq 0.03
  \; ,
  \label{expansion}
\end{equation}
as well as the small $|U_{e 3}|$ play a key role in this argument.
The  experimental setup must be chosen in such a way that the
$\delta_x$-dependent correction in (\ref{reduced}) induced by the
matter effect is suppressed. The product $|U_{e 3}^*U_{\mu 3}|$,
which is one side of the triangle, can be measured in studies of the
$\nu_{\mu} \to \nu_e$ transitions driven by $\Delta m_{32}^2$. For
this channel, in vacuum:
\begin{equation}
  P_{\mu e} = 4 |U_{e 3}^*U_{\mu 3}|^2 \sin^2
  \frac{\Delta m_{32}^2 L}{4E} + \Delta P_{\mu e} \; ,
  \label{emuchan}
\end{equation}
where the correction $\Delta P_{\mu e}$ is due to non-zero $\Delta
m_{21}^2$. In general, due to the matter effect it is not  possible
to write the transition probability in the simple form of equation
(\ref{emuchan}) with $\Delta P_{\mu e}\ll P_{\mu e}$.  The probability
can however be reduced to the form (\ref{emuchan}) in two limiting
cases \cite{Farzan:2002ct}:
\begin{itemize}
  \item The low-energy limit $E \ll E_{23}^R$ ($E_{23}^R \sim 6$ GeV
        is the resonance energy corresponding to the $2-3$ splitting)
        for which the matter correction is small;
  \item The short-baseline limit where `vacuum mimicking' condition
        is satisfied \cite{Akhmedov:2000cs,Yasuda:2001va}.
\end{itemize}
Unfortunately, neither of the proposed state-of-the-art long-baseline
experiments, No$\nu$A and T2K fulfill these requirements.  As
illustrated in figure 5 of \cite{Huber:2006vr}, the transition
probability for these setups is sensitive not only to
$|U_{e3}||U_{\mu 3}|$ but also to the value of the CP-violating
phase. Thus, the $|U_{\mu 3}U_{e 3}^*|$ side should be determined by
separately by measuring the values of $|U_{\mu 3}|$ and $|U_{e 3}|$.

The four elements in equation (\ref{elements}) can be determined, in
principle, as follows:
\begin{enumerate}
  \item $|U_{e3}|$ can be measured by reactor experiments with a
        typical baseline of $\sim 1$~km.
        In these experiments the matter effect is negligible and the
        survival probability for $\bar{\nu}_e$ can be written:
        \begin{equation}
          P_{e e} = 1 - 4 (1 - |U_{e3}|^2)
          |U_{e3}|^2 \sin^2 \frac{\Delta m_{23}^2 L}{4E} +
          \Delta P_{e e}~.
          \label{probab-ee}
        \end{equation}
        The relative correction is small,
        $\Delta P_{e e}/(1-P_{e e}) < 2$ \%, so $|U_{e 3}|$ can be
        determined with {\cal O}(1)\% accuracy \cite{Farzan:2002ct}.
        Experimental errors in the measurement of $P_{ee}$ will
        dominate over $\Delta P_{e e}$.
        If $\theta_{13}$ is close to the present upper bound, the next
        generation of reactor experiments will be able to measure it
        with a relative error of {\cal O}($10)$\%
        \cite{He:2006ud,Huber:2006vr,Cao:2005mh}.
        The uncertainty in $A_{1,2}$ (defined in equations (\ref{measure1})
        and (\ref{measure2})) arising from $\Delta |U_{e3}|$
        can be evaluated as:
        \begin{equation}
          \Delta A_1=\left[|U_{\mu 3}|-|U_{\mu 2}|
                   \frac{|U_{e3}|}{|U_{e2}|}\right] \Delta |U_{e3}| \; \ {\rm and} \
                   \Delta A_2=\left[|U_{\mu 3}|+|U_{\mu 2}|
                   \frac{|U_{e3}|}{|U_{e2}|}\right] \Delta |U_{e3}|
                   \; .
        \end{equation}
        Taking $|U_{e 3}|\simeq 0.15$ and
        $\frac{\Delta |U_{e3}|}{|U_{e3}|}\simeq 10 \%$ yields
        $\Delta A_{1,2}/A_{1,2}\simeq 20\%$.
  \item The element $|U_{\mu 3}|$ can  be measured in long-baseline
        $\nu_{\mu}$-disappearance experiments (one of the main
        motivations of which is to measure the mixing angle
        $\theta_{23}$ with high precision).
        For T2K and No$\nu$A, the matter effect cannot be neglected
        since $\Delta m_{23}^2/E \sim \sqrt{2}  G_Fn_e$, however effects
        due to the solar mass splitting are unimportant.
        To an
        approximation of {\cal O}$(\Delta m_{12}^2L/2E)\sim 0.01$, the survival probability can be treated in the
        two-neutrino oscillation framework.
        The value of $\sin \theta_{23}$ (thus
        $|U_{\mu 3}|=s_{23}c_{13}$) can be extracted with accuracy of
        $4\%$ or better \cite{Ayres:2004js};
  \item The values of $|U_{e1}|$ and $|U_{e 2}|$ can be obtained from
        the solar-neutrino data.
        Since the energy of solar neutrinos is low, to a good
        approximation, the matter effect on $|U_{e3}|$ can be neglected.
        Moreover,
         the solar-neutrino conversion driven by $\Delta m_{31}^2$
         produces only an averaged oscillation effect ($1\ll \Delta m_{31}^2 L/ E_\nu$).
        In this case the survival probability reads
        \cite{Lim:1987yd,Smirnov:1992ss}:
        \be
          P_{e e} = (1 - |U_{e3}|^2)^2
          P_2 (\tan^2\theta_{12}, \Delta m_{21}^2) + |U_{e3}|^4 \; ,
          \label{prob-sun}
        \ee
        where:
        \be
          \tan^2\theta_{12} = \frac{|U_{e2}|^2}{|U_{e1}|^2} \; ,
          \label{tan-sun}
        \ee
        and $P_2$ is the two-neutrino survival probability determined
        from the solution of the two-neutrino-evolution equation with
        the oscillation parameters $\tan^2\theta_{12}$, $\Delta
        m_{21}^2$ and the effective potential $(1 - |U_{e3}|^2)V_e$.
        Analysis of solar-neutrino data alone cannot yield an uncertainty in
        $\sin^2 \theta_{12}$ better than 19\% at the 3$\sigma$ C.L.
        \cite{Bandyopadhyay:2004cp}, even if the $pp$-flux data
        with a 3\% uncertainty is included.

        For a reactor experiment with a large baseline ($\sim 100$~km)
    such as KamLAND, $P_{ee}$ is given by:
        \be
          P_2 = 1 - \frac{4|U_{e1}|^2|U_{e2}|^2}{(1 - |U_{e3}|^2)^2}
          \sin^2 \frac{\Phi_{12}}{2}~.
          \label{p2}
        \ee
        This opens up the possibility of measuring the value of
        $\sin \theta_{12}$ with high precision.
        Although the current experiment, KamLAND, cannot reach a
    precision better than 18\% \cite{Bandyopadhyay:2004cp}, if
   a reactor experiment with a baseline of 60 km,  an exposure of
        $\sim 60$~GWkTy and a systematic error of 2\% is constructed,
        $\sin^2 \theta_{12}$ can be measured with an uncertainty of
    6\% at 3$\sigma$.
        Moreover, as shown in \cite{Strumia:2001gi}, combining data
    from KamLAND and Borexino will allow the uncertainty on
        $\sin^2 2 \theta_{12}$ to be reduced to 5\%.
        Then, using the measured value of $|U_{e1}|,\ |U_{e2}|$
        and the normalisation condition,
        $|U_{e1}|^2 + |U_{e2}|^2 = 1 - |U_{e3}|^2$, $|U_{e1}|$ and
    $|U_{e2}|$ can be determined separately.
        In this way, $|U_{e1}|$ can be determined with relatively high
        precision leading to $(\Delta A_{1,2}) < A_{1,2}/5$ for
        $\delta\simeq 90^\circ$.
        Such an uncertainty is small enough to establish the CP-violation; and
  \item The determination of $|U_{\mu1}|$ (and/or $|U_{\mu2}|$) is the
        most challenging part of the method.
        Note that in contrast to $|U_{e3}^* U_{\mu 3}|$, it is not
    possible to measure the combinations $|U_{e1}^* U_{\mu 1}|$ and
    $|U_{e2}^* U_{\mu2}|$, directly from oscillation experiments.
        Indeed, in vacuum the $\nu_{\mu} \to \nu_e$ transition
    probability is determined by the product
        $Re\left[U_{\mu1}^* U_{e1} U_{\mu2} U_{e2}^*\right]$ which
        depends not only on the absolute values of the matrix elements
    but also on their phases.
        Therefore one has to resort to the possibility of separately
    measuring $|U_{\mu1}|$ and $|U_{\mu2}|$.
        In fact, it is sufficient to measure a combination of
    $|U_{\mu1}|$ and $|U_{\mu2}|$ which differs from the
    normalisation condition (equation (\ref{normalisation})).
        This requires an experiment sensitive to the $\Delta m_{12}^2$
    splitting which appears usually as a sub-dominant mode.
        To suppress the leading effect and the interference of the
    leading and sub-leading modes, the oscillations driven by
    $\Delta m_{23}^2$ should be averaged out.
        This condition necessitates the following experimental
    configuration:
        \begin{itemize}
          \item The energy of the beam  should be low: $E < 1$ GeV;
                and 
          \item The baseline should be large,
                $L\gg  4E_\nu/\Delta m_{32}^2$.
        \end{itemize}
        Moreover, to avoid suppression of the sub-dominant mode  the
    baseline, $L$,
    should be of the order of the  oscillation length associated with the
    (1 - 2) splitting, $L \geq 2000$~km.
        Realisation of such a set-up is very challenging.
        The requirements listed above are fulfilled by the
   proposed Brookhaven to Homestake long-baseline experiment
    \cite{Beavis:2002ye}.
\end{enumerate}

In summary, if $|U_{e3}|$ saturates the present bound ($|U_{e3}|\sim
0.1$), the results of the next generation of reactor experiments can
be combined with the measurements of the $\nu_\mu$-survival
probability in T2K and/or No$\nu$A to determine $|U_{e3}|$ and
$|U_{\mu 3}|$, and consequently the third side of the
$e\mu$-triangle, with the required precision. Moreover, by combining
the results of KamLAND and Borexino, or alternatively by
constructing a reactor experiment with $L\sim 60$~km and an exposure
of 60 GWkTy, one can determine $|U_{e1}|$ and $|U_{e2}|$  with
sufficient accuracy to reconstruct the unitarity triangle. The main
obstacle to the construction of the unitarity triangle is the
precision measurement of $|U_{\mu 1}|$ and $|U_{\mu 2}|$ which
requires a very-long-baseline experiment with a setup similar to the
proposed Brookhaven to Homestake project \cite{Beavis:2002ye}.

The reconstruction of the unitarity triangle using measurements of
neutrino mixing in matter has been considered in
\cite{Zhang:2004hf,Xing:2005gk,Xing:2003ez}.
The area of the triangle will also be proportional to the CP-violating
phase $\delta$.
However to extract information on the neutrino parameters, several
different setups with distinct beam energies and matter densities are
required which correspond to different `triangles in matter'.

\subsubsection{Leptonic unitarity triangle and future experiments}

\begin{figure}
  \centering
  \includegraphics[bb=56 49 552 487,clip=true,width=0.6\linewidth]{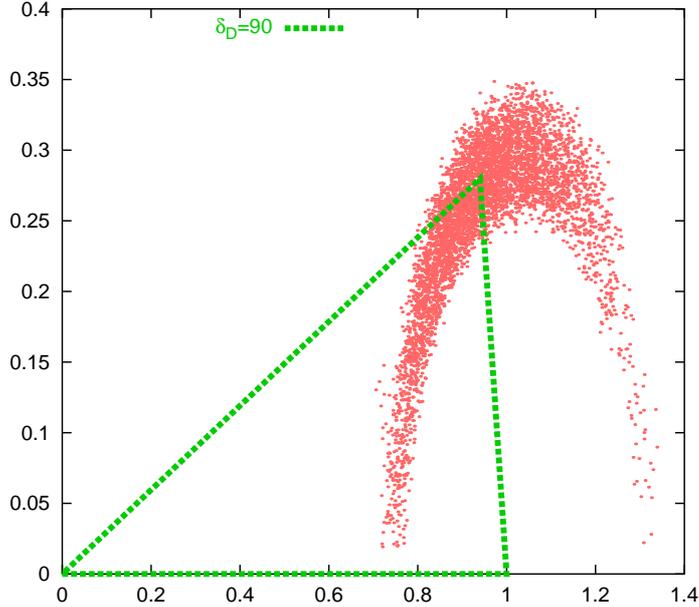}
  \caption{ The
    $e\mu$-triangle taking $s_{13}=0.15$,
    $\delta=90^\circ$ and the best fit values $s_{23}=0.67$ and $s_{12}=0.56$.
The  $|U_{e1}U_{\mu 1}^*|$ side is normalised to one.
    Each scatter point represents the possible position of vertex
    when the moduli of mixing matrix elements
    pick up
    random values around $|U_{e1}|=0.74$, $|U_{\mu 1}|=0.42$, $|U_{e 3}|=0.15$ and $|U_{\mu 3}|=0.67$ (which corresponds
     to the above mixing parameters) within the following uncertainty ranges:
    $\Delta|U_{e1}|/|U_{e1}|=5\%$;
    $\Delta|U_{\mu 1}|/|U_{\mu 1}|=10\%$;
    $\Delta |U_{e3}|/|U_{e3}|=10\%$; and
    $\Delta |U_{\mu 3}|/|U_{\mu 3}|=3\%$.
  }
  \label{trianglefuture}
\end{figure}

Figure (\ref{trianglefuture}) illustrates the possibility of
establishing CP-violation by the triangle method for the realistic
uncertainties
 which can be achieved after the next generation of experiments.
The shown triangle corresponds to $s_{13}=0.15$, $\delta_D=90^\circ$
and the current best fit for the rest of the mixing parameters
($s_{23}=0.67$ and $s_{12}=0.56$). The horizontal side
($|U_{e1}U_{\mu 1}^*|$) is normalised to one. The scatter points
show the position of the vertex of the triangle, when the moduli of
the mixing matrix elements take random values  around the central
points $|U_{e1}|=0.74$, $|U_{\mu 1}|=0.42$, $|U_{e 3}|=0.15$ and
$|U_{\mu 3}|=0.67$ (which correspond to the mixing parameters above)
within the future uncertainty ranges. To produce the scatter 
points, we have taken $\Delta|U_{e1}|/|U_{e1}|=5\%$, $\Delta
|U_{e3}|/|U_{e3}|=10\%$ and $\Delta |U_{\mu 3}|/|U_{\mu 3}|=3\%$
which correspond to the accuracy achievable respectively by combined
KamLAND and Borexino data analysis \cite{Strumia:2001gi}, reactor
experiments \cite{Huber:2006vr}, and T2K/No$\nu$A
\cite{Ayres:2004js}. We have taken  an optimistic accuracy of
$\Delta |U_{\mu 1}|/|U_{\mu 1}|=10\%$.
From this figure we observe that   if the
value of $s_{13}$ is close to its present upper bound and $\delta_D$
is maximal, the uncertainties outlined above will be small enough to
establish CP-violation.

The measurements described above can be complemented by those that
can be made at the proposed super-beams and at a Neutrino Factory.
While $|U_{\mu 1}|$ and $|U_{\mu 2}|$ can not be determined by these
experiments, the triangle method can be considered as an alternative
for resolving the eight-fold degeneracies which are encountered in
the conventional methods of searching for the CP-violating phase
\cite{Donini:2004hu,Burguet-Castell:2005pa,Burguet-Castell:2002qx,Huber:2005ep}.
For example, none of the setups we have suggested to reconstruct the
triangle is sensitive to ${\rm sgn}(\Delta m_{31}^2)$ thus the
triangle method can serve to resolve the ${\rm sgn}(\Delta
m_{31}^2)$ degeneracy.

\subsubsection{Beyond three neutrinos}

Deviation of the mixing matrix from unitarity may originate from
violation of the universality of weak interactions due, for example,
to mixing of neutrinos with heavy neutral leptons. This  affects not
only neutrino oscillations in vacuum and in matter but also the
leptonic decays, for example through the existence of
lepton-flavour violating decays \cite{Antusch:2006vw}. The discovery
of sterile neutrinos and their mixing with active neutrinos would
imply violation of unitarity for active-neutrino mixing. In the case
of four light neutrinos, the mixing can be represented in the form of
quadrangles. In this case, the number of Dirac CP-violating phases
will increase to three. A classification of the unitarity
quadrangles in the four-neutrino mixing scheme is given in
\cite{Guo:2002vm}. Relations between the areas of the unitarity
quadrangles and the re-phasing invariants of CP and T violation have
been established. Also quadrangles in matter were studied in
\cite{Zhang:2004hf}.

\subsubsection{Constraints on unitarity}

Neutrino oscillations constitute evidence for physics beyond the
Standard Model.
If new physics exists, it can manifest itself through unitarity
violation in the Standard Model couplings, the complete theory being
unitary and probability-conserving.
In the quark sector, the search for deviations from unitarity of the
CKM matrix is considered as a sensitive way to search for physics beyond
the Standard Model.
In the lepton sector unitarity violations can arise both from new
physics at low energy -- as in the case of the hypothetical existence
of additional light sterile neutrinos -- or at high energy -- as in
the case of the canonical see-saw mechanism
\cite{Minkowski:1977sc,Gell-Mann:1980vs,Yanagida:1979as,Mohapatra:1979ia},
where light neutrino masses are generated through mixing with heavy,
singlet, fermionic states.
In the following we will discuss both possibilities.

To study mixing among active and sterile neutrinos, we consider the
Standard Model field content plus $N_s$ sterile neutrinos.
The complete $(3+N_s)$-dimensional mixing matrix is unitary, but
the $3\times 3$ PMNS matrix is not since it is a sub-matrix.
Probability conservation in the complete theory implies:
\begin{equation}
  \label{Psterile}
  \sum_{y=e,\mu,\tau} P(\nu_x\to\nu_y)=1-P(\nu_x\to\nu_s)\, ,
\end{equation}
where $P(\nu_x\to\nu_s)\equiv P_{x s}$ is the oscillation probability
into sterile neutrinos.
Since neutral currents are sensitive to this sum, in principle a
neutral-current measurement alone would be sufficient to determine
$P_{x s}$.
However, in a realistic detector mis-identifications of charged
current and neutral current events, together with systematic
uncertainties on neutrino interaction cross sections, complicate the
analysis.
In reference \cite{Barger:2004db} the sensitivity of neutral-current
measurements to the sterile content of a neutrino beam in a
long-baseline oscillation experiment is studied.
The performance that can be expected of the present and next
generation of experiments (K2K, MINOS, and T2K) at $3\sigma$
sensitivity and the 90\%~C.L. exclusion limits for the sterile
oscillation probability will be of order $0.10-0.15$.

To date, deviations from unitarity coming from the additional light
neutrinos have been discussed.
However, similar deviations can be generated by the presence of heavy
neutrinos.
This is the case, for instance, in the see-saw mechanism, where $N_R$
right-handed neutrinos, with heavy Majorana masses, are added to the
Standard Model.
As before, the complete $(3+N_R)$-dimensional mixing matrix is
unitary, while the $3\times 3$ sub-matrix is not.
The main difference with the light-neutrino case is that the mixing
between the light and heavy states is minimal because the mass
difference is so large.
This case has been studied by many authors, both in general
frameworks in which heavy fermions have been added to the Standard
Model lagrangian \cite{Langacker:1988ur,Nardi:1991rg,Nardi:1994iv} and
in the specific neutrino context
\cite{Langacker:1988up,Bilenky:1992wv,Czakon:2001em,Bekman:2002zk}.
In particular, in reference \cite{Czakon:2001em} CP violation in
presence of non-unitarity induced by heavy neutrinos has been
considered and an enhancement of the effect has been observed.

Deviations from unitarity can also be studied in an effective-theory
approach, without the need for the introduction of new fermionic
states.
This can be done as long as the new physics resides at energies much
larger than the electroweak scale, such that heavy fields can be
integrated out.
The low-energy effective lagrangian will generally contain corrections
to the Standard Model couplings and a tower of non-renormalisable
higher-dimensional operators suppressed by powers of the large energy
scale, both of which can result in deviations from unitarity in the
mixing matrices.
In reference \cite{Antusch:2006vw} deviations from unitarity of
the leptonic mixing matrix are studied, in a minimal framework dubbed
MUV (Minimal Unitarity Violation).
In the MUV scheme sources of non-unitarity are allowed only in those
terms of the Standard Model lagrangian involving neutrinos and only
three flavours are considered, as in the standard case.
It is always possible to go to a basis, the mass
basis, where kinetic terms are diagonal and normalised and neutrino
masses are diagonal too.
Here the whole effect of new physics is encoded in the non-unitarity
of the leptonic mixing matrix.
In this framework, and taking a completely general mixing matrix, a
large set of neutrino data, including oscillations and decays, is
analysed, in order to see up to what point the measured elements of
the mixing matrix arrange themselves in a unitary pattern.

The starting point is the Standard Model lagrangian, where the PMNS
matrix is replaced by a generic matrix $N$, which relates the mass and
flavour basis: $\nu_\alpha = N_{\alpha i} \nu_i$.
Since $N$ is not unitary, and since the mass basis is still
orthonormal, the flavour basis is no longer orthogonal, and this gives
rise to new physical effects. The oscillation probability now reads:
\begin{equation}
  \label{prob}
  P_{\nu_\alpha \nu_\beta}(E,L)
  \equiv |\langle\nu_{\beta}|\nu_{\alpha}(L)\rangle|^2\,=\,
  \frac{|\sum_i N^*_{\alpha i}\,e^{i\,P_i\,L}\,N_{\beta i}|^2}
  {(NN^\dagger)_{\alpha\alpha}(NN^\dagger)_{\beta\beta}} \, .
\end{equation}
This formula is formally identical to the standard one, apart
from a normalisation factor in the denominator.
However, due to the non-unitarity of $N$, the oscillation probability
at $L=0$ is not zero, a phenomenon referred to as the `zero-distance'
effect:
\begin{equation}
  \label{zerodist}
  P_{\nu_\alpha \nu_\beta}(E, L=0)
  \propto |(NN^\dagger)_{\beta\alpha}|^2 \, .
\end{equation}
The zero-distance effect, and the fact that oscillations in matter
become non-diagonal, are the unique consequences of the non-unitarity
of $N$ on the phenomenology of neutrino oscillations.
The non-unitarity of $N$ also has consequences in other sectors.
Since the electroweak couplings are modified, interactions involving
the $W$ and $Z$ bosons are now sensitive to the elements of $N$.
However, since it is not possible to tag experimentally neutrino mass
eigenstates, in contrast to the quark sector, electroweak decays can
only be used to determine sums of products of mixing-matrix elements.
This information is extremely relevant in the determination of the
moduli of the matrix elements.

The number of parameters required to specify $N$ (9 moduli and 4 or 6
phases, depending on the Dirac or Majorana nature of neutrinos) is
larger than in the unitary case.
It is presently possible only to determine the moduli since all
positive oscillation signals to date correspond to disappearance
modes.
The elements of the `$e$-row' can be constrained using data from
CHOOZ \cite{Apollonio:2002gd}, KamLAND \cite{Araki:2004mb}, and SNO
\cite{Aharmim:2005gt}, together with the information on
$\Delta m^2_{23}$ resulting from an analysis of K2K data
\cite{Ahn:2002up}.
In contrast, less data is available that may be used to constrain the
elements of the $\mu$-row.
Data from K2K and SuperKamiokande \cite{Ashie:2005ik} can be used to
determine $|N_{\mu 3}|$ and the combination
$|N_{\mu1}|^2+|N_{\mu2}|^2$.
Putting all this information together, the following allowed $3\sigma$
ranges are obtained for the moduli of the elements of the leptonic
mixing matrix:
\bea
  |N| = \left( \begin{array}{ccc}
  0.76-0.89 & 0.45-0.66 & <0.37 \\
  \big[\,\sqrt{|N_{\mu1}|^2+|N_{\mu2}|^2}= & 0.57-0.86\,\big] & 0.57-0.86 \\
  ? & ? & ? \end{array}
  \right) \, .
  \label{N_osc}
\eea
Notice that using only oscillation experiments, and without assuming
unitarity, only half of the elements can be determined.
However, some information is also available from NOMAD
\cite{Astier:2001yj}, KARMEN \cite{Eitel:2000by}, BUGEY
\cite{Declais:1994su}, and the near detector at MINOS \cite{MINOS}.
These experiments exploit the zero-distance effect (equation
(\ref{zerodist})) to provide constraints on $NN^\dagger$.
Combining this information with equation (\ref{N_osc}), $|N_{\mu1}|$
and $|N_{\mu2}|$ can be disentangled.

In order to constrain all the elements of the mixing matrix,
other data have to be considered.
The decay widths for $W$ and $Z$ bosons are given by:
\begin{equation}
  \label{decays}
  \Gamma (W \rightarrow \ell_\alpha \nu_\alpha ) =
  \frac{G_F M_W^3}{6 \sqrt{2} \pi} (N N^\dagger)_{\alpha\alpha}
  \quad {\rm ; and} \quad
  \Gamma (Z \rightarrow \mbox{invisible} ) =
  \frac{G_F M_Z^3}{12 \sqrt{2} \pi}
  \sum_{\alpha,\beta} |(N N^\dagger)_{\alpha\beta}|^2 \, .
\end{equation}
These relations allow the diagonal elements of $N N^\dagger$ to be
constrained.
Additional information can be obtained from ratios of the rate of
decay of leptons, the $W$ boson and the electroweak decays of pions.

The off-diagonal elements of $N N^\dagger$ can be constrained using
rare charged-lepton decays such as
$l_\alpha \rightarrow l_\beta \gamma$.
The non-unitarity of $N$ forbids the GIM cancellation of the constant
term, and the branching ratio is approximated very accurately by:
\begin{eqnarray}
  \frac{\Gamma (\ell_\alpha \rightarrow \ell_\beta \gamma)}
  {\Gamma (\ell_\alpha \rightarrow \nu_\alpha \ell_\beta\overline{\nu}_\beta)}
  = \frac{100 \alpha}{96 \pi}
  \frac{| (N N^\dagger)_{\alpha\beta}|^2}
  {(N N^\dagger)_{\alpha\alpha}(N N^\dagger)_{\beta\beta}} \, .
\end{eqnarray}
Performing a global fit to all these electroweak data, the following
values are obtained at the $90\%$ CL:
\begin{eqnarray}
  |N N^\dagger | \approx
  \left( \begin{array}{ccc}
 0.994\pm  0.005   & < 7.0 \cdot 10^{-5}  &   < 1.6 \cdot 10^{-2}\\
 < 7.0 \cdot 10^{-5}   &  0.995 \pm  0.005 &  < 1.0 \cdot 10^{-2}  \\
 < 1.6 \cdot 10^{-2}   &  < 1.0 \cdot 10^{-2}  &  0.995 \pm 0.005
  \end{array}\right)
  \, .
  \label{nndag}
\end{eqnarray}
Similar bounds can be inferred for $N^\dagger N$ proving that, in the
MUV scheme, unitarity in the lepton sector is experimentally confirmed
from data on weak decays with a precision better than $5\%$.

The elements of the mixing matrix obtained from the analysis of
neutrino-oscillation experiments, equation (\ref{N_osc}), can now be
combined with the unitarity constraints obtained from weak decays,
equation (\ref{nndag}).
The resulting mixing matrix in the MUV scheme is:
\bea
  |N| = \left(
    \begin{array}{ccc}
0.75-0.89 & 0.45-0.65 & <0.20 \\
0.19-0.55 & 0.42-0.74 & 0.57-0.82 \\
0.13-0.56 & 0.36-0.75 & 0.54-0.82
    \end{array}
  \right) \, .
  \label{N_dec}
\eea
All the elements are now significantly constrained to be rather close
to those stemming from the usual unitary
analysis \cite{Gonzalez-Garcia:2004jd}:
\bea
  |U| = \left(
    \begin{array}{ccc}
      0.79-0.88 & 0.47-0.61 & <0.20 \\
      0.19-0.52 & 0.42-0.73 & 0.58-0.82 \\
      0.20-0.53 & 0.44-0.74 & 0.56-0.81
    \end{array}
  \right) \, .
  \label{U}
\eea

In the future, improvements in the measurements of the matrix elements
are expected, as well as improvements in the unitarity tests. 
On the one hand, the exploration of the appearance channels at future
facilities, such as super-beams
\cite{Itow:2001ee,Ayres:2004js,Gomez-Cadenas:2001eu,Campagne:2004wt}, 
beta-beams~\cite{Zucchelli:2002sa}, and the Neutrino Factory
\cite{Geer:1997iz,DeRujula:1998hd}, will permit the testing of the 
$\tau$-row directly and the measurement of the phases of the mixing
matrix, which up to now are completely unknown. 
On the other hand, improvements
on the unitarity bounds are expected from the experiments looking for
$\mu\rightarrow e\gamma$, but also from the bounds which can be
obtained at a Neutrino Factory. In particular, since the bounds on
rare $\tau$ decays are not likely to improve much, an improvement on
the bounds on $(NN^\dagger)_{e\tau}$ and $(NN^\dagger)_{\mu\tau}$
could be obtained with an OPERA-like detector placed at a short
baseline (100m) from a Neutrino Factory beam.

%% file: 04-BSnuM/04-05-Nonstandard-Interaction/def_temp.tex
\newcommand{\beqa}{\begin{eqnarray}}
\newcommand{\eeqa}{\end{eqnarray}}
\newcommand{\no}{\nonumber}
\newcommand{\AReq}[1]{equation (\ref{Eq:AR:#1})}
\newcommand{\AReqs}[1]{equations (\ref{Eq:AR:#1})}
\newcommand{\ARdm}[1]{{\Delta m^2_{#1}}}

%% file: 04-BSnuM/04-05-Nonstandard-Interaction/04-05-Nonstandard-Interaction.tex
\subsection{Non-standard interactions}

Neutrino oscillation experiments probe lepton-flavour
non-conservation, an effect which is not present in the Standard
Model.
In the Standard Model, the lepton sector exhibits a $U(1)^3$ flavour
symmetry, i.e. electron, muon, and tau numbers are conserved
guaranteeing that there are no leptonic flavour transitions. 
Neutrino masses break the symmetry $U(1)^3$; completely in the case of
Majorana neutrino masses, or down to $U(1)_L$ in the case of Dirac
masses.
However, neutrino masses are not the only way
in which the $U(1)^3$ symmetry may be broken. 
Non-standard interactions (NSIs) can also break the $U(1)^3$ symmetry
and generate flavour transitions. 
The dependence of the neutrino-oscillation signal on source-detector
distance and neutrino energy may be used to distinguish between the
various possibilities.

Any interaction that cannot be diagonalised simultaneously with the
weak interaction and the charged-lepton mass matrix breaks the
$U(1)^3$ leptonic-flavour symmetry. 
A simple example is a new effective four Fermi-interaction that, in the
basis where the charged-lepton-mass matrix and the $W$ interaction are
diagonal, is of the form $u \bar d \nu_e \bar\nu_\mu$.
Such an interaction allows the $e-\mu$ transition even for massless
neutrinos. 
Terms that break the lepton-flavour symmetry also generate flavour
transition in processes that involve charged leptons, for 
example, $\tau \to \mu \gamma$. 
Thus, in principle, such processes can be used to probe the same
physics as neutrino oscillation experiments. 
When the only source of flavour breaking is neutrino mass, the effect
in charged lepton processes is tiny due to the leptonic GIM
mechanism. 
For example, the amplitude for $\tau \to \mu \gamma$ is suppressed by
$m_3^2/m_W^2$ and thus BR($\tau \to \mu \gamma)\sim 10^{-50}$ which is
out of reach. 

The situation is different with NSIs. 
Here, the effect in charged lepton processes can be relatively large. 
The amplitude of the flavour transition in both the neutrino and the
charged-lepton sectors are expected to be of the same order. 
Since experiments with charged leptons are in principle easier than
those with neutrinos, it might seem that neutrino oscillation
experiments will not be sensitive to NSIs. 
However, in oscillation experiments the effect of the NSI amplitude
can be enhanced by interference with the standard oscillation
amplitude \cite{Gonzalez-Garcia:2001mp,Ota:2001pw}, an enhancement
that is not present in charged-lepton processes. 
Roughly speaking, if the new physics amplitude is small, and
parametrised by a small parameter $\varepsilon$, then the effect in
oscillation experiments is $O(\varepsilon)$ while for charged leptons
it is $O(\varepsilon^2)$. 
This enhancement makes the `probing power' of neutrino oscillation
experiments larger than one might naively expect.

Any neutrino-oscillation experiment can be divided into three
phases: production; propagation; and detection. 
NSIs can affect any of these phases. 
In the following we consider the production and detection processes
that are relevant to Neutrino Factories; an appearance experiment
where neutrinos are produced in the process 
$\mu^+\to e^+\nu_\alpha\bar\nu_\mu$ and detected by the processes 
$\nu_\beta d\to\mu^- u$ and $\nu_\beta d\to\tau^- u$ and
anti-neutrinos are produced and detected by the corresponding
charge-conjugate processes. 
A new interaction of the form $\mu \bar e \nu_\tau \bar \nu_{\mu}$
would affect oscillation experiments that use neutrinos produced in
muon decay. 
Similarly, interactions of the form $\mu \bar\nu_e u \bar d$ would
affect the detection processes.

The effect on the propagation can come in two ways. 
In vacuum oscillations, it comes from flavour-violating wave-function 
normalisation; non-diagonal kinetic terms arise, which cannot be
diagonalised simultaneously with the interaction of the $W$ 
boson. 
Such effects are likely to be relatively small and are not discussed
further \cite{Ota:2005et}. 
The effect on propagation in matter can be large.
For example, an interaction of the form 
$e \bar e \nu_\tau \bar \nu_\mu$ can generate $\mu-\tau$ transitions 
when neutrinos travel through a medium that contains electrons, such
as the Earth or the Sun. 

While NSIs can affect any of the three phases, they do not
necessarily affect them all.
The flavour structure of the new interactions that affect each phase
are different. 
Consider the case of interactions that involve two quarks and two
leptons; this kind of interaction affects both the detection and the
propagation. 
Yet, at detection the interaction is charged current while during
propagation the relevant interaction is neutral current. 
In many new-physics models these interactions are related, but this is
not automatic. 
Purely leptonic interactions affect the production and propagation. 
Yet, in the production the charged leptons are the electron and the
muon, while in propagation in matter both are electrons.
In section \ref{YG:subsec} we concentrate on effects due to new 
physics in production or detection. In section \ref{SubSect:AR} NSIs
effects on the propagation are discussed.

\subsubsection{Non-standard interactions in production and detection} 
\label{YG:subsec}

Consider a model-independent parameterisation of new-physics effects
on production and detection processes in neutrino oscillation
experiments
\cite{Grossman:1995wx,Gonzalez-Garcia:2001mp,Ota:2001pw,Ota:2002na,Ota:2005et}.
New physics in the source or the detector may be parameterised using
two sets of four-fermion couplings: $(G^s_{\rm NP})_{\alpha\beta}$;
and $(G^d_{\rm NP})_{\alpha\beta}$, where $\alpha,\beta=e,\mu,\tau$.
Here $(G^s_{\rm NP})_{\alpha\beta}$ refers to processes in the source
where a flavour eigenstate $\nu_\beta$ is produced in conjunction with
an incoming charged lepton, $\alpha^-$, or an outgoing $\alpha^+$. 
$(G^d_{\rm NP})_{\alpha\beta}$ refers to processes in the detector
where an incoming $\nu_\beta$ produces an $\alpha^-$.
While the $SU(2)_L$ gauge symmetry requires that the four-fermion
couplings of the charged current weak interactions be proportional to
$G_F\delta_{\alpha\beta}$, new interactions allow couplings with
$\alpha\neq\beta$. 
Phenomenological constraints imply that the new interaction is
suppressed with respect to the weak interaction, i.e.: 
$|(G_{\rm NP}^s)_{\alpha\beta}|\ll G_F$; and 
$|(G_{\rm NP}^d)_{\alpha\beta}|\ll G_F$.

In the SM, neutrino interactions have a Dirac $(V-A)(V-A)$ structure.
Admitting non-standard interactions, massless neutrinos can have
either the SM Dirac structure or a $(V-A)(V+A)$ structure. 
The effects of interactions of the form $(V-A)(V+A)$ at production or
detection are suppressed by ratios of charged-lepton masses and are
therefore very small and will be neglected \cite{Ota:2001pw}. 

In an appearance experiment where neutrinos are produced in the
process $\mu^+\to e^+\nu_\alpha\bar\nu_\mu$ and detected by the
process $\nu_\beta d\to\ell^- u$ and anti-neutrinos are produced and
detected by the corresponding charge-conjugate processes, the relevant
couplings are $(G_{\rm NP}^s)_{e\beta}$ and $(G_{\rm
NP}^d)_{\mu\beta}$. 
It is convenient to define small dimensionless quantities
$\varepsilon^{s,d}_{\alpha\beta}$ as follows: 
\beqa
  \label{YG:defeps}
  \varepsilon^s_{e\beta}&\equiv& \frac{(G_{\rm NP}^s)_{e\beta}}{
  \sqrt{|G_F+(G_{\rm NP}^s)_{ee}|^2+|(G_{\rm NP}^s)_{e\mu}|^2
  +|(G_{\rm NP}^s)_{e\tau}|^2}} \; {\rm ; and} \no \\
  \varepsilon^d_{\mu\beta}&\equiv& \frac{(G_{\rm NP}^d)_{\mu\beta}}{
  \sqrt{|G_F+(G_{\rm NP}^d)_{\mu\mu}|^2+|(G_{\rm NP}^d)_{\mu e}|^2
  +|(G_{\rm NP}^d)_{\mu\tau}|^2}}.
\eeqa
The assumption $|\varepsilon^{s,d}_{\alpha\beta}|\ll1$ means that
leading-order (linear) effects only need be considered.
The leading effects from flavour-diagonal couplings are proportional
to $\varepsilon$
(flavour-diagonal)$\times\varepsilon$(flavour-changing) and can
therefore be  neglected.

Non-zero values of $\varepsilon^{s,d}_{\alpha\beta}$ can be generated
if the three-by-three mixing matrix of the S$\nu$M is not unitary.
For example, suppose that there exists a fourth neutrino-mass
eigenstate $\nu_h$ which is heavy.  
If $m_h\gg m_\mu$, so that this mass eigenstate cannot be produced in
muon decay, then:
\beq
  \varepsilon^{d*}_{\ell e}+\varepsilon^s_{e\ell}\ 
  \to\ 
  -N_s^2U_{eh}U_{\ell h}^*.
  \label{YG:replac}
\eeq
where $\ell=\mu,\tau$ and $U_{eh}$ ($U_{\ell h}$) is the mixing
between the heavy neutrino mass eigenstate and the electron ($\ell$)
neutrino and $N_s$ is a normalisation factor given by:
\beq
  \label{YG:defNs}
  N_s=(|U_{e1}|^2+|U_{e2}|^2+|U_{e3}|^2)^{-1/2}=(1-|U_{eh}|^2)^{-1/2}.
\eeq
If there are many heavy states, equations (\ref{YG:replac}) and
(\ref{YG:defNs}) must be modified to include an implicit summation
over $h$.

The expression for the transition probability in neutrino-oscillation
experiments may now be written as a function of the mixing-matrix 
parameters and the new-physics parameters. 
For simplicity, consider a two-generation framework (expressions for
the three-flavour case can be found in \cite{Gonzalez-Garcia:2001mp}).
The state ($\nu_e^s$) that is produced in the source in conjunction
with an $e^+$ and the state ($\nu_\mu^d$) that is tagged by $\mu^-$
production in the detector may be written in terms of the mass
eigenstates as follows:
\beq
  |\nu_e^s\rangle\!=\!\sum_i\left[U_{e i}+\varepsilon^s_{e\mu}U_{\mu i}
  \right]\!|\nu_i\rangle,
  \quad
  |\nu_\mu^d\rangle\!=\!\sum_i\left[U_{\mu i}+\varepsilon^d_{\mu e}U_{e i}
  \right]\!|\nu_i\rangle.
\eeq
The transition probability, 
$P_{e\mu}=|\langle\nu_\mu^d|\nu_e^s(t)\rangle|^2$, where $\nu_e^s(t)$
is the time-evolved state that was purely $\nu_e^s$ at time $t=0$, is
then:
\beq
  \label{YG:probem}
  P_{e\mu}=\left|\sum_i e^{-iE_it}\left[U_{e i}U^*_{\mu i}
  +\varepsilon^s_{e\mu}|U_{\mu i}|^2+\varepsilon^{d*}_{\mu e}|U_{e i}|^2
  \right]\right|^2.
\eeq
The results will be presented in terms of the following parameters:
\beq
  \Delta m^2_{ij}\equiv m_i^2-m_j^2,\qquad
  \Delta_{ij}\equiv\Delta \frac{m_{ij}^2}{2E}, \qquad
  x_{ij}\equiv\frac{\Delta_{ij}\,L}{2} \; .
  \label{YG:defDelx}
\eeq
In the small-$x$ limit, $P_{e \mu}$ may be expanded to second
order in $x\equiv x_{12}$ and 
$\varepsilon\equiv \varepsilon^{d*}_{\mu e}+\varepsilon^s_{e\mu}$.  
In a basis in which the two-generation mixing matrix is real and is
parameterised by one angle $\theta$, the expression for $P_{e \mu}$
may be written:
\beq \label{YG:twogen}
  P_{e\mu}=x^2 \sin^2 2 \theta
  - 2 x \sin 2\theta \,\Im(\varepsilon) + |\varepsilon|^2\,.
\eeq
The first term is the S$\nu$M piece, while the second and third terms
arise only in the presence of new physics. 
The last term, which is a direct new-physics term, does not require
oscillations and is very small.
The second term is the most interesting one as it is an interference
term between the direct new-physics amplitude and the S$\nu$M
oscillation amplitude. 
There are two points to emphasise regarding this term:
\begin{enumerate}
  \item It is linear in $\varepsilon$, and for $x\gg \varepsilon$ it
        is larger than the direct new physics term: the interference
        increases the sensitivity to the new physics; and
  \item The interference is CP violating. This can be understood from 
        the fact that it is linear in $t$, namely it is T odd. In
        order for it to be CPT even it must also be CP odd.
\end{enumerate}

The interference term in equation (\ref{YG:twogen}) is CP violating and
its effect can be sought through measurements of $P_{e\mu}$ and the
transition probability of the CP-conjugate process, $P_{\bar
e\bar\mu}$. 
A CP transformation of the Lagrangian takes the elements of the mixing
matrix and the $\varepsilon$-terms into their complex conjugates. 
It is then straightforward to obtain the transition probability for
anti-neutrino oscillations. 
It is interesting to define the CP asymmetry, $A_{\rm CP}={P_-/ P_+}$,
where $P_\pm=P_{e\mu}\pm P_{\bar e\bar\mu}$. 
The CP-conserving rate $P_+$ is dominated by the S$\nu$M and is given
by $P_+=8x_{31}^2|U_{e3}U_{\mu3}^*|^2$.  
CP violation within the S$\nu$M ($P_-^{\rm S \nu M}$) is suppressed by
both the small value of $|U_{e3}|$ and the small mass-squared
difference $\Delta m^2_{21}$.  
For short distances ($x_{21},x_{31}\ll1$) it is further
suppressed since $P_-^{\rm S \nu M} \propto L^3$. 
The new-physics term ($P_-^{\rm NP})$ does not suffer from the last
two suppression factors, it does not require three generations, and it
has a different dependence on the distance, $P_-^{\rm NP}\propto L$. 
$P_-^{\rm S \nu M}$ and $P_-^{\rm NP}$ may be written:
\beq
  \label{YG:Acpsm}
  A_{\rm CP}^{\rm SM}=-2x_{21}\Im\left(\frac{U_{e2}U_{\mu2}^*}{
  U_{e3}U_{\mu3}^*}\right), \qquad
  A_{\rm CP}^{\rm NP}=-\frac {1}{x_{31}}\Im\left(
  \frac{\varepsilon}{U_{e3}U_{\mu3}^*}\right).
\eeq
The apparent divergence of $A_{\rm CP}^{\rm NP}$ for small $L$ is due
to the approximations that have been used. 
Specifically, there is an ${\cal O}(|\varepsilon|^2)$ contribution to
$P_+$ that is constant in $L$, namely $P_+={\cal O}(|\varepsilon|^2)$
for $L\to0$. 
In contrast, $P_-=0$ in the $L\to0$ limit to all orders in
$|\varepsilon|$. 

Equation (\ref{YG:Acpsm}) leads to several interesting
conclusions:
\begin{enumerate}
  \item It is possible that, in CP-violating observables, the
        new-physics contributions compete with, or even dominate over,
        the S$\nu$M ones in spite of the weakness of the interactions
        ($|\varepsilon|\ll1$);
  \item The different distance dependence of $A_{\rm CP}^{\rm SM}$ and
        $A_{\rm CP}^{\rm NP}$ will allow, in principle, an unambiguous
        distinction to be made between new-physics contributions of
        the type described here and the contribution from lepton
        mixing; and
  \item The $1/L$ dependence of $A_{\rm CP}^{\rm NP}$ suggests that
        the optimal baseline to observe CP violation from new physics
        is shorter than the one optimised for the S$\nu$M.
\end{enumerate}
Since long-baseline experiments involve the propagation of neutrinos
through the Earth, it is important to understand how matter effects
affect these results.
If a constant matter-density is assumed, then the matter contribution
to the effective $\nu_e$ mass, $A=\sqrt{2}G_F N_e$, where $N_e$ is the
electron density, is constant. 
In general, any new interaction also generates a new non-diagonal
contribution to the effective neutrino-mass matrix. 
Yet, since the new-physics effects are small, it is possible to
treat the effect of new-physics at production or detection and the
effect of new physics in the propagation separately.

The transition probability in matter is obtained by replacing the
mass-squared differences, $\Delta_{ij}$, and mixing angles, $U_{\alpha
i}$, with their effective values in matter, $\Delta_{ij}^m$ and
$U_{\alpha i}^m$.  
Considering only the two-generations and taking the small-$x$ limit as
before, the parameters $x^m$ and $\theta^m$ may be defined by:
\beq
  x^m=\frac{B}{\Delta} x, \qquad
  \sin 2\theta^m=\frac{\Delta}{B} \sin 2\theta,
\eeq
where $B=\Delta-A$. 
From equation (\ref{YG:twogen}) it is clear that matter effects cancel
at lowest order in $x$. 
Therefore, taking one higher order in $x$, the transition probability
in matter, $P^m$, may be written:
\beq
  P^m=P^v(1 \pm O(x^2))\,,
\eeq
where $P^v$ is the oscillation probability vacuum.
Since matter in  the Earth is not CP symmetric, its effect enters
the oscillation formula for neutrinos and anti-neutrinos with opposite
signs. 
Therefore, in contrast to the case of vacuum oscillation, $P_-$ will
receive contributions from terms which would be CP conserving in
vacuum and therefore $A_{CP}$ will be non-zero even if there are no
CP-violating amplitudes.
In particular, a fake asymmetry can be related to the real part of
$\varepsilon$.  

The matter-related contribution to $P_-$ may be denoted by
$P_-^m\equiv P_-(A)-P_-(A=0)$ and, since the leading contributions to
$P_+$ are the same as in the vacuum case, the matter-related
contribution to $A_{\rm CP}$ may be written 
$A_{\rm CP}^m\equiv P_-^m/P_+$.
The asymmetries for three neutrino generations in the small $x_{31}$
limit assuming $|x_{12}/x_{13}| \ll |U_{e3}|$ are:
\beq
\label{YG:asmmat}
  (A_{CP}^m)^{\rm SM}=\frac{2}{3}x_{31}^2\left(\frac{A}{\Delta_{31}}\right),
  \qquad
  (A_{CP}^m)^{\rm NP}=
  \frac{A}{\Delta_{31}}\Re
  \left({\varepsilon^{d*}_{\mu e}-\varepsilon^s_{e\mu}}
  \frac{U_{e3}}{U_{\mu3}^*}\right)\,.
\eeq
(more general results can be found in \cite{Gonzalez-Garcia:2001mp}).  
In equations (\ref{YG:Acpsm}) and (\ref{YG:asmmat}):
\begin{enumerate}
  \item Each of the four contributions has a different dependence on
        the distance. 
        In the short-distance limit the asymmetries may be written:
        \beq
          (A_{\rm CP}^m)^{\rm SM}\propto 
          L^2,\qquad A_{\rm CP}^{\rm SM}\propto L,\qquad
          (A_{\rm CP}^m)^{\rm NP}\propto L^0,\qquad 
          A_{\rm CP}^{\rm NP}\propto 1/L.
        \eeq
        Thus, it is possible, in principle, to distinguish between the
        various contributions;
  \item If the phases of the $\varepsilon$s are of order 1, then the
        genuine CP asymmetry will be larger (at short distances) than
        that due to the matter effect; and
  \item The search for CP violation in neutrino oscillations will
        allow us to constrain both $\Re(\varepsilon)$ and
        $\Im(\varepsilon)$. 
\end{enumerate}

A detailed study of the sensitivity of a future Neutrino Factory has
not been carried out.
Estimates indicated that $|\varepsilon| \sim 10^{-4}$ can be probed in
a future Neutrino Factory \cite{Gonzalez-Garcia:2001mp,Ota:2001pw}.  
Of course, it is interesting to search for such effects without any
specific new-physics framework in mind. 
In the following, however, we give several examples of specific 
new-physics models where large effects, $|\varepsilon| > 10^{-4}$, are 
possible \cite{YG:prep}. 

Consider first left-right symmetric (LRS) models.  
These models are defined by extending the symmetry of the Standard
Model to include right-handed electroweak interactions as follows:
\beq
  G_{\rm LRS}=SU(3)_C\times SU(2)_L\times SU(2)_R\times U(1)_{B-L}
  \times D,
  \label{YG:glrs}
\eeq
where $D$ is a discrete symmetry that requires, among other
constraints, $g_L=g_R$.  
Such models contain a scalar particle ($\Delta_L$) with quantum
numbers $\Delta_L(3,3,1)_{-2}$.
The couplings of $\Delta_L$ to leptons are given by:
\beqa
  {\cal L}_{\Delta_L}&=& f\bar L^ci\sigma_2\vec\sigma L\cdot\vec\Delta_L
  +{\rm h.c.}\no\\
  &=&-\sqrt{2} f_{ij}\Delta_L^0\bar\nu_i^c P_L\nu_j
  + f_{ij}\Delta_L^-\left(\bar \ell_i^c P_L\nu_j+\bar\ell_j^c P_L\nu_i\right)
  +\sqrt{2} f_{ij}\Delta_L^{--}\bar\ell_i^c P_L\ell_j+{\rm h.c.},
  \label{YG:delyuk}
\eeqa
where the $3\times3$ matrix $f$ is symmetric in flavour space,
$f_{ij}=f_{ji}$. 
The tree-level exchange of the $\Delta_L$ scalars lead to the following
four-fermion vertices:
\beqa
  \Delta_L^{--}-{\rm exchange}:& &{f_{ij}f_{kl}^*\over m_{--}^2}
  (\bar\ell_k\gamma^\mu P_L \ell_i)(\bar\ell_l\gamma_\mu P_L \ell_j),\no\\
  \Delta_L^{-}-{\rm exchange}:& &2{f_{ij}f_{kl}^*\over m_{-}^2}
  (\bar\ell_k\gamma^\mu P_L \ell_i)(\bar\nu_l\gamma_\mu P_L \nu_j),\no\\
  \Delta_L^{0}-{\rm exchange}:& &{f_{ij}f_{kl}^*\over m_{0}^2}
  (\bar\nu_k\gamma^\mu P_L \nu_i)(\bar\nu_l\gamma_\mu P_L \nu_j).
  \label{YG:delfie}
\eeqa
Effective couplings, $\varepsilon^s_{e\alpha}$, are induced in the
decays $\mu^+\to e^+\nu_\alpha\bar\nu_\mu$, with $\alpha=\mu$ or
$\tau$, through $\Delta_L^-$ exchange in equation (\ref{YG:delfie}).
(Note that the outgoing anti-neutrino must be a muon neutrino in order
for the interference to take place.)  
Such contributions are proportional to:
\beq
  2{f_{e\mu}f_{\mu\alpha}^*\over m_{-}^2}
  (\bar\mu\gamma^\mu P_L e)(\bar\nu_\alpha\gamma_\mu P_L \nu_\mu).
  \label{YG:lrssou}
\eeq
Inspection of equation (\ref{YG:lrssou}) indicates that an appropriate
definition of the LRS-induced coupling that is relevant to muon decay
is:
\beq
  {(G_\Delta)_{e\alpha}\over\sqrt2}={f_{e\mu}f_{\mu\alpha}^*\over2m_{-}^2}
  \quad \Longrightarrow \quad
  \varepsilon^s_{e\alpha}\equiv{(G_\Delta)_{e\alpha}\over G_F}=4
  {f_{e\mu}f_{\mu\alpha}^*\over g^2}{m_W^2\over m_{-}^2}.
  \label{YG:epslrs}
\eeq

Bounds on $\varepsilon^s_{e\alpha}$ can be obtained from charged-lepton
decays. 
If the $\Delta_L$-scalar is heavy, the mass-squared splittings among
its members, which break electroweak symmetry, are small and motivate
the approximation $m_-\approx m_{--}$. 
Then, using data from $\mu\to e\gamma$ and from $\tau\to\mu \mu e$ to
update tables 3 and 4 in \cite{Cuypers:1996ia}, one obtains:
\beq
  \varepsilon^s_{e\mu}\lsim 2\times10^{-5}, \qquad \mbox{\rm and} \qquad
  \varepsilon^s_{e\tau}\lsim  2\times10^{-3},
  \label{YG:etaulrs}
\eeq
indicating that $\varepsilon^s_{e\tau}$ can be large. 
Yet, it seems that models that saturate the bound have no particular
motivation. 
In generic models $\varepsilon^s_{e\tau}$ is related to the ratio of
the neutrino mass to the weak scale and thus is tiny. 
Of course, it may be possible to find models in which
$\varepsilon^s_{e\tau}$ is not related to the smallness of the
neutrino masses and is naturally large. 

Supersymmetric (SUSY) models without R-parity also contain scalars
with couplings to charged and neutral fermions \cite{Martin:1997ns}.
The couplings of the scalars $\tilde E_i(1,1)_{1}$, where $i=1,2,3$
to leptons are given by:
\beq
  {\cal L}_{\tilde E_i}= \lambda\bar L^ci\sigma_2 L\tilde E+{\rm h.c.}\no\\
  =\lambda_{ijk}\tilde E_k^-
  \left(\bar \ell_i^c P_L\nu_j-\bar\ell_j^c P_L\nu_i\right)+{\rm h.c.}.
  \label{YG:tilyuk}
\eeq
The $\lambda_{ijk}$ couplings are anti-symmetric in the flavour indices
$i,j$, $\lambda_{ijk}=-\lambda_{jik}$ and, in particular,
$\lambda_{eei}=\lambda_{\mu\mu i}=0$. 

Tree-level exchange of the $\tilde E_i^-$ scalars leads to the
following four fermions vertices:
\beq
2{\lambda_{ijm}f_{klm}^*\over m_{E_m}^2}
(\bar\ell_k\gamma^\mu P_L \ell_i)(\bar\nu_l\gamma_\mu P_L \nu_j).
\label{YG:tilfie}
\eeq
The contributions from $\tilde E_i^-$ exchange in equation
(\ref{YG:tilfie}) to the decays $\mu^+\to e^+\nu_\alpha\bar\nu_\mu$,
with $\alpha=\mu$ or $\tau$, are proportional to:
\beq
  2{\lambda_{e\mu i}\lambda_{\mu\alpha i}^*\over m_{E_i}^2}
  (\bar\mu\gamma^\mu P_L e)(\bar\nu_\alpha\gamma_\mu P_L \nu_\mu)
  \quad \Longrightarrow \quad
  \varepsilon^s_{e\alpha}\equiv{(G_{\tilde E})_{e\alpha}\over G_F}=4
  {\lambda_{e\mu i}\lambda_{\mu\alpha i}^*\over g^2}{m_W^2\over m_{E_i}^2}.
  \label{YG:epsrpv}
\eeq
Due to the anti-symmetry of the $\lambda_{ijm}$ couplings,
$\varepsilon^s_{e\mu}=0$.  
Tables 3 and 4 in \cite{Cuypers:1996ia} show that universality gives
the strongest bound on $\varepsilon^s_{e\tau}$:
\beq
  \varepsilon^s_{e\tau}\lsim 6\times10^{-2}.
  \label{YG:boundRP}
\eeq 

In general, only weak constraints on the values of the $\lambda_{ijk}$
couplings in R-parity violating SUSY models can be obtained. 
In particular, the upper bound given above can be saturated in a
generic model. 
The $\lambda$ couplings, however, contribute to neutrino masses (see,
for example, \cite{Grossman:2003gq}. 
Unless there is fine-tuning, the bounds on neutrino masses imply
$\varepsilon^s_{e\tau}\lsim 10^{-3}$. 
Thus, large effects are possible even without fine-tuning. 
Of course, the bound in (\ref{YG:boundRP}) can be saturated naturally
in models with extra structure, such as horizontal symmetries
\cite{Banks:1995by}. 

NSIs arising in supersymmetric models with R parity were studied in
\cite{Ota:2001pw,Ota:2005et}.
Measurements of charged-lepton decays allow strong limits to be placed
on NSI in SUSY models with R parity, implying that the relevant
couplings are small.
This class of model will not be discussed further here.

Finally, consider RS-type models \cite{Randall:1999ee} with right
handed neutrinos in the bulk \cite{Grossman:1999ra}.  
In such models bulk singlets are introduced with dimension-five mass
terms. 
When these mass terms are of the order of the fundamental scale, the
zero modes have very small couplings to the standard doublet neutrinos
that are confined to the Planck brane. 
Thus, exponentially small Dirac neutrino masses are generated. 
In addition to the zero modes, the higher Kaluza-Klein modes 
couple to the doublet neutrinos. 
However, their wave functions are not small at the visible brane. 
Thus, their dimension-four Yukawa couplings ($Y_5$) are not
particularly small, and large active-heavy mixing is expected.  
As a result of this mixing the effective $3\times 3$ mixing matrix is
not unitary, and this non-unitarity is equivalent to a new interaction in
production or detection. 

In order to have a viable model it is necessary to assume that the
$Y_5$, are small.  
Note that this is a mild fine-tuning as the most natural values for
these Yukawa couplings are $O(1)$.  
In this case, the mixing-matrix elements can be expanded in the small
mixing angles and we have \cite{Grossman:1999ra}:
\beq
  |U_{i\alpha}|^2 \approx \frac{1}{2 c_\alpha+1}\,\frac{v_0^2
  |Y_5^{i\alpha}|^2}{k^2} \,,
\eeq
where $v_0$ and $k$ are fundamental mass parameters of the theory and
$c_\alpha \equiv m_{bulk}^\alpha/k$ such that $m^\alpha_{bulk}$ are
the bulk masses of the singlet fermions.
In order to get neutrino masses in the range indicated by experiments,
the parameter $c_\alpha$ has to be in the range of $1.1$ to $1.5$.
Without any further input it seems natural to assume that all the mass
parameters $v_0$, $k$ and $m^\alpha_{bulk}$ take their naive values,
and therefore:
\beq \label{YG:yye}
  |U_{i\alpha}| \sim |Y_5^{i\alpha}|,
\eeq
up to coefficients of order unity. 
Since, by assumption, $|Y_5^{i\alpha}| \ll 1$, equation
(\ref{YG:replac}) yields:
\beq
  \varepsilon_{\ell e}\equiv \varepsilon^{d*}_{\ell e}+\varepsilon^s_{e\ell}
  \sim  Y_5^{e\alpha}Y_5^{*\ell\alpha}.
\eeq
The light-heavy mixing angles can be bounded from several processes
\cite{Grossman:1999ra,Kitano:2000wr}. 
The invisible width of the $Z$ leads to the constraint:
\beq
  |U_{eh}U_{\ell h}|\lsim10^{-2}.
  \label{YG:uucon}
\eeq
Limits on the decays $\mu \to e \gamma$ and $\tau \to e \gamma$ lead
to the following constraints \cite{Tommasini:1995ii}:
\beq
  |U_{eh}U_{\tau h}|\lsim10^{-2}, \qquad
  |U_{eh}U_{\mu h}|\lsim10^{-4} \; ;
  \label{YG:uucon-2}
\eeq
indicating that large effects are allowed for the tau case. 
For the muon channel the effects are not large but may still be
observable. 

Turning to the theoretical expectation for the mixing angles, naively,
it might be expected that the light-heavy mixing should be of order
unity. 
Yet, the Yukawa couplings may be rather small. 
Even so, the model seems to be more attractive for larger $Y_5$ and
therefore for large mixing angles. 

\subsubsection{Non-standard interactions in propagation}
\label{SubSect:AR}

\paragraph{Parameters and limits:}
\label{SubSect:AR:theory}

Non-standard neutrino interactions induced by new physics (NP) not yet
observed at accelerator experiments presumably arise at
scales $\LNP$ much larger than the typical energy involved in future
long-baseline experiments, $E\ll\LNP$. 
At such energies the non-standard effects are conveniently described
by effective interactions (operators) with dimension ($D$) 5 or more
(in energy).
The couplings of such operators involve inverse powers of the scale of
the new physics that generates them. 
The effect of such operators at lower scales is suppressed by
powers of $E/\LNP$, where $E$ is the typical energy of the experiment,
so that it is only necessary to take into account the lowest
dimensional interactions. 
The classic example is the Fermi interaction describing weak
interactions at scales lower than the weak scale $\LEW$. 
This four-fermion interaction has dimension 6 and its coupling
$2\sqrt{2}G_F\sim1/\LEW^2$ involves two inverse powers of the scale
$\LEW$ at which the operator is generated, which makes the weak
interactions weak at $E\ll\LEW$.  

The power of this `effective' description of high-energy interactions
is that: the effect of the most general high energy physics can be
conveniently parameterised in terms of a (finite) set of operators
only involving light fields, so that the knowledge of the physics
above $\LNP$ is not required; and the experimental identification of
the operators actually present at low energy provides important
information on the physics above $\LNP$. 
Indeed, weak interactions were first parameterised in terms of generic
four-fermion interactions. 
Unveiling the `V-A' (left-handed) structure of those interactions was
then crucial to the understanding of the renormalisable theory
underlying them (the SM).

At present the only available firm evidence of a non-renormalisable
remnant of higher energy physics is the $D = 5$ operator responsible
for neutrino masses and mixings :
\begin{equation}
  \label{Eq:AR:numasses}
  \frac{h_{ij}}{2\Lambda_L}(L_iH)(L_jH) ,
\end{equation}
where $L_i$, $i=1,2,3$ are the lepton doublets, $H$ is the
Higgs-doublet, and $\Lambda_L$ is the lepton-number-violation scale at
which the operator is generated. 
Once the Higgs gets a vacuum expectation value (vev), 
$\vev{H} = (0,v)^T$, that operator gives rise to Majorana neutrino
masses $m^\nu_{ij} = -h_{ij} v^2/\Lambda_L$, which forces $\Lambda_L$
to be near $10^{15}$~GeV for $h\sim1$, not very far from the
unification scale. 
The evidence for the existence of the operator in
\AReq{numasses} is very strong, as the understanding of neutrino masses
it provides is solid and general (the see-saw mechanism is just one
example of a high-energy mechanism giving rise to such an operator
\footnote{
  The operator in \AReq{numasses} accounts for essentially all
  high-energy mechanisms that generate neutrino masses. 
  The only possible alternative is that the neutrino masses originate
  at or below the weak scale. 
  The classic example is a Dirac mass term in the presence of an
  exactly conserved (at the perturbative level) lepton number. 
  This possibility is less appealing because it needs tiny Yukawa
  couplings for the neutrinos of all the three families. 
  The smallness of such Yukawas can however in turn be justified in
  terms of new symmetries appropriately broken
  \cite{Chikashige:1980ui,Gelmini:1980re,Chacko:2003dt,Davoudiasl:2005ks,Pascoli:2005zb} 
  or extra-dimensional mechanisms
  \cite{Dienes:1998sb,Arkani-Hamed:1998vp,Dvali:1999cn,Mohapatra:1999zd,Mohapatra:1999af,Barbieri:2000mg,Lukas:2000fy,Lukas:2000wn,Mohapatra:2000wn,Lukas:2000rg}.
}
).
However, such an operator has no significant effect on the
neutrino-matter interaction in long-baseline experiments, as it is
associated to the superheavy scale $\Lambda_L$. 
In order for new physics to have a measurable effect on the neutrino
interactions in matter, a new effective interaction has to be
associated to a scale not too much higher than the scale of the
physics giving rise to the Standard Model interactions (matter
effects), $\LEW\sim G_F^{-1/2}$. 
At present there is no firm evidence at all of operators
generated at such scale (which explains the variety of theoretical
models available for the physics accessible at the LHC). 
In the following therefore, a general parameterisation of the possible
operators relevant for neutrino interaction with ordinary matter is
used.

Consider only those operators that arise at a scale much lower than
$\Lambda_L$ for which lepton number is conserved.
The relevant interaction is then:
\begin{equation}
  \label{Eq:AR:L}
  \sum_{
    \substack{f=e,u,d \\ \alpha,\beta = e, \mu, \tau}}
    4\frac{G_F}{\sqrt{2}}
    \bar\nu_{\alpha L} \gamma^\mu \nu_{\beta L} \left(
    \epsilon^{f_L}_{\alpha\beta}\,\bar f_L \gamma_\mu f_L +
    \epsilon^{f_R}_{\alpha\beta}\,\bar f_R \gamma_\mu f_R \right) \;.
\end{equation}
Since the scale at which this interaction arises is supposed to be not
too far from the electroweak scale, its coupling may be parameterised
by $G_F\epsilon$, where $\epsilon\sim(\LEW/\LNP)^2$. 
\AReq{L} holds in a basis in which the kinetic terms are
canonical and the charged-fermion masses are diagonal. 
The effect of the coherent forward scattering induced by
\AReq{L} on neutrino propagation in an ordinary, neutral, unpolarised
medium is encoded in the parameters
\cite{Wolfenstein:1977ue,Nunokawa:1997dp,Bergmann:1999rz}:
\begin{equation}
  \label{Eq:AR:eab}
  \epsilon = \sum_{f=e,u,d}\frac{n^f}{n_e} \epsilon^f = \epsilon^e + 2\epsilon^u + \epsilon^d + \frac{n_n}{n_e}
  (2\epsilon^d + \epsilon^u) \;,
\end{equation}
where $\epsilon^f = \epsilon^{f_L}+\epsilon^{f_R}$, $n_f$ is the
number density of the fermion $f$ in the medium crossed by the
neutrinos ($n_n$ for the neutron), and the flavour indices have been
omitted. 
In Section \ref{SubSect:AR:propagation} the signatures of the new
interactions in terms of the $\epsilon_{\alpha\beta}$ parameters will
be discussed, independent of their origin. 
Constraints on the parameters are discussed here, focussing on the
non-flavour-diagonal couplings.

A model-independent limit on $\epsilon_{\mu\tau}$ can be inferred from
atmospheric-neutrino data
\cite{Fornengo:1999zp,Gonzalez-Garcia:2004wg}. 
The limit is obtained on the hypothesis that the NP interactions only
involve down quarks, $|\etu^d|< 0.013$ at 90\% C.L., corresponding to
$|\etu|\lesssim 0.4$. 
A recent combined analysis of Super-Kamiokande, K2K and MINOS
data \cite{Friedland:2006pi} also provides a bound on $\ete$. 
In the limit in which $\epsilon_{ee} = \epsilon_{\tau\tau} = 0$, the
analysis gives $|\ete|\lesssim 0.5$. 
The latter limit could improve with future MINOS data.
In \cite{Berezhiani:2001rs} the limit
$|\epsilon^{e_L,e_R}_{\alpha\beta}|\leq 0.53$ at 99\%C.L is obtained 
from the $e^+e^-\to\nu\bar\nu\gamma$ cross section measurement at
LEP. 
A stronger limit on $\etu$ from neutrino-scattering experiments,
$|\etu|< 0.1$, is found in \cite{Davidson:2003ha}. 
The latter also considers the limits from charged-lepton effects
induced by loops involving the vertex in \AReq{L}, which
gives, in particular, $|\eue| < 2\cdot 10^{-3}$.  

Stronger bounds can be obtained by relating the
$\epsilon_{\alpha\beta}$ parameters to operators involving the charged
leptons. 
The description of the effect of NSIs in neutrino propagation,
\AReq{L}, can be obtained in two steps. 
First, the general effective description just below the scale $\LNP$,
but above the weak scale, is written in terms of operators symmetric
under the SM gauge group. 
Then, the operators are run to the weak scale and matched with the
effective description below $\LEW$ in terms of the operators in
\AReq{L}. 
The presence of the intermediate, $\SUW$ symmetric step is relevant as
it relates the neutrino interactions in \AReq{L} to the
interactions of their $\SUW$ charged lepton partners. 
However, this relation is complicated by the fact that $\SUW$ is
broken. 
It is, in fact, possible to conceive of new physics affecting
neutrinos but not charged leptons, see below.
The amount of $\SUW$ breaking that can be tolerated is in turn bound
by electroweak-precision tests performed at LEP. 

Consider first the case in which $\SUW$ breaking is neglected and the
operators in \AReq{L} originate from $\SUW$ invariant
operators. 
Then, the experimental bounds on charged-lepton processes
imply \cite{Bergmann:1997mr,Bergmann:1999pk,Bergmann:2000gp}
\globallabel{Eq:AR:limits}:
\begin{align}
  & \epsilon^e_{e\mu}  \lesssim 10^{-6} &
  & \epsilon^e_{\mu\tau}  \lesssim 3\cdot 10^{-3} &
  & \epsilon^e_{e\tau}  \lesssim 4\cdot 10^{-3} \mytag \\
  & \epsilon^{u,d}_{e\mu}  \lesssim 10^{-5} &
  & \epsilon^{u,d}_{\mu\tau}  \lesssim 10^{-2} &
  & \epsilon^{u,d}_{e\tau}  \lesssim 10^{-2} \; . \mytag
\end{align}
For example, the extension of the MSSM including three singlet, chiral
neutrino fields (giving rise to a supersymmetric see-saw) can generate
large misalignment between leptons and sleptons, in turn inducing
non-standard interactions through one-loop diagrams involving the
sleptons. 
$\SUW$ breaking is negligible in this case, so that the strong
constraints in \AReq{limits} hold and suppress the effects in
neutrino propagation \cite{Ota:2005et}. 

These limits can be evaded by taking into account $\SUW$ breaking.
The extent to which the latter relaxes the limits depends on how the
operator in \AReq{L} is generated and how $\SUW$ breaking
enters. 
A general treatment should in principle be based on the most general
effective lagrangian at the EW scale, including the effective
contribution to the kinetic terms, along the lines of 
reference \cite{Berezhiani:2001rs}. 
Such a general analysis is not available, but it is clear that the
$\SUW$ symmetric limit is considerably weakened. 
This is supported by the analysis in
\cite{Bergmann:1999pk,Bergmann:2000gp}, where the case in which the 
operator in \AReq{L} is induced by the exchange of new heavy
bosons is considered. 
The effect of $\SUW$ breaking on the masses of such heavy bosons can
relax the bounds from the charged-lepton sector in
\AReq{limits} by a factor of seven without a conflict with the
electroweak-precision data. 
It is even possible to generate the neutrino operator in
\AReq{L} without giving rise to any charged-lepton effects if the new
physics (e.g. warped or flat extra-dimensions \cite{DeGouvea:2001mz})
induces the operator:
\begin{equation}
  \label{Eq:AR:peculiar}
  4\frac{G_F}{\sqrt{2}} \ve_{\alpha\beta} (H L_\alpha)^\dagger i\hat\partial(H
  L_\beta) \;.
\end{equation}
The latter contributes to the neutrino wave function, but not to that
of the charged lepton.
The neutrino kinetic term must therefore be brought back to the
canonical form by means of a non-unitary rotation. 
When acting on the standard Fermi interaction, the latter is rotated,
inducing extra contributions in the form in \AReq{L} but
leaving the charged-lepton sector completely unaffected.
The $\ve$ parameters are therefore constrained mostly by neutrino
experiments which give \cite{DeGouvea:2001mz} 
$|\ve_{e\mu}| < 0.05$, $|\ve_{e\tau}| < 0.1$, 
$|\ve_{\mu\tau}| < 0.013$.  
The couplings in \AReq{L} generated through this mechanism
are \cite{Campanelli:2002cc} \globallabel{Eq:AR:Zren}:
\begin{gather}
  \epsilon^e_{\alpha\beta} = -\frac{1}{2} \Bigr(\ve_{\alpha
  e}\delta_{\beta e} + \ve_{e\beta}\delta_{\alpha e}\Bigl) +
  \Bigl(\frac{1}{2} - 2\sin^2\theta_W\Bigr)\ve_{\alpha\beta} \mytag \\
  \epsilon^u_{\alpha\beta} =
  -\Bigl(\frac{1}{2}-\frac{4}{3}\sin^2\theta_W\Bigr)\ve_{\alpha\beta}  \qquad
  \epsilon^d_{\alpha\beta} =
  \Bigl(\frac{1}{2}-\frac{2}{3}\sin^2\theta_W\Bigr)\ve_{\alpha\beta} \;.
\end{gather}
so that the relevant parameters for the effects in neutrino
propagation are:
\begin{equation}
  \label{Eq:AR:1}
  \epsilon_{\alpha\beta} = -\frac{1}{2}\bigl(\ve_{\alpha
  e}\delta_{\beta e}+\ve_{e\beta}\delta_{\alpha e}\bigr)
  +\frac{1}{2}\frac{n_n}{n_e} \ve_{\alpha\beta} \;.
\end{equation}
Note that $\eue$, $\etu$, $\ete$ are, in this case, suppressed by the
relatively small factor $(n_n/n_e-1)/2$. 
As a consequence, the bounds on $\etu$, $\ete$ are stronger than the
ones from \AReq{limits}, despite the fact that bounds from
the charged-lepton sector are, in this case, essentially irrelevant.
The bound on $\epsilon_{e\mu}$ is, in contrast, weaker.

Finally, the limits on the impact of $\SUW$ breaking can be further
weakened if the effect on the precision observables of each source
of $\SUW$ breaking is considered separately or if it is assumed that
the Higgs is light.
In principle, the effects of two or more corrections (including the
effect of a Higgs that is heavier than expected) on the SM fit to
precision observables could in fact compensate each other, thus
allowing stronger $\SUW$ breaking effects \cite{Davidson:2003ha}.

\paragraph{Effects on neutrino propagation}
\label{SubSect:AR:propagation}

The possibility that new physics affects the neutrino transitions
observed in
solar~\cite{Wolfenstein:1977ue,Bergmann:1997mr,Bergmann:2000gp,Berezhiani:2001rt},
atmospheric~\cite{Ma:1997ww,Fornengo:1999zp,Bergmann:1999pk},
LSND~\cite{Bergmann:1998ft}, and supernova~\cite{Mansour:1997fi}
experiments has been widely studied in the literature since the
seminal paper of Wolfenstein \cite{Wolfenstein:1977ue}.
The effects of NSIs at production and decay are quite different from
those that arise in the propagation between source and detector.
This should make them relatively easy to disentangle. 
In fact, due to the geometrical $L^{-2}$ suppression, the effects at
production and detection are best studied at a smaller baseline
$L$~\cite{Gonzalez-Garcia:2001mp}, whereas in the case of new
interactions with matter the $L^{-2}$ suppression is compensated (up
to a certain $L$) by the development of the oscillation. 
Moreover, the possibility of a peculiar growth with the neutrino
energy opens up, which would give rise to a noticeable signature
\cite{Campanelli:2002cc}. 

As in ordinary matter the rate of incoherent scattering is
negligible, the effect of standard and non-standard interactions only
shows up through the coherent forward-scattering effect.
Such an effect is conveniently accounted for by the MSW
potential term in the neutrino Schroedinger equation. 
The potential induced by $\nu_\alpha f \rightarrow \nu_\beta f$
forward scattering ($\alpha,\beta=e,\mu,\tau$, $f=e,u,d$) induced by
the effective interaction in \AReq{L} can be parameterised as
$V_{\alpha\beta} = \epsilon_{\alpha\beta}V =
\sqrt{2}\epsilon_{\alpha\beta} G_F N_e$, where $V = \sqrt{2}G_F N_e$
is the MSW potential induced by the standard charged-current
interactions, $N_e$ is the electron number density and
$\epsilon_{\alpha\beta} = \epsilon^*_{\beta\alpha}$ are the parameters
defined in \AReq{eab}. 
In turn, the standard and non-standard MSW potentials can be reabsorbed
in an energy-dependent redefinition of the neutrino mass-squared matrix,
$M^2\to \Meff^2 +\text{universal terms}$, where:
\begin{equation}
  \label{Eq:AR:MM}
  M^2_\text{eff} = U
  \begin{pmatrix}
    0 & 0 & 0 \\
    0 & \ARdm{21} & 0 \\
    0 & 0 & \ARdm{31}
  \end{pmatrix} U^\dagger
  + 2 E V
  \begin{pmatrix}
    1+\epsilon_{ee} &
    \epsilon_{e\mu} &
    \epsilon_{e\tau}  \\
    \epsilon^*_{e\mu} &
    \epsilon_{\mu\mu} &
    \epsilon_{\mu\tau}  \\
    \epsilon^*_{e\tau} &
    \epsilon^*_{\mu\tau} &
    \epsilon_{\tau\tau}
  \end{pmatrix} \; ,
\end{equation}
$E$ is the neutrino energy, $U$ is the PMNS mixing matrix in the usual
parameterisation, and the flavour-universal terms do not play a role.

The possibility of observing the effect of non-universal diagonal
terms, $\epsilon_{\alpha\alpha}$, has been considered
in~\cite{Honda:2006gv}. 
Such terms could arise, for example, from a violation of universality
in $Z\nu\bar\nu$ couplings, in particular a correction to the
$Z\nu_e\nu_e$ coupling, to be compensated in the bound from the $Z$
invisible decay width by a corresponding correction to the
$Z\nu_\tau\nu_\tau$ coupling.  
The effect of a non-universality at the level of 1\% would amount in
first approximation to an energy-dependent shift in the effective
value of the $\theta_{23}$. 
The possibility of observing such a shift depends on the true value of
$\theta_{23}$ angle.  
The oscillation probability is proportional to $\sin^22\theta_{23}$,
which is almost insensitive to a 1\% shift in $\theta_{23}$, for
$\theta_{23} = \pi/4$.  
On the other hand,  the shift might have a chance to be observed for
example if $\sin^22\theta_{23} = 0.92$. 
This would require an experiment with neutrinos above the resonant
energy and therefore a large enough baseline ($L\sim
10000\,\text{km}$), in such a way that the first oscillation peak is
approached.
Most analyses concentrate on the possibility of observing the effect
of off-diagonal terms $\eue$, $\etu$, $\ete$, and in particular on
$\etu$ and $\ete$, since the bound on $\eue$ is too strong for it to
play any role.

Earlier work on the effect of non-vanishing $\etu$, $\ete$ on
$\nu_\mu\to\nu_\tau$, $\nu_e\to\nu_\tau$
oscillations \cite{Gago:2001xg,Huber:2001zw} assumed the presence of a
$\tau$ detector with an efficiency $\eta\sim 0.3$ for observing
$\nu_\tau$, $\bar\nu_\tau$. 
Moreover, the analyses were performed at fixed values of the
oscillation parameters, in particular $\theta_{13}$ and the
CP-violating phase $\delta$.
Later work \cite{Huber:2001de,Huber:2002bi} carried forward the
analysis by: focussing on the effect of $\ete$ in $\nu_e\to\nu_\mu$
transitions, which can be detected through the easier wrong-sign-muon
signal at a less ambitious muon detector; and by letting $\theta_{13}$
and $\delta$ vary.  
The latter might in fact have to be determined by the same experiments
sensitive to $\ete$, introducing an additional uncertainty on $\ete$.
At the same time, if the effect of a non-vanishing $\ete$ is taken
into account, the sensitivity to the oscillation parameters could
worsen. 
This is indeed the case if one considers the $\nu_e\to\nu_\mu$
transitions only. 
The wrong-sign-muon signal, however, is also due to $\nu_e\to\nu_\tau$
transitions producing a $\tau$ that then decays into a muon. 
This is important because the spectrum of the $\nu_e\to\nu_\tau$
transition can have a very peculiar behaviour in the presence of a
sizable $\ete$. 
Such a behaviour on the one hand enhances the $\nu_e\to\nu_\tau$
transitions at high energy (thus making the $\nu_e\to\nu_\tau$
contribution to the wrong-sign-muon signal important, sometimes even
predominant) on the other hand it allows the effects of $\theta_{13}$
and $\ete$ to be disentangled \cite{Campanelli:2002cc}. 

The presence of $\ete$ effects can reduce the sensitivity of
$\nu_e\to\nu_\mu$ transitions to $\theta_{13}$.
This is because an expansion in the small $\theta_{13}$ and $\ete$
parameters gives $P(\nu_e\to\nu_\mu) \approx A s^2_{13} + B s_{13}\ete
+ C\ete^2$, where $A,B,C$ depend on the baseline, on the energy, and on
the channel (neutrinos or anti-neutrinos) and $s_{13} =
\sin\theta_{13}$. 
The total rate of $\nu_e\to\nu_\mu$-induced wrong sign muon events
(obtained by convoluting $A,B,C$ with the energy dependence of fluxes,
cross-sections, efficiencies, etc.) then corresponds to an ellipse in
the $s_{13}$--$\ete$ plane, so that $s_{13}$ and $\ete$ cannot be
disentangled by a single total-rate measurement only. 
It also turns out that using the spectral information as well does not
help very much, as $\ete$ does not modify the spectrum of
$\nu_e\to\nu_\mu$ transitions significantly.
On the other hand, combining measurements in the neutrino and
anti-neutrino channels and combining measurements at different baselines
helps to reduce the degeneracy, but the sensitivity to $\theta_{13}$
is still reduced by one order of magnitude~\cite{Huber:2001de}. 
The situation is even worse in the presence of new-physics effects in
the production process. 
In this case the effect of a given $\theta_{13}$ (including
its energy and baseline dependence) can be faked in both the neutrino
and anti-neutrino channels by a proper combination of the NP parameters
controlling the exotic production process and the matter
effects \cite{Huber:2002bi}. 
A near detector, only sensitive to new effects at production, might
help in this case.

The degeneracy can be resolved by taking into account
the contribution of $\nu_e\to\nu_\tau$ transitions to the
wrong-sign-muon signal.
While the spectrum of $\nu_e\to\nu_\mu$ transitions is not
significantly affected by $\ete$, it turns out that  the spectrum of
$\nu_e\to\nu_\tau$ transitions can be.
In order to have an intuitive picture of the basic features of the
latter, and in general of all the $\epsilon_{\alpha\beta}$
parameters, consider first the approximation $\ARdm{21}=0$,
which is meaningful in the range of energies of interest. 
In this limit the mixing angle in vacuum $\theta_{12}$ becomes
un-physical. 
We can also consider a phase convention for the neutrino fields in
which the phases only appear in the $\epsilon$
parameters
\footnote{
  While in the absence of non-standard interactions
  the CP-violating phase also becomes un-physical in the $\ARdm{12}=0$
  limit, in the general case it does not. 
  In fact, the phase re-definition necessary to rotate $\delta$ away from
  the mixing matrix also acts on the non-diagonal new interactions. 
  Therefore, if the $\epsilon$ parameters are real to start with, they
  acquire a phase $\delta$ once $\delta$ has been rotated away from the
  mixing matrix in vacuum. 
  The phase has just moved from the mixing matrix in vacuum to the
  epsilon parameters.
}. 
\AReq{MM} then becomes:
\begin{equation}
  \label{Eq:AR:MM2}
  \Meff^2 = \ARdm{31}
  \begin{pmatrix}
    s^2_{13} + (E/\Eres)(1+\epsilon_{ee}) &
     s_{13}/\sqrt{2} + (E/\Eres) \epsilon_{e\mu} &
     s_{13}/\sqrt{2} + (E/\Eres) \epsilon_{e\tau} \\
     s_{13}/\sqrt{2} + (E/\Eres) \epsilon^*_{e\mu} &
     1/2 + (E/\Eres) \epsilon_{\mu\mu} &
     1/2 + (E/\Eres) \epsilon_{\mu\tau} \\
     s_{13}/\sqrt{2} + (E/\Eres) \epsilon^*_{e\tau} &
     1/2 + (E/\Eres) \epsilon^*_{\mu\tau} &
     1/2 + (E/\Eres) \epsilon_{\tau\tau}
   \end{pmatrix} \; ,
\end{equation}
where we also set $\theta_{23}=\pi/4$, $\cos\theta_{13}$, and
$\cos2\theta_{13}=1$. 

The neutrino effective masses and mixings follow from the
diagonalisation of $\Meff^2$. 
In the limit in which the non-standard interactions are switched off,
$\epsilon_{\alpha\beta}\to 0$, the usual expressions for the neutrino
masses and mixings in the presence of matter are recovered,
characterised by a resonant energy $\Eres$:
\begin{equation}
\label{Eq:AR:Eres}
\Eres \simeq 10{\rm~GeV} \fracwithdelims{(}{)}{\ARdm{31}}{2.5\cdot
10^{-3}{\rm~eV}^2}\fracwithdelims{(}{)}{1.65\,\mathrm{g}\cm^3}{\rho\, Y_e}
  \; ,
\end{equation}
where $\rho$ is the matter density and $Y_e$ is the number of
electrons per baryon in matter $n_e/n_B$. 
In particular, the characteristic suppression of the $\theta_{13}$,
$\theta_{12}$ mixing angles at energies higher than the resonance
energy is recovered. 
This is because in the $E/\Eres\gg1$ limit the large diagonal MSW term
in $(\Meff^2)_{11}$ is enhanced, which suppresses the mixing. 
In particular, $\sin^22\theta^m_{13}\sim \sin^22\theta_{13} (\Eres/E)^2$, so that the transition probabilities decrease with $E^2$:
\globallabel{Eq:AR:he1}
\begin{align}
  P(\nu_e\rightarrow\nu_\tau) & \sim \Big(\frac{\Eres}{E}\Big)^2
  \cos^2\theta_{23}\sin^2 2\theta_{13} \sin^2\frac{LV}{2}
  \mytag \; {\rm ; and} \\
  P(\nu_e\rightarrow\nu_\mu) & \sim \Big(\frac{\Eres}{E}\Big)^2
  \sin^2\theta_{23}\sin^2 2\theta_{13} \sin^2\frac{LV}{2} \; . \mytag
\end{align}
Note also that the mass-squared difference in matter, 
$\dmm{31}\sim 2EV$ grows with energy, canceling the $1/E$ dependence
in the oscillating term of the probability.

The situation is completely different in the presence of non-standard,
non-diagonal interactions; at least at very large energies
$E\gg\Eres$. 
The non-diagonal elements now also get a contribution that grows with
energy.
As a consequence, at sufficiently large energy, matter effects will
dominate in all the entries of $\Meff^2$.
Then: the effective mixing angles will be determined by matter effects
only, thus providing a determination of the $\epsilon_{\alpha\beta}$
ratios; and the mixing angles become energy independent, i.e.\
they do not suffer from the high-energy suppression anymore.
This is also true for the transition probabilities, the leading
terms of which are given in the large $E/\Eres$ limit by the following
simple expressions:
\globallabel{Eq:AR:he2}
\begin{align}
  \label{Eq:AR:he}
  P(\nu_e\rightarrow\nu_\tau) & \sim 4\left| \epsilon_{e\tau} +
    \frac{\Eres}{E} c_{23}s_{13} \right|^2 \sin^2 \frac{LV}{2} \mytag
    \\
  P(\nu_e\rightarrow\nu_\mu) & \sim4\left| \epsilon_{e\mu} +
    \frac{\Eres}{E} s_{23}s_{13} \right|^2 \sin^2 \frac{LV}{2} \;,
    \mytag
\end{align}
where the leading $\Eres/E$ correction to the energy-independent
amplitudes have been included.
The oscillation probability reaches a constant value
$4|\epsilon|^2\sin^2(LV/2)$ at high energies.

The behaviour of the transition probabilities at sufficiently high
energy is therefore drastically different in the presence of
non-standard flavour-changing interactions. 
Note also that at a Neutrino Factory, the energy independent
transition probability would be enhanced by the growth with energy of
the neutrino flux and of the neutrino cross section, thus giving rise
to a striking growth of the signal with energy. 
Of course, the interest of this observation depends on how large the
`sufficiently high' energy at which the energy-enhanced non-standard
effect dominates. 
This in turn depends on the entry of the $\Meff^2$ matrix under
consideration. 
In order for the NP effects to emerge in the ``atmospheric'' 23 block
in \AReq{MM2}, the $E/\Eres$ enhancement must be very large,
as the new $(E/\Eres)\etu$ effect competes with 1/2 and the limits on
$\etu$ are relatively severe. 
The situation is more promising in the 12 and 13 entries, where the
vacuum matrix element is suppressed by $s_{13}/\sqrt{2}$, so that the
new effect has a better chance to emerge. 
Particularly promising is the 13 entry, as values of $\ete$ as large
as 0.1 or more are not excluded (the limits on $\eue$ are the 
most stringent). 
Note that due to the large $\nu_\mu$--$\nu_\tau$ mixing, a large $\ete$
would also affect the $\nu_e\rightarrow\nu_\mu$ transitions, but would not
give rise, in this case, to an energy enhancement. 
This is because the large $\theta_{23}$ mixing communicating the
effect of $\ete$ to the $\nu_e\to\nu_\mu$ transition takes place at
the atmospheric mass-squared difference $\ARdm{31}$, which in the
large $\Eres/E$ limit is subleading compared to the other mass-squared
difference, $2EV$. 
This is also confirmed by \AReq{he2} ($\ete$ does not affect
$P(\nu_e\rightarrow\nu_\mu)$ at the leading order in $\Eres/E$). 

The $\ete$ term exceeds the standard term at energies $E\gtrsim\ENP =
|s_{13}/(\sqrt{2}\,\epsilon)|\Eres$. 
The regime in which the new effects are comparable to the standard ones
is therefore within the reach of a machine producing neutrinos of
energy $E_\nu$ such that: 
\begin{equation}
  \label{Eq:AR:threshold}
  |\epsilon| \gtrsim \frac{|s_{13}|}{\sqrt{2}}
   \frac{\Eres}{E_\nu} \; .
\end{equation}
At higher energies the non-standard effects start to dominate, and the
transition probability becomes constant in energy. 
For example, at a machine producing neutrinos with an energy of
$50$~GeV, the new effects are at least comparable to the standard ones
if $|\ete|\gtrsim 0.007 (|s_{13}|/0.05)$. 
Recall that $|s_{13}|=0.05$ corresponds to 
$\sin^2 2\theta_{13} = 10^{-2}$, a value not very far from the present 
bound and well within the typical sensitivity of a Neutrino Factory.

To investigate the sensitivity to CP-violating phases, define 
$\epsilon = |\epsilon|e^{i\phi}$.
The phase convention being considered is one in which the $\epsilon$'s
are the only complex parameters. 
In an alternative convention, in which the $\delta$ phase has not been
reabsorbed in $\epsilon$, the physical phase would be $\delta - \phi$.
\AReq{he2} shows that in the high-energy limit, the
probabilities depends on $\cos\phi$. 
This dependence is different in the neutrino and anti-neutrino channels,
as the matter effects in the anti-neutrino channel have opposite sign. 
As a consequence, $\cos\phi$ could be determined together with
$|\ete|$. 
The absolute value $|\ete|$ could in fact be determined in the 
high-energy regime, in which $|\ete|$ dominates the transition
amplitude. 
$\cos\phi$ could then be determined in the $E_\nu\sim\ENP$ regime in
which the interference between the standard and non-standard terms is
maximal. 
If $\cos\phi>0$, the two terms would interfere constructively in the
neutrino channel and destructively in the anti-neutrino one, while if
$\cos\phi<0$, the two terms would be destructive for neutrinos and
constructive for anti-neutrinos. 
The previous considerations hold of course provided that $\ENP>\Eres$,
or $|s_{13}/(\sqrt{2}\epsilon)|>1$. 
For $\ENP\lesssim\Eres$ the cancellation is spoiled by the $\ARdm{31}$
terms. 

If the condition in \AReq{threshold} is met in at least a
portion of the neutrino spectrum, the $\nu_\tau$ spectrum shows a
surprising enhancement at high energy. 
Direct $\tau$ detection is challenging and would require a very
granular detector for $\tau$ identification.
On the other hand, a coarse detector with only
muon-charge-identification capability would not miss the peculiar
feature of the signal, since the $\nu_e\to\nu_\tau$ channel
contributes to the wrong-sign muon spectrum through $\tau\to\mu$ decay
(B.R. $\approx 17\%$). 
Moreover, the unequivocal departure from the MSW prediction represents
a clean signal and allows the effect to be separated from standard
oscillations or from corrections due to the NSIs at production or
detection.
A detector capable of distinguishing electron-like from
neutral-current-like events would also be sensitive to the large
increase of the latter due to hadronic tau decays. 

Consider now a specific, favourable, case with oscillation parameters 
$\theta_{23}=\pi/4$, $\ARdm{31}=3\times 10^{-3}\eV^2$,
$\ARdm{21}=0\eV^2$ and $\sin^2 2\theta_{13}=0.001$ (the smaller the
value of $\sin^2 2\theta_{13}$, the more visible the new physics
effects). 
As for the $\epsilon$ parameters, the effect of $\eue$ on oscillation
probabilities is negligible, given the bounds discussed above. 
An $\etu$ at the experimental bound could give rise to non-negligible
effects~\cite{Gago:2001xg} but not to the high-energy enhancement we
are focussing on. 
We therefore set both $\eue = 0$ and $\etu = 0$ and we choose $\ete =
0.07$.

\begin{figure}
  \begin{center}
  \includegraphics[width=8cm]{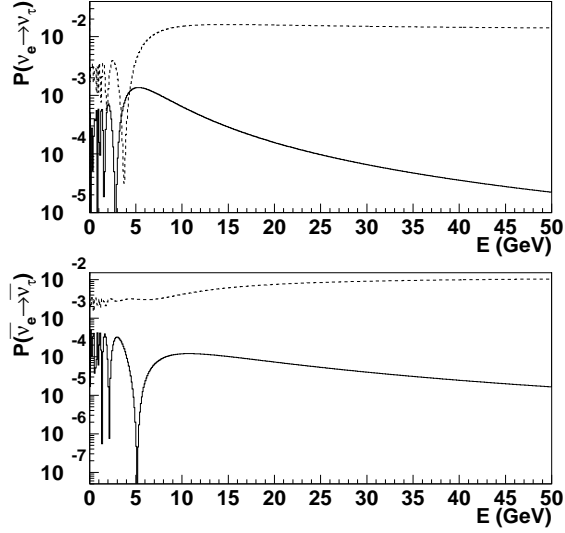}
  \end{center}
  \caption{
    The $\nu_e\rightarrow\nu_\tau$ and
    $\bar\nu_e\rightarrow\bar\nu_\tau$ oscillation probabilities in
    the standard case (full line) and in the presence of new physics
    (dashed line), for $\sin^2 2\theta_{13}=0.001$ and $\ete=0.07$.
    Adapted with kind permission of the Physical Review from figure 2 in
    reference \cite{Campanelli:2002cc}.
    Copyrighted by the American Physical Society.
  }
  \label{Fig:AR:probet}
\end{figure}

The oscillation probability in matter in the standard case is compared
to the oscillation probability in the presence of new physics in figure
\ref{Fig:AR:probet}.
While the standard oscillation probability decreases like $1/E_\nu^2$,
in the presence of new physics the probability reaches a constant
value at high energies larger than 10~GeV or so.
The difference is striking at high energy. 
For anti-neutrinos, the same behaviour is observed at high energy, but
a difference is noted at energies $E\sim\ENP$ or below. 
There, the two terms in the amplitude in equation (\ref{Eq:AR:he2}a) are
comparable and their relative sign is opposite for neutrinos and
anti-neutrinos. 
As in the present example $\ete>0$, a suppression of the probability
in the anti-neutrino channel is clearly visible.
The difference between the two CP-conjugated channels at $E\sim\ENP$
represents a powerful tool to constrain the phase of $\ete$. 
Note also that the behaviour at small $E$ strongly depends on the
$\ARdm{21}=0$ assumption, which has been kept for purposes of
illustration.

\begin{figure}
  \begin{center}
  \includegraphics[width=8cm]{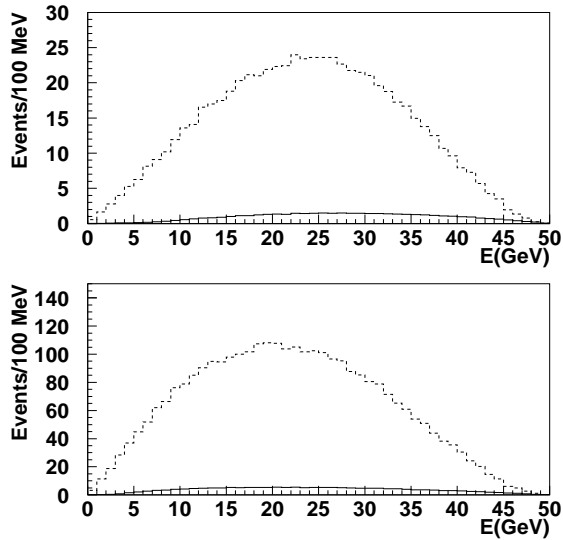}
  \end{center}
  \caption{
    Spectrum of wrong-sign muon events in a neutrino factory as
    described in the text in the case of $\mu^-$ (upper plot) and
    $\mu^+$ (lower plot) circulating in the storage ring.
    The full histogram corresponds to the standard case, the dashed
    histogram to the presence of new interactions.
    Taken with kind permission of the Physical Review from figure 4 in
    reference \cite{Campanelli:2002cc}.
    Copyrighted by the American Physical Society.
  }
  \label{Fig:AR:events-3}
\end{figure}

Consider now a Neutrino Factory with $10^{21}$ muon decays and a
40~kton detector with only muon identification capabilities, located
at a distance of 3000 km from the accelerator.  
Given the significant enhancement of the $\nu_e\to\nu_\tau$ transition
probability at high energy in the present example, we expect the
effect to be visible in the wrong-sign muon spectrum due to
$\tau\to\mu$ decays. 
The effect is indeed manifest in figure \ref{Fig:AR:events-3}, where
the spectrum of wrong-sign muon events in the standard case is
compared to the spectrum in the presence of new physics. 
The large difference between the two cases is essentially due to
$\tau$ decays. 
The wrong-sign muon signal due to $\nu_e\to\nu_\mu$
oscillations is in this case sub-leading in most of the energy range
and is significant only at intermediate energies
\cite{Huber:2001de,Huber:2002bi}. 

\subsubsection[Constraints from non-oscillation neutrino experiments]
{Constraints on non-standard interactions from
non-oscillation neutrino experiments}
\label{davidson}

In this section, bounds on NSIs arising from experiments in which
Standard Model parameters have been determined are presented
\cite{Davidson:2003ha,Berezhiani:2001rs,Barger:1991ae}.
These experiments include short-baseline neutrino experiments with
which $\sin^2 \theta_W$ was measured, LEP, and experiments used to
measure weak decays. 
There are also constraints from oscillation and astrophysical
experiments, which will be discussed in section \ref{friedland}. 
The four-fermion operators considered are of the form
$(\bar{\nu}_{\alpha} \gamma {\nu}_{\beta}) (\bar{f} \gamma f)$, where
$f$ is an electron or a first-generation quark. 
These operators differ from those of section \ref{YG:subsec}, in that
they have two neutrino legs (of possibly different flavour), and the 
remaining two legs are first-generation fermions of the same type.

Consider non-standard, neutral current, neutrino interactions
of the form of equation (\ref{Eq:AR:L}):
\beq
  {\cal L}_{eff}^{NSI} = -
  \sum_{P,f,\alpha, \beta} \varepsilon^{fP}_{\alpha \beta} 2 \sqrt{2} G_F
  (\bar{\nu}_\alpha \gamma_{\rho} L \nu_\beta) (\bar{f} \gamma^{\rho}P f)
  \; ,
  \label{SD-eps}
\eeq
where $f$ is a first-generation SM fermion ($e, u$ or $d$),
$\alpha$, $\beta$ are lepton flavour indices, and $P = L$ or $R$.   
The phase convention is such that $\varepsilon^{fP}_{\alpha \beta}$ is
real (CP violation in the new interactions in included in sections
\ref{YG:subsec} and \ref{friedland}. 
See also \cite{Gago:2001xg,Gonzalez-Garcia:2001mp}).
As in equation (\ref{YG:defeps}), non-standard interactions are normalised
as a perturbation away from  $G_F \rightarrow G_F(1 + \varepsilon)$.
However, the indices used here on $\varepsilon^{fP}_{\alpha \beta}$
differ from equation (\ref{YG:defeps}): $P = L$ [$R$] is allowed
in equation (\ref{SD-eps}), which gives NSI of the form $(V-A)(V-A)$
[$(V-A)(V+A)$], and the fermions $f$ are here restricted to be first
generation of the same flavour. 
So, for instance, we do not constrain the interaction discussed in
section \ref{YG:subsec}, for example equation (\ref{YG:lrssou}), because
it changes the flavour of the charged lepton. 

The four-fermion vertices of equation (\ref{SD-eps}) can be generated by
operators of dimension six, eight, and higher \cite{Berezhiani:2001rs},
with increasing powers of the Higgs-doublet vacuum-expectation value
(vev).
Due to Standard Model gauge symmetries, if equation (\ref{SD-eps})
arises at dimension six, then a 
$(\bar{\ell}\gamma \nu)(\bar{f} \gamma f^{`})$ operator arises with a
coefficient of the same order \cite{Bergmann:1999pk}.   
As discussed in section \ref{SubSect:AR}, charged-lepton physics
imposes tight constraints on the coefficients of such dimension-six
operators. 
However at dimension eight, an operator as in equation (\ref{SD-eps})
can appear at tree level without any charged-lepton counterpart
\cite{Berezhiani:2001rs}; the constraints summarised in this section
apply in this case. 
Notice that at dimension eight, $\varepsilon \propto v^4/\Lambda^4$,
where $v$ is the Higgs vev and $\Lambda$ the scale of new physics
The bounds presented below have been derived on the assumption that
only one operator is present at a time; the limits can be relaxed
when several NSIs are considered simultaneously
\cite{Davidson:2003ha}.

Non-standard interactions involving $\nu_e$ or $\nu_\mu$ and either
electrons or first-generation quarks, can be constrained by
neutrino-scattering data.
Such interactions would contribute to the neutral-current
cross section, in neutrino-beam experiments which determine 
$\sin^2 \theta_W$ by comparing the neutral-current and charged-current
event rates. 
Neutrino-flavour-diagonal NSIs interfere with the SM amplitudes, so
they contribute linearly.
The flavour changing $\varepsilon^{fP}_{\alpha \beta}$, 
$\alpha \neq \beta$, contribute quadratically, as in equation
(\ref{YG:twogen}).
Bounds are obtained from the CHARM \cite{Dorenbosch:1986tb}, CHARM II
\cite{Vilain:1994qy}, LSND \cite{Auerbach:2001wg}, and the NuTeV
\cite{Zeller:2001hh} experiments by requiring that the Standard Model
+ NSI contribution fit within the 90\% C.L. experimental result.  
The Standard Model parameters are taken from other precision data, and
the constraints are listed in table \ref{SD1}.  
NuTeV's results disagree with the Standard Model prediction, so in the
table, the NSIs which could fit this discrepancy have non-zero values. 
If the NuTeV result is supposed to have some other explanation, this
nonetheless gives an estimate of the sensitivity of the NuTeV data to
NSI.
\begin{table}
  \caption{
  \label{SD1}
    Current  90 $\%$ CL limits, that can be set on the coefficients $2
    \sqrt{2} G_F \varepsilon$ of four fermion vertices involving two
    neutrinos and two first generation fermions.
    See equation (\ref{SD-eps}) for the definition of $\varepsilon$.   
    The limits marked with an asterisk, $\as$, arise at one loop and
    are inversely proportional to $\log(\Lambda/m_W)$, taken $\gsim 1$. 
    The superscript $L,R$ of $\varepsilon$ is the chiral projector 
    $P= \{L,R \}$ in the operator.
  }
\begin{center}
\begin{tabular}{||  c | c | c ||} \hline \hline
  vertex& current limits& experiment\\
\hline \hline
 $(\bar{e} \gamma^{\rho} P {e} ) (\bar{\nu}_{\tau} \gamma_{\rho} L   \nu_\tau )$
& $|\varepsilon^{eP}_{ {\tau} {\tau} }|< 0.5$
& $(Z \rightarrow \bar e e )\as$  \\ \hline
 $(\bar{u} \gamma^{\rho} P {u} ) (\bar{\nu}_{\tau} \gamma_{\rho} L   \nu_\tau ) $
& $|\varepsilon^{uL}_{ {\tau}\tau }| < 1.4 $ ,  $|\varepsilon^{uR}_{ {\tau}\tau }| < 3 $
& $(Z \rightarrow \bar \nu \nu)\as$  \\ \hline
 $(\bar{d} \gamma^{\rho} P {d} ) (\bar{\nu}_{\tau} \gamma_{\rho} L   \nu_\tau )$
& $|\varepsilon^{dL}_{ \tau\tau}| < 1.1$,  $|\varepsilon^{dR}_{ \tau\tau}| < 6$
& $(Z \rightarrow \bar \nu \nu)\as$  \\ \hline
%
 $(\bar{e} \gamma^{\rho} P {e} ) (\bar{\nu}_{\mu} \gamma_{\rho} L   \nu_\mu )$
& $|\varepsilon^{e P}_{ \mu \mu}| < 0.03 $ & CHARM II \\ \hline
$(\bar{u} \gamma^{\rho} P {u} ) (\bar{\nu}_{\mu} \gamma_{\rho} L   \nu_\mu )$
& $\varepsilon^{uL}_{ {\mu} \mu } = -0.0053 \pm 0.0032 $
, $ |\varepsilon^{uR}_{{\mu} \mu }| < 0.006 $  & NuTeV \\
   \hline
 $(\bar{d} \gamma^{\rho} P {d} ) (\bar{\nu}_{\mu} \gamma_{\rho} L  \nu_\mu )$
& $\varepsilon^{dL}_{ {\mu} \mu} = 0.0043 \pm 0.0026 $ ,
   $|\varepsilon^{dR}_{ {\mu} \mu }| < 0.013 $  & NuTeV \\
   \hline
%
 $(\bar{e} \gamma^{\rho} P {e} ) (\bar{\nu}_{e} \gamma_{\rho} L \nu_e
   )$
& $-0.07 < \varepsilon^{eL}_{ {e} e} <0.1 $
, $-1 < \varepsilon^{eR}_{
   ee} <0.5 $
& LSND  \\\hline
 $(\bar{u} \gamma^{\rho} P {u} ) (\bar{\nu}_{e} \gamma_{\rho} L \nu_e
   )$
& $-1 < \varepsilon^{uL}_{ ee } < 0.3$
, $ -0.4 < \varepsilon^{uR}_{ ee } < 0.7 $
& CHARM
 \\ \hline
 $(\bar{d} \gamma^{\rho} P {d} ) (\bar{\nu}_{e} \gamma_{\rho} L \nu_e
   )$
& $ | \varepsilon^{dL}_{ ee}| < 0.3$
, $  |\varepsilon^{dR}_{ee}| < 0.5$  & CHARM \\
   \hline \hline
 $(\bar{e} \gamma^{\rho} P {e} ) (\bar{\nu}_{\tau} \gamma_{\rho} L \nu_\mu )$
&
 $| \varepsilon^{eP}_{\tau
  \mu}| < 0.4$  & $( \tau \rightarrow  \mu \bar{e} e )\as$ \\
 & $|  \varepsilon^{eP}_{ \tau
  \mu}| < 0.1$  &    CHARM II \\ \hline
 $(\bar{u} \gamma^{\rho} P {u} ) (\bar{\nu}_{\tau} \gamma_{\rho} L \nu_\mu )$
  & $|  \varepsilon^{uP}_{{\tau}
  \mu}| < 0.05   $
 & NuTeV \\ \hline
 $(\bar{d} \gamma^{\rho}  P{d} ) (\bar{\nu}_{\tau} \gamma_{\rho} L \nu_\mu )$
  & $|  \varepsilon^{dP}_{{\tau}
  \mu}| < 0.05   $
 & NuTeV \\ \hline
 $(\bar{e} \gamma^{\rho} P {e} )  (\bar{\nu}_{\mu} \gamma_{\rho} L \nu_e )$  &
   $| \varepsilon^{eP}_{ {\mu}
  e}| <  5 \times 10^{-4}$  &$( \mu \rightarrow  3 e )\as$  \\
  \hline
 $(\bar{u} \gamma^{\rho} P {u} ) (\bar{\nu}_{\mu} \gamma_{\rho} L \nu_e )$
 &
$| \varepsilon^{uP}_{ {\mu}
  e}| < 7.7\times 10^{-4} $ &$({\rm Ti} \mu \rightarrow {\rm  Ti} e )\as$  \\\hline
 $(\bar{d} \gamma^{\rho} P {d} )  (\bar{\nu}_{\mu} \gamma_{\rho} L \nu_e )$ &
$| \varepsilon^{dP}_{ \mu e
 }| < 7.7\times 10^{-4} $
  &$({\rm Ti} \mu \rightarrow  {\rm Ti} e )\as$  \\
   \hline
%
 $(\bar{e} \gamma^{\rho} P {e} )
   (\bar{\nu}_{\tau} \gamma_{\rho} L \nu_e )$  &
  $| \varepsilon^{eP}_{ {\tau}
  e}| < 0.8 $
 &  $( \tau \rightarrow  e \bar{e} e )\as$ \\
 &$|  \varepsilon^{eL}_{{\tau}e
  }| < 0.4, |  \varepsilon^{eR}_{ {\tau} e }| < 0.7   $
 &  LSND  \\
\hline
 $(\bar{u} \gamma^{\rho} P {u} )
   (\bar{\nu}_{\tau} \gamma_{\rho} L \nu_e )$  &
   $| \varepsilon^{uP}_{ {\tau}
  e}| < 0.7 $   &$( \tau \rightarrow  e \pi)\as$
    \\
  &$|  \varepsilon^{uP}_{{\tau}e
  }| < 0.5,  $
 &  CHARM \\ \hline
 $(\bar{d} \gamma^{\rho} P {d} )
   (\bar{\nu}_{\tau} \gamma_{\rho} L \nu_e )$ &
   $| \varepsilon^{dP}_{\tau e}
 | < 0.7 $  &$( \tau \rightarrow  e \pi)\as$
    \\
  &$|  \varepsilon^{dP}_{{\tau}e
  }| < 0.5,  $
 &  CHARM \\
 \hline
  \hline
\end{tabular}
\end{center}
\end{table}

Bounds on the interactions in equation (\ref{SD-eps}) can also be
obtained from radiative corrections. 
$W$ exchange between $\bar{\nu}$ and $\nu$ or $f$ will generate
effective interactions 
$(\bar{\ell}_\alpha \gamma_{\rho} L \ell_\beta) (\bar{f} \gamma^{\rho}P f)$ 
or
$(\bar{\ell}_\alpha \gamma_{\rho} L \nu_\beta) (\bar{f} \gamma^{\rho}L f^{`})$,
where $\ell$ is a charged lepton. 
This Standard Model loop transforms the non-standard neutrino
interaction to a charged-lepton interaction of strength 
$c \times 2 \sqrt{2} G_F \varepsilon^{fP}_{\alpha \beta}$, where:
\begin{equation}
  \label{SD-c}
  c \simeq \frac{\alpha }{4\pi s^{2}_{W}}\ln \left(
  \frac{\Lambda}{m_{W}}\right)
  \approx 0.0027 \, ,
\end{equation}
and $\Lambda$ is a new-physics scale which may conservatively be taken
to be $\sim$ TeV. 
Charged-lepton data can therefore constrain these NSI, even if the NSIs
do not involve charged leptons at tree level. 

The experimental bounds on $\mu \leftrightarrow e$ flavour change from
the charged-lepton sector (e.g. $\mu \rightarrow 3e$, 
$\mu \rightarrow e$ conversion on titanium) are very strong.  
Despite the loop-suppression factor of equation (\ref{SD-c}), they give
significant constraints on NSIs involving $\nu_\mu $ and $\nu_e$: 
$\varepsilon \lsim 10^{-3}$, see table \ref{SD1}.
The constraints from flavour-changing $\tau$ decays are weaker,
$\varepsilon \lsim 1$.
The upper limits on the $\tau$-decay branching ratios may improve in
the future; the limits in table \ref{SD1} scale as $\sqrt{BR}$, and
are calculated from 
$BR( \tau \rightarrow  \pi e ) =
BR( \tau \rightarrow  \mu \bar{e} e ) = 1.9 \times 10^{-7}$, and
$BR( \tau \rightarrow  e \bar{e} e ) = 2.0 \times 10^{-7}$.

Constraints on the non-standard interactions 
$(\bar{\nu}_\tau \gamma_{\rho} L \nu_\tau) (\bar{f} \gamma^{\rho}P f)$ 
can be obtained from their loop contribution to $Z$ decay. 
If the $Z$ decays to $\bar{\nu}_\tau \nu_\tau$, which then become
$\bar e e$ via the NSI $\varepsilon_{\tau \tau}^{eP}$, this
contributes to the decay $Z \rightarrow \bar{e} e$. 
Or if the $Z$ decays to $q \bar{q}$ ($q$ = $u$ or $d$), which become
$\bar{\nu}_\tau \nu_\tau$ via $\varepsilon_{\tau \tau}^{uP}$ or 
$\varepsilon_{\tau \tau}^{dP}$, this contributes to the invisible
width of the $Z$.  
The $Z$ decay branching ratios were measured at LEP to a precision
$\sim \alpha_{em}/\pi$, and support the global fits to Standard Model
parameters. 
This gives constraints of order $\varepsilon^{fP}_{\tau \tau} \sim 1$; 
see table \ref{SD1}. 
Better bounds on $\varepsilon_{\tau \tau}^{fP}$ can be found in
section \ref{friedland}.

\subsubsection{Oscillation experiments as probes of the NSI}
\label{friedland}

The effective low-energy operators induced by non-standard
interactions may appreciably modify the neutrino forward-scattering
amplitude on electrons and nucleons, as a result affecting neutrino
oscillations in matter.   
This makes neutrino-oscillation experiments a valuable low-energy tool
in searching for physics beyond the Standard Model. 
As the precision of neutrino-oscillation experiments increases, they
may begin to be regarded on the same footing as the existing precision
low-energy tools, such as the measurements of $K-\bar{K}$ mixing,
searches for flavour violating $\mu$ and $\tau$ decays, etc. 
In this section, a review of the sensitivity the existing
neutrino-oscillation experiments, including solar, reactor,
atmospheric, and accelerator neutrinos, to NSIs is presented.

\paragraph{NSI and oscillations: generalities}

Regardless of their origin, at the low energies relevant to neutrino
oscillations, NSIs are described by the effective low-energy,
four-fermion Lagrangian:
\begin{eqnarray}
  L^{NSI} = - 2\sqrt{2}G_F (\bar{\nu}_\alpha\gamma_\rho\nu_\beta)
  (\epsilon_{\alpha\beta}^{f\tilde{f} L}\bar{f}_L \gamma^\rho
  \tilde{f}_L + \epsilon_{\alpha\beta}^{f\tilde{f}
  R}\bar{f}_R\gamma^\rho \tilde{f}_{R})+ h.c. 
  \label{eq:lagNSI}
\end{eqnarray}
Here $\epsilon_{\alpha\beta}^{f\tilde{f} L}$
($\epsilon_{\alpha\beta}^{f\tilde{f} R}$) denotes the strength of
the NSI between the neutrinos $\nu$ of flavours $\alpha$ and
$\beta$ and the left-handed (right-handed) components of the
fermions $f$ and $\tilde{f}$.

Not all of these parameters impact neutrino oscillations in matter.
The propagation effects of NSI are, first of all, only sensitive to
$\epsilon_{\alpha\beta}^{f\tilde{f}}$ when there is no flavour change
of the background particle, $f=\tilde{f}$, as processes that change
the flavour of the background fermion do not add up coherently
\cite{nunu1}. 
Henceforth, the notation 
$\epsilon_{\alpha\beta}^{ffP}\equiv\epsilon_{\alpha\beta}^{fP}$ will
be used.
Secondly, only the vector component of the NSI enters,
$\epsilon_{\alpha\beta}^{f}\equiv\epsilon_{\alpha\beta}^{fL}+\epsilon_{\alpha\beta}^{fR}$,
with no sensitivity to the axial component. 
Therefore, the propagation and production/detection effects are
sensitive to different combinations of the NSI parameters, and hence
the corresponding measurements are complementary.

The matter piece of the oscillation Hamiltonian can be written (up
to an irrelevant overall constant) as:
\begin{equation}
H_{\rm mat}^{3\times3}=\sqrt{2}G_F n_e\begin{pmatrix}
1+\epsilon_{ee} & \epsilon_{e\mu}^\ast & \epsilon_{e\tau}^\ast \\
\epsilon_{e\mu} & \epsilon_{\mu\mu}& \epsilon_{\mu\tau}^\ast \\
\epsilon_{e\tau} & \epsilon_{\mu\tau} & \epsilon_{\tau\tau} \\
\end{pmatrix},
\label{eq:ham}
\end{equation}
where $n_e$ is the number density of electrons in the medium. The
epsilons here are the sum of the contributions from electrons
($\epsilon^{e}$), up quarks ($\epsilon^{u}$), and down quarks
($\epsilon^{d}$) in matter: $\epsilon_{\alpha\beta}\equiv
\sum_{f=u,d,e}\epsilon_{\alpha\beta}^{f}n_f/n_e$. Hence, unlike in
the standard case ($\epsilon_{\alpha\beta}=0$), the NSI-matter
effects depend on the chemical composition of the medium, not
only on the electron density, $n_e$.

The idea that non-standard neutrino interactions modify neutrino
oscillations in matter has been around for many years. 
It is already clearly spelled out in the seminal paper by Wolfenstein
\cite{Wolfenstein:1977ue} and has been elaborated by many authors
(\cite{Valle:1987gv,Roulet:1991sm, Guzzo:1991hi} and many others).
While in the 1980's and 1990's the focus was mainly on NSI as an
alternative to oscillations, in recent years the focus has shifted to
using neutrino-oscillation data to measure neutrino interactions. 

Because of the tight bounds on the parameters $\epsilon_{e\mu}$
and $\epsilon_{\mu\mu}$ (see Sect.~\ref{davidson}), it makes sense
to set them to zero while considering neutrino oscillations.
Moreover, the parameters $\epsilon_{\mu\tau}$ will also be set to
zero. 
This parameter was shown to be constrained
($\epsilon_{\mu\tau}<10^{-1}$) by the two-flavour analysis of the
atmospheric neutrino data \cite{Fornengo:2001pm}. 
Although a full 3-flavour analysis including $\epsilon_{\mu\tau}$ is
yet to be done, there are arguments that suggest that the two-flavour
bound may survive the generalisation to three flavours (unlike the
corresponding bound on $\epsilon_{e\tau}$, see
Sect.~\ref{sect:NSIatm}). 
Thus, only the effects of $\epsilon_{ee}$, $\epsilon_{e\tau}$,
and $\epsilon_{\tau\tau}$ will be considered. 
Even with this reduction, the parameter space of the problem is quite
large: different assignments of the diagonal and off-diagonal NSI to
electrons, and $u$ and $d$ quarks yield different dependences of the
oscillation Hamiltonian on the chemical composition and different
detection cross sections. 

\paragraph{NSI and solar neutrinos}

It is well known that the standard solar-neutrino analysis can be
done with only two neutrino states: $\nu_e$ and $\nu_\mu'$, where
the latter is a linear combination of $\nu_\mu$ and $\nu_\tau$
(The effect of the third state is to multiply the two-neutrino
survival probability by $\cos^4\theta$. 
See, e.g. \cite{Goswami:2004cn,Fogli:2005cq} for recent data
analyses.) 
This reduction involves performing a rotation in the $\mu-\tau$
sub-space by the atmospheric angle $\theta_{23}$ and taking  the first
two columns/rows of the mixing matrix. 
The vacuum-oscillation Hamiltonian then takes the usual form:
\begin{eqnarray}
  H_{\rm vac}^{2\times2} = \left(
  \begin{array}{rr}
    -\Delta \cos 2\theta & \Delta \sin 2\theta \\
    \Delta \sin 2\theta & \Delta \cos 2\theta
  \end{array} 
  \right),
  \label{eq:VAC_convention}
\end{eqnarray}
where $\Delta\equiv \Delta m^2/(4 E_\nu)$ and $\Delta m^2$ is the
mass splitting between the first and second neutrino mass states:
$\Delta m^2\equiv m^2_2-m^2_1$.

It turns out (quite fortunately and unlike the atmospheric
neutrino case, see section \ref{sect:NSIatm}) that the two-neutrino
reduction of the solar-neutrino analysis holds even when the
matter interactions become non-standard. 
The corresponding matter contribution to the two-neutrino oscillation
Hamiltonian can be written (once again, up to an irrelevant overall
constant) as:
\begin{eqnarray}
  H_{\rm mat}^{NSI} = \frac{G_F n_e}{\sqrt{2}}
  \left(\begin{array}{cc}
    1+\epsilon_{11} & \epsilon^\ast_{12} \\
    \epsilon_{12}   & -1-\epsilon_{11}
  \end{array} 
  \right),
  \label{eq:ceciconv}
\end{eqnarray}
where the quantities $\epsilon_{ij}$ ($i=1,2$) depend on the
original epsilons and on the rotation angle $\theta_{23}$:
\begin{eqnarray}
\epsilon_{11}=\epsilon_{ee} - \epsilon_{\tau\tau}
\sin^2\theta_{23},~~~~~~~~ \epsilon_{12}=-2\epsilon_{e\tau} \sin
\theta_{23}.
 \label{eq:NSIdef2}
\end{eqnarray}
In equation (\ref{eq:NSIdef2}) small corrections of order 
$\sin \theta_{13}$ or higher have been neglected. 
Equation (\ref{eq:NSIdef2}) shows that the flavour-changing-NSI effect
in solar-neutrino oscillations comes from $\epsilon_{e\tau}$, while
the flavour-preserving-NSI effect comes from both $\epsilon_{ee}$ and
$\epsilon_{\tau\tau}$.

A useful parameterisation is:
\begin{eqnarray}
  H_{\rm mat}^{NSI} = \left(
  \begin{array}{cc}
    A \cos 2\alpha & A e^{-2i\phi} \sin 2\alpha \\
    A e^{2i\phi} \sin 2\alpha  & -A \cos 2\alpha
  \end{array} 
  \right).
  \label{eq:MAT_convention}
\end{eqnarray}
Here the parameters $A=A(x)$, $\alpha$ and $\phi$ are defined as
follows:
\begin{eqnarray}
  \tan 2\alpha = |\epsilon_{12}|/(1+\epsilon_{11}) ,~~~~~
  2\phi=Arg(\epsilon_{12}),~~~~~
  A= G_F n_e \sqrt{[(1+\epsilon_{11})^2+|\epsilon_{12}|^2]/2}~.
  \label{eq:NSIdef}
\end{eqnarray}
In the absence of NSIs, $A=G_F n_e/\sqrt{2}$, $\alpha=0$, and the
Hamiltonian (equation (\ref{eq:MAT_convention})) reduces to its standard 
form. 
The effect of $\alpha$ is to change the mixing angle in the medium of
high density from $\pi/2$ to $\pi/2-\alpha$. 
The angle $\phi$ (related to the phase of $\epsilon_{e\tau}$) is a
source of CP violation. 
Solar-neutrino experiments, just like terrestrial-beam experiments
\cite{Gonzalez-Garcia:2001mp,Campanelli:2002cc}, are sensitive to its
effects \cite{Friedland:2004pp}, while the atmospheric neutrinos are
not (section \ref{sect:NSIatm}). 

\begin{figure}
  \centering
  \includegraphics[width=0.77\textwidth]{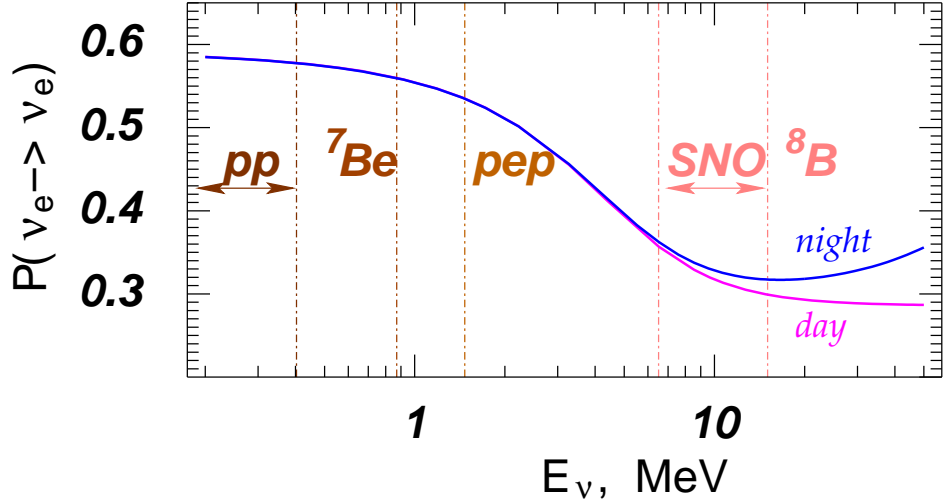}
  \caption{The electron neutrino survival probability and the day/night
  asymmetry as a function of energy for  the LMA solution.}
  \label{fig:PeeLMA}
\end{figure}

To understand the basic physics of the sensitivity of solar neutrinos
to NSI, first consider the electron-neutrino survival probability
$P_{ee}$ for the LMA-I solution in the standard case (no NSI). 
As shown in figure \ref{fig:PeeLMA}, $P_{ee}$ varies across the
solar-neutrino spectrum.  
On the low end ($pp$ neutrinos), it approaches
$\cos^4\theta+\sin^4\theta$. 
This is nothing but the (averaged) vacuum-oscillation value
$1-\sin^22\theta/2$. 
The low-energy solar neutrinos essentially are not affected by the
presence of matter, even at the production point in the core 
($\Delta m^2/2 E_\nu\gg\sqrt{2}G_Fn_e(r)$ for all $r$). 
On the high-energy end ($^8$B neutrinos), the survival probability
approaches $\sin^2\theta$: the Hamiltonian at the production point is
dominated by the matter term.

Between these two extremes lies the transition region where the matter
potential at the production point and the kinetic terms guiding vacuum
oscillations are comparable. 
It is natural to expect that this is the part of the solar-neutrino
spectrum that would be most sensitive to the non-standard neutrino
interactions.

\begin{figure}
  \centering
  \includegraphics[width=0.47\textwidth]{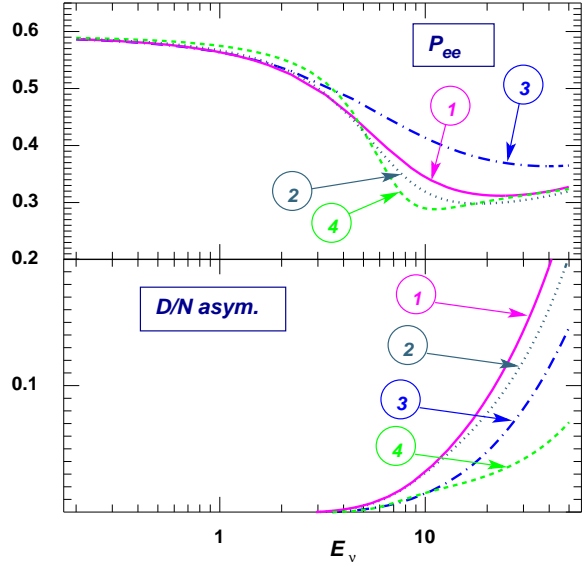}
  \caption{The electron neutrino survival probability and the day/night
  asymmetry as a function of energy for  $\Delta m^2=7\times 10^{-5}$ eV$^2$,
  $\tan^2\theta=0.4$ and several representative values of the NSI parameters: (1)
  $\epsilon_{11}^{u}=\epsilon_{11}^{d}=\epsilon_{12}^{u}=\epsilon_{12}^{d}=0$;
  (2) $\epsilon_{11}^{u}=\epsilon_{11}^{d}=-0.008$,
  $\epsilon_{12}^{u}=\epsilon_{12}^{d}=-0.06$;
  (3) $\epsilon_{11}^{u}=\epsilon_{11}^{d}=-0.044$,
  $\epsilon_{12}^{u}=\epsilon_{12}^{d}=0.14$;
  (4) $\epsilon_{11}^{u}=\epsilon_{11}^{d}=-0.044$,
  $\epsilon_{12}^{u}=\epsilon_{12}^{d}=-0.14$.
  Recall that the parameters in equation (\ref{eq:NSIdef2}) equal
  $\epsilon_{ij}=\epsilon^u_{ij}n_u/n_e+\epsilon^d_{ij}n_d/n_e$.
    Taken with kind permission of Physical Letters from figure 1 in
    reference \cite{Friedland:2004pp}.
    Copyrighted by Elsevier B.V.
}
  \label{fig:PeeAdn}
\end{figure}

Figure \ref{fig:PeeAdn} confirms these expectations. 
It shows that the behaviour of $P_{ee}$ in the transition region varies
considerably with $\epsilon_{e\tau}$,  both in amplitude and sign. 
Values of the order of $10^{-1}$ per quark can have a significant
effect. 
In fact, some of the parameter space can already be excluded as the
distortion of the spectrum at SNO would be unacceptably large. 
As an example, points with $\epsilon_{11} =0$ and
$\epsilon^u_{12}>0.14 $ are unacceptable at $90\%$ C.L. (here
$\epsilon^u_{\alpha\beta}=\epsilon^d_{\alpha\beta}$ is assumed)
\cite{Friedland:2004pp}. 
At the same time, possibilities such as curve 2 or 4 in the figure cannot
presently be excluded. 
Clearly, an excellent way to probe this part of the parameter space
would be to perform a high-statistics measurement of the
$^8$B-neutrino spectrum in the regime of low energies ($<6$ MeV). 

Note also that the day/night-asymmetry effect also changes in the
presence of NSI. 
In particular, for certain values of the NSI parameters, the day/night
asymmetry can be significantly reduced, as is clearly demonstrate by
curve 4 in the bottom panel of figure \ref{fig:PeeAdn}. 
In this case, the LMA-0 solution, characterised by 
$\Delta m^2\sim (1-2)\times 10^{-5}$ eV$^2$ and normally excluded by
the solar data, becomes allowed. 
One way to obtain this solution is by choosing NSI such that the angle
$\alpha$ (defined in equation (\ref{eq:MAT_convention})) becomes close
to $\theta$ \cite{Friedland:2004pp}. 
A choice can be made that is consistent with the atmospheric-neutrino
constraints. 
Another way is by choosing the flavour-preserving NSI to cancel the
standard matter term in the Earth \cite{Guzzo:2004ue}. 
The MSW effect in the Sun still happens in this scenario, because the
Sun has a different chemical composition than the Earth. 
Lastly, we note that it is even possible to obtain a solution for
$\theta>\pi/4$, the so-called LMA-D region \cite{Miranda:2004nb} (in
the `dark side' \cite{deGouvea:2000cq,Friedland:2000cp}). 
This requires quite large NSIs so that the sign of the matter effect
in the Sun is reversed.
For technical details, including approximate analytical expressions
for $P_{ee}$ and the day/night asymmetry, see
\cite{Friedland:2004pp}.

\paragraph{NSI and atmospheric neutrinos}
\label{sect:NSIatm}

On very general grounds, one expects the atmospheric neutrinos to be a
very sensitive probe of NSI. 
The reason is the remarkable agreement between the Super-Kamiokande
atmospheric-neutrino data and the predictions of the standard
$\nu_\mu\rightarrow\nu_\tau$ oscillation scenario. 
The agreement is non-trivial: with only two parameters, $\Delta
m_{atm}^2$ and $\theta_{23}$, it is possible to fit all presently
available Super-Kamiokande data, spanning five orders of magnitude in
energy, $E_\nu$, and three orders of magnitude in baseline, $L$. 
It may be expected that the introduction of non-standard
neutrino-matter interactions would change the oscillation pattern, 
breaking this beautiful fit.

Since the vacuum-oscillation Hamiltonian depends on the combination
$\Delta m^2/E_\nu$, while the non-standard matter potential,
$\sqrt{2}\epsilon_{\alpha\beta}G_F n_f$, is energy independent, the
high-energy part of the data-set is generally expected to be most
sensitive to non-standard interactions.  
The data in question are the stopping and through-going muon samples
\cite{Engel:1999zq} and these should be first examined for NSI
effects.

A simple estimate of the sensitivity could be obtained as follows. 
At very high energies, $E_\nu\gtrsim 50-100$ GeV, the vacuum-oscillation
length, $\sim 4\pi E_\nu/\Delta m^2$, becomes greater than the size of
the Earth.  
The standard oscillation mechanism predicts no oscillations for these
neutrinos. 
If the $\epsilon_{\mu\tau}$ NSI is present, it will drive oscillations
of the highest energy muon neutrinos, in conflict with the data. 
The simple criterion then is that the corresponding oscillation length
in matter, $\sim \pi (\sqrt{2}\epsilon_{\mu\tau}G_F n_e)^{-1}$ be
greater than the Earth's diameter. 
That yields $\epsilon_{\mu\tau}\lesssim 0.1$. 
Detailed two-neutrino ($\nu_\mu,\nu_\tau$) numerical analysis
\cite{Fornengo:2001pm} yields $\epsilon_{\mu\tau}\lesssim 0.08-0.12$
\footnote{
  Notice the difference in normalisations: our epsilons are normalised
  per electron, while \cite{Fornengo:2001pm} gives epsilons per $d$
  quark, resulting in a factor of $\sim4$ apparent difference
}.

With $\epsilon_{\tau\tau}$, the argument is slightly different. 
At the highest energies, where vacuum oscillations are not
operational, $\epsilon_{\tau\tau}$ has no effect. 
The effect appears at lower energies where vacuum oscillations are
predicted to occur: $\epsilon_{\tau\tau}$ introduces diagonal
splitting thus decreasing the effective mixing angle. 
Thus, one needs to compare $\sqrt{2}\epsilon_{\tau\tau}G_F n_e$ and
$\Delta m^2/2E_\nu$ at $E_\nu\sim20-30$ GeV, the highest energy at
which an oscillation minimum is expected to occur for neutrinos
traveling through the center of the Earth. 
This yields $\epsilon_{\tau\tau}\lesssim 0.2$, once again in
reasonable agreement with the numerical two-neutrino analysis
\cite{Fornengo:2001pm}. 

Clearly, these are very strong bounds; if they were to extend to
$\epsilon_{e\tau}$, the NSI effects on solar neutrinos discussed in
the previous sub-section would be excluded. 
It turns out, however, that this is not the case: when the analysis is
properly extended to three flavours, one finds that very large values
of both $\epsilon_{e\tau}$ and $\epsilon_{\tau\tau}$ are still allowed
by the data. 

\begin{figure}
  \centering
  \includegraphics[width=0.45\textwidth]{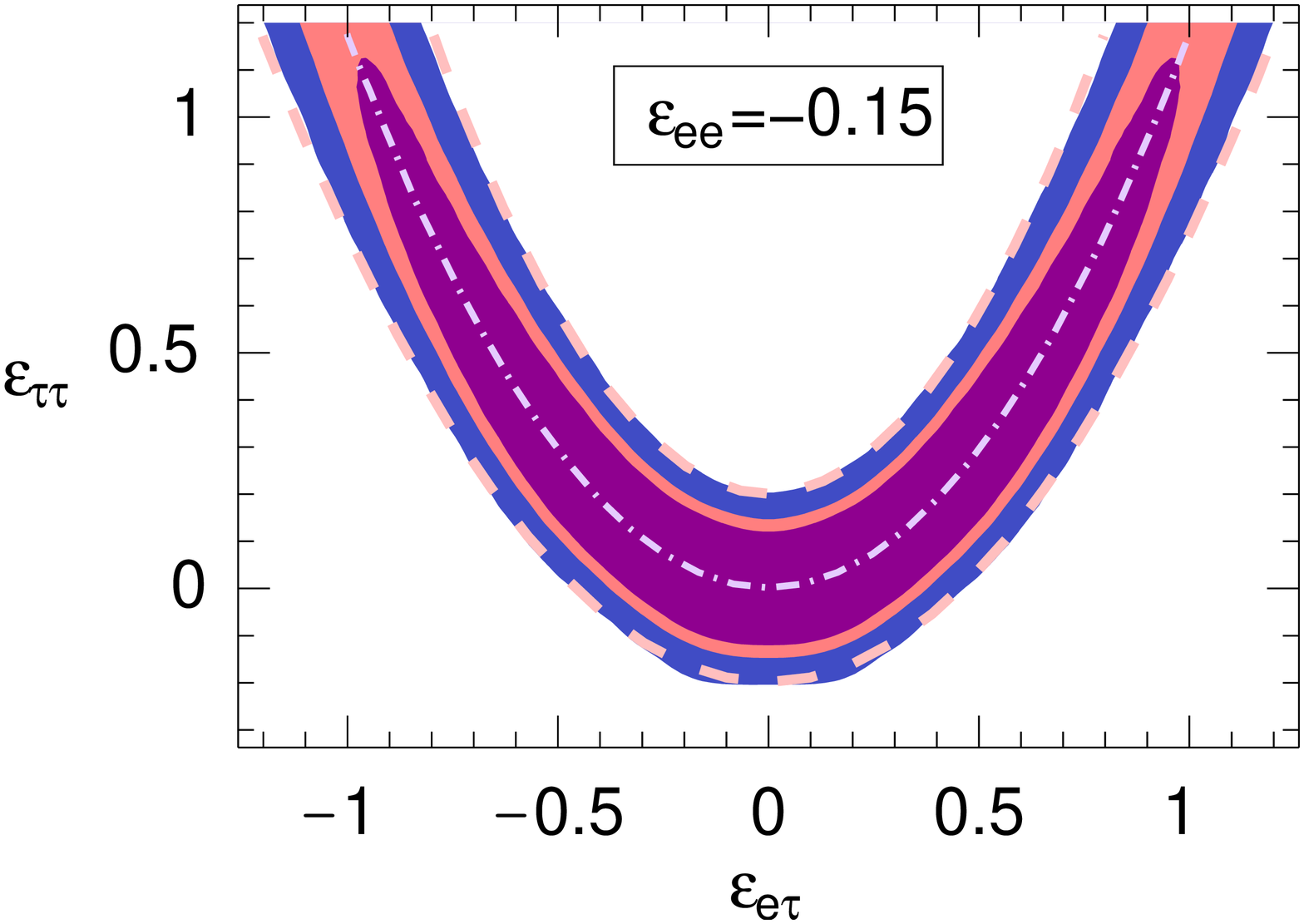}
  \includegraphics[width=0.49\textwidth]{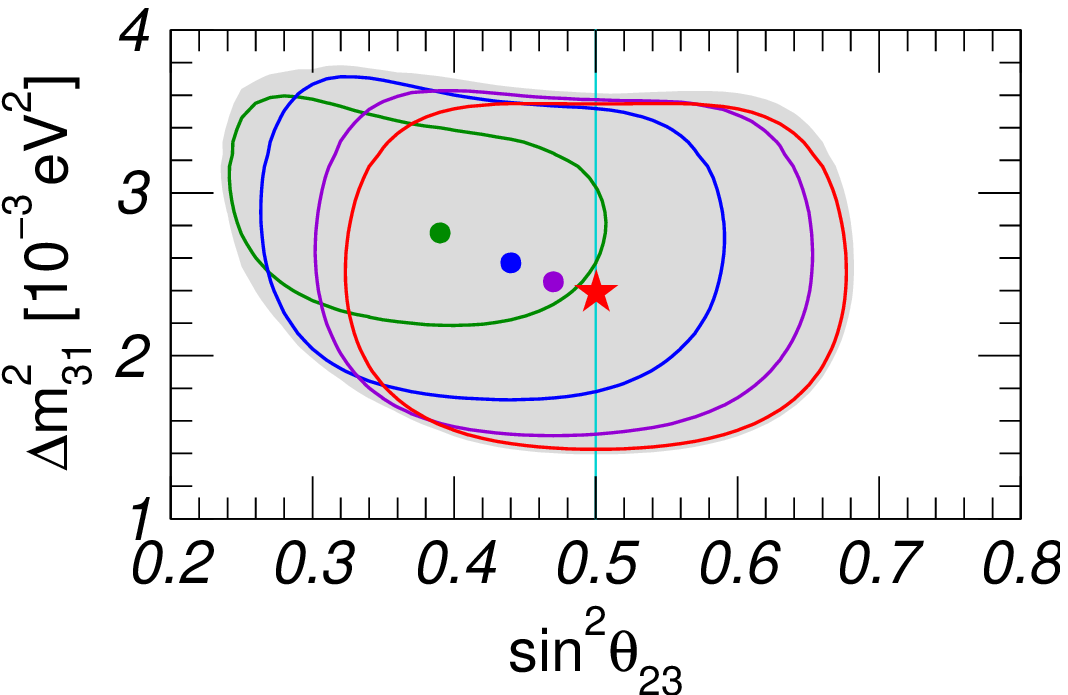}
  \caption{
    \textit{Left panel}: A 2-D section $(\epsilon_{ee}=-0.15)$ of the
    allowed region of the NSI parameters (shaded).
    The results are presented for $\Delta m^2_\odot=0$,
    $\theta_{13}=0$, and marginalised over $\theta$ and $\Delta m^2$. 
    The dashed contours indicate our analytical predictions. 
    See text for details. 
    \textit{Right panel}: The effect of the NSI on the allowed region
    and best-fit values of the oscillation parameters.
    Both figures taken with kind permission of the Physical Review from figures 1 and 2 in
    reference \cite{Friedland:2004ah}.
    Copyrighted by the American Physical Society.
  }
  \label{fig:scan}
\end{figure}

This surprising result is illustrated in the left panel of figure
\ref{fig:scan} (taken from \cite{Friedland:2004ah}), which shows a 2-D
slice of the allowed region in the 3-D parameter space of
$\epsilon_{ee}$, $\epsilon_{e\tau}$, and $\epsilon_{\tau\tau}$.  
Order-one values for both $\epsilon_{e\tau}$ and $\epsilon_{\tau\tau}$
are allowed, in other words, the NSI can be as large as, or even
larger than, the Standard Model neutrino interactions.

The contours presented in the left panel of figure \ref{fig:scan} have
been obtained by marginalising over $\Delta m_{atm}^2$ and
$\theta_{23}$.
The right panel of the figure shows what happens to the oscillation
parameters as one moves along the parabolic direction of the allowed
region away from the origin: the mixing angle becomes less than
maximal, while the mass splitting increases.  
The good fit to the data is maintained at the expense of changing the
oscillation parameters away from their standard values.

Both the shape of the allowed NSI region and the shift of the best-fit
oscillation parameters can be understood physically. 
The allowed region is reasonably well described by the equations:
\begin{eqnarray}
  \label{eq:width}
  |1+\epsilon_{ee}+\epsilon_{\tau\tau}-
  \sqrt{(1+\epsilon_{ee}-\epsilon_{\tau\tau})^2+4 |\epsilon_{e\tau}|^2}|
  \lesssim 0.4,\\
  \label{eq:width2}
  \cos^2 \beta \gtrsim \tan^2 \theta_{min}, \hskip 0.5truecm
  \cos^2 \beta \geq \left[\frac{2 \Delta m^2_{max}}{\Delta m^2_m}-1 \right]^{-1},
\end{eqnarray}
where:
\begin{eqnarray}
  \tan 2\beta &\equiv& 2 |\epsilon_{e\tau}|/(
  1+\epsilon_{ee}-\epsilon_{\tau\tau}); \\
  \label{tanb2eta}
  \Delta m^2_{m} &\equiv& \Delta m^2 \left[
  (\cos 2\theta (1+ \cos^2\beta) - \sin^2\beta)^2/4 + (\sin2\theta
  \cos\beta)^2\right]^{1/2};
\label{deltam2m}
\end{eqnarray}
and $\theta_{min}$ and $\Delta m^2_{max}$ denote the smallest mixing
and the largest mass splitting allowed by the low-energy data, 
$E \lesssim 1$~GeV, which are not affected by NSI. 
The derivation and discussion of these results are found in
\cite{Friedland:2004ah,Friedland:2005vy,Friedland:2006pi}. 
Under the conditions of equations (\ref{eq:width}) and (\ref{eq:width2}),
the high-energy atmospheric muon neutrinos undergo oscillations into a
state that is a linear combination of $\nu_e$ and $\nu_\tau$, instead
of purely into $\nu_\tau$ as in the standard case. 
This fact, however, is unobservable because at the energies in
question only the muon data is available. 
The low-energy neutrinos undergo `normal' vacuum oscillations, since
for them the vacuum-oscillation terms still dominate the Hamiltonian.

Notice that only the absolute value of $\epsilon_{e\tau}$
enters equations (\ref{eq:width}) to (\ref{deltam2m}). 
Unlike solar neutrinos, for $\theta_{13}=0$ atmospheric neutrinos are
completely insensitive to the phase of this parameter, which can be
explicitly seen also in figure \ref{fig:scan}. 
For $\theta_{13}\ne0$ there is some sensitivity, but the effect is
small \cite{Friedland:2005vy}. 

\paragraph{Combined analysis of the atmospheric and K2K data}
\label{sect:atmK2K}

\begin{figure}
  \centering
  \includegraphics[width=0.8\textwidth]{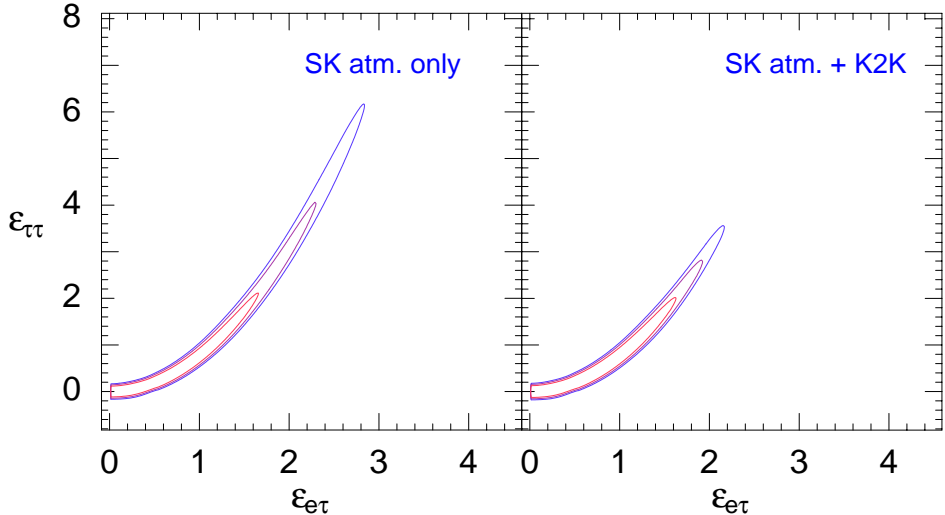}
  \caption{The role of K2K in constraining the allowed region of the
    NSI allowed by the atmospheric neutrino analysis. The value
    $\epsilon_{ee}=0.3$ was chosen.
    Taken with kind permission of Physical Review from figure 4 in
    reference \cite{Friedland:2005vy}.
    Copyrighted by the American Physical Society.
  }
  \label{fig:atmk2kimpact}
\end{figure}

Although K2K by itself is not sensitive to the effects of the
intervening matter because its baseline is too short (see
section \ref{sect:MINOS_firstdata}), the addition of the K2K
oscillation data to the Super-Kamiokande atmospheric data does
restrict the allowed NSIs. 
The reason behind this seemingly counter-intuitive result is that K2K,
by measuring the `true' vacuum oscillation parameters, restricts the
range over which these parameters could be varied to compensate for
the effects of the NSI, as described above. 
A typical impact of adding the K2K dataset is illustrated in
figure \ref{fig:atmk2kimpact}. 

\begin{figure}
  \centering
    \rotatebox{90}{
    \begin{minipage}{10cm}
    \includegraphics[width=0.9\textwidth]{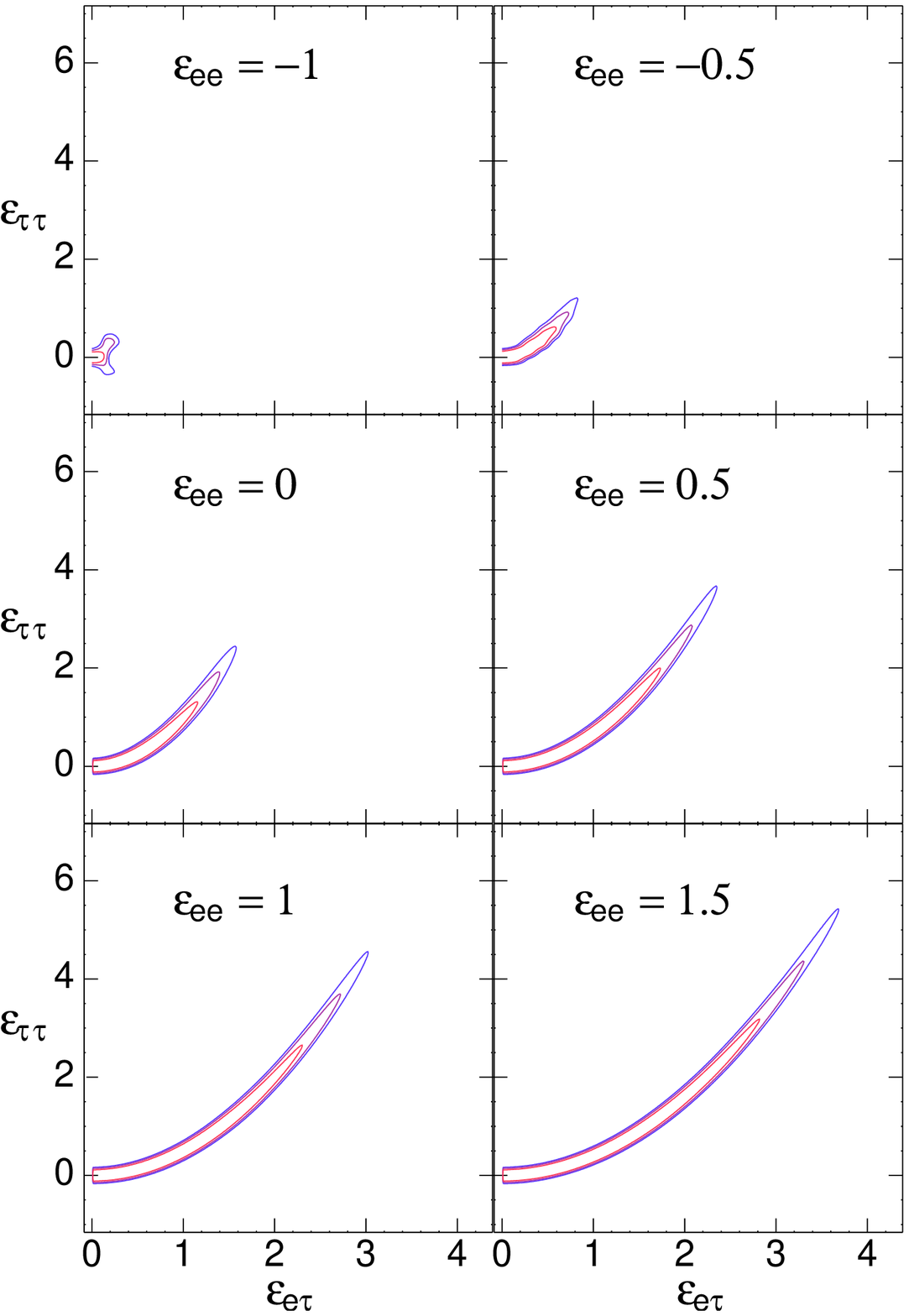}
  \caption{Ranges of the NSI parameters
    allowed by the combined analysis of the atmospheric and K2K data
    in the case of the inverted mass hierarchy. 
    Taken with kind permission of Physical Review from figure 1 in
    reference \cite{Friedland:2005vy}.
    Copyrighted by the American Physical Society.
    }
  \label{fig:atmk2k6panels}
  \end{minipage}
  \hspace{10mm}
  \begin{minipage}{10cm}
    \raisebox{10mm}{\includegraphics[width=0.9\textwidth]{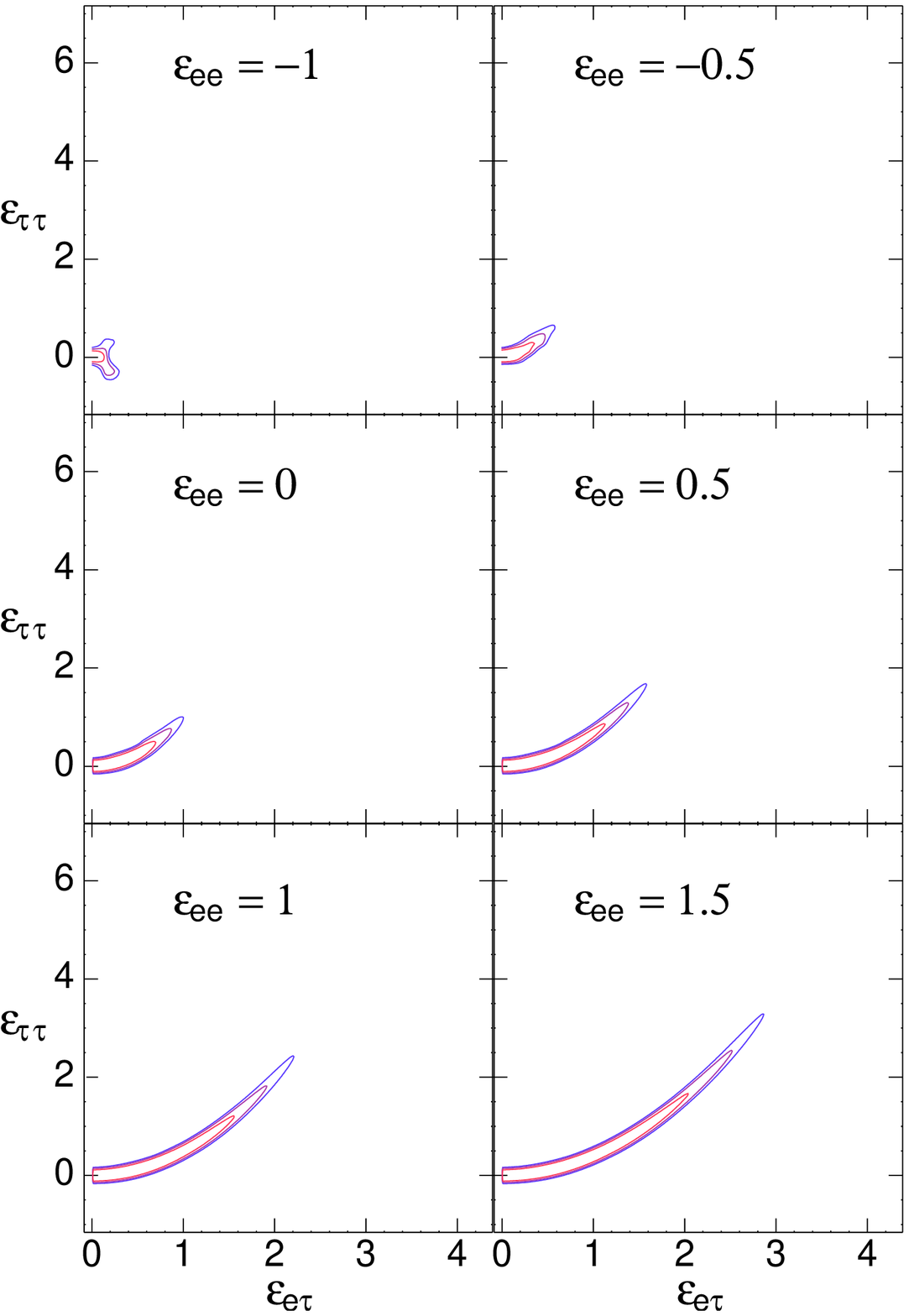}}
  \caption{Same as figure \ref{fig:atmk2k6panels} for normal mass hierarchy.
    Taken with kind permission of Physical Review from figure 6 in
    reference \cite{Friedland:2005vy}.
    Copyrighted by the American Physical Society.
  }
  \label{fig:atmk2k6panelsnormalhierarchy}
  \end{minipage}
  }
\end{figure}

Figure \ref{fig:atmk2k6panels} shows the ranges of the NSI parameters
allowed by the combined analysis of the atmospheric and K2K data. 
The different panels show sections of the 3-D region by contours of
constant $\epsilon_{ee}$. 
As before, in figure \ref{fig:scan}, the contours have been derived for 
$\theta_{13}=0$, $\Delta m_{21}^2=0$ and marginalised over
$\theta_{23}$ and $\Delta m_{23}^2$. 
Since the results are symmetric around $\epsilon_{e\tau}=0$, only
positive values of this parameter are shown. 
The mass hierarchy is assumed to be inverted.

The same analysis, repeated for the case of normal mass hierarchy, is
shown in figure \ref{fig:atmk2k6panelsnormalhierarchy}. 
The difference between the two hierarchies is a sub-leading effect that
is not described by equations (\ref{eq:width}) to (\ref{deltam2m}). 
Figures \ref{fig:atmk2kimpact}, \ref{fig:atmk2k6panels},
\ref{fig:atmk2k6panelsnormalhierarchy} have been adapted from
reference \cite{Friedland:2005vy}.

\subsubsection{The role of MINOS}

\paragraph{MINOS: first data release}
\label{sect:MINOS_firstdata}

The first question to address is whether NSIs can directly impact
the neutrino oscillations observed by MINOS. 
To do this one has to compare the quantity 
$l_{ref} = (\sqrt{2} G_F n_e \epsilon)^{-1}$, characterising the NSI 
matter effect, with the baseline of the experiment. 
For the average density of the continental crust, 
$(\sqrt{2} G_F n_e)^{-1}\simeq 1.9 \cdot 10^{3} $~km; 
this number is nearly an order of magnitude greater than the baseline
of K2K, 250 km, ensuring that K2K measures essentially the vacuum
oscillation parameters. 
The situation for MINOS is less clear-cut: with the baseline of
735~km, it is sensitive to matter effects, although at the
sub-dominant level. 

\begin{figure}
  \centering
  \includegraphics[width=0.87\textwidth]{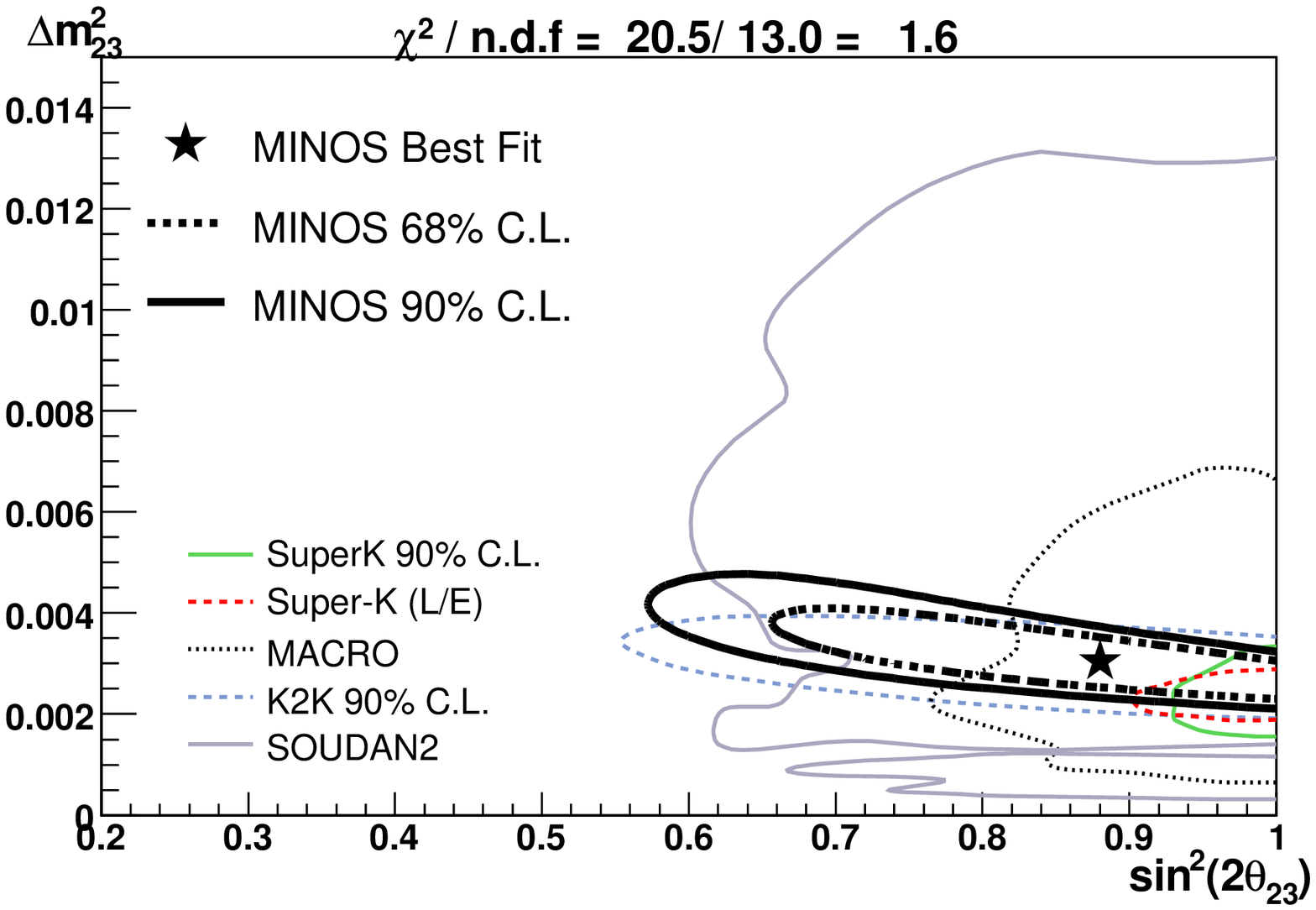}
  \caption{Neutrino oscillation parameters inferred from the analysis
    of the first MINOS data release. Other
    experiments are also included for comparison.
  Taken with kind permission of the MINOS Collaboration from \cite{MINOSI}. 
  }
  \label{fig:MINOSfigure}
\end{figure}

At the low-statistics stage ($0.97\times 10^{20}$ protons on target,
`MINOS~I'), the subdominant matter effects at MINOS can be neglected. 
In this approximation, MINOS simply measures the vacuum oscillations
parameters just as K2K does (see section \ref{sect:atmK2K}). 
It turns out, however, that MINOS~I does not add anything to
constraining the NSI parameters.  
This can be understood from figure \ref{fig:MINOSfigure}: the MINOS~I
dataset has very poor sensitivity in the direction in which the
oscillation parameters $\Delta m^2$ and $\theta$ (here 
$\theta\equiv\theta_{23}$) change to compensate for the effects of the
NSI (c.f. figure \ref{fig:scan}, right panel). 

\begin{figure}
  \centering
  \includegraphics[width=0.47\textwidth]{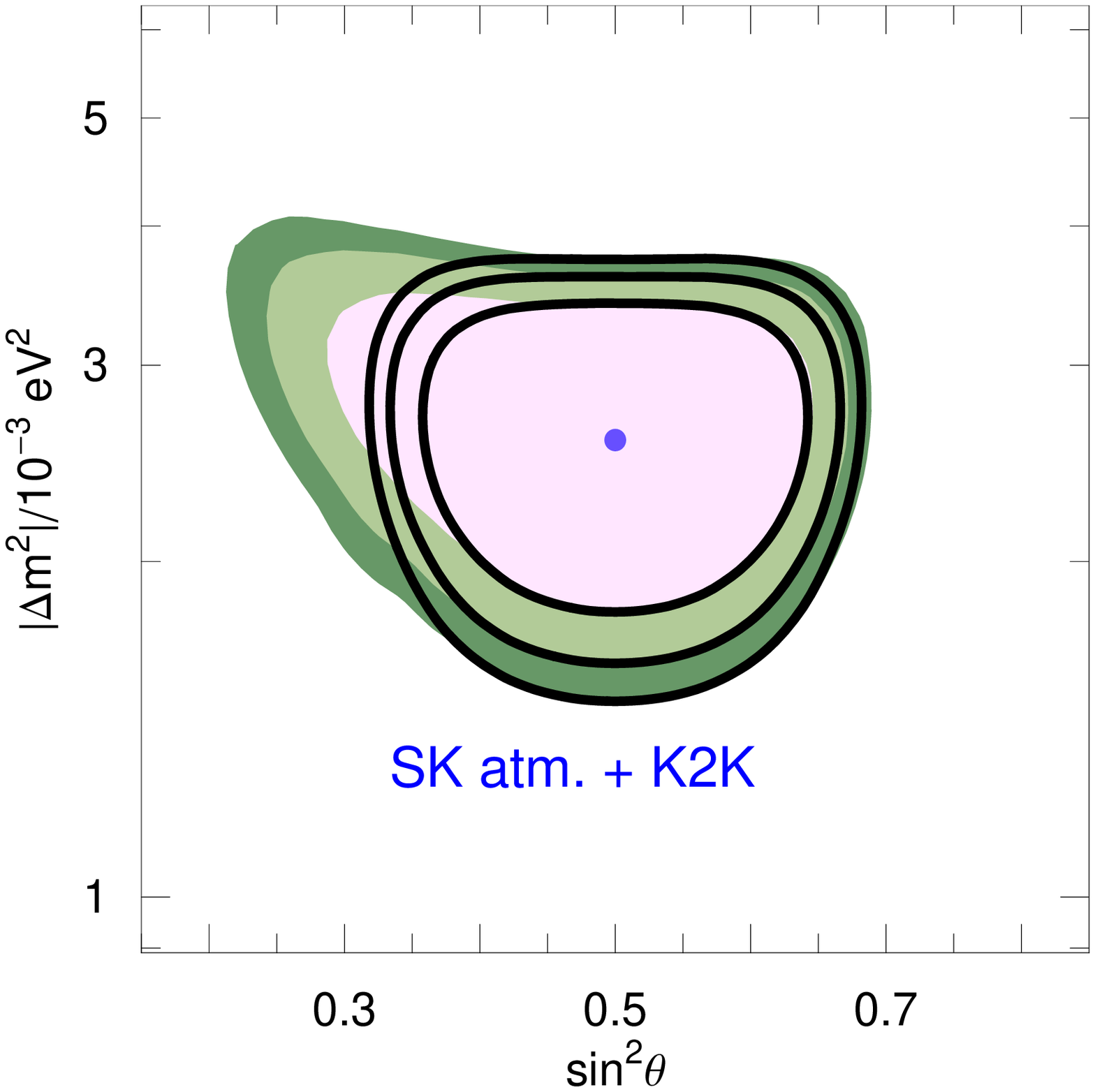}
  \includegraphics[width=0.47\textwidth]{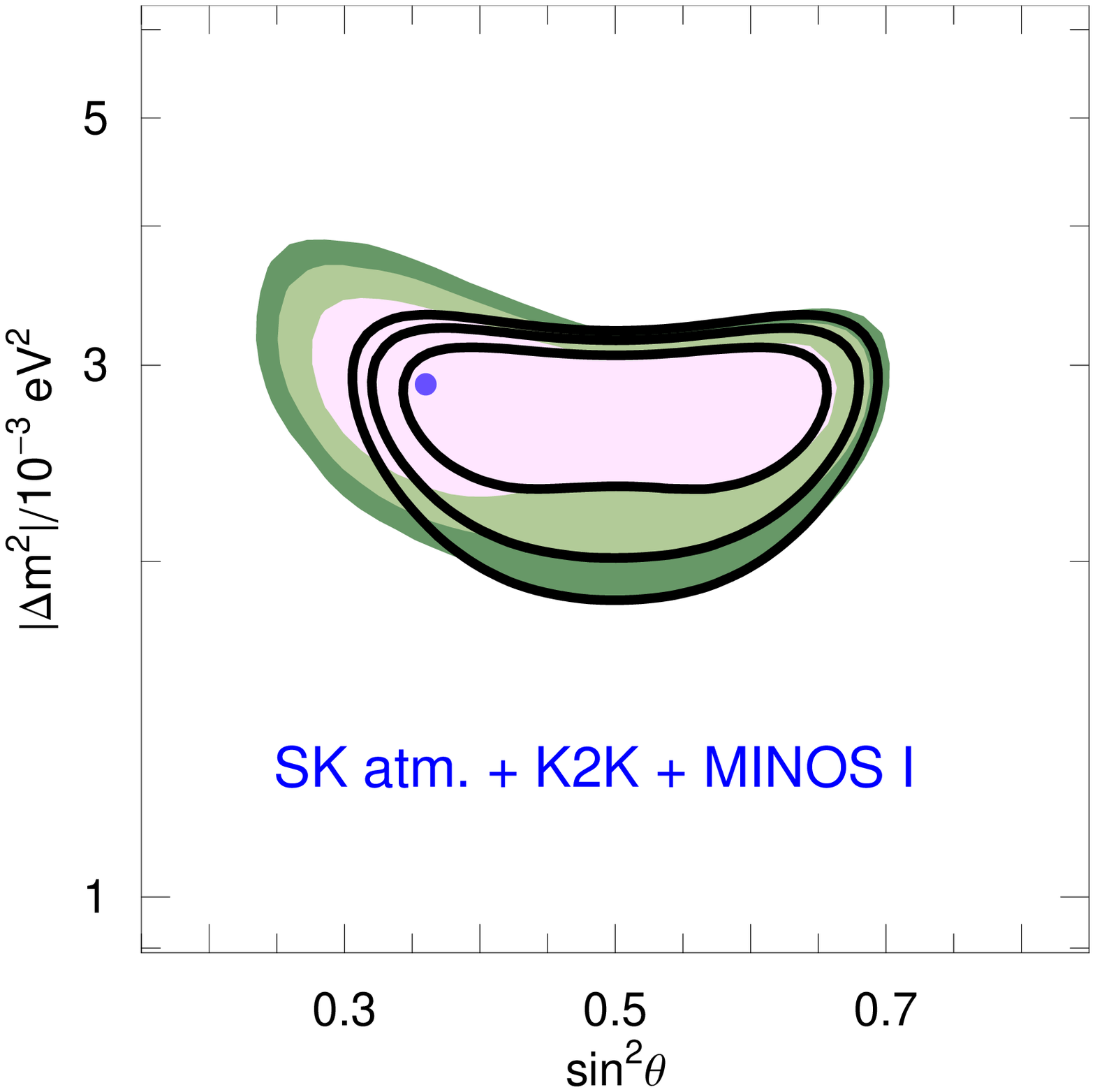}
  \caption{Regions in the space $\Delta m^2 - \sin^2 \theta$ allowed by the
  global fit before (\textit{left panel}) and after (\textit{right panel}) the MINOS
  results, with purely standard interactions (contours) and with NSI
  (filled areas).  For both cases we plot the regions allowed at 95\%,
  99\% and $3\sigma$ confidence levels for 2 degrees of freedom.  We
  have marginalised also over the sign of $\Delta m^2$ and took
  $-1\leq \epsilon_{ee} \leq 1.6$, motivated by one of the accelerator
  bounds (see \cite{Friedland:2004ah}).
    Both figures adapted with kind permission of Physical Review from figure 1 in
    reference \cite{Friedland:2006pi}.
    Copyrighted by the American Physical Society.
  }
  \label{fig:greenfigure}
\end{figure}

Indeed, the results of a detailed numerical fit, shown in figure
\ref{fig:greenfigure} confirms this. 
The part of the allowed region in the oscillation-parameter space that
arises because of the effect of the NSI (the part of the coloured
region outside of the black contours) remains upon the addition of the
data from MINOS~I, implying that the NSI effect can still be
compensated by the change of $\Delta m^2$ and $\theta$. 
The fits shown in figures \ref{fig:atmk2k6panels} and
\ref{fig:atmk2k6panelsnormalhierarchy} are basically unchanged by  
the addition of MINOS data \cite{Friedland:2006pi}. 
An updated dataset with $1.27\times 10^{20}$ protons on target has
been recently released \cite{Michael:2006rx}\footnote{
{\bf Note added}:
Figure \ref{fig:MINOSfigure} is the preliminary result by the MINOS group, and
the analysis described here was made as of September 2006.
See reference \cite{Adamson:2008zt} for the updated result with $3.36*10^{20}$POT.}.

\paragraph{MINOS: projections for the future}

\begin{figure}
  \centering
  \includegraphics[width=0.35\textwidth]{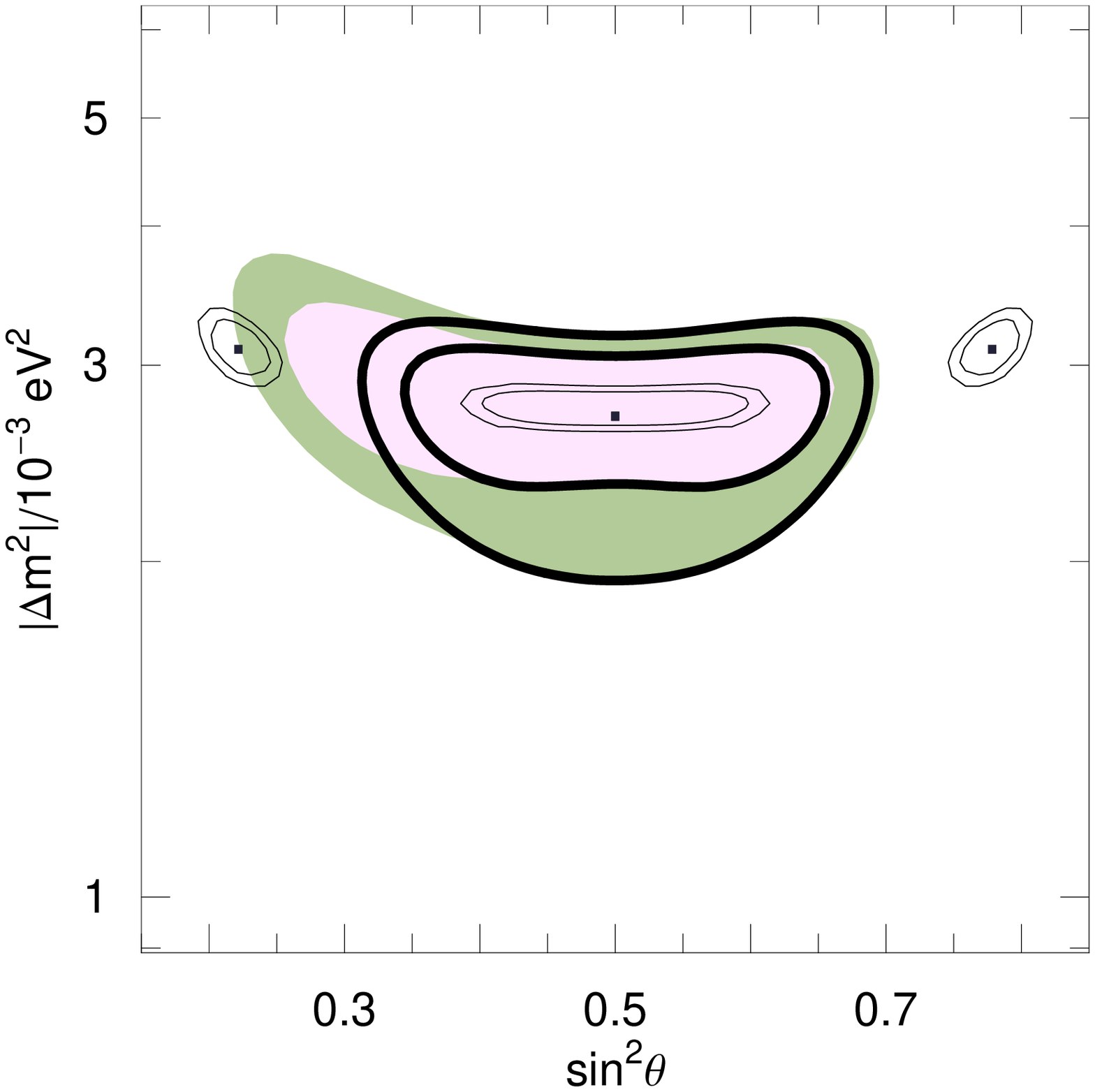}
  \includegraphics[width=0.55\textwidth]{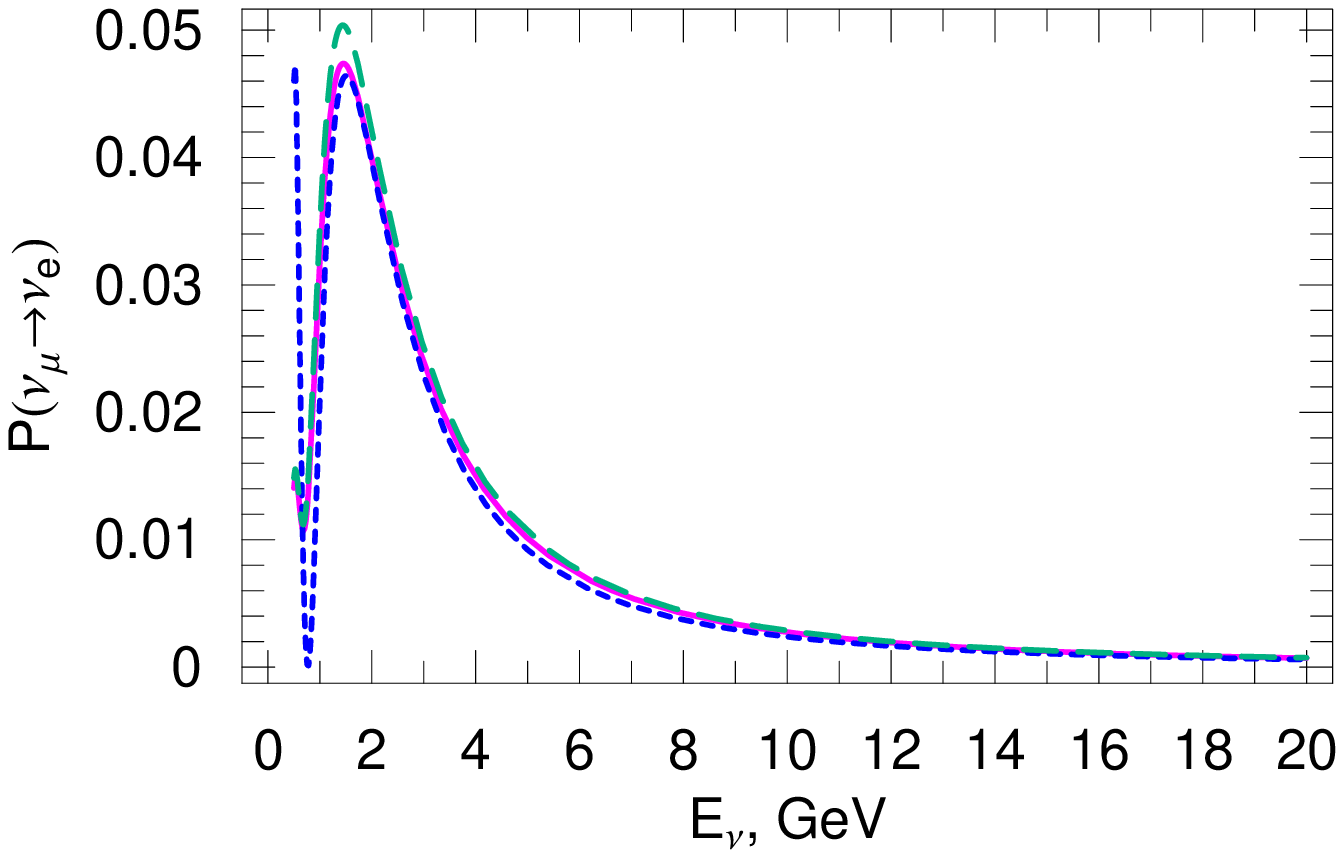}
  \caption{\textit{Left panel}: Results of fits to simulated MINOS
    data with high statistics of $25 \times 10^{20}$ protons on target
    (thin contours). The ``data'' were simulated for two sets of NSI
    and ``true'' oscillation parameters: (i) no NSI, $\sin^2
    \theta=0.5$ and $ \Delta m^2=2.7 \times 10^{-3}$, (ii)
    $\epsilon_{ee}=0$, $\epsilon_{\tau\tau}=0.81$,
    $\epsilon_{e\tau}=0.9$, $\sin^2 \theta=0.27$ and $ \Delta m^2=3.1
    \times 10^{-3}$. The fits were done in both cases in the
    assumption of no NSI; 90\% and 99\% C.L.  regions are shown.
    For reference, also shown are the regions allowed currently by all
    the data combined, at 90\% and 99\% C.L.  with (filled area) and
    without NSI (thick contours), as in Fig.  \ref{fig:greenfigure}.
    \textit{Right panel}: Conversion probability
    $P(\nu_\mu\rightarrow\nu_e)$ as a function of energy for (i)
    $\sin^2 2\theta_{13}=0.07$, $\Delta m_{23}=2.5\times10^{-3}$
    eV$^2$, $\sin^2\theta_{23}=1/2$ and standard neutrino interactions
    (short-dashed curve), vs.  (ii) $\sin^2 2\theta_{13}=0$, $\Delta
    m_{23}=2.9\times10^{-3}$ eV$^2$, $\sin^2\theta_{23}=0.36$ and
    $\epsilon_{ee}=0$, $\epsilon_{e\mu}=0.9$,
    $\epsilon_{\tau\tau}=0.81$ (solid curve). The NSI and
    $\theta_{13}$ effects are nearly completely degenerate.
    Both figures taken with kind permission of Physical Review from figure 3 and4 in
    reference \cite{Friedland:2006pi}.
    Copyrighted by the American Physical Society.
}
  \label{fakefits}
\end{figure}

The situation is expected to improve significantly as MINOS collects
more data. 
Figure \ref{fakefits} (left panel) shows the projected sensitivity of
MINOS with a data set corresponding to $25\times10^{25}$ protons on
target. 
Two scenarios, one corresponding to no NSI and one to large NSI (see
the caption), are considered.  
In the second scenario, the experiment would measure oscillation
parameters that are incompatible with those found from the atmospheric 
data under the assumption of the standard interactions. 
This incompatibility would indicate the need for new physics. 
The point $\epsilon_{ee}=\epsilon_{e\tau}=\epsilon_{\tau\tau}=0$ would
be excluded with confidence level (C.L.) higher than 99\%. 
By the same token, in the first scenario, the compensation mechanism 
between the NSI and the vacuum parameters would be significantly
constrained. 

Similar results are obtained with a more modest increase of the MINOS
statistics, to $16 \times 10^{20}$ instead of $25 \times 10^{20}$
protons on target. With this intermediate increase, the point 
$\epsilon_{ee}=\epsilon_{e\tau}=\epsilon_{\tau\tau}=0$ in the second
scenario would lie inside the 99\% C.L. contour, but outside the 95\%
C.L. contour.

MINOS will also be able to search for flavour-changing NSI effects
using the matter-induced conversion $\nu_\mu\rightarrow\nu_e$ 
\cite{Kitazawa:2006iq,Friedland:2006pi}. 
Schematically, this conversion can be viewed in two steps:
\begin{equation}\label{eq:twostep}
    \nu_\mu\stackrel{\Delta_{23},\theta_{23}}{\longrightarrow}\nu_\tau
    \stackrel{\epsilon_{e\tau}}{\longrightarrow}\nu_e.
\end{equation}
The first step has already been observed by MINOS, with the largest
conversion happening in the lower energy part of its spectrum 
($1.5-2$~GeV). 
Correspondingly, $\nu_e$ production according to
equation (\ref{eq:twostep}) is also expected to peak at low energy. 
The conversion probability $P(\nu_\mu\rightarrow\nu_e)$ as a function of
energy is shown in figure \ref{fakefits} (right panel). 
One can see that the probability indeed peaks at low energies and,
moreover, the effects of the NSI and $\theta_{13}$ are nearly
completely degenerate \cite{Huber:2001de}. 
Thus, if the conversion is observed, it will be necessary to break the
degeneracy by some other means. 

\paragraph{Summary}

In summary, the least constrained NSI parameters, $\epsilon_{ee}$,
$\epsilon_{e\tau}$, and $\epsilon_{\tau\tau}$ are presently being 
probed by both solar- and atmospheric-neutrino experiments. 
Solar neutrino experiments, by themselves, already exclude some parts
of the parameter space allowed by accelerator-based scattering
experiments. 
At the same time, the available data leaves a lot of possibilities
open.  
This is because the electron-neutrino survival probability as a
function of energy is presently measured well only above the SNO/SK
threshold of about 6 MeV.  
The crucial part of the spectrum below 5-6
MeV, where the transition from the matter-dominated to the vacuum
oscillation regime occurs, is measured very poorly. 
This situation should change in the next decade, as Borexino, KamLAND
(solar measurement), and other experiments come on line.

We have seen that atmospheric neutrinos, contrary to naive
expectations, also allow large NSI, comparable to, or even exceeding,
the strength of the Standard Model interactions. 
This happens because
the effects of the NSI can be compensated by changing the oscillation
parameters. 
This degeneracy is somewhat ameliorated, but not
eliminated, by the inclusion of the K2K data. 
Moreover, the first
data released by MINOS does not eliminate this degeneracy.
Again, this situation is expected to be
significantly improved in the future, as MINOS collects more data.

On the theoretical side, a lot of work on the implications of the
current data on NSI remains to be done. 
For example, a combined study of the atmospheric- and solar-neutrino
data has not yet been performed.

\subsubsection[Complementarity of long and short baseline experiments]
{Complementarity of long- and short-baseline experiments
for non-standard interactions}

The combination of long- and short-baseline experiments is effective
in distinguishing the oscillations due to $\theta_{13}$ and those due
to the NSI.
To see this, consider for simplicity the two-flavour scenario where
the oscillation probability can be expressed analytically.
The Hamiltonian for this case is:
\begin{eqnarray}
  U
  \left(
  \begin{array}{cc}
    0 & 0\\
    0 & \frac{\Delta m^2}{2E}
  \end{array}
  \right)
  U^\dagger
  +
  A
  \left(
    \begin{array}{cc}
      1+ \epsilon_{ee} & \epsilon_{e\tau}\\
      \epsilon_{e\tau} & \epsilon_{\tau\tau}
    \end{array}
  \right),
  \label{2f-NSImass}
\end{eqnarray}
where $A\equiv\sqrt{2}G_Fn_e$.
The effective mass-squared difference $\Delta m^2_M$, the mixing
$\theta_M$ and the oscillation probability $P(\nu_e\to\nu_\tau)$ at
distance $L$ in matter are given by:
\begin{eqnarray}
  \left( \frac{\Delta m^2_ML}{4E} \right)^2
  &=&
  \left( \frac{\Delta m^2L}{4E} \cos{2\theta}
  - \frac{AL}{2} (1+\epsilon_{ee}-\epsilon_{\tau\tau}) \right)^2
  +\left( \frac{\Delta m^2L}{4E} \sin{2\theta}
  + AL\epsilon_{e\tau} \right)^2 \; ;
  \label{2f-mass}\\
  \sin{2\theta_M}
  &=&
  \frac{ \Delta m^2 \sin{2\theta} + 4E A \epsilon_{e\tau} }
      { \Delta m^2_{M} } \; {\rm ; and}
  \label{2f-ang}\\
  P(\nu_e\to\nu_\tau)&=&\sin^2{2\theta_M}\sin^2
  \left(\frac{\Delta m^2_ML}{4E}\right).
  \label{2f-pro}
\end{eqnarray}
To have a large value of the oscillation probability
$P(\nu_e\to\nu_\tau)$, large values for both $\sin^2{2\theta_M}$ 
and $\sin^2\left(\Delta m^2_ML/4E\right)$ are required.
Equation (\ref{2f-pro}) implies: that the effect of the new physics in
$\sin^2\left(\Delta m^2_ML/4E\right)$ appears in a form
$AL(\epsilon_{ee}-\epsilon_{\tau\tau})$ or $AL\epsilon_{e\tau}$, so a
large deviation of $\Delta m^2_ML/4E$ from the standard value $\Delta
m^2L/4E$ requires that $AL\epsilon_{\alpha\beta}$ be non-negligible
irrespective of the neutrino energy $E$; and that, for the experiments
with $|\Delta m^2|L/E\simeq {\cal O}(1)$, multiplying by $L$ both the
numerator and the denominator of equation (\ref{2f-ang}), to obtain a
non-trivial new-physics contribution to the mixing angle $\theta_M$
again demands that $AL\epsilon_{\alpha\beta}$ be non-negligible.
These conditions imply that the baseline length has to be relatively
large for the new-physics effect to affect both of the factors in the
oscillation probability, since $A$ can be roughly estimated as
$A\simeq$1/(2000km) with $\rho\simeq$3g/cm$^3$. 
These features hold also in the case with three flavours.

\begin{figure}
    \begin{tabular}{cc}
      \resizebox{80mm}{!}
{\includegraphics
{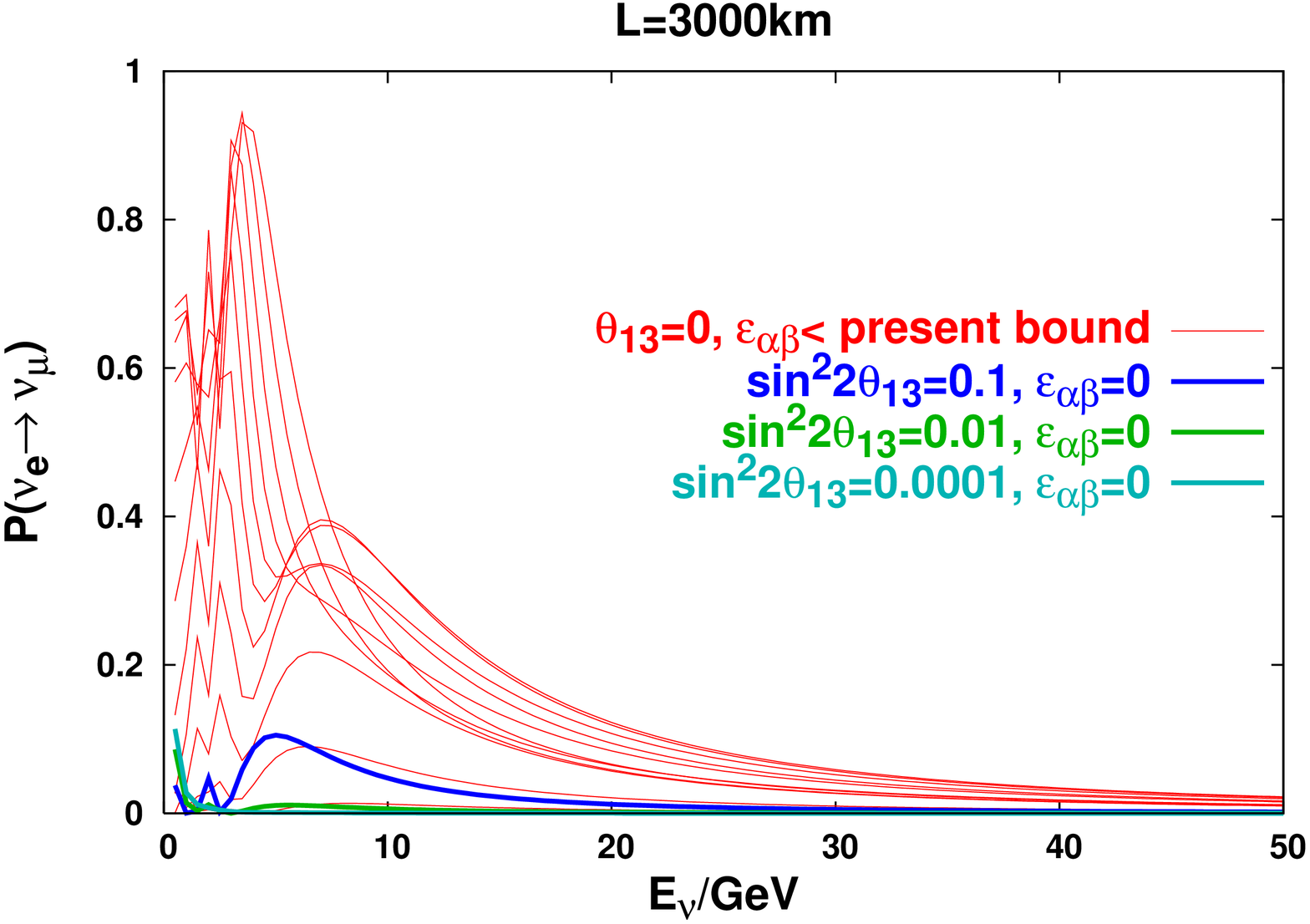}}&
\hspace*{-30mm}
      \resizebox{80mm}{!}
{\includegraphics
{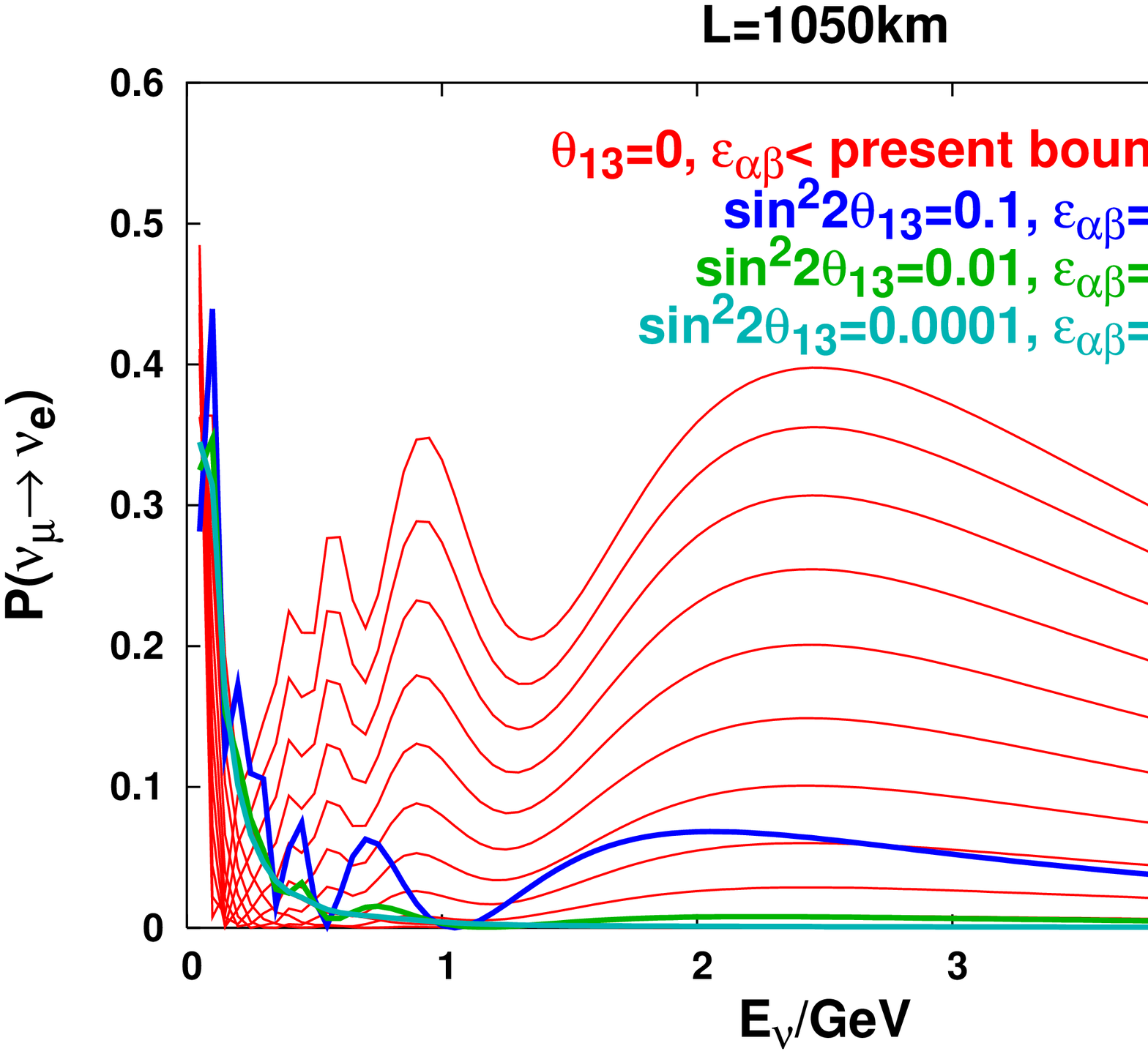}} \\
\\
\\
\\
\\
\\
      \resizebox{80mm}{!}
{\includegraphics
{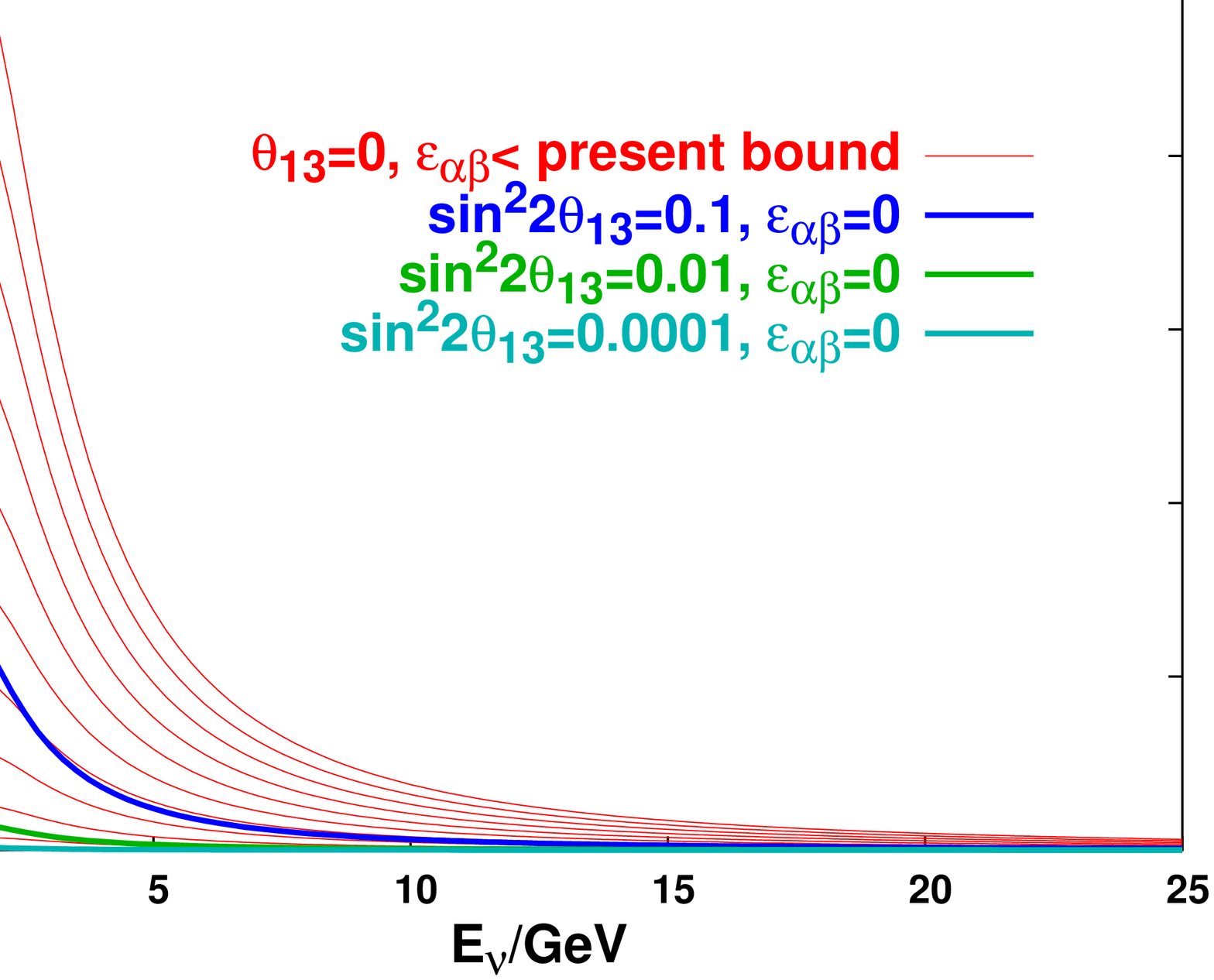}}&
\hspace*{-30mm}
      \resizebox{80mm}{!}
{\includegraphics
{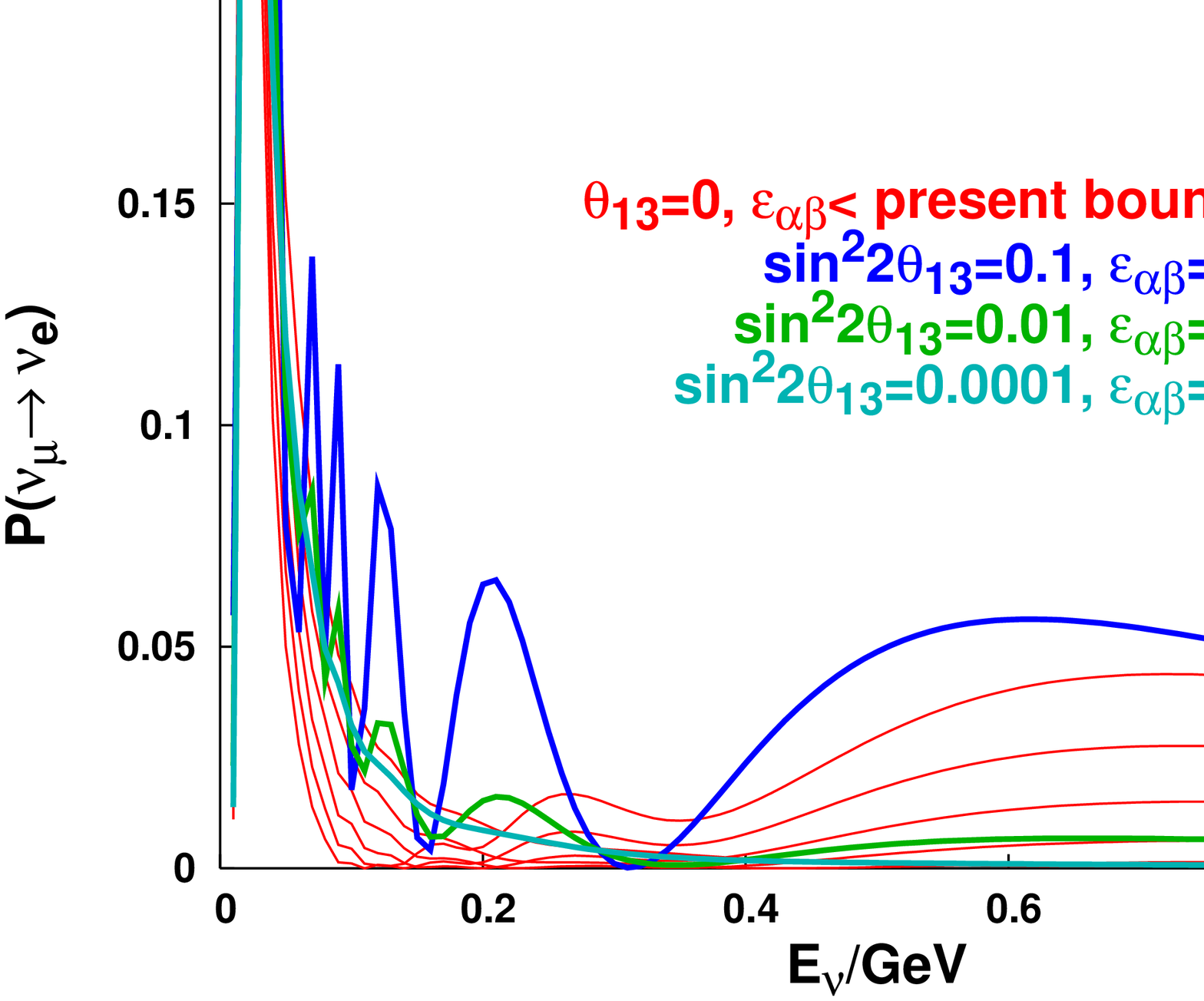}} \\
\\
\\
\\
\\
\\
      \resizebox{80mm}{!}
{\includegraphics
{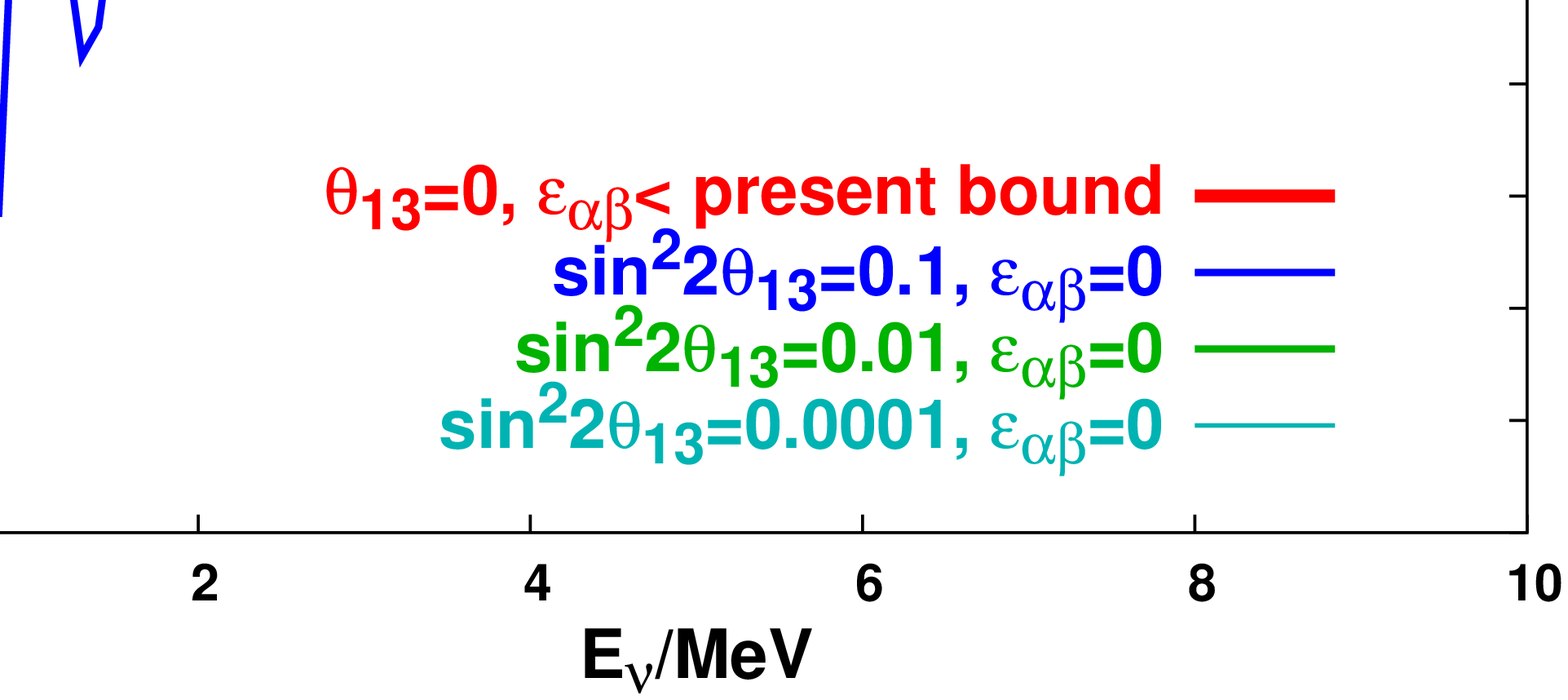}}
    \end{tabular}
\vspace*{-40mm}
\caption{Effects of $\theta_{13}$ versus those of the
NSI~\cite{yasuda-iss3,Kitazawa:2006iq}:
The oscillation probabilities with (red lines)
or without (blue, green and light blue lines) the NSI are plotted for
typical baseline lengths.  The red lines are plotted for various
values of $\epsilon_{ee}, \epsilon_{e\tau}, \epsilon_{\tau\tau}$
which are in the allowed region obtained by \cite{Friedland:2005vy}.}
  \label{longvsshort}
\end{figure}

The present and future generation of neutrino-oscillation experiments 
are designed mainly to probe neutrino oscillations with the
atmospheric-neutrino mass-squared difference $|\Delta
m^2_{\text{atm}}|\simeq 2.5\times10^{-3}$eV$^2$ and the typical
neutrino energy, $E$, of each experiment satisfies 
$|\Delta m^2_{\text{atm}}|L/E\simeq {\cal O}(1)$.
The baseline lengths, $L$, of these experiments, however, are quite
different and, when $\epsilon_{\alpha\beta}\sim{\cal O}(1)$,
only the experiments for which $AL$ is non-negligible will have
sensitivity to new physics.
Reactor experiments, for which $AL\ll 1$, are insensitive to
$\epsilon_{\alpha\beta}$.  
On the other hand, a reactor experiment has the advantage of having no
backgrounds due to new physics in measurements of the standard
oscillation parameters. 
For the T2K experiment, $AL\simeq 3/20$, so it has potential to
see the new physics effect.
MINOS, NOvA, T2KK, and a Neutrino Factory, since $AL$ is larger, have
greater potential to see the signal of $\epsilon_{\alpha\beta}$
\cite{Kopp:2007ne,Kopp:2007mi}.  
These effects can be seen in figure \ref{longvsshort}.

%% file: 05-Performance/05-Performance.tex
\section{Performance of proposed future long-baseline neutrino
         oscillation facilities} 
\label{Sect:Performance}
\renewcommand\thesubsection   {\thesection.\arabic{subsection}}
\renewcommand\thesubsubsection{\thesubsection.\arabic{subsubsection}}
\renewcommand\theparagraph    {\thesubsubsection.\arabic{paragraph}}
\renewcommand\thesubparagraph {\theparagraph.\arabic{subparagraph}}

\input 05-Performance/05-01-Introduction/05-01-Introduction

\input 05-Performance/05-02-Super-Beams/05-02-Super-Beams

\input 05-Performance/05-03-Beta-Beams/05-03-Beta-Beams

\input 05-Performance/05-04-Neutrino-Factory/05-04-Neutrino-Factory

\input 05-Performance/05-05-Comparisons/05-05-Comparisons

%% file: 05-Performance/05-01-Introduction/05-01-Introduction.tex
\subsection{Introduction}
\label{SubSect:Perf:Intro}

\begin{figure}
  \begin{center}
    \epsfig{file=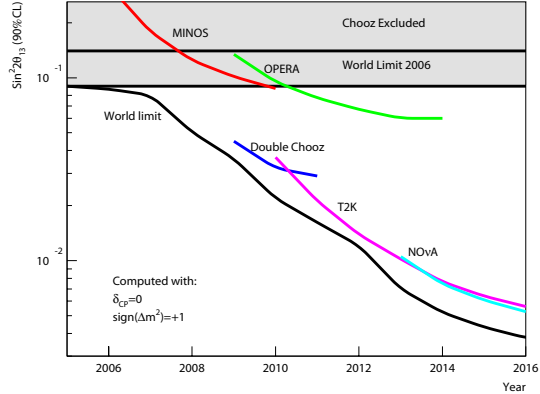,width=7cm}
  \end{center}  
  \caption{
    Projected evolution of the world limit on $\sin^2 2 \theta_{13}$
    at 90\% CL.
    The anticipated impact of the MINOS, OPERA, T2K, and NO$\nu$A
    long-baseline neutrino-oscillation experiments are shown together
    with that of the Double Chooz reactor-neutrino experiment are
    shown.
    Taken with kind permission of the Editor from the Proceedings of the
    Neutrino Telescopes 2007 \cite{Mezzetto:2007zz}.
    For a discussion of these experiments see section
    \ref{SubSect:SnuM}.
  } 
  \label{Fig:Th13vsTime}
\end{figure}

The precision with which the parameters of the Standard Neutrino Model
have been determined in fits to neutrino-oscillation data is shown in
figure \ref{fig:allowedareassolplusatm} and summarised in table
\ref{tab:globalallowedrange}.
Over the coming decade, the various long-baseline, reactor, solar, and
atmospheric neutrino experiments that are in operation or in
preparation will improve upon these results.
In particular, the strong push to determine the small mixing angle
will yield a measurement of $\theta_{13}$ if 
$\sin^2 \theta_{13} > 0.01$ and a substantially improved limit
otherwise.
Figure \ref{Fig:Th13vsTime} shows the evolution of the upper limit on 
$\sin^2 \theta_{13}$ that may be expected based on the performance
claimed for the various experiments \cite{Mezzetto:2007zz}.
The sensitivity to the small mixing angle improves significantly as
the data from each of the new experiments becomes available.
By around 2016, the rate of improvement in the sensitivity of the
neutrino-oscillation programme slows down and a new generation of
high-flux facilities is required.

The new facility must offer the best possibility of observing 
leptonic-CP violation and of determining the mass hierarchy 
(${\rm sgn} \Delta (m^2_{32})$). 
The optimisation of the facility depends on the value of
$\theta_{13}$.
If $\theta_{13}$ is large 
(such that $\sin^2 2 \theta_{13} \gtrsim 0.01$)
then it will have been measured, albeit with poor precision.
In this case, the high-sensitivity facility is required to offer the
best sensitivity to $\delta$ and (${\rm sgn} \Delta (m^2_{32})$).
If $\theta_{13}$ is small 
(such that $\sin^2 \theta_{13} \lesssim 0.01$) it is unlikely to have
been measured and the facility will, in addition, be required to have
the best possible sensitivity to $\theta_{13}$.

At the same time, the new facility must aim at providing measurements
of sufficient precision to inform the development of the theory of the
physics of flavour.
The status of the theoretical description of flavour is discussed in
detail in section \ref{Sect:NPandtheSNuM}.
Grand-unified theories typically provide relationships between the
neutrino-mixing parameters and those of the quarks. 
For such relationships to be tested requires that the precision with
which the neutrino-mixing parameters are determined matches that with
which the quark-mixing parameters are known.
At present the quark-mixing parameters are known at the percent
level. 
This sets the standard; the high-precision neutrino-oscillation
programme must deliver measurements of the neutrino-oscillation
parameters at the percent level.
To achieve this goal requires high-energy electron- and muon-neutrino
beams and highly sensitive neutrino-detection systems.

Three types of facility have been proposed to provide the neutrino
beams required to serve the high-sensitivity programme.
The Neutrino Factory gives the best performance over most of the
parameter.
Second-generation super-conventional-beam experiments may be an
attractive option in certain scenarios.
A beta-beam \cite{Zucchelli:2002sa}, in which electron neutrinos (or
anti-neutrinos) are produced from the decay of stored radioactive-ion
beams, in combination with a second-generation super-beam, may be
competitive with the Neutrino Factory.
The purpose of this chapter is to evaluate the physics performance of
a second-generation super-beams, a beta-beam facility, and the Neutrino
Factory and to present a critical comparison of their performance.

\subsubsection{Definition of observables}
\label{Sect:ObservablesDefined}

The observables that will be examined in sections
\ref{SubSect:Perf:SB}, \ref{SubSect:Perf:BetaBeam}, and
\ref{SubSect:Perf:NF}, and compared in section
\ref{SubSect:Perf:CompComb}, are defined below:
\begin{itemize}
  \item{\it Number of degrees of freedom:}
    The number of degrees of freedom that are used to convert
    $\Delta\chi^2$-values into confidence levels must be clearly
    defined.
    In the literature several different approaches can be found, for
    example: in \cite{Donini:2004iv,Donini:2005rn} the CP-violation
    discovery potential is defined as the smallest (largest) value of
    ``true'' $|\delta|$ (as a function of ``true'' $\theta_{13}$) for
    which the $3\sigma$ contour in the $( \theta_{13},\delta )$
    plane of any of the degenerate solutions reaches either
    $\delta = 0$ or $\delta = \pi$; while in \cite{Huber:2006wb}, the
    $\Delta\chi^2$ is marginalised over all parameters except
    $\delta$ and one degree of freedom is used.
    For definiteness, unless otherwise stated, we will use one degree
    of freedom throughout;
  \item{\it $\theta_{13}$-sensitivity and $\theta_{13}$ discovery
        potential:} 
    The $\theta_{13}$-sensitivity as a function of ``true''
    $\delta$ is the largest value of $\theta_{13}$ that fits the
    ``true'' value $\theta_{13} = 0$, after marginalisation over all
    parameters other than $\theta_{13}$, once all possible wrong
    choices of ${\rm sgn} ( \Delta m^2_{31} )$ and of the
    $\theta_{23}$-octant are taken into account.

    For the $\theta_{13}$ discovery potential, data are simulated for
    non-vanishing ``true'' $\theta_{13}$ and a given ``true''
    $\delta$.
    After marginalisation over all parameters other than $\theta_{13}$
    and taking into account all possible wrong choices of 
    ${\rm sgn} (\Delta m^2_{31} )$ and of the $\theta_{23}$-octant,
    if $\Delta \chi^2 (\theta_{13} = 0)\ge 9$, the ``true''
    $\theta_{13}$ is ``discovered at 3$\sigma$'';
  \item{\it CP discovery potential and sensitivity to maximal
        CP-violation:} 
    To obtain the $\delta$-discovery potential, data are simulated for
    ``true'' $\delta$ different from 0 and $\pi$ and a given ``true''
    $\theta_{13}$.  
    After marginalisation over all parameters other than $\delta$ and
    taking into account all possible wrong choices of 
    ${\rm sgn} ( \Delta m^2_{31} )$ and of the $\theta_{23}$-octant,
    if $\Delta \chi^2 (\delta = 0)$ and $\Delta \chi^2 (\delta = \pi)$
    are both larger than $9$, computed with respect to the absolute
    $\chi^2$ minimum, the ``true'' $\delta$ is ``discovered at
    3$\sigma$''.  

    Sensitivity to maximal CP-violation, refers to the possibility that
    a ``true'' $\delta = \pm \pi /2$ from $\delta = 0$ or 
    $\delta = \pi$ at a given CL as a function of some other parameter
    \cite{Cervera:2000kp,Huber:2006wb};
  \item{\it Sensitivity to the sign of the atmospheric mass difference:}
    We have sensitivity to the ``true'' mass hierarchy if, when
    performing an hypothesis test, after marginalisation over all
    parameters and taking into account all possible choices of the
    $\theta_{23}$-octant, we can exclude the wrong hierarchy at a
    given CL. 
    The procedure is to draw a contour in the ``true''
    ($\theta_{13},\delta$) plane for the mass hierarchy under
    consideration. 
    In most cases, a ``true'' normal hierarchy will be discussed,
    since the inverted hierarchy gives qualitatively similar
    results.
    Note that, for the ``true'' inverted hierarchy anti-neutrinos are 
    matter enhanced, thus compensating for the smaller cross-section
    with respect to neutrinos (see, for example, reference
    \cite{Huber:2005jk});
  \item{\it $\theta_{23}$--non-maximality discovery potential and
        sensitivity to the $\theta_{23}$-octant:} 
    Data are simulated for ``true'' $\theta_{23}$ different from
    $\pi/4$ and a given ``true'' $\Delta m^2_{31}$. 
    After marginalisation over all parameters but $\theta_{23}$, and
    taking into account all possible wrong choices of the sign of
    $\Delta m^2_{31}$, if $\Delta \chi^2 (\theta_{23} = \pi/4)\ge 9$,
    the corresponding deviation from maximality is ``discovered at 
    3$\sigma$''.

    If $\theta_{23} \neq \pi/4$, we have sensitivity to the ``true'' 
    $\theta_{23}$-octant if, when performing an hypothesis test, after
    marginalisation over all parameters and taking into account all
    possible choices of the mass hierarchy, we can exclude the wrong
    octant at a given CL.
    The procedure is to draw a contour in the ``true''
    ($\theta_{13},\delta$) plane for the ``true'' octant under
    consideration;
  \item{\it Precision on $\theta_{13}$ and $\delta$:}
    The precision on $\theta_{13}$ ($\delta$) is the projection of the
    (marginalised) $\Delta\chi^2$ onto the $\stheta$ ($\delta$) axis
    at a given CL. 
    Remember that, for different choices of the hierarchy and of the
    $\theta_{23}$-octant, several solutions can arise.
    In section \ref{SubSect:Perf:CompComb}, we also show our
    results as two-parameter contours in the ($\stheta,\delta$) plane
    for a set of ``true'' input pairs; and
  \item{\it Precision on $\Delta m^2_{31}$ and $\sin^2\theta_{23}$:} 
    The precision on $\Delta m^2_{31}$ ($\theta_{23}$) is the
    projection of the (marginalised) $\Delta\chi^2$ onto the 
    $\Delta m^2_{31}$ ($\sin^2 \theta_{23}$) axis at a given CL.
    Remember that, for different choices of the hierarchy, several
    solutions can arise.

    We will, in some cases, refer to the ``Fraction of (true)
    $\delta$'' (or the ``CP-fraction'').
    This is the fraction of the $\delta$-parameter space, 
    i.e. of ($0 < \delta < 2 \pi$) over which a facility has
    sensitivity to a given observable.
    For a graphical explanation of this procedure,  see e.g. figure 3
    of reference \cite{Barger:2006vy}. 
\end{itemize}

%% file: 05-Performance/05-02-Super-Beams/05-02-Super-Beams.tex
\subsection{The physics potential of super-beams}
\label{SubSect:Perf:SB}

\subsubsection{The super-beam concept}
\label{Sect:SuperBeamDef}

Conventional neutrino beams from $\pi$-decay have, up to now,
mainly been tuned for the study of $\nu_\mu$ disappearance
\cite{Ahn:2006zz,Michael:2006rx} or $\nu_\mu \rightarrow \nu_\tau$
appearance \cite{Kodama:1999hg}.
Such beams can be optimised for $\nu_{\mu} \rightarrow \nu_e$
searches.
The design of such a facility, producing high intensity, low energy
$\nu_\mu$ and $\bar{\nu_\mu}$ beams, requires the development of new,
high-power, proton accelerators delivering more intense proton beams
on target. 
In the following, a super-beam is taken to be a conventional neutrino
beam driven by proton driver with a beam power in the range 2~--5~MW.

The technology required for the super-beam is a development of that
used today in long-baseline neutrino-oscillation experiments.
Compared to beta-beam facilities or the Neutrino Factory, super-beams
have the advantage that the required technology is relatively well
known. 
The neutrino beam contains the dominant neutrino flavour
($\nu_\mu$ if the capture system focuses $\pi^+$ into the decay
channel) together with a small but unavoidable admixture of
$\bar{\nu}_\mu$, $\nu_e$ and $\bar{\nu}_e$.
The presence of $\nu_e$ and $\bar{\nu}_e$ in the primary beam limits
the super-beam sensitivity to $\nu_\mu \to \nu_e$ oscillations.
The intrinsic $\nu_e$ contamination, which grows with increasing
neutrino energy, must therefore be kept as low as possible. 
One way to achieve this is to arrange that the neutrino-beam axis is
tilted by a few degrees with respect to the vector pointing from the
source to the far detector (an off-axis beam).
The kinematics of the two-body $\pi$-decay ensures that all pions
above a given momentum produce neutrinos of similar energy at a given 
angle $\theta \ne 0$, with respect to the direction of the parent
pion.
The off-axis technique yields a low-energy beam of neutrinos with a
small energy spread.
Such neutrino beams have several advantages over the corresponding
broad-band on-axis beams; the narrow-band low-energy beam allows
energy cuts to be applied to reduce backgrounds and allows the $L/E$
of the  experiment to be tuned to the oscillation maximum.
However, the off-axis neutrino flux is significantly smaller than the
on-axis flux. 
Another way of reducing the $\nu_e$ background is
to design a beam line configuration where the contribution
by $K^+$ and $K^0$ results to be suppressed.

\subsubsection{T2K and T2HK }

The T2K facility consists of a conventional neutrino beam driven by
30~GeV protons from the J-PARC proton synchrotron at a beam power of
0.75~MW. 
The neutrino beam will illuminate the Super-Kamiokande detector at a
baseline of $L = 295$~km.
The facility is presently under construction, data taking is scheduled
to start at the end of 2009 \cite{Itow:2001ee}.
In the first year, the number of `protons-on-target' (pot) is
expected to be $\sim 10\%$ of the design value.
The T2K neutrino beam off-axis angle has been chosen
to be $2.5^\circ$ to maximize the
sensitivity of the experiment to $\theta_{13}$.

\begin{figure}
  \vspace{-0.5cm}
  \begin{center}
  \begin{tabular}{cc}
\hspace{-0.3cm}\epsfig{file=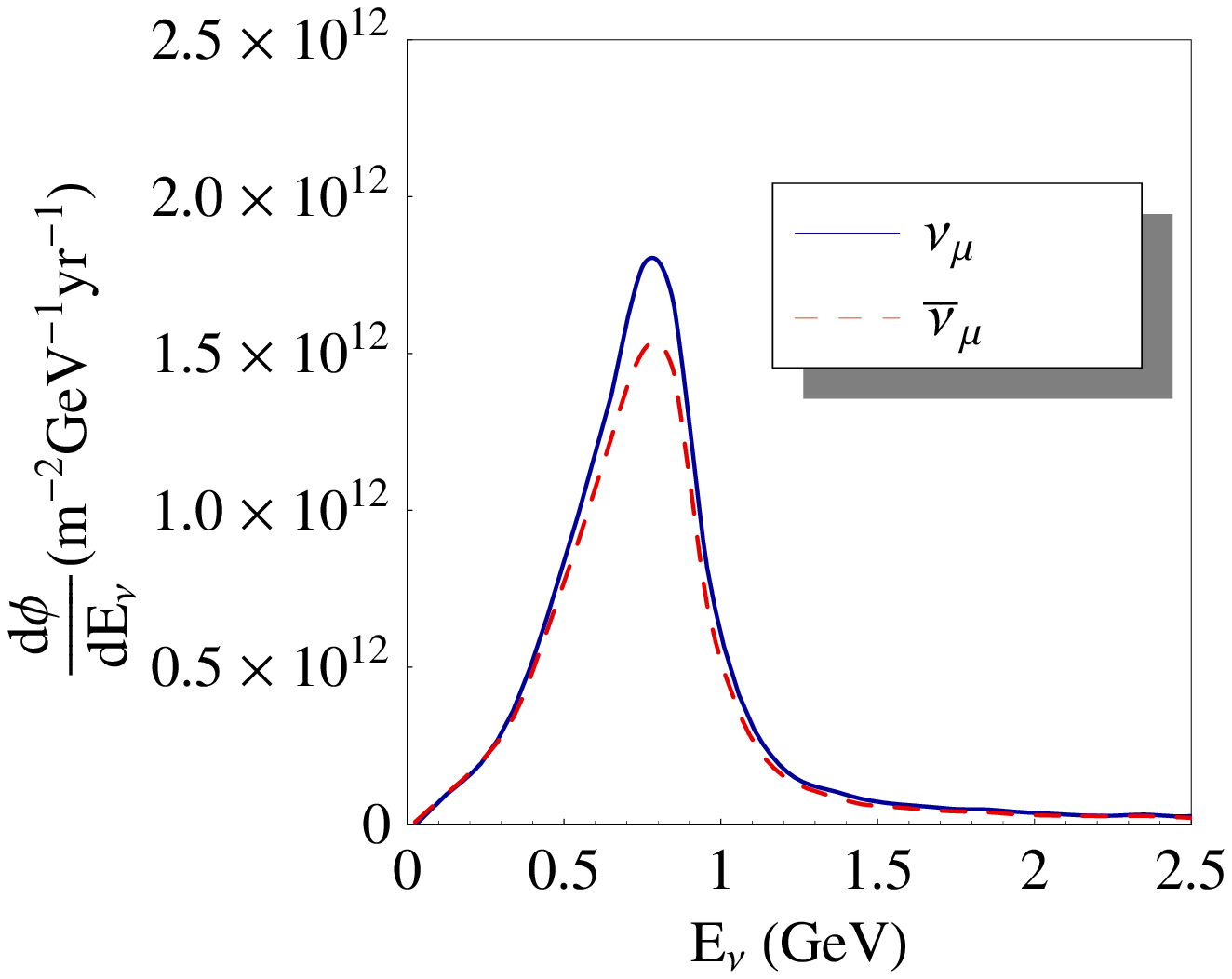,width=8.0cm}&
\hspace{-0.3cm}\epsfig{file=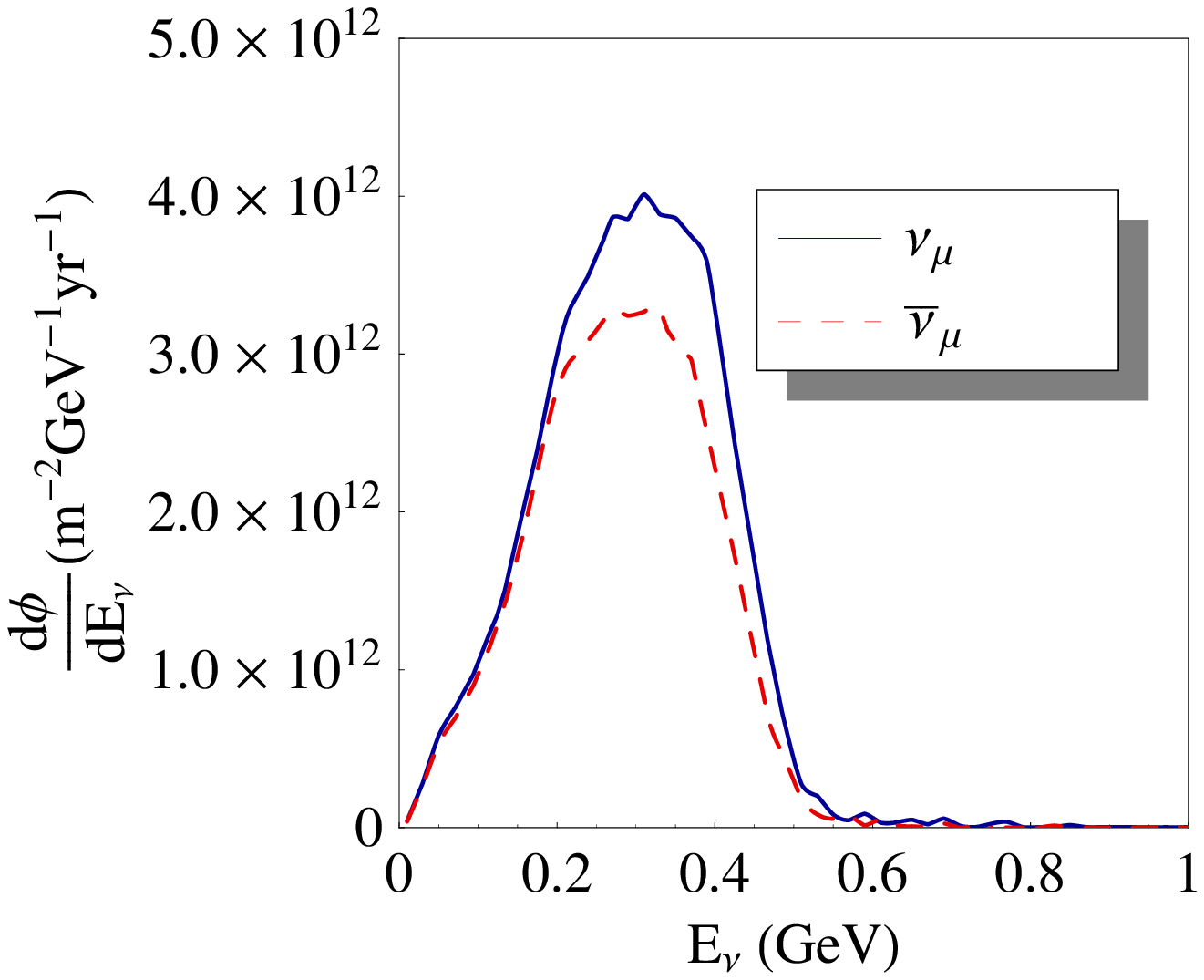,width=8.0cm} \\
    \end{tabular}
  \end{center}
  \caption{
    Left panel: T2HK fluxes at the Kamioka location (295 km baseline); 
    Right panel: SPL fluxes at the Fr\'{e}jus location (130~km
    baseline, proton beam energy 3.5 GeV).
  } 
  \label{fig:fluxes}
\end{figure}

An upgrade to the power of the J-PARC proton synchrotron to provide a
4~MW, 50~GeV proton beam is planned.
This, together with the construction of a mega-Tonne (Mton) class,
water \v Cerenkov detector (Hyper-Kamiokande) could provide enough
events to compete with beta-beam and Neutrino Factory facilities
if the mixing parameters are favourable.
This upgraded version of T2K, T2HK or T2K-II, is considered below. 
Figure \ref{fig:fluxes}(left) shows the neutrino fluxes expected at
Hyper-Kamiokande assuming a $2^\circ$ off-axis angle.

It has been proposed to exploit the J-PARC neutrino beam with a second
100 Kton \cite{Hagiwara:2005pe,Hagiwara:2006vn} or
0.5Mton \cite{Ishitsuka:2005qi,Kajita:2006bt}
water \v Cerenkov detector in Korea.
The second detector would be placed
at a $0.5^\circ$ \cite{Hagiwara:2005pe,Hagiwara:2006vn} or
at a $2.5^\circ$ \cite{Ishitsuka:2005qi,Kajita:2006bt}
off-axis angle for a baseline of $L = 1000$~km.
This combination of two baselines would give significant sensitivity
to the neutrino-mass hierarchy, reducing the degeneracy problem
present in searches for leptonic CP-violation (see section
\ref{sec:emu}).

\subsubsection{The SPL}

In the CERN super-beam project
\cite{Autin:2000mn,Mezzetto:2003mm,Gomez-Cadenas:2001eu,Apollonio:2002en},
the planned 4~MW Super-conducting Proton Linac (SPL) will deliver a
2.2~GeV proton beam on a mercury target to generate an intense
$\pi^+$ ($\pi^-$) beam focused by a magnetic horn into a short decay
tunnel. 
As a result, an intense $\nu_{\mu}$ ($\bar{\nu}_\mu$) beam will be
produced, providing a flux 
$\phi \sim 3.6 {\cdot} 10^{11} \nu_{\mu}$/year/m$^2$ 
($2.3 {\cdot} 10^{11} \bar{\nu}_{\mu}$/year/m$^2$), with an average
energy of 0.27~(0.25)~GeV aimed at a Mton-class, water \v Cerenkov
detector at the Modane laboratory in the Fr\'{e}jus area (a baseline
of $L = 130$~km).
The $\nu_e$ contamination from kaons will be suppressed by threshold
effects and the resulting $\nu_e/\nu_{\mu}$ ratio ($ \sim 0.4 \%$)
will be known to within $2\%$.

New developments show that the SPL potential could be improved by
raising the SPL energy to 3.5 GeV \cite{Campagne:2004wt}, to produce
more copious secondary mesons and to focus them more efficiently. 
This seems feasible if state-of-the-art RF cavities are used in place
of the LEP cavities assumed in the 2.2~GeV design
\cite{Garoby:2005se}.
In this upgraded configuration the neutrino flux could be increased by
a factor of 3 with respect to the 2.2 GeV configuration, with a
slightly higher energy of 0.28 GeV. 
The fluxes that the 3.5~GeV configuration will produce are shown in
figure \ref{fig:fluxes}(right).

\subsubsection{NO$\nu$A}

The NO$\nu$A experiment was proposed recently at FNAL to measure
$\nu_{\mu} \rightarrow \nu_e$ oscillations with a sensitivity 10
times better than MINOS \cite{Ayres:2004js}.
It consists of an upgraded NuMI Off-Axis neutrino beam with 
$E_{\nu} \sim 2$~GeV and a $\nu_e$ contamination of less than 0.5\%.
The baseline is $L=810$~km with the detector sited 12~km 
($\sim 0.85^{\circ}$) off-axis.
If approved, the experiment could start data taking in 2013.
The NuMI target will receive a 120~GeV/c proton beam with an expected
intensity of $6.5 {\cdot} 10^{20}$ pot/year.
The beam will be measured at a near and at a far detector, both
`totally active' liquid-scintillator detectors.
With and a five-year run and a detector mass $\sim 15$~Kton, NO$\nu$A
will achieve a sensitivity to $\sin^2 2 \theta_{13}$ comparable to
that which T2K can achieve.
The long baseline allows NO$\nu$A to make a measurement of 
$|\Delta m_{31}^2|$.

The possibility of exploiting NO$\nu$A together with a second
detector at a different baseline to determine the mass hierarchy has
been discussed \cite{MenaRequejo:2005hn,Mena:2005ri}.  
The potential of the increased data volume provided by the NuMI beam
instrumented with yet larger detectors, or detectors with larger
detection efficiency, in conjunction with a possible NuMI upgrade has
been studied \cite{Mena:2005sa}.
However, as yet there is no well-developed proposal for a NO$\nu$A
upgrade that is able to compete with other second generation
super-beams such as T2HK or the SPL.

\subsubsection{Wide-band super-beam}

A wide-band beam has been proposed, sited at BNL and serving a very
long baseline experiment
\cite{Diwan:2003bp,Diwan:2004bt,Barger:2006vy,Diwan:2006qf}.
In this proposal, the 28~GeV AGS would be upgraded to 1~MW and a
neutrino beam with neutrino energies in the range $0 - 6$ GeV could be
sent to a Mton water \v Cerenkov detector at the Homestake mine at a
baseline of 2540~km.

Wide-band beams possess the advantages of a higher on-axis flux and a
broad energy spectrum.
The latter allows the first and second oscillation nodes in the
disappearance channel to be observed, providing a strong tool to solve
the degeneracy problem. 
On the other hand, experiments served by wide-band beams must
determine the incident neutrino energy with good resolution and
eliminate the background from high energy tail of the spectrum. 

Upgrades to the FNAL main injector after the end of the Tevatron
programme are also under study and could provide a similar wide-band
neutrino beam.
The baseline in this case would be 1290~km. 
\footnote{
  Since the ISS concluded, the proposal to site a Deep Underground 
  Science and Engineering Laboratory (DUSEL) at the Homestake Mine in 
  South Dakota has been approved.
  A proposal to site a neutrino detector with a fiducial mass in
  excess of 100~kTonnes at DUSEL illuminated by a beam from FNAL is
  under discussion.
}
In the following, the flux obtained using 28~GeV protons and a 
200~m long decay tunnel will be used. 
For details of this spectrum see reference \cite{Diwan:2004bt}. 

The combination of channels and spectral information of a long
baseline wide-band beam experiment offers a promising means of solving
parameter degeneracies. 
However, the very long baseline decreases the event rate at the far
detector and reduces the sensitivity of the experiment to
$\theta_{13}$ and CP-violation; the sensitivity of the experiment to 
$\theta_{13}$ and $\delta$ is somewhat smaller than that of T2HK or
the SPL. 
Therefore, the following sections will focus on the performance of T2HK
and the SPL.
The performance of the wide-band beam will be discussed when
considering the determination of the mass hierarchy, where the long
baseline means that the wide-band beam out-performs T2HK and the SPL.
The wide-band beam is a very interesting option to search for leptonic
CP-violation, solving most of the degeneracies, if $\theta_{13}$ is
large enough, i.e. $\sin^2 2\theta_{13} > 5\times 10^{-3}$
($\theta_{13}>2^\circ$).

\subsubsection{Physics at a super-beam facility}

The first generation of neutrino super-beams, T2K and NO$\nu$A,
will study the $\nu_{\mu} \rightarrow \nu_e$ channel which is
sensitive to $\theta_{13}$ and $\delta$.
The experiments will start by running in neutrino mode.
This has the advantage that a large data set can be accumulated
relatively rapidly since the neutrino cross section is larger than the
anti-neutrino cross section.
Neutrino running alone, however, implies that the experiments
have no sensitivity to $\delta$. 
A second generation of upgraded super-beams, such as T2HK or the SPL,
could follow.
The extremely large data sets provided by these experiments would
yield sensitivity to much smaller values of $\theta_{13}$.
These experiments could also search for CP-violation by running with
anti-neutrinos, if $\theta_{13}$ is large enough. 
In the rest of this section, the sensitivity to $\theta_{13}$ and
$\delta$ of this second generation of super-beams will be
considered.

The search for small $\theta_{13}$ in the 
$\nu_{\mu}\rightarrow \nu_e$ channel suffers from parametric
degeneracies (see section \ref{sec:emu}). 
To alleviate this problem, and to improve significantly the
measurement of the atmospheric parameters $\theta_{23}$ and 
$\Delta m^2_{31}$ at these facilities, it is extremely useful to study
$\nu_{\mu}$ disappearance as well. 
Such measurements are also of importance in order to establish whether
$\theta_{23}$ is maximal in order to discriminate between different
mass models. 
The maximal-mixing-exclusion potential of the various super-beams will
therefore also be investigated below.

\subsubsection{The Water \v Cerenkov Detector}

 For small values of $\theta_{13}$, a very large data set is required
for the sub-leading $\nu_{\mu} \rightarrow \nu_e$ oscillation to be
observed.
The water \v Cerenkov is an ideal detector for this task since it is
possible to construct a detector of very large fiducial mass in which
the target material is also the active medium.
The \v Cerenkov light is collected by photo-detectors distributed over
the surface of the detector; the cost of instrumenting the detector,
therefore, scales with the surface area rather than the fiducial
mass.
Mton-class, water \v Cerenkov detectors are therefore ideal when
charge identification is not required and have been chosen for T2HK,
the SPL, and the wide-band beam long-baseline experiment. 
Such a device could also be the ultimate tool for proton-decay
searches and for the detection of atmospheric, solar, and supernov\ae
neutrinos.

Charged leptons are identified through the detection of \v Cerenkov
light in photo-multiplier tubes (PMTs) distributed around the vessel. 
The features of the \v Cerenkov rings can be exploited for particle
identification. 
A muon scatters very little in crossing the detector, therefore, the
associated \v Cerenkov ring has sharp edges. 
Conversely, an electron showers in the water, producing rings with
`fuzzy' edges. 
The total measured light can be used to give an estimate of the lepton
energy, while the time measurement provided by each PMT allows the
lepton direction and the position of the neutrino interaction vertex
to be determined. 
By combining all this information, it is possible to reconstruct the
energy, the direction, and the flavour of the incoming neutrino. 
It is worth noting that the procedure discussed above is suitable only
for quasi-elastic events ($\nu_l n \rightarrow l^- p$). 
Indeed, for non-quasi-elastic events more particles are present in the
final state that are either below the \v Cerenkov threshold or are
neutral, resulting in a poor measurement of the total event energy.
Furthermore, the presence of more than one particle above threshold
produces more than one ring, spoiling the particle identification
capability of the detector. 

The water \v Cerenkov is a mature technology that has been
demonstrated to be cost effective and to give excellent performance at
low neutrino energies. 
A detector with a fiducial mass as large as 20 times that of
Super-Kamiokande could be built and would be an optimal detector for
neutrino beams with energies around or below 1 GeV \cite{Jung:1999jq}.
There are three different proposals for such a detector, each of them
exploited by a different super-beam. 
Hyper-Kamiokande \cite{Itow:2001ee} could be located at the Kamioka
mine, at a distance of 295~km from J-PARC facility in Tokai. 
MEMPHYS \cite{deBellefon:2006vq}, in the Fr\'{e}jus area, could 
receive the SPL beam produced 130~km away at CERN. 
The wide-band beam produced at BNL (FNAL) could aim at a detector in
the Homestake mine \cite{Diwan:2006qf} at 2540~km (1290~km).

\subsubsection{Backgrounds and efficiencies}

In a conventional super-beam experiment, the search for 
$\nu_\mu \rightarrow \nu_e$ ($\bar{\nu}_\mu\rightarrow\bar{\nu}_e$)
is complicated by the $\nu_e(\bar{\nu}_e)$ contamination of the beam.
In a water \v Cerenkov detector, the appearance, $\nu_e(\bar{\nu}_e)$,
signal is detected by exploiting the high efficiency and high purity
of the detector in identifying electrons and muons in low multiplicity
interactions.
In addition to the $\nu_e(\bar{\nu}_e)$ contamination of the beam, the
main sources of background are the charged-current interactions of
$\nu_\mu(\bar{\nu}_\mu)$ and the production of $\pi^0$s in
neutral-current interactions. 
Even though the performance of water \v Cerenkov detectors is very
well studied, there are few analyses of the efficiencies and
backgrounds expected in the various super-beams considered here.

For T2HK, there is only the study reported in the letter of intent
\cite{Itow:2001ee}.
The expected signal- and background-event rates for the 
$\nu_\mu \rightarrow \nu_e$ channel are presented in table 2 of
reference \cite{Itow:2001ee}.
The expected efficiencies and fractional backgrounds have been
extracted for several analyses from this table
\cite{Huber:2002mx,Huber:2002rs,Huber:2004ug,Huber:2004gg,Huber:2005ep,Donini:2005db,Campagne:2006yx}.  
The signal efficiency, assumed to be constant, is 0.505. 
The various contributions to the background ($N^{bg}$), from the
$\nu_e(\bar{\nu}_e)$ contamination in the beam ($N_e^{CC}$), $\pi^0$
production in neutral-current events ($N^{NC}$), and $\nu_\mu$
charged-current interactions ($N_\mu^{CC}$ have the following weights:
\begin{equation}
  N^{bg} = 7.5 \cdot 10^{-2}N_e^{CC} + 
           5.6\cdot10^{-3} N^{NC}    + 
           3.3\cdot 10^{-4} N_\mu^{CC}
\end{equation}
The same efficiencies and backgrounds have been assumed for the
$\bar{\nu}_\mu \rightarrow \bar{\nu}_e$ channel since no further
information on this channel is available.
The efficiencies and backgrounds are assumed to be flat since no
energy dependence is presented. 
This is only an approximation and a more detailed description in terms
of migration and background matrices as in
\cite{Burguet-Castell:2003vv,Burguet-Castell:2005pa} would be
desirable.
For the spectral information, 20 bins of 40~MeV between 0.4~GeV and
1.2~GeV have been considered, convoluted with a Gaussian with 
$\sigma = 85$~MeV to account for the Fermi motion as in reference
\cite{Huber:2002mx}. 

The situation is very similar in the case of the SPL. 
The only available study is that of reference
\cite{Gomez-Cadenas:2001eu}, from which flat efficiencies and
backgrounds can be extracted. 
The efficiencies quoted in \cite{Gomez-Cadenas:2001eu} are 0.707 for
the $\nu_\mu \rightarrow \nu_e$ channel and 0.671 for 
$\bar{\nu}_\mu \rightarrow \bar{\nu}_e$.  
The backgrounds, in a notation consistent with that used above, are:
\begin{eqnarray}
  N^{bg} & = & 4.4 \cdot 10^{-1}N_e^{CC} + 2.7\cdot10^{-3} N^{NC} +
  4.6\cdot 10^{-4} N_\mu^{CC} \; ; {\rm and} \\
  \bar{N}^{bg} & = & 6.8 \cdot 10^{-1} N_e^{CC} + 4.4\cdot10^{-4} N^{NC} +
  1.3\cdot 10^{-3} N_\mu^{CC} \; .
\end{eqnarray}
These numbers have been used in several different studies
\cite{Donini:2005db,Campagne:2006yx,Donini:2004hu,Donini:2004iv,Donini:2005rn}. 
The SPL 3.5 GeV fluxes will be used in the following, as computed in
\cite{Campagne:2004wt}.  The energy
reconstruction is modelled with migration matrixes following
the work of \cite{Donini:2005db}.
For both the SPL and T2HK, a 440 Kton fiducial mass for the detector 
and 10 years running time have been assumed. 
The running time has been divided between the neutrino and
anti-neutrino mode in such a way as to produce a roughly equal number
of events for each channel. 
For both experiments, the rather optimistic value of 2\% has been
adopted for the systematic uncertainty.
The less optimistic case of 5\% systematic uncertainty is also
presented.
These errors are assumed to be uncorrelated between the various signal
channels (neutrinos and anti-neutrinos), and between the signal and
background samples.

For the wide-band beam long-baseline experiment, migration matrices
for both the signal and the background channels have been computed
\cite{Barger:2006vy} from a Monte Carlo simulation from the detector
described in \cite{chiaki}.
Following reference \cite{Barger:2006vy}, a 300 Kton fiducial mass
detector, 5~years neutrino running with 1~MW proton-beam power, and
5~years anti-neutrino running with a proton-beam power of 2~MW have
been considered.

\subsubsection{The super-beam performance}

In the following, the performance of T2HK and the SPL super-beams is
presented in terms of the $\theta_{13}$ and the CP-violation discovery
potential, the sensitivity of the facility to the maximality of
$\theta_{23}$, the mass hierarchy, and the octant of $\theta_{23}$.
The precision with which the atmospheric parameters can be measured is
also presented. 
To simulate the `data', the following set of `true values' for the
oscillation parameters are adopted: 
\begin{equation}\label{eq:default-params}
  \begin{array}{l@{\qquad}l}
    \Delta m^2_{31} = +2.5 \times 10^{-3}~\mathrm{eV}^2\,; &
    \sin^2\theta_{23} = 0.5\,;\\
    \Delta m^2_{21} = 8.0 \times 10^{-5}~\mathrm{eV}^2 \,;&
    \sin^2\theta_{12} = 0.3 \,;
  \end{array}
\end{equation}
and we include a prior knowledge of these values with a $1 \sigma$
accuracy of 5\% for $\theta_{12}$ and $\Delta m^2_{21}$. 
$\theta_{23}$ and $\Delta m^2_{31}$ can be measured by these
experiments and have been left free in the fits. 
These values and accuracies are motivated by recent global fits to
neutrino oscillation data \cite{Fogli:2005cq,Maltoni:2004ei}, and they
are always used except where explicitly stated otherwise.

\subsubsection{The $\theta_{13}$ discovery potential}
\label{sec:th13}

If the first generation of super-beam experiments do not demonstrate
that $\theta_{13}$ is non-zero, then the second generation facility
will be required to have a significantly improved sensitivity to
this parameter.
To assess the sensitivity of the proposed second-generation
super-beams to $\theta_{13}$, the following definition of the
discovery potential is used.
Data are simulated for a non-zero `true' value of $\stheta$ and for a
given true value of $\delta$.
If the $\Delta\chi^2$ of the fit to these data with $\theta_{13} = 0$
(marginalised over all parameters except $\theta_{13}$) is larger than
9, the corresponding true value of $\theta_{13}$ is taken to be
`discovered' at $3 \sigma$.  
In other words, the $3\sigma$-discovery limit as a function of the
true $\delta$ is given by the true value of $\stheta$ for which
$\Delta\chi^2(\theta_{13}=0) = 9$.  
In general, tests must also be made for degenerate solutions in
sign($\Delta m^2_{31}$) and the octant of $\theta_{23}$.

\begin{figure}
  \begin{center}
    \includegraphics[width=0.9\textwidth]{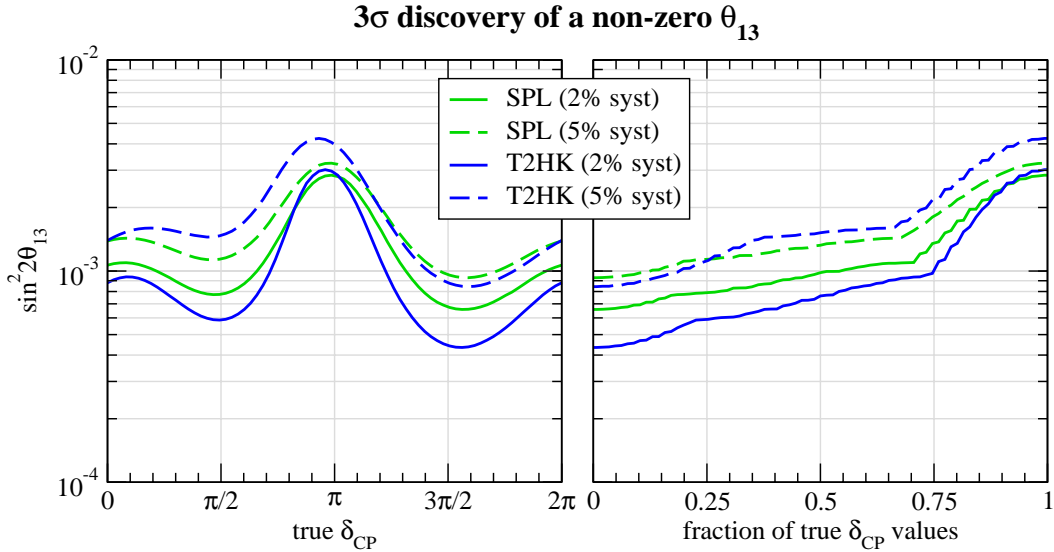}
  \end{center}
  \caption{
    $3\sigma$ discovery sensitivity to $\stheta$ for the SPL and T2HK
    as a function of the true value of $\delta$ (left panel) and as a
    function of the fraction of all possible values of $\delta$ (right
    panel). 
    Solid (dashed) lines are for 2\% (5\%) systematic errors. 
    Adapted with kind permission of the Journal of High Energy Physics 
    from figure 9 in reference \cite{Campagne:2006yx}.
    Copyrighted by SISSA.
  }
  \label{fig:th13SB}
\end{figure}

The discovery limits for the SPL super-beam and for T2HK are shown in 
figure \ref{fig:th13SB}. 
The performance of the two facilities is rather similar, and a
discovery potential down to $\stheta \simeq 4\times 10^{-3}$ is within
reach for all possible values of $\delta$. 
For certain values of $\delta$ (around $\delta = \pi/2$ or
$3\pi/2$) the sensitivity is significantly improved, and discovery
limits below $\stheta \simeq 10^{-3}$ are possible for a large
fraction of all possible values of $\delta$.
The wide-band beam long-baseline experiment has a slightly lower
sensitivity ranging from $\stheta \simeq 2\times 10^{-3}$ to 
$\stheta \simeq 5\times 10^{-3}$ (see figure 5 of reference 
\cite{Barger:2006vy}). 

Figure \ref{fig:th13SB} also illustrates the effect of systematic
uncertainties on the $\theta_{13}$ discovery reach.  
The lower (solid) boundary of the band for each experiment corresponds
to a systematic error of 2\%, whereas the upper (dashed) boundary is
obtained for a systematic uncertainty of 5\%. 
These uncertainties include the (uncorrelated) normalisation
uncertainties on the signal as well as the background; the
dominant uncertainty is the uncertainty on the background. 
For the SPL, systematic uncertainties have a rather small impact on
the sensitivity, whereas for the larger data set acquired by T2HK, the
limit is more strongly affected.

\subsubsection{CP-violation discovery potential}
\label{sec:CPV}

If $\theta_{13}$ is shown to be non-zero, then it becomes important to
assess the leptonic CP-violation (CPV) discovery potential
quantitatively, i.e. to assess the extent to which the proposed
second-generation super-beam experiments can establish that 
$\delta$ differs from $0$ or $\pi$. 
The CPV-discovery potential is evaluated as follows.
Simulated data sets were produced for a range of assumed `true' values
of $\stheta$ and $\delta$. 
These data were then fitted using the CP-conserving values 
$\delta = 0$ and $\delta = \pi$, all other parameters being
marginalised and the sign and octant degeneracies being taken into
account.
If no fit with $\Delta \chi^2 < 9$ is found, CP conservation can be
excluded at $3\sigma$ confidence level for the chosen values of
$\delta^\mathrm{true}$ and $\stheta^\mathrm{true}$.

\begin{figure}
  \begin{center}
    \includegraphics[width=0.65\textwidth]{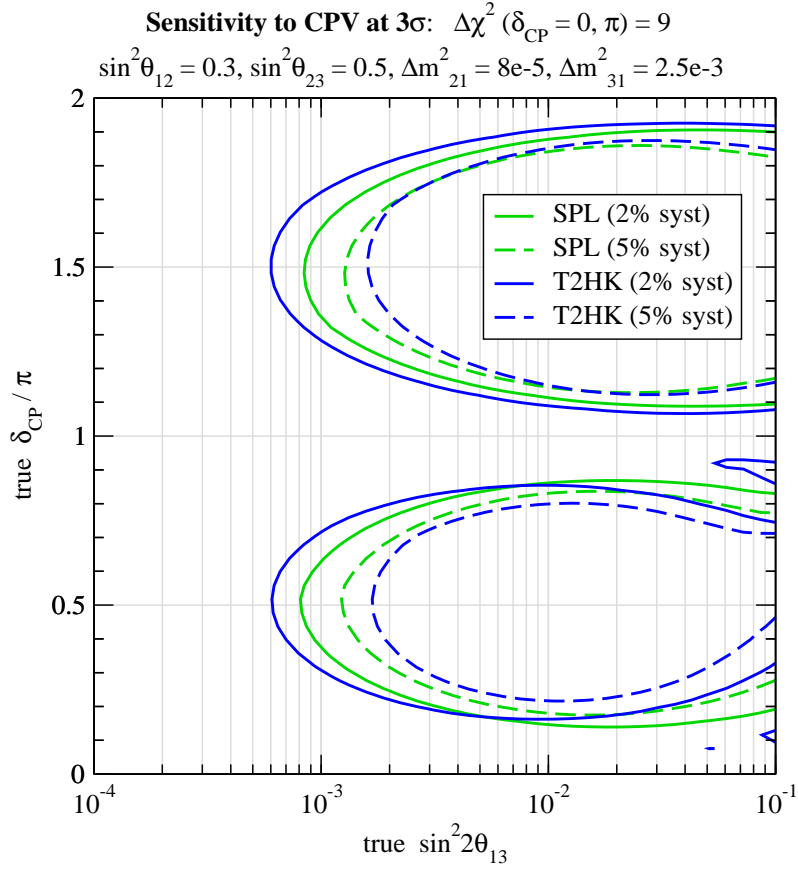}
  \end{center}
  \caption{
    CPV discovery potential the SPL, and T2HK: for parameter values
    inside the ellipse-shaped curves CP conserving values of $\delta$
    can be excluded at $3\sigma$ $(\Delta\chi^2>9)$.
    Solid (dashed) lines are for 2\% (5\%) systematic errors.
    Adapted with kind permission of the Journal of High Energy Physics 
    from figure 11 in reference \cite{Campagne:2006yx}.
    Copyrighted by SISSA.
  }
  \label{fig:CPV}
\end{figure}

The CPV discovery potential for the SPL super-beam, and for T2HK is
shown in figure \ref{fig:CPV}. 
As in the case of the $\theta_{13}$ discovery potential, the
performance of the two facilities is comparable.
For an assumed systematic uncertainty of 2\%, maximal CPV (for
$\delta^\mathrm{true} = \pi/2, \, 3\pi/2$) can be discovered at
$3\sigma$ down to $\stheta \simeq 6\times 10^{-4}$ for T2HK, and
$\stheta \simeq 8\times 10^{-4}$ for the SPL super-beam.
The CPV discovery potential of the wide-band long-baseline super-beam
would be limited to $\stheta \simeq 4\times 10^{-3}$ (see the right
panel of figure 7 in reference \cite{Barger:2006vy}). 
The best sensitivity to CPV is obtained for $\stheta \gtrsim 10^{-2}$, 
where, for a systematic uncertainty of 2\%, CPV can be established for 
75\% of all possible values of $\delta$.
The figure shows the expected performance for systematic uncertainties
of 2\% and 5\%. 
Again, T2HK is more strongly affected by the systematic uncertainties,
out-performing the SPL super-beam for a 2\% uncertainty but being
out-performed by it for a 5\% uncertainty. 

The sensitivity maximum around $\stheta \simeq 10^{-2}$ can easily be
understood from the oscillation probability. 
The interference term that allows the measurement of $\delta$ is
suppressed by $\sin2\theta_{13}$ and $\Delta m^2_{21} L/4E$ (see, for
example equation (7) of reference \cite{Cervera:2000kp}). 
There are two other leading terms in the probability, one suppressed 
by $\sin^22\theta_{13}$ and the other suppressed by $(\Delta
m^2_{21}/4E)^2$. 
For $\sin2\theta_{13} \simeq \Delta m^2_{21} L/4E$, the three terms in
the oscillation probability will be of the same order of magnitude and
the CP-violation signal will not be hidden by the other two terms.
On the other hand, if $\sin2\theta_{13}$ becomes too large or too
small, one of the two CP-conserving terms dominates the interference
term resulting in a loss of sensitivity. 
Indeed, for experiments built at the first peak of the atmospheric
oscillation, $\sin2\theta_{13} \simeq \Delta m^2_{21} L/4E$ for
$\stheta \simeq 10^{-2}$. 
If the experiment operates at the second oscillation peak the larger
$\Delta m^2_{21} L/4E$ will shift the maximum of the CP-violation
sensitivity to larger values of $\stheta$, as can be seen 
in the right panel of figure 3 of reference \cite{Donini:2006dx}.

\subsubsection{Maximal $\theta_{23}$ exclusion potential}

\begin{figure}
  \vspace{-0.5cm}
  \begin{center}
    \hspace{-1cm} \epsfxsize8.25cm\epsffile{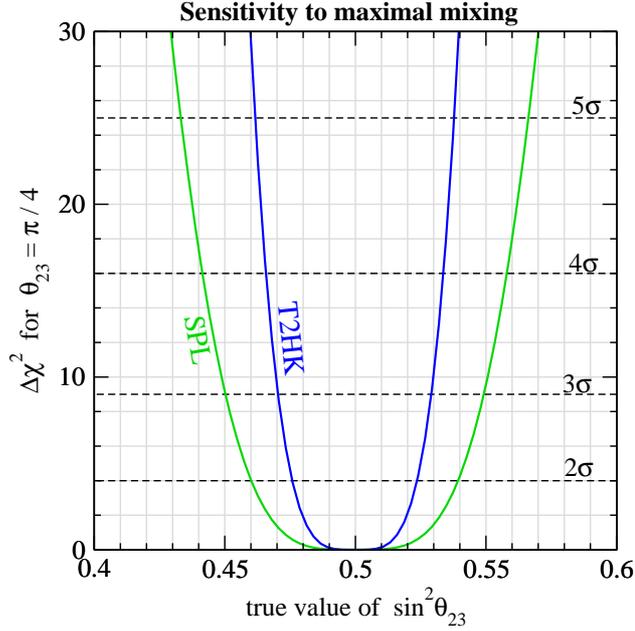}
  \end{center}
  \caption{
    3$\sigma$ maximal mixing exclusion potential for the SPL and
    T2HK. 
    The $\Delta\chi^2$ for maximal $\theta_{23}$ is shown as a
    function of the true value of $\sin^2\theta_{23}$. 
  }
  \label{fig:sensit2k}
\end{figure}

Experiments able to study the $\nu_\mu \rightarrow \nu_\mu$
oscillation can address the issue of the maximality of $\theta_{23}$
which is crucial to discriminate between different models of neutrino
mass.
The potential to exclude maximal $\theta_{23}$ has been computed in
the following way: data are simulated for different true values of
$\sin^2 \theta_{23}$, if the $\Delta\chi^2$ of the fit to these data
with $\sin^2\theta_{23} = 0.5$ (marginalised over all parameters
except $\sin^2\theta_{23}$) is larger than 9, then maximal mixing can
be excluded at 3$\sigma$.
Figure \ref{fig:sensit2k} shows that both T2HK and the SPL super-beam
can measure at $3\sigma$ any deviation from maximal mixing larger than
10\%. 
However, T2HK, with its better spectral information, out-performs
the SPL, going down in sensitivity to deviations of 6\% from maximal
mixing. 
The importance of energy resolution in the disappearance channel to
exclude maximal mixing is discussed in reference
\cite{Donini:2005db}.

\subsubsection{Sensitivity to the atmospheric parameters}
\label{sec:atm}

\begin{figure}
  \begin{center}
    \includegraphics[width=0.55\textwidth]{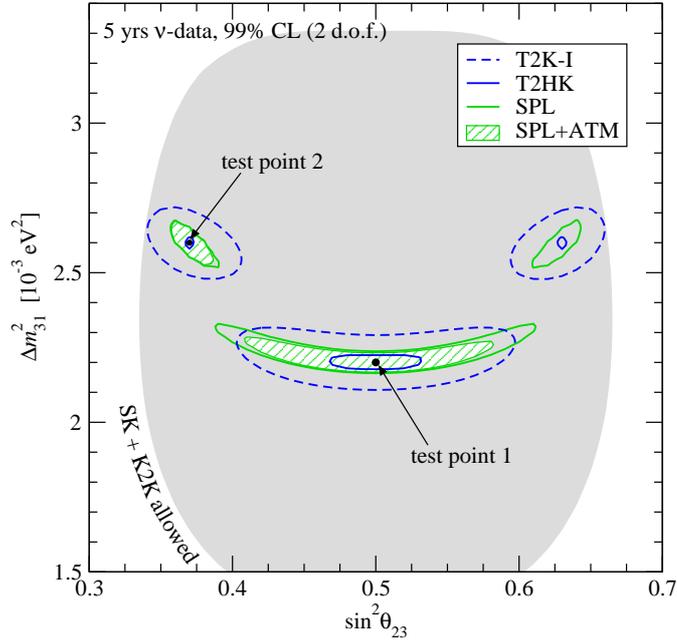}
  \end{center}
  \caption{
    Allowed regions of $\Delta m^2_{31}$ and $\sin^2\theta_{23}$ at
    99\%~CL (2 d.o.f.)  after 5~yrs of neutrino data taking for the
    SPL, T2K, and T2HK projects, and the combination of the SPL with
    5~years of atmospheric-neutrino data in the MEMPHYS detector. 
    For the true parameter values we use 
    $\Delta m^2_{31} = 2.2\, (2.6) \times 10^{-3}~\mathrm{eV}^2$ and 
    $\sin^2\theta_{23} = 0.5 \, (0.37)$ for the test point 1 (2), and
    $\theta_{13} = 0$ and the solar parameters as given in equation
    (\ref{eq:default-params}).  
    The shaded region corresponds to the 99\%~CL region from present
    SK and K2K data \cite{Maltoni:2004ei}. 
    Taken with kind permission of the Journal of High Energy Physics 
    from figure 8 in reference \cite{Campagne:2006yx}.
    Copyrighted by SISSA.
  } 
  \label{fig:atm-params}
\end{figure}

The $\nu_\mu$ disappearance channel available in super-beam
experiments allows the atmospheric parameters $|\Delta m^2_{31}|$ and
$\sin^2\theta_{23}$ to be determined precisely (see, e.g., references
\cite{Donini:2005db,Antusch:2004yx,Minakata:2004pg} for recent
analyses).
Figure \ref{fig:atm-params} illustrates the improved precision with
which these parameters will be determined in future super-beam
experiments.
The figure shows the allowed regions at 99\%~CL for T2K, the SPL, and
T2HK, where, in each case, five years of neutrino data are assumed. 
Table \ref{tab:atm-params} gives the corresponding relative accuracies
at 3$\sigma$ for $|\Delta m^2_{31}|$ and $\sin^2\theta_{23}$.
\begin{table}
  \begin{center}
    \begin{tabular}{lcrrr}
      \hline\noalign{\smallskip}
        & True values  & T2K & SPL & T2HK \\
      \noalign{\smallskip}\hline\noalign{\smallskip}
      $\Delta m^2_{31}$   & $2.2\cdot 10^{-3}$ eV$^2$ & 4.7\% & 3.9\% & 1.1\% \\
      $\sin^2\theta_{23}$ & $0.5$                     & 20\%  & 22\%  & 6\%   \\
      \noalign{\smallskip}\hline\noalign{\smallskip}
      $\Delta m^2_{31}$   & $2.6\cdot 10^{-3}$ eV$^2$ & 4.4\% & 3.0\% & 0.7\% \\
      $\sin^2\theta_{23}$ & $0.37$                    & 8.9\% & 4.7\% & 0.8\% \\
      \noalign{\smallskip}\hline
    \end{tabular}
  \end{center}
  \caption{
    Accuracies at $3\sigma$ on the atmospheric parameters 
    $|\Delta m^2_{31}|$ and $\sin^2\theta_{23}$ for 5 years of
    neutrino data from T2K, SPL, and T2HK for the two test points
    shown in figure \ref{fig:atm-params} 
    ($\theta^\mathrm{true}_{13} = 0$). 
    The accuracy for a parameter $x$ is defined as 
    $(x^\mathrm{upper} - x^\mathrm{lower})/(2 x^\mathrm{true})$, 
    where $x^\mathrm{upper}$ ($x^\mathrm{lower}$) is the upper (lower)
    bound at 3$\sigma$ for 1~d.o.f.\ obtained by projecting the
    contour $\Delta \chi^2 = 9$ onto the $x$-axis. 
    For the accuracies for test point~2 the octant-degenerate solution
    is neglected.
  }
  \label{tab:atm-params}
\end{table}

From the figure and the table it is evident that T2K and T2HK
are very good at measuring the atmospheric parameters, only a
modest improvement is possible with SPL with respect to T2K.
T2HK provides excellent sensitivity to these parameters: for 
test-point~2, for example, sub-percent accuracies are obtained at 
3$\sigma$. 
The disadvantage of the SPL with respect to T2HK is the limited
spectral information. 
Because of the lower beam energy, nuclear Fermi motion is a severe
limitation for energy reconstruction in the SPL super-beam, whereas in
T2HK the somewhat higher energy allows an efficient use of spectral
information in quasi-elastic events. 
The effect of spectral information on the disappearance measurement is 
discussed in detail in reference \cite{Donini:2005db}.

For test point~1 (maximal mixing for $\theta_{23}$), rather poor
accuracies are obtained for $\sin^2\theta_{23}$ for T2K and the SPL
($\sim20\%$), and only $6\%$ for T2HK. 
The reason is that in the disappearance channel $\sin^22\theta_{23}$
(rather then $\sin^2\theta_{23}$) is measured.
This translates into rather large errors for $\sin^2\theta_{23}$ if
$\theta_{23} = \pi/4$~\cite{Minakata:2004pg}.
For the same reason it is difficult to solve the octant degeneracy.
It can be seen that for test point~2, with a non-maximal value of
$\sin^2\theta_{23} = 0.37$, the degenerate solution is still present
around $\sin^2\theta_{23} = 0.63$ in each of the three experiments.

\subsubsection{Sensitivities to the mass hierarchy and the $\theta_{23}$ octant}

\begin{figure}
  \begin{center}
    \includegraphics[width=0.5\textwidth]{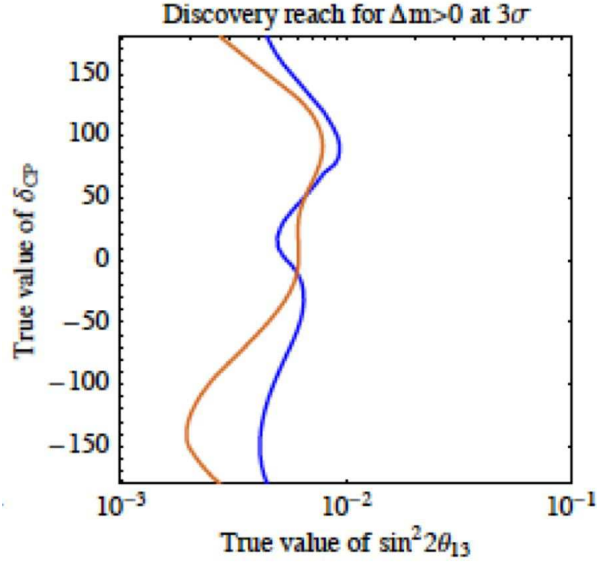}
  \end{center}
  \caption{
    Sensitivity of the wide band beam long baseline experiment to the
    mass hierarchy at $3\sigma$ $(\Delta\chi^2 = 9)$ as a function of
    the true values of $\sin^22\theta_{13}$ and $\delta$. 
    The blue (dark) curves are for $L=1300$ km and the red (light)
    curves for $L = 2500$ km. 
    The figure is taken from reference \cite{Diwan:2006qf}.
  } 
  \label{MH1300}
\end{figure}

The determination of the mass hierarchy is a secondary goal for
super-beams such as T2HK and the SPL which have too short a baseline
to exploit the matter effects required to solve these degeneracies. 
Indeed, the sensitivity to the hierarchy of T2HK is limited to some
favourable values of $\delta$, while the SPL has no sensitivity
whatsoever. 
However, the long baseline of the wide-band beam experiment gives
significant sensitivity to the mass hierarchy.
In figure \ref{MH1300}, the discovery potential for a normal mass
hierarchy is shown for two different baselines: 2500~km, roughly the
baseline between BNL and Homestake; and 1300~km the distance between
FNAL and Homestake. 
It can be seen that, for the 1300~km baseline, 
if $\sin^2 2 \theta_{13} > 10^{-2}$, the mass hierarchy can be
measured for any value of $\delta$.
The sensitivity is further increased to 
$\sin^2 2 \theta_{13} > 8\times 10^{-3}$ for the longer baseline. 

As was shown above, neither experiment is sensitive to the octant of
$\theta_{23}$. 
However, as pointed out in references
\cite{Peres:2003wd,Gonzalez-Garcia:2004cu}, atmospheric-neutrino data
may allow the octant of $\theta_{23}$ to be determined. 
If 5~years of atmospheric-neutrino data in MEMPHYS are added to the
SPL super-beam data, the degenerate solution for test point~2 can be
excluded at more than $5\sigma$, as can be seen in figure
\ref{fig:atm-params}, and hence the octant degeneracy is solved in
this example. 
Of course, this way of measuring the octant works even better if
atmospheric data taken with Hyper-Kamiokande are combined with T2HK
data, see below. 

\subsubsection{Combination with atmospheric neutrino measurements}
\label{sec:atmospherics}

Combining atmospheric-neutrino events to the long-baseline
neutrino-beam data is an attractive method of resolving degeneracies
\cite{Huber:2005ep}. 
If $\theta_{13}$ is sufficiently large, Earth matter effects in
multi-GeV, $e$-like atmospheric-neutrino events are sensitive to the
mass hierarchy
\cite{Petcov:1998su,Akhmedov:1998ui,Bernabeu:2003yp}. 
Moreover, sub-GeV, $e$-like events provide sensitivity to the octant of
$\theta_{23}$ \cite{Kim:1998bv,Peres:2003wd,Gonzalez-Garcia:2004cu}
due to oscillations driven by $\Delta m^2_{21}$ (see also reference
\cite{Kajita} for a discussion of atmospheric neutrinos in the context
of Hyper-Kamiokande).
Following reference \cite{Huber:2005ep}, the potential of the various
second-generation super-beam experiments is investigated with the
combined beam- and atmospheric-neutrino data set below.
A general three-flavour analysis of atmospheric data is performed
based on reference \cite{Gonzalez-Garcia:2004cu} and references
therein. 
Fully-contained $e$-like and $\mu$-like events (further divided into
sub-GeV $p_l<400$~MeV, sub-GeV $p_l > 400$~MeV, and Multi-GeV events)
are included. 
In addition, partially-contained $\mu$-like events, stopping 
muons, and through-going muons are considered. 
Each of these data samples is divided into 10 zenith angle bins.

\begin{figure}
  \begin{center}
    \includegraphics[width=0.9\textwidth]{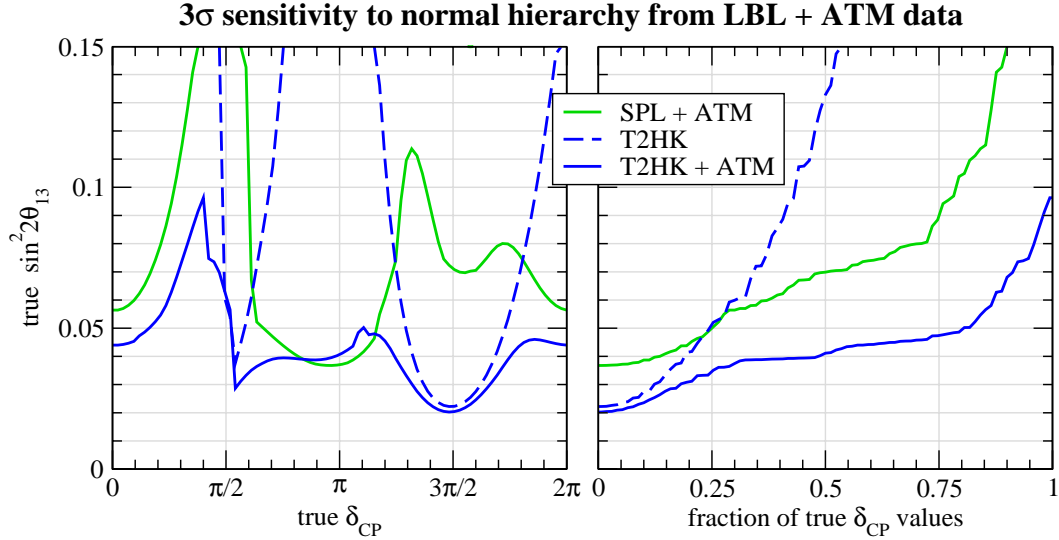}
  \end{center}
  \caption{
    Sensitivity to the mass hierarchy at $3\sigma$ 
    $(\Delta\chi^2 = 9)$ as a function of the true values of
    $\sin^22\theta_{13}$ and $\delta$ (left panel), and
    the fraction of true values of $\delta$ (right panel). 
    The solid curves are the sensitivities from the combination of the
    super-beams and atmospheric neutrino data, the dashed curves
    correspond to super-beam data only. 
    Adapted with kind permission of the Journal of High Energy Physics 
    from figure 16 in reference \cite{Campagne:2006yx}.
    Copyrighted by SISSA.
  }
  \label{fig:hierarchy}
\end{figure}

Figure \ref{fig:hierarchy} shows how the combination of
atmospheric plus long-baseline yields sensitivity to the sign of
$\Delta m^2_{31}$. 
For the long-baseline data alone, the SPL super-beam has no
sensitivity (because of the very small matter effects that arise in
the relatively short baseline) and the sensitivity of T2HK 
depends strongly on the true value of $\delta$. 
However, by including data from atmospheric neutrinos the mass 
hierarchy can be determined at the $3\sigma$~CL provided
$\sin^22\theta_{13} \gtrsim 0.05-0.09$ for the SPL, and
$\sin^22\theta_{13} \gtrsim 0.03-0.05$ for T2HK. 
Both experiments have the worst sensitivity around 
$\delta = \pi/2$, where the enhancement of the neutrino signal
and the suppression of the anti-neutrino signal typical of the normal
hierarchy is masked by the opposite effect of the CP-violating phase.
Here, T2HK would only be able to exclude an inverted hierarchy if
$\sin^22\theta_{13} \gtrsim 0.1$ and the SPL loses its sensitivity
altogether. 
Conversely there are maximums of the sensitivity around 
$\delta = 3\pi/2$, where $\delta$ enhances the neutrino
signal and suppresses that of the anti-neutrino. 
Comparing figure \ref{fig:hierarchy} with figure \ref{MH1300} it is
clear that, even when combined with atmospheric data, the sensitivity
of T2HK and the SPL to the mass hierarchy is rather poor, being
out-performed by the longer baseline wide-band beam experiment by an
order of magnitude. 

\begin{figure}
  \begin{center}
    \includegraphics[width=0.55\textwidth]{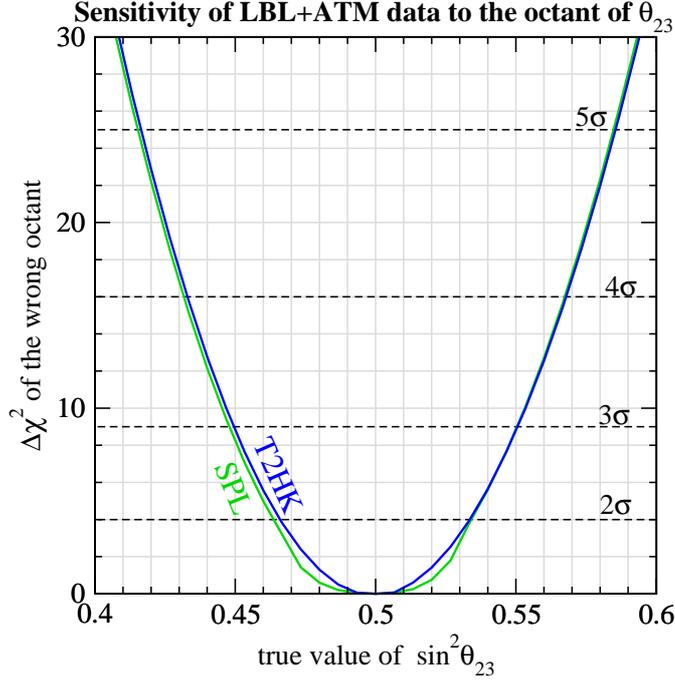}
  \end{center}
  \caption{
    $\Delta\chi^2$ of the solution with the wrong octant of
    $\theta_{23}$ as a function of the true value of
    $\sin^2\theta_{23}$. 
    A true value of $\theta_{13} = 0$ has been assumed. 
    Adapted with kind permission of the Journal of High Energy Physics 
    from figure 17 in reference \cite{Campagne:2006yx}.
    Copyrighted by SISSA.
  }
  \label{fig:octantSB}
\end{figure}

Figure \ref{fig:octantSB} shows the potential of atmospheric plus
long-baseline data to exclude the octant-degenerate solution. 
Since this effect is based mainly on oscillations driven by
$\Delta m^2_{21}$, there is very good sensitivity even for
$\theta_{13} = 0$; a non-zero value of $\theta_{13}$ improves the
sensitivity in most cases \cite{Huber:2005ep}. 
From the figure one can see that both experiments can identify
the true octant at 
$3\sigma$ for $|\sin^2\theta_{23} - 0.5| \gtrsim 0.05$.

\subsubsection{Super-Beam associated with a beta-beam}

A beta-beam could exploit the intense proton driver required to drive
a super-beam and both facilities could illuminate the same far
detector.
The SPL in particular could be complemented by a low-$\gamma$
beta-beam in the CERN design (see section
\ref{SubSect:Perf:BetaBeam}).
It is therefore interesting to study possible complementarities
between the two facilities. 
The main difference between the two neutrino beams is the different
initial neutrino flavour, $\nu_e$ ($\bar{\nu}_e$) for a beta-beam and 
$\nu_{\mu}$ ($\bar{\nu}_\mu$) for a super-beam. 
This implies that at the near detector all relevant cross-sections can
be measured. 
In particular, the near detector exposed to the beta-beam will
measure the cross section for the SPL appearance search, and vice
versa.
If both experiments run with neutrinos and anti-neutrinos the following 
transition probabilities can be measured: $P_{\nu_e\to\nu_\mu}$, 
$P_{\bar\nu_e\to\bar\nu_\mu}$, $P_{\nu_\mu\to\nu_e}$, and
$P_{\bar\nu_\mu\to\bar\nu_e}$. 
Tests of the T and CPT symmetries would thus be possible, in addition
to CP-violation, since matter effects are very small because of the
relatively short baseline. 

\begin{figure}
  \begin{center}
    \includegraphics[width=0.9\textwidth]{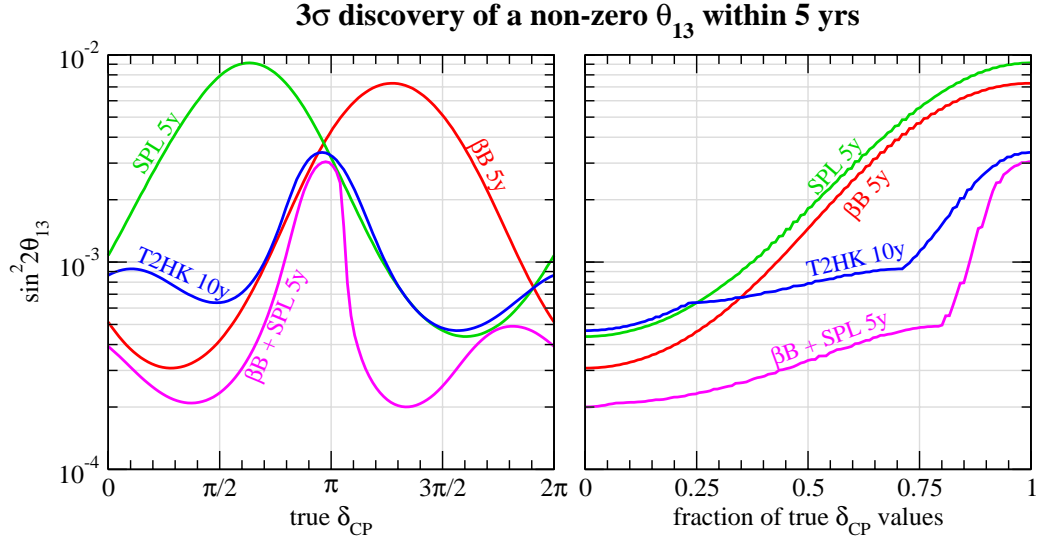}
  \end{center}
  \caption{
    Discovery potential of a finite value of $\stheta$ at $3\sigma$
    $(\Delta\chi^2>9)$ for 5~yrs neutrino data from \BB, SPL, and the
    combination of \BB\ + SPL compared to 10~yrs data from T2HK (2~yrs
    neutrinos + 8~yrs anti-neutrinos).  
    Taken with kind permission of the Journal of High Energy Physics 
    from figure 14 in reference \cite{Campagne:2006yx}.
    Copyrighted by SISSA.
  }
  \label{fig:th13-5yrs}
\end{figure}

\begin{figure}
  \begin{center}
    \includegraphics[width=0.6\textwidth]{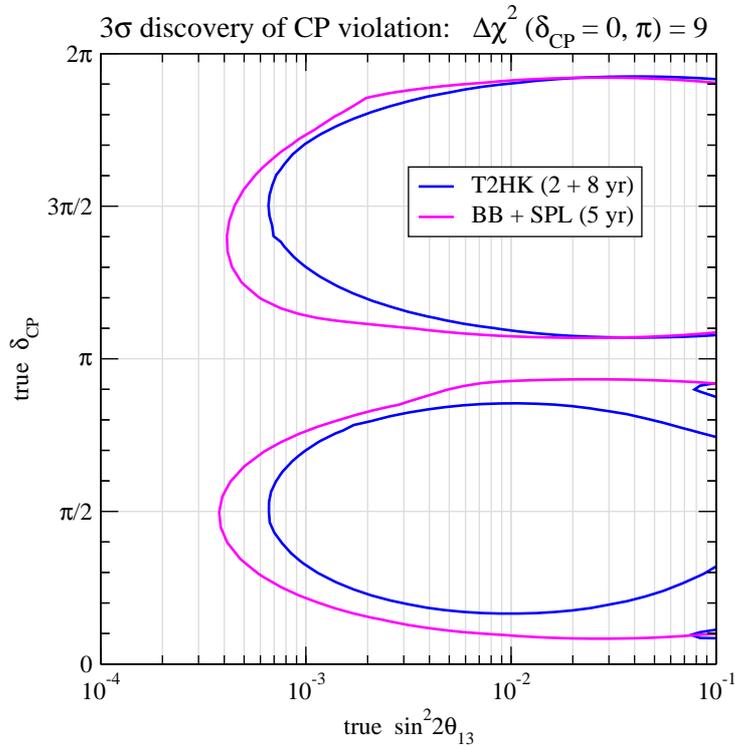}
  \end{center}
  \caption{
    Sensitivity to CPV at $3\sigma$ $(\Delta\chi^2>9)$ for combining
    5~yrs neutrino data from \BB\ and SPL compared to 10~yrs data from
    T2HK (2~yrs neutrinos + 8~yrs anti-neutrinos).  
    Taken with kind permission of the Journal of High Energy Physics 
    from figure 15 in reference \cite{Campagne:2006yx}.
    Copyrighted by SISSA.
  }
  \label{fig:CP-5yrs}
\end{figure}

However, if CPT symmetry is assumed, the beta-beam channels are
redundant: the only gain in combining the two facilities is an
increase in the size of the data set which does not help to solve the
degeneracies \cite{Donini:2004hu}. 
Nevertheless, this also means that in principle all information
can be obtained from neutrino data alone because of the relations
$P_{\bar\nu_e\to\bar\nu_\mu} = P_{\nu_\mu\to\nu_e}$ and
$P_{\bar\nu_\mu\to\bar\nu_e} = P_{\nu_e\to\nu_\mu}$. 
This implies that (time consuming) anti-neutrino running can be
avoided. 
This is illustrated in figures \ref{fig:th13-5yrs} and
\ref{fig:CP-5yrs}.
In figure \ref{fig:th13-5yrs} the $\theta_{13}$ discovery potential is
shown for 5~years of neutrino data from the $\gamma = 100$ beta-beam
and the SPL super-beam. 
Luminosities of $5.8 \cdot 10^{18}$ ($2.2 \cdot 10^{18}$) decays per
year for $^6$He ($^{18}$Ne) have been assumed. 
From the left panel it can be seen how each experiment plays the role
an anti-neutrino run would have played the single-facility case. 
Combining these two data sets results in a slightly better sensitivity
than 10 years (2$\nu$+8$\bar\nu$) of T2HK data.
In addition, figure \ref{fig:CP-5yrs} shows that the combination is
also effective in searching for CPV, 5 years of neutrino data from the
beta-beam and the SPL leads to a better sensitivity than 10 years of
T2HK alone. 

\subsubsection{Super-Beam associated with the Neutrino Factory}

As described in the section \ref{SubSect:Perf:NF}, the Neutrino
Factory suffers acutely from the degeneracy problem because its energy
and baseline are such that it operates far from the oscillation
maximum.
With its high energy and long baseline, the Neutrino Factory is ideal
to tackle the problem of the sign degeneracies through matter
effects. 
However, the fact that the oscillation peak occurs in the lowest
energy bin with relatively low efficiency, causes the intrinsic
degeneracy to spoil its sensitivity to CP-violation.  
Super-beams, with a completely different $L/E$ and operating at the
first oscillation maximum, do not suffer as badly from this
degeneracy.
On the other hand, the short baselines and lower energies favoured by
super-beams strongly limit their ability to solve the sign degeneracy
by exploiting matter effects. 
The combination of data from these two facilities can therefore be a
very effective tool to solve the degeneracy problem.
Furthermore, the intense pion beam that would produce the muons
required for the Neutrino Factory beam might also be exploited as a
super-beam source.
Indeed, the 2.2 GeV SPL beam was originally conceived and
optimised as the first stage of a Neutrino Factory project.
Thus, in a Neutrino Factory a super-beam comes `for free'. 
A Mton class water \v Cerenkov detector would still be needed to fully
exploit its potential, though.

Detailed studies of the ability to solve the eightfold degeneracy by 
combining the Neutrino Factory and the SPL super-beam can be found in 
references \cite{Burguet-Castell:2002qx,Donini:2003kr,Mena:2005ek}. 
An impressive synergy between the two facilities is found, lifting
all the degenerate solutions for large fractions of the parameter
space. 
However, a more detailed study fully including the systematics in the
considered detectors is still required.

%% file: 05-Performance/05-03-Beta-Beams/05-03-Beta-Beams.tex
\subsection{The physics potential of beta-beam facilities}
\label{SubSect:Perf:BetaBeam}

A beta-beam \cite{Zucchelli:2002sa} is produced from boosted,
radioactive-ion decays and therefore is a pure $\nu_e$ or
$\bar{\nu}_e$ beam. 
The flavour transitions that can, in principle, be studied in this
facility are:
\begin{eqnarray}
  &\nu_e \rightarrow  \nu_\mu \nonumber ~~~~~\nu_e \rightarrow \nu_e ~~~~~
  \nu_e \rightarrow \nu_\tau \nonumber \\
  &\bar{\nu}_e \rightarrow  \bar{\nu}_\mu \nonumber ~~~~~\bar{\nu}_e \rightarrow \bar{\nu}_e ~~~~~
  \bar{\nu}_e \rightarrow \bar{\nu}_\tau. \nonumber 
\end{eqnarray}
There are three variables that determine the properties of the
facility: the type of ion used, and in particular the
the end-point kinetic energy of the electron in the $\beta$-decay,
$E_0$; the relativistic $\gamma$ (energy divided by mass) of the ion;
and the baseline, $L$. 
Once these parameters are fixed, the neutrino flux can be calculated
precisely since the kinematics of $\beta$ decay is very well known. 
In the laboratory frame, the neutrino flux, $\Phi^{\rm lab}$, is given
by \cite{Burguet-Castell:2003vv}:  
\begin{eqnarray}
  \left.{d\Phi^{\rm lab}\over dS dy}\right|_{\theta\simeq 0} 
  \simeq {N_\beta \over \pi L^2} {\gamma^2 \over g(y_e)} y^2 (1-y)
  \sqrt{(1-y)^2 - y_e^2} \; ; 
\end{eqnarray}
where $N_\beta$ is the number of ion decays per unit time, $m_e$ is
the mass of the electron, $dS$ is the element of solid angle,
$0 \, \leq \, y={E_\nu \over 2 \gamma E_0} \, \leq \, 1-y_e$, and  
$y_e=m_e/E_0$; and:
\begin{eqnarray}
  g(y_e)\equiv {1\over 60} \left\{ \sqrt{1-y_e^2} (2-9 y_e^2 - 8 y_e^4) + 15 y_e^4 \log\left[{y_e \over 1-\sqrt{1-y_e^2}}\right]\right\}.
\end{eqnarray}
Note that the shape of the flux, and in particular the average
(anti-)neutrino energy, is essentially constant for a particular 
$\gamma E_0$ and that, if the number of decaying ions and the baseline
are kept fixed, the flux increases with $\gamma$.

\subsubsection{Beta-beam setups}
\label{sec:setups}

The choice of isotope is a compromise between production yield, 
$E_0$, and lifetime. 
Isotopes should be sufficiently long-lived to avoid strong
losses in the acceleration phase, but must decay fast enough to
generate a neutrino beam of sufficient flux.
Lifetimes of the order of $1$~s are considered reasonable.

The following isotopes have been identified as good candidates:
$^6{\rm He}$ with $E_0= 3506.7$~keV to produce $\bar{\nu}_e$ and
$^{18}{\rm Ne}$ with $E_0=3423.7$~keV to produce $\nu_e$
\cite{Zucchelli:2002sa}. 
More recently two ions with larger $E_0$ have been also considered:
$^8{\rm Li}$ ($E_0= 12.96~$MeV) and $^8{\rm B}$ ($E_0=13.92~$MeV)
\cite{Rubbia:2006pi,Donini:2006dx}. 
At the same $\gamma/L$, the neutrino beams produced by the ions 
${\rm Li}/{\rm B}$ are typically at three to four times more energetic
than those of ${\rm He}/{\rm Ne}$. 

Optimisation of the $\gamma$ factor and the baseline should take
into account the following physics requirements: 
\begin{itemize}
  \item 
    $L / \langle E_\nu\rangle$ should be near the first atmospheric
    maximum so that oscillation signals are as large as possible.
    For a particular ion, this means that $L$ and $\gamma$ have a
    constant ratio and therefore that the neutrino flux is constant; 
  \item 
    The neutrino energy should be above $\mu$-production threshold;
  \item 
    The neutrino energy should be large enough for a measurement of the
    spectral distortion to be used to resolve the intrinsic
    degeneracy;
  \item 
    The baseline should be as long as possible to all the mass
    hierarchy to be determined through the observation of matter
    effects; and
  \item 
    Event rate: increasing $\gamma$ at fixed ion flux increases the
    neutrino energy and therefore the number of events since the
    neutrino cross sections increase with energy.
\end{itemize}
All these requirements point in the same direction: increasing the
$\gamma$ factor as much as possible and tuning the baseline to sit near
the atmospheric-oscillation peak.
Practical issues will lead to constraints on the maximum $\gamma$
that can be achieved.
If an existing accelerator infrastructure was developed to host a
beta-beam, the $\gamma$s which could be achieved for He and Ne are: 
\begin{itemize}
  \item{\it CERN-SPS:}  
    $\gamma_{{\rm He}} = 150, \gamma_{{\rm Ne}} = 250$;
  \item{\it Refurbished SPS:}
    $\gamma_{{\rm He}} = 350, \gamma_{{\rm Ne}} = 580$;
  \item{\it Tevatron:} 
    $\gamma_{\mathrm He} = 350, \gamma_{\mathrm Ne} = 580$; and
  \item{\it LHC:} 
    $\gamma_{\mathrm He} \sim 2500, \gamma_{\mathrm Ne} \sim 4000$.
\end{itemize}
The $\gamma$s that could be achieved for ${\rm Li}/{\rm B}$ are
$\gamma_{{\rm Li}/{\rm B}} = 8/9 \gamma_{{\rm He}/{\rm Ne}}$.   

For  $\gamma_{{\rm He}}=150$, bending magnets of 5 T and a useful
decay length of 36\%, the decay ring length is $\sim 6 \, 880$~m. 
If $\gamma$ is increased and the bending magnets are the same, the
decay ring should be scaled proportionally to maintain the same
fraction of useful ion decays.
The ions in the decay ring should be kept in small bunches in order to
keep the machine duty-cycle small; this is required to keep the
background from atmospheric neutrinos at a negligible level (see the
discussion in section \ref{sec:atmos}).

An appropriate long baseline site is also required. 
To reduce the background from cosmic muons, an underground location is
preferable.
Therefore, an additional constraint for the choice of baseline would
be the availability of an appropriate site, preferably with an
existing and underground laboratory. 
It should be noted that a detector for a beta-beam could be versatile
enough to allow other data samples to be studied (for example:
atmospheric neutrinos; supernova neutrinos; etc.). 
Fortunately a number of alternatives exist that roughly match the
$\gamma$s noted above.

\subsubsection{The low-energy beta-beam: LE$\bbph$}

A low-energy beta-beam, with average neutrino energies in the sub-GeV
range, matches the distance from CERN to the Modane laboratory in the
Frejus tunnel, $L=130$~km.
The nice feature of this option is that the appropriate $\gamma$
could be achieved with the present CERN SPS. 
In the first proposal along these lines
\cite{Mezzetto:2003ub,Bouchez:2003fy}, a $\gamma_{{\rm He}} = 60,
\gamma_{{\rm Ne}} =100$ was chosen so that the baseline would sit near
the first atmospheric peak. 
It was then realised, in reference \cite{Burguet-Castell:2003vv}, that
this was not optimal. 
The new standard choice is $\gamma_{He} = \gamma_{Ne} =100$
\cite{Mezzetto:2005ae}.
In reference \cite{Burguet-Castell:2005pa}, a scan in $\gamma$ was
performed for this baseline, assuming a fixed ion flux, and the
optimal $\gamma$ was found to be $\gamma \geq 90-100$ and with
little improvement for larger $\gamma$. 
The average neutrino energy is $\sim 0.4$~GeV, a little above the
atmospheric peak at the CERN-Frejus baseline. 

\subsubsection{High-energy beta-beams: HE$\bbph$}

Neutrino beams with average energies in the $1-1.5$~GeV range could reach the atmospheric peak at $L\sim 700$~km,  matching the 
distance between CERN-Canfranc,  CERN-Gran-Sasso or Fermilab-Soudan. 
Such a beam could be achieved in two ways:
\begin{itemize}
\item[(a)] by using more powerful accelerators, such as a refurbished SPS or the Tevatron to increase $\gamma_{{\rm He}} = \gamma_{{\rm Ne}} =350$\cite{Burguet-Castell:2003vv}
\item[(b)] by using higher $E_0$ ions such as ${\rm Li}/{\rm B}$ at moderate 
$\gamma\sim 100$ that could be achieved also with the Fermilab Main Injector,  but increasing significantly the number of decaying ions to compensate for the loss of flux \cite{Rubbia:2006pi, Rubbia:2006zv} 
\end{itemize}

Even higher energy beta-beams have also been considered
\cite{Burguet-Castell:2003vv,Huber:2005jk}. 
If it were possible to accelerate the ions in LHC without significant
additional losses, it would be possible to produce a beam with
$\gamma= {\cal O}(1000)$.
In this case, with a baseline of a few thousand kilometers, better
sensitivity to matter effects and the sign of $\Delta m^2_{31}$ 
would be achieved \cite{Burguet-Castell:2003vv,Huber:2005jk}. 
The performance of such a setup will be presented below. 
However, such an increase in $\gamma$ looks rather far-fetched at
present and it is more likely that a greenfield scenario for the
beta-beam would end up providing a higher intensity of ions
\cite{Rubbia:2006zv} rather than larger boosts.

\subsubsection{Ion production and $\nu$ fluxes}

The only detailed studies on ion-production and acceleration performed
up to now have concentrated on using the ISOLDE technique for ion
production and the CERN PS and SPS for acceleration
\cite{Autin:2002ms}.
The EURISOL beta-beam group baseline assumes $\gamma=100$ and a flux
corresponding to $2.9\times 10^{18}$ ${\rm He}$ and $1.1\times 10^{18}$
${\rm Ne}$ decays per year \cite{EurisolDS}.
The goal is to achieve this performance without assuming modifications
to the present CERN accelerators.
No study has yet been performed for the ${\rm Li}/{\rm B}$ option, so
the ion flux assumed in this case should be considered as a goal.
Since the ion production system would be common, it is reasonable to
assume that a refurbished SPS could be used to accelerate ions to
higher $\gamma$ without further loses. 
More ions must be stored in the decay ring at higher $\gamma$ since
the ion lifetime is dilated, this may limit the neutrino flux.
On the other hand, at higher $\gamma$ the duty cycles that have been
used in the baseline scenario to reduce the atmospheric background can
be relaxed.
Therefore, the fluxes noted above will be used for both the low- and
the high-$\gamma$ setups.

\begin{figure}
  \begin{center}
    \epsfig{file=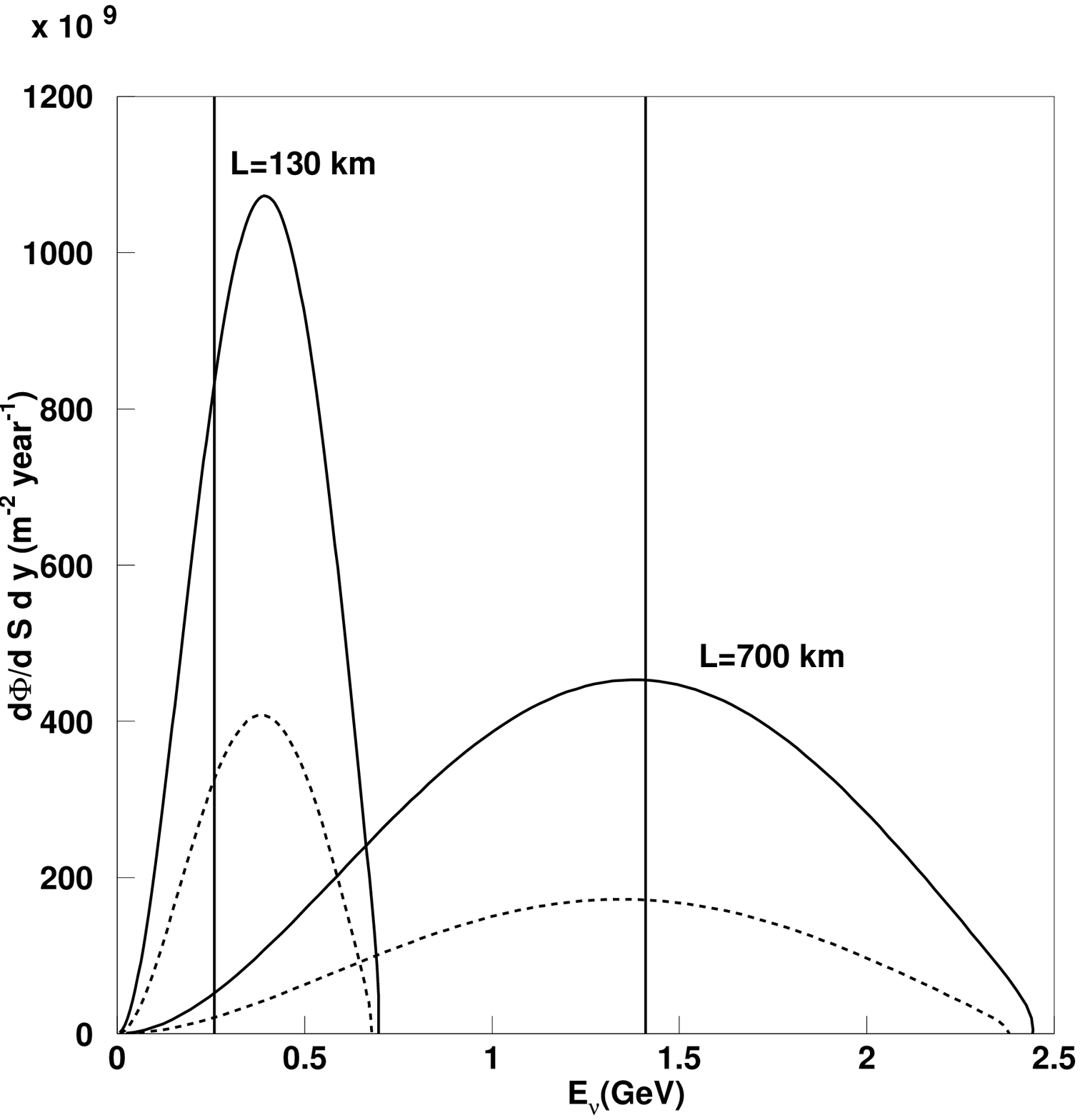,width=7cm}
    \epsfig{file=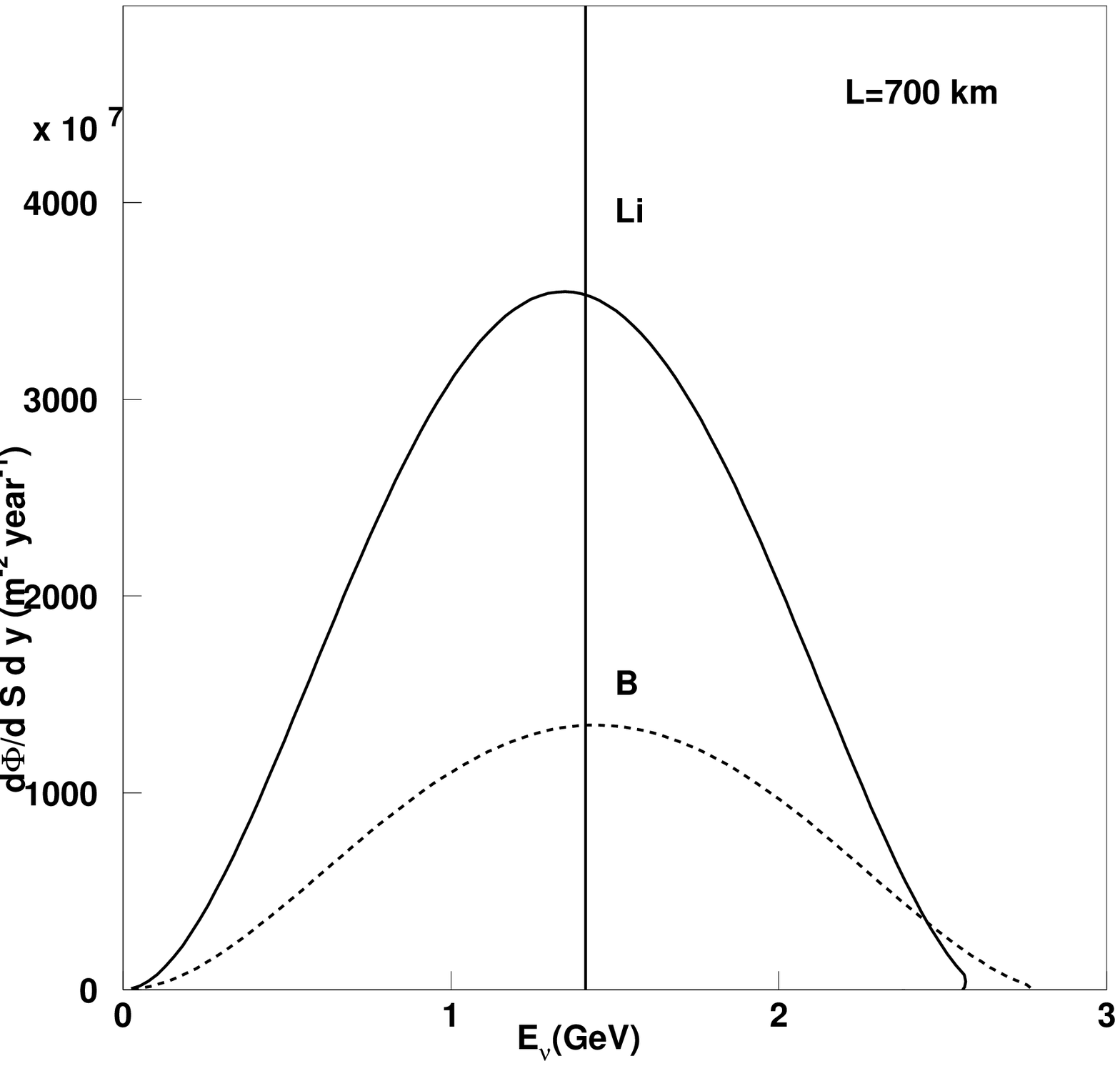,width=7cm}
  \end{center}  
  \caption{
    $\bar{\nu}_e$ (solid) and $\nu_e$ (dashed) fluxes as a function of
    the neutrino energy for He and Ne at $\gamma=100, 350$ (left) and
    for Li and B at $\gamma=100$ (right), assuming the number of
    decaying ions to be the standard one in all cases. 
    The vertical lines correspond to the energy position of the
    atmospheric peak for $\Delta m^2_{13}$ = 0.0025~eV$^2$. 
  } 
  \label{fig:bbfluxes}
\end{figure}

The neutrino fluxes at the detector location for the LE$\beta\beta$
and the HE$\beta\beta$ and the standard ion fluxes are shown in
figure \ref{fig:bbfluxes}. 
As explained above, the shape of the flux depends only on the
combination $2 \gamma E_0$, which defines the end-point of the
spectrum and therefore it is rather similar for the two HE$\beta\beta$
options. 
On the other hand, the absolute flux depends on the combination
$(\gamma/L)^2$ and is therefore smaller for lower $\gamma$ as can be
seen by comparing the left and the right plots of
figure \ref{fig:bbfluxes}; although they are very similar in shape,
they differ by a factor 10 in absolute value. 
The properties of the various beta-beam setups are summarised in table
\ref{tab:bbparam}.
\begin{table}
  \begin{center}
    \begin{tabular}{cccccc}
      \hline
      Ion & $\gamma$ & $L(km)$ & ${\bar\nu}_e$ CC & $\nu_e$ CC & $\langle E_\nu \rangle (GeV)$\\
      \hline
      He/Ne & 100 & 130 & 28.9 & 32.8 & 0.39/0.37 \\
      He/Ne & 350 & 700 & 62.0 & 55. & 1.35/1.3 \\
      Li/B  & 100 & 700 &  5.0 & 4.9 & 1.3/1.4 \\
      \hline
    \end{tabular}
  \end{center}
  \caption{
    Number of charged-current events per \hbox{kton-year} and average
    neutrino energy, in the absence of oscillation, for the different
    options and a number of decaying ions of 
    $N_{He/{\rm  Li}}=2.9\times 10^{18}~year^{-1}$ and
    $N_{Ne/B}=1.1\times 10^{18}~{year}^{-1}$.
  }
  \label{tab:bbparam}
\end{table}

Given the fact that proposals for new techniques by which the ion
yield may be increased \cite{Rubbia:2006zv} have not yet been fully
exploited, and on the assumption that a number of improvements to the
present PS and SPS at CERN are likely to occur in the LHC-upgrade
programme, it does not seem unreasonable to consider a greenfield
scenario in which the number of ions is increased up to a factor 10
with respect to the baseline defined above.
We will consider the reach of such an aggressive facility in section
\ref{Sect:HighFlux}.

\subsubsection{ Detector technology}

The golden signals at a beta-beam facility are: a muon from the
appearance channel; and an electron from the disappearance channel.
The silver channel ($\tau$ production) is not open for most of the
setups considered and has not been studied in any detail. 

Since the beam, at source, is a pure flavour eigenstate, the principal
uncertainties in the measured oscillation probabilities arise from
uncertainties in the background rates and the precision with which the
efficiencies can be determined. 
The main requirements for an optimal detector are, therefore, good
particle identification (i.e. $\mu/e/\pi$ separation) and good
neutrino-energy resolution.
Several types of massive detector can be optimised to identify muons
and electrons in the GeV range. 

The fact that the beta-beam produces a pure $\nu_e$ (or $\bar{\nu_e}$)
beam means that the golden (muon appearance) channel is free from
the beam-generated 'wrong-sign muon' background that is present at the
Neutrino Factory.
This means that it is not necessary to magnetise the beta-beam
detector; a significant advantage that the beta-beam has over the
Neutrino Factory.
Since magnetisation is not required, a very massive, water \v Cerenkov
detector is an appropriate technology choice for the beta-beam. 
Such a detector has a broad physics potential beyond oscillation
physics: proton decay; detection of neutrinos from supernovas; etc. 
It is hard to imagine that one can achieve megaton detector masses
with a different type of technology.

Detectors that have been considered for the beta-beam to date include:
\begin{itemize}
  \item 
    A 500~kton fiducial water \v Cerenkov
    \cite{Mezzetto:2003ub,Bouchez:2003fy,Burguet-Castell:2003vv};
  \item 
    A 50~kton NOvA-like detector \cite{Huber:2005jk}; and
  \item 
    A 40~kton Iron calorimeter \cite{Donini:2006tt}.
\end{itemize}
We will give some details of the performance of the first two
options. 
Very recently a liquid argon TPC has also been discussed in the
context of the beta-beam in reference \cite{Rubbia:2006zv}. 
We refer to this work for details.  

\subsubsection{Water \v Cerenkov}

\begin{figure}
  \begin{center}
    \epsfig{file=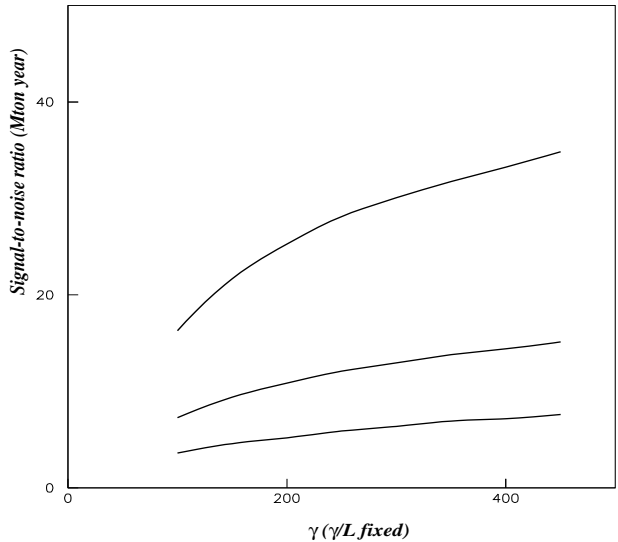,width=9cm,height=8cm} 
  \end{center}
  \caption{
    Signal to noise (i.e. square root of the signal plus background
    events add in quadrature for the different energy bins) for the   
    appearance signal from ${\rm Ne}$ in units of one Mton-year as a
    function of $\gamma$, holding $\gamma/L\simeq 0.5$ fixed. 
    The three curves correspond to $\theta_{13}=8^\circ, 3^\circ$ and
    $1^\circ$.
  }
  \label{fig:wclimit}
\end{figure}

Water \v Cerenkovs are optimised to search for the quasi-elastic (QE),
charged current (CC) events; it is not possible to measurement
the hadronic energy and therefore it is possible to reconstruct the
neutrino energy only for QE events. 
Figure \ref{fig:wclimit} shows the signal-to-noise ratio for a
megaton-year exposure as a function of $\gamma$ (for fixed $\gamma/L
\sim 0.5$) for a neutrino beam from ${\rm Ne}$ decays. 
The signal-to-noise ratio is defined in each energy bin (seven bins
are considered in all setups between 200MeV and the end-point), 
the results for all bins are then added in quadrature. 
The expected improvement with $\gamma$ slows down above 
$\gamma \sim 400$, because events at higher energies are likely to
give more than one ring and therefore are not likely to be selected,
while the background-selection efficiency continues to increase.
The figure shows that there is little benefit from increasing $\gamma$
above 300--400 using a water \v Cerenkov \cite{Burguet-Castell:2005pa}. 
On the other hand, for lower $\gamma$ this technology is probably close
to optimal given the large mass that one could envisage for this type
of detector.   

Detailed Monte Carlo studies of a Super-Kamiokande-like detector have
been performed to quantify the efficiencies and backgrounds for the
$\mu$-appearance signal. 
The signal selection cuts are essentially three: 
\begin{itemize}
  \item 
    Single ring, contained events; and
  \item 
    $\mu$-like ring;
  \item 
    Delayed ring: Michel electron from $\mu$-decay.
\end{itemize}
The energy resolution for QE events is quite good, mainly limited by
Fermi motion. 
However, the contamination from non-QE events, which increases with
energy, introduces a shift between the true and reconstructed
neutrino energies. 
In order to take into account this fact properly, migration matrices
for efficiencies and backgrounds that allow for the `migration' from
true to reconstructed neutrino energy are used as first described in
reference \cite{Burguet-Castell:2003vv}.
In the analysis presented below, reconstructed energy bins of
$100~$MeV for the LE$\beta\beta$ and $200~$MeV for the HE$\beta\beta$
will be considered. 

The main source of background are neutral current (NC) events with one 
positively-charged pion being produced through the $\Delta$ resonance.
Negatively-charged pions are very much suppressed by the delayed-ring
cut, because of the large absorption cross section for negative pions.
For the HE$\beta\beta$ setup, multi-pion events are also a significant
source of background.  
In third place, a few charged current (CC) events, in which the
electron ring goes undetected and a single pion is mis-identified as a
muon, also survive the selection cuts. 
A more detailed analysis would be needed to see whether the presence
of a low-energy electron in these events could be revealed by means of
a more sophisticated reconstruction algorithm.

\begin{figure}
  \begin{center}
    \epsfig{file=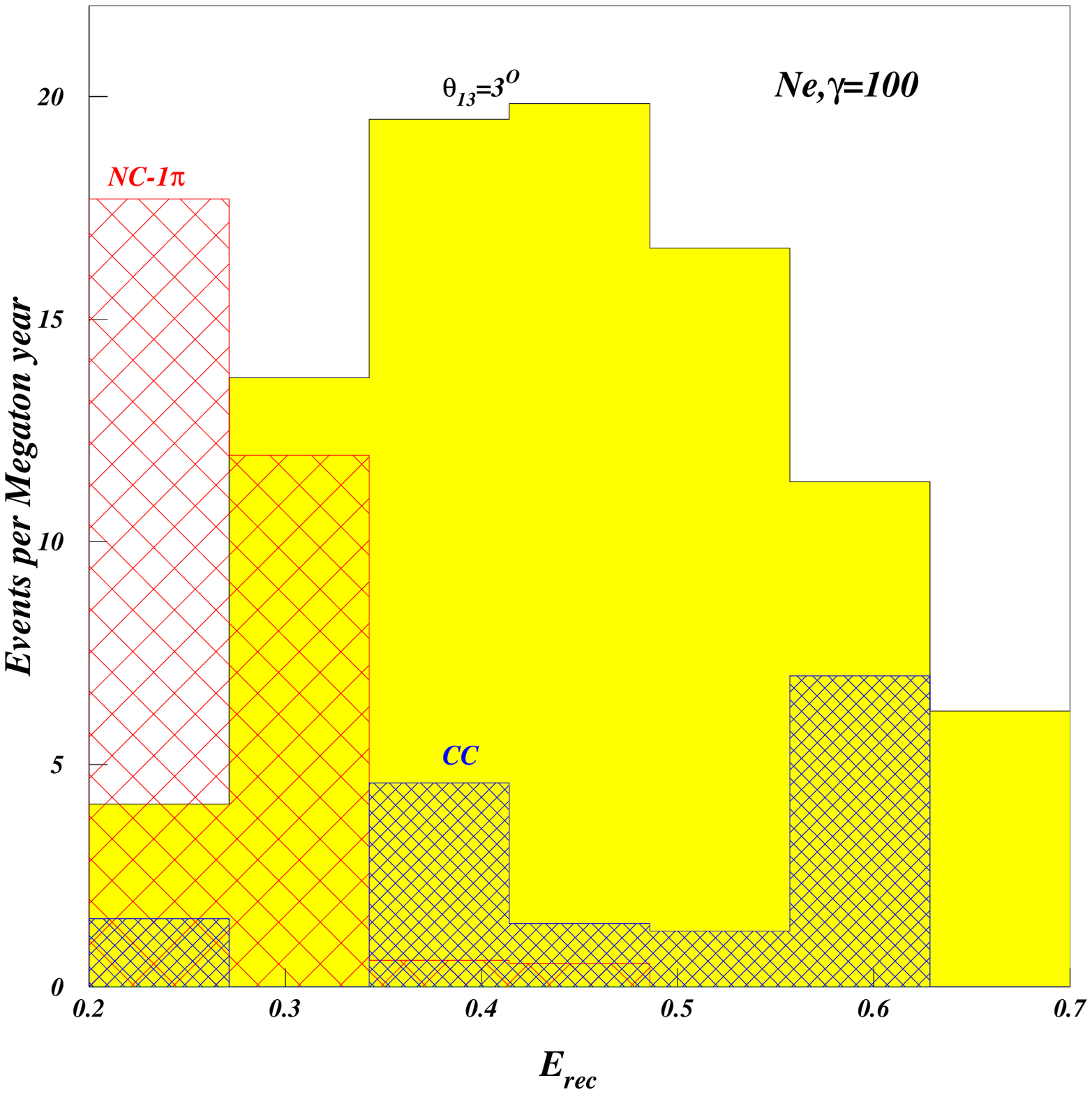,width=7.cm} 
    \epsfig{file=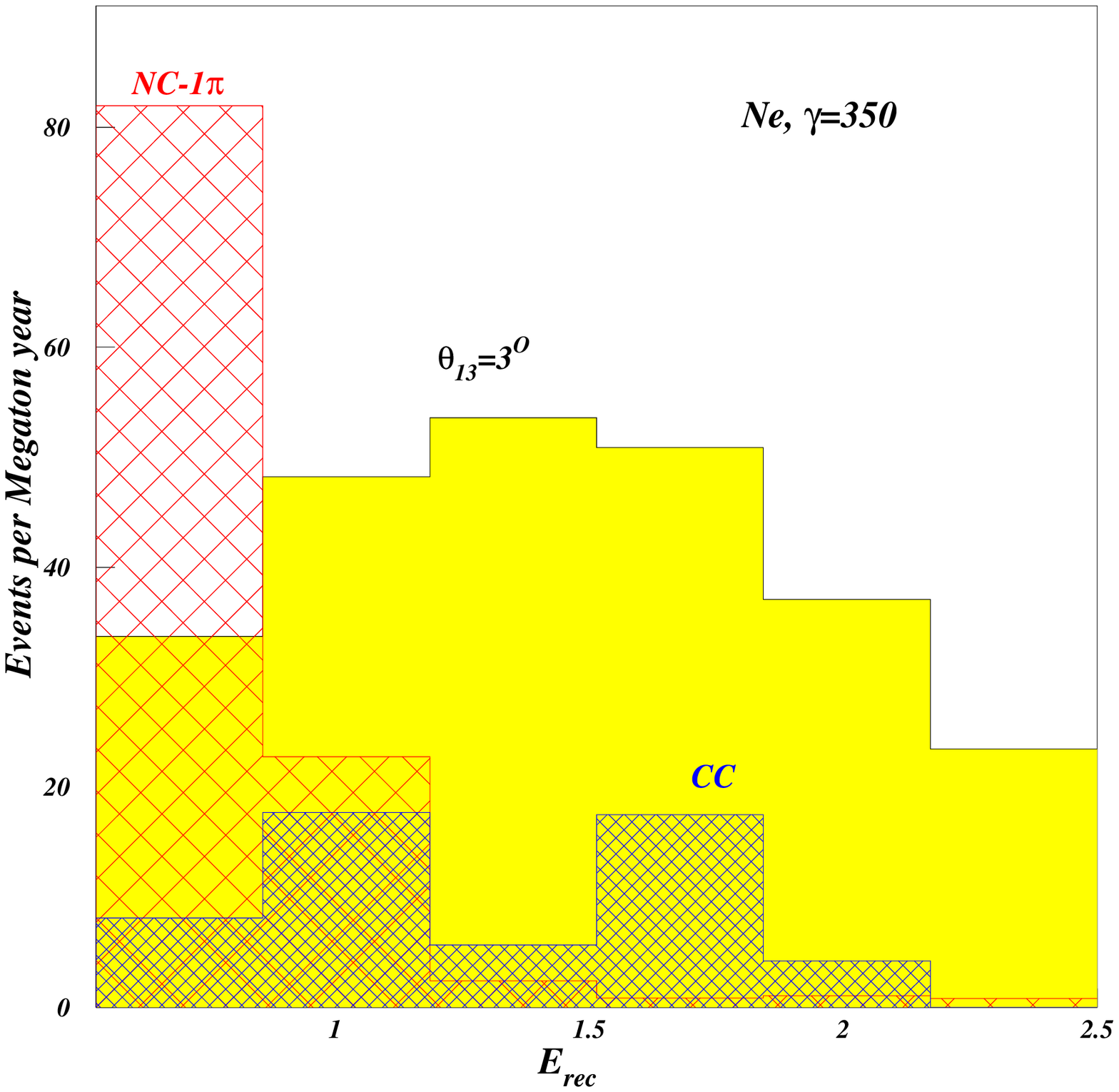,width=7.cm} 
  \end{center}
  \caption{
    Reconstructed energy in a water Cerenkov per Megaton year for
    signal  and NC and CC background (hatched) for the LE setup (left)
    and the HE one (right). The true values assumed are
    $\theta_{13}=3^\circ$ and $\delta=90^\circ$.
  } 
  \label{fig:signal_bg_bb}
\end{figure}

Figure \ref{fig:signal_bg_bb} shows a comparison of the expected
signal for $\theta_{13}=3^\circ$ and $\delta=\pi/2$ (near the
sensitivity reach of T2K-I) together with the different background
contributions for each setup. 
The level of NC background is rather large, especially for the HE
setup, but, owing to the very different kinematics of QE and NC
events, the reconstructed neutrino energy for the NC events is
strongly peaked at much lower values \cite{Burguet-Castell:2003vv},
making this background easily distinguished from the signal.

In the comparison plots which follow, a global normalisation
uncertainty of $2\%$ will be considered.
This normalisation uncertainty is taken to include the
fiducial-volume uncertainty.
In addition, a $1\%$ uncertainty in the ratio of neutrino to
anti-neutrino cross sections, an optimistic assumption if the present
knowledge of this ratio is taken into account.
A dedicated neutrino cross-section-measurement programme using a near
detector at the same facility would be required to reach such a
precision.
For the disappearance signal the uncertainty on the global
normalisation is the most important, so we neglect background
uncertainties and considered only the normalisation error.

\subsubsection{NO$\nu$A-like detector}

A totally-active, liquid-scintillator detector (TASD) \'a la NO$\nu$A
has been considered in \cite{Huber:2005jk} for $\gamma \geq 500$. 
The main advantage of this technology is that the neutrino energy can
be reconstructed for non-QE events, which become dominant at higher
energies, as well as for QE events.
On the other hand, it may be difficult to build a detector of this
type with a mass much larger than a few tens of kilotons. 
A fiducial mass of $50~$kton will be assumed.

The detector performance has been studied in the NO$\nu$A proposal.
Since the detector has been proposed for the conventional NUMI beam,
the study considered only efficiencies and backgrounds for $e$-like 
events.
These efficiencies and backgrounds are summarised in table
\ref{tab:nova}. 
While assuming the same efficiencies and backgrounds for the $\mu$
signal might be conservative, as argued in \cite{Huber:2005jk}, the
physics is quite different and a detailed study of this detector for
the beta-beam is essential for a reliable comparison to other
technologies to be made.
\begin{table}
  \begin{center}
    \begin{tabular}{lcccc} 
      \hline
      &\multicolumn{2}{c}{Appearance}&\multicolumn{2}{c}{Disappearance}\\
      &$\nu$&$\bar\nu$&$\nu$&$\bar\nu$\\
      \hline
      Signal efficiency&0.8&0.8&0.2&0.2\\
      Background rejection&0.001&0.001&0.001&0.001\\
      Signal error&2.5\%&2.5\%&2.5\%&2.5\%\\
      Background error&5\%&5\%&5\%&5\%\\ \hline
    \end{tabular}
  \end{center}
  \caption{
    The signal efficiencies and background rejection respectively and
    the systematical errors for the various signals and backgrounds
    used in \cite{Huber:2005jk}.
  } 
  \label{tab:nova}
\end{table}

The energy resolution is assumed to be a Gaussian with a width of
3$\%/\sqrt{E}$ for muons and 6$\%/\sqrt{E}$ for electrons and the
background is conservatively assumed to have the same energy spectrum
as the signal. 

\subsubsection{Atmospheric backgrounds}
\label{sec:atmos}

A very important source of background for all detector technologies
are atmospheric-neutrino events. 
A detector like Super-Kamiokande will record approximately 
120 $\nu_{\mu}+\bar{\nu}_{\mu}$ interactions per kiloton-year
(including the disappearance of $\nu_{\mu}$ into $\nu_{\tau}$).

There atmospheric background may be reduced in two ways. 
Firstly, the energy is often poorly reconstructed for these events
since they come from all directions while the signal comes from the
direction of the beam. 
The cut on the reconstructed energy to be within the range of energies
produced by the beta-beam significantly reduces the background without
affecting the signal efficiency. 
Secondly, selecting events for which the reconstructed 
neutrino direction is consistent with the beam also preferentially
selects beam-induced events. 
While the neutrino direction cannot be measured directly, it is
increasingly correlated with the observable lepton direction at high
energies.  
A directional cut is more effective as $\gamma$ increases, but is
never perfectly efficient. 
For a similar signal efficiency, background rejection for the
HE$\beta\beta$ was estimated in reference
\cite{Burguet-Castell:2005pa} to be a factor three better than for the
LE$\beta\beta$.

\begin{figure}
  \begin{center}
    \epsfig{file=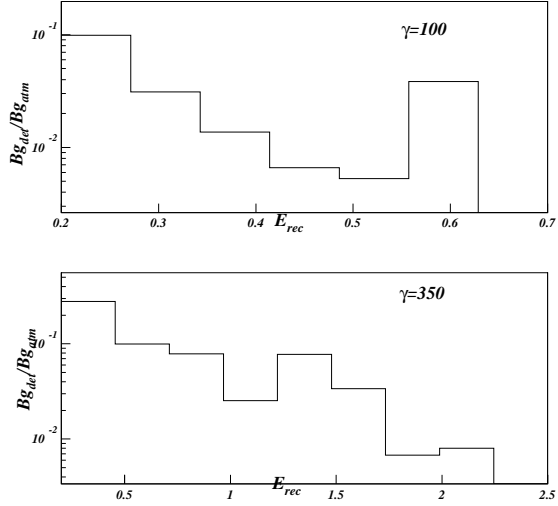,width=8cm} 
  \end{center}
  \caption{
    Ratio of $\nu$ background events coming from the detector
    misidentification to those coming from atmospheric neutrinos for
    the LE (up) and HE (down) setups. 
    The statistics corresponds to a megaton year ($10^7$ sec).
  }
  \label{fig:atm}  
\end{figure}

Without imposing any directional cut, we show the ratio of the
detector to atmospheric backgrounds in reconstructed-energy bins for
the LE and HE setups in figure \ref{fig:atm}.
Since the atmospheric background can be measured with very good
accuracy, the systematics associated to its subtraction are very small
and therefore it would be sufficient if this ratio could be made of
${\cal O}(1)$.  
Such a rejection factor can be achieved by timing the parent ion
bunches. 
It was estimated \cite{Bouchez:2003fy} that a rejection factor of $5
\times 10^{-5}$ per bunch is feasible with bunches 10~ns in length. 
As already indicated in reference \cite{Burguet-Castell:2005pa}, this
rejection power is an more than sufficient given the ratios shown in
figure \ref{fig:atm} which indicate that a global rejection of
$10^{-2}$ is probably sufficient for the LE option and could be even
relaxed further for the HE option. 
In the analysis presented below, the atmospheric background is assumed
to be negligible.  

\subsubsection{Analysis of performance and optimisation}

The following `standard' setups will be compared directly:
\begin{itemize}
  \item 
    LE$\beta\beta$ with a 500~Mton fiducial Water \v Cerenkov;
  \item 
    HE$\beta\beta$ with a 500~Mton fiducial Water \v Cerenkov: HE-a;  and
  \item 
    HE$\beta\beta$ with a 50~kton fiducial TASD: HE-b.
\end{itemize}
Further details of the experimental analyses can be found in
references
\cite{Burguet-Castell:2005pa,Mezzetto:2005ae,Huber:2005jk,Campagne:2006yx}.
The performance of these setups will be compared assuming the S$\nu$M
and using the following central values for the known oscillation parameters:
\begin{eqnarray}
  &\sin^2 \theta_{23} = 0.44  & \Delta m^2_{13} = +2.5 \times 10^{-3} eV^2  \nonumber\\ 
  & \sin^2 \theta_{12}= 0.3  & \Delta m^2_{12} = 0.8 \times 10^{-4} eV^2 \, .
  \label{eq:true}
\end{eqnarray}
All the plots labeled $ISS 2006$ assume these `true' values, however
this is not the case for all plots shown below.

\subsubsection{Sensitivity to $\theta_{13}$}

\begin{figure}
  \begin{center}
    \epsfig{file=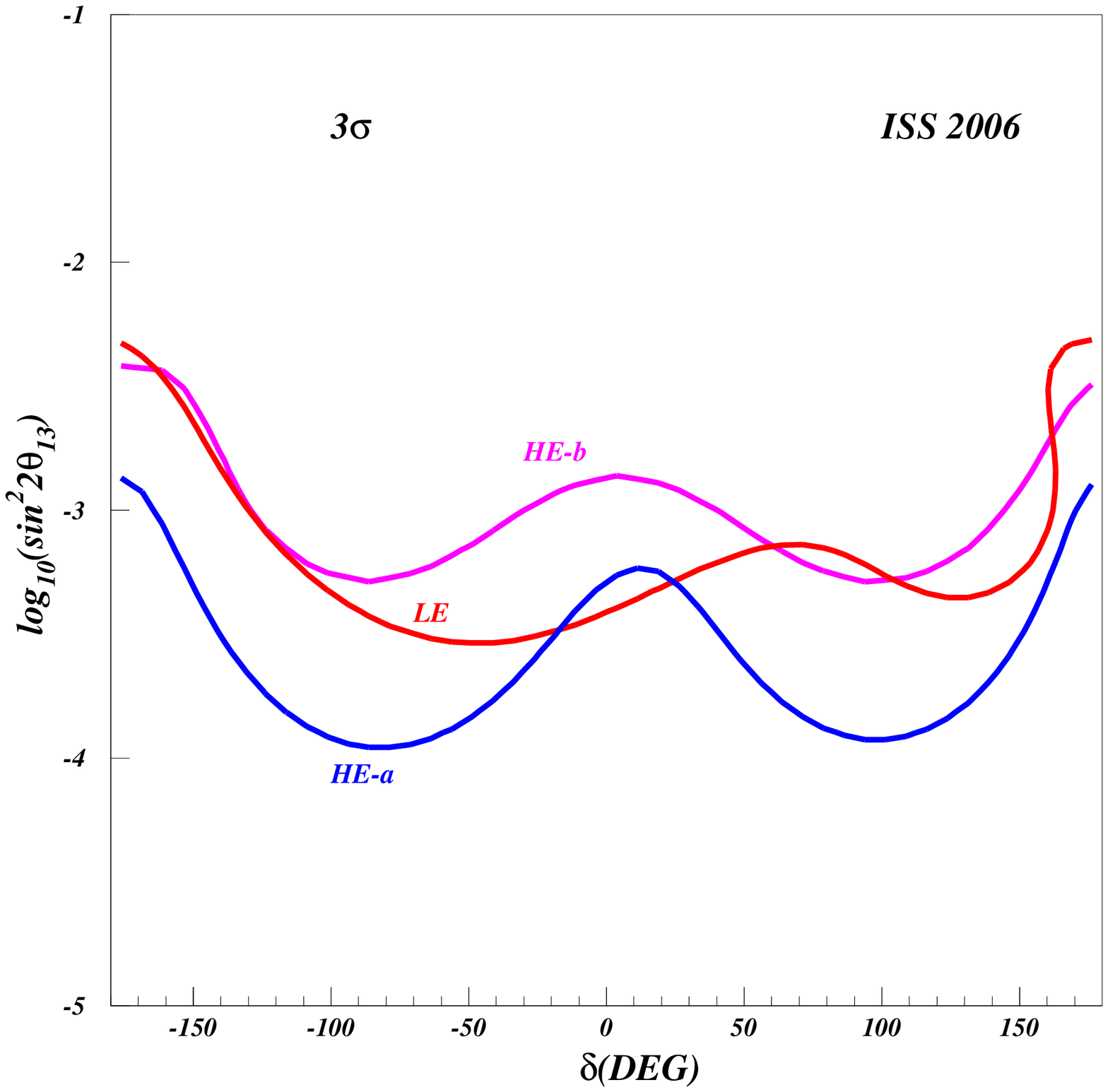,width=7cm}
    \epsfig{file=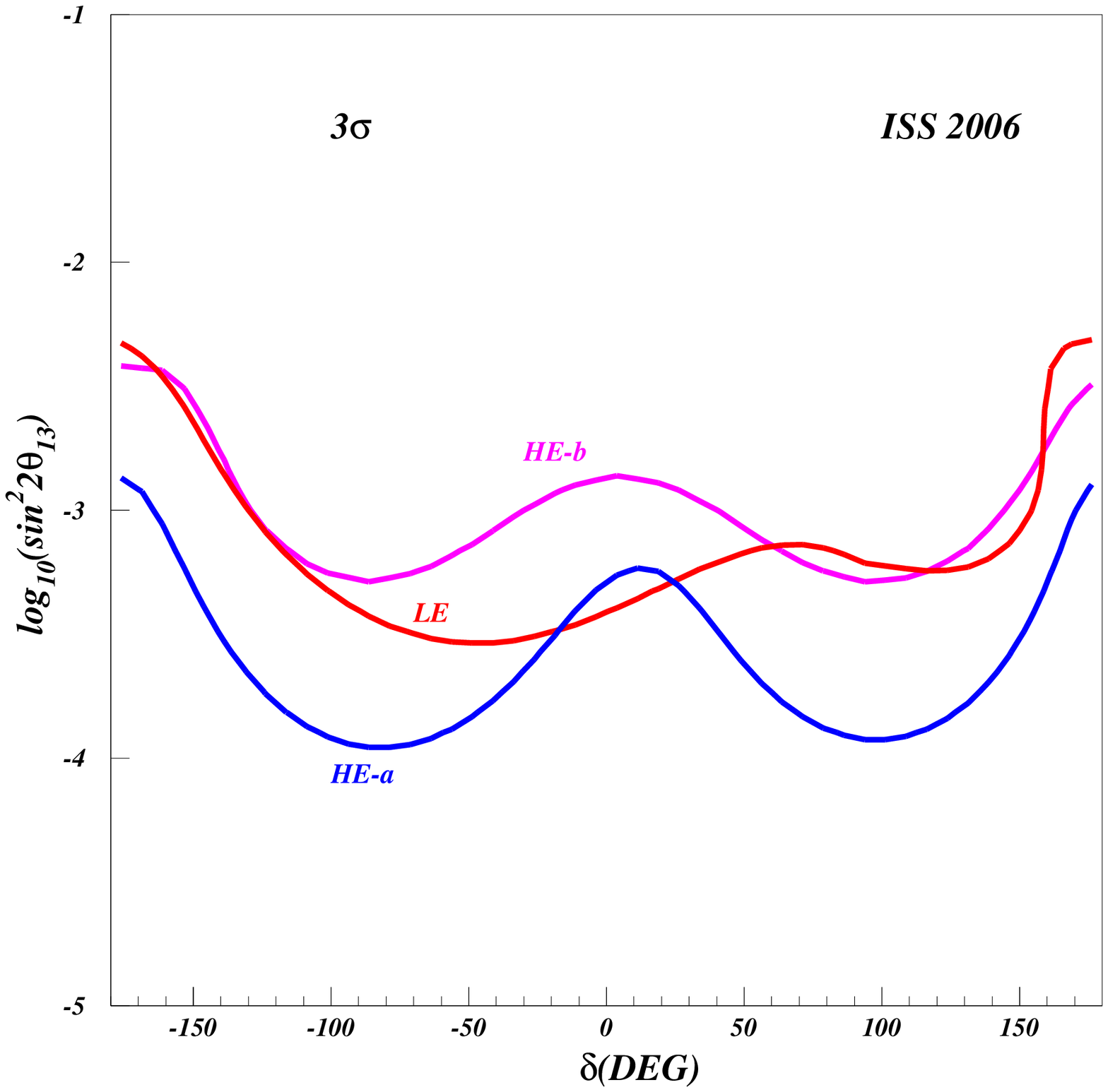,width=7cm} 
  \end{center}
  \caption{
    3$\sigma$ sensitivity to $\theta_{13}$ for the LE$\bbph$,
    HE$\bbph$-a, HE$\bbph$-b. The left plot does not include the
    discrete ambiguities and the right one does.
  } 
  \label{fig:th13}
\end{figure}

In figure \ref{fig:th13} we compare the sensitivity to
$\theta_{13}\neq 0$ for the LE$\bbph$ and three HE$\bbph$ options
using three types of detector. 
On the left plot only the intrinsic degeneracy is included, while
the right plot also takes into account discrete ambiguities.
Comparison of the left and right panels indicates that the effect of
the discrete ambiguities is quite small.
For the HE options, the bigger mass yields improved sensitivity as
expected, while the LE option with a 500~kton detector slightly
out-performs the HE-b option with a detect for which the fiducial mass
is a factor 10 smaller.

\subsubsection{Sensitivity to CP violation}

\begin{figure}
  \begin{center}
    \epsfig{file=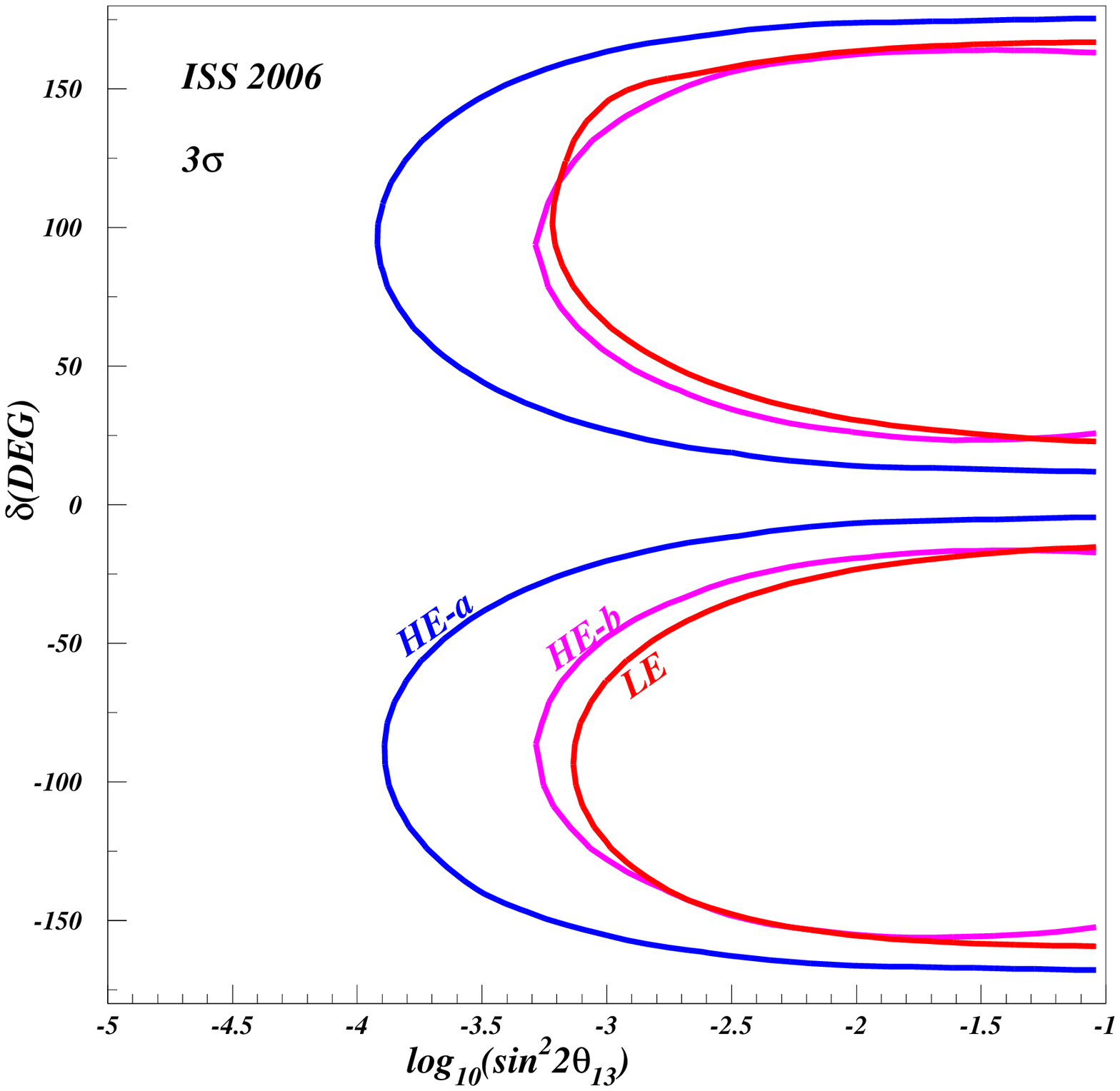,width=7cm}  
    \epsfig{file=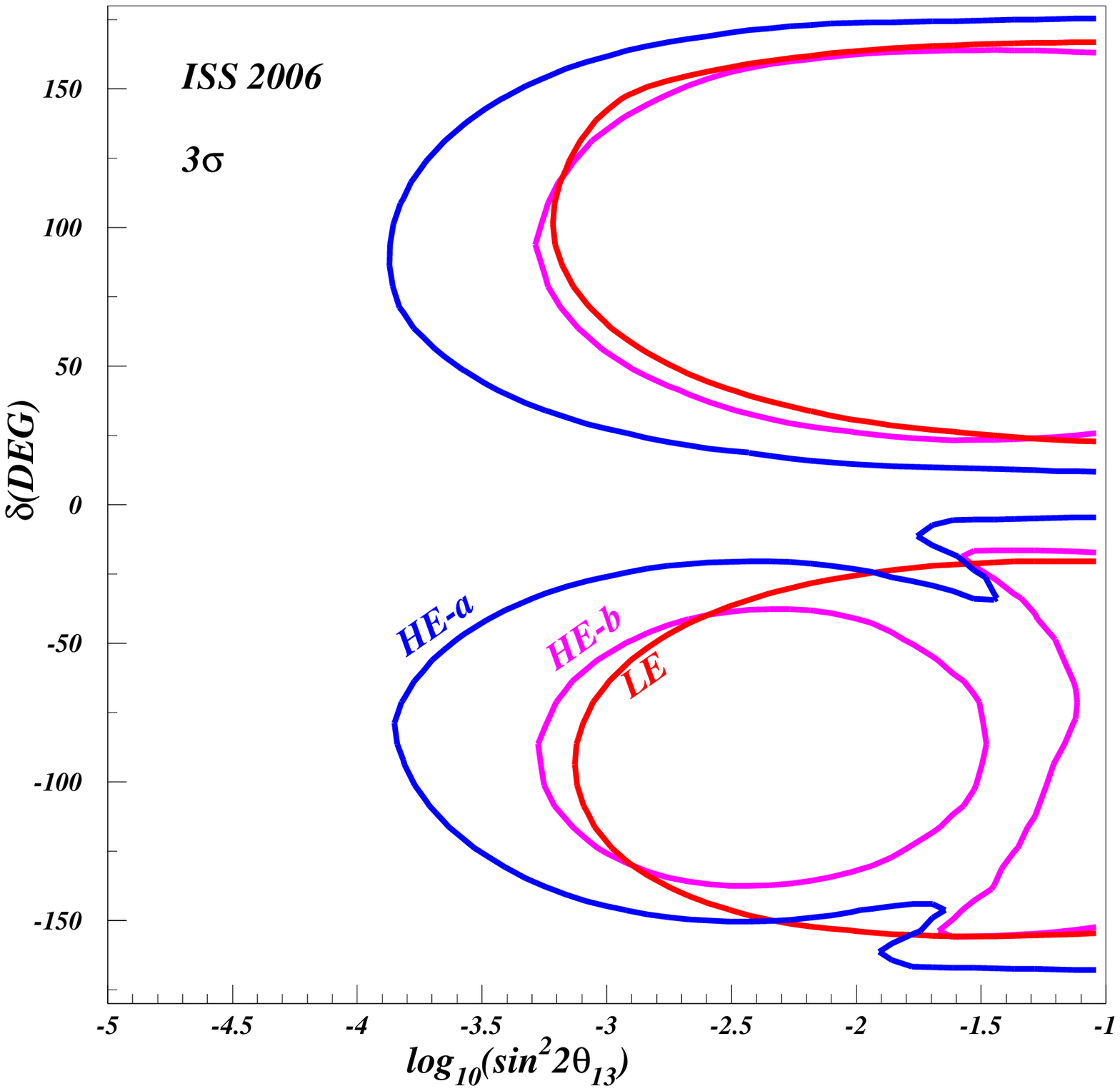,width=7cm} 
  \end{center}
  \caption{
    3$\sigma$ sensitivity to $\theta_{13}$ for the LE$\bbph$, 
    HE$\bbph$-a, HE$\bbph$-b  neglecting the discrete ambiguities
    (left) and including them (right). 
  } 
  \label{fig:cpv}
\end{figure}

The sensitivity to CP violation for the same setups is compared in
figure \ref{fig:cpv}.
Again, the discrete ambiguities are not present in the left panel, but
are taken into account in the the right panel. 
In the case of CP-violation sensitivity, the HE-a option out-performs
the others, while the LE option is similar or slightly worse than
HE-b.
The sign ambiguity is directly responsible for the loss of sensitivity
in a band at negative $\delta$ for the HE setups. 
This is a well-known effect that has also been observed in T2HK
analyses (see for example  reference \cite{Ishitsuka:2005qi}).
A combination with another experiment/measurement to resolve the
correlation between $\delta$ and the hierarchy is necessary. 
The different alternatives by which this can be done have not yet been
explored.

\subsubsection{Sensitivity to the discrete ambiguities}

\begin{figure}
  \begin{center}
    \epsfig{file=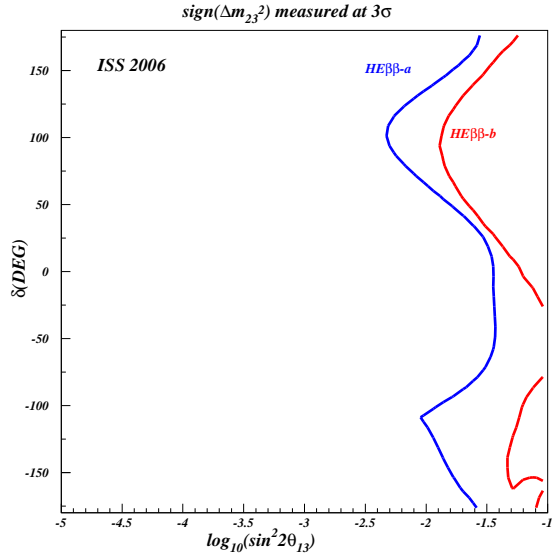,width=8cm}
  \end{center}
  \caption{
    3$\sigma$ sensitivity to the normal mass hierarchy assuming for
    the HE$\bbph$-a and HE$\bbph$-b setups.
  } 
  \label{fig:sign}
\end{figure}

The sensitivity to the sign of $\Delta m^2_{23}$ for the HE options is
compared in figure \ref{fig:sign} (the sensitivity of the LE option is
not shown since the short baseline makes fives it little or no
sensitivity).
The HE-a option again out-performs the HE-b option.
The dependence on $\delta$ is very strong.  
Only for values of $\sin^2 2 \theta_{13} > 0.03$ can the normal
hierarchy be established at 3$\sigma$ for any value of $\delta$. 
A significant improvement can be made by combining the beta-beam data
with atmospheric-neutrino data, this will be discussed in the next
section. 

The sensitivity to the octant of $\theta_{23}$ is extremely weak for
the choice we have made of $\theta_{23}$ for all the setups.
However, as we will see below this does not interfere with the
measurement of $\theta_{13}$ and $\delta$.

\subsubsection{Measurement of $\theta_{13}$ and $\delta$}

\begin{figure}
  \begin{center}
    \epsfig{file=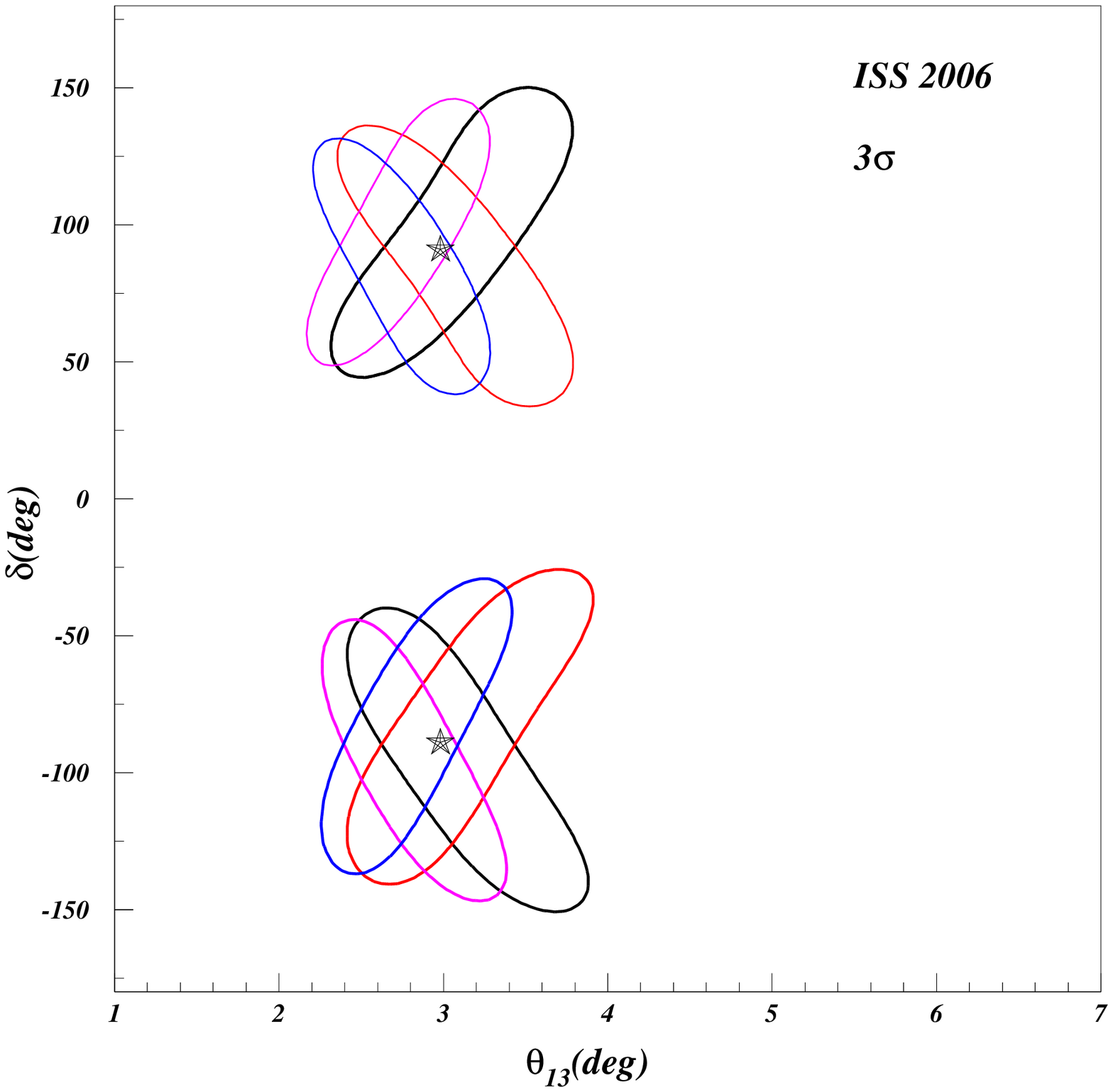,width=7cm}  
    \epsfig{file=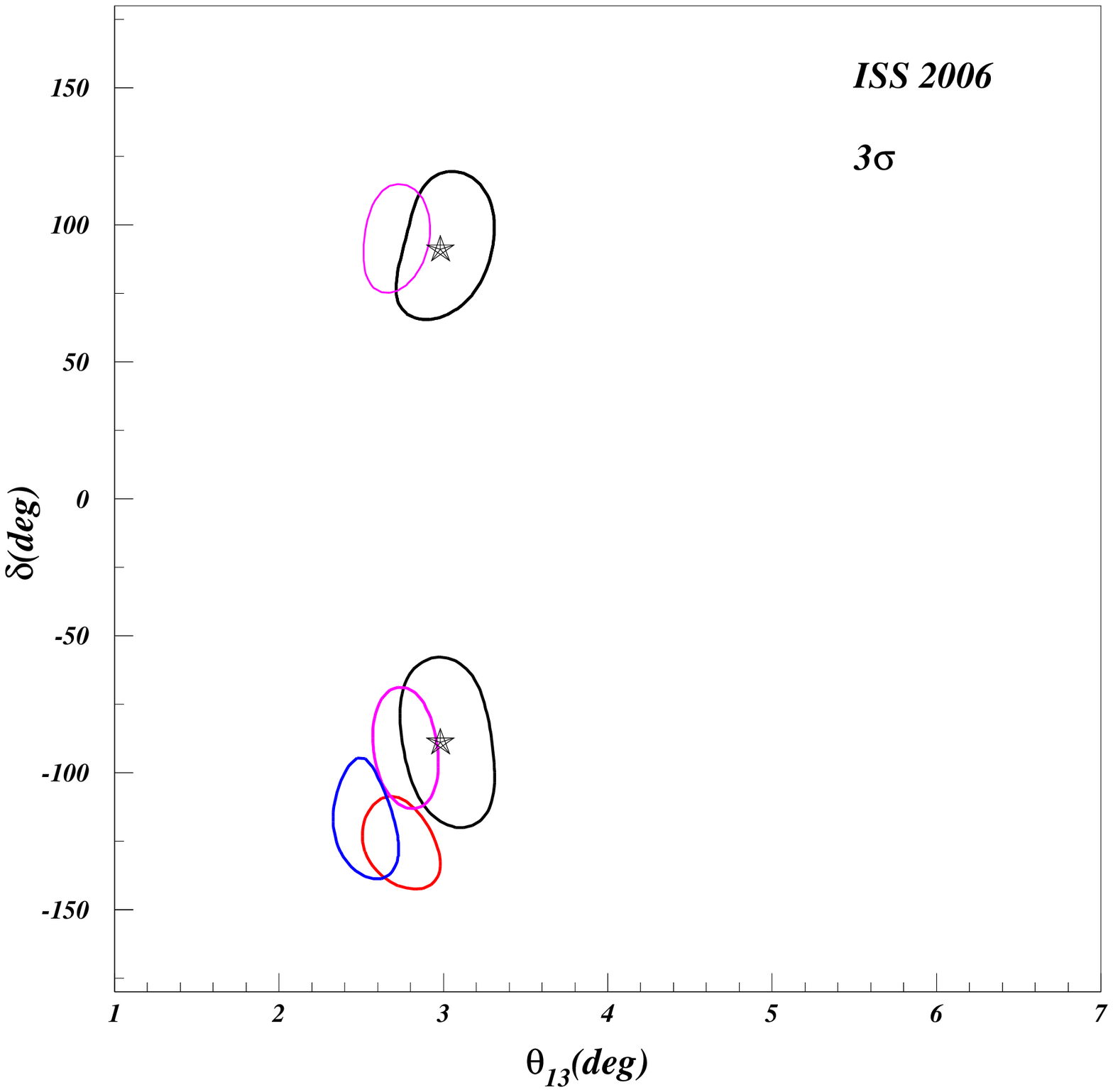,width=7cm}  \\
    \epsfig{file=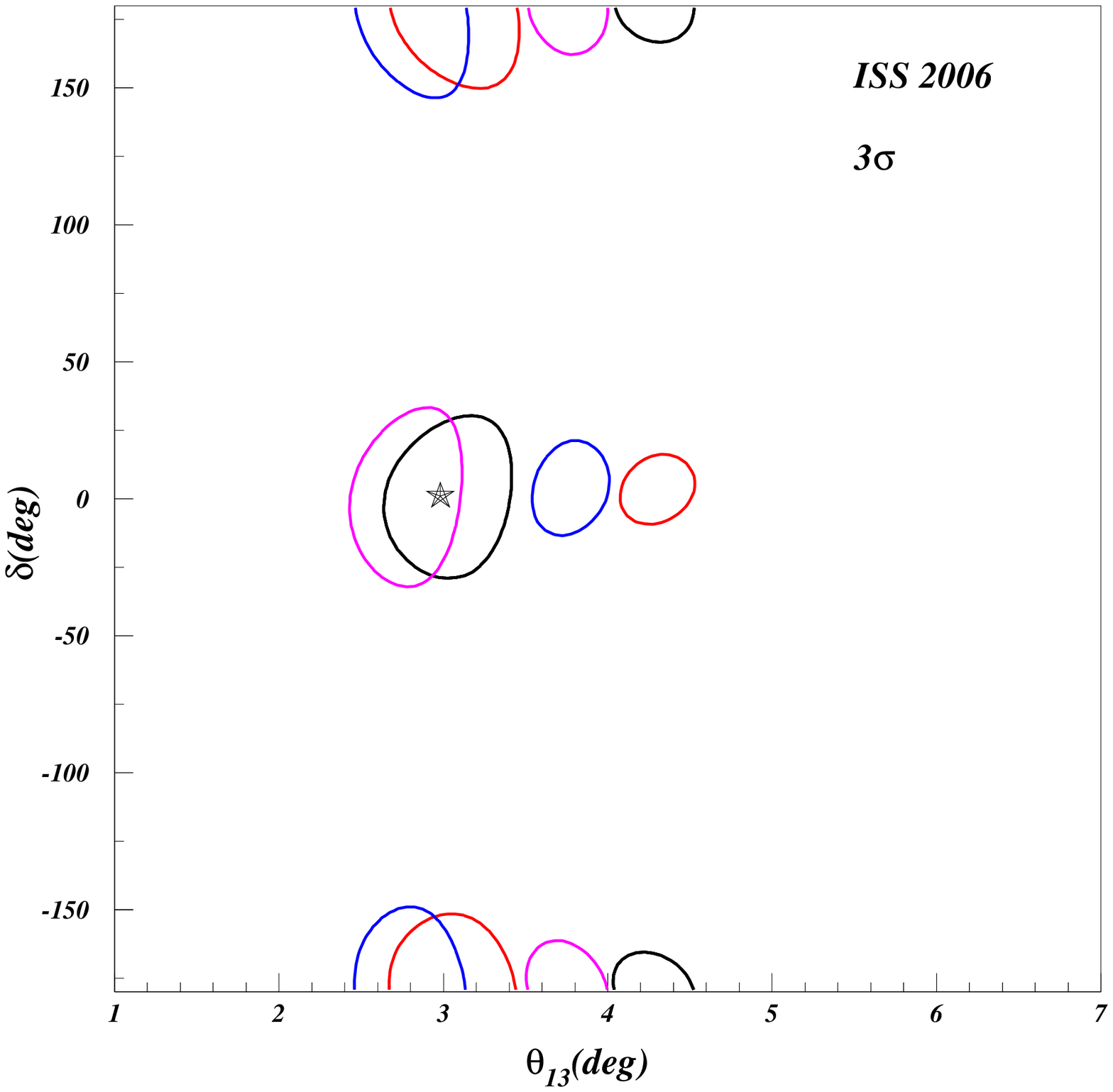,width=7cm}  
    \epsfig{file=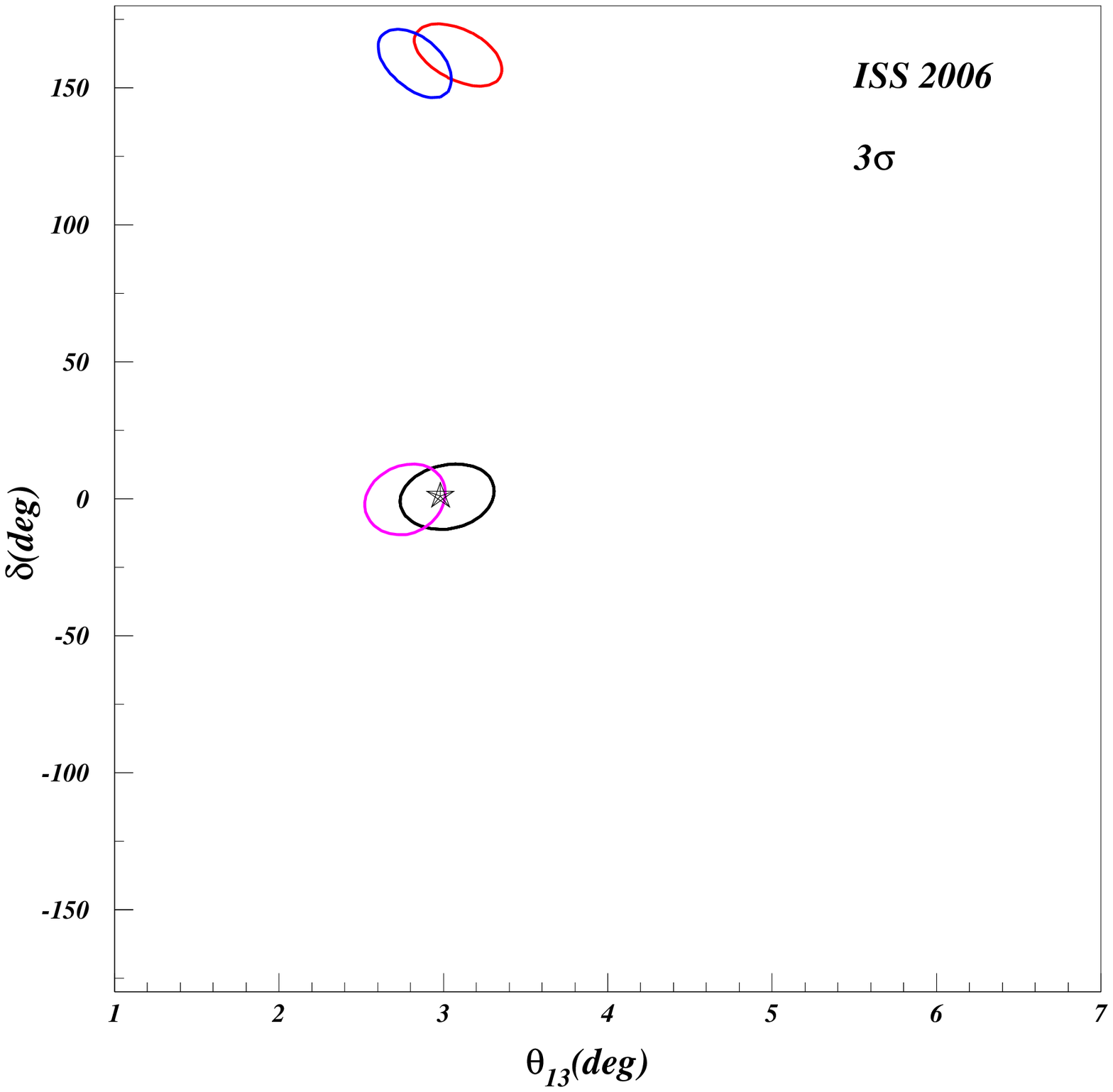,width=7cm}  \\
  \end{center}
  \caption{
    3$\sigma$ CL contours obtained in the LE$\bbph$ (left) and
    HE$\bbph$-a setup (right) for three values of the true parameters
    indicated by the stars.
    The solid black ellipses show the the intrinsic degeneracy, the
    pink ellipses the octant degeneracy, the red ellipses show the
    mass-hierarchy degeneracy, and the blue ellipse is the combined 
    mass-hierarchy/octant degeneracy.
  }
  \label{fig:fits}
\end{figure}

The results from fitting the appearance and disappearance signals to
extract the parameters $(\theta_{13}$ and $\delta)$, for the true
values indicated by the stars ($\theta_{13}=3^\circ$,
$\delta=90,-90,0$) are shown in figure \ref{fig:fits}.
The left panels correspond to the LE setup and the right panels to the
HE-a option.  
The uncertainties on $\delta$ and $\theta_{13}$ are significantly
larger for the LE$\bbph$ and in particular the eight-fold degeneracy
is fully present in this case.
The intrinsic degeneracy is resolved for the HE-a setup for all values
of $\delta$.
The octant degeneracy remains in all cases for the HE-a setup, while
the hierarchy and mixed degeneracies are resolved for
$\delta=90^\circ$. 

\subsubsection{Towards an optimal beta-beam setup}

While the sensitivity of the setups considered above to CP violation
and $\theta_{13}$ is comparable to the sensitivity that may be
achieved at the Neutrino Factory, the ability to resolve the discrete
degeneracies is rather limited.
A number of ideas have been considered to improve the physics reach of
beta-beams, particularly as regards the discrete ambiguities, these
ideas will be discussed below.

\subsubsection{Combination with atmospheric data}

Any large detector that could be used for a beta-beam can provide 
more precise measurements of the atmospheric-neutrino flux. 
This is certainly the case for the water \v Cerenkov considered in
setups LE$\beta\beta$ and HE$\beta\beta$-a for which the fiducial mass
considered is twenty times larger than that of Super-Kamiokande. 
Also in the case of a much smaller detector, such as a magnetised iron
calorimeter, the measurement of the neutrino and anti-neutrino fluxes
could add valuable information on the oscillation parameters
\cite{Agafonova:2000xm}. 

The physics potential that results from the combination of these
measurements with those in a long-baseline experiment were first
studied in reference \cite{Huber:2005ep}, where the case of the T2HK
super-beam was considered. 
More recently, the same analysis has been performed for the
LE$\beta\beta$ \cite{Campagne:2006yx}. 
In both cases, it has been found that for sufficiently large values of
$\theta_{13}$, the combination with atmospheric data is extremely
helpful in resolving the discrete degeneracies related to the mass
hierarchy and the $\theta_{23}$ octant. 
As we have seen, the LE$\beta\beta$ setup has no sensitivity to
either, while the HE$\beta\beta$ options have some, $\delta$-dependent,
sensitivity.

\begin{figure}
  \begin{center}
    \epsfig{file=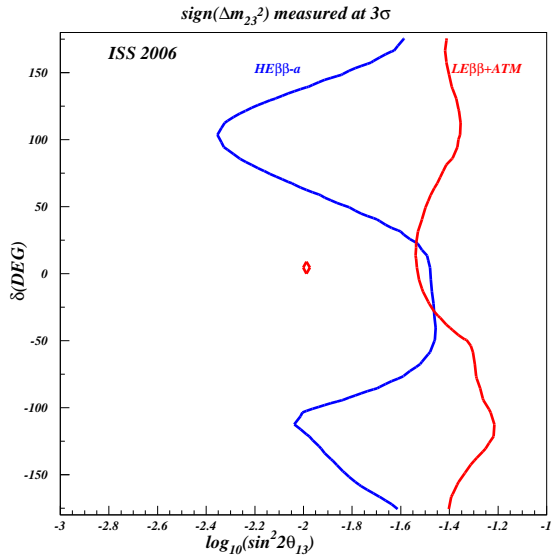,width=8cm}
  \end{center}
  \caption{
    3$\sigma$ sensitivity to the normal mass hierarchy assuming for
    the HE$\bbph$-a and the LE$\bbph$ in combination with atmospheric
    data.
  }
  \label{fig:signatm}
\end{figure}

The regions in which the sgn$\Delta m^2_{31}$ can be established at
3$\sigma$ by combining atmospheric-neutrino data with the
LE$\beta\beta$ and the HE-a setup are shown in figure \ref{fig:signatm}.
The combination of the LE setup with atmospheric data results in a
significant sensitivity to the hierarchy, although the combination
does not improve the sensitivity of the  HE-a setup.
The combination with atmospheric data is also possible for all
HE setups, this analysis has not yet been done.
It is expected, however, that including the atmospheric data in this
case will improve the sensitivity to the sgn$\Delta m^2_{31}$ for
those values of $\delta$ for which the sensitivity is poor and to
improve the sensitivity to CP violation in the negative $\delta$
region in the right panel of figure \ref{fig:cpv}. 

Concerning the octant ambiguity, the combination of a LE$\bbph$ with
atmospheric data has been shown not to improve the sensitivity that
can be achieved using the atmospheric data alone
\cite{Campagne:2006yx}.
It will be interesting to see whether, in the case of the HE$\bbph$
for which the sensitivity to the octant is better
\cite{Burguet-Castell:2005pa}, the situation changes and there is some
improvement as is in the case for the combination of the atmospheric
data with other super-beams such as T2HK or the SPL.

\subsubsection{An associated super-beam}

In the first beta-beam scenario considered at CERN, the complex also
included a conventional neutrino beam, the SPL super-beam, using the
same baseline (CERN-Frejus) and detector (water \v Cerenkov) (see
section \ref{SubSect:Perf:SB}.
The advantage of having the two types of beam, is that,
in addition to CP-conjugate transitions, T-conjugate and CPT-conjugate
transitions could also be measured \cite{Mezzetto:2003ub}. 
In particular, the comparison of the $\nu_e \rightarrow \nu_\mu$ 
and $\nu_\mu \rightarrow \nu_e$ oscillation probabilities is a T-odd
observable and is therefore sensitive to $\delta$. 

Besides the theoretical interest of these measurements in the search
for new physics, the determination of $\delta$ through such a T-odd
measurement is advantageous from the experimental point of view
because several systematic uncertainties would be cancelled. 
For example, the error on the Earth matter density is not relevant for
this measurement. 
 
The first analysis of the performance of a super-beam and beta-beam
combination \cite{Mezzetto:2003ub,Bouchez:2003fy} did not include
spectral information and in this situation, given that the two
$\langle E_\nu\rangle/L$ are very similar, it was clear that
degeneracies, in particular the intrinsic, one would remain
\cite{Donini:2004hu}. 
Later, the spectral information has been included and it has been
shown that the SPL on its own is able to resolve the intrinsic
degeneracy \cite{Campagne:2006yx}, however this is not the  
case for the LE$\beta\beta$ as we have seen. 

\begin{figure}
  \begin{center}
    \epsfig{file=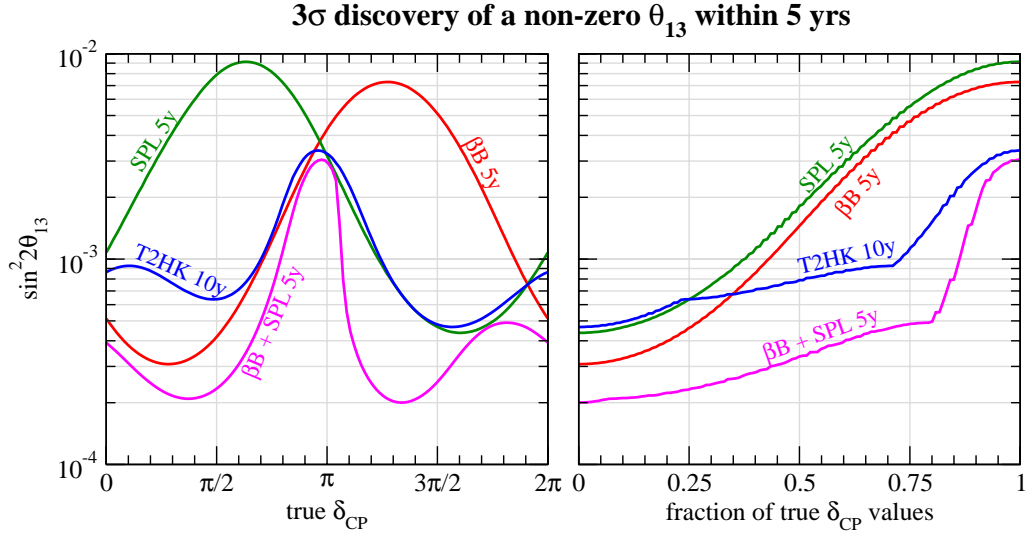,width=14cm}
  \end{center}
  \caption{
    3$\sigma$ sensitivity to $\sin^2 2\theta_{13}$ for 5 years of run
    with a LE$\bbph$ (with twice the standard ion flux) and the SPL
    super-beam, compared with a 10 year run with T2HK (2 years
    neutrinos and 8 anti-neutrinos. 
    Figure taken from reference \cite{Campagne:2006yx}. 
    The parameters are not the same as those in equation
    (\ref{eq:true}).
    Taken with kind permission of the Journal of High Energy Physics 
    from figure 14 in reference \cite{Campagne:2006yx}.
    Copyrighted by SISSA.
  } 
  \label{fig:bbspl}
\end{figure}

A real synergy of both types of experiment has been explored recently
\cite{Campagne:2006yx}. 
The idea is to use only neutrino runs in both beams, which has the
advantage that the cross sections are larger than for anti-neutrinos.
In figure \ref{fig:bbspl}, the sensitivity to $\theta_{13}$ for this
combination with a five-year run of both the SPL and beta-beam is
shown to outperform a ten-year run of T2HK.  

\subsubsection{Combination of different ions}

In reference \cite{Donini:2006dx}, a combination of the beams produced
by the four ions ${\rm He}/{\rm Ne}/{\rm Li}/{\rm B}$ (the
`alternating-ion' scenario) with a $\gamma$ below the present SPS
limit has been considered.  
In this case, the baseline chosen was $630$~km (CERN-Canfranc), which
corresponds to the first atmospheric peak for the ${\rm Li}/{\rm B}$
beam at $\gamma\sim 100$, while the ${\rm He}/{\rm Ne}$ beam is close
to the second peak for a similar $\gamma$. 
The ion fluxes are assumed to be the standard ones for 
${\rm Li}/{\rm B}$ as for $He/Ne$, so the total neutrino and
anti-neutrino fluxes from all the ions are similar, but the shapes are
quite different since the end-point values, $E_0$, differ (see section
\ref{sec:setups}).

\begin{figure}
  \begin{center}
    \epsfig{file=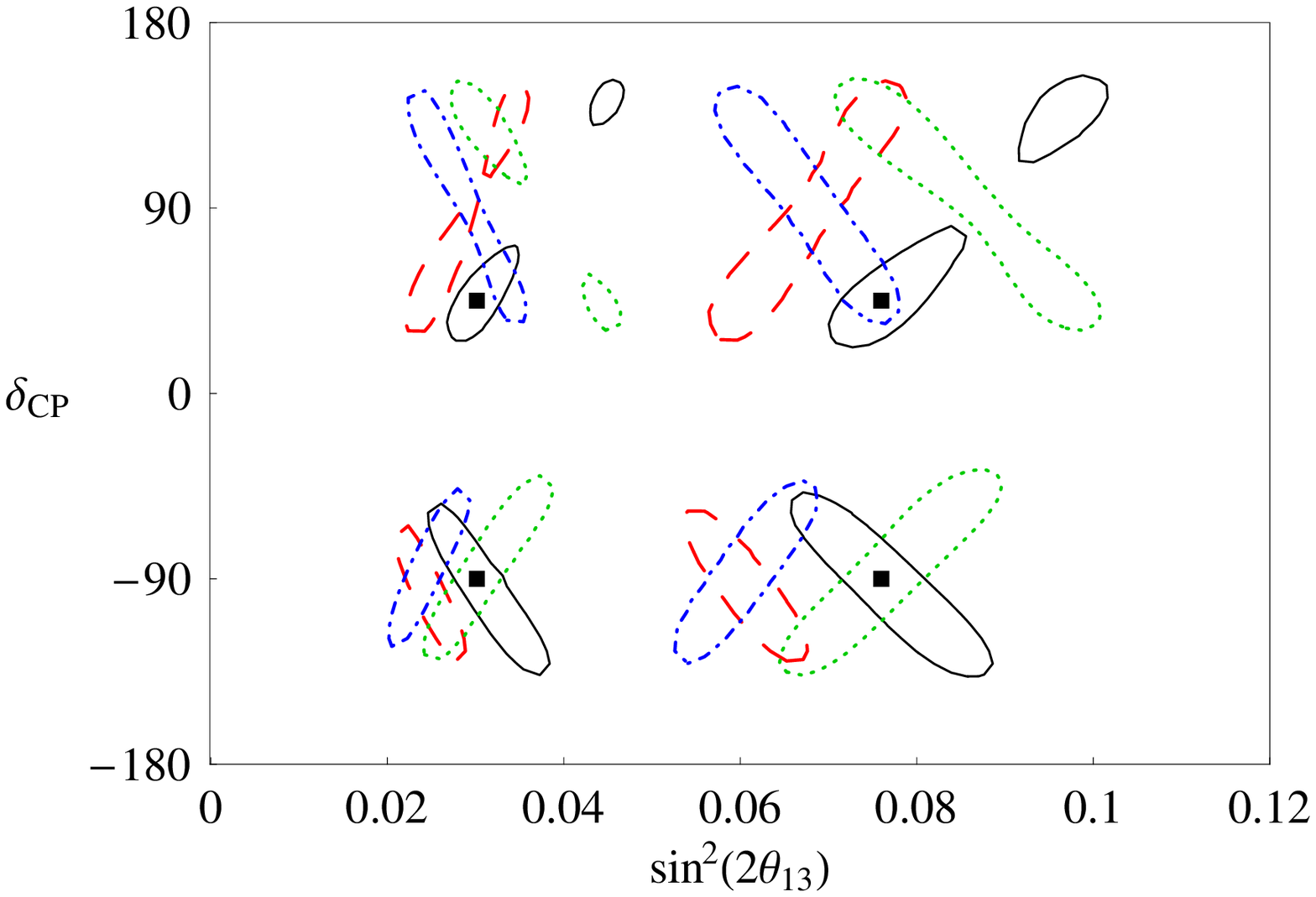,width=7.5cm} 
    \epsfig{file=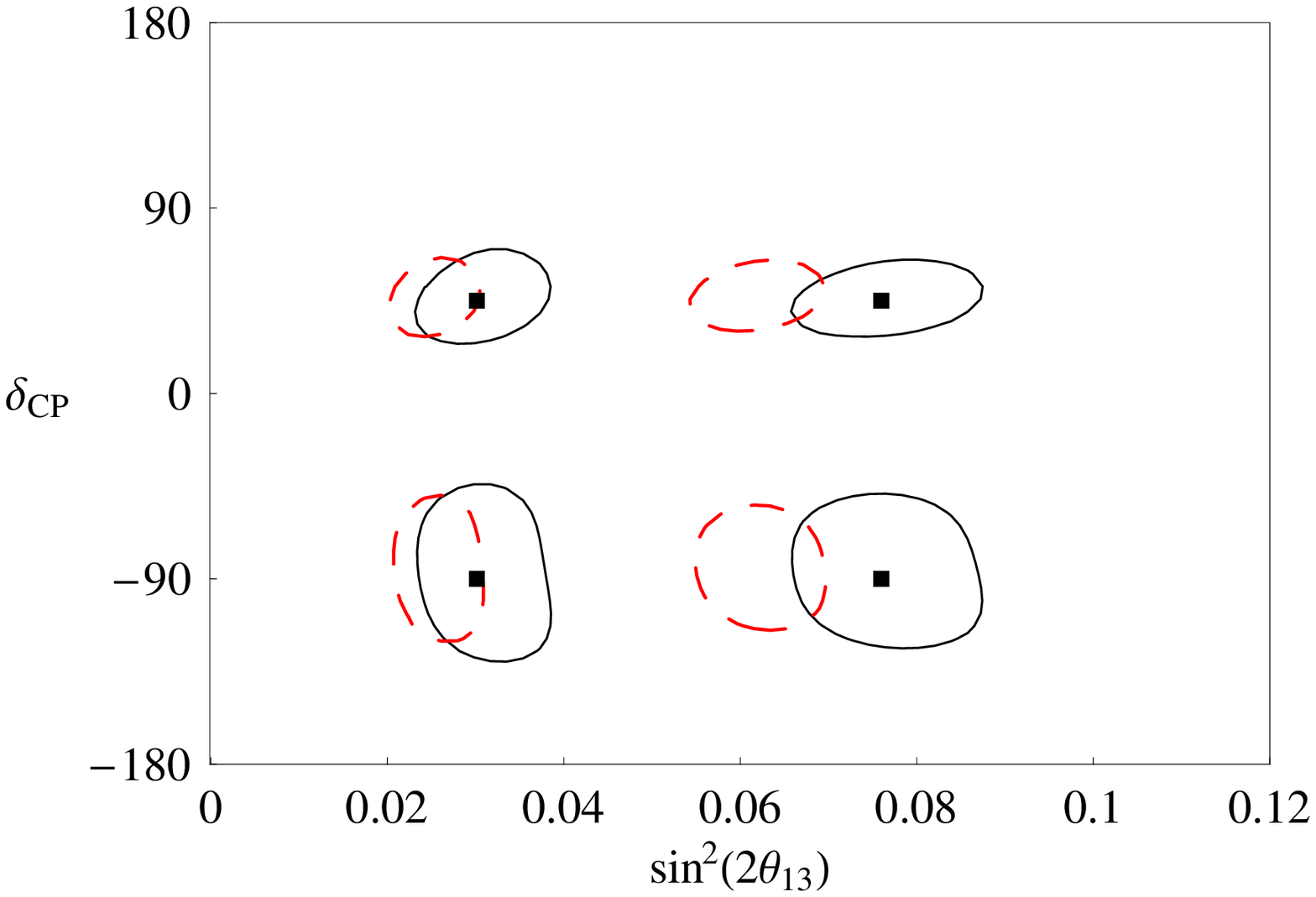,width=7.5cm}
  \end{center}
  \caption{
    90$\%$ contours for a LE$\bbph$ with $\gamma=120$ (left) and an
    alternating ion scenario (right) both at $L=630~km$
    (CERN-Canfranc). The true parameters are denoted with a thick
    black square. Figure taken from reference
    \cite{Donini:2006dx}. The parameters are not the same as those in
    equation (\ref{eq:true}).
    Taken with kind permission of Physical Letters from figure 2 in
    reference \cite{Donini:2006dx}.
    Copyrighted by Elsevier B.V.
  } 
  \label{fig:ioncomb}
\end{figure}

The main advantage of this combination over the LE$\beta\beta$ setup
is the use of two different $L / \langle E_\nu\rangle$ which is a very
powerful way of resolving degeneracies. 
In particular, the intrinsic degeneracy which severely limits a precise
determination of $\delta$ in the LE$\beta\beta$ setup is absent in the
combination.
This, however, is at the expense of having larger statistical
uncertainties due to the smaller flux. 
Fits for $(\theta_{13},\delta)$ for a LE$\bbph$ with a slightly larger
$\gamma=120$ at $L=130~km$ and the combination of four ions at
$L=630~km$ are compared in figure \ref{fig:ioncomb}.
The eight-fold degeneracy of the former is reduced to a two-fold
degeneracy in the latter at $90\%$CL, only the octant ambiguity
remains unresolved. 
The sensitivity of this combination to the hierarchy has also been
shown to be very significant and much less dependent on $\delta$ than
in the case of the HE-a setup.
We therefore conclude that this combination outperforms the
LE$\beta\beta$ if $\theta_{13}$ is not too small, within the reach of
T2K phase I.  
 
\subsubsection{Higher $\gamma$ ?}

\begin{figure}
  \begin{center}
    \epsfig{file=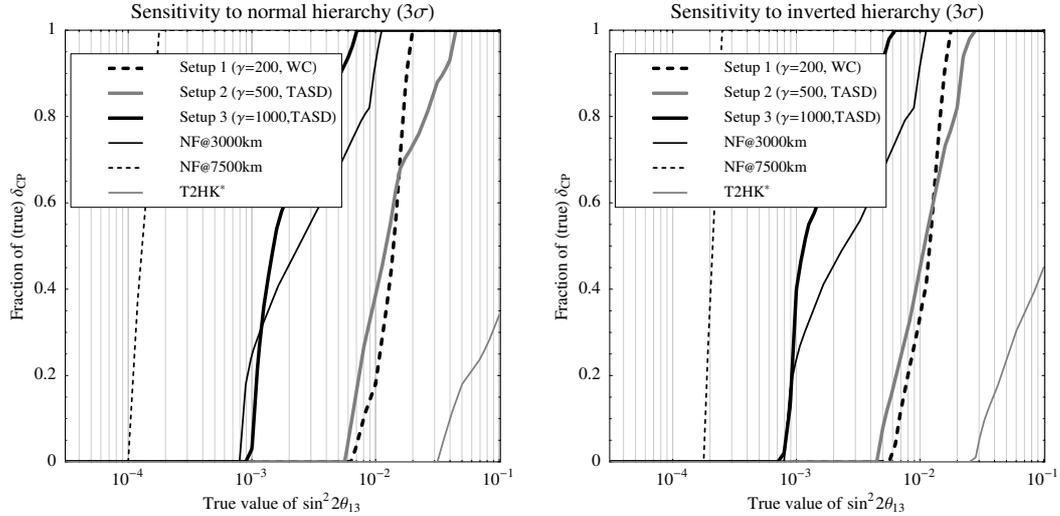,width=15cm}
  \end{center}
  \caption{
    3$\sigma$ sensitivity to the normal mass hierarchy for different
    beta-beam setups including one with $\gamma=1000$ combined with
    the same detector as in the HE-b setup discussed in the text. 
    The Neutrino Factory setups are also included for comparison. 
    Figure taken from reference \cite{Huber:2005jk}. 
    The parameters are not the same as those in equation (\ref{eq:true}).
    Taken with kind permission of the Physical Review from figure 12 in
    reference \cite{Huber:2005jk}.  
    Copyrighted by the American Physical Society.
  } 
  \label{fig:vhebb}
\end{figure}

The possibility of using more powerful accelerators such as the LHC to
achieve even higher $\gamma$ has also been discussed. 
The increase in $\gamma$ allows smaller detectors, optimised for
events in the multi-GeV range, to be considered. 
The physics potential of a very high $\gamma$ beta-beam with 
$\gamma \geq 1000$, but assuming the same ion flux, has been
considered in \cite{Burguet-Castell:2003vv,Huber:2005jk}. 
In the first reference a $\sim 50$kton idealised scintillator detector
was assumed, while in the second the NO$\nu$A type detector discussed
above was considered.
The data-sample size is therefore improved very much with respect to
the HE$\bbph$-a setup since the gain in $\gamma$ is compensated by a
decrease in the detector mass.
The sensitivity to sgn$\Delta m^2_{31}$ is compared for three
beta-beams setups and the Neutrino Factory in figure \ref{fig:vhebb}.
The conclusion of these studies is that going to such high $\gamma$
improves the sensitivity to the hierarchy and therefore resolves the
correlation of $\delta$ with sgn$\Delta m^2_{31}$, so improving the
sensitivity to CP violation.

A related idea has been proposed more recently in
\cite{Agarwalla:2006vf}. 
The goal is to arrive to the magic baseline ($L \sim 7 \, 500$~km
\cite{Huber:2003ak} using $^8{\rm B}$ and $^8{\rm Li}$ ions with
$\gamma$ in the range 250--500; this could be achieved with a
refurbished SPS at CERN pointing at the Indian Neutrino Observatory
(INO) where a large magnetised iron calorimeter in the 50-100~kton
range (ICAL) could be used as the far detector.  
For details of the potential of this setup see
\cite{Agarwalla:2006vf}.

\subsubsection{Higher fluxes ?}
\label{Sect:HighFlux}

The standard ion fluxes that have been used in this study are based on
the CERN design for a LE$\bbph$, using the present CERN SPS and PS and
requiring a duty cycle of a few $10^{-3}$. 
In the present design, a large fraction of the ions being produced are
lost in the acceleration process and it is likely that a refurbished
PS or SPS could eliminate some of the present losses.
The refurbishing of these old machines is likely to be required to
serve the LHC programme and, therefore, it is likely that further
optimisations to increase the neutrino flux can be considered.
Furthermore, an entirely new approach to producing the required
unstable ions, using ionisation cooling, has been recently proposed in
reference \cite{Rubbia:2006pi}. 
Although there remain many details to work out, this novel approach
offers the possibility of increasing the ion-production yield by
several orders of magnitude. 
The physics reach of such a beta-beam with such high fluxes would be
outstanding and it is therefore of the upmost importance to explore
possible optimisations that could be achieved with realistic
improvements in the accelerators or/and the ion-production technique.

\subsubsection{Monocromatic $e$-capture beams}

Triggered by the beta-beam concept, a different type of neutrino beam
has been proposed in reference \cite{Bernabeu:2005jh}. 
The idea is to produce neutrinos from boosted ions that undergo an
$e$-capture transition, that is an atomic electron is captured by a
proton, anti-neutrino beams cannot be produced this way. 
Kinematically it is a two-body decay and therefore the neutrino energy
is well-defined and given by the difference between the initial and
final nuclear mass energies minus the excitation energy of the
final-state nucleus.
Such transitions are usually dis-favoured, but there are a few nuclei
(see table \ref{tab:nuclei}) for which the decay rate is significant.
\begin{table}
  \begin{center}
    \begin{tabular}{cccc}
      \hline
      Decay                      & $T_{1/2}$ &  $E_\nu$ (keV) & EC/$\beta^+$ (\%) \\
      \hline
      $^{148}$Dy $\to$ $^{148}$Tb       &  3.1 m   &  2062     &   96/4      \\
      $^{150}$Dy $\to$ $^{150}$Tb       &  7.2 m   &  1397     &   99.9/0.1  \\
      $^{152}$Tm $2^-$ $\to$ $^{152}$Er &  8.0 s   &  4400     &   45/55     \\
      $^{150}$Ho $2^-$ $\to$ $^{150}$Dy &   72 s   &  3000     &   77/33    \\
      \hline
    \end{tabular}
  \end{center}
  \caption{\it Decay properties of some rare-earth nuclei.}
  \label{tab:nuclei}
\end{table}

The neutrino flux can easily be shown to be \cite{Bernabeu:2005jh}:
\begin{equation}\label{master}
\frac{d^2N_\nu}{ dS dE}
= \frac{1}{\Gamma} \frac{d^2\Gamma_\nu}{dS dE} N_{ions}
\simeq \frac{\Gamma_\nu}{\Gamma} \frac{ N_{ions}}{\pi L^2} \gamma^2
\delta{\left(E - 2 \gamma E_0 \right)},
\end{equation}
where $\gamma$ is the boost factor of the parent ion, $E_0$ is the
neutrino energy in the laboratory frame, and $\Gamma_\nu/\Gamma$ is
the e-capture branching fraction. 
The neutrino energy in the detector will be peaked at $2\gamma E_0$
and the requirement that the neutrino energy be reconstructed
accurately in the detector can be relaxed.
One can easily disentangle the different oscillation parameters by
performing counting experiments at different values of $\gamma$. 

As in the case of the beta-beam, a possible implementation of the
concept would involve the use of EURISOL to produce the unstable ions,
the SPS to accelerate them, and a decay ring. 
However, to allow electron capture to occur, we need to keep one
electron bound to the ion's nucleus, partly ionised particles have a
short vacuum life-time (even in a very good vacuum collisions 
with the few remaining atoms suffice to cause them rapidly to lose the
remaining electron).
The ion that has been proposed as optimal is $^{150}$Dy, which could
be accelerated in the CERN SPS up to a maximum $\gamma$ of 195.   

The main advantage of an electron-capture beam over a `conventional'
beta-beam, for a similar number of ion decays, is that all the
intensity is peaked at the energy(ies) of interest. 
In a beta-beam the broad spectrum implies that many neutrinos will be
produced at energies for which the dependence on $\delta$ is less
pronounced, and/or the cross section is too low. 
It is also a excellent tool to discriminate against backgrounds of
various types.

\begin{figure}
  \begin{center}
    \includegraphics[width=10cm]{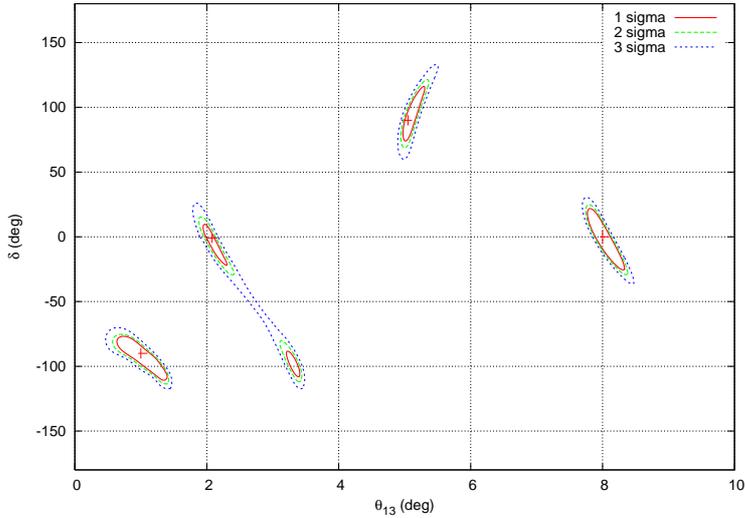} 
  \end{center}
  \caption{
    Combined fits of $\theta_{13}$ and $\delta$ for different central
    values of the parameters. 
    The known parameters are not the same as those of
    equation (\ref{eq:true}). 
    Figure taken from \cite{Bernabeu:2005kq}.
  }
  \label{fig:fits_ec}
\end{figure}

Even though no realistic study of the expected ion flux has yet been
performed, an analysis of the performance of such a beam assuming an
intensity of $10^{18}$ ions/year has been presented in
\cite{Bernabeu:2005jh,Bernabeu:2005kq}.
Using a $440$~kton fiducial mass water \v Cerenkov located at a
distance of 130~km (CERN-Frejus baseline), 5 years running time for
each of $\gamma=195$ and $\gamma=90$, the precision in the
determination of $\theta_{13}$ and $\delta$ that could be obtained is
illustrated in figure \ref{fig:fits_ec}. 
Due to the lack of a CP-conjugate observable, such as one would have
with an anti-neutrino beam, all sensitivity to the CP phase is
lost at a given $\gamma$. 
However, it is remarkable that the measurement of the oscillation
probabilities at two  energies results in a significant sensitivity to
$\delta$.
An alternative, that would improve the sensitivity to $\delta$, would
be to combine an $e$-capture beam with a standard beta-beam from
$^6$He using the same detector.   

%% file: 05-Performance/05-04-Neutrino-Factory/05-04-Neutrino-Factory.tex
\subsection{Optimisation and physics potential of a Neutrino Factory 
            oscillation experiment}
\label{SubSect:Perf:NF}
%
%
\input 05-Performance/05-04-Neutrino-Factory/05-04-00-Neutrino-Factory
%
%
\input 05-Performance/05-04-Neutrino-Factory/05-04-01-Neutrino-Factory-setup
%
%
\input 05-Performance/05-04-Neutrino-Factory/05-04-02-Neutrino-Factory-golden
%
%
\input 05-Performance/05-04-Neutrino-Factory/05-04-03-Neutrino-Factory-clones
%
%
\input 05-Performance/05-04-Neutrino-Factory/05-04-04-Neutrino-Factory-best
%
\input 05-Performance/05-04-Neutrino-Factory/05-04-05-Neutrino-Factory-lowE

%% file: 05-Performance/05-04-Neutrino-Factory/05-04-00-Neutrino-Factory.tex
In a Neutrino Factory \cite{Geer:1997iz,DeRujula:1998hd} muons are
accelerated from an intense source to energies of several tens of GeV
and injected into a storage ring with long straight sections. 
The muon decays $\mu^+ \rightarrow e^+ \, \nu_e \, \nubarmu$ and
$\mu^- \rightarrow e^- \, \nubare \, \nu_\mu$ provide a very well
known flux of neutrinos with energies up to the muon energy itself.
Neutrino Factory designs have been proposed in Europe
\cite{Autin:1999ci,Gruber:2002tn}, the US
\cite{MuColl,Finley:2000cn,Ozaki:2001bb,Alsharoa:2002wu,Zisman:2003bh},
and Japan \cite{Kuno:2001tb}.  
The conclusion of these studies is that an accelerator complex capable
of providing $\sim 10^{21}$ muon decays per year can be built.   
One of the most striking features of the Neutrino Factory is the
precision with which the characteristics of all components of the beam
would be known. 
The following effects were considered in reference
\cite{Blondel:2004ae}:
\begin{itemize}
  \item 
    Beam polarisation: with a polarimeter, the beam energy and energy
    spread can be measured and the degree to which the polarisation
    dependence of the neutrino flux affects the measured rates can be
    tested to high precision \cite{Blondel:2000vz}; 
  \item 
    Beam divergence and radiative corrections in muon
    decay \cite{Broncano:2002hs}; 
  \item 
    Absolute normalisation of the flux to be obtained from a beam
    monitor; and
  \item 
    Absolute cross-section normalisation using the inverse-muon-decay
    reaction, $\nu_{\mu} e^- \rightarrow \mu^- \nu_e$, in the near
    detector.  
    In principle, a normalisation of fluxes and cross-sections with a
    precision of $10^{-3}$ can be achieved. 
\end{itemize}

Some of these features should also be present for a beta-beam, and for
any facility in which a stored beam of well-defined optical properties
is used to produce neutrinos. 
This is an important difference with respect to super-beams for which
the precision with which the neutrino and anti-neutrino cross sections
and fluxes are known is determined by the degree to which the 
particle-production spectra are known.

Twelve oscillation processes can be studied using the Neutrino Factory 
which and store beams of both positive and negative muons (see table
\ref{tab:procs}).
Neutrinos produced from the decay of positive and negative muons 
must not be confused.
The required separation can be achieved by running the two polarities
in turn or by careful timing if the two polarities are stored
simultaneously. 
In order to take full advantage of this flavour-richness, the optimal
detector should be able to perform both appearance and disappearance
experiments, providing lepton identification and charge
discrimination. 
\begin{table}
  \begin{center}
    \begin{tabular}{|c|c|c|}
      \hline
      $\mu^+ \rightarrow e^+\nu_e\nubarmu$ &  $\mu^- \rightarrow e^-\nubare$  & \\
      \hline
      $\nubarmu \rightarrow \bar \nu_\mu$  & $\nu_\mu \rightarrow \nu_\mu$   & disappearance \\
      $\nubarmu \rightarrow \bar \nu_e$   & $\nu_\mu \rightarrow \nu_e$     & appearance (challenging) \\
      $\nubarmu \rightarrow \bar \nu_\tau$ & $\nu_\mu \rightarrow \nu_\tau$  & appearance (atm. oscillation) \\
      $\nu_e \rightarrow \nu_e$        & $\bar \nu_e \rightarrow \bar \nu_e$   & disappearance \\   
      $\nu_e \rightarrow \nu_\mu$      & $\bar \nu_e \rightarrow \bar \nu_\mu$  & appearance: ``golden'' channel \\  
      $\nu_e \rightarrow \nu_\tau$     & $\bar \nu_e \rightarrow \bar \nu_\tau$ &  appearance: ``silver'' channel \\ 
      \hline
    \end{tabular}
    \label{tab:procs}
  \end{center}
  \caption{
    Oscillation processes in a Neutrino Factory.
  }
\end{table}

The search for $\nu_e \rightarrow \nu_{\mu}$ transitions (the `golden
channel') \cite{Cervera:2000kp} appears to be particularly attractive
at the Neutrino Factory.
It can be studied in appearance mode, by looking for muons with
charge opposite to that of the stored muon beam (`wrong-sing muons'),
thus strongly reducing the dominant background (`right-sign muons'). 
The wrong-sign-muon channel yields an impressive sensitivity to
$\sin^2\theta_{13}$ and sensitivity to the leptonic CP-violating phase,
$\delta$, down to very small values of $\theta_{13}$ 
\cite{Cervera:2000kp,Burguet-Castell:2001ez,Huber:2003ak}. 
For example, with two 40 Kton MINOS-like magnetised-iron detectors at
two different baselines, exposed to beams of both polarity and
$10^{21}$ muon decays, it will be possible to explore $\theta_{13}$
down to $\sin^2 2 \theta_{13} \ge 1 \times 10^{-5}$ 
($\theta_{13} \ge 0.1^\circ$) and to measure $\delta$ for most 
of the parameter space \cite{Burguet-Castell:2001ez}. 
The relatively high energy of the neutrinos produced through the decay
of high-energy stored muons implies that baselines of several
thousand kilometers are needed for Neutrino Factory experiments.  
For such baselines, CP asymmetries are dominated by matter effects
\cite{Freund:2000ti,Geer:2000pu,Donini:1999jc} that can be used to
determine unambiguously sign($\Delta m^2_{31}$) for large enough
$\theta_{13}$. 

The determination of ($\theta_{13},\delta$) at the Neutrino Factory is
not free of ambiguities; up to eight different regions of the
parameter space can fit the same experimental data in the
$(\theta_{13},\delta)$ plane.
In order to solve these ambiguities, a single experimental measurement
for a single neutrino beam is not enough. 
One possible solution to this problem is to combine detectors looking 
for `golden' muons at different baselines (i.e., different
$L/E$).
A second possibility is to make use of the rich flavour content of the
Neutrino Factory beam. 
The $\tau$ appearance channel (`silver channel')
\cite{Donini:2002rm,Autiero:2003fu} has been advocated as a powerful
means of resolving ambiguities, if a detector capable of $\tau$
identification can be used.
This can readily be understood since the ${\delta}$-dependence of
the silver and the golden channel are different, while the dependence
of the two channels on matter effects and ${\theta_{13}}$ is similar.
On the other hand, the $\nu_\mu$-disappearance channel is rather
effective for large values of $\theta_{13}$ in measuring the
$\theta_{23}$ octant \cite{Donini:2005db}.
A detector capable of measuring the charge of the electrons has been
shown to allow the resolution of ambiguities by
separating the events into several classes (right-sign muons,
wrong-sign muons, electrons, and neutral currents) and performing a
fine energy binning down to low energies.  
Such a possibility was first studied assuming the feasibility of a
magnetised liquid-argon detector \cite{Bueno:2000fg}, and recently
updated in reference \cite{Huber:2006wb}.  
R\&D efforts for a liquid-argon detector embedded in a magnetic field  
are ongoing \cite{Ereditato:2005yx} (the first curved tracks were
recently observed in a 10 litre liquid-argon TPC embedded in magnetic 
field \cite{Badertscher:2005te}). 
A third possibility is an improved detector (with a much lower muon
energy threshold) to look for `golden muons' solving at the same
time all the degeneracies. 

This section is organised as follows: in section \ref{sec:NFsetup} the
`standard' Neutrino Factory setup is introduced and the different
detectors are described; section \ref{sec:golden} contains a review of
the performance of the magnetised iron detector located at 
$L \sim 3000$~km from the source (i.e., the `standard' setup) and of
the problems that must be faced; in section \ref{sec:clones} possible
improvements to this setup are considered combining detectors at
different baselines, channels (following table \ref{tab:procs}) and
improving the `standard' detector; in section \ref{sec:best} the main
characteristics that are needed to use the Neutrino Factory at its
best are addressed. 

%% file: 05-Performance/05-04-Neutrino-Factory/05-04-01-Neutrino-Factory-setup.tex
\subsubsection{The Neutrino Factory setup}
\label{sec:NFsetup}

In the following, the `standard' Neutrino Factory refers to a facility 
in which a 50~GeV stored-muon beam delivers a luminosity of 
$1\times 10^{21}$ muon decays per year.
The total luminosity per muon polarity is taken as a given,
independent of, for example, the specific choice of proton-driver beam
power, storage-ring geometry, or the time spent running with a
particular polarity.
Notice that in the literature several different options have been
considered for each of these \cite{Apollonio:2002en}. 
Three detectors of different technologies, each specifically optimised
to detect a particular signal, have been considered.

\paragraph{Magnetised Iron Detector (MID): the `golden channel'}
\label{sec:mid}

\noindent
The most important signal at the Neutrino Factory is the 
`golden channel', i.e. the appearance channel $\nu_e \to \nu_\mu$.  
The signal is tagged by `wrong-sign muons', muons in the detector with
charge opposite to that of the muons in the storage ring.  
In order to extract the signal from the dominant source of background,
i.e. non-oscillated $\bar \nu_\mu$ (giving rise in the detector to a
huge number of `right-sign muons'), a magnetised detector is required.
This requirement represents the most important difference between the
detectors adopted for super-beam and beta-beam facilities and those
needed to take full advantage of the Neutrino Factory. 
As a consequence, large, water \v Cerenkov detectors are dis-favoured
and medium size magnetised detectors must be considered.

\begin{figure}
  \begin{center}
    \mbox{\epsfig{file=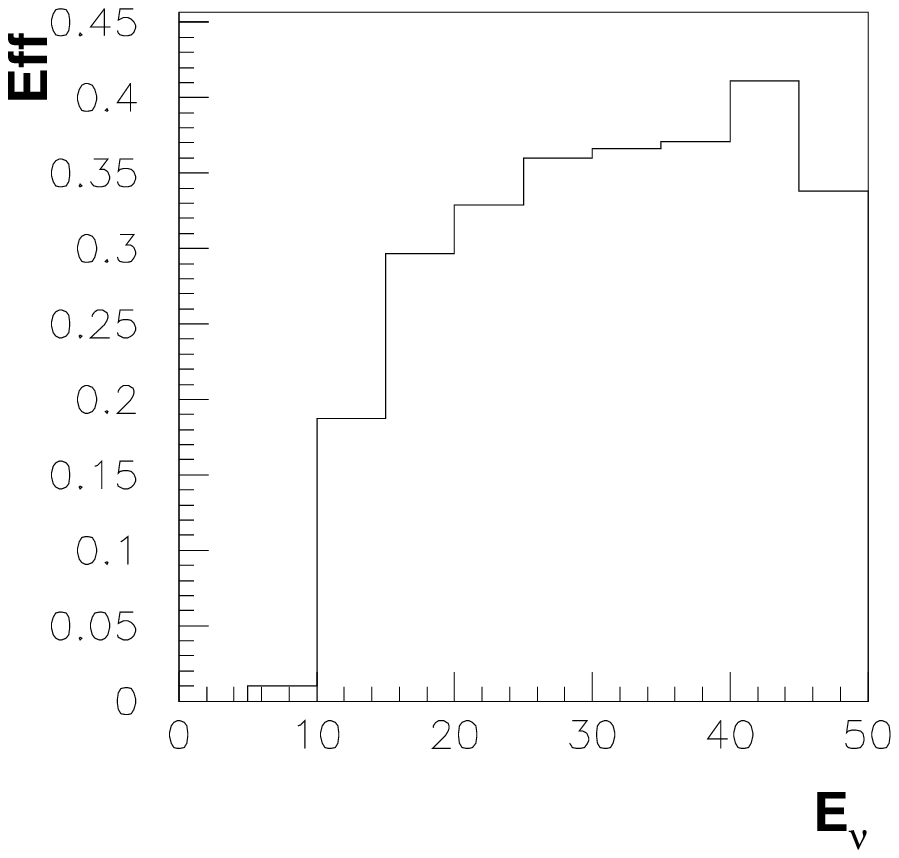,width=6cm} 
    \epsfig{file=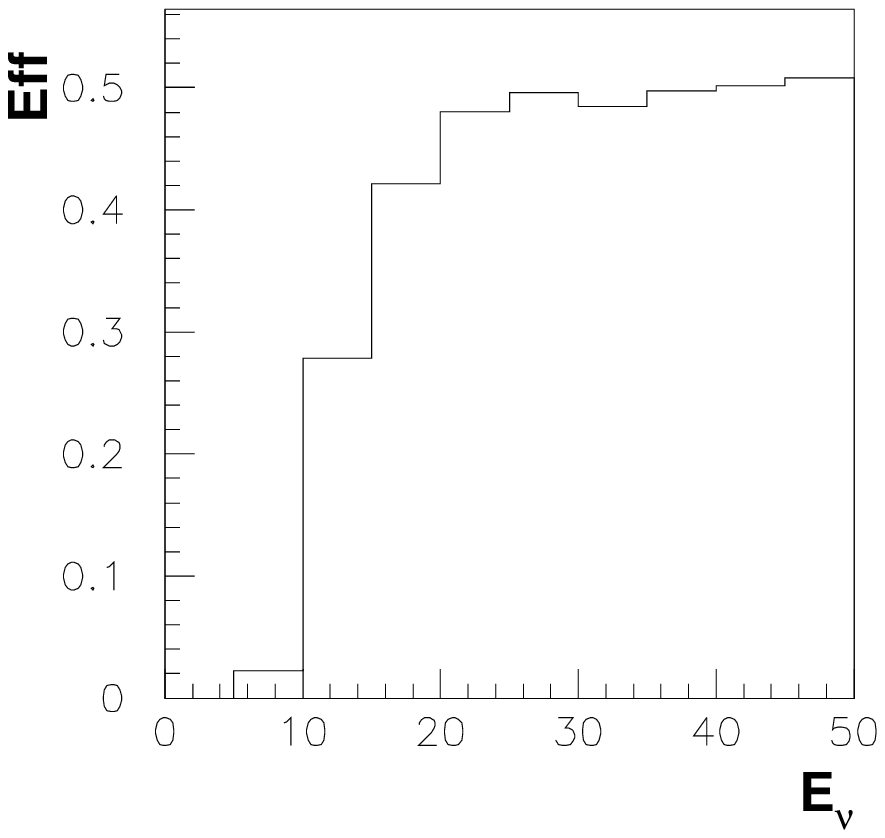,width=6cm}} \\
  \end{center}
  \caption{
    Signal efficiency at the magnetised iron detector for $\mu^+$
    (left panel) and $\mu^-$ (right panel) as a function of the
    neutrino energy.
    Taken with kind permission of Nuclear Physics B from figure 18
    in reference \cite{Cervera:2000kp}.
    Copyrighted by Elsevier Science B.V.
  }
  \label{fig:effmid}
\end{figure}

The reference detector, a 50~Kton magnetised iron calorimeter of the
MINOS type, was optimised in reference \cite{Cervera:2000vy} for the
study of $\nu_e \to \nu_\mu$ oscillations.
Tight kinematic cuts were applied to decrease the dominant and
sub-dominant backgrounds (right-sign muons and charmed-meson decays).
Such cuts, although strongly reducing the background, have the
dis-advantage that a significant proportion of the signal with neutrino
energy below 10 GeV is removed.
This can be seen in figure \ref{fig:effmid}, where the efficiencies of
the golden channel in the magnetised iron detector with MINOS-like
performance is shown. 
Measurements of the energy spectrum below 10 GeV, however, have been
shown to be extremely important; the first oscillation peak for $L\sim
3000$~Km lies precisely in this energy range. 
For this reason, the Neutrino Factory is the single facility
considered in this report most affected by degeneracies.
The measurement of the spectrum both below and above the oscillation
maximum has been shown to be crucial in the solution of many of the
parametric degeneracies that compromise the
($\theta_{13},\delta$)-measurement. 
The improvement in performance obtained by increasing the signal
efficiency for neutrino energies below 10~GeV is considered in section
\ref{sec:det}.

Different treatments of the energy response of the detector can be
found in the literature.
For example, in reference \cite{Cervera:2000kp} the energy resolution
was assumed to be $0.2 \times E_\nu$.
The effect of this finite energy resolution was taken into account by
grouping the events in five bins of width $\Delta E_\nu = 10$~GeV. 
This approach is quite conservative especially at low energy where 
most of the oscillation signal is found.
In reference \cite{Huber:2006wb} a finer binning was adopted, with a
more detailed treatment of resolution effects.
There are several differences between the treatment in reference
\cite{Huber:2006wb} and that of reference \cite{Cervera:2000kp}.
First, the energy response of the detector is modeled by folding the
raw-event distribution with a Gaussian resolution kernel of width
$\sigma_E=0.15 \times E_\nu$; in this way, the results become
independent of the bin width, provided that the binning is fine
enough.
Secondly, 43 bins of variable $\Delta E_\nu$ were considered in the
energy range $E_\nu \in [1,E_\mu]$ GeV.
The bins were defined as follows: 18 bins of 
$\xi \times 500 \, \mathrm{MeV}$; 10 bins of 
$\xi \times 1 \, \mathrm{GeV}$; and 15 bins of 
$\xi \times 2 \, \mathrm{GeV}$ from the lowest to the highest energy,
where $\xi = (E_{\mu}-1)/49$ is an overall scale factor ($\xi=1$
corresponding to the `standard' 50~GeV Neutrino Factory). 
The fast oscillations that arise at low energies and that can lead to
`aliasing' effects are averaged, at the probability level, over a
width of  150~MeV \cite{Huber:2004ka,Huber:2007ji} for muon energies up to 100 GeV
and baselines up to 9000 Km.
This procedure has been tested, i.e. it has been verified that the
$\chi^2$-values do not change if a finer binning is chosen or if a
different averaging procedure at the probability level is used.

The different procedures that have been adopted in the literature to
account for the energy response of the detector could make, in
principle, a significant difference in the results.
This is especially true at the Neutrino Factory, where the parametric
degeneracies play such a big role. 
Indeed, some of the degeneracies are energy dependent and can be
solved if the detector considered has good energy resolution.
For this reason the results obtained with the various binning
procedures and efficiencies used in references \cite{Cervera:2000kp}
and \cite{Huber:2006wb} have been compared using GLoBES
\cite{Huber:2004ka,Huber:2007ji}.
A point at $\sin^22\theta_{13}=10^{-3}\,\, [\theta_{13} = 1^\circ]$
was chosen for this comparison since at these intermediate values the
impact of degeneracies is, in general, largest.  
For the true solution the results obtained using the different
procedures could not be distinguished.
In contrast, for the intrinsic degeneracy a $\sim30\%$ difference in
the $\Delta\chi^2$ was observed.
Such a difference, however, would also arise in other modification of
the setup, such as, for example, the assumption of a different
low-energy muon threshold or efficiency.
The degeneracy is, nonetheless, always present at the 5$\sigma$
level.
Therefore, at the current level of accuracy in the detector
simulation, there is no qualitative difference.
In the context of improved detector simulations that will become
available, however, it will be extremely important to describe
accurately the detector response.

The results shown in below will be based on the treatment of
references \cite{Huber:2002mx,Huber:2006wb}, unless otherwise stated. 
For the wrong-sign muon signal, flat efficiencies of 0.45 (neutrinos)
and 0.35 (anti-neutrinos) for energies in the range 
$E_\nu \in [20,50]$~GeV are assumed.  
A linear rise of the efficiencies from the lower threshold (between 0
at $4$~GeV) to their final value at $20$~GeV is assumed. 
The energy resolution is treated as described above.  
The relatively high neutrino-energy threshold is the result of
optimising for the purest possible sample of wrong-sign muons, thus
selecting events with the highest possible energy. 
Indeed, the lower the muon energy, the higher the likelihood to
mis-identify the muon charge or the nature of the event (charged
current versus neutral current) becomes.  
As the average muon energy will decrease with the neutrino energy, we
model the total, fractional background with the function 
$\beta E_\nu^{-2}$, where $\beta =10^{-3}$. 
Integrating from 4~GeV to 50~GeV, we fix the weight factor, $\beta$,
by matching roughly the total fractional background obtained in
reference \cite{Cervera:2000vy}.  
The fractional backgrounds considered in the following are:
$5\times10^{-6}$ of the neutral current events; and 
$5\times10^{-6}$ of the right-sign muon events.

It must be noted that this detector can also be used to look for
$\nu_\mu \to \nu_\mu$ dis-appearance, providing a very good measurement
of the atmospheric parameters $\theta_{23}$ and $\Delta m^2_{31}$ and
giving some handle on the `octant degeneracy'.
For the right-sign muon sample there is no need to determine
accurately the charge of the muon, since wrong-signs muons 
constitute only a negligible fraction of the sample. 
Therefore, the efficiencies and thresholds reported by MINOS
\cite{Ables:1995wq}, and a signal efficiency of 0.9 starting at 1~GeV
are used. 
The backgrounds in this case are $10^{-5}$ of all neutral-current
events and all wrong sign muon events. 
The latter are added directly to the signal. 

For both channels we use a 2.5\% systematic error on the signal and
a 20\% systematic error on the background normalisation.
A different magnetised-iron calorimeter detector for a Neutrino Factory 
experiment has been described in reference \cite{Indumathi:2004pn}.

\paragraph{Emulsion Cloud Chamber (ECC): the `silver channel'}
\label{sec:ecc}

\noindent
To soften the parametric degeneracy problem, it has been proposed to
take advantage of the `silver channel', i.e. $\nu_e \to \nu_\tau$
oscillations \cite{Donini:2002rm}.
This is a unique feature of the Neutrino Factory, where the average
neutrino energy is high enough to produce $\tau$ CC events; not even
the highest $\gamma$ beta-beam discussed above can look for this
signal.
The signal can be tagged looking for wrong-sign muons in coincidence
with a $\tau$-decay vertex, to distinguish them from golden channel
wrong-sign muons.
Therefore, a detector with muon-charge identification and vertex
reconstruction is needed. 
Two technologies have been considered in the literature: liquid-argon
detectors \cite{Bueno:2000fg} and emulsion-cloud-chamber (ECC)
techniques. 
The latter has been extensively studied for the OPERA detector that is
under construction at the Gran Sasso laboratory, and a dedicated
analysis to use this technique at the Neutrino Factory has been
published in reference \cite{Autiero:2003fu}. 
In \cite{Autiero:2003fu}, a 5~Kton ECC was considered and a
detailed study of the main sources of background performed. 
In the following, the ECC will be considered as the standard detector
to study the silver channel. 

The various backgrounds to the silver-channel signal are presented
in table \ref{tab:silverbckg}. 
The ECC detector is assumed to have a fiducial mass of 5~Kton as in
reference \cite{Autiero:2003fu}.  
In addition, an overall signal efficiency of approximately 10\%,
chosen to reproduce the signal-event numbers from table 4 in reference
\cite{Autiero:2003fu}, is assumed.
The background rejection factors are taken from reference
\cite{Autiero:2003fu} as well and are summarised in table
\ref{tab:silverbckg}.
\begin{table}
  \begin{center}
    \begin{tabular}{lr} \hline
      Background source & Rejection factor \\ \hline
      Neutrino induced charm production & $10^{-8} \, \times \, (N_{CC}(\nu_e)+N_{CC}(\nu_\mu))$ \\ 
      Anti-neutrino induced charm production & $3.7 \cdot 10^{-6} \, \times \, N_{CC}(\bar{\nu}_\mu)$ \\ 
      $\tau^+\rightarrow\mu^+$ decays & $10^{-3} \, \times \, N_{CC}(\bar{\nu}_\tau)$ \\ 
      $\mu$ matched to hadron track & $7 \cdot 10^{-9} \, \times \, N_{CC}(\bar{\nu}_\mu)$ \\ 
      Decay-in-flight and punch-trough hadrons & $6.97 \cdot 10^{-7} \, \times N_{NC}\, +$ \\
      & $+ \, 2.1 \cdot 10^{-8} \, \times \, N_{CC}(\nu_e)$\\ 
      Large-angle muon scattering & $10^{-8} \, \times \, N_{CC}(\nu_\mu)$ \\ \hline
    \end{tabular}
  \end{center}
  \caption{
    The background sources and rejection factors for the silver
    channel measurement in the $\mu^+$-stored phase. 
    From reference \cite{Autiero:2003fu}.
  }
  \label{tab:silverbckg}
\end{table}

The energy resolution is assumed to be $20\% \times E$, implemented as
in \cite{Huber:2006wb}.
It is further assumed that silver-channel data-taking only occurs
when $\mu^+$ are stored (running with $\mu^-$ will produce very few
silver events, due to the $\bar \nu_\tau N$ cross-section
suppression).
A 15\% systematic uncertainty on the signal and a 20\% systematic
uncertainty on the background normalisation are assumed.

Notice that this detector can also be used to look for 
$\nu_\mu \to \nu_\tau$ appearance, i.e. precisely the purpose for
which it is being built in the framework of the CNGS experiment. 
This channel can be useful to measure the atmospheric parameters, as
well as the $\nu_\mu$ dis-appearance channel discussed above. 
Other possible $\tau$ decay channels, such as decay into electrons or
into hadrons, have not been considered as they would need a dedicated 
analysis and a totally different detector.

\paragraph{Liquid Argon Detector (LAr): the `platinum' channel}
\label{sec:lar}

\noindent
In addition to the channels discussed above, 
$\nu_{\mu}(\bar{\nu}_{\mu})\rightarrow\nu_e(\bar{\nu}_e)$ oscillations
can be also observed at a Neutrino Factory. 
This channel, the `platinum channel' (since the observation of a small
number of events can be extremely valuable) is the T-conjugate of the
golden channel. 
It is also its CP-conjugate, albeit with different matter effects.  
Combined with the golden channel, the platinum channel will help to
resolve many of the correlations and degeneracies.

\begin{figure}
  \begin{center}
    \epsfig{file=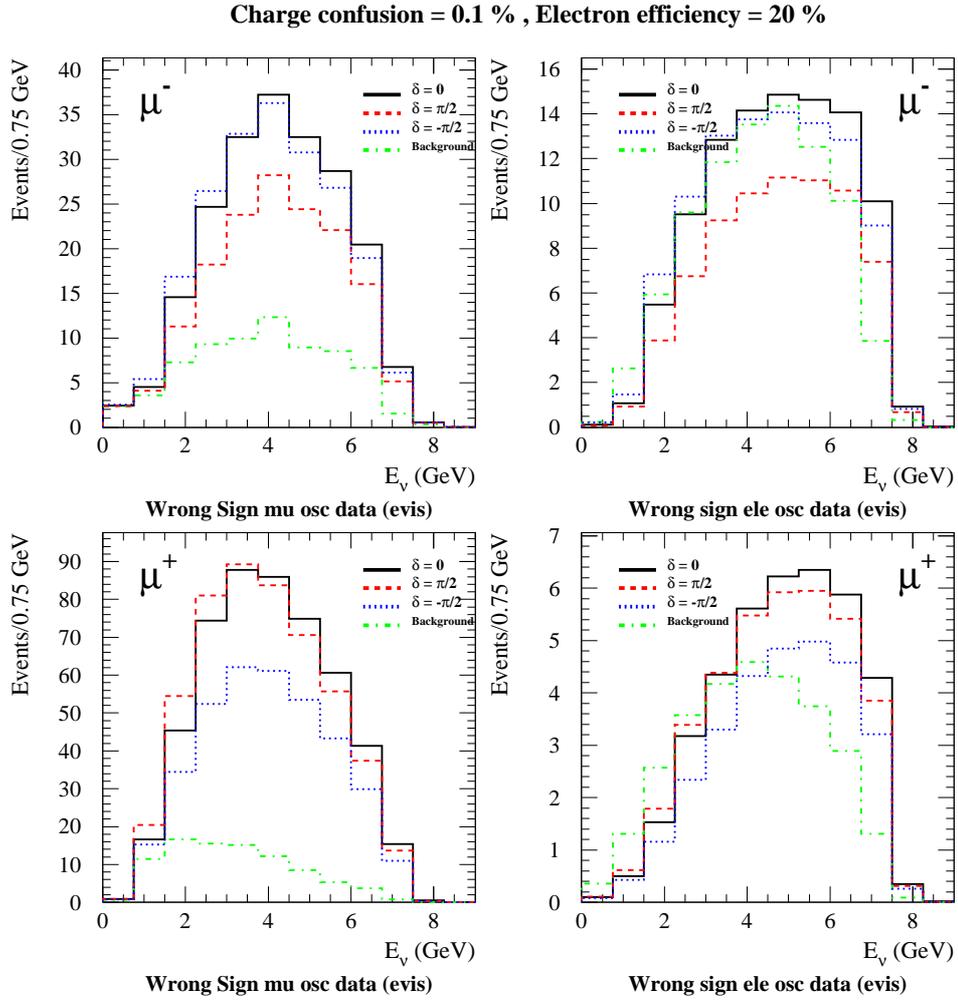,width=14cm}
  \end{center}
  \caption{
    Visible energy distribution for wrong-sign muons
    (left panel) and wrong-sign electrons (right panel) normalised to
    $10^{21}$ muon decays.  
    The electron efficiency $\epsilon_e$ is assumed to be 20\% and
    charge confusion probability is set to 0.1\%. 
    Three sets of curves are represented, corresponding to
    $\delta=+\pi/2$ (dashed line), $\delta=0$ (full line) and
    $\delta=-\pi/2$ (dotted line).
    The background contribution from tau decays is also shown.
    The other oscillation parameters are $\Delta m^2_{32}=3\times
    10^{-3}\ \rm eV^2$, $\Delta m^2_{21}=1\times 10^{-4}\ \rm eV^2$,
    $\sin^2 \theta_{23} = 0.5$, $\sin^2 \theta_{12} = 0.5$ and 
    $\sin^2 2\theta_{13} = 0.05$. 
    Taken with kind permission of Nuclear Physics B from figure 18
    in reference \cite{Bueno:2001jd}.
    Copyrighted by Elsevier Science B.V.
  }
  \label{fig:effargon}
\end{figure}

To take advantage of this channel, a detector that can identify the
charge of the electrons (to reduce the dominant background from
non-oscillated $\nu_e$) is required. 
Electron charge identification has so far only been studied for a
magnetised liquid-argon TPC \cite{Rubbia:2001pk}. 
A lower-energy detection threshold of 0.5~GeV was applied.
In reference \cite{Rubbia:2001pk} it was pointed out that electrons
and positrons of higher energy tend to shower early, which means
that the track is short and the curvature is hardly measurable.
Therefore, there may be an upper energy threshold above which it is
no longer possible to measure the electron charge.
The efficiency and the dominant backgrounds to the platinum channel in
a liquid-argon detector are shown in figure \ref{fig:effargon}.

In reference \cite{Huber:2006wb}, the $\nu_e$-appearance performance of the
MINOS detector (which has been estimated in reference \cite{NUMI714})
was adopted. 
An extra background corresponding to 1\% of the non-oscillated $\nu_e$
was added to take into account the difference between the Neutrino
Factory beam and the NuMI beam. 
The fractional background of this analysis is in agreement with that
of reference \cite{Rubbia:2001pk}.  

The `standard' platinum-channel detector is assumed to be liquid-argon
TPC with a fiducial mass of 15 Kton. 
The signal efficiency, which is assumed to be energy independent, is
taken to be 20\% \cite{Rubbia:2001pk}, and the background-rejection
factors are summarised in table \ref{tab:platinumbckg}.  
Furthermore, the energy resolution is assumed to be 
$15 \% \times E_\nu$. 
The upper threshold for the electron/positron-charge identification
(CID) is assumed to be 7.5~GeV. 
The CID background is assumed to be $1\%$ \cite{Rubbia:2001pk} and the
other backgrounds are taken from~reference \cite{NUMI714}.

In some of the figures of sections \ref{sec:clones} and \ref{sec:best},
however, the impact of an `improved' platinum-channel detector is
discussed. 
The `improved' detector has the following characteristics: a 50~Kton
fiducial mass; an energy-independent signal efficiency of 40\%;
background-rejection factors as given in reference \cite{NUMI714}
extrapolated to higher energy; and CID background as for the standard
setup. 
The CID upper energy threshold is varied continuously from 7.5~GeV to
50~GeV.
The performance of this improved detector are labeled in the figures
`platinum$^*$'.

Both for platinum and platinum$^*$ detectors a 2.5\% systematic error
on the signal and a 10\% systematic error on the background
normalisation have been assumed.
\begin{table}
  \begin{center}
    \begin{tabular}{lr} \hline
      Background source & Rejection factor \\ \hline 
      Muon dis-appearance  & $10^{-3} \, \times \, N_{CC}(\nu_\mu) \, (N_{CC}(\bar{\nu}_\mu))$ \\ 
      Tau appearance & $5 \cdot 10^{-2} \, \times \, N_{CC}(\nu_\tau) \, (N_{CC}(\bar{\nu}_\tau))$ \\  
      Neutral current reactions & $10^{-2} \, \times \, N_{NC}$ \\ 
      Wrong sign electron/positron & $10^{-2} \, \times \, N_{CC}(\bar{\nu}_e) \, (N_{CC}(\nu_e))$ \\ \hline
    \end{tabular}
  \end{center}
  \caption{
    The background sources and rejection factors for the platinum
    channel measurement for the $\mu^-$-stored phase, while the
    brackets refer to the $\mu^+$-stored phase.
    The numbers, besides the background from electron/positron CID,
    are taken from reference \cite{NUMI714}.
  }
  \label{tab:platinumbckg} 
\end{table}

%% file: 05-Performance/05-04-Neutrino-Factory/05-04-02-Neutrino-Factory-golden.tex
\subsubsection{Physics potential of the golden channel}
\label{sec:golden}

In this section, the physics potential of the standard golden-channel
detector is presented.
The $\nu_\mu \to \nu_\mu$ dis-appearance channel will be included in
this section as well.
Through this channel, an independent measurement of the atmospheric
parameters is possible. 
This serves to reduce significantly the impact of the uncertainties 
induced by uncertainties in these parameters in the
($\theta_{13},\delta$) measurement \cite{Huber:2002mx,Donini:2005rn}.

Results will be presented following the definitions given in section
\ref{sec:pic}.
Most of the figures in this section are taken from reference
\cite{Huber:2006wb}, where the following input (or `true') values were
used:
\begin{eqnarray}
  \Delta m^2_{31}   &= 2.2^{+1.1}_{-0.8}\cdot10^{-3}\,\mathrm{eV}^2 \; ; \quad\sin^2\theta_{23}&=0.5^{+0.18}_{-0.16} \; ; \nonumber \\
  \Delta m^2_{21}   &= 8.1^{+1.0}_{-0.9}\cdot10^{-5}\,\mathrm{eV}^2 \; ; \quad\sin^2\theta_{12}&=0.3^{+0.08}_{-0.07} \; ; \\
  \sin^2\theta_{13} &= 0^{+0.047}_{-0}                              \; ; \quad \delta        &=0^{+\pi}_{-\pi} \; . \nonumber
  \label{equ:params}
\end{eqnarray}
The ranges represent the current 3$\sigma$ allowed ranges (from
reference \cite{Maltoni:2004ei} (see also references
\cite{Fogli:2003th,Bahcall:2004ut,Bandyopadhyay:2004da}), both choices
of $\mathrm{sgn}(\ldm)$ are allowed. 
A 5\% additional uncertainty on $\sdm$ and $\theta_{12}$ from solar
experiments at the time that data from the Neutrino Factory becomes
available is assumed \cite{Bahcall:2004ut}. 
Matter-density uncertainties at the level of 5\%, uncorrelated between
different baselines, have been included
\cite{Geller:2001ix,Ohlsson:2003ip}.
Whenever discussing the  octant degeneracy, the `true value' of the
atmospheric angle has been fixed to $\theta_{23} = 0.44$ (or $0.56$)
\cite{Fogli:2005cq}. 

The precision with which many of the S$\nu$M observables can be
measured strongly depend on the true values of $\stheta$ and
$\delta$.
Hence, the results are presented in terms of two-dimensional plots in
the $(\theta_{13}, \delta)$ plane.   
Each point in the plot corresponds to a different input
($\theta_{13},\delta$) pair. 
Notice that, in all plots, the $\Delta \chi^2$ is marginalised over
the external atmospheric and  solar parameters, as well as over the
matter density, to take into account fully the correlations among the
various parameters.   

The requirements for the optimisation of the standard Neutrino Factory
are summarised in table \ref{tab:requirements}. 
There are two very important results. 
No baseline performs optimally for all the observables considered, a
`shorter' baseline $L \sim 3 \, 000 - 5 \, 000$~km is needed to
provide good sensitivity to CP-violation and for the precise
determination of the leading atmospheric parameters; a longer baseline, 
$L \simeq 7 \, 500$~km, is required to give optimal sensitivity to
$\stheta$, the mass hierarchy, and for the disentanglement of
degeneracies in the CP-violation measurements. 
For the muon energies, we find that $E_{\mu} \gtrsim 20$~GeV is
sufficient for most applications, and $E_{\mu} \sim 40$~GeV should be
on the safe side.
Therefore, we find that the main challenge for a Neutrino Factory will
be the baseline, which can affect the physics potential much more than
a muon energy lower than previously assumed.
\begin{table}[t]
  \begin{center}
    \begin{tabular}{lrrr}
      \hline
      Performance indicator & $L$ [km] & $E_{\mu}$ [GeV] \\
      \hline
      {\bf Three-flavour effects:} \\
      $\stheta$ sensitivity & $\sim 7 \, 500 $ (``magic baseline'') & 20-50 \\
      Mass hierarchy sensitivity & $\gtrsim 6 \, 000 $ & 20-50 \\
      Max. CP violation sensitivity & $\sim 3 \, 000 - 5 \, 000$ & $>$ 30 \\
      \hline
      {\bf Leading atmospheric parameters:} & \\
      $\ldm$ precision & $ \gtrsim  3 \, 000$ & $\gtrsim 40$\\
      Deviation from maximal mixing ($\theta_{23}$) & $\gtrsim 3 \, 500 + 50 \cdot E_{\mu} /\mathrm{GeV}$ & $\gtrsim 20$ \\
      \hline
      {\bf Optimisation for large $\boldsymbol{\stheta}$:} \\
      Mass hierarchy sensitivity & $> 1 \, 000$ & $>10$\\
      CP violation sensitivity ($\Delta \rho = 1\% \, \bar{\rho}$)  &  $\sim 1 \, 500 - 5 \, 500$ & 20-50  \\
      CP violation sensitivity ($\Delta \rho = 5\% \, \bar{\rho}$) &  $\sim 1 \, 500 - 2 \, 000$ & 20-50  \\
      & $\sim 4 \, 500 - 5 \, 500$ & 20-40  \\
      \hline
    \end{tabular}
  \end{center}
  \caption{
     Requirements for the near-optimal performance of our `standard
     Neutrino Factory' (one individual experiment) for 
     $\ldm = 0.0022$~{eV}$^2$ for different performance indicators.
  }
  \label{tab:requirements} 
\end{table}

\paragraph{$\theta_{13}$-sensitivity}

\begin{figure}
  \begin{center}
    \begin{tabular}{cc}
      \hspace{-1cm} \raisebox{2mm}{\epsfxsize8.25cm\epsffile{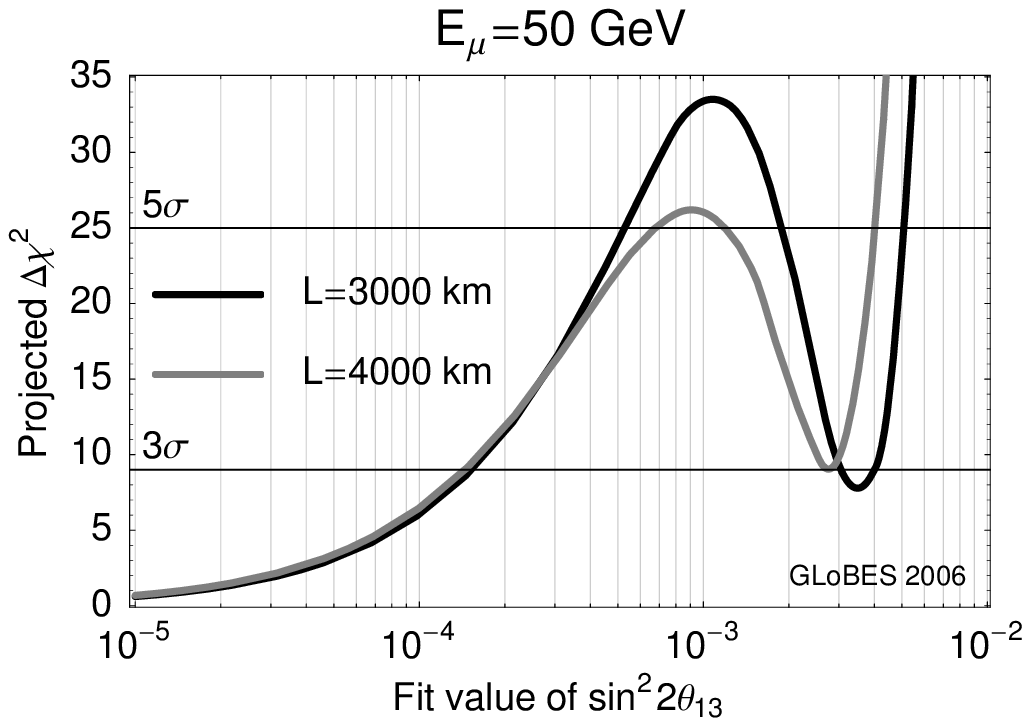}} & 
      \hspace{-0.5cm} \epsfxsize6.25cm\epsffile{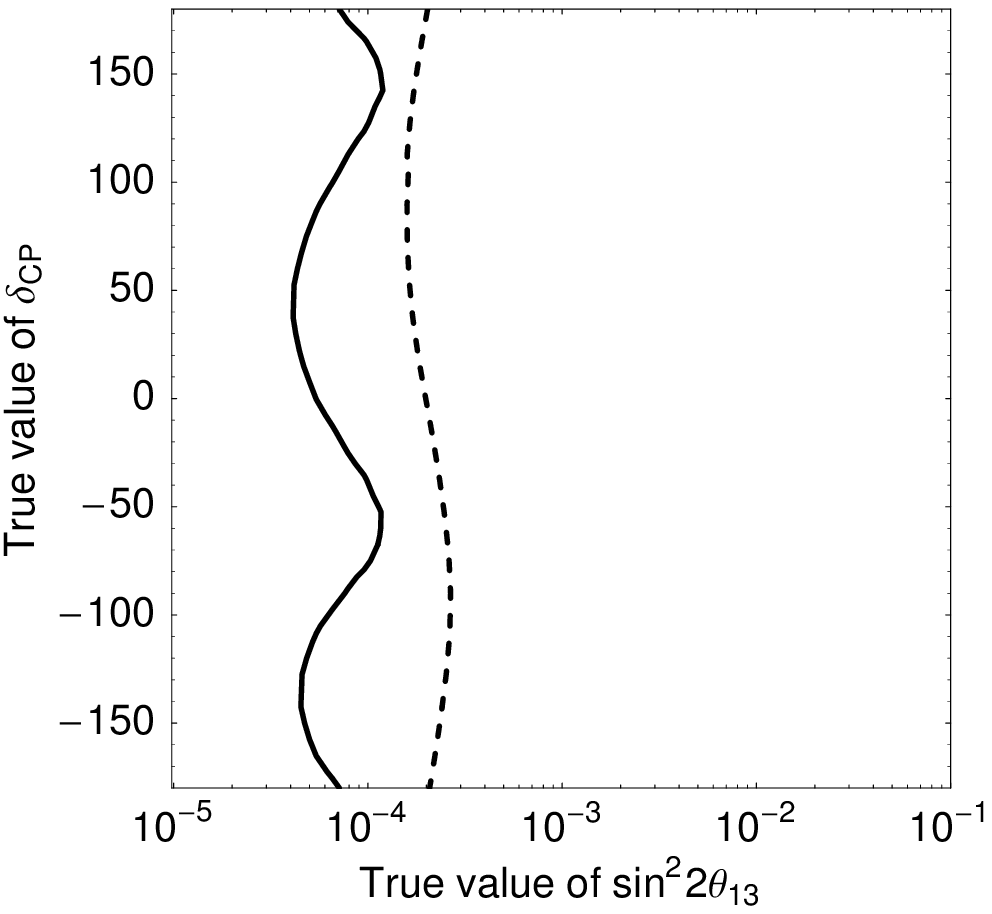} \
    \end{tabular}
  \end{center}
  \caption{
    Left panel:
    Projected $\Delta \chi^2$ for the $\stheta$
    sensitivity as a function of the fit value of $\stheta$ for 
    $E_\mu = 50$~GeV and two different baselines as given in the plot
    legend (includes degeneracies).
    Right panel: 3$\sigma$ $\theta_{13}$-discovery; solid lines refer
    to the $L = 4000$~km Neutrino Factory; dashed lines to the 
    $L = 7500$~km Neutrino Factory.
    Left panel taken with kind permission of the Physical Review from 
    figure 4 in reference \cite{Huber:2006wb}.
    Copyrighted by the American Physical Society.
  }
  \label{fig:sensi}
\end{figure}

\noindent
The $\Delta \chi^2$ function, marginalised over all parameters other
than $\theta_{13}$, for a fit to data from the golden-channel
detector under the conditions described above is shown in figure
\ref{fig:sensi}.
The left panel of figure \ref{fig:sensi} shows that the $\Delta \chi^2$
function has two minima, the first corresponding to the input value
$\theta_{13} = 0$ and the second for 
$\sin^2 2 \theta_{13} \ge 10^{-3}$ (the intrinsic degeneracy).  
If there is no signal (hypothesis $\stheta=0$), the degeneracy will
worsen the $\stheta$ sensitivity, since a fake solution with a
relatively large $\stheta$ will still be consistent with $\stheta=0$.
Therefore, it is not possible to exclude rather large $\stheta$
values.
It must be stressed that results at 3$\sigma$ are strongly dependent
on small changes in the luminosities, the external parameters or the
setup.  
It can be seen in the figure how, for $L = 4000$~km, the 
$\Delta \chi^2$ at the second minimum increases and, at 3$\sigma$, the
$\theta_{13}$-sensitivity improves by one order of magnitude with
respect to the case of $L = 3000$~km.
However, at 5$\sigma$ the degeneracy is still present for both
baselines: these two cases will therefore be interpreted as
qualitatively similar, as in fact they are.
In the right panel of figure \ref{fig:sensi} the $\stheta$ discovery
potential for $L=4000$~km and $L=7500$~km at $3\,\sigma$ is shown.
Although the performance is slightly worse for the  longer baseline,
the $\delta$-dependence is much weaker.

\begin{figure}
  \begin{center}
    \includegraphics[width=0.9\textwidth]{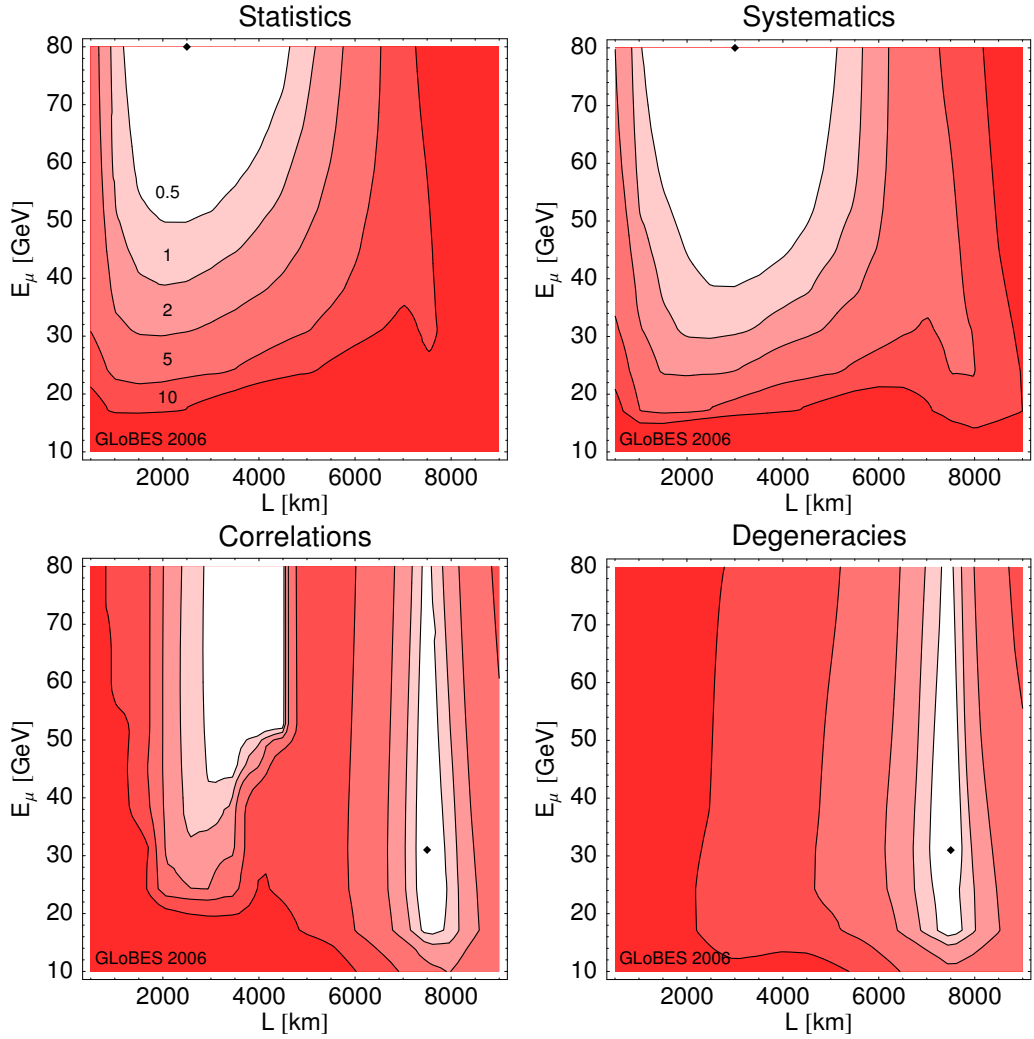}
  \end{center}
  \caption{
    Sensitivity to $\stheta$ ($5 \sigma$) relative to the optimum
    (white) within each plot.   
    The different panels correspond to successively taking into
    account statistics, systematics, correlations, and degeneracies.
    The different contours represent the region within a factor of
    0.5, 1, 2, 5, and 10 above the maximal sensitivity in each plot.
    The maximal sensitivities are $\stheta < 1.4 \cdot 10^{-5}$
    (statistics), $2.8 \cdot 10^{-5}$ (systematics), 
    $2.4 \cdot 10^{-4}$ (correlations), and $5.0 \cdot 10^{-4}$
    (degeneracies), obtained at the energies and baselines marked by
    the diamonds.
    Taken with kind permission of the Physical Review from 
    figure 3 in reference \cite{Huber:2006wb}.
    Copyrighted by the American Physical Society.
  }
  \label{fig:t13sens} 
\end{figure}

Figure \ref{fig:t13sens} shows the $\stheta$ sensitivity at $5 \sigma$
as a function of the baseline $L$ and the parent muon energy $E_\mu$. 
The different panels correspond to taking into account, successively,
statistical uncertainties, systematic uncertainties, correlations, and
degeneracies.
The different contours represent the region within a factor of 0.5, 1,
2, 5, and 10 above the maximum sensitivity in each plot.
The maximum sensitivity (obtained for the energies and baselines
marked by the diamonds) are: $\stheta < 1.4 \cdot 10^{-5}$
(statistics), $2.8 \cdot 10^{-5}$ (systematics), $2.4 \cdot 10^{-4}$
(correlations), and $5.0 \cdot 10^{-4}$ (degeneracies), respectively.

When statistical and systematic uncertainties only are considered
(i.e., $\delta$ is fixed to the value for which we get maximum
sensitivity), figure \ref{fig:t13sens}(upper row), baselines from 
$1\, 000$ to $ 4 \, 000$~km with as much muon energy as possible give
the best sensitivity.
However, when correlations and degeneracies are taken into account,
the benefit of the `magic baseline' \cite{Huber:2003ak} becomes more
apparent.
At the magic baseline all dependence on $\delta$ cancels and many
of the degeneracies disappear `by magic', thus improving the $\stheta$
sensitivity.
This happens for $V = \sqrt{2} G_F n_e = 2 \pi / L$, or, in terms of
the constant matter density $\rho$, for approximately two nucleons per
electron, equivalent to:  
\begin{equation}
  L_{\mathrm{magic}} \, [\mathrm{km}]  \simeq 32 \, 726 \, \, 
    \frac{1}{\rho \, [\mathrm{g/cm^3}]} \, .
\end{equation}
Numerically, it can be shown to be closer to 
$L_{\mathrm{magic}} \sim 7 \, 250$~km for a realistic PREM 
\cite{Dziewonski:1981xy} profile by
minimising the $\delta$-dependence in the appearance
rates.
At this distance, the optimal muon energy need not to be higher than 
40~GeV (or even 30~GeV).
The reason for this is that the $\stheta$ term in the appearance
probability does not drop as a function of the baseline at the mantle
resonance energy.
Therefore, matter effects prefer lower energies, whereas higher muon
energies imply higher event rates and a relative decrease of events at
the mantle resonance.
The optimum is determined by a balance between these two factors.
The magic baseline has two obvious drawbacks: the event rate is
reduced by the large distance; and it does not allow for a CP
measurement.

\paragraph{CP-discovery potential}

\begin{figure}
  \begin{center}
    \begin{tabular}{cc}
      \hspace{-0.5cm} \epsfxsize6.25cm\epsffile{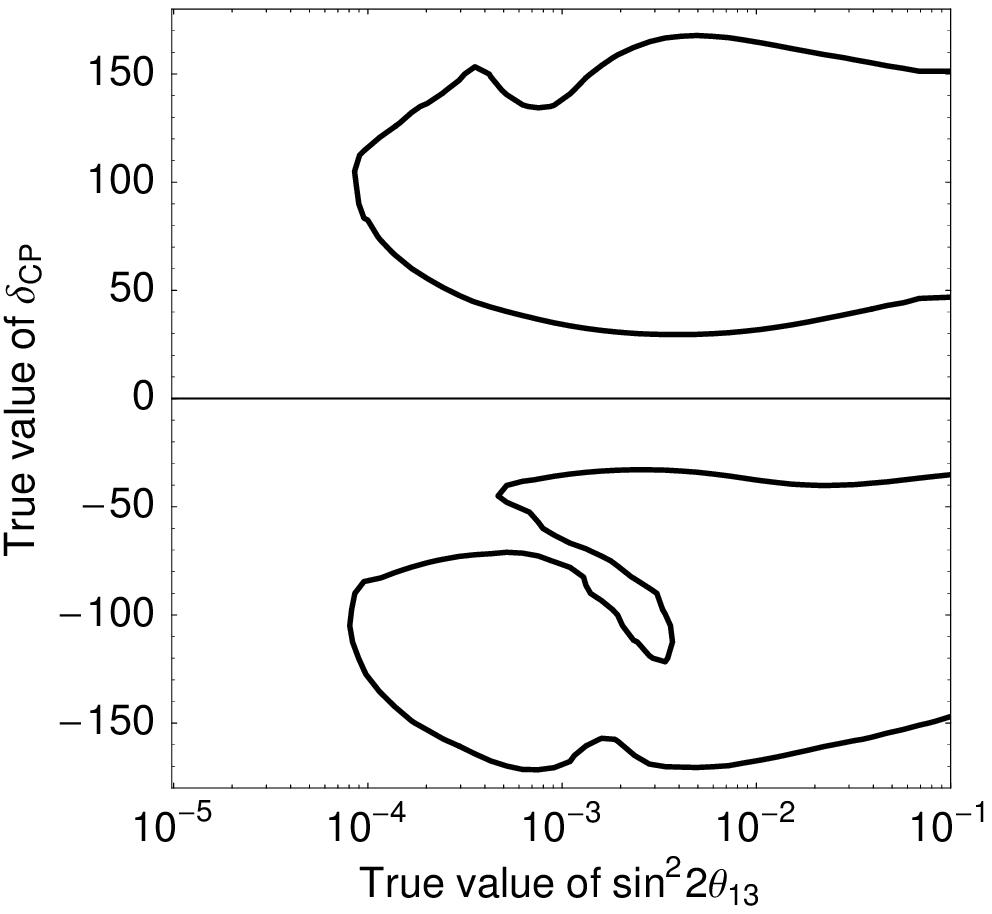} &
      \hspace{-1cm} \raisebox{1mm}{\epsfxsize8.25cm\epsffile{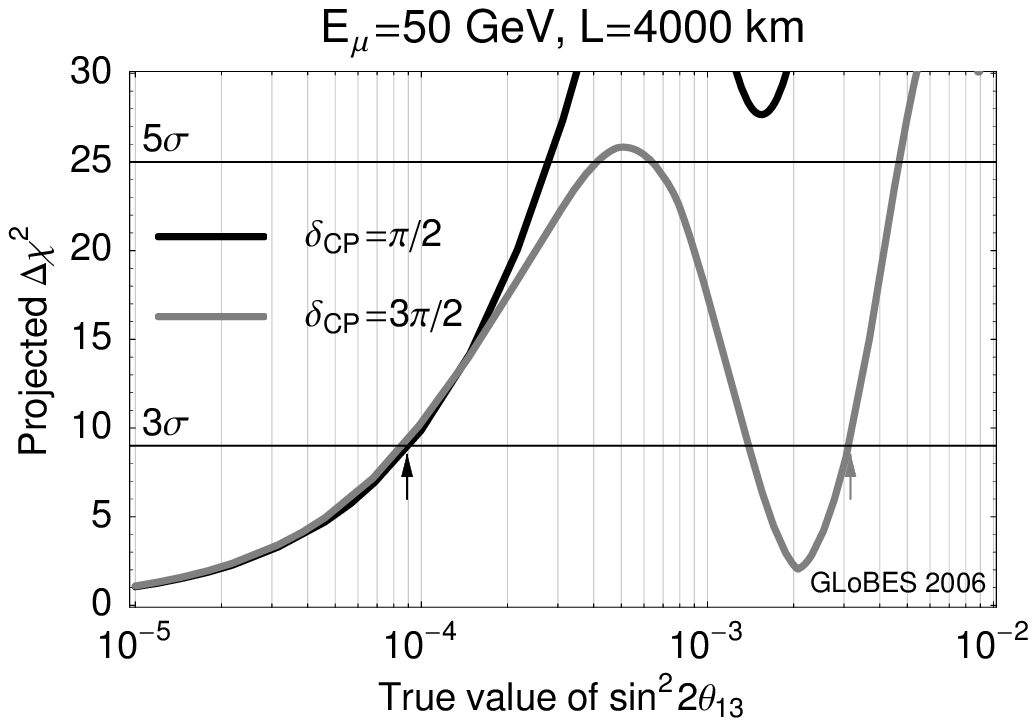}} \
    \end{tabular}
  \end{center}
  \caption{
    Left panel: 3$\sigma$ CP-discovery potential for the 50~GeV
    Neutrino Factory at $L = 4000$ km. 
    Right panel (Figure taken from reference \cite{Huber:2006wb}):
    Projected $\Delta \chi^2$ for the wrong choice of the
    hierarchy, computed for maximal CP-violation, $\delta=\pi/2$ and
    $\delta=3 \pi/2$, as a function of the true value of $\stheta$. 
    The arrow represents the smallest $\stheta$ for $\delta=\pi/2$
    (grey arrow) and $\delta=3 \pi/2$ (black arrow) above which
    CP-violation can be found for any value of $\stheta$.
    Right panel taken with kind permission of the Physical Review from 
    figure 4 in reference \cite{Huber:2006wb}.
    Copyrighted by the American Physical Society.
  }
  \label{fig:cpdid}
\end{figure}

\noindent
Figure \ref{fig:cpdid} (left panel) shows the CP-discovery potential
for the standard Neutrino Factory defined above for a baseline of $L =
4000$~km.  
No CP-discovery potential has been evaluated for the Neutrino Factory
and a baseline of 7000 km; due to matter effects and the choice of the
baseline (close to the magic baseline), the sensitivity to
$\delta$ vanishes.  
The Neutrino Factory with a baseline of 4000~km is not as good as one
would expect from its $\theta_{13}$-sensitivity.
This may be explained as follows: as a general rule, for small values of
$\theta_{13}$ the degeneracies flow toward $\delta = 0^\circ$ and
$|\delta| = 180^\circ$ (see references \cite{Donini:2003vz} and
\cite{Burguet-Castell:2002qx}), thus mimicking a non-CP violating
phase.
Especially problematic is the case where the data is fitted with
inverted mass hierarchy, in this case it is possible to fit the data
with $\delta=\pi$ for intermediate true values of
$\stheta\sim10^{-3}\,\,[1^\circ]$, the so called
$\pi$-transit \cite{Huber:2002mx}. 
Due to a `parametric conspiracy' between the chosen energy and
baseline and the matter effects, at the Neutrino Factory the typical 
value of $\theta_{13}$ for which this happens is much larger than at
the SPL and or at T2HK.  
Therefore, although from the statistical point of view the Neutrino
Factory would certainly out-perform both the SPL and T2HK, in practice
for small values of $\theta_{13}$ a CP-violating phase will be
difficult to distinguish from a non--CP-violating one, if the sign-
and octant-degeneracies are not solved.

Figure \ref{fig:cpdid} (right panel) shows the $\Delta \chi^2$ for
the wrong choice of the mass hierarchy, computed for maximal
CP-violation (i.e. true $\delta/2 = \pi$ or $3 \pi/2$), as a function of
the true $\theta_{13}$.
The $\Delta \chi^2$ is marginalised over all parameters other then
$\delta$ and computed for fitted $\delta = 0$ or $\pi$. 
Sensitivity to maximal CP-violation is then represented (for a fixed
$L$ and $E_\mu$) by the region of true $\theta_{13}$ for which 
$\Delta \chi^2$ is bigger than a given (1 dof) confidence level. 
For $\theta_{13} \to 0$, it becomes more and more difficult to
distinguish CP-violation from CP-conservation. 
However, it can be seen that for $\delta=3 \pi/2$ a second minimum
appears both at 3$\sigma$ and at 5$\sigma$ for larger $\theta_{13}$,
not present for $\delta = \pi/2$. 
This is the $\pi$-transit that was noted before. 
If the mass hierarchy is unknown, no sensitivity to maximal
CP-violation is possible if $\stheta$ lies in this region.

The largest (rightmost) $\stheta$ value for which $\Delta \chi^2 \geq 9$
(or 25) represents the smallest $\stheta$ for which it is possible
unambiguously to observe maximal CP-violation, although the
sensitivity may be restored at lower values of $\stheta$. 
This value of $\stheta$ is labeled by an arrow in the figure.
Conservatively, this value is taken as the benchmark for the
$(L,E_\mu)$ optimisation.
Figures will be presented at 3$\sigma$ only since the results do not
depend on the chosen confidence level.
 
\begin{figure}
  \begin{center}
    \includegraphics[width=0.9\textwidth]{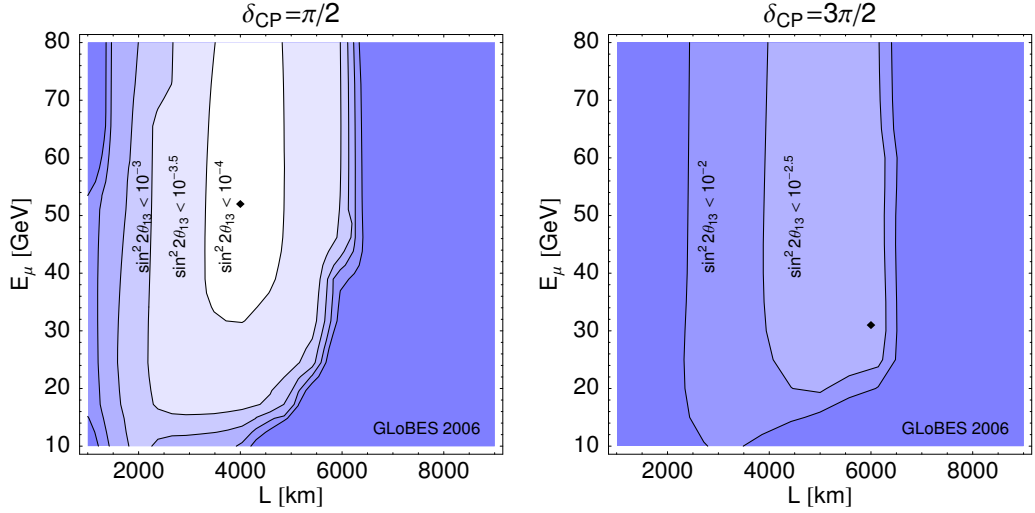}
  \end{center}
  \caption{
    Sensitivity to maximal CP-violation ($\delta=\pi/2$ or 
    $3 \pi/2$) for a normal ``true'' mass hierarchy as a function of
    $L$ and $E_\mu$. 
    The sensitivity is given as maximal reach in $\stheta$ at the 
    $3 \sigma$ 1 dof CL including correlations and degeneracies.
    The minima, marked by the diamonds, are 
    $\stheta = 8.8 \cdot 10^{-5}$ (left panel) and $\stheta = 1.3
    \cdot 10^{-3}$ (right panel).
    Taken with kind permission of the Physical Review from figure
    5 in reference \cite{Huber:2006wb}.
    Copyrighted by the American Physical Society.
  }
  \label{fig:cpel} 
\end{figure}

Figure \ref{fig:cpel} shows the sensitivity to maximal CP violation
(as defined above) for the two different choices of $\delta$. 
For $\delta=\pi/2$, we find the optimal performance at about 
$3 \, 000 - 5 \, 000$~km for $E_{\mu} \gtrsim 30$~GeV, whereas larger
energies are not required. 
When $\delta = 3\pi/2$, the absolute $\stheta$ reach is rather poor,
once again the most conservative value of $\stheta$ above which
CP violation can be determined has been chosen. 
In this case, degeneracies affect the CP-violation performance. 
It has been demonstrated in reference \cite{Huber:2004gg} that the
magic baseline can be used to solve these degeneracies in the third
and fourth quadrants of $\delta$.
If $\delta$ turned out to be in this region, to improve the
sensitivity, a second baseline is needed to solve the sign-degeneracy,
thus alleviating the effects of the $\pi$-transit.

\paragraph{Sensitivity to the mass hierarchy}
\label{sec:hierarchy}

\noindent
The $\nu_e \to \nu_\mu$ oscillation probability in matter depends on
the sign of $\Delta m^2_{31}$.  
A change of this sign is equivalent to a CP transformation, that is,
interchanging the probability of neutrinos and anti-neutrinos.  
Thus, matter effects themselves induce a non-vanishing CP-odd
asymmetry.
The maximum sensitivity to the sign of $\Delta m^2_{31}$, using the
PREM matter-density profile, is expected at a baseline 
${\cal O}(7000)$~km. 
The asymmetries from different energy bins, however, peak at slightly
different baselines. 
Therefore, spectral information can be used to improve the measurement
of the sign of $\Delta m^2_{31}$.

\begin{figure}
  \begin{center}
    \hspace{-1cm} \epsfxsize6.25cm\epsffile{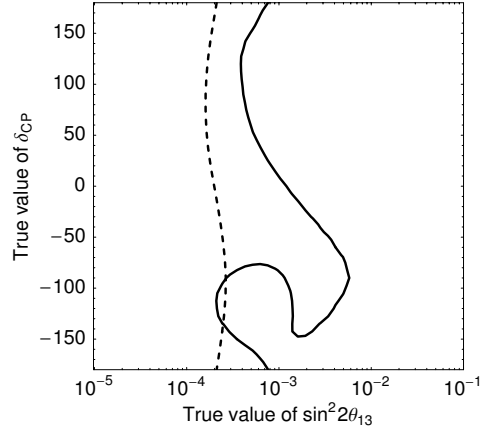} 
  \end{center}
  \caption{
    3$\sigma$ sensitivity to the sign($\Delta m^2_{31}$) for normal
    `true' mass hierarchy. 
    Solid (dashed) lines refer to the $L = 4000$~km ($7500$~km)
    Neutrino Factory.
  }  
  \label{fig:signo}
\end{figure}

The discovery potential for the normal `true' mass hierarchy is shown
at the $3\sigma$ confidence level in figure \ref{fig:signo}, evaluated
for baselines of $L=4000$~km and $L=7500$~km. 
The sensitivity of the short and the long baselines are identical for
$\delta\simeq-110^\circ$.
For this particular parameter set it is also possible to lift the
degeneracies at the short baseline. 
Compared to figure 19 of reference \cite{Donini:2005db}, 
the better sensitivity of the short baseline for 
$\delta \sim 100^\circ$ depends on the efficiency function used, see
section \ref{sec:mid}. 
For all other values of $\delta$, the longer baseline has a better
sensitivity. 

\begin{figure}
  \begin{center}
    \includegraphics[width=\textwidth]{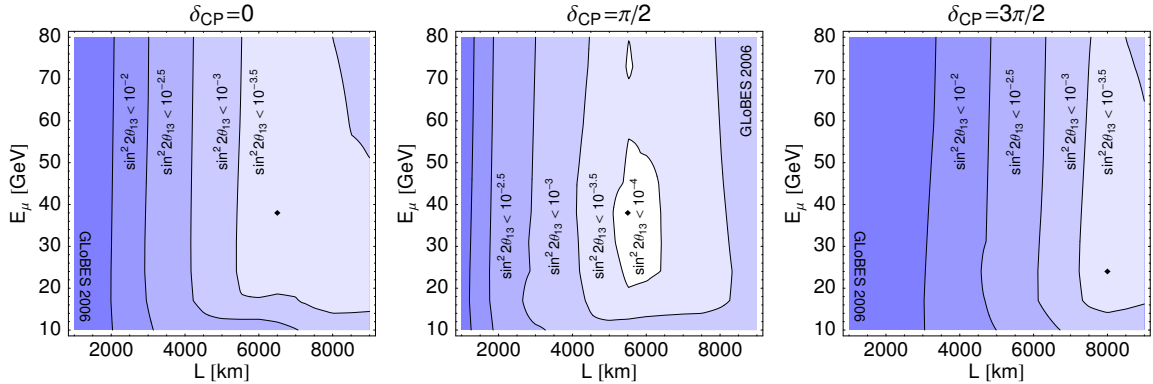}
  \end{center}
  \caption{
    Sensitivity to a normal `true' mass hierarchy for different values
    of $\delta$ (plot labels) as a function of $L$ and $E_\mu$. 
    The sensitivity is given as the maximal reach in $\stheta$ at 
    $3 \sigma$ including correlations and degeneracies. 
    The minima, marked by the diamonds, are 
    $\stheta = 1.8 \cdot 10^{-4}$ (left panel), 
    $\stheta = 6.7 \cdot 10^{-5}$ (middle panel), and 
    $\stheta = 1.6 \cdot 10^{-4}$ (right panel). 
    Taken with kind permission of the Physical Review from figure
    6 in reference \cite{Huber:2006wb}.
    Copyrighted by the American Physical Society.
  }
  \label{fig:signel} 
\end{figure}

The normal mass-hierarchy sensitivity reach in $\stheta$ as a function
of baseline and parent muon energy is shown in figure \ref{fig:signel}
for different values of $\delta$. 
The mass-hierarchy sensitivity increases with the baseline because of
the matter effects.   
This means that for very small true values of $\stheta$, a very long
baseline is required.
The muon energy is of secondary interest, as long as it is larger than
about $20 \, \mathrm{GeV}$.
In fact, for $\delta=\pi/2$ or very long baselines 
$L > 8\, 000$~km, a muon energy larger than $50$~GeV is dis-favoured
because of the matter resonance at lower energies.  
In all cases, the magic baseline $L \simeq 7\, 500$~km is near the
optimum. 
For certain values of $\delta$, there are `gaps' in the $\stheta$
axis for which no unambiguous measurement of the hierarchy is
possible, corresponding to a second minimum in $\Delta \chi^2$, as 
was the case in figure \ref{fig:cpdid} (right panel).  
In figure \ref{fig:signel}, such gaps occur for $\delta = 3\pi/2$. 
In this case, only the largest value of $\stheta$ above which
mass-hierarchy sensitivity can be achieved unambiguously is shown. 
Therefore, figure \ref{fig:signel} (right panel) actually shows the
ranges for the `gapless' determination of the mass hierarchy. 
Thus, for very long baselines $L \gtrsim 7\, 500$~km, the mass
hierarchy can be determined over the full range of $\stheta$. 
Note that, in this case, such a baseline itself allows the degeneracies
to be solved.

\paragraph{Measurement of the atmospheric parameters}

\noindent
Except for any suppressed three-flavour effects, a Neutrino Factory
will be useful for the precision measurement of the leading
atmospheric parameters $\ldm$ and $\theta_{23}$.
For simplicity, the case in which the true $\stheta=0$ is considered
in this section, because $\stheta>0$ yields complicated correlations
in the dis-appearance channel (cf., equation (33) in reference
\cite{Akhmedov:2004ny}).
Results are presented for both hierarchies as a function of 
$|\Delta m^2_{31}|$ (see section \ref{sec:pic} and reference
\cite{Donini:2005db} for a discussion of the subject).
The solution for the inverted hierarchy, depending on the definition
of the large mass-squared splitting, always differs somewhat from the
original solution. 
However, there is no qualitative difference to the best-fit solution
for $\stheta=0$.

The $\nu_\mu$ dis-appearance channel is extremely useful for the
determination of the atmospheric-neutrino parameters $\ldm$ and
$\sin^2\theta_{23}$.
An impressive accuracy can be attained, even with the standard setup.
However, a better precision can be achieved with a lower muon
identification threshold.
This can be achieved by loosening the kinematic cuts needed for a good
muon charge identification.
With no-CID, low-energy bins have a much higher efficiency, 
which in turn maximises the oscillatory signal. 
The price that must be paid is that neutrino and anti-neutrino rates
have to be added in this case, which is not a major problem for the
dis-appearance channel \cite{Freund:2000ti}.

\begin{figure}
  \begin{center}
    \includegraphics[width=\textwidth]{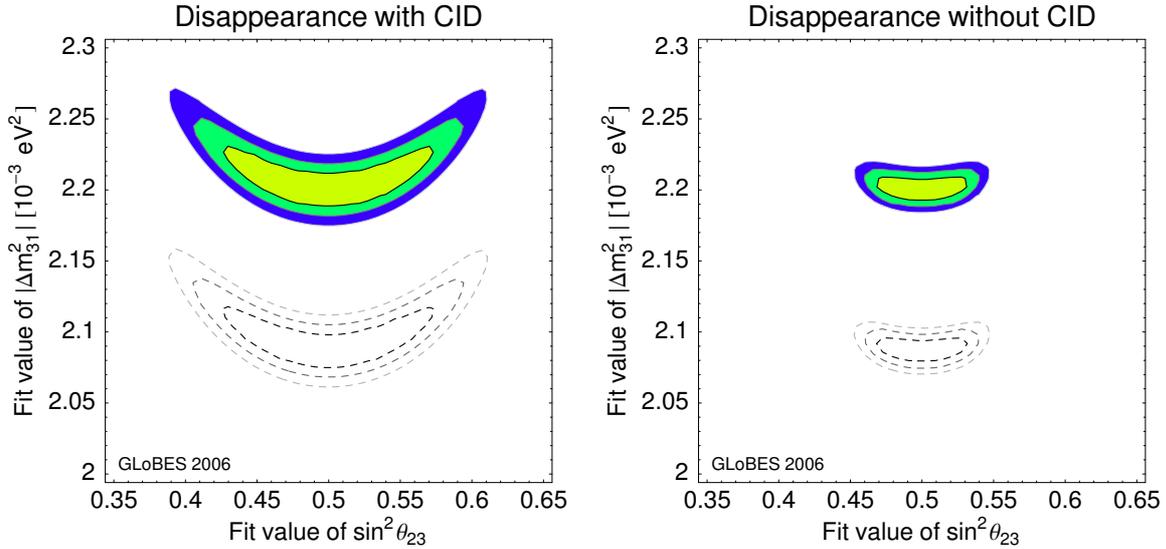}
  \end{center}
  \caption{
    Comparison of ($\ldm$-$\theta_{23}$)--precision between CID
    (left panel) and no CID (right panel) in the dis-appearance channel
    including all correlations ($1\sigma$, $2 \sigma$, $3\sigma$,
    2~d.o.f., $\stheta=0$). 
    The appearance information is added as usual with CID.
    Dashed curves correspond to the inverted hierarchy solution. 
    Taken with kind permission of the Physical Review from figure
    1 in reference \cite{Huber:2006wb}.
    Copyrighted by the American Physical Society.
  }
  \label{fig:cid} 
\end{figure}

The raw data set is therefore split into two samples: the first with
charge identification (CID), used for the appearance channel and 
modeled accordingly to reference \cite{Huber:2002mx}; the second without
charge identification (no-CID). 
In this case the MINOS efficiencies and thresholds from
references \cite{Ables:1995wq,Huber:2004ug} are used. 
Note that this implies two different energy-threshold functions. 
The fact that there are almost no events below about 
$4$~GeV in the appearance channel is appropriately
modelled.
For details of the shape of the appearance channel threshold function,
the efficiencies, and the model of the energy resolution, see appendix
B.2 of reference \cite{Huber:2002mx}.
By comparing the two panels of figure \ref{fig:cid}, it can be seen
that it is extremely helpful not to use the CID information in  
the dis-appearance channel (cf. reference \cite{deGouvea:2005mi}).

\begin{figure}
  \begin{center}
    \includegraphics[width=0.9\textwidth]{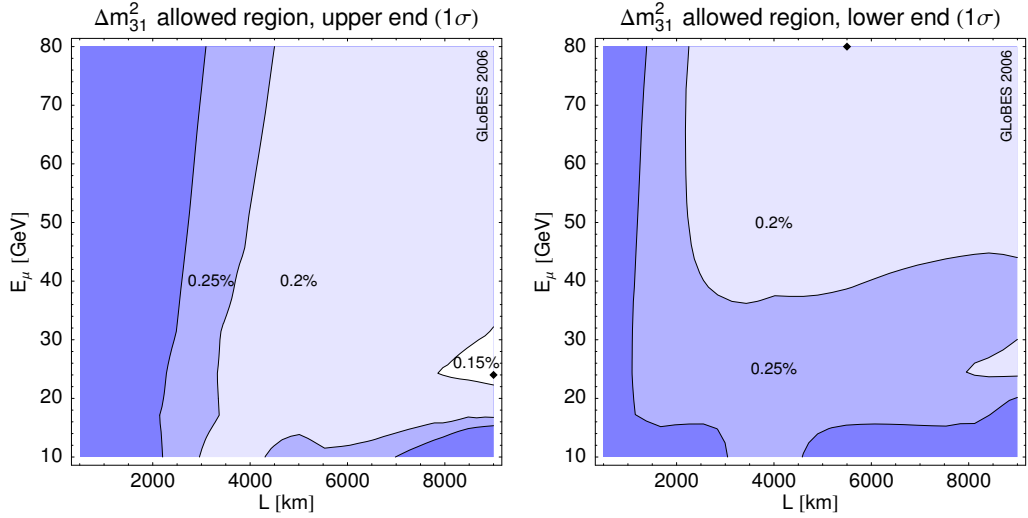}
  \end{center}
  \caption{
    Relative precision on $\ldm$ (at $1 \sigma$) as a function of $L$
    and $E_\mu$, including all parameter correlations for a normal
    mass hierarchy and $\stheta=0$.
    The upper end (left panel) and lower end (right panel) of the
    allowed region are given separately because the $\Delta \chi^2$ is
    quite asymmetric.  
    The minima, marked by the diamonds, occur at 0.14\% (left panel)
    and 0.18\% (right panel).
    Taken with kind permission of the Physical Review from figure
    7 in reference \cite{Huber:2006wb}.
    Copyrighted by the American Physical Society.
  } 
  \label{fig:dmel} 
\end{figure}

Figure \ref{fig:dmel} shows the relative precision on $\ldm$ as a
function of $L$ and $E_\mu$ (at $1\sigma$ CL for 1 degree of freedom),
including all parameter correlations, for a normal `true' mass
hierarchy.  
The upper end (left panel) and lower end (right panel) of the allowed
region are given separately, because the $\Delta \chi^2$ is quite
asymmetric in many cases. 
The first oscillation maximum can be found at:
\be
  L_{\mathrm{max}} \sim \left ( 564 \, \frac{E}{\mathrm{GeV}} \right ) \mathrm {km} \, , 
  \label{eq:lmax}
\ee
which explains the optimum observed for 
$E_{\mu} \sim 10$~GeV at about $3 \ 500$~km (remember that the
mean neutrino energy is somewhat below $E_{\mu}$).  
For $E_\nu \geq 2$~GeV (below this energy no significant rate is
observed), $L \gtrsim 1 \, 000$~km is a necessary condition to be able
to disentangle $\theta_{23}$ from $\ldm$. 
If $L \ll L_{\mathrm{max}}$, $\theta_{23}$ and $\ldm$ are highly
correlated. 
The separate analysis of the dataset without CID yields an extremely
good (compared to, e.g., reference \cite{Freund:2001ui}) relative
precision on $\ldm$ of the order of $0.2 \%$ for 
$L \gtrsim 3 \, 000$~km and $E_{\mu} \gtrsim 40$~GeV. 
This comes from the ability to resolve the oscillation maximum at low
energies for long enough baselines and large enough data sets because
of the lower threshold and the higher overall efficiency of the no-CID
dis-appearance channel sample.
Although the total rate decreases for longer baselines, more
oscillation maxima can be resolved. 
Note that we have included sufficiently many bins at low energies to
incorporate these effects.

\begin{figure}
  \begin{center}
    \includegraphics[width=0.9\textwidth]{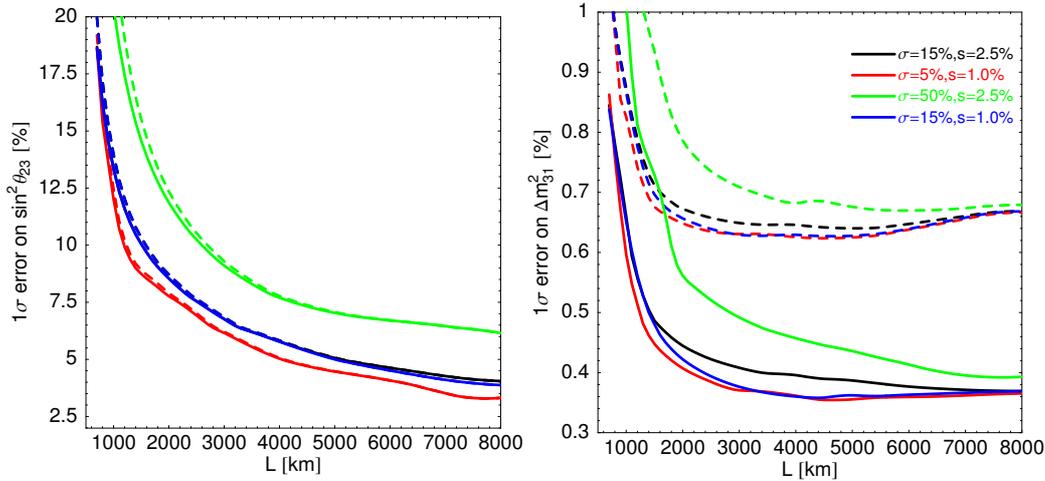}
  \end{center}
  \caption{
    The relative $1\,\sigma$ (full width) errors on
    $\sin^2\theta_{23}$ (left panel) and $\ldm$ (right panel)
    as a function of the baseline. 
    The result is shown for various combinations of energy resolution
    $\sigma$ and systematic error $s$.
    The solid (dashed) lines assume an uncertainty on $\sdm$ and
    $\theta_{12}$ of 5\% (10\%). 
    The results are computed for $\stheta\equiv0$.
    Taken with kind permission of the Physical Review from figure
    14 in reference \cite{Huber:2006wb}.
    Copyrighted by the American Physical Society.
  }
  \label{fig:atmprec} 
\end{figure}

It has been shown in reference \cite{deGouvea:2005mi} that the
energy resolution has a significant influence on the accuracy on the
leading parameters. 
In figure \ref{fig:atmprec} the relative 1$\sigma$ (full width)
errors on $\sin^2\theta_{23}$ (left panel) and $\ldm$ (right panel) as
a function of the baseline are shown. 
The different coloured lines correspond to different values of the
energy resolution, $\sigma$, and the normalisation error of the signal,
$s$.
Interestingly, the signal error seems to be quite unimportant. 
The energy resolution, on the other hand, has a relatively large
impact, especially at the shorter baselines. 
The dashed lines show the effect of increasing the uncertainty on the
solar parameters to $10\%$ (instead of $5\%$), the increased
uncertainty leads to a considerable deterioration in precision. 
Irrespective of the error on the solar parameters and the energy
resolution, longer baselines are preferred, especially for
$\sin^2\theta_{23}$.

\paragraph{Sensitivity to maximal $\theta_{23}$ and the octant-discovery potential}

\begin{figure}
\vspace{-0.5cm}
  \begin{center}
    \begin{tabular}{cc}
      \hspace{-1cm} \raisebox{5mm}{\epsfxsize7.5cm\epsffile{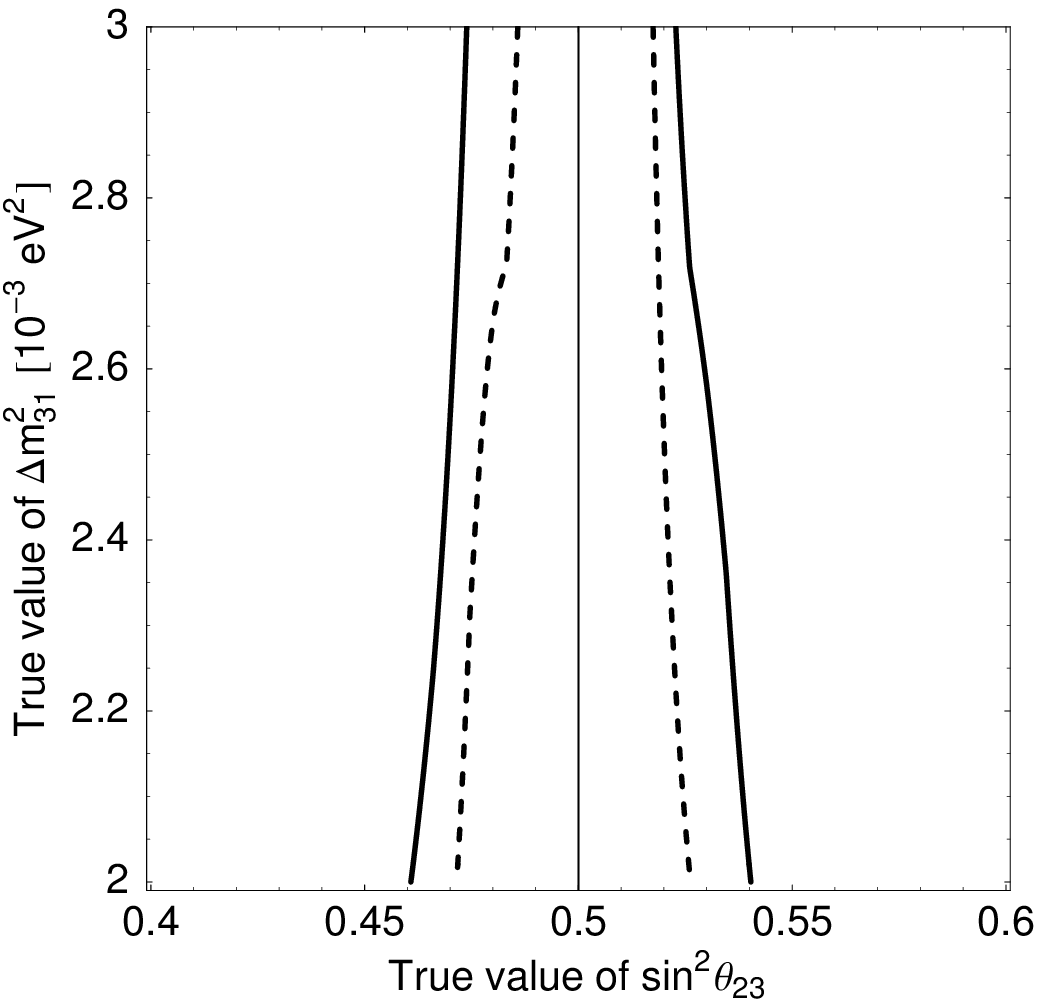}} & 
      \hspace{-0.5cm} \epsfxsize8.25cm\epsffile{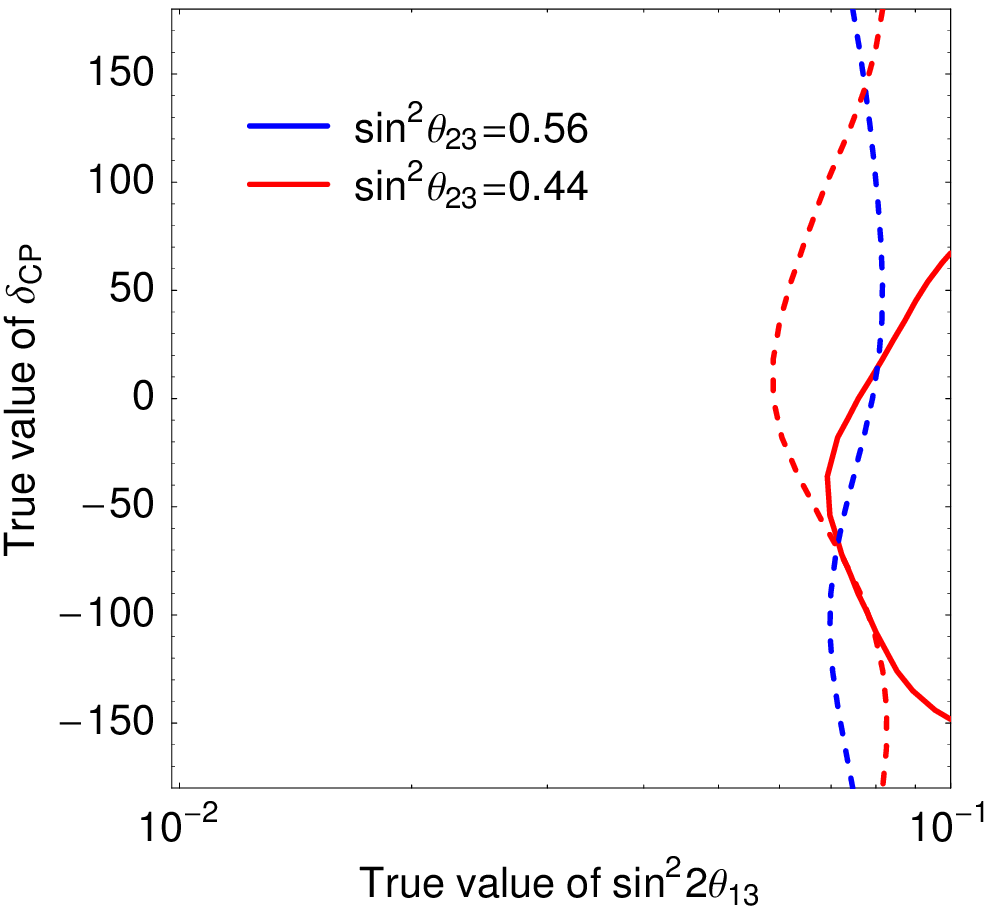} \
    \end{tabular}
  \end{center}
  \vspace{-\baselineskip}
  \caption{
    Left panel: 3$\sigma$ sensitivity to deviations from maximal
    $\theta_{23}$;.
    Right panel: 3$\sigma$ sensitivity to the $\theta_{23}$-octant.  
    Solid (dashed) lines refer to the $L = 4000$~km ($L = 7500$~km)
    Neutrino Factory.
  } 
  \label{fig:octant}
\end{figure}

\noindent
A natural explanation for maximal mixing ($\theta_{23}=\pi/4$) might
involve a new symmetry between $\nu_\mu$ and $\nu_\tau$.
Therefore, the degree to which $\theta_{23}$ differs from $\pi/4$ is a
powerful tool to discriminate between different neutrino-mass models
\cite{Altarelli:2003vk,Antusch:2004yx}.
Figure \ref{fig:octant} (left panel) shows the sensitivity to
deviations from maximal $\theta_{23}$.
The curves have been computed for $\theta_{13}=0$. 
Deviations as small as 10\% of $\sin^2 \theta_{23}$ from maximal
mixing could be established at the $L = 4000$~km baseline for certain
values of $\Delta m^2_{31}$.
A better sensitivity may be obtained for $L = 7500$~km, however, in this
case, the energy and baseline match the first oscillation peak in
matter.  
The sensitivities reached are at the level of 
$\sin^2 \theta_{23} \in [0.45-0.48]$ almost independent of the value
of $\Delta m^2_{31}$, which means that deviations from maximal mixing
of the order of 4\% could be established. 

Notice that, although this sensitivity is rather good, in general it
is very difficult to determine the octant in which the atmospheric
angle lies. 
It is quite difficult to break the $\theta_{23} \to \pi/2-\theta_{23}$
symmetry induced by the leading term in the transition probability;
the sub-leading terms that could help in lifting this degeneracy are
strongly correlated.
However, for $\theta_{13} \neq 0$, full advantage can be taken of
matter effects in the muon-neutrino dis-appearance signal.
The Neutrino Factory shows a certain (limited) capability to solve this
degeneracy, irrespective of the baseline and the value of $\delta$. 
The ability to resolve the octant is shown in figure \ref{fig:octant}
(right panel); longer baselines perform better and it is easier to
resolve the octant if the true $\theta_{23}<\pi/4$.

\begin{figure}
  \begin{center}
    \includegraphics[width=0.4\textwidth]{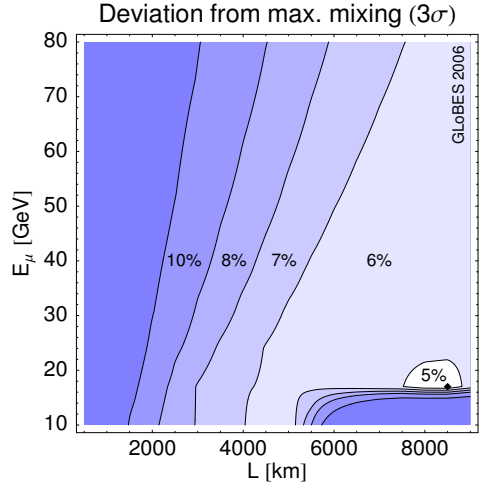}
  \end{center}
  \caption{
    Sensitivity to deviations from maximal mixing for a normal mass
    hierarchy and $\stheta=0$ as a function of $L$ and $E_\mu$.
    The sensitivity is given as relative deviation of 
    $\sin^2 \theta_{23}$ from $0.5$ in per cent at $3 \sigma$ 
    including all parameter correlations, where only the upper
    branch $\sin^2 \theta_{23} >0.5$ is taken into account.  
    The minimum, marked by a diamond, is at 4.2\%. 
    Taken with kind permission of the Physical Review from figure
    8 in reference \cite{Huber:2006wb}.
    Copyrighted by the American Physical Society.
  }
  \label{fig:devmaxel} 
\end{figure}

Figure \ref{fig:devmaxel} shows the sensitivity to deviations from
maximal mixing for a normal mass hierarchy and $\stheta=0$ as a
function of $L$ and $E_\mu$. 
The sensitivity is given as the relative deviation of 
$\sin^2 \theta_{23}$ from $0.5$ in per cent at $3 \sigma$,  including
all parameter correlations.
Note that only the upper branch $\sin^2 \theta_{23} >0.5$ is taken
into account, because there is hardly any sensitivity to the
$(\theta_{23}, \pi/2 - \theta_{23})$ ambiguity \cite{Fogli:1996pv} and 
the problem is symmetric around $\theta_{23} = \pi/4$. 
A very similar qualitative and quantitative behaviour is found to that
reported in reference \cite{Freund:2001ui}. 
However, the low-energy performance for very long baselines 
($L \gtrsim 6 \, 000$~km) is significantly improved as the
efficiencies at lower energies are better when including $\nu_\mu$
dis-appearance data without CID.
Most importantly, it is very hard to improve the sensitivity to
deviations from maximal mixing with the standard setup, probably
because of the rather large normalisation uncertainties that have been
assumed. 
In particular, T2HK could achieve a similar quantitative performance
\cite{Antusch:2004yx}.  

\paragraph{Optimisation for large $\stheta$}

\noindent
Consider now large values of $\stheta$, $\stheta \simeq 0.1$
$[\theta_{13} \simeq 9^\circ]$, which means that it will be measured
at the next generation of super-beam experiments.
It is well known that for large $\stheta$, matter-density uncertainties
affect the precision measurements of $\stheta$ and $\delta$ (see,
e.g., references \cite{Huber:2002mx,Ohlsson:2003ip}).
Therefore, it is an interesting question whether the optimisation
changes for large $\stheta$ depending on the matter-density
uncertainty, and if the performance of conventional techniques can be
exceeded.

\begin{figure}
  \begin{center}
    \includegraphics[width=0.9\textwidth]{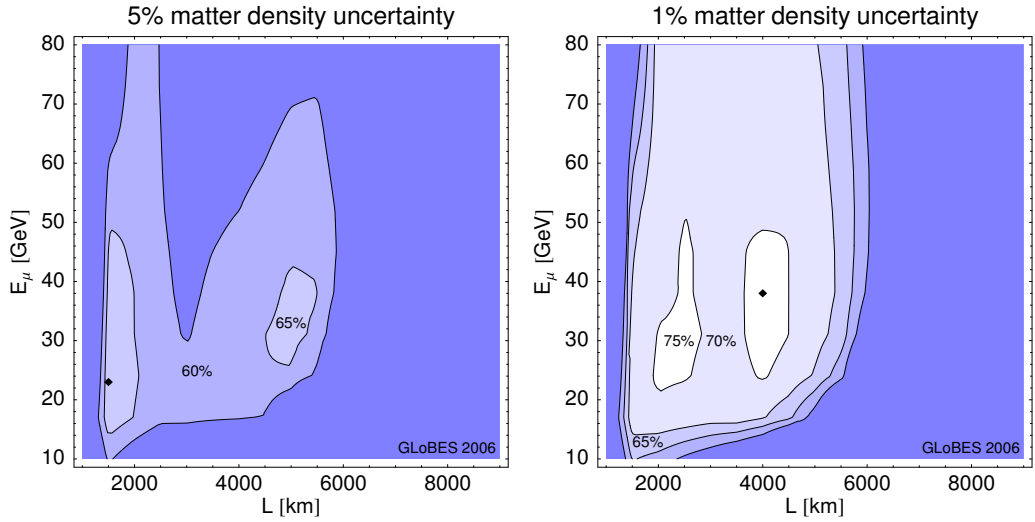}
  \end{center}
  \caption{
    Fraction of (true) $\delta$ as function of $L$ and $E_\mu$ for
    the measurement of CP-violation for $\stheta=0.1$ and a normal
    mass hierarchy ($3\sigma$, including all parameter correlations
    and degeneracies). 
    The left panel corresponds to a matter density uncertainty of
    $5\%$, and the right panel to a matter density uncertainty of
    $1\%$.  
    The maxima, marked by the diamonds, are at 68\% (left) and 77\%
    (right). 
    Taken with kind permission of the Physical Review from figure
    9 in reference \cite{Huber:2006wb}.
    Copyrighted by the American Physical Society.
  } 
  \label{fig:lthetael} 
\end{figure}

For the mass hierarchy, the optimisation is hardly affected by the
matter-density uncertainty. 
As a general rule, the mass hierarchy can be determined for all values
of $\delta$ for $L \gtrsim 1\, 000$~km almost independent of muon
energy.
Discovery of non-vanishing $\stheta$ is possible independent of
$\delta$.
The CP-violation discovery potential is shown in figure
\ref{fig:lthetael} as a function of $L$ and $E_\mu$, for $\stheta=0.1$
and a normal mass hierarchy, for two different values of the matter
density uncertainty: 5\% (left panel); and 1\% (right panel). 

The maximum achievable CP-fraction depends on the matter-density
uncertainty, and is only marginally affected by the choice of baseline
for baselines between $1 \, 500$ and $5 \, 500$~km for small
matter-density uncertainty.
A very peculiar behaviour of for larger matter-density uncertainty
can be observed in the left panel of figure \ref{fig:lthetael}.
A matter density uncertainty of $\sim 5\%$ is more realistic with the
present level of understanding
\cite{Geller:2001ix,Ohlsson:2003ip,Pana}.
In particular, the combination $L=3 \, 000$~km and $E_{\mu}=50$~GeV,
which is often considered, performs especially badly. 
It is not trivial to explain this loss of sensitivity.
First, smaller muon energies are preferred since matter density
uncertainties hardly affect the leading $\sin^2 2 \theta_{13}$-term
close to the matter resonance (which is acting as a background to the
$\delta$ measurement; see figure 3 of reference
\cite{Ohlsson:2003ip}). 
Second, shorter baselines are preferred since matter effects are
smaller there and, therefore, the impact of density uncertainties is
reduced. 
Third, there is a second maximum for $L \simeq 5 \, 000$~km, where the 
CP-asymmetric term is enhanced for $E \sim 10$~GeV, equation
(\ref{eq:lmax}); remember that the mean neutrino energy is
considerably below the muon energy).
These factors together cause the structure in the left panel of figure 
\ref{fig:lthetael}.

Comparison of figure \ref{fig:lthetael} (right panel) with figure
\ref{fig:cpel}(left) shows that for small values of the matter-density
uncertainty, the `usual' optimisation for CP violation is
qualitatively recovered.
The optimal performance for small matter density uncertainties is
reached in a wide range of $L$ and $E_{\mu}$.

%% file: 05-Performance/05-04-Neutrino-Factory/05-04-03-Neutrino-Factory-clones.tex
\subsubsection{Solving degeneracies}
\label{sec:clones}

Various solutions have been proposed to reduce the parametric
correlations and degeneracies observed in the simultaneous measurement
of $\theta_{13}$ and $\delta$ at the Neutrino Factory. 
The design of the magnetised iron detector used to measure
golden-channel wrong-sign muons and the tight kinematic cuts
applied to reduce the background, result in very few events being
collected in the energy region below 10~GeV. 
This region, however, is where the first oscillation peak lies for
neutrinos produced in the decay of 50~GeV muons. 
Unfortunately, having sufficient statistics above and below the
oscillation peak has been shown to be the key to solve many of the
parametric degeneracies. 
This is why Neutrino Factory experiments suffer from this problem
more than super-beam or beta-beam experiments, for which the results
are limited by statistics.

The different methods that may be used to resolve the correlations and
degeneracies will be discussed under three headings:
\begin{itemize}
  \item 
    Combining data collected using several magnetised iron
    detectors of the type described in section \ref{sec:mid} placed at
    a number of baselines;
  \item 
    Combining different channels collected using the detectors
    described in sections \ref{sec:ecc} and \ref{sec:lar}; and
  \item 
    Improving the performance of the detector described in section
    \ref{sec:mid}. 
\end{itemize}

\paragraph{Combining baselines}
\label{sec:baselines}

\begin{figure}
  \begin{center}
    \includegraphics[width=8cm]{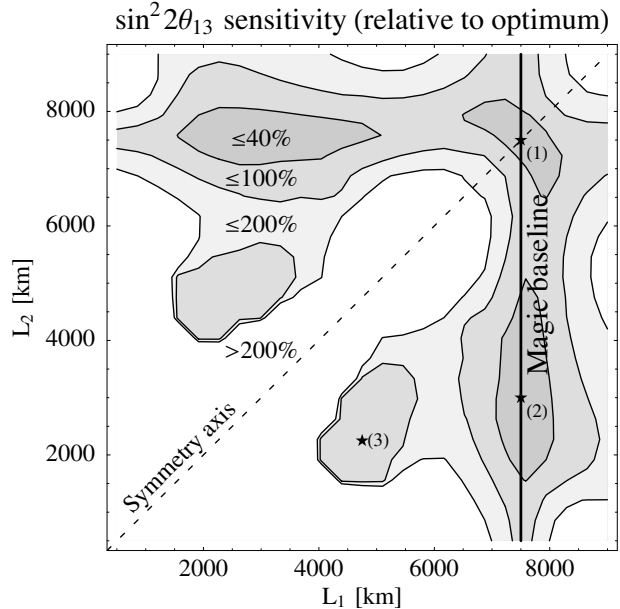}
  \end{center}
  \caption{
    The $\sin^2 2 \theta_{13}$ sensitivity limit relative to the
    optimum value of $5.9 \cdot 10^{-5}$ at 
    $L_1=L_2 \simeq 7 \, 500$~km.
    It is plotted at the $3 \sigma$ confidence level as function of
    the baselines $L_1$ and $L_2$ heading from the 50~GeV Neutrino
    Factory towards two 25~Kton detectors. 
    The sensitivity limit includes full correlations and
    degeneracies. 
    The true parameters for this figure are $\Delta
    m^2_{31}=3\cdot10^{-3}\,\mathrm{eV}^2$, $\theta_{23}=\pi/4$,
    $\Delta m^2_{21}=7\cdot10^{-5}\,\mathrm{eV}^2$ and
    $\sin^2\theta_{12}=0.28$.
    Taken with kind permission of Physical Review from figure 1 in
    reference \cite{Huber:2003ak}.
    Copyrighted by the American Physical Society.
  }
  \label{fig:2base}
\end{figure}

\noindent
The first option to resolve the degeneracies is to combine golden-muon
signals from experiments located at different baselines.
It was recognised very early in the literature that certain types
of correlations are less pronounced if data from different baselines
are analysed together, see for example
\cite{Cervera:2000kp,Burguet-Castell:2001ez}.
It turns out that the most useful additional baseline is around
$L\sim7500$~km, the magic baseline \cite{Huber:2003ak}. 
At this distance, matter effects completely suppress any three-flavour
effect and allow for an unambiguous measurement of $\stheta$ and of
the mass hierarchy (see also section \ref{sec:hierarchy}). 
In figure \ref{fig:2base}, the sensitivity to $\stheta$ is shown for a
Neutrino Factory with two baselines as a function of the two
baselines.
Clearly, the combination of $L_1\sim 3\, 000$~km and $L_2=7500$~km has a
very good performance (star labeled `(2)'). 
The other possible choice, i.e. putting all the detector mass at
$L=7500$ km (star labeled `(1)'), is very good for $\stheta$
measurements but would have no sensitivity to CP-violation at 
all. 
The third possible solution `(3)' is fine tuned, as shown in
figure 2 of \cite{Huber:2003ak}.

The combination, $L \sim 3\, 000$~km and $L \sim 7\, 000$~km, is very
effective; it allows for a clean measurement of $\stheta$ and of the
sign of $\Delta m_{31}^2$ at the magic baseline and for a good
measurement of $\delta$ at the shorter baseline, where
the ($\theta_{13},\delta$) correlation is strongly reduced because
$\theta_{13}$ is already constrained by the magic-baseline data. 
A second detector at 3\, 000 km in combination with the first at or
around the magic baseline significantly improves the
$\sin2\theta_{13}$ sensitivity, by about an order of magnitude
(the results do not change significantly if the detector is anywhere
between 3\, 000~km to 5\, 000~km). 
The sensitivity is not strongly affected by the exact value of the
location of the first detector in the range 6\, 000~km to 9\, 000~km
either \cite{Gandhi:2006gu}.

\paragraph{Combining channels}
\label{sec:channels}

\begin{enumerate}
 \item {\it The silver channel:}  
  In reference \cite{Donini:2002rm} it was noticed that muons
  arising from  $\tau$ decay when $\tau$s are produced via a 
  $\nu_e \to \nu_\tau$ transition show a different 
  $(\theta_{13}, \delta)$ correlation from those coming from 
  $\nu_e \to \nu_\mu$ transitions.  
  By using an Emulsion Cloud Chamber (ECC) capable of $\tau$-decay
  vertex recognition, it is possible to use the complementarity of the
  information from $\nu_e \to \nu_\tau$ and from $\nu_e \to \nu_\mu$
  to solve the intrinsic degeneracy.
  The relatively small mass of the ECC, the small $\nu_{\tau}$-nucleon
  cross section, the small $\tau \to \mu$ branching ratio, and the
  decay-vertex requirement, cause the statistical significance of the
  silver channel to be much lower than that of the golden channel.
  Silver muons, in combination with golden muons, are also extremely
  helpful in dealing with the $[\theta_{23},\pi/2-\theta_{23}]$
  ambiguity, since the leading term in $P(\nu_e \to \nu_\tau)$ is
  proportional to $\cos^2 \theta_{23}$, whereas the analogous term in
  $P(\nu_e \to \nu_\mu)$ is proportional to $\sin^2 \theta_{23}$. 
  However, the sensitivity of the silver/golden channel combination to
  the $\theta_{23}$-octant strongly depends on the value of
  $\theta_{13}$.

  The addition of the silver-channel data does not affect the
  golden-channel baseline optimisation.
  The golden channel suffers significantly from degeneracies at the 
  $4 \, 000$~km baseline, in particular for true
  $\delta=3\pi/2$. For $\stheta\sim 3 \times 10^{-3}$, the
  sensitivities to maximal CP-violation and to the mass hierarchy are
  lost, and a sensitivity gap appears. 

  For a golden channel setup fixed to $\mathrm{E_\mu=50\,GeV}$ and
  $\mathrm{L_{MID}=4\, 000\,km}$, the optimal ECC baseline to close the
  sensitivity gap is found between 2500~and~5\, 000~km.
  It will therefore be assumed in the following that the ECC detector
  is located at the second golden-channel detector baseline
  (3\, 000~km).

 \item {\it The $\nu_\mu \to \nu_\tau$ channel:}
  Using the ECC detector, it is possible to disentangle the (dominant) 
  $\nu_\mu \to \nu_\tau$ appearance oscillation from the $\nu_\mu$
  disappearance channel.  
  This oscillation probability, which is the main goal of the CNGS
  experiments, is extremely sensitive to the atmospheric parameters
  $\theta_{23}$ and $\Delta m^2_{31}$.
  In principle, it could be used to complement the information from
  the $\nu_\mu$ disappearance channel in the MID detector to solve the
  octant-degeneracy.

  A detailed study of this channel at the ECC is lacking. 
  A preliminary analysis shows that the performance of this channel is
  similar to the $\nu_\mu$ disappearance channel at the MID detector.

 \item {\it The platinum channel:}
  The platinum channel is the T-conjugate of the golden channel.
  Therefore, the $(L,E_\mu)$-optimisation of the platinum channel is
  the same as that of the golden channel.
  It will be assumed that the platinum-channel detector should be
  sited at the same baseline as the golden-channel detector.
 
  As for the silver channel, the platinum channel is strongly limited
  by statistics. 
  In the left panel of figure \ref{fig:pltresh}, $\Delta \chi^2$ (with 
  respect to the absolute $\chi^2$ minimum) values at the
  intrinsic-degeneracy and sign-degeneracy minima are shown as a
  function of the upper electron/positron-charge identification
  threshold for $\sin^2 2 \theta_{13} = 2.5 \times 10^{-3}$.  
  Recall that, for the standard  platinum-channel detector
  \cite{Rubbia:2001pk}, the upper threshold has been fixed to
  7.5~GeV. 
  As can be seen in the figure, for a 15~Kton magnetised liquid-argon
  (LAr) detector the $\Delta \chi^2$ at the degenerate solutions does
  not change a lot.  
  This is a severe limitation of this channel due to the small
  size of the available data set.
  On the other hand, for a 50~Kton magnetised LAr detector, both
  degeneracies are lifted if the upper electron CID threshold is
  increased above 30~GeV.  
  In this case, the sensitivity gap can be closed completely.  
  This means that only with both a significant increase in the
  detector mass and in the electron CID threshold this channel can
  help in solving degeneracies for low $\theta_{13}$. 
  The 15~Kton LAr detector, as well as the 5~Kton ECC detector, can
  contribute for intermediate values of $\theta_{13}$ but not for such
  low values. 
\begin{figure}
  \begin{center}
    \raisebox{1mm}{\includegraphics[width=0.39\textwidth]{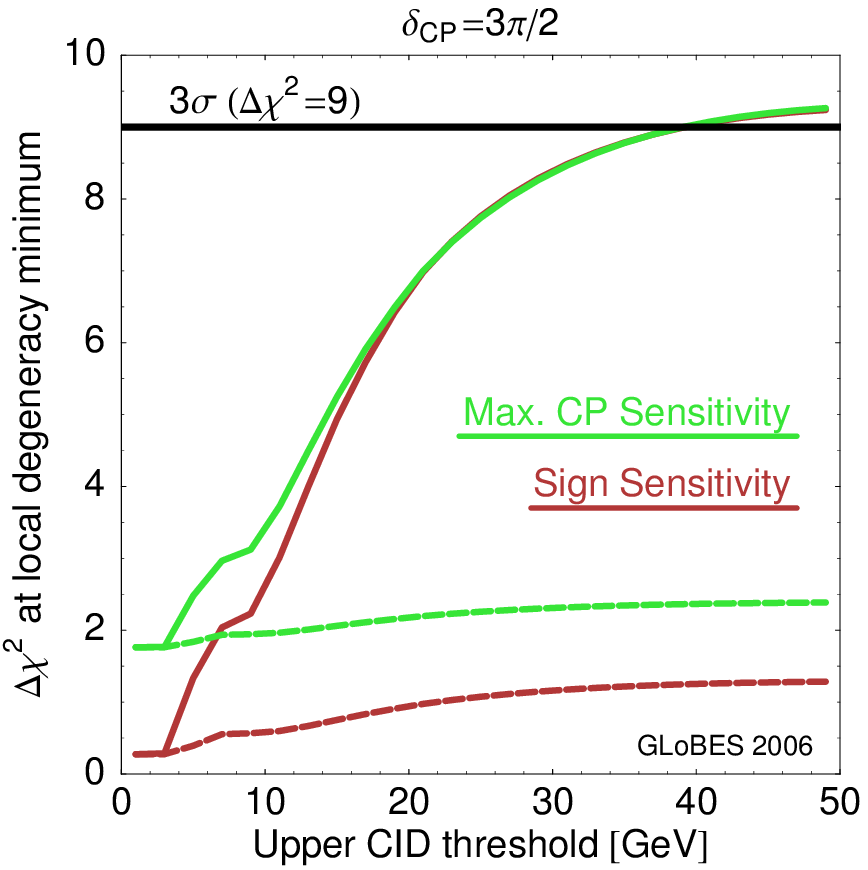}}
    \hspace{0.1cm}
    \includegraphics[width=0.41\textwidth]{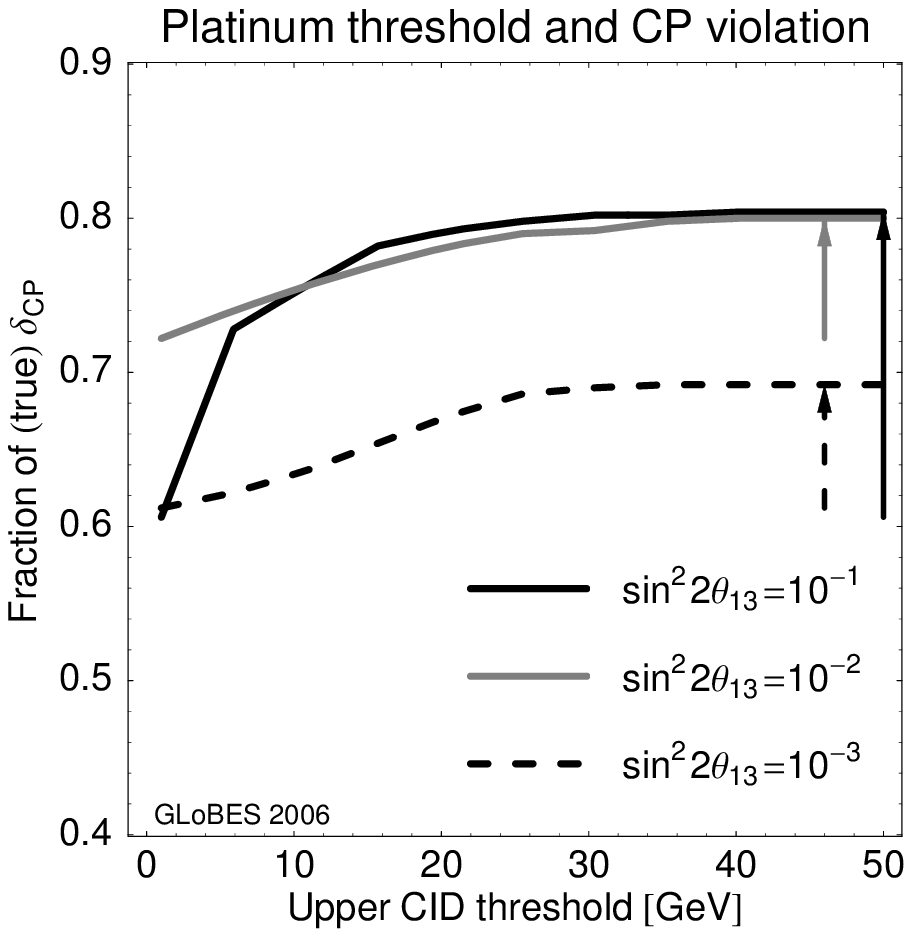}
  \end{center}
  \caption{
    Left panel: Dependence of $\Delta \chi^2$ value at the intrinsic-
    and sign-degeneracy minima (light gray/green and dark gray/red,
    respectively) on the upper electron CID threshold, for input
    values $\sin^2 2 \theta_{13} = 2.5 \times 10^{-3}$
    (i.e. $\theta_{13} \sim 1.5^\circ$) and $\delta = 3 \pi/2$.
    The baseline is assumed to be 4\, 000~km and $E_\mu = 50$ GeV.
    Solid (dashed) lines stand for the improved (standard) platinum
    detector, with 50 (15) Kton mass and 40\% (20\%) efficiency.  
    Right panel: The fraction of (true) $\delta$ for which
    CP-violation can be discovered at $3 \sigma$ as a function of the
    upper electron CID threshold (for a normal mass hierarchy), 
    combining the 50 Kton golden detector at $L = 4\, 000$~km and the
    improved 50~Kton platinum detector with 40\% efficiency, for
    $E_\mu = 50$~GeV. 
    Different curves refer to different values of $\stheta$.  
    The arrows refer to the improvement in the physics potential by
    using the platinum channel.
    Taken with kind permission of the Physical Review from figures
    16 and 17 in reference \cite{Huber:2006wb}.
    Copyrighted by the American Physical Society.
    Figures taken from reference \cite{Huber:2006wb}.
  } 
  \label{fig:pltresh}
\end{figure}
 
  The platinum channel may be used to solve degeneracies both for
  intermediate and large values of $\stheta$.
  In figure \ref{fig:pltresh}(right), the CP-fraction for which
  CP-violation can be discovered as a function of the upper CID
  threshold is shown. 
  The dependence of the discovery potential on the threshold is
  relatively shallow for $\stheta \lesssim 10^{-2}$, whereas for
  larger $\stheta$ a 6~GeV upper threshold can increase the
  CP-fraction by about 10\%. 
  This means that if $\stheta$ turns out to be large, a relatively low
  upper threshold could be acceptable.
  However, if it is intended to use the platinum-channel detector as a
  degeneracy-solver, the threshold will need to be as high as 20 to
  30~GeV. 

 \item {\bf The $\nu_e$ disappearance channel}
  While only electron-neutrino (and anti-neutrino) appearance has been
  considered in this section, one could also think about implementing
  the $\nu_e$ disappearance channel. 
  The impact of this channel for $\stheta=0.1$ has been tested and
  some improvement has been observed, though it is not as beneficial
  as the platinum-appearance channel.
  If one cannot achieve CID to the assumed level, the
  $\nu_e$-disappearance channel alone without CID can provide useful
  information. 
  The $\nu_e$-disappearance channel with CID performs worse than
  that without CID (as was the case for the $\nu_\mu$-disappearance
  channel).
  The $\nu_e$-disappearance is not considered further in the rest of
  this section.
  If $\nu_e$ detection is eventually implemented, the
  disappearance-channel data should be exploited as well. 
  The main issue determining its usefulness is, of course, how well
  systematic uncertainties can be controlled. 
\end{enumerate}

{\bf \noindent Combination of the additional channels}

\noindent
The combination of the data from the additional channels 
with golden muons is now discussed in terms of the three
observables: sensitivity to $\stheta$; sensitivity to maximal
CP-violation; and sensitivity to the mass hierarchy.

The relative contribution to the physics reach can be roughly
understood by looking at the statistical significance of the various
options.  
To this end, in table \ref{tab:channelevents} the signal and
background event rates (as well as the signal over the square root of
the background) for two specific points in the parameter space,
representing two conceptually different cases, $\stheta= 10^{-1}
[\theta_{13} = 9^\circ]$ or 
$\stheta=3 \times 10^{-3} [\theta_{13} = 1.6^\circ]$ are presented.
For $\stheta=10^{-1}$, the golden channel suffers from the matter
density uncertainties.  
For $\stheta= 3 \times 10^{-3}$, on the other hand, the golden channel
suffers from degeneracies.
In both cases, additional channels could improve the Neutrino Factory
performance (but are limited by the size of the data set). 
It can be seen from table \ref{tab:channelevents} that the golden
channel deserves its name, having the largest statistical significance
for both values of $\stheta$.
This is due to the fact that muons are relatively straightforward to
detect and easy to distinguish from backgrounds.  
The silver channel has a much lower statistical weight and a
relatively high background contamination. 
The event rates for the silver channel are also given at a ECC
detector baseline of 732~km, the distance from CERN to Gran Sasso 
where the OPERA detector will be located.
No data are shown for the $\mu^-$-stored phase, see reference
\cite{Autiero:2003fu}.
It can be seen that the variation of the baseline does not have a big
impact on the total rates.  
Notice that the size of the platinum-channel data set is larger when
the Neutrino Factory operates in $\mu^-$-polarity, when the golden
channel is weaker because of the matter effect suppression. 
Thus, it acts as an anti-neutrino mode without matter-effect
suppression. 
\begin{table}
  \begin{center}
    \begin{tabular}{lrrr} \hline
      $\sin^22\theta_{13}=10^{-1}$ & Signal & Background & S/$\mathrm{\sqrt{B}}$ \\ \hline 
      Golden & 31000 (6000) & 39 (73) & 5000 (700) \\ 
      Silver & 210 (--) & 32 (--) & 37 (--) \\  
      Silver@732km & 260 (--) & 110 (--) & 25 (--) \\
      Platinum & 4 (120) & 140 (110) & 0.3 (11) \\ 
      $\mathrm{(Golden)_{\mathrm{MB}}}$ & 5100 (340) & 9 (17) & 1700 (83) \\ \hline
    \end{tabular} 

    \vspace*{0.3cm}

    \begin{tabular}{lrrr} \hline
      $\sin^22\theta_{13}=3 \times 10^{-3}$ & Signal & Background & S/$\mathrm{\sqrt{B}}$ \\ \hline 
      Golden & 1900 (450) & 39 (72) & 300 (53) \\ 
      Silver & 3 (--) & 33 (--) & 0.5 (--) \\  
      Silver@732km & 1.7 (--) & 110 (--) & 0.2 (--) \\
      Platinum & 1 (5) & 170 (110) & 0.08 (0.5) \\ 
      $\mathrm{(Golden)_{\mathrm{MB}}}$ & 200 (10) & 9 (17) & 67 (2.4) \\ \hline
    \end{tabular}
  \end{center}
  \caption{
    The (rounded) event rates in the $\mu^+$ ($\mu^-$)-stored phase
    for the golden channel and the standard silver and platinum
    channels at a baseline of 4\, 000~km and for 
    $\mathrm{E_\mu= 50\,GeV}$. 
    For comparison reasons, we also give the golden channel event
    rates at the magic baseline ($L = 7500$~km) and the silver channel
    event rates at $L = 732$~km.
    The upper table is calculated for $\sin^2 2\theta_{13}=10^{-1}$
    and the lower table for $\sin^22\theta_{13}=3 \times 10^{-3}$.
    The remaining oscillation parameters are fixed as in
    equation (\ref{equ:params}), with $\delta=0$.
  } 
  \label{tab:channelevents}
\end{table}

The performance of the golden channel can also be improved by a second
detector at the magic baseline, as was stressed in section
\ref{sec:baselines}. 
Therefore, the golden-channel event rates are also given at the magic
baseline for comparison.    
Despite the strong reduction in the neutrino flux, there are still a
very large number of golden muons, and the signal-to-background ratio
is still much better than for the silver or platinum channels.  
It may therefore be expected that additional channels will only be
useful in those regions of the parameter space where the performance
of the Neutrino Factory is strongly affected by either degeneracies or
correlations (i.e., for intermediate $\theta_{13}$).

For $\stheta=3 \times 10^{-3}$, the combination of the silver or
platinum channel with the golden channel data is comparable, with a 
slightly better impact of the golden/silver combination on the
sensitivity to the mass hierarchy.  
For $\stheta=10^{-1}$, however, the golden/platinum combination has a
rather larger margin of improvement with respect to the golden/silver
combination.  
The reason for this lies in the $\tau$ production threshold which
suppresses the most useful silver events around the first oscillation
maximum.  
Thus, an increase in the size of the silver data set is not helpful.
On the other hand, if one can go beyond the 15~Kton magnetised LAr
detector and increase the upper electron CID threshold up to 30~GeV,
the CP-discovery potential of the Neutrino Factory is significantly
improved.
 
\begin{figure}
  \begin{center}
    \includegraphics[width=0.4\textwidth]{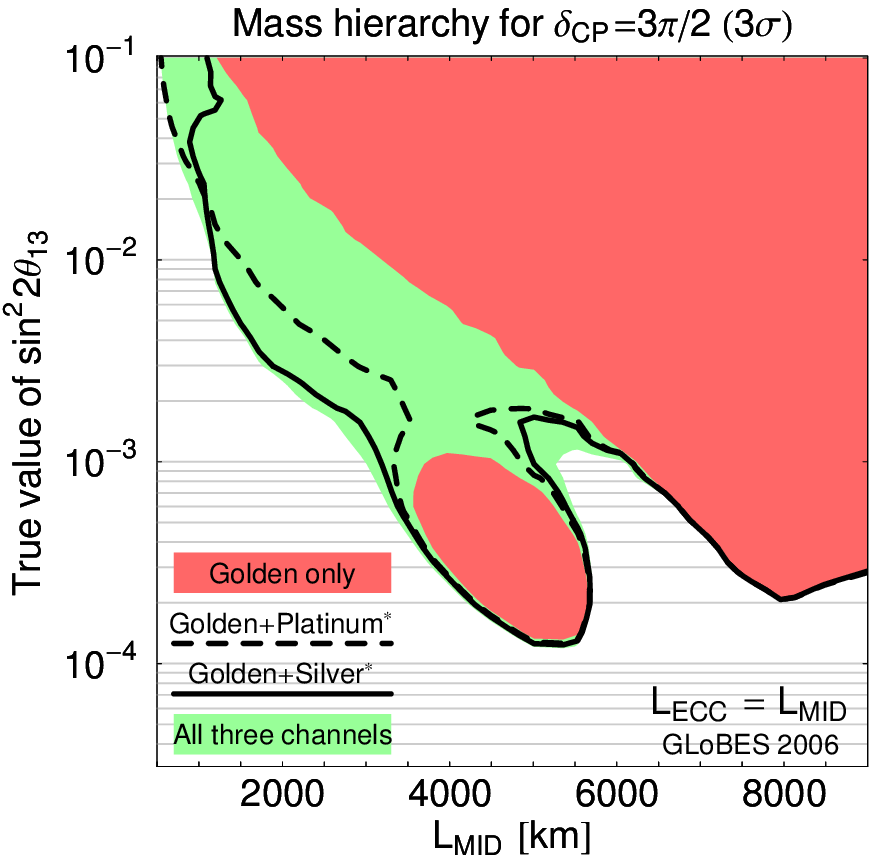} \hspace{0.1cm.eps}
    \includegraphics[width=0.4\textwidth]{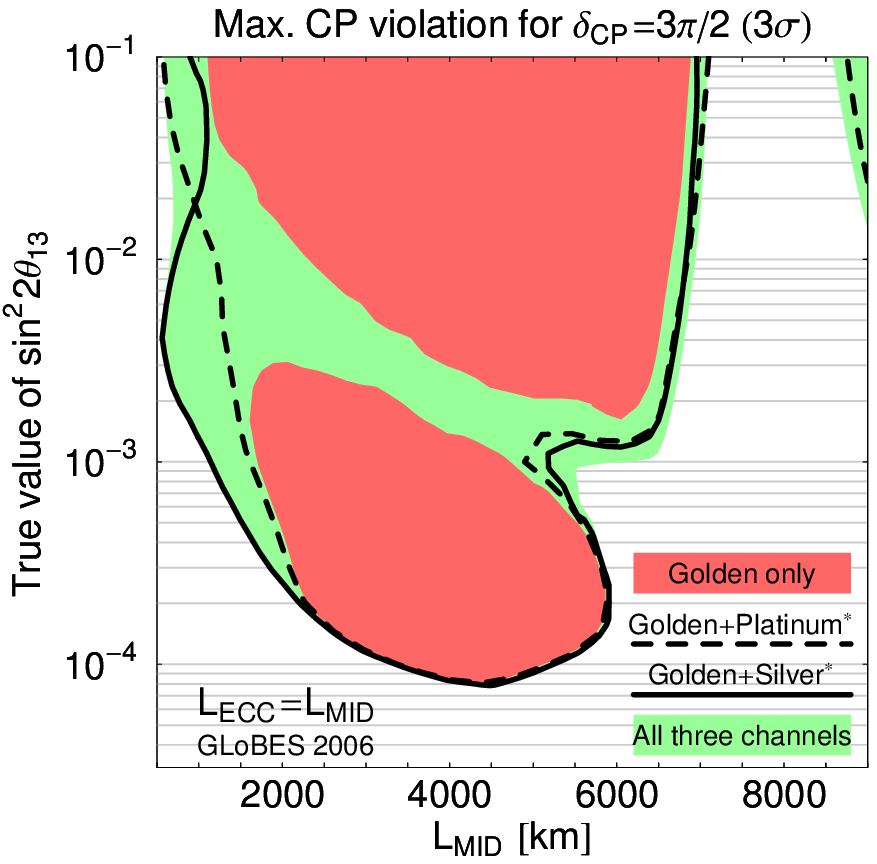}
  \end{center}
  \caption{
    The sensitivity to mass hierarchy (left pane) and to maximal
    CP-violation (right panel) at $3\sigma$ for the combination of
    different channels as given in the plot legends, for
    $\delta=3\pi/2$.  
    All correlations and degeneracies are taken into account.  
    Taken with kind permission of the Physical Review from figure
    20 in reference \cite{Huber:2006wb}.
    Copyrighted by the American Physical Society.
  }
  \label{fig:ChannelsMHCP}
\end{figure}

Although the additional channels do not improve significantly the
$\theta_{13}$ sensitivity of the Neutrino Factory, they help in
solving some of the degeneracies. 
This is shown in figure \ref{fig:ChannelsMHCP}, where the sensitivity
to the mass hierarchy (left panel) and to maximal CP-violation (right
panel) are presented for different combinations of golden, silver, and
platinum channels as a function of the (common) baseline.
The plots refer to $\delta=3 \pi/2$, a value for which the
degeneracy problem is severe. 
Notice that the plots (taken from reference \cite{Huber:2006wb}) show
golden data combined with data from the silver$^*$ and platinum$^*$ detectors.
The latter refers to the 50~Kton upgrade of the platinum detector
described in section \ref{sec:lar}.
The former to the silver detector with a data set 5 times as large as
that assumed above.
Since solving the degeneracies for intermediate $\theta_{13}$ does not
rely significantly on the statistical weight of the data, the results
shown would not change much using standard silver and platinum
detectors.
 
For the sensitivity to the mass hierarchy figure
\ref{fig:ChannelsMHCP} (left panel), the additional silver- and
platinum-channel data can indeed improve the sensitivity and close the
sensitivity gap between the dark shaded regions in a large range of
baselines. 
The $4 \, 000$~km baseline with channel combination becomes as good
as the magic baseline to measure the mass hierarchy, for 
$\delta \approx 3\pi/2$.   
It has been checked that the impact of the additional channels is
small for $\delta=0$ and negligible for $\delta=\pi/2$.
 
For the sensitivity to maximal CP-violation figure
\ref{fig:ChannelsMHCP} (right panel), it can also
be seen that the combination of silver and/or platinum channels with
the golden one completely closes the degeneracy gap. 
For $\mathrm{L\approx4\, 000\,km}$ and $\delta = 3 \pi/2$,
CP-violation can be determined unambiguously for $\sin^2 2
\theta_{13}$ as small as $10^{-4} [\theta_{13} = 0.3^\circ]$.  
It has been checked that the impact of the additional channels is
negligible for $\delta = \pi/2$ for baselines around $4 \,
000$~km, since the effect of degeneracies is small for that specific
value of $\delta$.
 
In addition to the baseline optimisation, the dependence of the
sensitivities on the energy of the stored muons can be studied.  
As far as the sensitivity to the normal mass hierarchy is concerned,
the variation of the maximal reach in (true) $\stheta$ is of minor
importance, and even improves slightly for the choice of smaller
muon energy. 
For the golden channel only, or golden and platinum channels combined,
the maximum is approximately reached for $E_\mu \sim 30\,\mathrm{GeV}$. 
A sensitivity-gap for $\sin^2 2 \theta_{13} \in [1,5] \times 10^{-3}$
cannot be cured by the golden channel alone, independent of
$\mathrm{E_\mu}$.
However, if combined with the silver or platinum channel, the
sensitivity-gap can be closed for parent energies $E_\mu \gtrsim
20\,\mathrm{GeV}$ (golden/platinum) or larger than about $E_\mu
\gtrsim 25\,\mathrm{GeV}$ (golden/silver).
For the platinum combinations (or all channels combined), the
additional information not only allows a lower energy neutrino beam to
be used, but also favours a lower parent energy of $E_\mu \sim
30\,\mathrm{GeV}$. 
On the other hand, when only the silver-channel data are used, the
$\tau$-production threshold disfavours low muon energies.  
For the sensitivity to maximal CP violation, qualitatively the same
results as for the mass hierarchy are obtained.

\paragraph{Improved detector}
\label{sec:det}

\noindent
A Neutrino Factory requires a large investment in accelerator R\&D and
infrastructure.
A joint optimisation of both accelerator and detector, however, has
been neglected so far and it is worth considering whether significant
gains in performance can be achieved with an increased emphasis on the
detector side of the experiment.  
The main problem is the lack of reliable performance predictions for
large magnetic detectors.  
The goal of this section is not to prove the feasibility of certain
detector properties or parameters, but to demonstrate the possible
gain in the physics reach if certain properties can be achieved.
Therefore, the following statements or assumptions about the detector
performance are not to be mistaken for a claim of feasibility, but
should be understood as desirable improvements; the extent to which
such performance can be achieved must be determined by extensive
R\&D.
Choices for the various factors affecting the detector performance
have been made with the intention that the assumptions are not too far
away from what may be possible \cite{ISSdetectorWG}.
However, the effect of varying the detector-performance assumptions
on the physics performance will be discussed in some cases.
These results may serve as guidelines to focus efforts in detector
R\&D.
They should be interpreted as indicating the `optimisation potential
of the detector', rather than as the `optimised detector' per se. 

\begin{enumerate}
 \item{\it Improved detector assumptions:}
 \label{sec:requirements}
  The main limitation of a Neutrino Factory compared to other neutrino
  facilities comes from the fact that the standard detector has a
  relatively high neutrino-energy threshold (necessary for muon charge
  identification), which makes the first oscillation maximum basically
  inaccessible (\cf, reference \cite{Cervera:2000vy}). 
  All measurements have to be performed in the high energy tail of the
  oscillation probability, off the oscillation maximum. 
  This is the reason why it is the facility most affected by the
  eightfold-degeneracy \cite{Barger:2001yr,Huber:2002mx}. 
  Amongst the possible solutions to this problem, the physics reach of
  a `better detector' has been considered \cite{Huber:2002mx}. 
  In the following, reference \cite{Huber:2002mx} is taken as a
  starting point and discuss improvements in the detection threshold
  and energy resolution. 

  Achieving a lower threshold probably requires a finer granularity of 
  the detector, i.e. , a higher sampling density in the
  calorimeter. 
  This should at the same time improve the energy resolution of the
  detector. 
  The energy resolution is parameterisation by 
  $\sigma_E \, [\mathrm{GeV}]=[\sigma \, \sqrt{E_\nu}+ 0.085 ]\,\mathrm{GeV}$ 
  with $\sigma = 0.15$ for the energy resolution (as compared to
  $\sigma_E=0.15 \, E_\nu$ in section \ref{sec:mid}, corresponding to
  $\sigma \simeq 0.5$), where the constant part models a lower limit
  from Fermi motion.
  For definiteness, the neutrino energy threshold is taken to be
  $1\,\mathrm{GeV}$ and a constant efficiency of $0.5$ is taken for
  all neutrino appearance events above threshold. 
  The background model assumes that the
  threshold will only affect events below the threshold, not
  events above, i.e. , there is down-feeding of background but no
  up-feeding. 
  The reason behind this assumption is that a mis-identified 
  neutral-current event should always have a reconstructed energy which is
  lower than the true energy, since there is missing energy in every
  neutral current event. 
  This setup of combined lower threshold, increasing background
  fraction, and better energy resolution will be called `optimal
  appearance'. 
  Similar numbers are quoted for the NO$\nu$A detector
  \cite{Ayres:2004js}.

  The following setups will be compared:
  \begin{enumerate}
    \item {\it Standard detector:} as described in section
          \ref{sec:mid};
    \item {\it Optimal appearance:} 
          $\sigma=15\%$, $\beta=10^{-3}$, full efficiency of 50\%
          already reached at $1 \, \mathrm{GeV}$;
    \item {\it Better threshold:} 
          Same as (b), but $\sigma=50\%$ as for (a); and
   \item {\it Better energy resolution:} Same as (b), but old
         threshold from (a).
  \end{enumerate}

  As before, it is assumed that the systematic uncertainty on the
  background is $20\%$ and the corresponding uncertainty for the
  signal is $s=2.5\%$ for all these setups.
  To a very good approximation, it is safe to say that varying $s$
  from $1\%$ to $5\%$ does not change the results at all. 
  On the other hand, the weight factor $\beta$ is only important so
  long as it does not become too large, but even an increase of a
  factor of 10 is not devastating.
  Note, however, that the error on the background is quite
  conservative compared to the numbers usually quoted for super-beams.
  Most certainly, the impact of an increased background will be
  strongly reduced by reducing this uncertainty. 
  For more details see figure 13 of \cite{Huber:2006wb}.

 \item{\it Impact on physics reach:}

  Changing the detector threshold by a significant amount certainly
  should affect the choice of the optimal baseline and muon energy.
  In the left panel of figure \ref{fig:th13le}, the sensitivity to
  $\stheta$ at $5\sigma$ is shown for the optimal detector as a
  function of the baseline and muon energy, including the effect of
  degeneracies.
  The maximal reach, marked by the diamond, is 
  $\stheta = 1.1 \cdot 10^{-4}$.
  It is reached for $L \sim 7\,500\,\mathrm{km}$ and
  $E_\mu=24\,\mathrm{GeV}$.  
  Compared to figure \ref{fig:t13sens} (lower right), a second maximum
  in the sensitivity is present at shorter baselines even when
  degeneracies are included. 
  Energies as low as $20\,\mathrm{GeV}$ work reasonably well for both
  baselines. 
  It is interesting to see whether the improvements are mainly due to
  the lower threshold or the better energy resolution.
  This is illustrated in figure \ref{fig:th13le}(right), where
  different combinations of lower threshold or better energy
  resolution are compared with the standard setup on the basis of the
  sensitivity to $\stheta$ (in this figure, $E_\mu$ is fixed to 
  $50 \, \mathrm{GeV}$).
  The main effect for the $\stheta$ sensitivity improvement clearly
  comes from a lower energy threshold, the better energy resolution
  playing a very minor role. 
  Note that the maximum in this figure occurs at around 
  $3\, 000 \, \mathrm{km}$ for the optimal detector because the muon
  energy has been fixed.
  A comparison to figure \ref{fig:th13le}(left) shows that this is not
  the global maximum in $(L,E_\mu)$-space.
  \begin{figure}
    \begin{center}
      \includegraphics[width=0.45\textwidth]{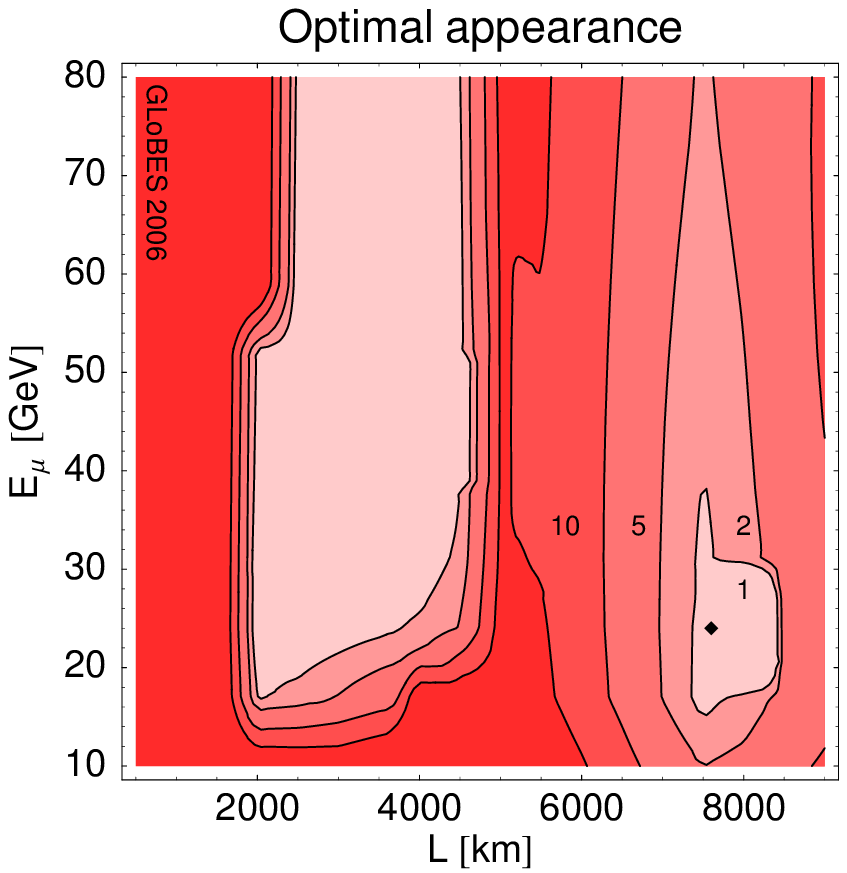}\hspace*{0.4cm}%
      \raisebox{-0.5cm}{\includegraphics[width=0.51\textwidth]{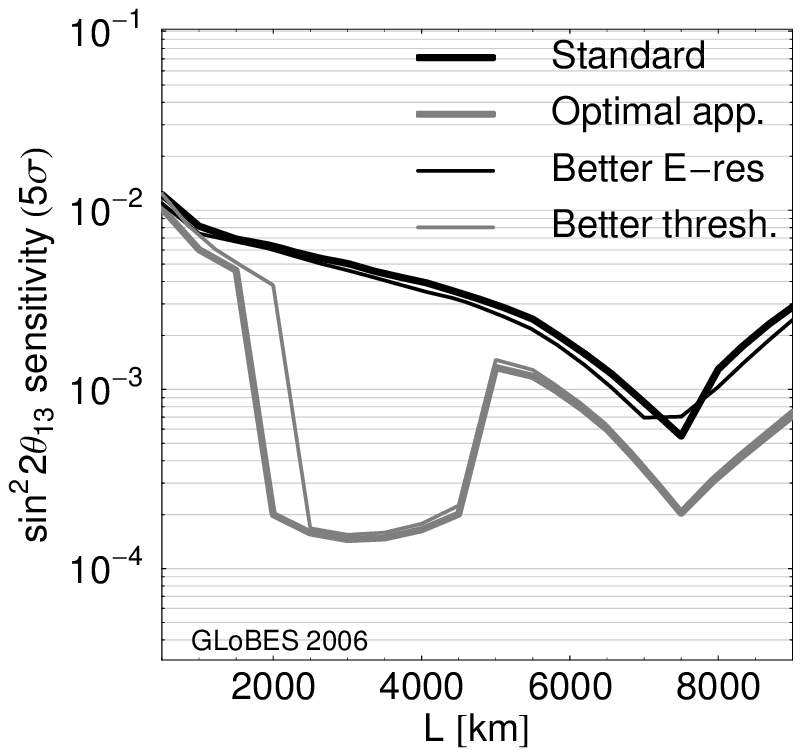}}
    \end{center}
    \caption{
      $\stheta$ sensitivity at $5\sigma$ for several improved detector
      options.  
      The left hand panel shows the $\stheta$ sensitivity as a function
      of baseline and muon energy relative to the maximal reach for the
      `optimal appearance' detector including degeneracies similar to
      figure \ref{fig:t13sens} (lower right).  
      The maximal reach, marked by a diamond, is $\stheta=1.1 \cdot
      10^{-4}$.
      The right hand panel shows the $\stheta$ sensitivity as a function
      of the baseline for different detector options (see plot legend)
      and fixed $E_\mu=50 \, \mathrm{GeV}$. 
      Note that the better energy resolution option uses a different
      background model, which leads to the crossing with the
      `standard' curve at $L \sim 7 \,500$~km.
      Taken with kind permission of the Physical Review from figure
      10 in reference \cite{Huber:2006wb}.
      Copyrighted by the American Physical Society.
    }
    \label{fig:th13le}
  \end{figure}

  The behaviour of the sensitivities to CP-violation and mass hierarchy
  is substantially the same, as is shown in figure
  \ref{fig:mhcpoptdet}
  In this figure $\delta=3 \pi/2$ was chosen, since for this
  value degeneracies have a larger impact than for $\delta = \pi/2$
  and any improvement is more obvious.
  The left panel shows the sensitivity to the mass hierarchy at
  $3\sigma$, where sensitivity is given within the shaded/marked
  areas.  
  The red (dark) shaded regions show the results for the standard
  detector whereas the blue (light) shaded regions show the result for
  the optimal setup. 
  Clearly, the accessible range in $\stheta$ improves and the
  constraints on the baseline become somewhat weaker for the better
  detector. 
  The difference between having only a better threshold (dashed line)
  and only a better energy resolution (solid line) is quite large. 
  The same happens for the sensitivity to CP-violation, figure
  \ref{fig:mhcpoptdet} (right panel).
  For all the sensitivities considered, large improvements come from a
  lower threshold, while the improved energy resolution makes only a
  minor contribution. 
  The choice of the optimal $L$ and $E_\mu$ seems to be essentially
  unaffected by a better detector.
  \begin{figure}
    \begin{center}
      \includegraphics[width=\textwidth]{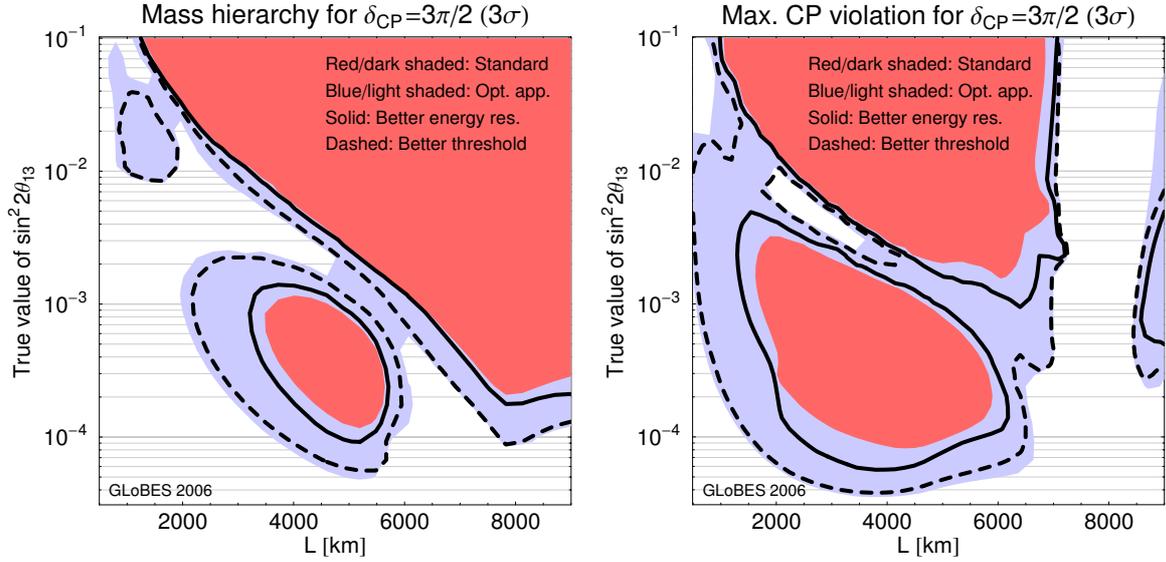}
    \end{center}
    \caption{
      The normal mass hierarchy (left panel) and CP-violation (right
      panel) sensitivities (at $3 \sigma$) as a function of baseline
      and true $\stheta$ for a normal hierarchy and 
      $\delta=3 \pi/2$, different detector options (see legend) and
      fixed $E_\mu=50 \, \mathrm{GeV}$.  
      Sensitivity is given in the shaded/enclosed regions.
      Taken with kind permission of the Physical Review from figure
      11 in reference \cite{Huber:2006wb}.
      Copyrighted by the American Physical Society.
    }
    \label{fig:mhcpoptdet}
  \end{figure}

  At this stage it is not clear how difficult it will be to push the
  threshold to lower values. 
  The previous sections have demonstrated that the measurement of
  $\delta$ is the most demanding for the detector. 
  Figure \ref{fig:thresh} shows the CP-violation discovery potential
  at 3$\sigma$ (depicted as the CP-fraction) for several low
  energy thresholds, for the optimal appearance
  detector.
  Lowering the threshold to 5 GeV is enough to resolve most of the
  degeneracies at intermediate $\theta_{13}$. 
  On the other hand, a significant gain is observed for large
  $\theta_{13}$ for thresholds below 5 GeV.
  \begin{figure}
    \begin{center}
      \includegraphics[width=0.5\textwidth]{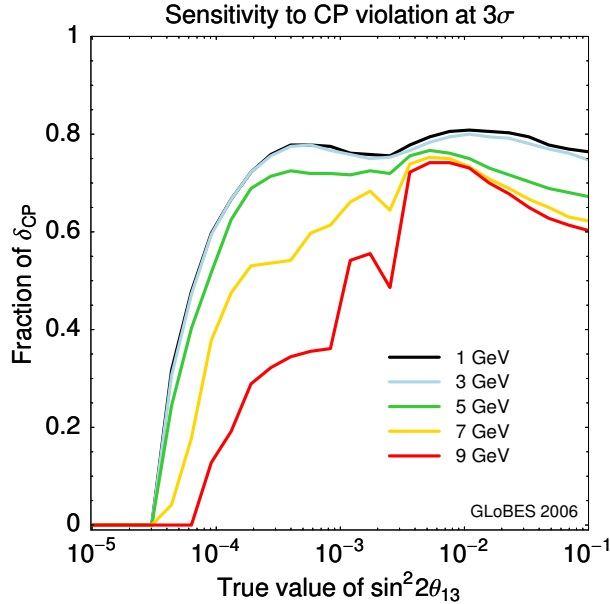}
    \end{center}
    \caption{
      CP discovery reach at $3\sigma$ for the optimal appearance
      detector at $L=4\, 000\,\mathrm{km}$ for various choices of the
      energy threshold as explained in the legend.
    }
    \label{fig:thresh} 
  \end{figure}

  One important issue in this context is the performance of a Neutrino
  Factory if $\stheta$ turns out to be large, $\stheta \sim 10^{-1}$. 
  There will be information regarding this case from reactor experiments
  by around 2010 (see references \cite{Ardellier:2004ui,Huber:2006vr}
  for Double Chooz). 
  Note that $\stheta$ discovery and mass hierarchy measurement are
  rather easy for large values of $\stheta$, which means that the
  optimisation is focused on the measurement of $\delta$.

  Figure \ref{fig:CPfracoptdet} shows the fraction of $\delta$ for
  which the sensitivity to CP violation is at or above the $3\sigma$
  level as a function of the baseline for $\stheta=10^{-1}$ and
  different combinations of experimental setup and matter-density
  uncertainty.
  For comparison the CP-fraction for which \JHFHK\ would be sensitive to
  CP-violation is shown; super-beams can be competitive for large
  $\theta_{13}$.
  In the left panel the results are shown for the canonical value of the
  matter-density uncertainty of $5\%$. 
  Clearly, the standard Neutrino Factory setup does not perform better
  than the super-beam. 
  The situation changes once better detectors are considered. 
  The optimal setup defined previously would yield a significant
  improvement over the super-beam for nearly all choices of the baseline
  above $1500\,\mathrm{km}$.  
  It also can be seen that the improvement comes from both the lower
  threshold and better energy resolution. 
  In this scenario, the detector performance is crucial in making the
  case for a Neutrino Factory.
  The right panel shows the result if the matter-density uncertainty
  could be reduced to $1\%$. 
  Quite obviously, this would further improve the performance of the
  Neutrino Factory.
  These results for the optimal detector hold for a lower muon energy
  around $20\,\mathrm{GeV}$ as well.
  \begin{figure}
    \begin{center}
      \includegraphics[width=\textwidth]{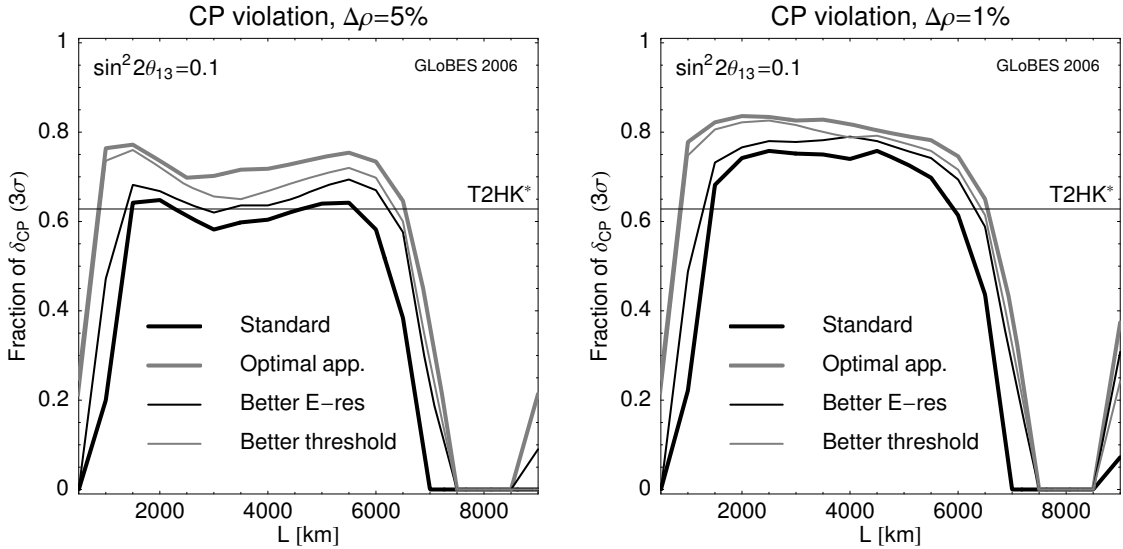}
    \end{center}
    \caption{
      The CP-fraction for the sensitivity to CP-violation (at
      $3\sigma$) for a normal hierarchy as function of baseline for
      different detector options (see legend) and 
      $E_\mu=50 \, \mathrm{GeV}$. 
      The left (right) panel corresponds to 5\% (1\%) matter density
      uncertainty.
      Taken with kind permission of the Physical Review from figure
      12 in reference \cite{Huber:2006wb}.
      Copyrighted by the American Physical Society.
    } 
    \label{fig:CPfracoptdet} 
  \end{figure}

  In the case of large $\stheta$, improving the detector energy
  resolution and energy threshold would allow a shorter baseline of
  about $1500\,\mathrm{km}$ and a muon energy of $20\,\mathrm{GeV}$ to
  be chosen, while the option 
  $4\, 000 \, \mathrm{km}$ at $50 \, \mathrm{GeV}$ does not mean a
  significant loss in sensitivity (depending on the matter-density
  uncertainty, the loss is about 5\% to 8\% in the CP-fraction).
  Furthermore, for one Neutrino Factory baseline only, it can be
  concluded that lower threshold, better energy resolution, and lower
  matter-density uncertainty would equally help to improve the
  performance. 
\end{enumerate}

%% file: 05-Performance/05-04-Neutrino-Factory/05-04-04-Neutrino-Factory-best.tex
\subsubsection{The optimal Neutrino Factory}
\label{sec:best}

The optimised setups from the previous sections are compared below.
The baseline and muon-energy optimisation is not discussed further, 
rather, these parameters are fixed from the earlier discussion. 
The muon energy is fixed, unless otherwise stated, to 
$E_{\mu} = 50 \, \mathrm{GeV}$.
Note that the matter-density uncertainty is assumed to be correlated
among all channels at the same baseline.

For the optimal baseline, CP-violation measurements favour a baseline
around $4 \, 000 \, \mathrm{km}$ (but baselines between 
$3 \, 000 \, \mathrm{km}$ and $5 \, 000 \, \mathrm{km}$ do not 
affect the sensitivity too much). 
For large values of $\stheta$, shorter baselines 
$L \gtrsim 1 \, 500 \, \mathrm{km}$ are possible as well. 
Note that the short baseline 
($L \lesssim 5 \,000 \, \mathrm{km}$) is affected by correlations and
degeneracies for small and intermediate values of $\stheta$, which
means that it has moderate $\stheta$ and mass hierarchy
sensitivities. 
In addition, this result has been tested for larger values of $\ldm$,
and it does not change significantly (whereas the absolute physics
potential increases).

As far as baseline upgrades are concerned, a degeneracy-solving 
baseline is necessary to improve the $\stheta$ sensitivity, the
$\stheta$ discovery reach, and the mass-hierarchy discovery reach. 
A baseline in the range $L \sim 7 \, 000 - 7 \,500 \, \mathrm{km}$  
(i.e., the magic baseline) can play this role, since the appearance
probability does not depend on $\delta$ at this distance and the 
intrinsic-degeneracy can be solved unambiguously independent of
the oscillation parameters, possibly over-estimated luminosities,
confidence level, etc. (see reference
\cite{Burguet-Castell:2001ez}).
Furthermore, matter effects are stronger than for the shorter
baseline, which means that the magic baseline is sensitive to
different physics, rather than being simply a luminosity upgrade. 
Moreover, it helps CP-violation measurements at large $\stheta$,
and can establish the MSW effect in the Earth even for
$\stheta=0$ \cite{Winter:2004mt}. 
Since this baseline is useful in all physics scenarios, one may want
to choose a Neutrino Factory setup with two such baselines 
from the very beginning.
The second baseline will be a major challenge from the engineering
point of view. 
However, the physics potential of this baseline is well established 
and the technical feasibility should be rather predictable.
In the plots of this section, the index `MB' refers to the magic
baseline.

For detector upgrades, an improvement of the golden-channel detector
is certainly the main objective. 
In particular, lowering the detection threshold will greatly improve
the physics potential in all physics scenarios and for both the 
mass-hierarchy and the CP-violating-phase measurements. 
It has been demonstrated that an improved detector would allow the use
of a lower parent-muon energy, $E_{\mu} \sim 20 \, \mathrm{GeV}$
instead of $E_{\mu} \sim 50 \, \mathrm{GeV}$, thus reducing the effort
on the accelerator side. 
The improvement of the detector with respect to energy resolution and
threshold should be possible.
Notice that an improved detector will not be able to solve all the
degeneracies on its own.

Between the various additional channels, the platinum channel
($\nu_\mu \to \nu_e$) will be very useful for large 
$\stheta \gtrsim 10^{-2}$ provided the electron-charge-identification
threshold can be increased up to $\sim 10 - 15 \, \mathrm{GeV}$ (see
the right panel of figure \ref{fig:pltresh}) and enough events can be
collected.
The reference 15~Kton magnetised LAr detector of reference
\cite{Rubbia:2001pk} is statistically limited and would not improve
the performance of the Neutrino Factory significantly.
Platinum-channel searches may be implemented in the golden detector
(thus allowing for a 50~Kton magnetised detector, something extremely
difficult for the liquid-argon technique), but the electron-neutrino
detection may not turn out to be technically possible at this level
and might be effective only at much lower energy. 
It must be noted that a detector looking for the platinum channel is 
complementary to the improved golden detector theoretically, since a 
different combination of CP-violation and matter effects would be 
measured and it would permit the measurement of T-violation.
However, it should be a secondary objective after improving the
golden-channel threshold. 

For intermediate values of $\theta_{13}$, the silver and platinum
channels give similar results as degeneracy-solvers, the former having
a slightly larger.
This channel, also, is statistically limited and any possible
improvement of the detector (mass, magnetisation of the emulsions,
better vertex-identification efficiency, etc.) would be extremely
helpful.  
Notice that the silver channel is interesting for applications such
as searches for physics beyond the S$\nu$M or deviations from maximal
mixing; the discussion below is restricted to the measurement of the
parameters of the S$\nu$M.

With the reference MID detector (with MINOS-like performance), the
muon energy of a Neutrino Factory should be in the range $40$~GeV to
$50 \, \mathrm{GeV}$ to be optimised for all measurements.
The muon energy may not have to be as high as $50 \,\mathrm{GeV}$ for  
neutrino-oscillation physics because of the matter resonance in the 
Earth's mantle. 
An improvement of the detection threshold could reduce the muon energy
to $20 \, \mathrm{GeV}$ while achieving excellent physics
sensitivities, and the physics scenario `large $\stheta$' may even
allow for lower energies.
Note that the use of the silver channel disfavours low muon energies,
i.e., $E_\mu$ should be $\sim 25 \, \mathrm{GeV}$ or greater.

\begin{figure}
  \begin{center}
    \includegraphics[width=0.5\textwidth]{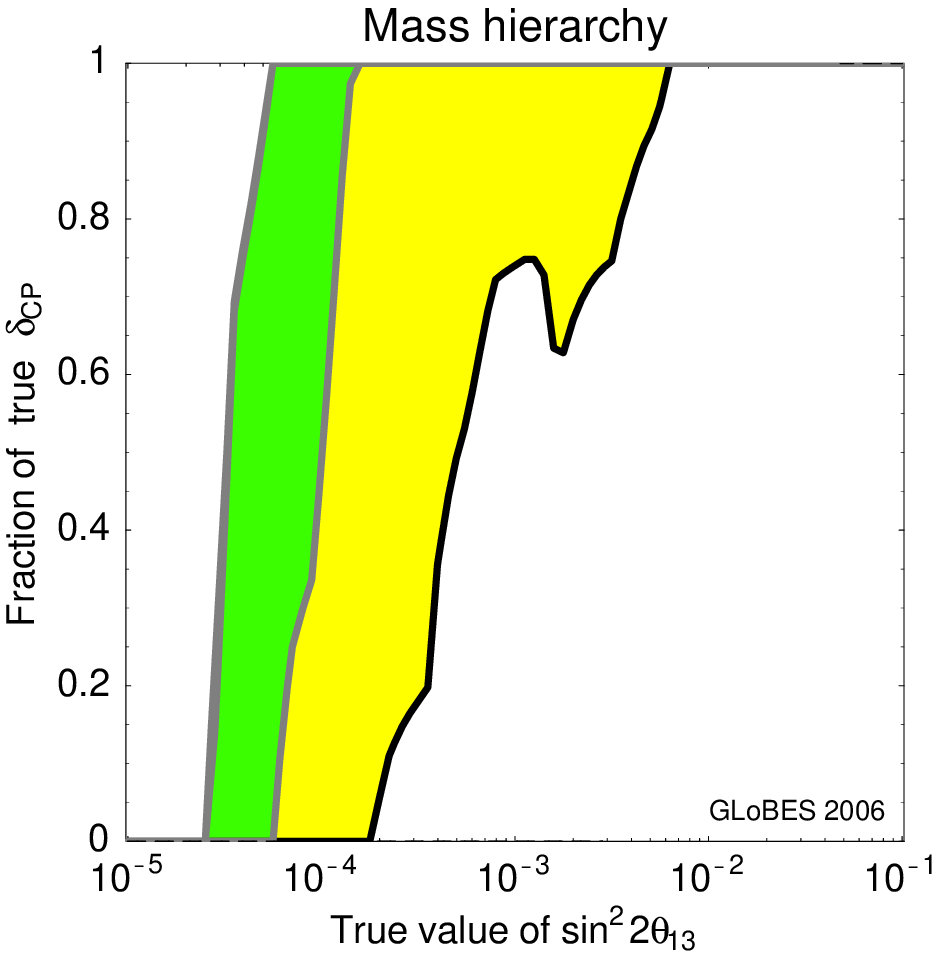}~\includegraphics[width=0.5\textwidth]{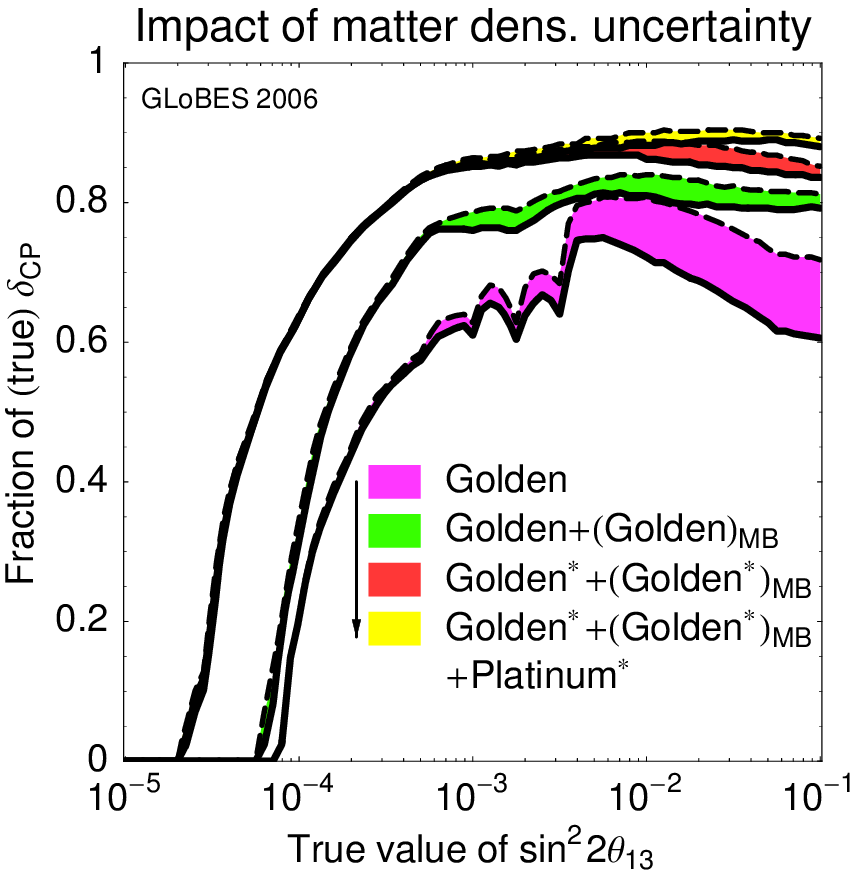}
  \end{center}
  \caption{
    Left panel: CP-fraction of  the sensitivity to the mass hierarchy at
    3$\sigma$.   
    The different shaded areas correspond to successively taking into
    account: 1) the magic baseline (yellow) and 2) an improved
    detector at $E_\mu=20\,\mathrm{GeV}$ (green).  
    Right panel: CP-discovery potential at 3$\sigma$.  
    The different lines correspond to successively taking into account  
    additional optimisations as given in the legend. 
    Solid (dashed) stands for a 5\% (2\%) matter density uncertainty.  
    Shaded areas represent the improvement potential with respect to
    the unknown matter density profile.  Notice that in going from
    Golden to Golden* the muon energy goes down from
    $E_\mu=50\,\mathrm{GeV}$ to $E_\mu=20\,\mathrm{GeV}$.
    Taken with kind permission of the Physical Review from figures
    23 and 24 in reference \cite{Huber:2006wb}.
    Copyrighted by the American Physical Society.
  }
  \label{fig:optsummary}
\end{figure}

The left panel of figure \ref{fig:optsummary} summarises the outcome
of this optimisation discussion by presenting the CP-fraction for the
sensitivity to the mass hierarchy, successively switching on the
magic baseline and the golden$^*$ improved detector.
One can easily read off the excellent combined potential for 
mass hierarchy and CP-violation of the Neutrino Factory below 
$\stheta \lesssim 10^{-2}$. 
Remember that none of the suggested improvements could be achieved
with a simple luminosity upgrade, i.e., adding mass to the
golden-channel detector. 

Finally, it is well known that the matter-density uncertainty is
important for $\stheta$ and $\delta$ measurements at large $\stheta$
(see, \eg , references \cite{Huber:2002mx,Ohlsson:2003ip} for the
relevant regions in parameter space). 
Since the magic baseline and the platinum channel extract the
information on $\stheta$ (and $\delta$) in a different way compared
to the golden channel, one may suspect that the correlation with the
matter density can be partially eliminated. 
The impact of the matter-density uncertainty on our optimisation
summary is shown in the right panel of figure
\ref{fig:optsummary}.  
For the $L=4 \, 000  \, \mathrm{km}$ baseline alone, it can be seen
that the impact of matter density uncertainties is rather large
(`Golden'). 
However, adding the magic baseline and (possibly) the platinum channel
reduce this dependence significantly.
This result is very interesting since in this case an improvement on
the knowledge of the matter density profile may not be necessary
anymore.
Nevertheless, note that a lower matter density uncertainty cannot
replace the detector, channel, and baseline improvements discussed in
this section.

In conclusion, the optimal Neutrino Factory setup for oscillation
parameter measurements has two baselines (at 
$L \sim 1500 -- 4000 \, \mathrm{km}$ and one at 
$L \simeq 7 \, 500 \, \mathrm{km}$, respectively), 
a `better' golden channel detector (with lower threshold and higher 
energy resolution) and a muon energy of 
$E_{\mu} \sim 25 \, \mathrm{GeV}$. 
This set of improvements exhausts the optimisation potential in most
of the parameter space.  
The only region where an additional gain may be achieved is for large
$\stheta\sim 10^{-1}$ (see section \ref{Sect:LowENF}). 
Here, adding a high-mass platinum-channel detector (with electron CID
capability) would decrease the impact of the matter density
uncertainty. 
If, for any reason, the long baseline cannot be implemented,
combination of the golden detector with a standard silver or
platinum detector (with a slight preference for the former) can
significantly improve the performance of the Neutrino Factory for
intermediate $\theta_{13}$.

As far as future Neutrino Factory R\&D is concerned, the ability to
operate two baselines as well as the lower detection threshold of the
golden detector are the most critical components to the optimised
physics potential. 
Furthermore, a better energy resolution of the golden-channel detector
would improve the physics potential further.

%% file: 05-Performance/05-04-Neutrino-Factory/05-04-05-Neutrino-Factory-lowE.tex
\subsubsection{Low-energy neutrino factory}
\label{Sect:LowENF}

In reference \cite{Geer:2007kn}, it has been suggested that a very low
energy Neutrino Factory, where the stored muons have an energy of
$4.12$ GeV, may be exciting if $\theta_{13}$ proves to be large
($\theta_{13} \geq 2^\circ$). 

The primary neutrino-oscillation channel at a Neutrino Factory
requires the identification of wrong-sign muons, and hence a detector
with excellent muon-charge identification. 
Early studies \cite{Albright:2000xi} based on a MINOS-like segmented 
magnetised detector suggested that, to reduce the charge
mis-identification rate to the $10^{-4}$ level while retaining a 
reasonable muon reconstruction efficiency, the detected muon needs to
have a minimum momentum of $\sim 5$~GeV.  
The analysis obtained a 50\% reconstruction efficiency for
charged-current neutrino interactions exceeding $\sim 20$~GeV.
This effectively places a lower limit of about 20~GeV on the desired
energy of the muons stored in the Neutrino Factory (see section
\ref{sec:golden}).
Recently, a refined analysis has shown that, with more sophisticated
selection criteria, high efficiencies ($> 80$\%) can be obtained for
neutrino interactions exceeding $\sim 10$~GeV, with efficiencies
dropping to $\sim 50\%$ by 5~GeV, motivating the proposed improvement
in the magnetised iron detector studied in section \ref{sec:det}. 
This new analysis suggests that a MINOS-like detector could be used at
a Neutrino Factory with energy less than 20~GeV, but probably not less
than 10~GeV.  

Therefore, to consider a lower energy Neutrino Factory, a finer
grained detector that enables reliable sign-determination with good
efficiency for muons of lower energy is needed.   
One way to achieve this could be to use a totally active, magnetised,
segmented detector, of the type proposed for the NO$\nu$A
detector \cite{Ayres:2004js} but within a large magnetic volume.
Initial studies seem to show that, for this technology, the muon
reconstruction efficiency is expected to approach unity for momenta
exceeding $\sim 200$~MeV/c, with a charge mis-identification lower
than $10^{-4}$ ($10^{-3}$) for momenta exceeding approximately
400~MeV/c (300~MeV/c).

Whether these numbers are realistic must be confirmed by further and
more detailed studies.
Nevertheless, with a magnetised far detector concept that makes it
possible to measure neutrino interactions down to about $0.8$ GeV, it
becomes interesting to consider a Neutrino Factory with a stored-muon
energy of a few GeV.  
In present designs for a 25~GeV Neutrino Factory \cite{Geer:2006tb},
there at least two acceleration stages are required to accelerate the
muons from $\sim 1$~GeV to 25~GeV.  
A Neutrino Factory for which the final muon energy is a few GeV would
require only one acceleration stage.

Present Neutrino Factory studies suggest that it would be reasonable
to expect, for a Neutrino Factory with (without) an ionisation-cooling
channel before the pre-accelerator, about $5 \times 10^{20}$ 
($3 \times 10^{20}$) useful positive muon decays per year and 
$5 \times 10^{20}$ ($3 \times 10^{20}$) useful negative muon decays
per year in a given straight section. 
In reference \cite{Geer:2007kn} it is assumed that the same luminosity
can be achieved for a 4.12~GeV Neutrino Factory.
Two setups have been considered: 
\begin{itemize}
  \item{\it Setup A:} 
    Five years data taking with $3 \times 10^{20}$ useful muon decays
    per muon polarity per year; or 
  \item{\it Setup B:}
    Ten years of data taking with $5 \times 10^{20}$ useful muon
    decays per muon polarity per year. 
\end{itemize}
In both cases, a 20~Kton fiducial mass, magnetised, totally active
NO$\nu$A-type detector is considered. 

Assuming the previous hypothesis on the neutrino flux and the far
detector size and performances, the physics potential of this setup
has been studied in reference \cite{Geer:2007kn} for two reference
baselines: $1280$ Km, the distance from Fermilab to Homestake, and
$1480$ Km, the distance from Fermilab to Henderson mine.  
Taking advantage of both the golden channel and of the 
$\nu_\mu \to \nu_\mu$ disappearance (but not of the silver channel,
since the neutrino energy is too low to produce taus), it has been
shown that: 
\begin{itemize}
  \item
    Maximal atmospheric neutrino mixing can be excluded at $99\%$~CL
    if $\sin^2 \theta_{23}<0.48$ ($\theta_{23}<43.8^\circ$); 
  \item 
    If $\theta_{23} \neq 45^\circ$, the $\theta_{23}$-octant is
    identified correctly at $99\%$ CL if $\theta_{13}> 1^\circ$ for
    Setup A and $\theta_{13}> 0.6^\circ$ for Setup B, independent of
    the value of the CP violating phase, $\delta$; 
  \item
    The neutrino-mass hierarchy is identified at the $95\%$ CL; and
  \item 
    The CP violating phase, $\delta$, is measured with a $95\%$ CL
    error lower than $20^\circ$, if $\sin^2 2 \theta_{13}>0.01$ (i.e.
    $\theta_{13}> 3^\circ$) assuming the more conservative exposure
    scenario.
\end{itemize}
All sensitivities are computed assuming $2$ degrees of freedom and a
$2\%$ overall systematic error. 
The statistical error is included, but no background has been
considered. 
Finally, the detector efficiency has been assumed to be $100\%$
above 0.8 GeV, and zero below this threshold.

%% file: 05-Performance/05-05-Comparisons/05-05-Comparisons.tex
\subsection{Comparisons}
\label{SubSect:Perf:CompComb}

The physics reach of second-generation super-beams, beta-beam
facilities, and the Neutrino Factory have been reviewed in detail in
the preceding sections. 
The purpose of this section is to make a quantitative comparison of
the discovery potential  (as defined in section \ref{sec:th13}) of the
three classes of facility for the three unknown quantities 
$\sin^2 2 \theta_{13}$, ${\rm sign} \Delta m_{31}^2$, and $\delta$.

The sensitivity of each of the proposed facilities depends on the
choice of a number of parameters; optimised parameter choices may
require R\&D programmes to be carried out successfully.
To assess the degree to which such R\&D programmes can improve the
physics reach, a `conservative' and an `optimised' set-up is assumed
for each facility; the discovery reach for each facility being
presented as a band, one edge of which corresponds to the conservative
parameter set, the other to the optimised parameter set.
For each setup, appearance and disappearance data taken using both
neutrino and anti-neutrino beams are considered.
In each case, the matter density is assumed to be known with an
uncertainty of 2\%. 
$\theta_{23}$ and $\Delta m^2_{31}$ were assumed to be known within 10\%,
whereas  $\theta_{12}$ and $\Delta m^2_{21}$ were assumed to be known
within 4\%. 
The conservative and optimised set-ups for each of the three types of
facility under consideration are summarised below. 
\begin{itemize}
  \item{Second-generation super-beams:}
    The three super-beam facilities considered, the SPL, T2HK, and the
    wide-band beam experiment, were defined in section
    \ref{Sect:SuperBeamDef}.
    The aspects of these facilities that are most important to the
    performance comparison are summarised below:
    \begin{itemize}
      \item{T2HK} is the proposed upgrade from the T2K experiment. 
        Here a proton-beam power of $4\,\mathrm{MW}$ has been assumed.
        A megaton class water Cherenkov detector with a fiducial mass
        of $440\,\mathrm{kt}$ at a baseline of $295\,\mathrm{km}$ has
        been assumed. 
        The running time assumed was 2 years for neutrinos and 8 years
        for anti-neutrinos (here, one year corresponds to
        $10^7\,\mathrm{s}$).  
        For more details see \cite{Campagne:2006yx};
      \item{SPL} is a CERN-based version of a superbeam. 
        A proton-beam power of $4\,\mathrm{MW}$ a megaton class water
        Cherenkov detector with a fiducial mass of $440\,\mathrm{kt}$
        at a baseline of $130\,\mathrm{km}$ have been assumed. 
        The running time assumed was 2 years for neutrinos and 8 years
        for anti-neutrinos.
        For more details see \cite{Campagne:2006yx}.
      \item{WBB} is the proposal originally put forward by BNL to use
        an on-axis, long baseline, wide-band neutrino beam pointed
        to illuminate a water Cherenkov detector. 
        Here, a proton-beam power of $1\,\mathrm{MW}$ has been assumed
        for neutrino running and a proton-beam power of
        $2\,\mathrm{MW}$ has been assumed for anti-neutrino running.
        The detector assumed was a water Cherenkov detector with a
        fiducial mass of $300\,\mathrm{kt}$ at a baseline of
        $1300\,\mathrm{km}$. 
        The running time assumed was 5 years for neutrinos and 5 years
        for anti-neutrinos. 
        For more details see \cite{Barger:2006vy,Diwan:2006qf}. 
\end{itemize}
    The optimised parameter set corresponds to the assumption of a
    total systematic uncertainty of 2\%. 
    The conservative parameter set assumes a total systematic
    uncertainty of 5\%;

\item{Beta-beam facilities:} The beta-beam facilities were defined in
  section \ref{sec:setups}.  The conservative option is taken to be
  the CERN baseline scenario with stored $^6$He and $^{18}$Ne beams at
  $\gamma=100$ serving a $440\,\mathrm{kt}$ (fiducial) water Cherenkov
  detector at a baseline of $130\,\mathrm{km}$. The running time
  assumed was is 5 years with $2.9\cdot10^{18}$ $^6$He decays per year
  and $1.1\cdot10^{18}$ $^{18}$Ne decays per year. 
  A systematic uncertainty of 2\% was assumed.
  For more details see \cite{Campagne:2006yx}.
  
  The optimised parameter set assumes stored $^6$He and $^{18}$Ne
  beams at $\gamma=350$ illuminating a $440\,\mathrm{kt}$ (fiducial)
  water Cherenkov detector at a baseline of $730\,\mathrm{km}$. The running
  time is 5 years with $2.9\cdot10^{18}$ $^6$He decays per year and
  $1.1\cdot10^{18}$ $^{18}$Ne decays per year. A systematic
  uncertainty of 2\% was assumed. 
  For more details see \cite{Burguet-Castell:2005pa}
  
\item{The Neutrino Factory:} The Neutrino Factory setups were defined
  in section \ref{sec:NFsetup}.  The conservative setup assumes
  $10^{21}$ useful muon decays per year and a stored muon-beam energy
  of 50~GeV. The running time is 4 years with $\mu^+$ and 4 years with
  $\mu^-$.  Neutrino events are recorded in a $50\,\mathrm{kt}$ golden
  detector (defined in section \ref{sec:mid}) at a baseline of 4000~km.
  This detector is assumed to have an appearance $\nu_\mu$ threshold
  rising linearly from 0 at $4\,\mathrm{GeV}$ to its final value at
  $20\,\mathrm{GeV}$. Systematic uncertainties of 2.5\% on the signal
  and 20\% on the background\footnote{The fact that the number of
  background events is small means that the large systematic
  uncertainty on the background-event rate has almost no impact on the
  performance}. 
  For more details see \cite{Huber:2002mx,Huber:2006wb}.
    
  The optimised setup assumes a 20~GeV stored muon beam delivering
  $10^{21}$ muon decays per year and baseline. The running time
  assumed was 5 years with $\mu^+$ and 5 years with $\mu^-$.  Neutrino
  interactions are recorded in two improved golden detectors, called
  golden*. Both have a mass of $50\,\mathrm{kt}$. One is placed at a
  baseline of 4000~km, the second at a baseline of 7500~km. The
  improved detector has a threshold of $1\,\mathrm{GeV}$, above which
  the effeciency is constant. Note, that the results bascially are
  unchanged if the threshold is raised to $3\,\mathrm{GeV}$, since
  there is only a very small neutrino flux between $1\,\mathrm{GeV}$
  and $3\,\mathrm{GeV}$. A systematic uncertainty of 2.5\% has been
  assumed. 
  For more details see \cite{Huber:2006wb}.
\end{itemize}

\begin{figure}
  \begin{center}
    \includegraphics[width=\textwidth]%
      {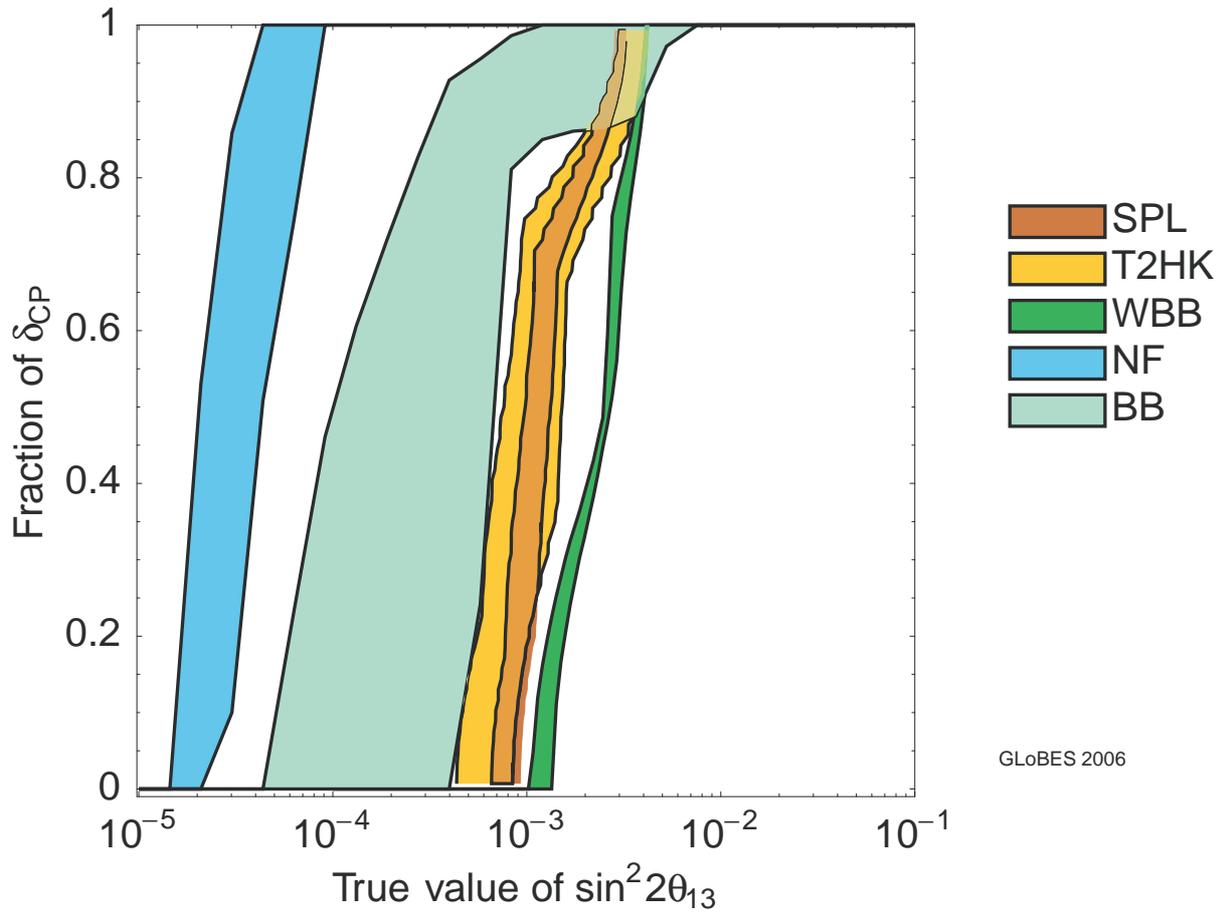}
  \end{center}
  \caption{
    The discovery reach of the various proposed facilities in 
    $\sin^2 2 \theta_{13}$.
    In the area to the right of the bands, $\sin^2 2 \theta_{13} = 0$
    can be excluded at the $3\sigma$ confidence level.
    The discovery limits are shown as a function of the fraction of
    all possible values of the true value of the CP phase $\delta$
    (`Fraction of $\delta_{\rm CP}$') and the true value of 
    $\sin^2 2 \theta_{13}$.
    The right-hand edges of the bands correspond to the conservative
    set-ups while the left-hand edges correspond to the optimised
    set-ups, as described in the text.
    The discovery reach of the SPL super-beam is shown as the orange
    band, that of T2HK as the yellow band, and that of the wide-band
    beam experiment as the green band.
    The discovery reach of the beta-beam is shown as the light green
    band and the Neutrino Factory discovery reach is shown as the blue
    band.
  }
  \label{Fig:Comp:Th13}
\end{figure}

Figure \ref{Fig:Comp:Th13} shows the discovery reach of the various
facilities in $\sin^2 2 \theta_{13}$.
The figure shows the fraction of all possible values of the true value
of the CP phase $\delta$ (`Fraction of $\delta_{\rm CP}$') for which
$\sin^2 2 \theta_{13} = 0$ can be excluded at the $3\sigma$ confidence
level as a function of the true value of $\sin^2 2 \theta_{13}$.
Of the super-beam facilities, the most sensitive is the T2HK with the
optimised parameter set.
The SPL super-beam performance is similar to that of T2HK, while the
performance of the WBB is slightly worse.
The limit of sensitivity of the super-beam experiments is 
$\sim 5 \times 10^{-4}$; for $\sin^2 2 \theta_{13} \gsim 10^{-3}$ the
super-beam experiments can exclude $\sin^2 2 \theta_{13} = 0$ at the
$3\sigma$ confidence level for all values of $\delta$.
The conservative beta-beam set-up has good sensitivity to 
$\sin^2 2 \theta_{13}$ for $\sin^2 2 \theta_{13} \sim 10^{-3}$, but
runs out of sensitivity for values of $\theta_{13}$ only just less
than the sensitivity limit of T2HK.
The optimised ($\gamma=350$) beta-beam has significantly better
performance, with a sensitivity limit of 
$\sin^2 2 \theta_{13} \gsim 5 \times 10^{-5}$.
Both the conservative and the optimised Neutrino Factory set-ups
have a significantly greater $\sin^2 2 \theta_{13}$ discovery reach;
the optimised set-up having a sensitivity limit of 
$\sim 1.5 \times 10^{-5}$.

\begin{figure}
  \begin{center}
    \includegraphics[width=\textwidth]%
      {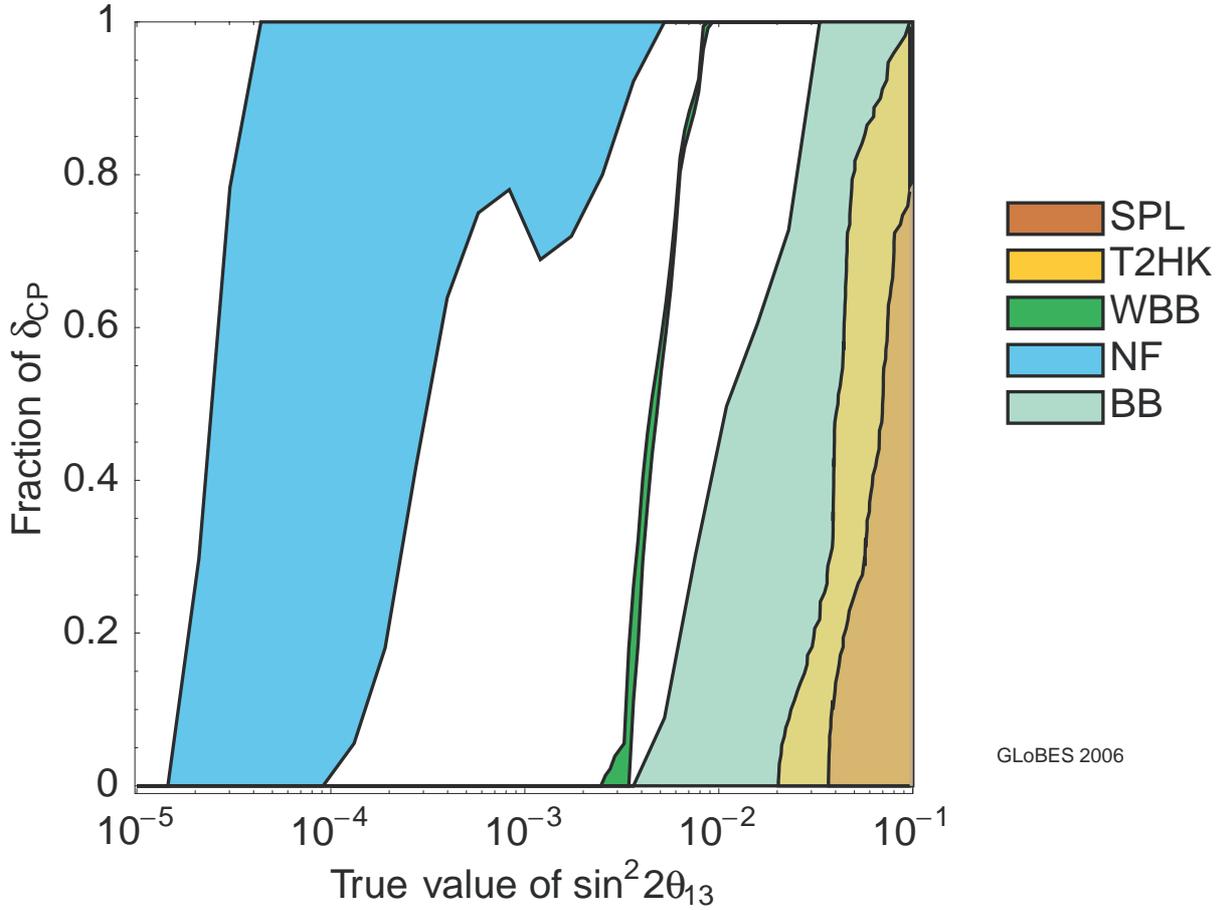}
  \end{center}
  \caption{
    The discovery reach of the various proposed facilities for the  
    discovery of the mass hierarchy.
    In the area to the right of the bands, sign$\Delta m^2_{31}$
    can be established at the $3\sigma$ confidence level.
    The discovery limits are shown as a function of the fraction of
    all possible values of the true value of the CP phase $\delta$
    (`Fraction of $\delta_{\rm CP}$') and the true value of 
    $\sin^2 2 \theta_{13}$.
    The right-hand edges of the bands correspond to the conservative
    set-ups while the left-hand edges correspond to the optimised
    set-ups, as described in the text.
    The discovery reach of the SPL super-beam is shown as the orange
    band, that of T2HK as the yellow band, and that of the wide-band
    beam experiment as the green band.
    The discovery reach of the beta-beam is shown as the light green
    band and the Neutrino Factory discovery reach is shown as the blue
    band.
  }
  \label{Fig:Comp:SgnDM13}
\end{figure}

Figure \ref{Fig:Comp:SgnDM13} shows the discovery reach of the various
facilities in sgn$\Delta m^2_{31}$.
The various bands shown in the figure have the same meaning as those
shown in figure \ref{Fig:Comp:Th13}; the discovery reach is again
evaluated at the $3\sigma$ confidence level.
Of the super-beam set-ups considered only the WBB has significant
sensitivity to the mass hierarchy with a sensitivity limit of 
$\sin^2 2 \theta_{13} \gsim 3 \times 10^{-3}$.
Of the beta-beam set-up only the optimised, $\gamma=350$ option with
the relatively long baseline of 730~km is competitive with the WBB,
having a comparable sensitivity limit.
The Neutrino Factory, benefitting from the long baseline, out-performs
the other facilities. 
The sensitivity limit of the conservative option being
$\sin^2 2 \theta_{13} \gsim 1.5 \times 10^{-4}$, while the sensitivity
limit of the optimised facility is
$\sin^2 2 \theta_{13} \gsim 1.5 \times 10^{-5}$.

\begin{figure}
  \begin{center}
    \includegraphics[width=\textwidth]%
      {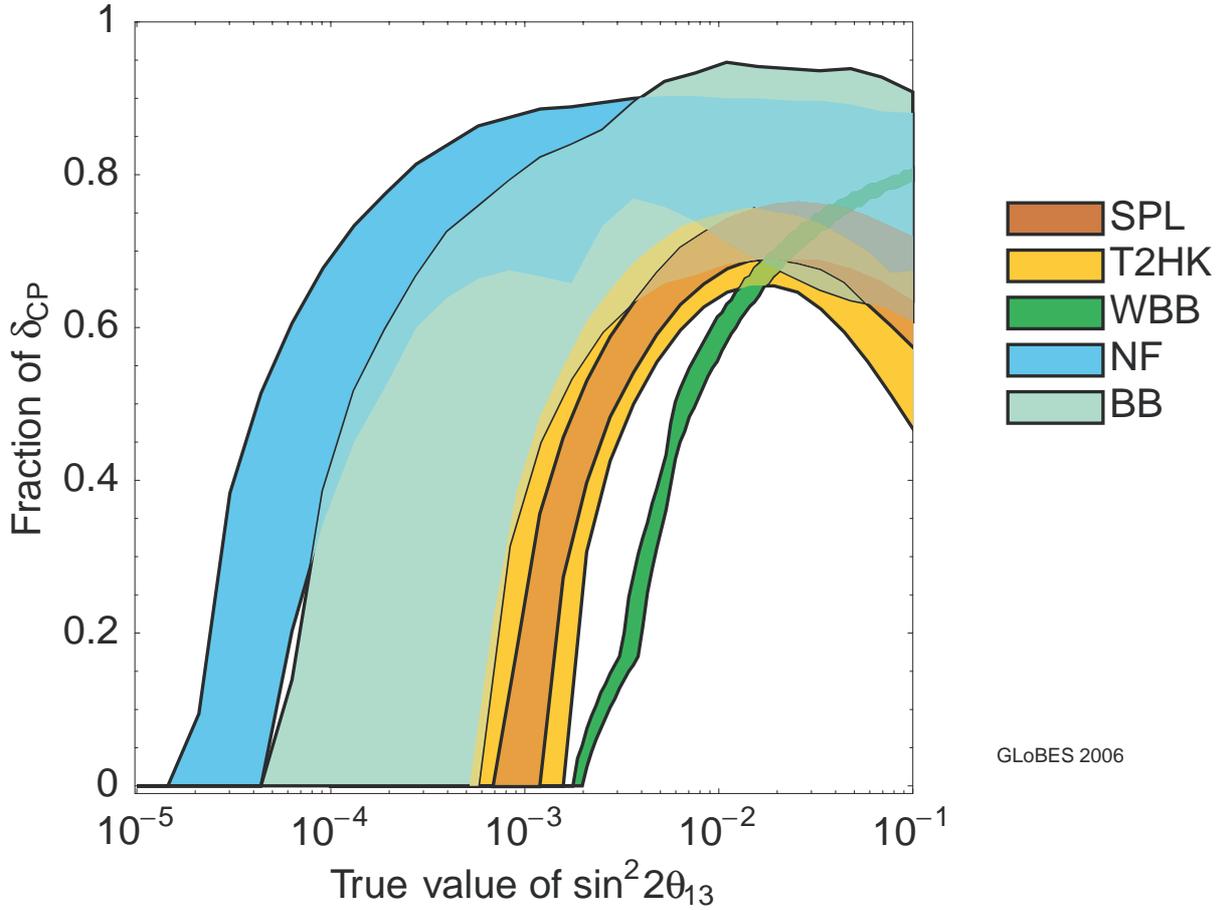}
  \end{center}
  \caption{
    The discovery reach of the various proposed facilities in 
    the CP phase $\delta$.
    In the area to the right of the bands, $\delta = 0$ and 
    $\delta = \pi$ can be excluded at the $3\sigma$ confidence level.
    The discovery limits are shown as a function of the fraction of
    all possible values of the true value of the CP phase $\delta$
    (`Fraction of $\delta_{\rm CP}$') and the true value of 
    $\sin^2 2 \theta_{13}$.
    The right-hand edges of the bands correspond to the conservative
    set-ups while the left-hand edges correspond to the optimised
    set-ups, as described in the text.
    The discovery reach of the SPL super-beam is shown as the orange
    band, that of T2HK as the yellow band, and that of the wide-band
    beam experiment as the green band.
    The discovery reach of the beta-beam is shown as the light green
    band and the Neutrino Factory discovery reach is shown as the blue
    band.
  }
  \label{Fig:Comp:CPV}
\end{figure}

Figure \ref{Fig:Comp:CPV} shows the discovery reach of the various
facilities in the CP phase $\delta$.
The various bands shown in the figure have the same meaning as those
shown in figure \ref{Fig:Comp:Th13}; the discovery reach is again
evaluated at the $3\sigma$ confidence level.
The T2HK and the SPL super-beams show a greater sensitivity to CP
violation for $\sin^2 2 \theta_{13} \sim 10^{-3}$ than the WBB
experiment.
However, the WBB experiment has sensitivity for a larger range
of values of $\delta$ that the other super-beam facilities considered
for $\sin^2 2 \theta_{13} \sim 10^{-1}$.
The performance of the conservative ($\gamma=100$) beta-beam is
comparable to that of the optimised T2HK experiment.
The optimised ($\gamma=350$) beta-beam shows considerably better
performance; a sensitivity limit of $\sim 4 \times 10^{-5}$ and a CP
coverage of around 90\% for $\sin^2 2 \theta_{13} \gsim 10^{-2}$.
For low values of $\theta_{13}$ ($\sin^2 2 \theta_{13} \lsim 10^{-4}$
the conservative Neutrino Factory performance is comparable with that
of the optimised beta-beam. 
For larger values of $\theta_{13}$, the CP coverage of the optimised
beta-beam is significantly better.
The optimised Neutrino Factory out-performs the optimised beta-beam
for $\sin^2 2 \theta_{13} \lsim 4 \times 10^{-3}$.
For larger values of $\theta_{13}$ the optimised beta-beam has a
slightly larger CP coverage.

In summary, for large values of $\theta_{13}$ 
($\sin^2 2 \theta_{13} \gsim 10^{-2}$),
the three classes of facility have comparable sensitivity; the best
precision on individual parameters being achieved at the Neutrino
Factory.
For intermediate values of $\theta_{13}$
($5 \times 10^{-4} \lsim \sin^2 2 \theta_{13} \lsim 10^{-2}$),
the super-beams are out-performed by the beta-beam and the Neutrino
Factory. 
For small values of $\theta_{13}$
($\sin^2 2 \theta_{13} \lsim 5 \times 10^{-4}$),
the Neutrino Factory out-performs the other options.
A significant amount of conceptual design work and hardware R\&D is
required before the performance assumed for each of the facilities can
be realised.
Therefore, an energetic, programme of R\&D into the accelerator
facilities and the neutrino detectors must be established with a view
to the timely delivery of conceptual design reports for the various
facilities.


%% file: 06-Alternatives/06-Alternatives.tex
\section{The potential of other alternatives}
\label{Sect:Alternatives}

\input 06-Alternatives/06-01-Solar-and-reactor/06-01-Solar-and-reactor

\input 06-Alternatives/06-02-Atmospheric/06-02-Atmospheric

\input 06-Alternatives/06-03-0nu2beta/06-03-0nu2beta

\input 06-Alternatives/06-04-Unitarity/06-04-Unitarity

%% file: 06-Alternatives/06-01-Solar-and-reactor/06-01-Solar-and-reactor.tex
\subsection{Solar- and reactor-neutrino experiments}

Possible future solar- and reactor-neutrino experiments are discussed
together in this section.
In addition, a comparative study of the sensitivity of these
experiments to $\ms$ and $\sss$ is presented.

\subsubsection {The Generic $pp$ experiment}

The present generation of solar- and reactor-neutrino experiments will
not be able to determine $\sss$ with an accuracy better than
10\%--15\%. 
To make a more precise measurement of $\sss$ in solar-neutrino
experiments it is necessary to make a precise measurement of the 
$pp$-neutrino flux \cite{Bahcall:2003ce}, 
sub-MeV solar-neutrino experiments (LowNu experiments) are therefore
being planned for the detection of the $pp$ neutrinos using either
charged-current reactions 
(LENS \cite{LENSWWWSite}, 
MOON \cite{MOONWWWSite}, 
SIREN \cite{Liubarsky:2000gq}) 
or electron-scattering
processes 
(XMASS \cite{Namba:2004nz}, 
CLEAN \cite{Nikkel:2005qj}, 
HERON \cite{Lanou:2005ku}, 
MUNU \cite{MUNUWWWSite}, 
GENIUS \cite{GENIUSWWWSite}) 
\cite{lownu}.  

\begin{figure}
  \begin{center}
    \includegraphics[width=11.0cm,height=9cm]
    {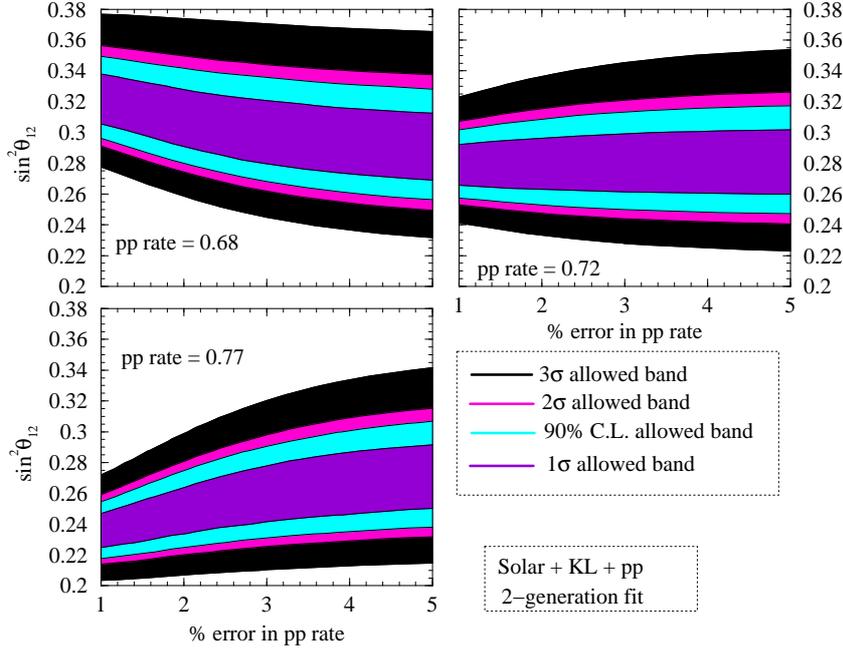}
    \caption{\label{Fig:SnuM:pp}
      Sensitivity plot showing the C.L. (1 dof) allowed range of
      $\sin^2\theta_{12}$ as a function of the error in $pp$ rate 
      for three different values of measured $pp$ rate.
      Adapted with kind permission of the Physical Review from figure 3
      in reference \cite{Bandyopadhyay:2004cp}.
    }
  \end{center}
\end{figure}

Figure \ref{Fig:SnuM:pp} shows the dependence of the sensitivity of
solar-neutrino measurements to $\sin^2 \theta_{12}$ on the precision
with which the $pp$ flux is known \cite{Bandyopadhyay:2004cp}.
The results are for a generic $\nu_e$-$e$ scattering experiment with a
threshold of 50 keV. 
The figure shows the two-generation allowed range of $\sss$ from the 
global analysis of \kl and solar data including the LowNu $pp$ rate,
as a function of the error in the $pp$ measurement.
Three illustrative $pp$ rates of 0.68, 0.72, and 0.77 are considered
and the experimental error in the $pp$ measurement is varied from 1\%
to 5\%. 
By adding the $pp$-flux data in the analysis, the error on
$\sin^2\theta_{12}$ reduces to 14\% (19\%) at 3$\sigma$ for a 1\%
(3\%) uncertainty in the measured $pp$ rate
\cite{Bandyopadhyay:2004cp}.  
Performing a similar three-neutrino oscillation analysis it is found
that, as a consequence of the uncertainty on $\sin^2\theta_{13}$, the
error on the value of $\sin^2\theta_{12}$ increases to 
17\% (21\%) \cite{Bandyopadhyay:2004cp}.  

\subsubsection{The SK-Gd reactor experiment}

A detector with the fiducial mass of the Super-Kamiokande (SK)
detector that was sensitive to reactor neutrinos would be able to make
a very precise measurement of $\theta_{12}$.
In view of this, there has been a proposal to dope SK with 
gadolinium (Gd) by dissolving 0.2\% of gadolinium chloride in the SK
water \cite{Beacom:2003nk}. 
SK receives the same reactor flux as KamLAND and, in principle, could
detect these reactor $\anue$ through inverse beta decay.
The inverse beta-decay process produces an electron and a neutron: the
electron produces \v Cerenkov light which can be detected; the neutron
must be detected through neutron capture.
Unfortunately, neutron capture on a proton releases a photon with an
energy of only 2.2~MeV, which can not be detected in SK.
The addition of Gadolinium circumvents this problem since 
neutron capture on gadolinium releases an 8~MeV $\gamma$ cascade
which is above the SK threshold.
With its 22.5 kton of ultra pure water, the SK detector offers a
target with $1.5\times 10^{33}$ free protons for the antineutrinos
coming from the various reactors in Japan. 
Therefore, for the same measurement period, the SK-Gd reactor
experiment may be expected to yield a data set roughly 43 times that
which can be provided by the \kl experiment.  

\begin{figure}
  \begin{center}
  \includegraphics[width=10.0cm,height=8cm]
  {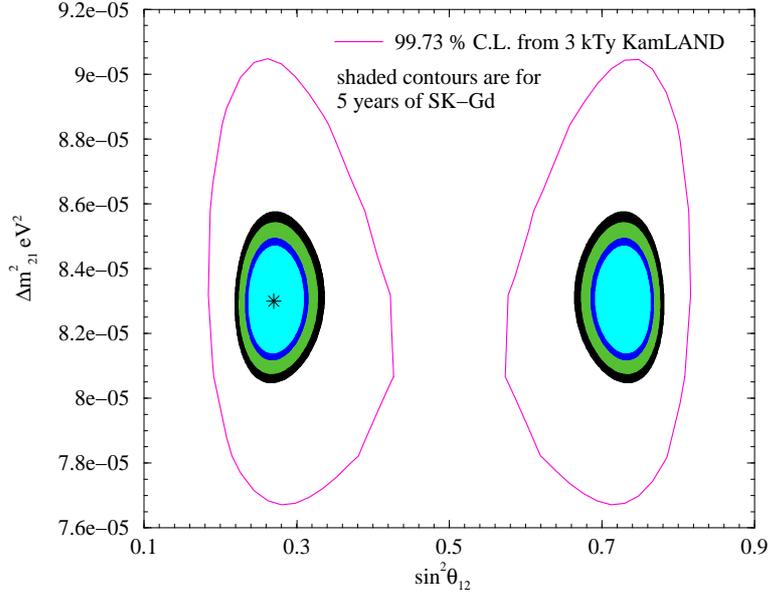}
    \caption{\label{Fig:SnuM:skgd}
      The 90\%, 95\%, 99\%, 99.73\% C.L. (2 dof) allowed regions in the
      $\ms-\sss$ plane from an analysis of prospective data, obtained in
      5 years of running of the SK-Gd detector. 
      The open contours shows the 99.73\% C.L. allowed areas expected
      from 3 kTy of \kl data. The definition of the C.L. correspond to a
      two parameter fit.  
      Taken with kind permission of Springer-Verlag GMBH from 
      reference \cite{Choubey:2005es}.
      Copyrighted by the Springer-Verlag GMBH.
    }
  \end{center}
\end{figure}

In \cite{Choubey:2004bf}, the reactor-$\anue$ data expected 
in the proposed SK-Gd detector is simulated for 
$\ms=8.3\times 10^{-5}$ eV$^2$, $\sss = 0.27$, and  
divided into 18 energy bins, with a visible-energy threshold
of 3~MeV and bin width of 0.5~MeV. 
The precision with which the parameters $\ms$ and $\sss$ can be
determined after a five-year exposure is shown in figure
\ref{Fig:SnuM:skgd} \cite{Choubey:2004bf}.
Also shown for comparison in the figure is the 99.73\% C.L. line
expected from a 3~kTy exposure of KamLAND. 
The precision expected from SK-Gd is superior in both $\ms$ and
$\sss$.
The \sig{} spread in $\ms$ and $\sss$ expected from five-years data taking in
SK-Gd would be at the level of 2\%--3\% and 18\% respectively
\cite{Choubey:2004bf}. 
This is to be compared with the corresponding spread of 6\% and 32\%
expected from 3~kTy of \kl data. 
Results for a similar experimental set-up in Europe and the
corresponding accuracy in the measurement of $\ms$ and $\sss$ has been
studied recently \cite{Petcov:2006gy}.

\subsubsection{The SPMIN reactor experiment}

\begin{figure}
  \begin{center}
  \includegraphics[width=11.0cm,height=9cm]
  {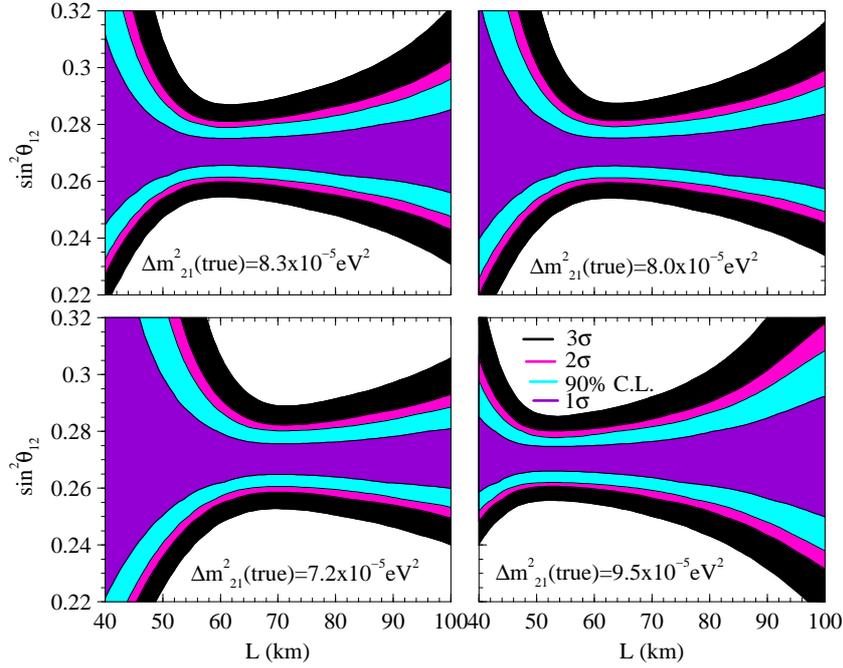}
    \caption{\label{Fig:SnuM:spmin}
      Sensitivity plots for the SPMIN reactor experiment showing the
      $1\sigma$,  $1.64\sigma$, $2\sigma$, and $3\sigma$ (1 dof) range of
      allowed values for $\sss$ as a function of the baseline $L$. 
      Taken with kind permission of the Physical Review from figure 9 in
      reference \cite{Bandyopadhyay:2004cp}.
      Copyrighted by the American Physical Society.
    }
  \end{center}
\end{figure}

The solar mixing angle could be measured with great accuracy in a
reactor experiment with the baseline tuned to the Survival Probability
MINimum (SPMIN) \cite{Bandyopadhyay:2003du}.
Figure \ref{Fig:SnuM:spmin} shows the $\sss$ sensitivity expected in a
reactor experiment as a function of the baseline $L$
\cite{Bandyopadhyay:2004cp}. 
The sensitivity has been evaluated on the assumption of a total
systematic uncertainty of 2\% and a data set corresponding to 73~GWkTy
(given as a product of reactor power in GW and the exposure of the
detector in kTy).
The true value of $\sss$ is assumed to be $0.27$ and the positron
spectrum that would be observed in the detector is simulated for four 
different assumed values of for $\ms$.  
The spectrum is thus simulated at each baseline and the range of
values of $\sss$ allowed by the experiment is plotted as a function of
the baseline. 
The baseline at which the band of allowed values of $\sss$ is
narrowest is the ideal baseline for the SPMIN reactor experiment. 
The figure confirms that this ideal baseline depends critically on the
true value of $\ms$.
The optimal baseline for $\ms=8.0(8.3)\times 10^{-5}$ eV$^2$ is 
63~km (60~km).  
At the optimal baseline, the SPMIN reactor experiment can achieve a
precision of $\sim 2(6)\%$ at $1\sigma(3\sigma)$ in the measurement of
$\sss$ \cite{Bandyopadhyay:2004cp,Minakata:2004jt}.

Figure \ref{Fig:SnuM:spmin} gives the impression that the optimal
baseline for a given value of $\ms$ is very well defined. 
However, note that in figure \ref{Fig:SnuM:spmin} $\ms$ was allowed 
to vary freely. 
The uncertainty in the $\ms$ measurement translates to extra
uncertainty in the $\sss$ measurement. 
If $\ms$ could be measured to a very high
precision in some other experiment, such as \kl or SK-Gd, then the
uncertainty in $\sss$ due to $\ms$ can be reduced significantly.
If $\ms$ was kept fixed, the choice of the baseline for the SPMIN
experiment becomes much broader \cite{Bandyopadhyay:2004cp}.

The measurement of $\sss$ is statistics limited making a large
exposure very important.
For example, the sensitivity to $\sss$ improves from 3(10)\% to 2(6)\%
at $1\sigma(3\sigma)$ as the exposure is increased from 20~GWkTy to
60~GWkTy.
The effect of systematics on the $\sss$ measurement can be checked by
repeating the analysis with a more conservative estimate of 5\% for
the systematic uncertainty. 
For $\ms({\rm true})=8.3 \times 10^{-5}$ eV$^2$, the spread in $\sss$
at $L=60$~km increases from 6.1\% to 8.6\% at $3\sigma$, as the
systematic error is increased from 2\% to 5\% . 
Finally, the impact of the error on $\theta_{13}$ on the precision of
$\sss$ is to increase the uncertainty in $\sss$ from 6.1\% to 8.7\% at
$3\sigma$, for $\ms({\rm true})=8.3 \times 10^{-5}$ eV$^2$ and
$L=60$~km \cite{Bandyopadhyay:2004cp}.

%% file: 06-Alternatives/06-02-Atmospheric/06-02-Atmospheric.tex
\subsection{Atmospheric neutrino experiments}
\label{sec:atm_future}

The effect of the sub-dominant terms in the Super-Kamiokande (SK)
atmospheric-neutrino data is not yet statistically significant.
However, the sub-dominant terms, if observed in a future high
statistics atmospheric-neutrino experiment, can be used to constrain:
the extent to which $\theta_{23}$ differs from $45^{\circ}$; the octant
in $\theta_{23}$ is to be found; and ${\rm sgn}(\Delta m_{31}^2)$.

Assuming that the matter-density is constant, the excess of
electron-type events in a water-\v Cerenkov experiment such as SK is
given by
\cite{Peres:1999yi,Peres:2003wd,Petcov:1998su,Akhmedov:1998ui,Bernabeu:2001xn,Bernabeu:2003yp}:
\bea
  \frac{N_e}{N_e^0} - 1 &\simeq& 
  \sin^22\theta_{12}^M
  \sin^2\left(\frac{(\ms)^M L}{4E}\right ) 
  \times (r\cos^2\theta_{23} - 1) \nonumber \\
  &+& \sin^22\theta_{13}^M
  \sin^2\left(\frac{(\ma)^M L}{4E}\right )
  \times (r\sin^2\theta_{23} - 1) \nonumber \\
  &+& \sin\theta_{23}\cos\theta_{23} ~ r ~ Re
  \bigg[A_{13}^* A_{12} \exp(-i\delta)\bigg]~,
  \label{eq:electronexcess}
\eea
where $L$ is the baseline, $E$ is the energy of the neutrino,
$r=N_e/N_\mu$, $N_e$ and $N_\mu$ being the number electron and muon
events respectively in the detector in the absence of oscillations, and
$\theta_{12}^M$, $\theta_{13}^M$, $(\ms)^M$ and $(\ma)^M$ are the mixing angle
and mass-squared differences in matter. 

The first term in equation (\ref{eq:electronexcess}) is the 
$\Delta m_{21}^2$-driven oscillation term -- which is more important
for the sub-GeV neutrino sample. 
Since $r\simeq 0.5$ in the sub-GeV regime, this term brings an 
excess (depletion) of sub-GeV electron events if $\theta_{23} < \pi/4$
($\theta_{23} > \pi/4$). 
It can thus be used to study the maximality and octant of
$\theta_{23}$ through the sub-GeV electron sample
\cite{Peres:1999yi,Gonzalez-Garcia:2004cu}.
The second term is the $\theta_{13}$-driven oscillation term. 
Being dependent on $\sin^2\theta_{23}$, this term goes in the opposite
direction to the first term. 
Therefore, for sub-GeV neutrinos, larger $\theta_{13}$ would imply
that the effect of the first term would be suppressed by this term. 
However, for multi-GeV neutrinos, there will be large matter effects
inside the earth and this is the dominant term for the electron
neutrinos. 
The $\sin^2\theta_{23}$ dependence of this term could then be used to
study the maximality and the octant of $\theta_{23}$ through the
multi-GeV electron sample \cite{Choubey:2006jk,Huber:2005ep}.
Since matter effects bring in sensitivity to the ${\rm sgn}(\Delta m_{31}^2)$, 
this term can be used to study the mass hierarchy.
The last term is the `interference' term \cite{Peres:2003wd}, which
depends on  $\delta$. 
The effect of this term could be to dilute the effect of the first
two terms and spoil the sensitivity of the experiment. 
However, being directly dependent on $\delta$, this term also
brings in some sensitivity to the CP phase itself
\cite{Peres:2003wd,Fogli:2005cq}. 

The depletion of muon events in the limit of $\Delta m_{21}^2=0$ 
is given by:
\bea
  1 - \frac{N_\mu}{N_\mu^0} &=& 
  (P^1_{\mu\mu} + P^2_{\mu\mu}) + (P^3_{\mu\mu})^\prime \sin^2\theta_{23}
  (\sin^2\theta_{23} - \frac{1}{r}) \, ,
\eea
where:
\bea
  P^1_{\mu\mu} &=& \sin^2 \theta^M_{13} {\sin^2 2\theta_{23}} \sin^2
  {\left[({A}+\Delta m_{31}^2) - {(\Delta m_{31}^2)}^M\right]L 
  \over 8E} \, ; \\
  P^2_{\mu\mu} &=& \cos^2 \theta^M_{13} {\sin^2 2\theta_{23}} \sin^2
  {\left [({A}+\Delta m_{31}^2 ) + {(\Delta m_{31}^2)}^M\right ]L \over 8E} 
  \, ; \\
  (P^3_{\mu\mu})^\prime 
  &=& \sin^2 2\theta^M_{13} \sin^2
  {{(\Delta m_{31}^2)}^M L\over 4E} \, ;
  \label{eq:muondeficit}
\eea
and $A=2\sqrt{2}G_FN_eE$ is the matter potential.
The approximation of a vanishing $\Delta m_{21}^2$ has been made in
equation (\ref{eq:muondeficit}) only for the sake of simplicity, since
the main sub-dominant effect in the muon-neutrino channel comes from
matter effects, which are large for multi-GeV neutrinos for
which the $\Delta m_{21}^2$ dependence is of less importance. 
The results presented in later sections have been obtained using
the full numerical solution of the three-generation equation.
For small values of $\theta_{13}$, matter effects are very small and
$P^{2}_{\mu\mu}$ is the dominant term in the survival probability.
Since this term depends on $\sin^2 2\theta_{23}$, in the absence of
matter effects, sensitivity to the $\theta_{23}$ octant is not
expected from experiments probing the $P_{\mu\mu}$ channel alone. 
However, if $\theta_{13}$ is not small, neutrinos which travel through 
large baselines suffer large matter effects. 
The mixing angle $\theta_{13}$ increases in matter and the third term
$(P_{\mu\mu}^3)^\prime$ becomes important as well. 
Since this term has a strong dependence on $\sin^2\theta_{23}$, rather
than $\sin^2 2\theta_{23}$, the $P_{\mu\mu}$ channel is expected to
develop sensitivity to the octant of $\theta_{23}$ in the presence of
large matter effects \cite{Choubey:2005zy}.
Also, by probing matter effects in the resultant muon signal,
the neutrino-mass hierarchy can be probed
\cite{Bernabeu:2001xn,Palomares-Ruiz:2004tk,Indumathi:2004kd,Gandhi:2004md,Gandhi:2004bj,Gandhi:2005wa,Petcov:2005rv}

\begin{figure}
  \begin{center}
    \includegraphics[width=14.0cm, height=8cm]
    {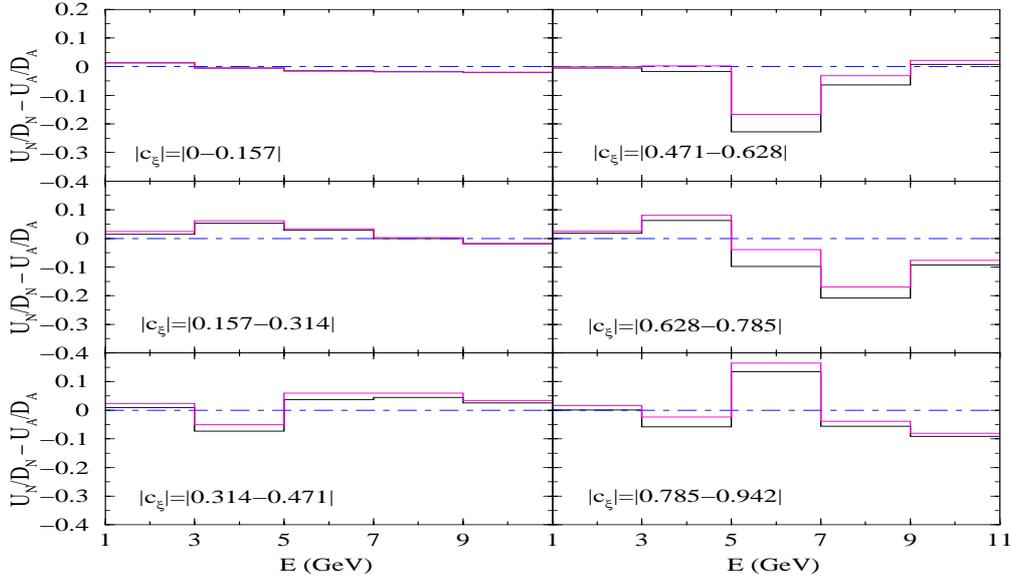}
  \end{center}
  \caption{
    The difference between the up-down ratio for the neutrinos
    ($U_N/D_N$) and anti-neutrinos ($U_A/D_A$) shown for the various
    energy and zenith-angle bins.
    The solid black and solid magenta lines are for
    neutrinos/anti-neutrinos travelling in matter with $\sa=0.5$ and
    0.36 respectively.
    Taken with kind permission of the Physical Review from figure 6 in
    reference \cite{Choubey:2005zy}.
    Copyrighted by the American Physical Society.
  }
  \label{Fig:SnuM:p22L}
\end{figure}

High-energy (multi-GeV) neutrinos are sensitive to matter effects.
Since upward-going neutrinos have a longer path length through matter
than downward-going neutrinos, matter effects may be studied by
evaluating the up-down asymmetry using multi-GeV atmospheric-neutrino
data.
In contrast to matter effects in the electron-neutrino-appearance
channel, the disappearance probability, $P_{\mu\mu}$, is a function of
$L$ and $E$. 
This is illustrated in figure \ref{Fig:SnuM:p22L} \cite{Choubey:2005zy}, 
which shows the difference between the ratio of upward-going to
downward-going muon events for atmospheric neutrinos ($U_N/D_N$) and
anti-neutrinos ($U_A/D_A$). 
The rate estimates have been made for a large magnetised-iron
detector, such as that proposed for the India-based Neutrino
Observatory (INO) \cite{Athar:2006yb}.
The normal mass hierarchy is assumed and the results are shown 
for different energy and zenith-angle bins. 
Since, for a given mass hierarchy, large matter effects appear 
either in the neutrino or in the anti-neutrino channel, the difference
in the ratios for neutrinos and anti-neutrinos gives the net matter
effect.
The figure indicates that the matter effect is largest for neutrinos
travelling $L\simeq 7000$~km with $E\sim 5$~GeV and that the net
matter effect changes sign with $L$ and $E$.
Thus, in order to see the matter effects it is necessary to bin the
data judiciously both in energy and zenith angle. 
The figure also shows that that $\Delta P_{\mu\mu}$ depends on the
value of $\theta_{23}$.

Magnetised-iron calorimeters are expected to have good energy and
zenith-angle resolution.
Therefore, fine binning would allow such detectors to observe matter
effects in the muon signal. 
The magnetic field which allows muon-neutrino induced events to be
distinguished from anti-muon-neutrino events enhances the sensitivity
of these detectors to matter effects since, as noted above, matter
effects appear either in the neutrino or the anti-neutrino channel.
Iron calorimeters have two principal disadvantages: the neutrino
energy threshold is relatively high, allowing for the detection of
multi-GeV neutrinos only; and electron-neutrino induced events can not
be detected.

Water \v Cerenkov detectors have the advantage that sub-GeV neutrinos can
be detected. 
However, the energy resolution is worse than that of an iron
calorimeter. 
For the results presented here, the data is binned in sub-GeV and
multi-GeV bins and therefore the matter effect in the $P_{\mu\mu}$
channel is largely averaged out. 
This averaging implies that only a very small residual matter effect
in the multi-GeV muon sample may be observed. 
However, matter effects in the $P_{\mu e}$ channel do not change sign
over most of the relevant range of $E$ and $L$ in the multi-GeV
regime.
Therefore, the multi-GeV electron sample has large matter effects and
can be used to study the deviation of $\theta_{23}$ from maximality
and the $\theta_{23}$ octant, as well as the mass hierarchy.

\subsubsection{Is the mixing angle $\theta_{23}$ maximal?}

\begin{figure}
  \raisebox{5mm}{\includegraphics[width= 5.2cm, height=5.4cm]{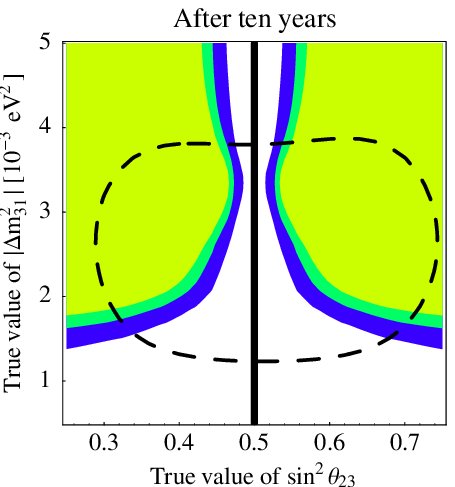}}
  \includegraphics[width= 5.2cm, height=5.4cm]{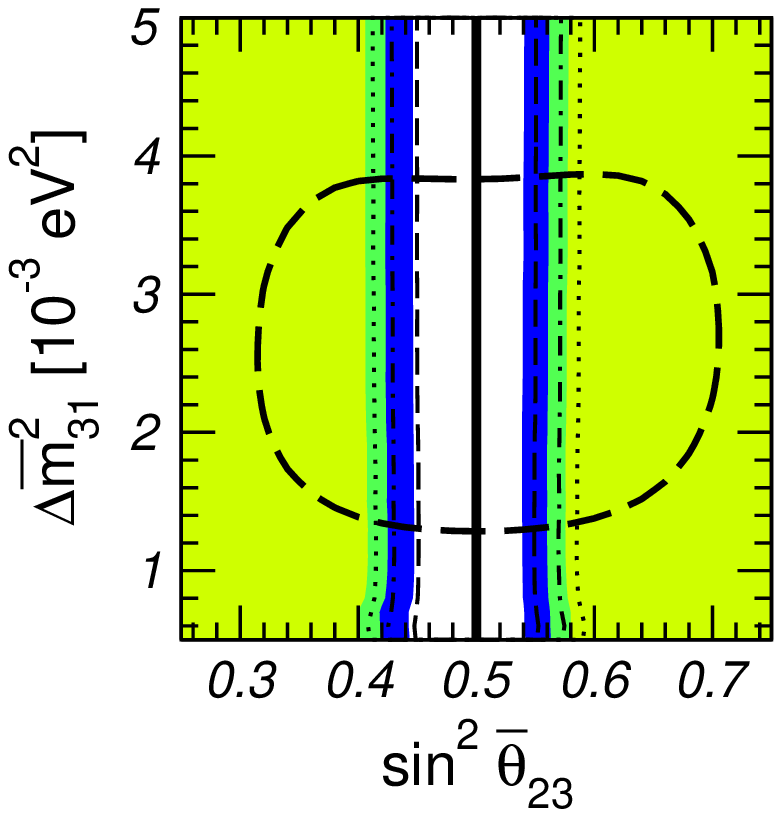}
 \raisebox{2mm}{ \includegraphics[width= 5.4cm, height=5.6cm]{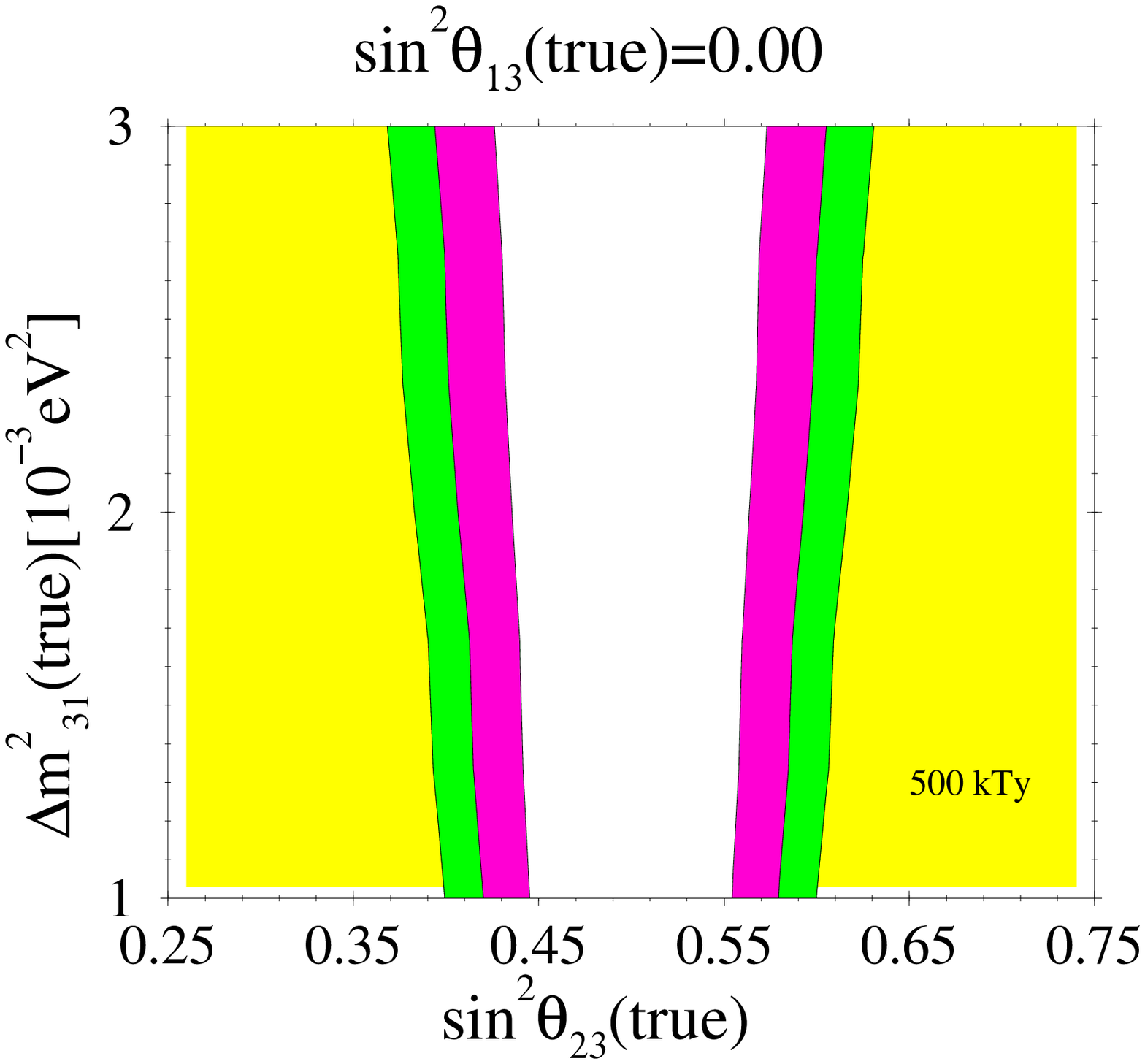}}
  \caption{
    The regions of $\mat$ and $\sat$ where maximal $\theta_{23}$
    mixing can be rejected at $1\sigma$ (inner bands), $2\sigma$
    (middle bands) and $3\sigma$ (outer bands) C.L. 
    The sensitivity expected: the left panel (taken from
    \cite{Antusch:2004yx}) shows the sensitivity expected from the
    combined data from the long base-line experiments. 
    The middle panel (taken from \cite{Gonzalez-Garcia:2004cu}) shows the
    sensitivity expected with atmospheric neutrinos in a megaton water 
    detector (SK50). 
    The right-hand panel (taken from \cite{Choubey:2006jk})
    shows the corresponding
    reach expected from 500 kTy atmospheric neutrino data in large
    magnetised-iron detectors.
    The true value of $\theta_{13}$ is assumed to be zero.
    Middle panel taken with kind permission of the Physical Review from
    figure 4 in reference \cite{Gonzalez-Garcia:2004cu}, copyrighted
    by the American Physical Society.
    Left panel taken with kind permission of the Physical Review from
    figure 1 in reference \cite{Antusch:2004yx}, copyrighted
    by the American Physical Society.
  }
  \label{fig:maximality}
\end{figure}

The measurement of both the magnitude and sign of the deviation of
$\sa$ from its maximum is of great importance. 
The deviation of $\sa$ from 0.5 may be quantified by defining 
$D \equiv \frac{1}{2} - \sa$.
At present, the best limit $|D|$ comes from the SK
experiment giving $|D| \leq 0.16$ at $3\sigma$ 
\cite{Ashie:2005ik}; the sign of $D$ is unknown at present.
The potential of atmospheric-neutrino experiments to test the
deviation of $\theta_{23}$ from maximality is shown in figure
\ref{fig:maximality}.
The figure also shows the sensitivity obtained by combining data from
the current and the next generation of long-baseline experiments.
The combined long-baseline data set includes five years of running
for each of the following: MINOS; ICARUS; OPERA; T2K; and NO$\nu$A.
The middle panel shows the sensitivity to $|D|$ of
atmospheric-neutrino experiments with water \v Cerenkov detectors with a
data set corresponding to an exposure of 4.6~Megaton-years.
The left panel shows the corresponding sensitivity of
atmospheric-neutrino data in large magnetised-iron detectors with an
exposure of 500-kiloton-years.
At $\Delta m_{31}^2$(true)$=2.5\times 10^{-3}$ eV$^2$, it should be 
possible to measure $|D|$ within 19\% and 25\% at $3\sigma$
with atmospheric neutrinos using water and iron detectors
respectively. 
This is slightly weaker than the sensitivity of the combined
long-baseline experiments, where it should be possible to measure
$|D|$ to within 14\% at $3\sigma$. 
However, note that all the results presented in figure
\ref{fig:maximality} have been obtained assuming that the true value
of $\theta_{13}$ was zero. 
For non-zero $\theta_{13}$, the presence of matter effects in the
$P_{\mu\mu}$ channel brings a marginal improvement in the sensitivity
of atmospheric-neutrino experiments using a magnetised-iron detector. 
For the megaton-water atmospheric-neutrino experiment, very large
matter effects in the $P_{\mu e}$ channel bring a significant
improvement in the determination of $|D|$, making this experiment
comparable to, or better than, the long-baseline experiments for
studying the deviation of $\theta_{23}$ from maximality
\cite{Choubey:2006jk}.

\subsubsection{Resolving the $\theta_{23}$ Octant Ambiguity} 
\label{sec:octant}

If the true value of $\theta_{23}$ is not $45^\circ$, then the
question of whether $\theta_{23} > \pi/4$ ($D$ positive) or 
$\theta_{23} < \pi/4$ ($D$ negative) arises. 
This ambiguity is generally regarded as the most difficult to
resolve.
As discussed above, the presence of matter effects in the
zenith-angle- and energy-binned atmospheric-$\numu/\anumu$ data 
opens up the possibility of probing the octant of $\theta_{23}$ in
magnetised-iron detectors \cite{Choubey:2005zy}.
On the other hand, atmospheric $\nue/\anue$ data in water \v Cerenkov
detectors could also give information on the octant of $\theta_{23}$,
both through the $\Delta m_{21}^2$-dependent sub-dominant term in the
sub-GeV sample \cite{Peres:1999yi,Gonzalez-Garcia:2004cu}, and through
the matter effect in the multi-GeV sample
\cite{Choubey:2006jk,Huber:2005ep}.
This, therefore, opens the possibility of combining 
atmospheric-neutrino data with data from long-baseline experiments to
resolve parameter degeneracies \cite{Huber:2005ep,Campagne:2006yx}.

\begin{figure}
\begin{center}
  \includegraphics[width=7.0cm, height=7.0cm]
  {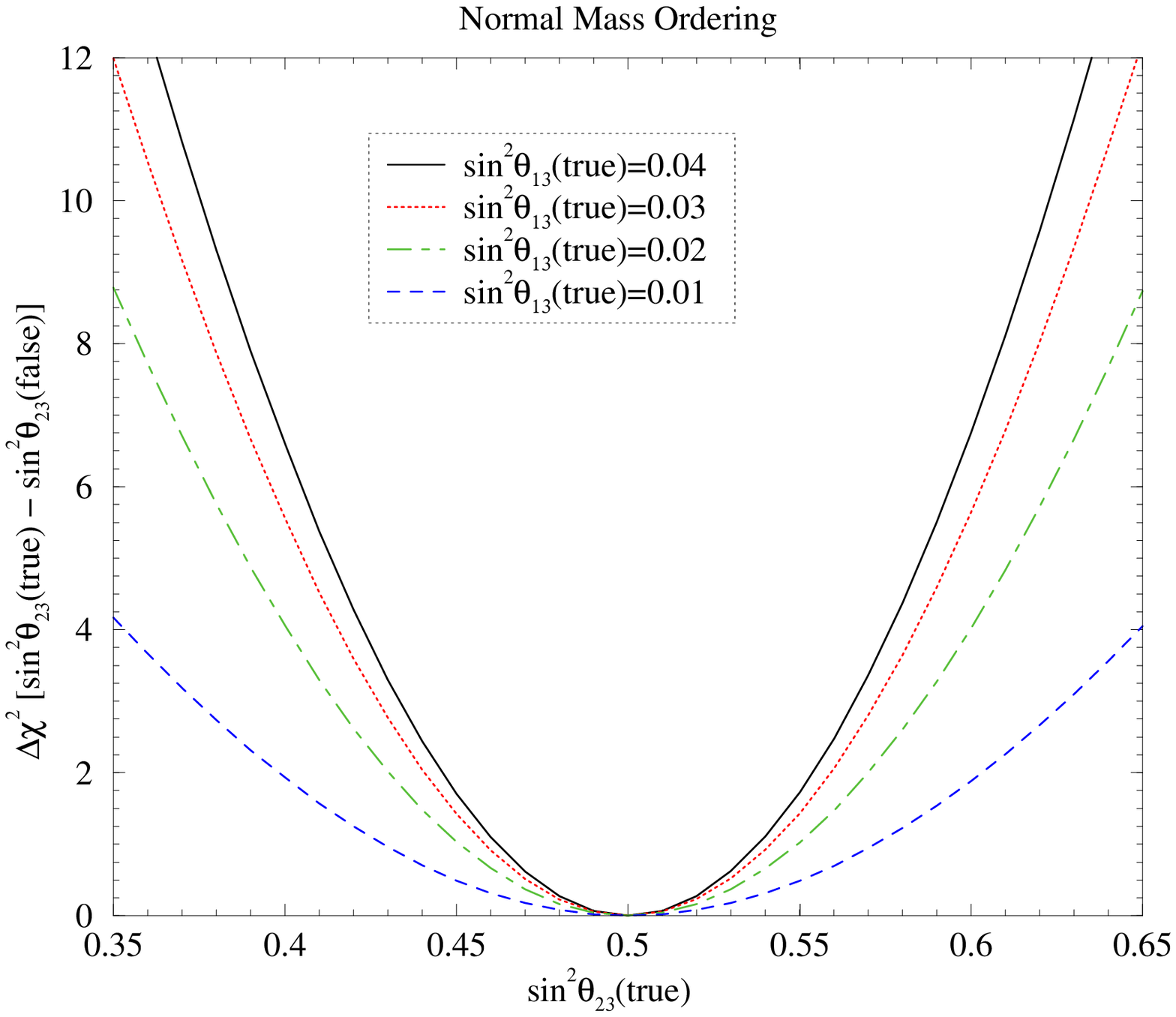}
\hspace{5mm}
  \includegraphics[width=7.0cm, height=7.0cm]
  {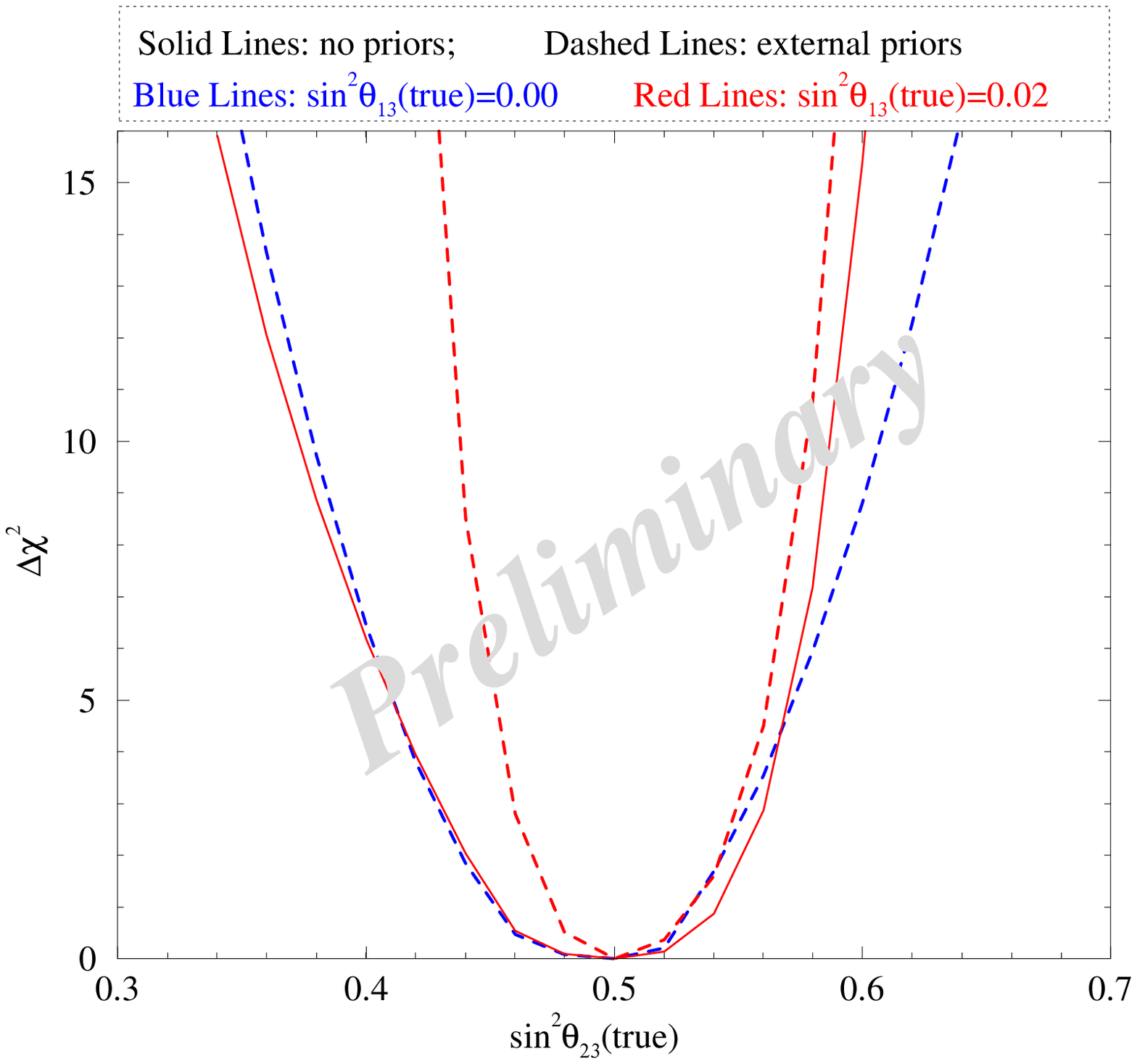}
  \caption{
    Plot showing the octant sensitivity as a function of
    $\sin^2\theta_{23}$(true), for an atmospheric neutrino experiment
    with large magnetised-iron calorimeter (left panel) and
    megaton water detector (right-hand panel).
      Taken from \cite{Choubey:2006jk}.
  }
  \label{fig:delchioctant}
  \end{center}
\end{figure}

In order to obtain the limiting value of $\sin^2\theta_{23}$(true) 
which could still allow for the determination of the sign of $D$ it is
convenient to define:
\begin{equation}
  \Delta \chi^2 \equiv \chi^2 (\sin^2\theta_{23} ({\rm true}),
  \sin^2\theta_{13} ({\rm true}), {\rm others(true)}) 
  - \chi^2(\sin^2\theta_{23} ({\rm
  false}),\sin^2\theta_{13}, {\rm others}),
  \label{Eq:chioctant}
\end{equation}
with $\sin^2\theta_{23}$(false) restricted to the wrong octant and
`others' comprising $\Delta m^2_{31}$, $\Delta m^2_{21}$,
$\sin^2\theta_{12}$, and $\delta$.  
These parameters, along with $\sin^2\theta_{13}$ as well as
$\sin^2\theta_{23}$(false), are allowed to vary freely in the fit.
The results of the fit are shown in figure \ref{fig:delchioctant} for
a 500-kiloton-year exposure in a large magnetised-iron calorimeter
(left panel) and a 4.6~Megaton-year exposure of a water \v Cerenkov
experiment (right-hand panel) \cite{Choubey:2006jk}.
For the magnetised-iron detector, the results are presented using four
different values of $\sin^2\theta_{13}$(true), assuming a normal mass
ordering.
For a given $\sin^2\theta_{13}$(true), the range of
$\sin^2\theta_{23}$(true) for which $\sin^2 \theta_{23}$(false) can be
ruled out with atmospheric neutrinos in magnetised-iron detector is
given in table \ref{tab:octant}. 
These results can be compared to the sensitivity that can be obtained
using a water \v Cerenkov detector, which is shown for normal mass
hierarchy in the right-hand panel of figure \ref{fig:delchioctant} and
reported in table \ref{tab:octant}.
The octant determination can be performed reasonably well even if
$\sin^2\theta_{13}$(true) was zero \cite{Gonzalez-Garcia:2004cu}.
However, if $\sin^2\theta_{13}$(true) is non-vanishing and reasonably
large, the octant sensitivity of this experiments becomes
significantly enhanced through earth matter effects appearing 
in the multi-GeV electron sample \cite{Choubey:2006jk,Huber:2005ep}.
\begin{table}
  \begin{tabular}{ccc}
    \hline
    & &  \cr
    Type of Experiment & $\sin^2\theta_{23}$(false) excluded
    at 3$\sigma$ if: & for \cr
    & &  \cr
    \hline
    & &  \cr
    &$\sin^2\theta_{23} ({\rm true}) < 0.402 \ {\rm or} \ > 0.592 $
    &$\sin^2\theta_{13} ({\rm true}) = 0.02$ \cr
    Magnetised-Iron (0.5 MTy)
    &$\sin^2\theta_{23} ({\rm true}) < 0.421 \ {\rm or} \ > 0.573$
    &$\sin^2\theta_{13} ({\rm true}) = 0.04$ \cr
    & &  \cr
    \hline
    & &  \cr
    & 
    $\sin^2\theta_{23} ({\rm true}) < 0.383 \ {\rm or} \ > 0.600$
    & $\scht=0.00$ \cr
    Water \v Cerenkov (4.6 MTy) & 
    $\sin^2\theta_{23} ({\rm true}) < 0.438 \ {\rm or} \ > 0.573$
    & $\scht=0.02$ \cr
    & &  \cr
    \hline
  \end{tabular} 
  \caption{
    A comparison of the potential of different experiments to rule out
    the wrong $\theta_{23}$ octant at \sig{} (1 dof). 
    The third column gives the condition on the true value of $\sch$
    needed for the $\theta_{23}$ octant resolution.
  }
  \label{tab:octant}
\end{table}

\subsubsection{Resolving the ambiguity in the neutrino-mass hierarchy}

\begin{figure}
  \begin{center}
    \includegraphics[width=16.0cm, height=6.0cm]
    {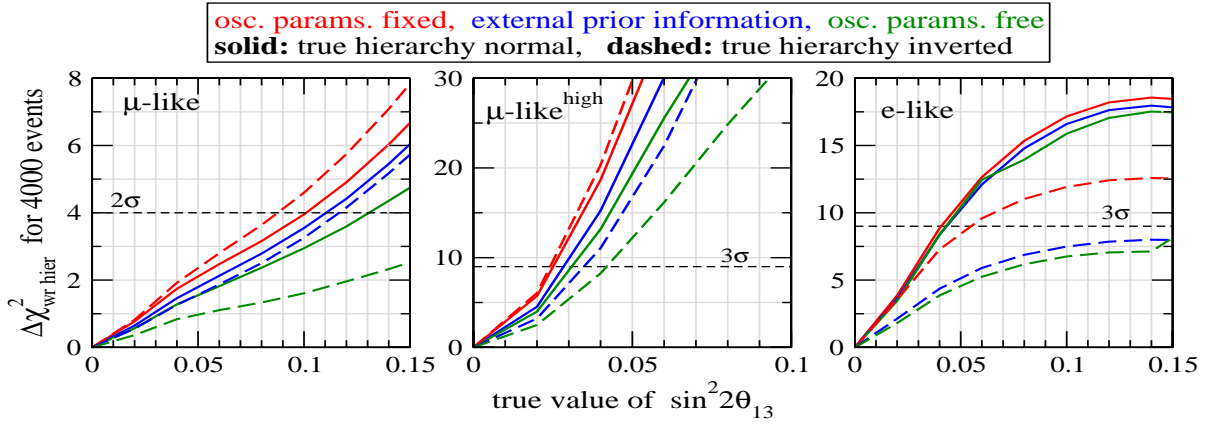}
  \end{center}
  \caption{
    $\Delta \chi^2$ for the wrong hierarchy as a function of 
    $\sin^22\theta_{13}$(true). See the text for the details.
  }
  \label{fig:hierarchyINO}
\end{figure}

Large matter effects in atmospheric neutrinos can be exploited to
probe the sign of $\Delta m_{31}^2$.
Figure \ref{fig:hierarchyINO} shows the sensitivity to 
${\rm sign}(\Delta m_{31}^2)$ that is expected in a magnetised-iron
calorimeter with 4000 observed upward going events
\cite{Petcov:2005rv}.
The simulation has been performed for both the normal and the inverted
hierarchy; the curves show the $\chi^2$, and hence the C.L., with
which the wrong hierarchy can be ruled out.
Fits have been carried out under the following conditions: all
parameters other than the mass hierarchy are fixed (red lines); 
external priors have been used for the oscillation parameters (blue
lines); and all oscillation parameters are allowed to vary freely in the
fit (green lines). 
The left panel is for muon events in a detector with 15\% energy
and $15^\circ$ zenith angle resolution, the middle panel is for muon
events with 5\% energy and $5^\circ$ zenith angle resolution, while
the right-hand panel is for electron events.
For vanishing $\theta_{13}$, the matter effects vanish giving
$\chi^2=0$. 
As $\theta_{13}$ increases, matter effects increase, thereby
increasing the sensitivity of the experiment to the hierarchy.
For a magnetised-iron calorimeter such as INO, where the energy
resolution is expected to be around 15\% and the zenith angle
resolution to be around $15^\circ$, the wrong hierarchy can be ruled
out at $\sim 2\sigma$ using the muon events, if
$\sin^22\theta_{13}$(true)$=0.1$ and $\sin^2\theta_{23}$(true)$=0.5$, 
and where the information from the other long-baseline experiments 
on the oscillation parameters have been included through the priors. 
Comparison of the left with the middle panel shows that the
sensitivity to the hierarchy increases if the detector resolution is
improved.  
Comparison of the left with the right-hand panel shows that the
sensitivity to the hierarchy increases if the detector is able to
detect electron-type events.
Of course, since matter effects increase with $\theta_{23}$, the
sensitivity to the hierarchy increases as the true value of
$\theta_{23}$ increases.

The sign of $\Delta m_{31}^2$ can be determined using the excess in
the multi-GeV electron sample that arises due to matter effects using
a water \v Cerenkov detector
\cite{Gandhi:2005wa,Huber:2005ep,Choubey:2006jk,Gandhi:2007td}.
The wrong hierarchy can be ruled by a 4.6 Megaton-year exposure
of such an experiment at more than 2$\sigma$ if
$\sin^22\theta_{13}$(true)$=0.1$ and $\sin^2\theta_{23}$(true)$=0.5$
\cite{Huber:2005ep,Choubey:2006jk}.
This is comparable to the sensitivity of the magnetised-iron detectors
discussed above. 
However, since water detectors use the excess in electron events for
multi-GeV neutrinos for which matter effects contribute to the
probability $P_{\mu e}$, the excess is also dependent on the CP phase
$\delta$.
If the value of $\delta$ is allowed to vary freely in the fit
then the sensitivity decreases appreciably \cite{Choubey:2006jk}.

%% file: 06-Alternatives/06-03-0nu2beta/06-03-0nu2beta.tex
\subsection{Neutrino Mass Hierarchy from Future $0\nu\beta\beta$ Experiments}

If neutrinos are Majorana particles, it may be possible to observe
the process $(A,Z) \rightarrow (A,Z-2) + 2 \, e^-~$, neutrinoless
double-beta decay (\obb). 
The effective mass that may be extracted, or bounded, in a \obb{}
experiment is given by the coherent sum: 
$\meff =  \left| \sum_i m_i \, U_{ei}^2 \right|$, 
where $m_i$ is the mass of the $i^{\rm th}$ neutrino mass state, the
sum is over all the light-neutrino mass states and $U_{ei}$ are the
matrix elements of the neutrino mixing matrix, i.e. \meff{} depends on
7 out of the 9 parameters contained in the neutrino-mass matrix.
In particular, the effective mass that may be extracted from \onbb
depends on the neutrino-mass spectrum. 
There have been a large number of papers written on the implications
of a future measurement of \meff{} (see for example
\cite{Petcov:2005yq}).
At present, the best limit on the effective mass is given by the
Heidelberg--Moscow collaboration $\meff \le 0.35\, z~{\rm eV}$, 
where $z(={\cal O}(1))$ indicates that there is an uncertainty 
in the value of the \nme{} (NME) involved in the \obb{} process
\cite{Klapdor-Kleingrothaus:2000sn}.
Several new experiments are running, under construction, or in the
planing phase \cite{Aalseth:2004hb}. 
It is reasonable, therefore, to expect that \meff{} will be probed
down to $\simeq 0.04$ eV and it is pertinent to ask if such a
measurement can help determine the neutrino-mass hierarchy.

\begin{figure}
  \begin{center}
    \includegraphics[width=16.0cm,height=8cm]
    {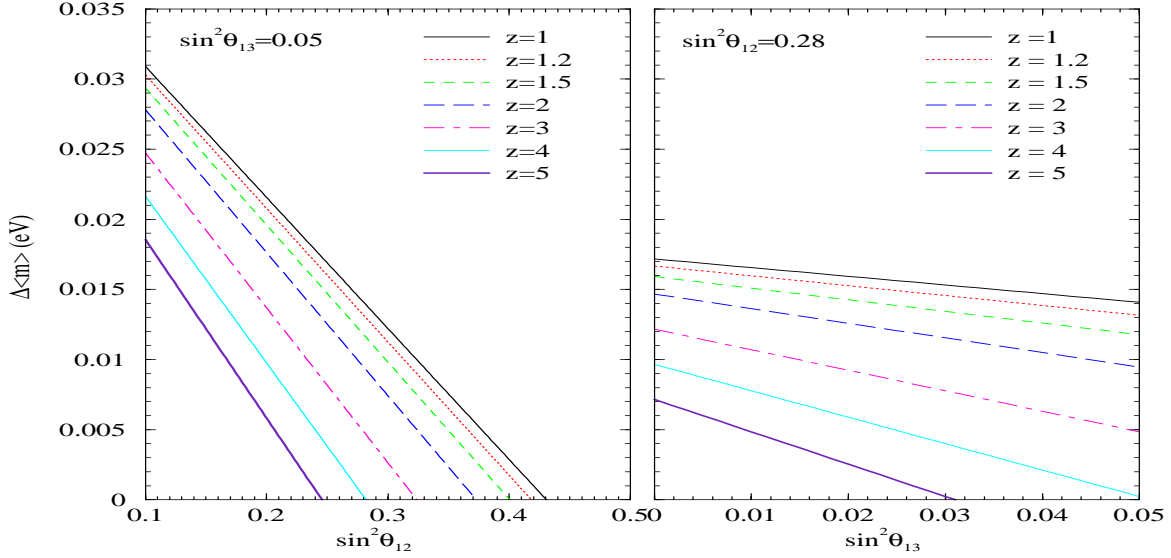}
  \end{center}
  \caption{
    The difference between the minimal value of \meff{} for IH 
    and the maximal value of \meff{} for NH for different $z$, as a
    function of $\sin^2 \theta_{12}$ (left--hand panel) and 
    $\sin^2 \theta_{13}$ (right--hand panel). 
        Taken with kind permission of Physical Review Letters from figure 4 in
        reference \cite{Willmann:1998gd}.  
        Copyrighted by the American Physical Society.
        Taken with kind permission of the Physical Review from figure 5 in
        reference \cite{Choubey:2005rq}.
        Copyrighted by the American Physical Society.
  }
  \label{Fig:SnuM:diffnme}
\end{figure}

For the normal-hierarchy (NH) scheme, for which $m_1 \ll m_2 \ll m_3$,
and assuming that $m_1$ can be neglected, the effective mass may be 
written:
\be
  \meffnh \simeq \left| \sqrt{\ms} \, \sss \, \csh + \sqrt{\ma} \, \sch \, 
  e^{2i(\beta - \alpha)} 
  \right| \; .
  \label{eq:meffnh}
\ee
For the inverted-hierarchy (IH) scheme, assuming that 
$m_3 \ll m_1 < m_2$, and neglecting $m_3$, the effective mass may be
written: 
\be
  \meffih \simeq \sqrt{|\ma|} \, \csh \, \sqrt{ 1 - \sin^22\theta_{12} 
  \, \sin^2
 \alpha} ~.
\label{eq:meffih}
\ee
Any positive signal for $\obb$ will be able to distinguish the IH
scheme from the NH scheme if the difference between the predicted
values for $\meff$ for the IH scheme and the NH scheme is larger than 
the error in the measured value of $\meff$. 
Among the most important errors involved is the one coming from the 
uncertainty in the value of the \nme. 
Figure \ref{Fig:SnuM:diffnme} shows the difference in the predicted 
values of $\meffnhmax$ and $\meffihmin$ taking into account the error
in the \nme{} \cite{Choubey:2005rq}. 
$\meffnhmax$ and $\meffihmin$ are the largest and smallest values for
\meff{} that are allowed, given the present knowledge of the
oscillation parameters, in the NH and IH scheme respectively.
This uncertainty is incorporated through the parameter $z$, which
gives the factor by which the nuclear matrix elements are uncertain
(see \cite{Choubey:2005rq} for the details).
It was argued in \cite{Choubey:2005rq} that, for a given mass
hierarchy, the uncertainty in the prediction of \meff{} coming from
the uncertainty in the allowed values of $\ma$ and $\ms$ can be
neglected since these parameters are expected to be measured with
very high accuracy in the immediate future.
Therefore, the major uncertainty in \meff{} will come from the
uncertainty on the values of $\sss$ and $\sch$. 
Figure \ref{Fig:SnuM:diffnme} shows the impact of the uncertainty in
the values of $\sss$ and $\sch$ on the sensitivity of the future
\obb{} experiments to the neutrino-mass hierarchy. 
The figure shows that for $\sch$ close to its current limit and
assuming $z = 2$, $\sss = 0.3$, and $\Delta \meff \simeq 0.01$~eV
it should be possible to determine the mass hierarchy if the
experimental uncertainty in \meff{} is less than 0.01~eV. 
The chances of determining the hierarchy is largest when $\sch=0$.
More importantly, while the dependence on $\sch$ is weak, the
sensitivity of the \obb{} experiments to the hierarchy is strongly
dependent on $\sss$. 
Therefore, a substantial reduction in the uncertainty on the allowed
values of $\sss$ is a prerequisite for the determination of the
neutrino-mass hierarchy using \obb{} experiments.

So far, the assumption that the lightest neutrino mass was close to
zero has been made.
If the lightest neutrino had a mass $m_0 \gtap 0.01$ eV, it would not
be possible to distinguish between the NH and IH schemes using \obb{}
measurements, the mass spectrum in that case would be
quasi-degenerate.
However, we could still use \obb{} to put a limit on the absolute
neutrino-mass scale. 
For a quasi-degenerate (QD) mass spectrum, with a common mass scale
$m_0$, the limit on the neutrino mass reads \cite{Choubey:2005rq}:
\be
  m_0 \le z \, \meff_{\min}^{\rm exp} \,  \frac{1 + \tan^2 \theta_{12}}
  {1 - \tan^2 \theta_{12} - 2 \, |U_{e3}|^2 } 
  \equiv z \, \meff_{\min}^{\rm exp} \,  f(\theta_{12}, \theta_{13})~.
  \label{eq:m0_lim}
\ee 
Currently, the uncertainty on $f(\theta_{12}, \theta_{13})$ is around
50\%, $1.9 < f(\theta_{12}, \theta_{13})< 5.6$.   
It is expected to reduce to $\sim$ 21\% ($\sim 9\%$) at $3\sigma$ 
if a low energy $pp$ solar-neutrino experiment (a reactor experiment
at the SPMIN) should be built. 
The uncertainty depends only a little on the value of $\theta_{13}$. 
From the current limit on the effective mass, 
$\meff \le 0.35 \, z$~eV, with the accepted value of $z \simeq 3$
and our current knowledge of $f(\theta_{12}, \theta_{13})$, we can set
a limit on $m_0$  of 5.6 eV, clearly weaker than the limit from
tritium beta decay experiments.
However, if $f(\theta_{12}, \theta_{13})$ was known with an
uncertainty of $20\%$, say  $2.7 < f(\theta_{12}, \theta_{13})< 4.0$,
then for $z \, \meff_{\min}^{\rm exp}=0.1$~eV the limit would become
$0.3 ~{\rm eV} \ltap m_0 \ltap 0.4$ eV. 
Of course, if there is no signal for \obb, but just an upper limit 
on $z \, \meff_{\min}$, the allowed range of $m_0$ will be replaced by
an upper limit corresponding to the largest value in the range.
The examples given above, indicate that, for the QD mass spectrum, a
measurement of, or a better constraint on, \meff{} will lead to a
stronger limit on the absolute neutrino mass scale than can currently
be obtained from direct kinematic searches.

%% file: 06-Alternatives/06-04-Unitarity/06-04-Unitarity.tex
\subsection{Astrophysical methods of  determining  the mixing parameters}
\label{icecube}

Measuring the fluxes of neutrinos from astrophysical sources can
help us to determine the mixing-matrix elements. 
The goal of this section is to discuss this method, focusing on the 
possibility of extracting $|U_{\mu 1}|$, arguably the most challenging 
element of the mixing matrix to measure. 
In section \ref{section-coherence}, we discussed the possibility of
using neutrino beams made up either of pure or incoherent mass
eigenstates to extract the moduli of the mixing-matrix elements
by studying their charged-current interactions. 
Astrophysical sources
can yield 
such beams through three classes of mechanisms: adiabatic
conversion; neutrino decay; and decoherence. The second and third
cases will be discussed in detail in sections \ref{unstable}
and \ref{stable}. Here we comment on the case of adiabatic
conversion which takes place for solar
neutrinos. Propagating from central regions of the Sun, the electron
neutrinos with energies $E > 10$ MeV are converted to a state which
nearly coincides with $\nu_2$ at the surface of the Sun. As a
result, by studying the charged-current interactions of the solar
neutrinos with $E>10$~MeV, we can determine $|U_{e2}|$.
Unfortunately, the energy of these neutrinos will be too small to
allow  muon production  at the detectors; so, they cannot be used to
extract $|U_{\mu 2}|$. However, there is a possibility that more
energetic neutrinos ($E\gg m_\mu$) may be produced inside the Sun:
if the dark matter is composed of Weakly Interacting Massive
Particles (WIMPs), over time they can be accumulated in the core of
the Sun. Thus, the WIMP-annihilation rate in the core of the Sun
will increase, giving rise to a relatively high energy flux (for a
recent review, see \cite{Halzen:2005ar}). As shown in
\cite{Farzan:2002ct}, for the low energy part of the spectrum
$E_\nu<5$~GeV, the transition probability in the Sun will be
adiabatic and therefore the oscillation probabilities will depend
only on the absolute values of the elements of $U_{PMNS}$. In
\cite{Farzan:2002ct}, it was suggested that the value of
$|U_{\mu 1}|$ could be derived by studying these
neutrinos. Unfortunately, because of 
the high energy-threshold of large-scale neutrino detectors, this
method does not seem to be feasible. There is another mechanism for
production of neutrinos with $E_\nu>1$~GeV inside the Sun:
cosmic-ray collisions in the Sun can give rise to `solar-atmospheric
neutrinos'. Recently in \cite{Fogli:2006jk}, it has been shown that
the oscillation probability of these neutrinos (after averaging over
neutrino and anti-neutrino channels) depends only on the absolute
values of the elements of the PMNS matrix. However, low statistics
(only $\sim$ten events in ICECUBE per year) render this an unsuitable
tool for the extraction of $|U_{\mu 1}|$.

\subsubsection{General remarks about astrophysical neutrinos}
\label{capability}

The  methods for extracting $|U_{\mu 1}|$  discussed here are based
on the flavour-identification capability of neutrino telescopes.
Since neutrino telescopes cannot distinguish between neutrino and
anti-neutrino; neutrino and anti-neutrino events will therefore
enter the same data sample.

In the energy range 1--100~TeV, a neutrino telescope can identify
two types of neutrino events: muon-track events; and shower-like
events. The muon-tracks originate from the charged current (CC)
interactions of $\nu_\mu$ ($\bar{\nu}_\mu$) as well as CC
interactions of $\nu_\tau$ ($\bar{\nu}_\tau$), with the subsequent
decay of the $\tau$ ($\tau^+$) to $\mu$ ($\mu^+$). Shower-like
events can be produced in three ways: neutral current (NC)
interactions of all the active neutrinos; CC interaction of $\nu_e$
($\bar{\nu}_e$); and CC interactions of $\nu_\tau$
($\bar{\nu}_\tau$) and the subsequent decay of $\tau$ ($\tau^+$)
through non-muonic decay modes. It is convenient to define the
ratio:
\begin{equation}
  R\equiv {{\rm muon-track \ events} \over {\rm shower-like \ events}}.
\end{equation}
Following  the above discussion, $R$ can be written as:
\begin{equation}
  R={ \int_{E_{th}^\mu}\left[
  \frac{dN_\mu^{CC}(F_{\nu_\mu},F_{\bar{\nu}_\mu})}{dE_\mu}+B
  \frac{dN_{\tau\to \mu}^{CC}(F_{\nu_\tau},F_{\bar{\nu}_\tau})}
  {dE_{\mu}}\right]\times R_\mu(E_\mu) dE_\mu \over T
  \int_{E_{th}^{sh}}^{E_{cut}} \left( \sum_\alpha
  \frac{dN^{NC}(F_{\nu_\alpha},F_{\bar{\nu}_\alpha})}{dE}+
  \frac{dN^{CC}(F_{\nu_e},F_{\bar{\nu}_e})}{dE}+ (1-B)
  \frac{dN^{CC}(F_{\nu_\tau},F_{\bar{\nu}_\tau})}{dE} \right)d E} \; ,
  \label{ratioR}
\end{equation}
where $R_\mu$ and $T$  are respectively the muon range and the
thickness of the detector, and $B \equiv {\rm Br}(\tau\to \mu \nu_\mu
\nu_\tau)$. $d N_\alpha^{CC}/dE$ and $d N_\alpha^{NC}/dE$ are
respectively the rates of CC and NC interactions of $\nu_\alpha$ and
$\bar{\nu}_\alpha$:
\begin{equation}
  \frac{dN_\mu^{CC}(F,\tilde{F})}{dE_\mu}=\int^{E_{cut}}\frac{ dF}{
  dE_{\nu} }\frac{d \sigma_{CC}}{d E_\mu} dE_{\nu}+
  \int^{E_{cut}}\frac{ d\tilde{F}}{ dE_{\bar{\nu}} }\frac{d
  \bar{\sigma}_{CC}}{d E_\mu} dE_{\bar{\nu}} \; ;
 \label{muCC}
\end{equation}
\be
  \frac{d N^{CC}(F,\tilde{F})}{dE} =
  \sigma_{CC} \frac{dF}{dE} + \bar{\sigma}_{CC}
  \frac{d\tilde{F}}{dE} \;   \label{CC}
\ee and \be
  \frac{d N^{NC}(F,
  \tilde{F})}{dE}  =\int_0^{E-E_{th}}\left[
  \frac{d\sigma_{NC}}{dE_{\nu_f}} \frac{dF}{dE} +
  \frac{d\bar{\sigma}_{NC}}{dE_{{\nu}_f}}
  \frac{d\tilde{F}}{dE}\right]dE_{\nu_f} \; .
  \label{NC}
\ee
Here $\sigma_{CC}$ and $\bar{\sigma}_{CC}$ are the charged-current
cross sections for $\nu$ and $\bar{\nu}$, and $d\sigma_{NC}/dE_{\nu_f}$
and $d\bar{\sigma}_{NC/dE_{\bar{\nu}_f}}$ are the partial cross
sections for $\nu(E) N \to \nu_f(E_{\nu_f}) +{\rm jet}$ and
$\bar{\nu}(E) N \to \bar{\nu}_f(E_{\bar{\nu}_f}) +{\rm jet}$,
respectively.
Finally,
\begin{eqnarray}
  \frac{dN_{\tau\to
  \mu}^{CC}(F,\tilde{F}) } {dE_{\mu}}=
           \nonumber
\end{eqnarray}
\be
  \int_0^{E_\tau}\int_0^{E_{cut}} f(E_\tau, E_\mu) \frac{ d
  {\sigma}_{CC}}{dE_\tau} \frac{dF}{d E_{{\nu}_\tau}} dE_{\nu_\tau}
  dE_\tau+\int_0^{E_\tau}\int_0^{E_{cut}} f(E_\tau, E_\mu) \frac{ d
  \bar{\sigma}_{CC}}{dE_\tau} \frac{d\tilde{F}}{d
  E_{\bar{\nu}_\tau}} dE_{\bar{\nu}_\tau} dE_\tau \; ,
  \label{tautomuCC}
\ee where $f(E_\tau, E_\mu)$ is the probability of the production of
a muon with energy  $E_\mu$ in the decay of a $\tau$ lepton with
energy  $E_\tau$.

The possibility of measuring $R$ using the ICECUBE experiment has been
studied in detail in \cite{Beacom:2003nh} and it has been found that
with $ E_\nu^2 d F_\nu/dE_\nu=10^{-7}~{\rm GeV}~{\rm cm}^2~{\rm sec}^{-1}$,
the ratio $R$ can be measured with 20\% accuracy after one year of
data taking.
Since the statistical error dominates, by increasing the data-taking
time to 10 years, the uncertainty would decrease to 7\%.

Notice that we have used the fact that, at this energy range
($E_{c.o.m}\gg m_\tau$), the cross sections are approximately equal
for all flavours. In the above formul\ae, $E_{th}^\mu$ ($\sim$1~TeV)
and $E_{th}^{sh}$ are respectively the thresholds for detecting
muon-track and shower-like events and $E_{cut}$ ($ \sim 100$~TeV) is
the energy above which neutrinos will be absorbed in the Earth.
Above $E_{cut}$, all the neutrinos will be absorbed in the Earth but
$\nu_\tau$ can re-appear as a result of the transitions $\nu_\tau \to
\tau \to \nu_\tau$. Of course, the final $\nu_\tau$ reaching the
detector will be less energetic than the original one, which can fake
a $\nu_\tau$ with energy less than $E_{cut}$. Consequently, the
ratio $R$ will turn out to be smaller than expected if this
phenomenon is not taken into account. In order to be able to extract
$|U_{\mu 2}|$ with the required precision, it will be necessary to
evaluate the correction due to such an effect. Estimating this
correction requires some knowledge of the energy spectrum for
$E>E_{cut}$ and is therefore model dependent.

Notice that before entering the detector, the upward-going neutrinos
pass through the Earth. However, this will not significantly change
the flavour composition because, for $E>1$~TeV, $\Delta m_{31}^2/2E
\ll \sqrt{2} G_F n_e$ and  the effective flavour mixing in the Earth
is therefore strongly suppressed.

\subsubsection{Unstable neutrinos arriving from cosmic distances}
\label{unstable}

In \cite{Farzan:2002ct}, the possibility of employing the decaying
neutrinos to derive the CP-violating phase has been proposed.
In a series of papers \cite{Beacom:2002vi,Beacom:2003zg}, the idea has
been further elaborated.
In the following, the results will be reviewed.

In the SM, neutrinos  are stable,
 however, in the framework of Majoron models, the rapid decay of
neutrinos may become a possibility
\cite{Gelmini:1980re,Chikashige:1980qk}:\footnote{
It was shown later in ref. \cite{Schechter:1981cv} that
the decays discussed in \cite{Gelmini:1980re,Chikashige:1980qk} are so much suppressed
that these decay modes are phenomenologically irrelevant.
However, majoron couplings are rather model-dependent, and
it is possible to contrive models where they are sizable enough to
lead to lifetimes of phenomenological interest.
For more on these issues, see \cite{Valle:1990pk,Pakvasa:2003zv}.
Here we simply assume that fast invisible decays of neutrinos
are possible, and ask ourselves whether such decay modes lead
to interesting consequences.}

\begin{equation}
  \nu_i \rightarrow \bar{\nu}_j +J \; ,
\end{equation}
where $\nu_i$ and $\nu_j$ are mass eigenstates, and $J$ is a Goldstone
boson called the Majoron.

If  the lifetime of the neutrinos in their rest frame is finite but
much larger than $\sim 10^{-3}$ sec, the solar and atmospheric
neutrinos will not undergo decay; however, neutrinos  from very
distant sources (i.e., the gamma-ray bursters, the Active Galactic
Nuclei, AGN, and supernov\ae) can decay before reaching the
detectors. At the detectors, the neutrino flux from the distant
sources  will be composed only of the lightest neutrinos,
$\nu_1$ and $\bar{\nu}_1$: $F_1$ and $F_{\bar{1}}$. Notice that we
have assumed that the ordering of the neutrino masses is normal:
$m_1<m_2<m_3$. As a result, regardless of the flavour composition at
the source, we expect that at the detector:
\begin{equation}
  dF_{\nu_e}/dE:dF_{\nu_\mu}/dE:dF_{\nu_\tau}/dE=
  |U_{e1}|^2:|U_{\mu1}|^2:|U_{\tau 1}|^2 \; ,
  \label{decayratio}
\end{equation}
and recalling that mixing matrices of neutrinos and anti-neutrinos
are the complex conjugate of of one-another, we have
\begin{equation}
  dF_{\bar{\nu}_e}/dE:dF_{\bar{\nu}_\mu}/dE:dF_{\bar{\nu}_\tau}/dE=
  |U_{e1}|^2:|U_{\mu1}|^2:|U_{\tau 1}|^2 \; .
  \label{decaybarratio}
\end{equation}
Notice that this result is independent of energy. Equation
(\ref{ratioR}) implies:
\begin{equation}
  R = \frac{|U_{\mu1}|^2 + B \xi_1 |U_{\tau 1}|^2}{\xi_2 +
  |U_{e1}|^2\xi_3 + (1- B) \xi_3 |U_{\tau 1}|^2} \; ,
\end{equation}
where (see equations (\ref{muCC}), (\ref{CC}), (\ref{NC}), and
(\ref{tautomuCC}) for definitions):
\begin{equation}
  \xi_1 = {\int_{E_{th}^\mu}\frac{dN_{\tau\to
  \mu}^{CC}(F_1,F_{\bar{1}})} {dE_{\mu}}\times R_\mu(E_\mu) dE_\mu
  \over\int_{E_{th}^\mu}\frac{dN_\mu^{CC}(F_{1},F_{\bar{1}})}
  {dE_\mu}R_\mu(E_\mu)dE_\mu} \; ;
\end{equation}
and:
\begin{equation}
  \xi_2={T \int_{E_{th}^{sh}}^{E_{cut}}  \sum_\alpha
  \frac{dN^{NC}(F_{1},F_{\bar{1}})}{dE} \over
  \int_{E_{th}^\mu}\frac{dN_\mu^{CC}(F_{1},F_{\bar{1}})}{dE_\mu}R_\mu(E_\mu)
  dE_\mu} \; ; 
\end{equation} 
and finally:
\be 
  \xi_3= {T
  \int_{E_{th}^{sh}}^{E_{cut}}\frac{dN^{CC}(F_{1},F_{\bar{1}})}{dE}
  d E \over
  \int_{E_{th}^\mu}\frac{dN_\mu^{CC}(F_{1},F_{\bar{1}})}{dE_\mu}R_\mu(E_\mu)
  dE_\mu} \; .
\end{equation}
In first approximation, $F_{\nu_e}:F_{\nu_\mu}:F_{\nu_\tau}\sim
0.6:0.15:0.15$, which significantly deviates from what is expected
in the case of stable neutrinos. Thus, by measuring $R$ with a
moderate precision, we can  test whether neutrinos are stable or
not. To extract $|U_{\mu 1}|$ precisely enough, higher accuracy in
the measurement of $R$ will be required. Though the  flux arriving
at the Earth is purely composed of $\nu_1$ and $\bar{\nu}_1$, for
extracting $R$ from the data the knowledge of the  dependence of the
neutrino flux on  energy, $F_1(E)$, is necessary since detection of
processes contributing to $R$ have different kinematics. As
discussed in \cite{Beacom:2003nh}, the spectrum of neutrinos can be
determined by measuring the total energy of muon-track events. The
accuracy with which the spectrum can be determined, as indicated in
\cite{Beacom:2003nh}, strongly depends on the overall shape of the
spectrum. Another limiting factor will be  the size of the data
sample which depends on the, as yet unknown, neutrino luminosity at
the source.

$\gamma$-ray bursts may be accompanied by a flux of energetic
($\sim 1$~TeV) neutrinos \cite{Meszaros:2001ms}. Taking the distance
of the $\gamma$-ray burster from the Earth to be of order $10^{28}$
cm, one finds that $\stackrel{(-)}{\nu_2}$  and
$\stackrel{(-)}{\nu_3}$ will decay before reaching the detectors if
their lifetimes in their rest frame, $\tau_{\nu_i}$, satisfy the
following inequality:
\begin{equation}
  \tau_{\nu_i} \stackrel {<} {\sim} 10^{16} {\rm sec} ~ \left({m_{\nu_i} \over
  E}\right) \left({L \over 10^{28}~{\rm cm}} \right).
  \label{ineq}
\end{equation}
In the case of a hierarchical spectrum $m_1\simeq 0$, $m_3 \sim
0.05$ eV and $m_2 \sim 0.009$ eV, from equation (\ref{ineq}) we find
that in order to have en route decay of $\nu_2$ and $\nu_3$ coming
from gamma ray bursters, their respective lifetimes have to be
shorter than 10 sec and 100 sec. For the quasi-degenerate spectrum
with $m_1\simeq m_2\simeq m_3= 0.1$ eV, the bound is weaker:
$10^3$~sec. 
Taking the coupling of the Majoron to neutrinos to be ${\cal
O}\left(10^{-6}\right)$ (corresponding to the bound from supernova
cooling considerations \cite{Farzan:2002wx}), we find the lifetime
of neutrinos in their rest frames to be of order of 1~sec for $m_\nu
= \left(\Delta m_{21}^2\right)^{1/2}$, which means neutrinos with
TeV-scale energies that come from cosmic distances can decay before
reaching the detectors, whereas  neutrinos with TeV-scale energies
produced inside our Galaxy will not have enough time to decay before
reaching the  Earth. If nature is so kind as to set the lifetime of
neutrinos in this range, the two methods described in this section
and the next may be combined to extract the value of $|U_{\mu 1}|$.
All these considerations are essentially at an `idea level' and
further study is necessary to see if useful information can be
obtained from the proposed measurements.

 The flux of neutrinos with TeV-scale energies from an
individual $\gamma$-ray burster at cosmological distance $z \sim 1$
produces $(10^{-1}-10)$ muons in 1~km$^3$-size detectors
\cite{Meszaros:2001ms}. Since these neutrinos are correlated in time
with the $\gamma$-ray bursts and coming from the same source, they
can be distinguished from background neutrino fluxes. The rate of
$\gamma$-ray bursts detectable on the Earth is $\sim 10^3$/year, so
the data sample is fairly large and useful information on mixings
may be obtained.

\subsubsection{Stable neutrinos and loss of coherence}
\label{stable}

Consider stable or meta-stable neutrinos produced by cosmological
sources. For example, consider again the neutrinos with $E\sim
1$~TeV accompanying the $\gamma$-ray bursts \cite{Meszaros:2001ms}
or TeV neutrinos from the center of our Galaxy ($L\sim {\rm  ~ 10~
kpc}$). For such neutrinos, the oscillation length is much smaller
than the distance from the source; i.e., $\Delta m_{21}^2 L/E
\gg 1$. As a consequence, the (anti-)neutrino beam will loose its
coherence and  the transition probability is therefore averaged out
as:
\begin{equation}
  P_{\alpha \beta}  = \bar{P}_{\alpha \beta} = \sum_{i} |U_{\alpha
  i}|^2|U_{\beta i}|^2 \; ,
  \label{easily}
\end{equation}
where $P_{\alpha \beta}$ and $\bar{P}_{\alpha \beta}$ are
respectively the probabilities of transitions $\nu_\alpha \to
\nu_\beta$ and $\bar{\nu}_\alpha \to \bar{\nu}_\beta$. To derive
(\ref{easily}) the fact that $|\bar U_{\alpha i}|=|U_{\alpha i}|$ has
been used. In particular:
\begin{equation}
  P_{\mu \mu} = \sum_{i} |U_{\mu i}|^4 = K_{\mu\mu} -
  2 |U_{\mu2}|^2 |U_{\mu 1}|^2 \; ,
  \label{mu-mu}
\end{equation}
and:
\begin{equation}
  P_{e\mu} = \sum_{i} |U_{\mu i}|^2|U_{e i}|^2  =
  K_{e\mu} -
  |U_{\mu2}|^2 (|U_{e 1}|^2 - |U_{e 2}|^2) \; ,
  \label{e-mu}
\end{equation}
where $K_{\mu\mu}$ and $K_{e\mu}$ are known functions of $|U_{e1}|$,
$|U_{e2}|$, $|U_{e3}|$, $|U_{\mu 3}|$ which do not depend on
$|U_{\mu1}|^2$ and $|U_{\mu2}|^2$. The probability $P_{ee}$ does not
depend on $|U_{\mu1}|^2$ and $|U_{\mu2}|^2$.

The probabilities in equations (\ref{easily}), (\ref{mu-mu}), and
(\ref{e-mu}) have the following properties which play a key role in the
calculations: $P_{\alpha \beta} = P_{\beta \alpha}$; the probabilities
for neutrinos and anti-neutrinos are equal; and the probabilities do
not depend on energy.

Let us assume that at the source
$F_{\nu_e}:F_{\nu_\mu}:F_{\nu_\tau}=w_e:w_\mu:w_\tau$.
After traveling long distances (${\Delta m_{21}^2 L/ 2 E} \gg 1$), the
flavour ratio will evolve into:
\begin{equation}
  F_{\nu_e}:F_{\nu_\mu}:F_{\nu_\tau}=
  \sum_{\alpha, i} w_\alpha |U_{\alpha i}|^2 |U_{ei}|^2:
  \sum_{\alpha, i} w_\alpha |U_{\alpha i}|^2 |U_{\mu i}|^2:
  \sum_{\alpha, i} w_\alpha |U_{\alpha i}|^2 |U_{\tau i}|^2.
  \label{stableratio}
\end{equation}
Thus, the ratio $R$ depends on $|U_{\mu 1}|$ and measuring this ratio,
the value of $|U_{\mu 1}|$ can, in principle, be derived
\cite{Farzan:2002ct} (see also
\cite{Serpico:2005sz,Serpico:2005bs,Winter:2006ce}).
However, this ratio strongly depends on the original flavour
composition.
Two different processes for neutrino production have been suggested
with different predictions for the flavour ratios:
\begin{enumerate}
  \item $\pi^{+}  \to \mu^{+} + \nu_{\mu}$, and then
        $\mu^+ \to e^+  + \nu_e + \bar{\nu}_{\mu}$; and the
        CP-conjugate of these processes.
        These processes yield $F^0_{\nu_e}:F^0_{\nu_\mu}:F^0_{\nu_\tau}=1:2:0$
        at the source; and
  \item Decay of the neutron: $n\to p+ e+\bar{\nu}_e$ which yields
        $F^0_{\nu_e}:F^0_{\nu_\mu}:F^0_{\nu_\tau}=1:0:0$.
\end{enumerate}
The two cases can be discriminated by a moderately accurate
measurement of $R$. However, as shown in \cite{Kashti:2005qa}, the
muon produced in pion decay can lose energy before it decays, which
in turn reduces the value of $F^0_{\nu_e}:F^0_{\nu_\mu}$ at the
source. Moreover, the two processes  can simultaneously be at work
which again will result in an unknown flavour ratio at the source.
In order to extract $|U_{\mu 1}|$ with an accuracy of 10\%, it is
necessary to know the original flux with a precision  better than
10\%. As discussed in \cite{Farzan:2002ct}, if
$F_{\nu_e}/F_{\nu_\mu}$ and $F_{\nu_\tau}/F_{\nu_\mu}$ are
separately measured, it will be possible to independently  extract
the original flavour ratio. Such information can be derived, if the
detector can discriminate between electronic showers (resulting from
the CC interactions of $\nu_e$ or the CC interaction of $\nu_\tau$
and the subsequent decay of the produced $\tau$ to the electron) and
hadronic showers (produced by the NC interaction of all neutrinos or
the CC interaction of $\nu_\tau$ and the subsequent `hadronic' decay
of the produced $\tau$). Although such a discrimination is in
principle possible but, in practice, it will be challenging
\cite{Beacom:2003nh}.

\subsubsection{Summary \label{conclusion}}

In this section, we have discussed the possibility of
extracting information on the mixing parameters by studying
neutrinos from astrophysical sources. As discussed in
section \ref{Subsubsect:triangleCP},    measuring the value of
$|U_{\mu 1}|$ is essential for reconstructing the unitarity triangle;
it is extremely challenging for the accelerator-based experiments to
measure $|U_{\mu 1}|$.
We have therefore focused on
deriving $|U_{\mu 1}|$ from the astrophysical-neutrino 
data in this section.

The flavour ratio of astrophysical neutrinos can be employed  to
measure $|U_{\mu 1}|$. In the case of 
stable neutrinos, the result would suffer from the uncertainty in 
the flavour composition of the flux at the source. We have 
argued that if neutrinos decay on their way with 
$\tau_\nu<10-10^3$~sec (depending on the neutrino-mass scheme) such 
an uncertainty would not affect the results. So, if there 
sources of sufficient luminosity to provide reasonable data samples,
astrophysical neutrinos can be considered a useful means of deriving
$|U_{\mu 1}|$ and thus reconstructing the unitarity triangle.

%% file: 07-Muon-Physics/07-Muon-Physics.tex
\section{Muon physics}
\label{Sect:MuonPhysics}

\input 07-Muon-Physics/07-01-Introduction/07-01-Introduction

\input 07-Muon-Physics/07-02-FMP/07-02-FMP

\input 07-Muon-Physics/07-03-LFV-theory/07-03-LFV-theory

\input 07-Muon-Physics/07-04-FV-exp/07-04-FV-exp

\input 07-Muon-Physics/07-05-normal-decay-WF/07-05-normal-decay-WF

\input 07-Muon-Physics/07-06-conclusions/07-06-conclusions

%% file: 07-Muon-Physics/07-01-Introduction/07-01-Introduction.tex
\subsection{Introduction}

Ever since the discovery of the muon, the study of its properties and 
decays have
contributed to a deeper understanding of Nature at the smallest distance
scale. Muon physics played a fundamental role in establishing the V--A
structure of weak interactions and the validity of quantum electrodynamics.  
Moreover, muon physics has not yet exhausted its potential and, indeed,
may provide crucial information regarding one of the most fundamental quests
in modern physics: the structure of the theory which lies beyond the
Standard Model of particle physics.  The present 3.4 standard deviation
difference between the measured
\cite{Bennett:2002jb,Bennett:2004pv,Bennett:2006fi}
and Standard Model \cite{Miller:2007kk} values of the muon 
anomalous magnetic moment,
$a_\mu = (g_\mu - 2)/2$, might be such an example.

A muon storage ring is an essential part of the Neutrino Factory idea, with
the primary aim of the machine being the study of neutrino
properties.  The Neutrino Factory is also 
an ideal place to study muon properties, since
they provide, necessarily, muon fluxes which are orders of magnitude larger
than that which can be obtained at present. For example, at the Paul
Scherrer Institut (PSI) beams of $10^8$~$\mu$/s are available.  At the
Japan Proton Accelerator Research Complex (J-PARC) the proposed muon
intensity for the PRISM experiment is $10^{11}$ to $10^{12}$~$\mu$/s.  
At a Neutrino Factory fluxes as large as $10^{13}$ to
$10^{14}$~$\mu$/s could be available. It is, therefore, imperative to 
understand how to take full advantage of these intense muon beams in order
to improve significantly on the reach of low-energy muon experiments.

Independent of whether or not $(g-2)_\mu$ is constraining, or pointing to,
new physics, it, along with the suite of muon experiments described below, 
will provide significant
information from the precision frontier that is complementary to that
expected from the Large Hadron Collider.  If charged
lepton-flavour-violation, or a permanent electric-dipole moment are 
observed, they will help
clarify our understanding of the information gained at the LHC.
If not observed, along with $(g-2)_\mu$,
 they will restrict possible interpretations of 
the new physics.  In order to proceed to significantly greater
sensitivities, the electric-dipole moment and lepton-flavour violating
experiments would greatly benefit from this new, very intense muon
source.

While precise measurements of the muon lifetime and Michel parameters
provide tests for the theory of weak interactions and its possible extensions,
one of the main interests in muon physics lies in the search for processes
that violate muon number, or the observation of a permanent muon
electric-dipole moment (EDM).
 The discovery of decays such as
$\mu^+ \to e^+ \gamma$, $\mu^+\to e^+e^-e^+$, or $\mu^-$--$e^-$
conversion in nuclei, or the observation of a muon EDM,
would be an indisputable proof of the existence of new
dynamics beyond the Standard Model.

Global symmetries (like individual lepton numbers), as opposed to local
symmetries, are considered not to be based on fundamental principles and are
expected to be violated by gravitational effects, in the strong regime, and,
more generally, by higher-dimensional effective operators which describe local
interactions originating from some unknown high-energy dynamics. Baryon number
conservation is another example of an abelian global symmetry of the
Standard Model, which can be broken by new-physics effects.

Atmospheric- and solar-neutrino experiments have provided strong
evidence for neutrino oscillations, which has now been confirmed by
terrestrial experiments at accelerators and reactors.
 This implies  violation of individual
lepton numbers ($L_i$) and, most likely, of total lepton number ($L$), which
is a first indication of physics beyond the Standard Model.
Current neutrino data indicate values of the neutrino masses corresponding
to non-renormalisable interactions at a scale $M\sim 10^{9-14}$~GeV. New
lepton-number violating dynamics at the scale $M$ cannot yield observable 
rates for rare muon processes, since the corresponding effects are suppressed
by $(m_{\mu}/M)^4$. The observation of muon-number violation in muon decays
would thus require new physics beyond that responsible for neutrino masses.
Theoretically, however, there is no reason why $L_i$ and $L$ would be
broken at the same energy scale. Indeed, in many frameworks, such as
supersymmetry, the $L_i$-breaking scale can be close to the weak scale.
In this case, muon processes with $L_\mu$ violation would occur with rates
close to the current experimental bounds.

It is also very important to stress that the information which can be
extracted from the study of rare muon processes is, in many cases, not
accessible to high-energy colliders. Take supersymmetry as an example. 
While the LHC can significantly probe slepton masses, it cannot compete
with muon-decay experiments in constraining the slepton mixing angles.

In the following section we discuss dipole moments, lepton-flavour
violation and other muon-decay experiments.  Many additional details
can be found in the excellent report of the CERN working 
group \cite{Aysto:2001zs} of 2001, on which this document is based.

%% file: 07-Muon-Physics/07-02-FMP/07-02-FMP.tex
\subsection{The Magnetic and Electric Dipole Moments of the Muon }

The electric- and magnetic-dipole 
moments have been an integral part of relativistic electron
(lepton) theory since Dirac's famous 1928
paper, in which he pointed out
that an electron in external electric and magnetic fields has 
``the two extra terms:
\begin{equation}
\frac{e \hbar }{ c}\left({\mathbf \sigma} , {\mathbf H} \right) 
+ i \frac{e \hbar}{ c}\rho_1 \left( {\mathbf \sigma} , {\mathbf E} \right) ,
\label{eq:dirac-dpm}
\end{equation}
\dots [which], when divided by the factor $2m$, can be regarded as the
additional potential energy of the electron due to its new
degree of freedom.\cite{Dirac:1928hu}''
These terms represent the magnetic-dipole (Dirac) moment and electric
dipole moment interactions with the external magnetic and electric
fields. 

In modern notation, the magnetic dipole moment (MDM) interaction becomes:
\begin{equation}
\bar u_{\mu}\left[ eF_1(q^2)\gamma_{\beta} +
\frac{i e}{ 2m_{\mu}}F_2(q^2)\sigma_{\beta \delta}q^{\delta}\right]
u_{\mu} \, ,
\end{equation}
where $F_1(0) = 1,$ and $F_2(0) = a_{\mu}$, the latter being the
anomalous (Pauli) moment.
The electric dipole moment (EDM) interaction is:
\begin{equation}\bar u_{\mu}
\left[\frac {i e}{ 2 m_{\mu}} F_2(q^2) - F_3(q^2)\gamma_5
\right]\sigma_{\beta \delta}
q^{\nu}u_{\mu} \, ,
\end{equation}
where $F_2(0) = a_{\mu}$, $ F_3(0) = d_{\mu}$, with:
\begin{equation}
d_{\mu} = \left( \frac{\eta} {2} \right) \left(\frac {e \hbar }{ 2 mc}\right) 
\simeq \eta \times 4.7\times 10^{-14}\ e\,{\rm cm}.
\label{eq:eta}
\end{equation}
This $\eta$, which is the EDM analogy to $g$ for the MDM,
 should not be confused with the Michel parameter $\eta$.

The existence of an EDM implies that both {\sl P} and {\sl T} are 
violated \cite{Purcell_50,Landau:1957tp,Ramsey_58}.  
This can be seen by considering the
non-relativistic Hamiltonian for a spin one-half
particle in the presence of both an electric and a magnetic field:
${\mathcal H} = - \vec \mu \cdot \vec B  - \vec d \cdot \vec E$.
The transformation properties of $\vec E$, $\vec B$, $\vec \mu$ and $\vec d$
are given in the table \ref{tb:edm}(a), and we see that while
$\vec \mu \cdot \vec B$ is even under all three,
$\vec d \cdot \vec E$ is odd under both {\sl P} and
{\sl T}. While parity violation has been observed in many weak processes,
direct {\sl T} violation has only been observed in the 
neutral-kaon system \cite{Angelopoulos:1998dv}.
In the context of CPT symmetry, an EDM implies CP
violation, which is allowed by the Standard Model for decays
 in the neutral-kaon and $B$-meson 
sectors.
\begin{table}
\begin{center}
 \begin{minipage}{0.95\textwidth}
  \begin{minipage}{0.27\textwidth}
     \begin{tabular}{cccc} \hline
      & {$\vec E$ } &{$\vec B$ }  &{$\vec \mu$ or  $\vec d$} \\
      \hline
      {\sl P} & - & + & + \\
      {\sl C} & - & - & - \\
      {\sl T} & + & - & - \\
      \hline 
      \\
      & & (a)  & \\
    \end{tabular}
  \end{minipage}
\hskip0.2in
  \begin{minipage}{0.05\textwidth}
    \begin{center}
      \begin{tabular}{ccc} \hline
   { Particle}  &{ Present EDM} & { Standard Model}  \\
                 & Limit ($e$~cm)           & Value ($e$~cm) \\
   \hline
   $n$ & {$2.9 \times 10^{-26}$ } (90\%CL)\cite{Baker:2006ts}  & {$10^{-31}$ }  \\
   \hline
   $e^-$  & {$\sim 1.6 \times 10^{-27 }$} (90\%CL)\cite{Regan:2002ta} &{$10^{-38}$ } \\
   \hline
   {$\mu$} &{$<10^{-18}$ } (CERN)\cite{Bailey:1978mn} & {$10^{-35}$ }\\
   & $\sim10^{-19}$ $^\dag$ (E821) \  & \\
   \hline
   $^{199}Hg$ & $ 2.1 \times  10^{-28}$  (95\%CL)\cite{Romalis:2000mg} \\
   \hline
   $^\dag$Estimated \\
      &  (b)  & \\
      \end{tabular}
     \end{center}
   \end{minipage}
\end{minipage}
 \end{center}
\caption{
  (a) Transformation properties of the magnetic and electric fields
      and dipole moments.
  (b) Measured limits on electric dipole moments, and their Standard
      Model values
}
\label{tb:edm}
\end{table}

The identification of new sources of CP violation
appears to be a crucial requirement for explaining the dominance 
of matter over anti-matter in the Universe. 
Permanent electric-dipole moments of fundamental particles
would violate both time reversal (T) and parity
(P) invariance, and with the assumption of
CPT conservation also the CP symmetry \cite{Landau:1957tp,Ramsey_58}.
The present limits from EDM searches are given in
table \ref{tb:edm}(b).

\begin{figure}
\begin{center}
\subfigure[ ]{\includegraphics[width=0.5\textwidth]{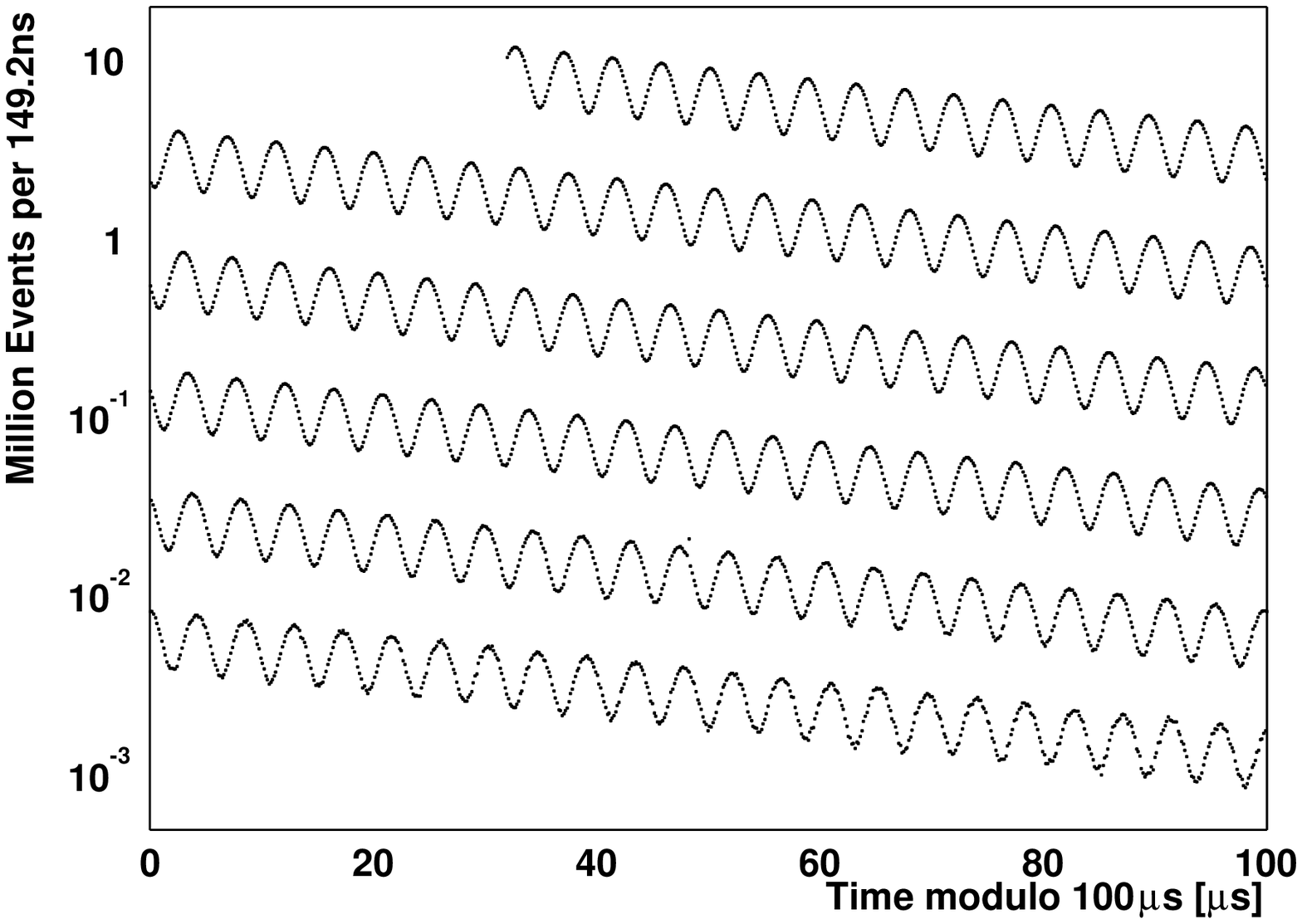}}
\subfigure[ ]{\includegraphics[width=0.4\textwidth]{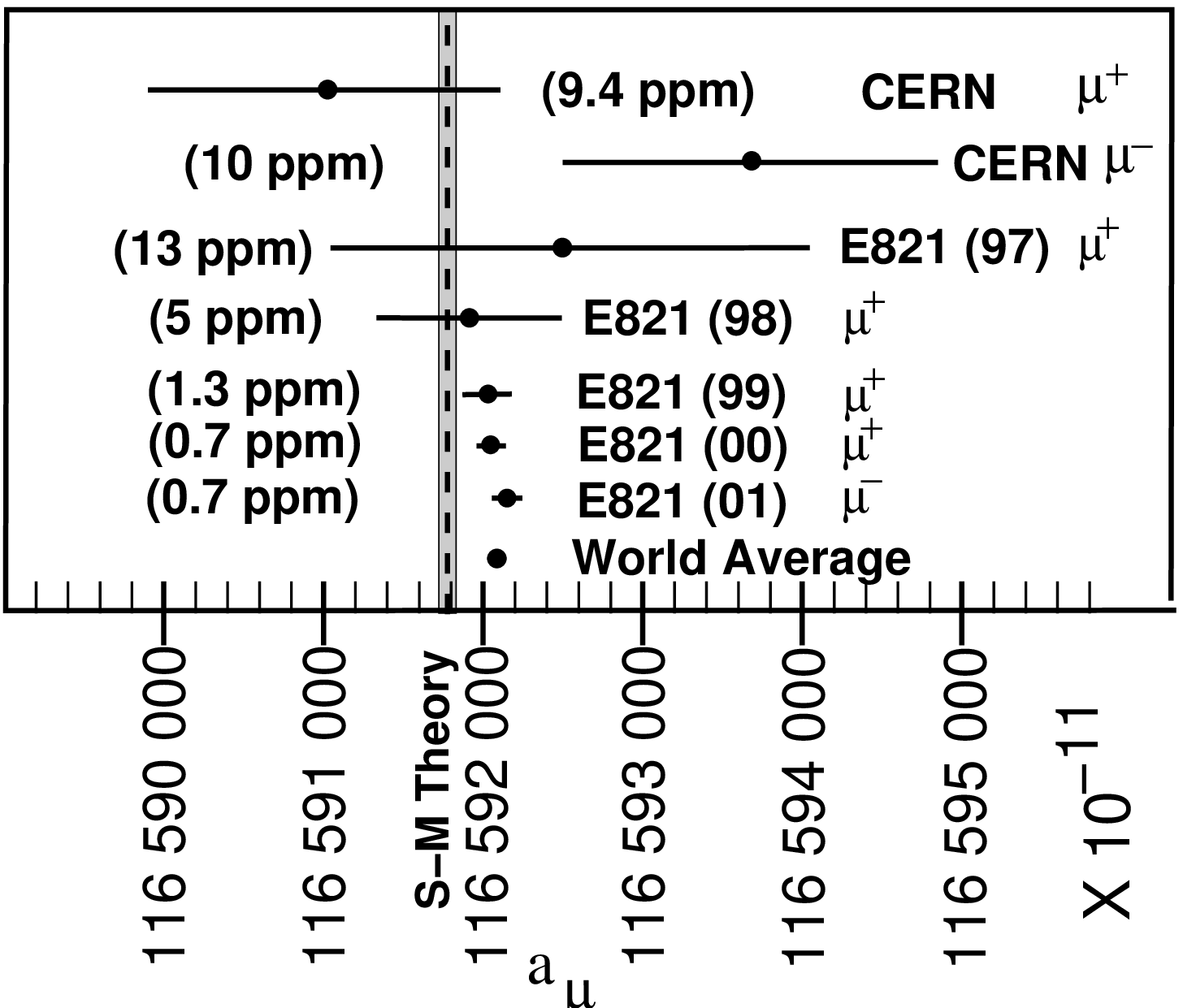}}
\caption{(a)The time spectrum of $3.6 \times 10^9$ 
electrons with energy greater than
1.8 GeV from the 2001 E821 data set \cite{Bennett:2004pv}. 
The diagonal ``wiggles'' displayed  modulo 100~$\mu$s result from
the muon spin precession in the storage ring.
    Adapted with kind permission of Physical Review from figure 2 in
    reference {Bennett:2006fi}.
    Copyrighted by the American Physical Society.
(b)Measurements of the muon anomaly, indicating the value, as well as
the muon's sign.  As indicated in the text, to obtain
the value of $a_{\mu^-}$ and the world average
 {\sl CPT} invariance is assumed. The theory value is taken from
reference \cite{Miller:2007kk}, which uses electron-positron
annihilation to determine the hadronic contribution.
\label{fg:wig-result}}
\end{center}
\end{figure}

The anomalous magnetic moment (anomaly) of 
the muon, $a_{\mu} \equiv g_\mu - 2$,
has a long history of constraining models of physics beyond the 
Standard Model.  It has now been measured to a relative precision of 
0.54 parts per million \cite{Bennett:2002jb,Bennett:2004pv,Bennett:2006fi}.
Muons are stored in a super-ferric storage ring,
and the spin difference frequency between the 
cyclotron frequency and the muon-spin-rotation frequency is given by:
\begin{equation}
\vec \omega_a = - {\frac {e}{m}}
\left[ a_{\mu} \vec B -
\left( a_{\mu}- {\frac  {1}{ \gamma^2 - 1}} \right) \vec \beta \times \vec E
\right].
\label{eq:tbmt}
\end{equation}
which is the frequency that the spin precesses relative to the momentum.
  At $\gamma= 29.3$ the electric field
used for vertical focusing does not contribute to the 
spin precession for a muon on the central orbit.  By counting high-energy
positrons as a function of time, one observes the muon lifetime modulated
by the $(g-2)$ precession, as shown in figure \ref{fg:wig-result}(a).
Both $a_{\mu^+}$ and $a_{\mu^-}$ were measured. Assuming
CPT invariance, the E821 collaboration obtained the 
 anomalous magnetic moment \cite{Bennett:2006fi}:
\begin{equation}
  a_\mu(\mathrm{Expt}) = 11\,659\,181.2(6.9) \times
  10^{-10}~~\mbox{(0.54\,ppm)} \, .
\label{eq:E821-amu}
\end{equation}
The total uncertainty includes a 0.46~ppm statistical uncertainty
and a 0.28~ppm systematic uncertainty, combined in quadrature.

\begin{figure}
\begin{center}
  \includegraphics[width=0.65\textwidth,angle=0]{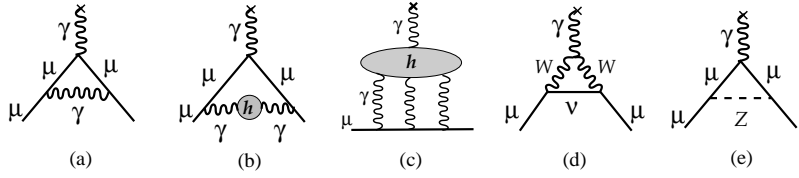}
\end{center}
  \caption{The Feynman graphs for: (a) Lowest-order QED (Schwinger) term;
 (b)Lowest-order hadronic contribution; (c) hadronic light-by-light
contribution;  (d)-(e) the lowest order electroweak $W$ and $Z$ 
contributions.  With the present 
limits on $m_h$, the contribution from the single Higgs loop is negligible.}
  \label{fg:radcor}

\end{figure}

The Standard Model theory value consists of well known QED and Weak
contributions, plus a hadronic contribution of about 60 ppm of $a_{\mu}$
which dominates the uncertainty on the Standard Model value. The
leading-order
contributions are shown diagrammatically in figure \ref{fg:radcor}. 

The hadronic contribution has been the source of substantial 
work \cite{Davier:2003pw,Davier:2004gb}, which continues 
to the present.  The lowest order
must be taken from $e^+e^-\rightarrow {\rm hadrons}$ using a
dispersion relation \cite{Miller:2007kk}:
\begin{equation}
a_{\mu}^{({\rm Had;1})}=\left(\frac{\alpha m_{\mu}}{ 3\pi}\right)^2
\int^{\infty} _{4m_{\pi}^2} \frac{ds }{ s^2}K(s)R(s);
\end{equation}
where:
\begin{equation}
R\equiv  \frac {\sigma_{\rm tot}(e^+e^-\to{\rm hadrons})} 
{ \sigma_{\rm tot}(e^+e^-\to\mu^+\mu^-)}\, ,
\end{equation}
and $K(s)$ is a known function \cite{Miller:2007kk}.
 The only assumptions here are analyticity and the optical
theorem.    Recently published data
on the hadronic cross
sections \cite{Aloisio:2004bu,Akhmetshin:2006wh,Akhmetshin:2006bx,Achasov:2006vp} 
 have significantly
reduced the uncertainty on the hadronic 
contribution \cite{Hagiwara:2006jt,Davier:2007ua}.
The present Standard Model value is \cite{Miller:2007kk}:
\begin{equation}
 a_{\mu}^{{\rm ( SM07)}} = 116\,591\,785 (61) \times 10^{-11}\, .
\label{eq:SM07}
\end{equation}
When compared with the experimental value in equation (\ref{eq:E821-amu})
one obtains 3.4 standard deviation
difference between experiment and 
theory \cite{Hagiwara:2006jt,Davier:2007ua,Miller:2007kk}.

It has been proposed that the hadronic contributions 
could also be determined from hadronic $\tau$-decay data, using the
conserved 
vector current (CVC) hypothesis\cite{Alemany:1997tn}.  
Such an approach can only give the
iso-vector part of the amplitude, 
i.e. the $\rho$ but not the $\omega$ 
intermediate states. In contrast, the  $e^+e^-$ annihilation
cross section contains both iso-vector and iso-scalar contributions,
with the cusp from $\rho-\omega$ interference as a dominant feature.
Since hadronic $\tau$ decay goes through
the charged-$\rho$ resonance, and $e^+e^-$ annihilation goes through
the neutral $\rho$, understanding
the isospin corrections is essential in this approach.  This use of 
the CVC can be checked by comparing the hadronic contribution to
$a_\mu$ obtained from each method.  Alternately, one can take
 the measured branching ratio for
 $\tau^- \rightarrow V^- \nu_\tau $, where $V$ is any vector final
state ({\it e.g.} $\pi^- \pi^0$) and compare it to that predicted using
CVC and $e^+e^-$ data,  applying all the appropriate 
isospin corrections. At present, neither comparison
gives a satisfactory result \cite{Davier:2007ua},
and the prescription of CVC with the 
appropriate isospin correction
 seems to have aspects that are not understood.
 Given two consistent
$e^+e^-$ data sets and the uncertainties inherent in the
 required isospin corrections to the $\tau$ data,
the most recent Standard Model
 evaluations do not use the $\tau$ data to
determine 
$a^{(\rm Had;1)}$ \cite{Hagiwara:2006jt,Davier:2007ua,Miller:2007kk}.
Additional $e^+e^-$ data are expected to become available
in the next year which should increase our confidence in the 
$e^+e^-$-based evaluation.

Since the muon anomaly results from virtual particles that couple to the 
muon, or photon, in principle it is sensitive to all such particles, not just
the known Standard Model particles. Thus the
muon anomaly is sensitive to a number of potential candidates 
for physics beyond the Standard Model \cite{Czarnecki:2001pv}, e.g.,
new particles that couple to the muon
such as the supersymmetric partners of the weak gauge 
bosons \cite{Martin:2002eu,Stockinger:2006zn}; muon
substructure, where the contribution depends on the substructure scale
$\Lambda$ as,
$\delta a_{\mu} (\Lambda_{\mu}) \simeq  {m^2_{\mu}}/{ \Lambda^2_{\mu}}$;
$W$-boson substructure;
and  extra dimensions \cite{Appelquist:2001jz}.

\begin{figure}
\begin{center}
  \includegraphics[width=0.6\textwidth,angle=0]{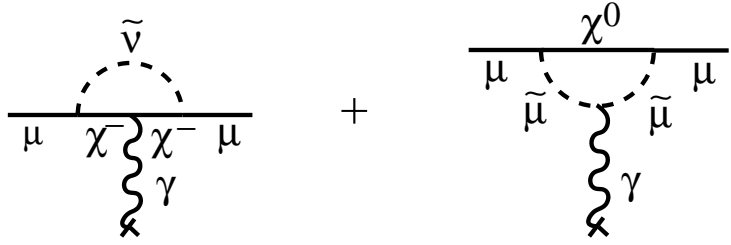}
\end{center}
  \caption{The lowest-order supersymmetric contributions to the muon 
anomaly. The $\chi$ are the superpartners of the Standard Model gauge bosons.
  \label{fg:susy}}

\end{figure}

The potential contribution from
supersymmetry has generated a lot of attention \cite{Martin:2002eu,Stockinger:2006zn},
with the relevant diagrams shown in figure \ref{fg:susy}. 
 A simple model
with equal masses \cite{Czarnecki:2001pv} gives
\begin{eqnarray}
 a_{\mu}^{({\rm SUSY})} 
 & \simeq &
\frac {\alpha(M_Z) }{ 8 \pi \sin^2 \theta_W} 
\frac {m^2_{\mu} }{ \tilde m^2}
\tan \beta \left( 1 -\frac {4\alpha }{ \pi}
\ln\frac{\tilde m }{ m_{\mu}}\right) \nonumber \\
 & \simeq &
({\rm sgn} \mu) \ 13 \times 10^{-10}\ \tan \beta\ 
\left( \frac {100\ {\rm GeV}  }{ \tilde m}\right)^2 \; ;
\end{eqnarray}
where $\tan \beta$ is the ratio of the two vacuum expectation values of the
two Higgs fields.  If the SUSY mass scale were known, 
then $ a_{\mu}^{({\rm SUSY})} $ would provide a clean way to determine
$\tan \beta$.

\begin{figure}
\begin{center}
\subfigure[ ] {\includegraphics[width=.3\textwidth]{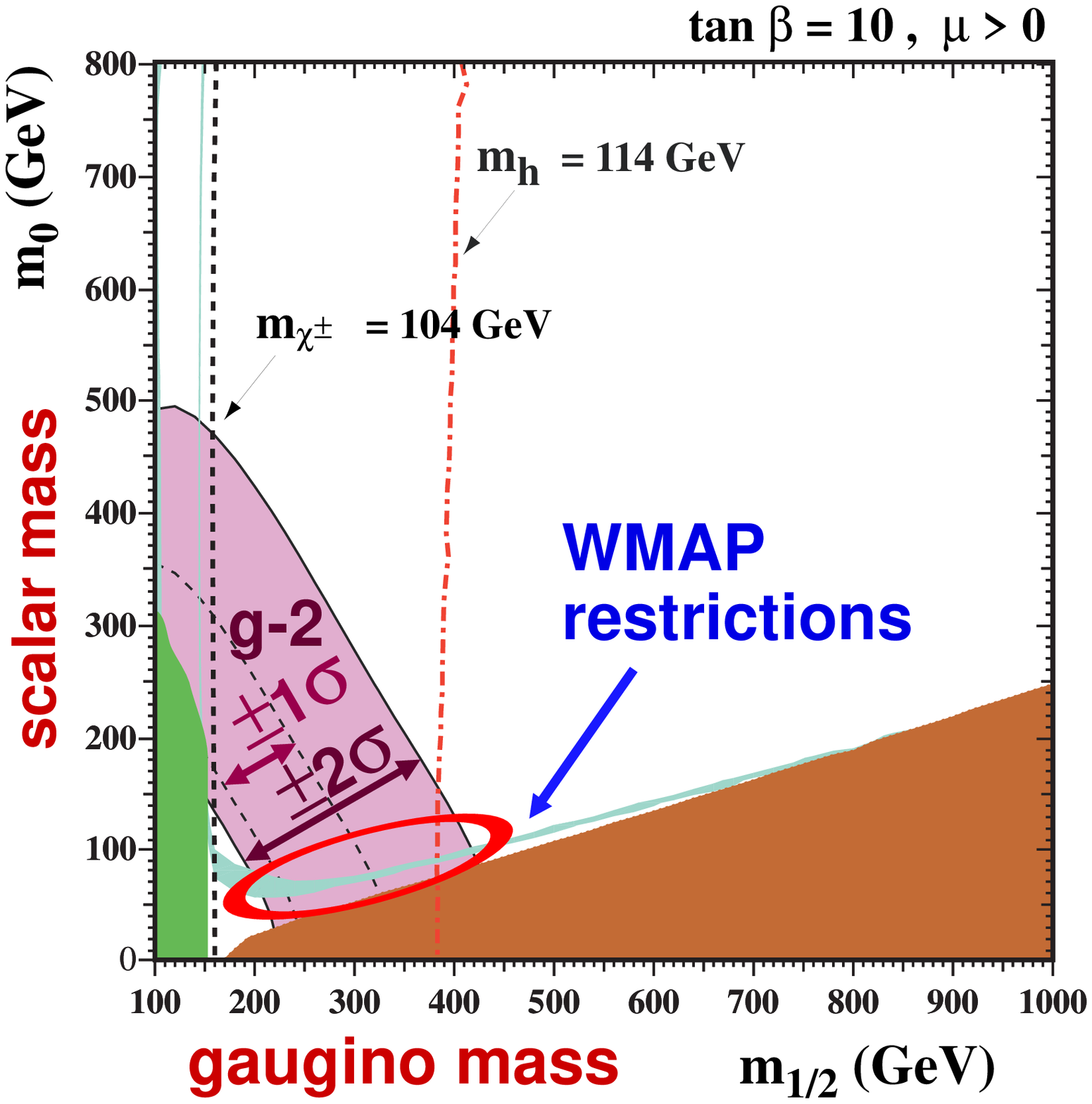}}
\subfigure[ ]{\includegraphics[width=.3\textwidth]
{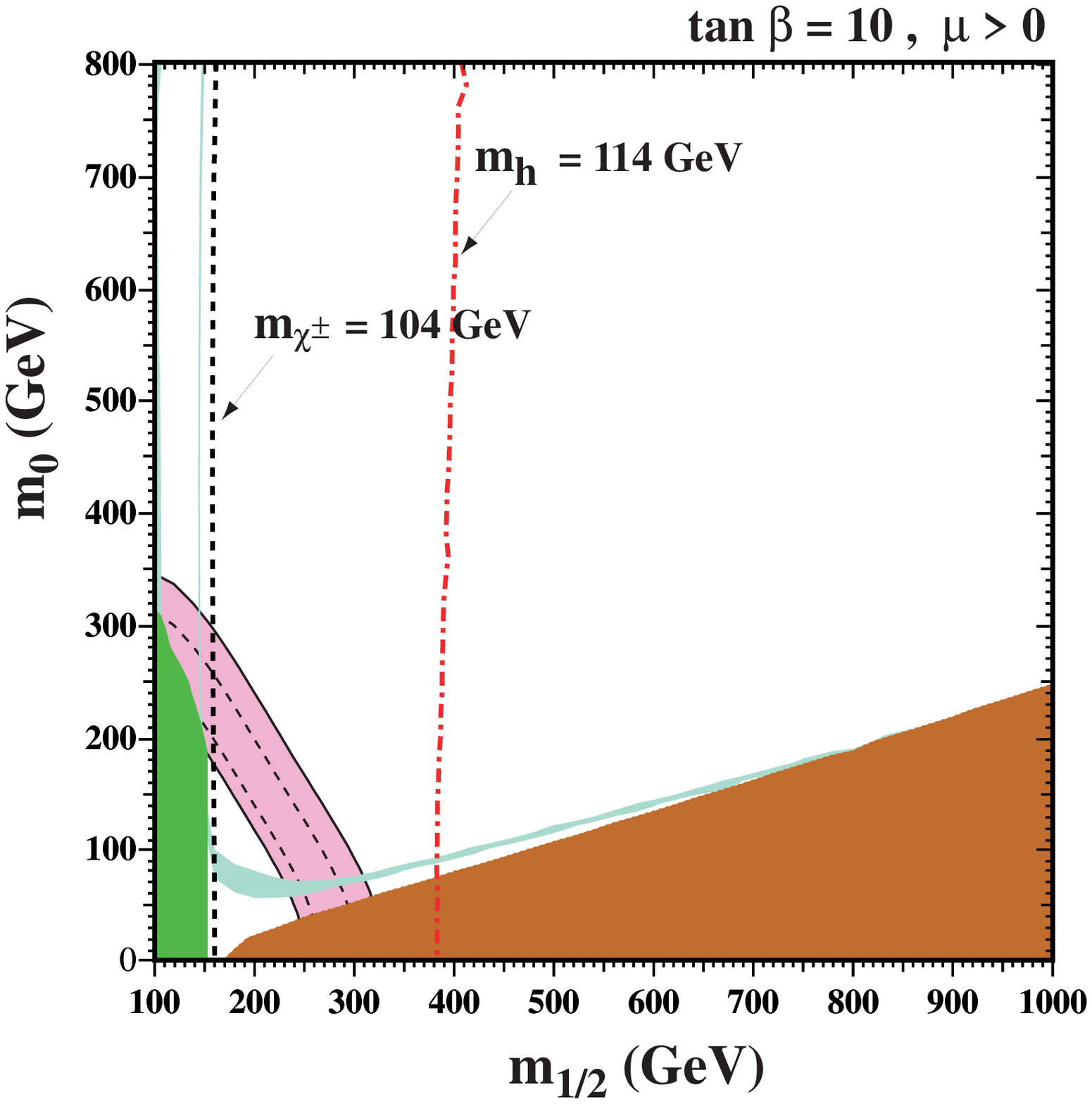}}
\subfigure[]{\includegraphics[width=.3\textwidth]
{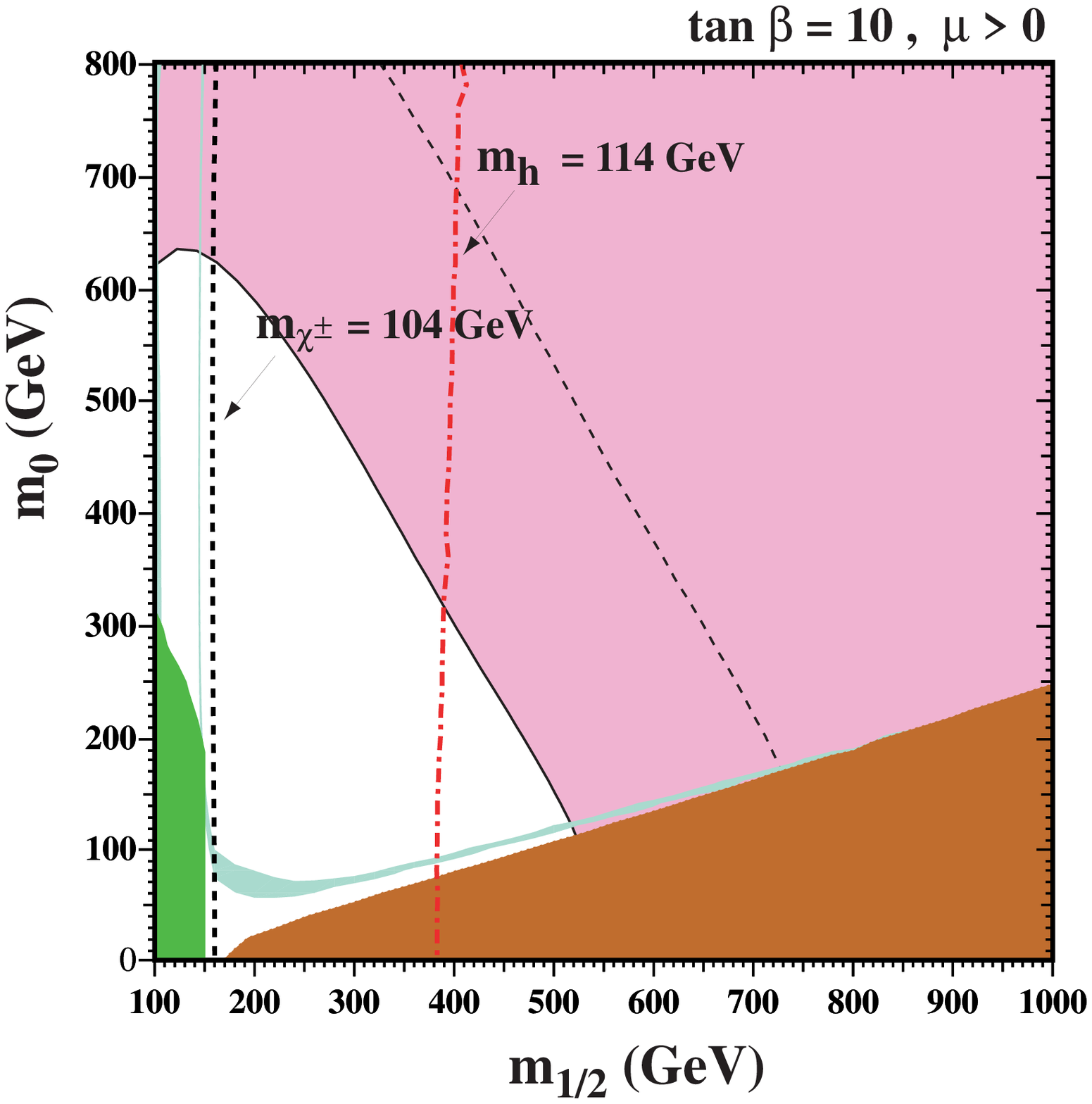}} 
\caption{
The $m_0$--$m_{1/2}$ plane of the CMSSM parameter space for
  $\tan\beta=10$, $A_0=0$, sign$(\mu)=+$. 
(a) The $\Delta a_{\mu}^{(\rm today)}=295(88)\times 10^{-11}$
 between experiment and Standard Model theory is from
reference \cite{Miller:2007kk}, see text. 
The brown wedge on the lower right is excluded by
the requirement the dark matter be neutral.  Direct limits on
 the Higgs and chargino $\chi^\pm$ masses are indicated by vertical
lines. Restrictions from the WMAP satellite data are shown as a 
light-blue line. The $(g-2)$ 1 and 2-standard deviation boundaries are shown
in purple.
The region ``allowed'' by WMAP and $(g-2)$  is indicated by the ellipse,
which is further restricted by the limit on $M_h$.
 (b) The plot with
 $\Delta a_{\mu}=295(39)\times 10^{-11}$, which assumes that in the
future both
the theory and experimental errors decrease to $22 \times 10^{-11}$.
(c) The same errors as (b), but $\Delta = 0$.
 (Figures courtesy of K. Olive)
\label{fg:dark}
}  
\end{center}
\end{figure}

\begin{figure}
\begin{center}
\subfigure[ ] {\includegraphics[width=.33\textwidth]{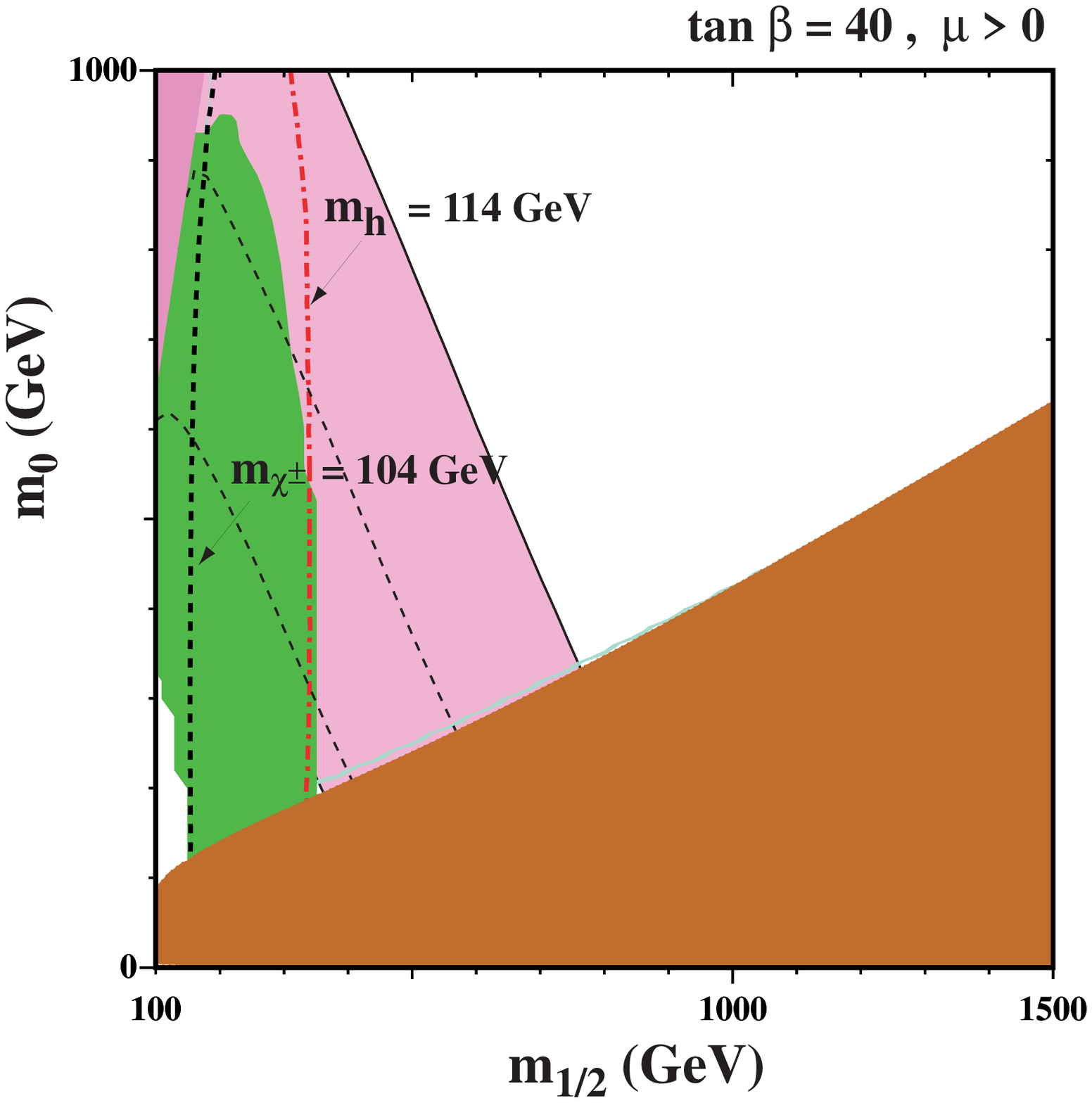}}
\subfigure[ ]
{\includegraphics[width=.33\textwidth]{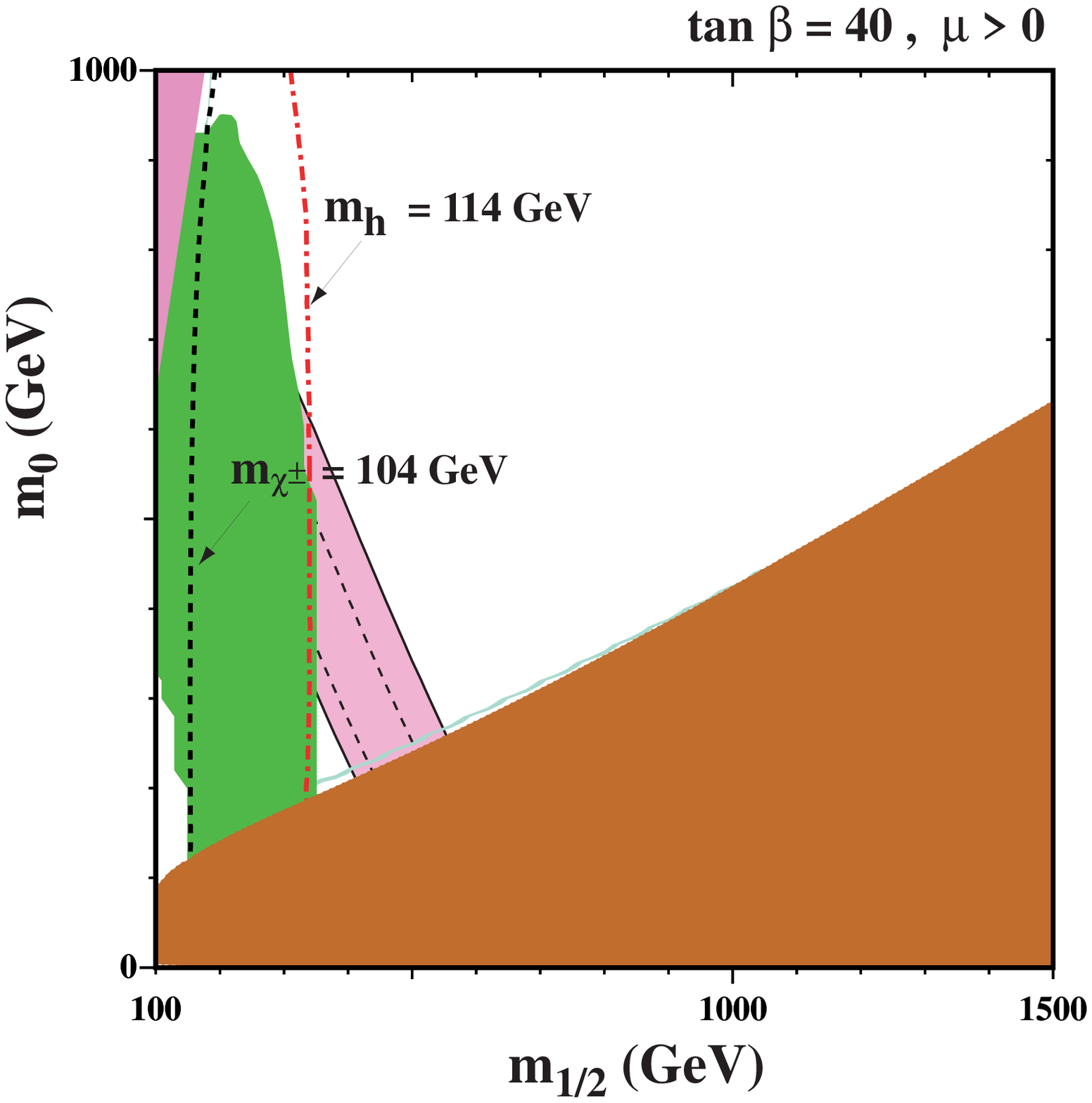}}\subfigure[]{\includegraphics[width=.33\textwidth]{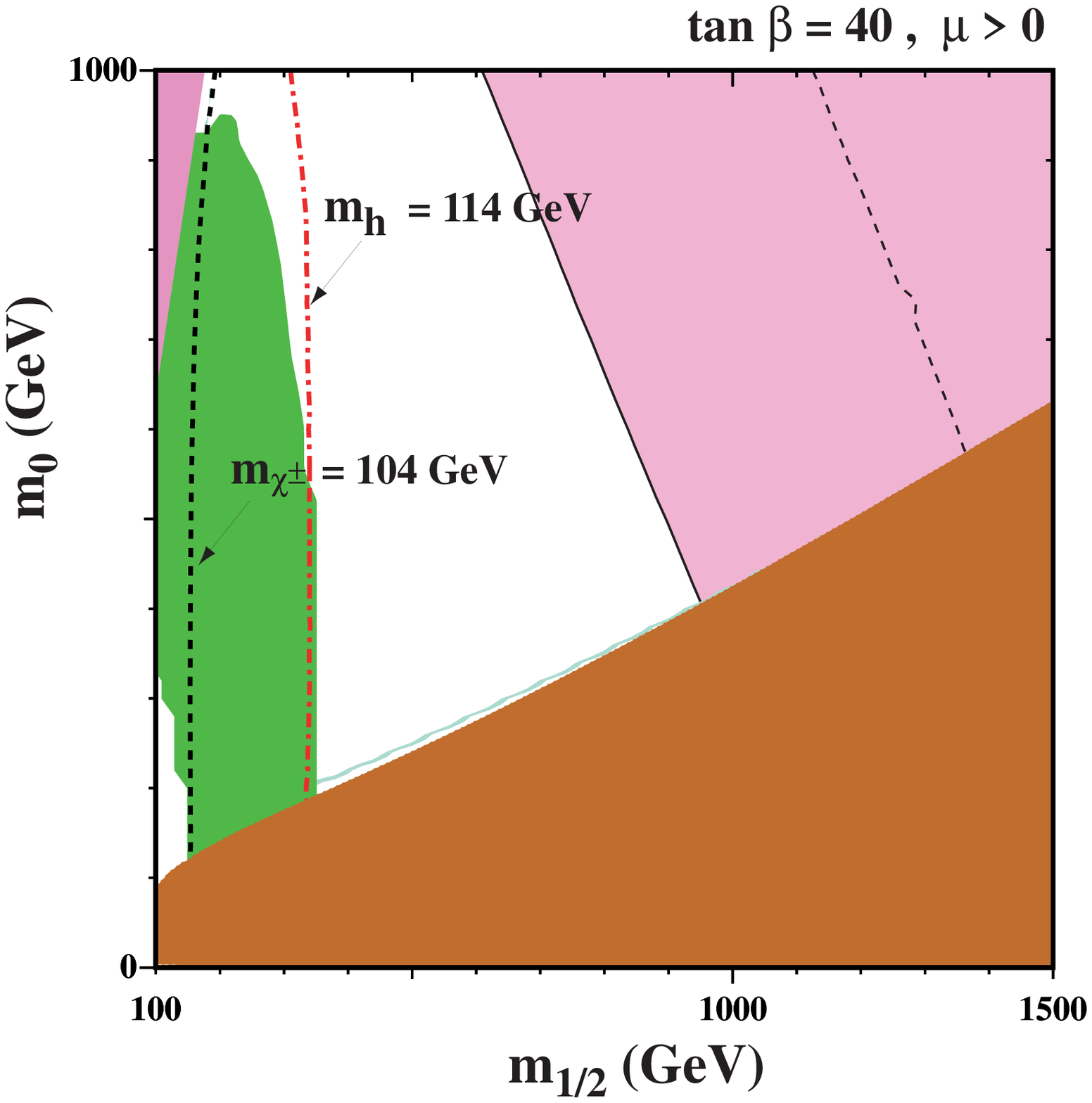}}
\caption{The CMSSM plots as above, but 
with $\tan \beta = 40$.
(a) As in figure \ref{fg:dark} but for $\tan \beta = 40$ (b) The plot with
 $\Delta a_{\mu}=295(39)\times 10^{-11}$, which assumes that  in the
future both
the theory and experimental errors decrease to $22 \times 10^{-11}$.
(c) The same errors as (b), but $\Delta = 0$.
 (Figures courtesy of K. Olive)
\label{fg:dark2}}
\end{center}
\end{figure}

One candidate for the cosmic dark matter is the lightest
supersymmetric partner, the neutralino, $\chi^0$ in figure
\ref{fg:susy}.
In the context of a constrained minimal
supersymmetric model (CMSSM), $(g-2)_\mu$ provides an orthogonal
constraint on dark matter \cite{Ellis:2003cw,Ellis:2005mb} from that provided by the 
WMAP survey, as can be seen in
figures \ref{fg:dark} and \ref{fg:dark2}.

The results from E821 at the Brookhaven AGS are interesting, if not
definitive.
Whatever the final interpretation of $a_\mu$ turns out to be, it
will constrain the theories
of physics beyond the Standard Model \cite{Hertzog:2007hz}.
This ability is clearly
demonstrated in figures \ref{fg:dark} and \ref{fg:dark2} (see
reference \cite{Hertzog:2007hz} for additional examples).
  An improved experiment is
possible at existing facilities, and does not need the ultra high flux of 
muons that would be available at the Neutrino Factory. However, it is clear
that the measurement needs to be further refined.

\begin{figure}
\begin{center}
  \includegraphics[width=.90\textwidth]{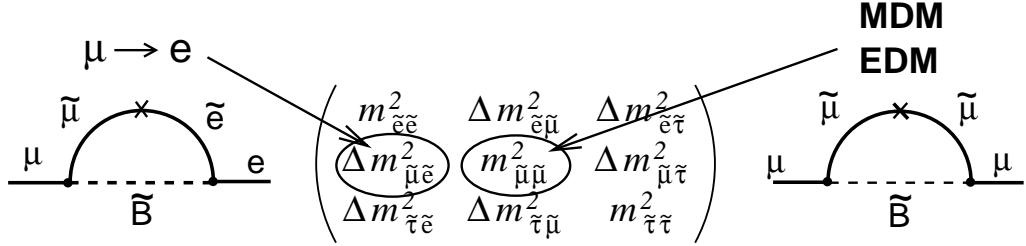}
\end{center}
  \caption{The supersymmetric contributions to the anomaly, and to
$\mu \rightarrow e$ conversion, showing the relevant slepton mixing matrix
elements. The MDM and EDM give the real and imaginary parts of
the matrix element respectively.
\label{fg:susymix}}
\end{figure}

While the MDM has a substantial Standard Model value, the Standard Model
 EDMs
for the leptons are immeasurably small and lie orders of magnitude below
the present experimental limits (see table \ref{tb:edm}).
Thus an EDM at a measurable level would signify physics 
beyond the Standard Model. 
SUSY models, and other dynamics at the TeV scale
do predict EDMs at measurable
levels \cite{Babu:2000dq,Feng:2001sq,Feng:2002wf,Ellis:2001yz,Ellis:2002xg}.  In the context of SUSY, the
EDM and MDM provide information on the diagonal matrix element of
the slepton mixing matrix, while muon flavour violation provides information
on the off-diagonal matrix element, as indicated in figure \ref{fg:susymix}.

If the muon possessed a permanent electric-dipole moment, the spin
precession formula (equation (\ref{eq:tbmt})) is modified by the
addition of a second term: 
\begin{equation}
\vec \omega =
 -\frac {e} {m} 
\left[ a_{\mu} \vec B -
\left( a_{\mu}- \frac{1}{ \gamma^2 - 1} \right) 
{  \frac {\vec \beta \times \vec E }{ c }}
\right] 
  +
\frac{e }{ m}\left[ \frac{\eta}{ 2} \left( \frac {\vec E }{ c} +
\vec \beta \times \vec B \right) \right] 
\label{eq:omegawedm}
\end{equation}
where
$d_{\mu} = (\eta/ 2) ({e \hbar / 2 m c }) \simeq \eta \times 4.7\times 
10^{-14} \ \ e{\rm  - cm}$
and 
$a_{\mu} = (g-2) / 2$.  For reasonable values of 
$\beta$, the motional electric field 
$\vec \beta \times \vec B$ is much larger than electric fields that can be
obtained in the laboratory and the two vector frequencies are orthogonal
to each other. 

A new idea optimises the EDM signal in a storage ring using
 the motional electric field in the rest frame of the muon
interacting with the EDM to cause spin motion \cite{Farley:2003wt}.
The dedicated experiment will use a new storage ring
operated with $\gamma < 29.3$, (e.g. $\gamma =5$,
$p_\mu$ = 500~MeV/c), so that
the $\vec \beta \times \vec E$ term in equation (\ref{eq:omegawedm})
does not vanish. This permits a radial electric field
to be used to stop the $(g-2)$ precession (i.e. make the first
bracketed term in equation (\ref{eq:omegawedm}) go to zero).
Then the spin will follow the momentum
as the muons go around the ring, except for any movement (out of plane)
arising from
an EDM.  Thus the EDM would cause a steady build-up of the spin out of the
plane with time.  Detectors would be placed above and below the storage 
region, and a time-dependent, up-down asymmetry 
$R= ( N_{\rm up} - N_{\rm down})/ ( N_{\rm up} + N_{\rm down} )$ 
would be the signal of an EDM.

Two muon EDM experiments are being discussed.   Adelmann
 and Kirsh \cite{Adelmann:2006ab} have proposed
 that a 
sensitivity of $5 \times 10^{-23}\,e-$cm 
could be achieved with a small storage ring at PSI.
An experiment to search for a permanent EDM of the muon 
with a design sensitivity of $10^{-24}$ $e$-cm has been presented 
to J-PARC as a letter of intent \cite{J-PARC-uEDM-loi}.
These sensitivities 
lie well within values predicted by some SUSY models
\cite{Babu:2000dq}.
For a dedicated muon EDM experiment, a sensitivity of 
$10^{-24}\ e$~cm requires the product of polarisation times 
detected decays to be $NP^2 = 10^{16}$, a flux only available at a
the front-end of a Neutrino Factory, or other high-power proton
accelerator. 
The sensitivity is limited by the muon flux, and it should be possible to
improve significantly on the sensitivity at a higher intensity facility
such as a Neutrino Factory.

If an EDM were to be discovered, one would wish to measure as 
many EDMs as possible to understand
the nature of the interaction.
The muon provides a unique opportunity to search for an EDM of a 
second-generation particle.
While naively the muon and electron EDMs
scale linearly with mass, in some theories the muon EDM is
greatly enhanced relative to linear scaling relative to the electron EDM
when the heavy neutrinos of the theory are 
non-degenerate.\cite{Babu:2000dq,Ellis:2001yz,Ellis:2002xg}

%% file: 07-Muon-Physics/07-03-LFV-theory/07-03-LFV-theory.tex
\subsection{Search for muon number violation}
\subsubsection{Theoretical considerations}
\label{LNV}

In the Standard Model (SM), muon number is exactly conserved. When neutrino
masses are added and neutrino oscillations take place, muon-number violating
processes involving charged leptons become possible as well. However, because
of the smallness of neutrino masses, the rates for these processes are
unobservable \cite{Petcov:1976ff,Bilenky:1977du,Cheng:1977sw,Marciano:1977wx,Lee:1977qz,Lee:1977ti}; for instance:
\begin{equation}
B(\mu \to e \gamma )=\frac{3\alpha}{32 \pi}\sum_i \left|
V_{\mu i}^*V_{ei} \frac{m_{\nu_i}^2}{M_W^2}\right|^2 \sim
10^{-60} \left|\frac{V_{\mu i}^*V_{ei}}{10^{-2}}\right|^2
\left( \frac{m_{\nu_i}}{10^{-2}~{\rm eV}}\right)^4\, .
\end{equation}
The observation of muon-number violation in charged muon decay would,
therefore, serve as an unambiguous sign of new physics and indeed, a number of
SM extensions may be probed sensitively by the study of rare muon decays. Here
we will concentrate on supersymmetric models and models with extra dimensions,
but it should be pointed out that
various other SM extensions also predict observable rates for the rare $\mu$
decays: models with new $\rm Z^\prime$ gauge bosons \cite{Bernabeu:1993ta};
leptoquarks \cite{Davidson:1993qk}; or Lorentz-invariance 
violation \cite{Coleman:1998ti,Bluhm:1999dx,Dehmelt:1999jh,Hughes:2001yk}.
For a review
on muon number violation, see reference \cite{Kuno:1999jp}.

\subsubsection{Model-independent analysis of rare muon processes}

Although a purely model-independent analysis based on effective operators
cannot make any prediction for the absolute rate of rare muon processes, it
can be very useful in determining the relative rates. We will compare the
rates for $\mu^+\to e^+\gamma$, $\mu^+\to e^+e^-e^+$, and $\mu^-$--$e^-$
conversion. In a large class of models, the dominant source of individual
lepton number violation comes from a flavour non-diagonal magnetic-moment
transition. Let us therefore consider the effective operator
\begin{equation}
{\cal L}=\frac{m_\mu}{\Lambda^2} {\bar \mu}_R \sigma^{\mu\nu}e_L
F_{\mu\nu}+{\rm h.c.} \label{magmom}
\end{equation}
This interaction leads to the following results for the branching ratios of
$\mu^+\to e^+\gamma$ ($B(\mu\to e\gamma)$) and $\mu^+\to e^+e^-e^+$ 
($B(\mu\to 3e)$), and for the rate of $\mu^-$--$e^-$ conversion in
nuclei normalised to the nuclear capture rate ($B(\mu N\to eN)$):
\begin{eqnarray}
B(\mu \to e\gamma)&=& \frac{3(4\pi)^2}{G_F^2\Lambda^4}\;, \label{meg}\\
\frac{B(\mu \to 3e)}{B(\mu \to e\gamma)}&=&\frac{\alpha}{3\pi}
\left( \ln\frac{m_\mu^2}{m_e^2}-\frac{11}{4}\right) =6\times 10^{-3}\;,
\label{m3e}\\
\frac{B(\mu N\to eN)}{B(\mu \to e\gamma)}&=& 10^{12}~B(A,Z)~
\frac{2G_F^2m_\mu^4}{(4\pi)^3\alpha}=2\times 10^{-3}~B(A,Z)\;.\label{m4e}
\end{eqnarray}
Here $B(A,Z)$ is an effective nuclear coefficient which is of order 1 for
elements heavier than aluminium ~\cite{Czarnecki:1998iz}. The logarithm in equation \ref{m3e}
is an enhancement factor for $B(\mu \to 3e)$, which is a consequence of the
collinear divergence of the electron-positron pair in the $m_e\to 0$ limit.
Nevertheless, because of the smaller phase space and extra power of $\alpha$,
$B(\mu \to 3e)$ and $B(\mu N\to eN)$ turn out to be suppressed with
respect to $B(\mu \to e\gamma)$ by factors of $6\times 10^{-3}$
and $(2$--$4)\times 10^{-3}$, respectively. 
See reference \cite{Aysto:2001zs} for further discussion.

Next, let us include an effective four-fermion operator which violates
individual lepton number: 
\begin{equation}
{\cal L}=\frac{1}{\Lambda_F^2}{\bar \mu}_L\gamma^\mu e_L {\bar f}_L
\gamma_\mu f_L +{\rm h.c.}\;, \label{op4}
\end{equation}
where $f$ is a generic quark or lepton. The choice of the operator in
equation \ref{op4} is made for concreteness, and our results do not depend
significantly on the specific chiral structure of the operator. First we
consider the case in which $f$ is neither an electron nor a light quark, and
therefore $\mu^+\to e^+e^-e^+$ and $\mu^-$--$e^-$ conversion occur only at the
loop level. Comparing the $\mu^+ \to e^+ \gamma$ rate in equation \ref{meg} with
the contributions from the four-fermion operator to $B(\mu \to 3e)$ and
$B(\mu N\to eN)$, we find:
\begin{eqnarray}
\frac{B(\mu \to 3e)}{B(\mu \to e\gamma)}&=&\frac{8\alpha^2N_f^2}{9(4\pi)^4}
\left(\frac{\Lambda}{\Lambda_F}\right)^4\left[ \ln \frac{{\rm max}
(m_f^2,m_\mu^2)}{M_F^2}\right]^2\;, \label{prim}\\
\frac{B(\mu N\to eN)}{B(\mu \to e\gamma)}&=& 10^{12}~B(A,Z)~
\frac{32G_F^2m_\mu^4N_f^2}{9(4\pi)^6}
\left(\frac{\Lambda}{\Lambda_F}\right)^4\left[ \ln \frac{{\rm max}
(m_f^2,m_\mu^2)}{M_F^2}\right]^2\;. \label{prim2}
\end{eqnarray}
Here $N_f$ is the number of colours of the fermion $f$ and $M_F$ is the 
heavy-particle mass generating the effective operators (typically $M_F$ is
much smaller than $\Lambda$ or $\Lambda_F$ because of loop factors and mixing
angles). The logarithms in equations \ref{prim} and \ref{prim2} correspond to
the anomalous dimension mixing of the operator in equation \ref{op4} with the
four-fermion operator generating the relevant rare muon
process~\cite{Raidal:1997hq}. If $\Lambda \sim \Lambda_F$, then the
contributions from the four-fermion operator are irrelevant, since the ratios
in equations \ref{m3e} and \ref{m4e} are larger than those in equations \ref{prim}
and \ref{prim2}. More interesting is the case in which the four-fermion 
operator in equation \ref{op4} is generated at tree level, while the
magnetic-moment transition in equation \ref{magmom} is generated only at one loop,
as in models with $R$-parity violation \cite{Hall:1983id,Barger:1989rk,Dreiner:1997uz} or with leptoquarks
\cite{Davidson:1993qk}. In this case, we expect:
\begin{equation}
\left(\frac{\Lambda}{\Lambda_F}\right)^4\simeq \frac{(4\pi)^3}{\alpha}\;.
\label{parg}
\end{equation}
If equation \ref{parg} holds and if we take $M_F\simeq 1$~TeV, then the ratios in 
equations \ref{prim} and \ref{prim2} become of order unity, so the
different rare muon processes have comparable rates.

Alternatively, if the fermion $f$ in equation \ref{op4} is an electron (or a
light quark), the effective operator can mediate $\mu \to 3e$ (or
$\mu^-$--$e^-$ conversion) at tree-level, and the corresponding process can
dominate over the others~\cite{deGouvea:2000cf}. For instance, we obtain:
\begin{equation}
\frac{B(\mu \to 3e)}{B(\mu \to e\gamma)}=
\frac{1}{12(4\pi)^2}\left(\frac{\Lambda}{\Lambda_F}\right)^4\;,
\end{equation}
for the case $f=e$.

In conclusion, the various rare muon processes are all potentially very
interesting. In the event of a positive experimental signal for muon-number
violation, a comparison between searches in the different channels and the use
of the effective-operator approach  will allow us to identify quickly
the correct class of models.

%% file: 07-Muon-Physics/07-04-FV-exp/07-04-FV-exp.tex
 \subsection{Experimental prospects}
\begin{figure}
\centerline{\includegraphics[width=10cm]{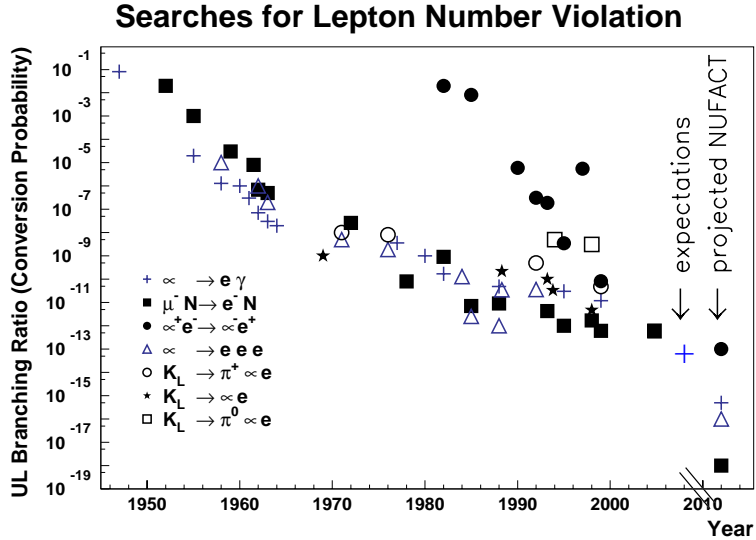}}
 \caption{Historical development of the 90\% C.L. upper
 limits (UL) on branching
ratios
 respectively conversion probabilities
 of muon-number violating processes which involve muons and kaons.
 Sensitivities expected for planned searches are indicated in the year
 2008 (see also reference \cite{Kuno:1999jp}). The projections for
 a neutrino factory (NUFACT) are also shown.
    Taken from \cite{Aysto:2001zs}.
 \label{fig:history}}
 \end{figure}

 The experimental sensitivities achieved during the past decades in tests
 of muon number conservation are illustrated in  figure \ref{fig:history},
and given in table \ref{tab:mulim}.
 Generally the tests were limited  by the intensities of the available
 $\mu$ and K beams, but in some cases detector limitations have played a role
 as well.
\begin{table}
\begin{minipage}{\textwidth}
\caption{Present limits on rare $\mu$ decays. \label{tab:mulim}}
\begin{tabular}{lrccc}
\hline
mode  &upper limit (90\% C.L.)  &year   &Exp./Lab.     &reference      \\

\hline
$\mu^+\to e^+ \gamma$                   &$1.2 \times 10^{-11}$
&2002   &MEGA / LAMPF           &\cite{Brooks:1999pu,Ahmed:2001eh}       \\
$\mu^+\to e^+ e^+ e^-$                  &$1.0 \times 10^{-12}$                  &1988   &SINDRUM I/ PSI         &\cite{Bellgardt:1987du}          \\
$\mu^+ e^- \leftrightarrow \mu^- e^+$   &$8.3 \times 10^{-11}$                  &1999   &PSI                    &\cite{Willmann_97,Willmann:1998gd}                     \\
$\mu^-\;$Ti$\;\to e^-$Ti                &$6.1 \times 10^{-13}$                  &1998   &SINDRUM II / PSI       &\cite{Wintz:1998rp}                    \\
$\mu^-\;$Ti$\;\to e^+$Ca$^*$            &$3.6 \times 10^{-11}$                  &1998   &SINDRUM II / PSI       &\cite{Kaulard:1998rb}                  \\
$\mu^-\;$Pb$\;\to e^-$Pb                &$4.6 \times 10^{-11}$                  &1996   &SINDRUM II / PSI       &\cite{Honecker:1996zf}                 \\
$\mu^-\;$Au$\;\to e^-$Au                &$7 \times 10^{-13}$                    &2006   &SINDRUM II / PSI       &\cite{Bertl:2001fu}                      \\
\hline
\end{tabular}
\end{minipage}
\end{table}

All recent results with $\mu^+$ beams were obtained with ``surface'' muon
 beams (see for instance~\cite{Pifer:1976ia}),
that consist of muons originating in the decay of
$\pi^+$'s that stopped at the surface of the pion-production target,
or ``sub-surface'' beams, in which the muons originate from the decays
of $\pi^+$ stopping just below the surface.  Because of the narrow
momentum spread, such beams are superior to conventional pion decay
channels in terms of muon stop density; they permit the use of
relatively thin (typically 10~mg/cm$^2$) foils to stop the beam; and
they offer the highest muon stop densities that can be obtained at
present. Such low-mass stopping targets are required for the ultimate
resolution in positron momentum and emission angle, photon yield, or
the efficient production of muonium in vacuum. 

In this section we indicate  how far experimental searches
could benefit from muon beam intensities which are 2--3 orders of
magnitude higher than are presently available. Further details can be
found in the CERN study of 2001\cite{Aysto:2001zs}. 

 \subsubsection{$\mu \rightarrow e \gamma$}

Neglecting the positron mass, the 2-body decay $\mu^+ \to e^+ \gamma$ 
of muons at rest is characterised by:
\begin{eqnarray*}
E_{\gamma}      &=&     E_e = m_{\mu}c^2/2 = 52.8~{\rm MeV} \; ;    \\
\Theta_{e\gamma}&=&     180^{\circ}                         \; {\rm ; and} \\
t_{\gamma}      &=&     t_e                                 \; .
\end{eqnarray*}
All $\mu \to e \gamma$ searches performed during the past three
 decades were limited by accidental coincidences between a positron
 from normal muon decay,
$\mu \to e \nu \overline{\nu}$, and a photon produced in the decay
 of another muon, either by bremsstrahlung or by $e^+ e^-$ annihilation
 in flight. This background
dominates the intrinsic background from radiative muon decay
 $\mu \to e \nu \overline{\nu} \gamma$.  Accidental $e\gamma$
 coincidences can be suppressed by testing
the three conditions listed above.
The vertex constraint resulting from the ability to trace 
back positrons and photons to an extended stopping target
 can further reduce background.
 Attempts have been made to suppress accidental coincidences
 by observing the low-energy positron associated with the photon, but 
with minimal success.

 The most sensitive search to date was performed by the MEGA
Collaboration at the Los Alamos Meson Physics Facility 
(LAMPF)\cite{Brooks:1999pu,Ahmed:2001eh}, which established
 an upper limit  (90\% C.L.) on $B_{\mu\rightarrow e \gamma}$ 
of $1.2 \times 10^{-11}$~\cite{Brooks:1999pu}.
 The MEG experiment~\cite{MEGprop} at PSI, aims at a single-event sensitivity
of $\sim 10^{-13} - 10^{-14}$, and  began commissioning
in early 2007.  
A surface muon beam is employed that reaches an intensity around
$5\times10^8\,\mu^+/s$.

A straightforward improvement of more than an order of
 magnitude in suppression of accidental background results from
 the DC beam at PSI, as opposed to the pulsed
LAMPF beam which had a macro duty cycle of 7.7\% . Another 
order-of-magnitude improvement is achieved by superb time resolution 
($\approx 0.15$~ns FWHM on $t_{\gamma}- t_e$).

\begin{figure}
\includegraphics*[width=\linewidth]{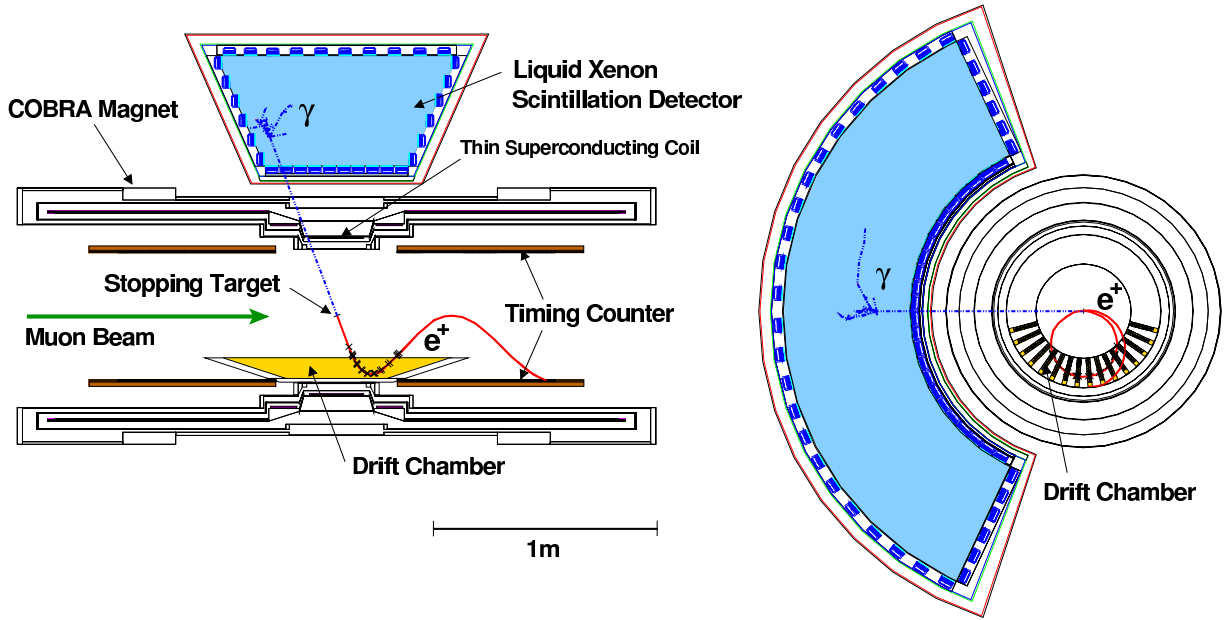}
\caption{Side and end views of the MEG setup. The magnetic field 
is shaped such that positrons are quickly swept out of the tracking
 region thus minimising the load on
the detectors. The cylindrical 0.8~m$^3$ single-cell LXe detector is
 viewed from all sides by by 846 PMTs immersed in the LXe allowing 
the reconstruction of photon energy,
time, conversion point and direction and the efficient rejection of
 pile-up signals.
    (Figure courtesy of T. Mori)
}
\label{fg:meg}
\end{figure}

The MEG setup is shown in figure \ref{fg:meg}.  The MEG spectrometer magnet
makes use of a unique ``COBRA''(COnstant Bending RAdius) design which
 results in a graded magnetic
field varying from 1.27~T at the centre to 0.49~T at both ends. This 
field distribution not only results in a constant projected bending 
radius for the 52.8~MeV positron,
for emission angles $\theta$ with $|\cos\theta| < 0.35$ , but also 
sweeps away positrons with low longitudinal momentum more effectively
than does a solenoidal field as used by MEGA.
This design significantly reduces the instantaneous rates in the drift
chambers.
\begin{figure}
  \begin{center}
    \includegraphics*[clip, trim=0 0 0 0, width=0.58\linewidth]%
      {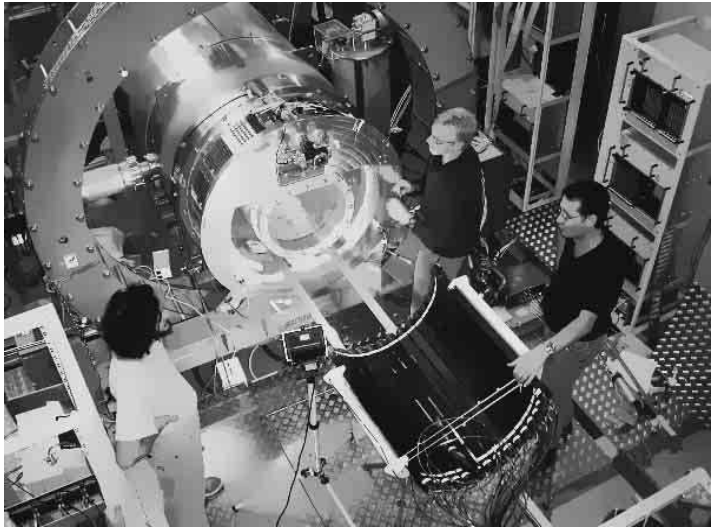}
  \end{center}
  \caption{
    Installing one of the timing counters into the COBRA magnet during
    the pilot run of the positron spectrometer at the end of 2006. 
    The large ring is one of two Helmholtz coils used to compensate
    the COBRA stray field at the locations of the photomultipliers of
    the LXe detector.\label{fig:MEG2}
    (Figure courtesy of T. Mori)
  }
\end{figure}
The drift chambers are made of 12.5~$\mu$m thin foils supported by
 C-shaped carbon-fibre
frames which are out of the way of the positrons. The foils have 
``vernier" cathode pads which permit the measurement of the trajectory 
coordinate along the anode wires with an
accuracy of about 500~$\mu$m.  
There are two timing counters at each end of the magnet 
(see figure \ref{fig:MEG2}), each of which consists of a layer 
of plastic scintillating fibres and 15 plastic
scintillator bars of $4\times 4\times 90$~cm$^3$. The fibres 
give hit positions along the beam axis and the bars measure positron
 timings with a precision of
$\sigma = 40$~ps. The counters are placed at large radii so only 
high-energy positrons reach them, giving a total rate of a few$\times
10^4$/s for each bar.

High strength aluminium-stabilised conductor is used to make the
magnet as thin  as 0.20$X_0$, so that 85\% of 52.8~MeV/$c$ gamma rays
traverse the magnet without interaction
before entering the gamma ray detector placed outside the magnet.
Whereas MEGA used rather inefficient pair spectrometers to detect
the photon, MEG developed a novel
liquid-xenon scintillation detector as shown in figure \ref{fg:meg}. By 
viewing the scintillation light from all sides, the electromagnetic 
shower induced
by the photon can be reconstructed which allows a precise measurement
 of the photon conversion point and direction~\cite{Mihara:2004dj}.
 Special PMTs that work at liquid-zenon (LXe)
temperature ($-110^\circ$C), persist under high pressures and are 
sensitive to the VUV scintillation light of LXe ($\lambda\approx 178$~nm) have been developed in
collaboration with Hamamatsu Photonics. To identify and separate
 pile-up efficiently, fast waveform digitising is used for all the
 PMT outputs. 

The performance of the detector was measured with a prototype detector.
 The results are shown in table \ref{tab:MEG}.
First data taking with the complete setup is scheduled for the
 second half of 2007. A sensitivity of ${\cal O}(10^{-13})$ 
for the 90\% C.L. upper limit in case no
candidates are found should be reached after two years.
\begin{table}
\begin{minipage}{\textwidth}
\caption{Performance of a prototype of the MEG LXe detector at $E_{\gamma}$=53~MeV. \label{tab:MEG}}
\begin{tabular*}{\textwidth}{@{\extracolsep{\fill}}llll}
\hline
&observable             &resolution ($\sigma$)  &\\
\hline
&energy                 &1.2\%                  &\\
&time                   &65~ps                  &\\
&conversion point       &$\approx$4~mm          &\\
\hline
\end{tabular*}
\end{minipage}
\end{table}

 As a next
 step it seems reasonable to consider experiments aiming at a sensitivity
of $10^{-15}$
 or better. However, it is not at all obvious how to reach
such levels of
 sensitivity without running into the background of accidental $e\gamma$
coincidences.
Surface-muon rates ten-times larger than those used by MEG already can
be achieved  today.
However, to exloit such rates, the background suppression would
have to be improved by two orders of magnitude.

Accidental background, $N_{\rm acc}$, scales with the detector
 resolutions as:
\begin{equation*}
N_{\rm acc} \propto \Delta E_e \cdot \Delta t 
 \cdot (\Delta E_{\gamma} \cdot \Delta \Theta_{e\gamma}
 \cdot \Delta x_{\gamma})^2 \cdot A^{-1}_T \;\; ,
\end{equation*}
with $x_{\gamma}$ the coordinate of the photon trajectory at
the target and $A_T$ the target area. Here, it has been assumed that
the photon can be traced back to the
target with an uncertainty that is small compared to $A_T$.
 Since the angular resolution is dictated by the positron
 multiple scattering in the target, this can
be written:
\begin{equation*}
N_{\rm acc} \propto \Delta E_e \cdot \Delta t 
 \cdot (\Delta E_{\gamma}  \cdot \Delta x_{\gamma})^2 
\cdot \frac {d_T}{A_T}\;\; ,
\end{equation*}
with $d_T$ the target thickness. When using a series of
 $n$ target foils each of them could have a thickness of
 $d_T/n$ and the beam would still be stopped.
Since the area would increase like $n \cdot A_T$ the 
background could be reduced in proportion with $1/n^2$:
\begin{equation*}
N_{\rm acc} \propto \Delta E_e \cdot \Delta t  
\cdot (\Delta E_{\gamma} \cdot \Delta x_{\gamma})^2 
\cdot \frac {d_T/n }{n \cdot A_T}\; ,
\end{equation*}
so a geometry with ten targets, 1~mg/cm$^2$ each,
 would lead to the required background suppression.

 The expected number $N_s$ of observed $\mu \rightarrow e \gamma$ 
decays can be written as:
 \begin{equation}
 N_s = R_{\mu} T \cfrac{\Omega}{4\pi} \, 
 \epsilon_e \epsilon_{\gamma} \epsilon_{cut} \,  
B_{\mu\rightarrow e \gamma} \;,
 \label{eqn1}
 \end{equation}
 where $R_{\mu}$ is the muon-stop rate, $T$ is the total measuring time,
$\Omega$ is
 the detector solid angle (we assume identical values for the photon and
the positron
detectors), $\epsilon_e$ and  $\epsilon_{\gamma}$ are the positron-
and photon-detection efficiencies, $\epsilon_{cut}$ is the efficiency
of the selection cuts. 
Selection cuts
 can be applied on the reconstructed positron energy ($E_e$), photon
energy ( $E_\gamma$ ),
 opening angle ($\theta_{e \gamma}$) and relative timing ($t_{e
\gamma}$).

 In the MEG experiment at PSI, the background is
 dominated by accidental coincidences of a positron from normal muon
decay and a
 photon which may originate in the decay $\mu^+ \rightarrow  e^+ \nu
 \overline{\nu} \gamma $ or may be produced by
an ${e}^+$ through
 external bremsstrahlung or annihilation in flight. In a DC beam the
number of accidental
 coincidences is given by:
 \begin{equation}
 N_b = {R_{\mu}}^2 f_e \epsilon_e f_\gamma \epsilon_\gamma   
 (\cfrac{\Omega}{4\pi})^2 \, \pi \cfrac{\Delta \theta_{e
\gamma}^2}{\Omega}
 \, 2\Delta t \, T\;,
 \label{eqn2}
 \end{equation}
 where $f_e$ ($f_\gamma$) is the $e^+$ ($\gamma$) yield per stopped muon
within the 
 selection window and $\Delta t$ is the cut applied on the $e^+ - \gamma$
time difference.
 For a non-DC beam, $N_b$ must multiplied by the inverse of the duty
cycle.

 Since the accidental background rises quadratically with the
 muon-stop rate, it will be
 even more problematic in future experiments using a higher beam
intensity.
 An experiment with $R_\mu = 10^{10} \mu/{\rm s}$ and all the other
quantities of
 equation (\ref{eqn1}) unchanged would yield one $\mu \rightarrow e \gamma$ 
event for $B_{\mu \rightarrow e \gamma}
= 10^{-16}$.
 However, the accidental background would increase to $10^4$ events. It
is obvious that better detector resolutions
and/or improved experimental concepts are required.

\subsubsection{$\mu^+ \to e^+e^+e^-$}

From an experimental point of view the decay $\mu\to 3e$ offers some 
important advantages compared to the more familiar $\mu \to e \gamma$
discussed in the previous section. 
The principal background is  from accidental coincidences
between positrons from normal muon decay and $e^+e^-$ pairs originating
 from photon conversions or scattering of positrons off atomic electrons
(Bhabha scattering).  Since the
 final state contains only charged particles, the setup may consist of 
a magnetic
spectrometer without the need for an electromagnetic calorimeter with
 its limited performance in terms of energy and directional resolution,
 rate capability,
and event definition in general.  On the other hand, of major concern are 
the high rates in the tracking system of a $\mu \to 3e$ setup which has
 to withstand the load of the full muon decay spectrum.

The present experimental limit, $B(\mu \to 3e) < 1 \times 10^{-12}$
\cite{Bellgardt:1987du}, was published in 1988. Since no new proposals
exist for this decay mode we shall analyse the prospects of an improved
experiment with the SINDRUM experiment as a point of reference. A
detailed description of the experiment may be found in
reference \cite{Bertl:1985mw}.

Data were taken during six months using a 25~MeV/$c$ sub-surface
beam. 
The beam was brought to rest with
a rate of $6 \times 10^6\, \mu^+$~s$^{-1}$ in a hollow 
double-cone foam target
(length 220~mm, diameter 58~mm, total mass 2.4~g). SINDRUM I
 is a solenoidal spectrometer with a relatively low magnetic
field of 0.33~T corresponding to a transverse-momentum threshold 
around 18~MeV/$c$ for particles crossing the tracking system. This
system consists of five cylindrical MWPCs concentric with the beam axis.
 Three-dimensional space points are found by measuring the charges
induced on cathode strips oriented $\pm 45^\circ$ relative to the sense
 wires. Gating times were typically 50~ns.
The spectrometer acceptance for $\mu \to 3e$ was 24\% of 4$\pi$ sr 
(for a constant transition-matrix element)
so the only place for a significant improvement in sensitivity
 would be the beam intensity. 
\begin{figure}
\centerline{\includegraphics[height=70mm]{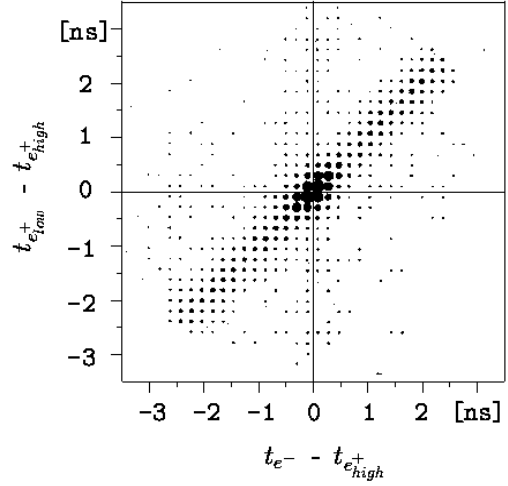}}
\caption{Relative timing of $e^+e^+e^-$
events. The two positrons are labelled according to the invariant mass
when combined with the electron. One notices a contribution of
correlated triples in the centre of the distribution. These events
are mainly $\mu \to 3e \nu \overline{\nu}$ decays. The concentration
of events along the diagonal is due to low-invariant-mass $e^+e^-$
pairs in accidental coincidence with a positron originating in the
decay of a second muon. The $e^+e^-$ pairs are predominantly due to
Bhabha scattering in the target.
    Taken from \cite{Aysto:2001zs}.
\label{SNDR_dt}}
\end {figure}

Figure~\ref{SNDR_dt} shows the time distribution of the recorded
 $e^+e^+e^-$ triples. Apart from a prompt contribution of correlated
triples one notices a dominant contribution from accidental coincidences 
involving low-invariant-mass $e^+e^-$ pairs. Most of these are explained by
Bhabha scattering of positrons from normal muon decay
 $\mu \to e \nu \overline{\nu}$. The accidental
background thus scales with the target mass, but it is not obvious how 
to reduce this mass  significantly below the 11~mg/cm$^2$
achieved in this search.

\begin{figure}
\centerline{\includegraphics[height=60mm]{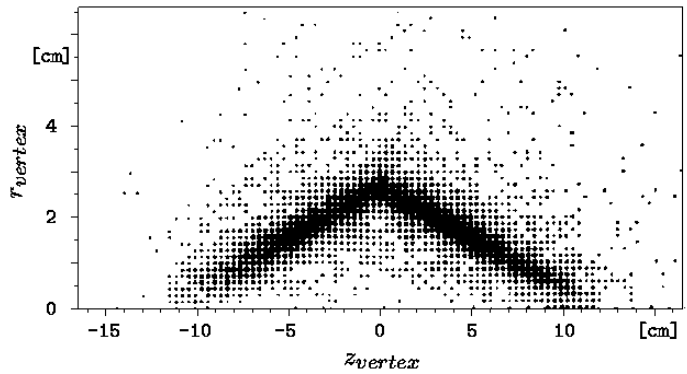}}
\caption{Spatial distribution of the vertex fitted to
prompt $e^+e^+e^-$ triples. One clearly notices the double-cone target.
    Taken from \cite{Aysto:2001zs}.
\label{SNDR_vtx}}
\end {figure}

Figure~\ref{SNDR_vtx} shows the vertex distribution of prompt events.
 One should keep in mind that most of the uncorrelated triples contain
$e^+e^-$ pairs coming from the target and their vertex distribution will
 thus follow the target contour as well. This 1-fold accidental
background is
suppressed by the ratio of the vertex resolution (couple of mm$^2$) and the
 target area. There is no reason, other than the cost of  the
detection system, not to choose a much larger target. Such an  increase might
 also help to reduce the load on the tracking detectors.
Better vertex resolution would help as well. At these low energies tracking
 errors are dominated by multiple scattering in the first
detector layer but it should be possible to gain by bringing it closer to 
the target.

\begin{figure}
\centerline{
\includegraphics[height=5.5cm]{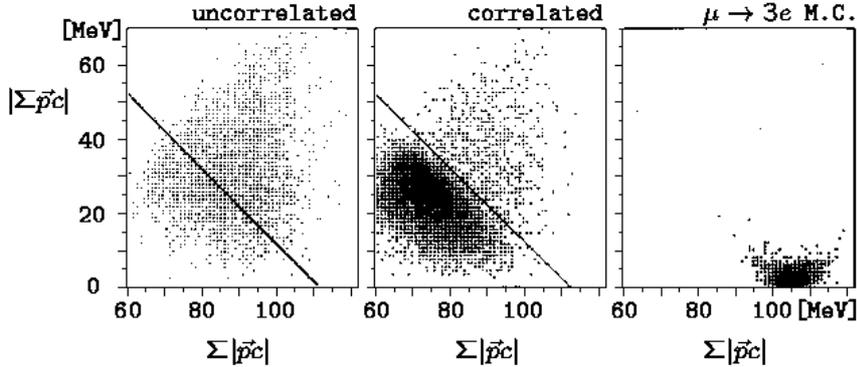}}
\caption{Total momentum versus total energy for three event
classes discussed in the text. The line shows the kinematic limit
(within resolution) defined by
$\Sigma |\vec{p}c| + |\Sigma \vec{p}c| \leq m_{\mu}c^2$ for any
muon decay. The enhancement in the distribution of correlated triples
below this limit is due to the decay $\mu \to 3e \nu \overline{\nu}$.
    Taken from \cite{Aysto:2001zs}.
\label{SNDR_ep}}
\end {figure}

Finally, figure \ref{SNDR_ep} shows the distribution of total momentum
versus total energy for three classes of events: (i) uncorrelated
$e^+e^+e^-$ triples; (ii) correlated  $e^+e^+e^-$ triples; and (iii)
simulated $\mu \to 3e$ decays. The distinction between uncorrelated
and correlated triples has been made on the basis of relative timing
and vertex as discussed above. 

What would a $\mu \to 3e$ set-up look like that would aim at a 
single-event sensitivity around $10^{-16}$, that would make use
 of a beam rate around
$10^{10}$~$\mu^+$/s? The SINDRUM I measurement was background-free
 at the level of $10^{-12}$ with a beam of $0.6  \times 10^{7}$~$\mu^+$/s.
 Taking into account that
background would have set in at $10^{-13}$, the increased stop rate would
 raise the background level to $\approx 10^{-10}$; so six orders of magnitude
in background reduction would have to be achieved. Increasing the target
 size and improving the tracking resolution should bring
two orders of magnitude from the vertex requirement alone. Since the 
dominant sources of background are accidental coincidences
between two decay positrons (one of which undergoes Bhabha scattering),
 the background rate scales with the momentum-resolution
squared. Assuming an improvement by one order of magnitude, i.e., from
 the $\approx 10\%$ FWHM obtained by SINDRUM I to $\approx 1\%$
for a new search, one would gain two orders of magnitude from the 
constraint on total energy alone. The remaining factor 100 would result
from the test on the collinearity of the $e^+$ and the $e^+e^-$ pair.

As mentioned in reference \cite{Bertl:1985mw}, a dramatic suppression
of background could be achieved by requiring a minimum opening angle
(typically 30$^\circ$) for both $e^+e^-$ combinations. Depending on
the mechanism for $\mu \to 3e$, such a cut might lead to a strong loss
in $\mu \to 3e$ sensitivity as well.

Whereas background levels may be under control, the question remains 
whether detector concepts can be developed that work at the high beam
rates proposed. A large modularity will be required to solve problems
 of pattern recognition. Also the trigger for data readout may turn
 out to be  a great challenge.

\subsubsection{$\mu \rightarrow e$ conversion} 

When negative muons stop in matter, they quickly get captured,  form
muonic atoms, and mostly reach the atomic ground ($1s$) state before decaying.
The main channels are muon decay in orbit (DIO):
\begin{equation}
\mu^{-} + (A,Z) \to e^{-} + \overline {\nu}_e + \nu_{\mu} + (A,Z) \; ;
\end{equation}
and nuclear
muon capture (NMC):
\begin{equation}
\mu^{-} + (A,Z) \to \nu_{\mu} + (A,Z-1) \; ;
\end{equation}
 where the final nucleus is likely to be in an excited state.

Because of the two-body final state,
neutrinoless 
$\mu^{-}-e^-$ conversion in muonic atoms:
\begin{equation}
\mu^{-}+(A,Z) \to e^{-}+(A,Z) \; ;
\end{equation}
with a nucleus of a mass number $A$ 
and an atomic number $Z$,
has the greatest potential for significant increases in sensitivity over
present limits; potentially by as many as six orders of magnitude.
The electrons produced in $\mu - e$ conversion are mono-energetic with
the energy: 
\begin{equation}
E_{\mu e}=m_{\mu}c^2-B_{\mu}(Z)-R(A)\;,
\end{equation}
where $B_{\mu}(Z)$ is the atomic binding energy of the muon, and
 R is the atomic recoil energy, for a muonic atom with atomic
 number $Z$ and mass number $A$.
In the lowest approximation $B_{\mu}(Z)\propto Z^2$ and $R(A)\propto A^{-1}$.

For conversions that leave the nucleus in its ground state, the nucleons act
coherently which boosts the conversion probability relative to the
rate of the dominant process of ordinary nuclear muon capture. 
 The electron is emitted with
energy $E_e \approx m_{\mu}c^2$, which coincides with the endpoint
of muon DIO, the only intrinsic physics background.
Since the energy distribution of muon decay in orbit falls steeply
above $m_{\mu}c^2/2$ the experimental set-up may have a large
signal acceptance and  the detectors can still  be protected
against the vast majority of decay and capture background events.

The muon-electron conversion probability,
$B_{\mu-e}$, varies as a function of $A$ and $Z$, and with the
probability that the nucleus stays in its ground state.
Calculations\cite{Wei59,Kosmas:1989pj,Kosmas:1990tc,Kitano:2002mt,Kosmas:2002if,Czarnecki:2002ac}
 predicted a steady rise of the
branching ratio until $Z \approx 30$, from which point it was expected
to drop again. For this reason most experiments were performed on 
medium-heavy nuclei. The nuclear-physics calculations predict
 the coherent fraction to be larger than
80\% for all nuclear systems \cite{Chiang:1993xz,Kosmas:1997ec}.

Muon decay in orbit (DIO) constitutes an intrinsic background source 
which can only be suppressed with sufficient electron-energy
resolution.
Energy distributions for DIO electrons have been calculated for a
number of muonic
atoms\cite{Shanker:1979ap,Herzog:1980my,Watanabe:1987su}.
The process
predominantly results in electrons with energy $E_{\rm DIO}$ below
 $m_{\mu}c^2/2$, the kinematic endpoint in free muon decay, with a
 steeply falling high-energy
component reaching up to $E_{\mu e}$. In the endpoint region the DIO
 rate varies as $(E_{\mu e}-E_{\rm DIO})^5$ and a resolution of
 $1-2\,$MeV (FWHM) is sufficient
to keep the DIO background under control. Since the DIO endpoint rises 
at lower $Z$, great care has to be taken to avoid low-$Z$
 contaminations in and around the target.

Another background source is due to radiative muon capture (RMC):
\begin{equation}
 \mu^-(A,Z) \to \gamma(A,Z-1)^*\nu_{\mu} \; ,
\end{equation}
after which the photon
 creates an $e^+e^-$ pair either
internally (Dalitz pair) or through $\gamma \to e^+e^-$-pair 
production in the target. The RMC endpoint can be kept below
 $E_{\mu e}$ for selected isotopes.

Most low-energy muon beams have large pion contaminations. Pions may
 produce background when stopping in the target through radiative pion
 capture (RPC) which
takes place with a probability of $O(10^{-2})$. Most RPC photons have
 energies above $E_{\mu e}$. As in the case of RMC, these photons may
 produce background
through $\gamma \to e^+e^-$ pair production. There are various strategies 
to cope with RPC background:
\begin{itemize}
\item
One option is to keep the total number of $\pi^-$ stopping in the target
 during the live time of the experiment below $10^{4-5}$. This can be 
achieved with the
help of a moderator in the beam, exploiting the range difference between
pions and muons of given momentum or with a muon storage ring exploiting
the difference in lifetime; and
\item
Another option is to exploit the fact that pion capture takes place in a
time-scale far below a nanosecond. The background can thus be suppressed
  by using a pulsed beam and selecting only delayed events.
\end{itemize}

Cosmic rays (electrons, muons, photons) are a copious source of electrons 
with energies around $\approx100\,$MeV. With the exception of
$\gamma \to e^+e^-$ in the target, these events can be recognised by
the presence of an incoming particle. Passive shielding and veto
counters above the detection system also help to suppress this
background.
\begin{figure}
\begin{center}
\includegraphics[height=\linewidth, angle=90]{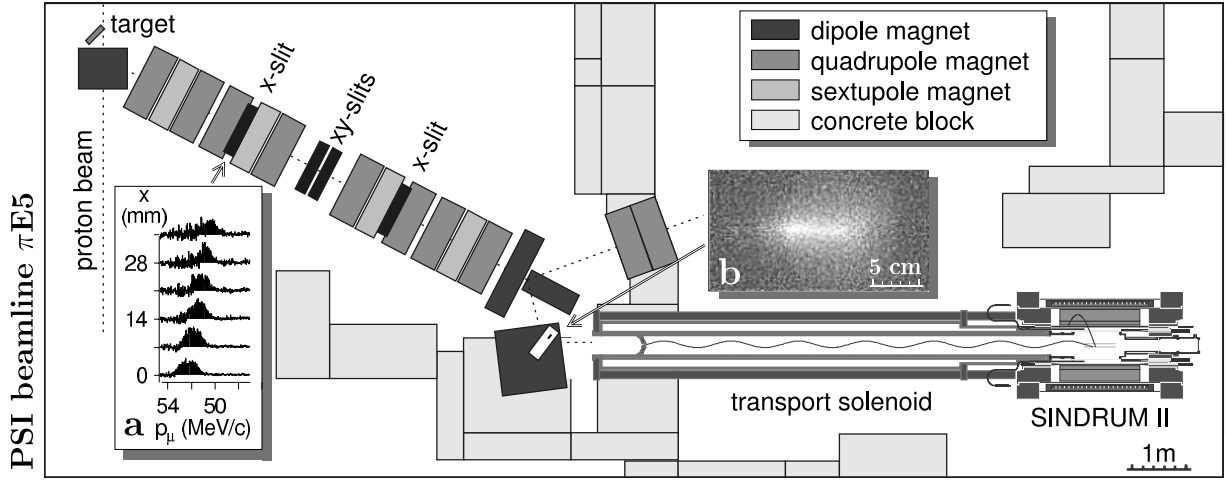}
\caption{Plan view of the SINDRUM II experiment. The
 $1\,$MW $590\,$MeV proton beam hits the $40\,$mm carbon production
 target (top left of the figure). The $\pi$E5
beam line transports secondary particles ($\pi, \mu, e$) emitted in the
 backward direction to a degrader situated at the entrance of a solenoid
 connected axially to
the SINDRUM~II spectrometer. Inset a) shows the momentum dispersion at 
the position of the first slit system.
Inset b) shows a cross section of the beam at the position of the beam
 focus.
    Taken with kind permission of the European Physical Journal from 
    figure 2 in reference \cite{Bertl:2001fu}.
    Copyrighted by Springer Berlin/Heidelberg. 
\label{fig:sindrum2}}
\end{center}
\end{figure}

\begin{figure}
  \begin{center}
    \includegraphics[clip, trim=20 210 0 0, width=0.60\linewidth]%
      {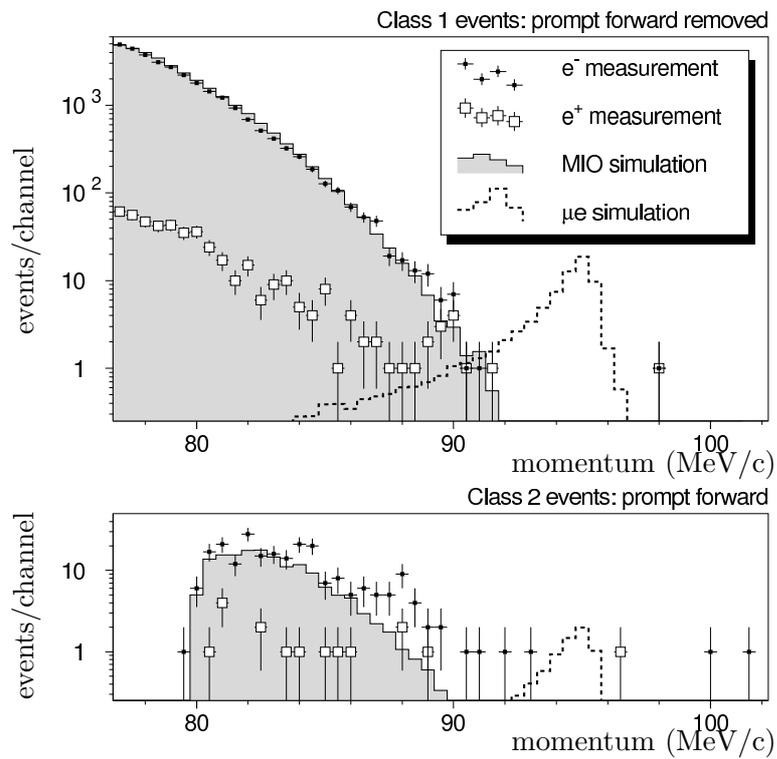}
    \begin{picture}(0,0)(0,0) 
      \put(-287,50){\rotatebox{89.99999}{events/channel}}
      \put(-100,-2){momentum (MeV/c)}
    \end{picture}
    \includegraphics[clip, trim=20 20 0 320, width=0.60\linewidth]%
      {07-Muon-Physics/Figures/etot_all.eps}
    \begin{picture}(0,0)(0,0) 
      \put(-287,50){\rotatebox[origin=c]{89.99999}{events/channel}}
      \put(-100,-2){momentum (MeV/c)}
    \end{picture}
  \end{center}
  \caption{
    Momentum distributions of electrons and positrons for two
    event classes described in the text. Measured distributions are
    compared with the results of simulations of muon decay in orbit
    and $\mu - e$ conversion. 
    N.b.decay in orbit is labelled ``MIO'' in this figure.
    Taken with kind permission of the European Physical Journal from 
    figure 11 in reference \cite{Bertl:2001fu}.
    Copyrighted by Springer Berlin/Heidelberg. 
  }
  \label{fig:sindrumgold}
\end{figure}

The present best limits (see table \ref{tab:mulim}) have been measured 
with the SINDRUM~II spectrometer at PSI. 
Most recently, a search was
 performed on a gold
target \cite{Bertl:2001fu}. 
In this experiment (see figure \ref{fig:sindrum2}),
pion suppression is based on the the fact that the range of pions is a
factor of two shorter than that of muons at the selected momentum
(52~MeV/c). 
A simulation using the measured range   
distribution shows that about one in $10^6$ pions cross an 8~mm thick CH$_2$ 
moderator. Since
these pions are relatively slow, 99.9\% of them decay before reaching the 
gold target which is situated some $10\,$m further downstream. As a
result, pion stops in the
target have been reduced to a negligible level. What remains are radiative
pion capture in the degrader and $\pi^- \to e^- \overline{\nu}_e$
decay-in-flight shortly before entering the degrader. The resulting
electrons may reach the target where they can scatter into the solid
angle acceptance of the spectrometer. 
${\cal O}$(10)
events are expected
with a flat energy distribution between 80 and $100\,$MeV. These events are
peaked in the forward direction and show a time correlation with the
cyclotron RF signal. To cope
with this background two event classes have been introduced based on the
values of polar angle and rf phase. Figure \ref{fig:sindrumgold} shows the
corresponding momentum distributions.
The spectra show no indication for $\mu - e$ conversion.
The corresponding upper limit:
\begin{equation}
B_{\mu e} \equiv 
\frac {\Gamma(\mu^-{\rm Au}\to e^-{\rm Au_{g.s.}}) }
{ \Gamma_{\rm  capture}(\mu^-{\rm Au})}
 < 7 \times 10^{-13} \;\;\; 90\%\; {\rm C.L. ;}
\label{eq:limit2}
\end{equation}
has been obtained with the help of a likelihood analysis of the
momentum distributions shown in figure \ref{fig:sindrumgold} taking
into account: muon decay in orbit; $\mu - e$ conversion; a
contribution taken from the observed positron distribution describing
processes with intermediate photons, such as radiative muon capture;
and a flat component from pion decay-in-flight or cosmic rays.

Based on a scheme originally developed during the eighties for the Moscow
Meson Factory \cite{Dzhilkibaev:1989zb}, $\mu e$-conversion experiments
are being considered both in the USA and in Japan. The key elements
are: 
\begin{itemize}
\item
A pulsed proton beam permits the removal of
 pion background by selecting events 
in a delayed time window. Proton extinction factors
between pulses of $\leq 10^{-9}$ are needed;
\item
A large acceptance capture solenoid surrounding the pion-production
target leads to a major increase in muon flux; and
\item
A bent solenoid transporting the muons to the experimental target results
 in a significant increase in momentum transmission compared to a conventional
quadrupole channel. A bent solenoid not only removes neutral particles and
 photons but also separates electric charges. 
\end{itemize}

While the MECO proposal at BNL~\cite{meco_proposal} was terminated,
an experiment with a similar design philosophy 
called mu2e  has been given stage one approval at Fermilab.
Significant studies have been carried out on the 
targeting and proton beam design, which require a re-configuring of the
antiproton complex to become a muon beam production area.
This would take place  after the Tevatron collider
 stops operation. 

\begin{figure}
\begin{center}
\includegraphics[width=0.8\linewidth]{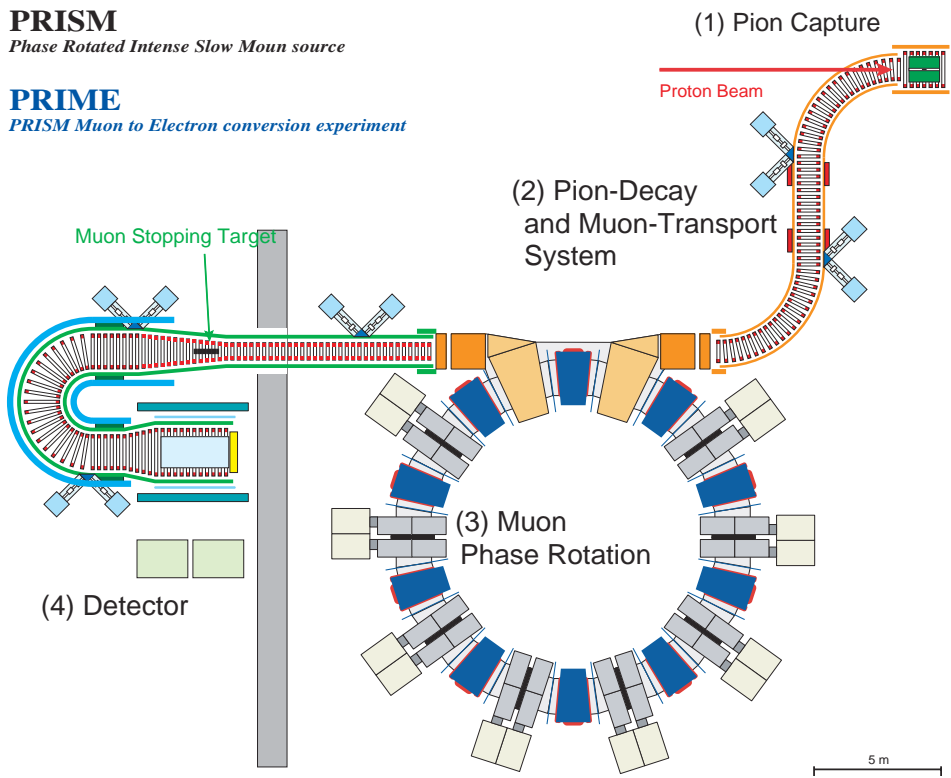}
\caption{Layout of PRISM/PRIME. The experimental target is situated at
 the entrance of the 180$^{\circ}$ bent solenoid that transports decay 
electrons to the
detection system. See text for further explanations.
Reproduced with the kind permission of the PRISM/PRIME collaboration.
\label{fig:PRISM}}
\end{center}
\end{figure}

Further improvements are being considered for an experiment at J-PARC.
To exploit fully the lifetime difference to suppress pion induced 
background, the separation has to
occur in the beamline rather than after the muon has stopped, since 
the lifetime of the muonic atom may be significantly shorter than the
 2.2~$\mu$s lifetime of the free muon. For
this purpose a muon storage ring, PRISM (Phase Rotated Intense Slow Muon
source, see figure \ref{fig:PRISM}), is being considered~\cite{prism03}
which makes use of
large-acceptance fixed-field alternating-gradient (FFAG) magnets. A portion
 of the PRISM-FFAG ring is presently under construction as an R\&D project.
As the name suggests the ring is also used to reduce the momentum spread of
 the beam 
(from $\approx$30~\% to $\approx$3~\%).
This is achieved by accelerating late muons
and decelerating early muons in RF electric fields. The scheme requires
 the construction of a pulsed proton beam~\cite{LOI26},
which is under consideration by the J-PARC Laboratory Management.
 The low momentum spread of the muons allows the use of a relatively
 thin target which is an essential ingredient for high resolution in the
positron momentum measurement 
with the PRIME detector\cite{PRIME}.
Table \ref{tab:mue_table} lists the $\mu ^-$ stop rates and single-event
sensitivities for the various projects discussed above.
\begin{table}
\begin{minipage}{\textwidth}
\caption{$\mu-e$  conversion searches. \label{tab:mue_table}}
\begin{tabular*}{\textwidth}{@{\extracolsep{\fill}}lllcccc}
\hline
project         &Lab            &status         &$E_p$ [GeV]    &$p_{\mu}$ [MeV/c]      &$\mu^-$ stops [s$^{-1}$]       &$S$
\footnote{value of $B_{\mu e}$ corresponding to an expectation of one observed event}   \\
\hline
{\scriptsize SINDRUM II}      &{\scriptsize PSI}            &{\scriptsize finished}       &0.6  &52$\pm$1   &~~~~$10^7$             &$2\times10^{-13}$      \\
{\scriptsize MECO}            &{\scriptsize BNL}            &{\scriptsize cancelled}      &8~~  &45$\pm$25      &~~~~$10^{11}$          &$2\times 10^{-17}$     \\
{\scriptsize mu2e}            &{\scriptsize FNAL}           &{\scriptsize scientific approval}    &8~~   &45$\pm$25      &0.6$\times 10^{10}$    &$4\times 10^{-17}$     \\
{\scriptsize PRISM/PRIME}     &{\scriptsize J-PARC}         &{\scriptsize under consideration}            &40~~  &68$\pm$3       &~~~~$10^{12}$          &$5\times 10^{-19}$     \\
\hline
\end{tabular*}
\end{minipage}
\end{table}

\subsubsection{Muonium-anti-muonium conversion}

Muonium is the atomic bound state of a positive muon and an electron.
For leptons, a spontaneous conversion of muonium ($\mu^+e^-$) into
anti-muonium 
($\mu^-e^+$)  would be completely analogous 
to the well known
${\rm K}^0-\overline{{\rm K}^0}$ oscillations in the
quark sector.  A search was suggested  in 
1957 by Pontecorvo \cite{Pontecorvo:1957cp,Feinberg_61} three years
before the atom was discovered 
by Hughes et al. \cite{Hughes_60,Hughes_90}.
The process could proceed
at tree level through bi-lepton exchange or through various 
loops. Predictions for the process exist in a variety of
speculative models including left-right symmetry, 
$R$-parity-violating supersymmetry, GUT theories, and several others
\cite{Herczeg:1992pt,Halprin:1982wm,Mohapatra:1991ij,Wong:1985pb,Halprin:1993zv,Fujii:1993su,Frampton:1997in,Pleitez:1999ix}. 

\begin{figure}
\begin {center}
\includegraphics[width=0.6\textwidth]{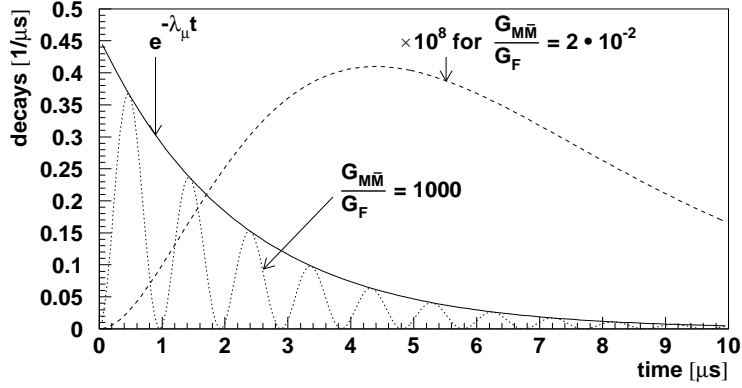}
\end {center}
\vspace*{-0.5cm}
\caption{Time dependence of the probability to observe an anti-muonium decay  
for a system which was initially in a pure muonium state.  
The solid line represents the exponential decay of muonium  
in the absence of a finite coupling. 
The decay probability as anti-muonium is given for a coupling strength of 
$G_{\rm M\overline{M}}   = 1000 G_F$ 
by the dotted line and for a coupling strength 
small compared to the muon decay rate (dashed line). 
In the latter case the maximum of the probability 
is always at about 2 muon lifetimes.  
Only for strong coupling   could 
several oscillation periods   be observed.
Taken with kind permission of Springer Berlin/Heidelberg
from figure 2 in reference \cite{Willmann_97}.
Copyrighted by the Springer Berlin/Heidelberg.
\label{mmbar_oscill}}
\end {figure}

Any possible coupling between muonium and its anti-atom  will give rise to
oscillations between them. For atomic s-states with 
principal quantum number, $n$,
a splitting of their
energy levels:
\begin{eqnarray} 
 \delta =
 {\displaystyle \frac{8 G_{\rm F}}{\sqrt{2} n^2 \pi a_0^3}
 \frac{G_{\rm M\overline{M}}}{G_{\rm F}} } \; ;
\end{eqnarray}
is caused,
where $a_0$ is the Bohr radius of the atom, $G_{\rm M\overline{M}}$
is the coupling constant in an effective four-fermion interaction  
and ${G}_{\rm F}$ is the weak interaction Fermi 
coupling constant.
For the ground state we have $\delta =   
1.5 \times 10^{-12} \, {\rm eV} \times (G_{\rm M\overline{M}}/G_{\rm F})$
which corresponds to 519 Hz for $G_{\rm M\overline{M}}=G_{\rm F}$.
An atomic system created at time $t = 0$ 
as a pure state of muonium can be expected to be observed 
in the anti-muonium state at a later time $t$ with a time dependent 
probability of:
\begin{eqnarray} 
\label{t_oscill} 
{ p_{\rm M\overline{M}}(t)} &= \sin^2\left(\frac{\delta \,
t}{2\,\hbar}\right) 
e^{-\lambda _\mu t}  
&\approx \left(\frac{\delta \, t}{2\,\hbar}\right)^2  e^{-\lambda
_\mu t}\;, 
\end{eqnarray} 
where $ \lambda _\mu = 1/\tau_\mu $ is the muon decay rate
(see figure \ref{mmbar_oscill}). 
The approximation is valid for a weak coupling as suggested by the  
known experimental 
limits on $G_{\rm M\overline{M}}$.
\begin{figure}
\begin {center}
\includegraphics[width=0.6\textwidth]{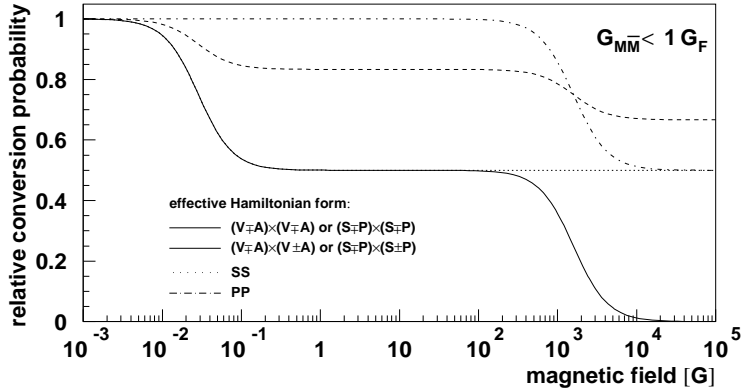}
\end {center}
\vspace*{-0.5cm}
\caption{The muonium to anti-muonium conversion probability 
          depends on external magnetic fields 
          and the coupling type. Independent 
          calculations were performed  by Wong and Hou 
          {\cite{Hou:1995pi}} and Horikawa and 
          Sasaki {\cite{Horikawa:1995ae}}. 
        Taken with kind permission of Springer Berlin/Heidelberg
        from figure 3 in reference \cite{Willmann_97}.
        Copyrighted by the Springer Berlin/Heidelberg.
\label{mbar_in_fields}}
\end{figure}

The degeneracy of corresponding states in the atom and its anti-atom 
is removed by external magnetic fields which can cause a suppression  
of the conversion and a reduction of the probability $p_{\rm
M\overline{M}}$. 
The influence of an external magnetic field depends on the interaction
type of 
the process. The reduction of the conversion 
probability has been calculated for all possible 
interaction types as a function of field strength  
(figure \ref{mbar_in_fields}) \cite{Hou:1995pi,Horikawa:1995ae}.  
In the case of an observation of the conversion process,   
the coupling type could be revealed by  measurements of 
the conversion probability at two different  
magnetic-field values. 
 
The conversion process is strongly suppressed for muonium 
in contact with matter
since a transfer of the negative  
muon in anti-muonium to any other atom is energetically 
favoured and breaks 
up the symmetry between muonium and anti-muonium by opening up  
an additional decay channel for the anti-atom only
\cite{Morgan:1970yz,morg_73} 
\footnote{In gases 
at atmospheric pressures the conversion probability  is approximately
five orders of magnitude smaller than in vacuum 
mainly  due to scattering of the atoms from gas molecules.  
In solids the reduction amounts to 10 orders of magnitude.}. 
Therefore any new sensitive experiment needs to employ  
muonium atoms in vacuum \cite{Willmann:1998gd,Willmann_97}. 

The most recent experiment, which was carried out at PSI, utilised 
a powerful signature in which the identification of both
constituents of the anti-atom  and their coincident detection
after its decay.
In this experiment, an energetic electron appears in the $\mu^-$
decay. 
The positron from the atomic shell remains with an 
average kinetic energy of 13.5~eV. 
The energetic particle could be observed in a 
magnetic, wire-chamber spectrometer and a position-sensitive
microchannel plate
(MCP) served as a detector for atomic shell positrons onto which these
particles could be transported in a guiding magnetic field after
post-acceleration in an electrostatic device. A clean vertex
reconstruction and
the observation of annihilation $\gamma$'s in a pure CsI detector
surrounding
the MCP were required in an event signature \cite{Willmann:1998gd,Willmann_97}.
Half a year of data-taking was carried out what is currently the most
intense
surface-muon source; the $\pi$E5 channel at PSI. The previous upper 
bound on the total conversion probability 
per muonium atom 
${  P}_{{\rm M\overline{M}}}= \int  p_{\rm M\overline{M}}(t) {\rm d}t$ 
was improved by more
than three  orders of magnitude and yielded an upper bound of 
${P}_{{\rm M} \overline{{\rm M}}}\leq 8.0 \times 10^{-11}/{ 
S}_{\rm
B}$. Here, a magnetic field correction ${S}_{\rm B}$ is
included which accounts for the 0.1 T magnetic
field in the experiment. 
${S}_{\rm B}$ is of order unity and 
depends on the type of the ${\rm M\overline{M}}$ interaction. 
For an assumed effective (V--A)$\times $(V--A)-type four-fermion interaction   the
quoted result   corresponds to an upper limit   for the coupling
constant of    
${G}_{{\rm M} \overline{{\rm M}}}\leq 3.0\times 10^{-3} {
G}_{\rm F}$  (90~\%~C.L.).
Several limits on model parameters 
were significantly improved, such as the mass of the  bi-leptonic
gauge boson, and some models were strongly disfavoured,
such as a certain $Z_8$ model with radiative mass generation 
and the minimal version of 331 models
\cite{Willmann:1998gd,Willmann_97}.
\begin{figure}
  \begin{minipage}{1.5in}  
  \centering{  
   \hspace*{-1.6cm}  
   \mbox{  
   \includegraphics[width=2in]{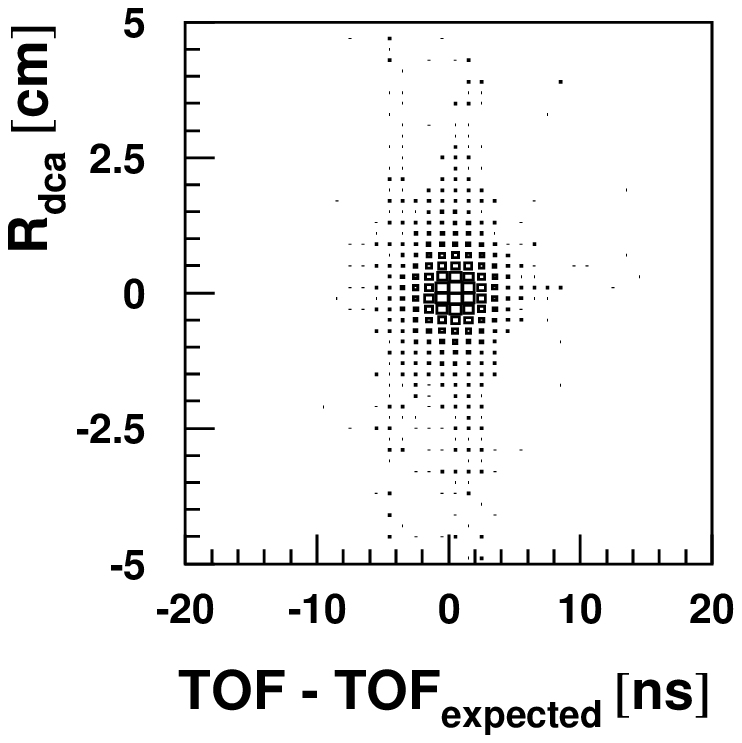}
         }  
             }  
  \end{minipage}     
 \hspace*{0.1cm}  
 \begin{minipage}{1.5in}  
    \centering{  
   \hspace*{.3cm}  
   \mbox{  
   \includegraphics[width=2in]{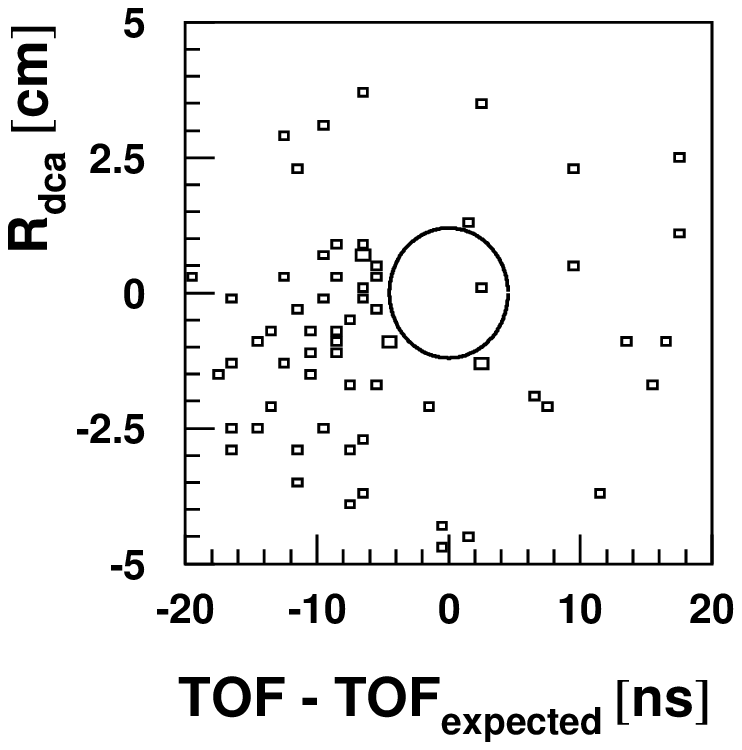}  
         }  
             }  
   \end{minipage}  
 \centering\caption 
        {The distribution of the distance of closest approach ($R_{dca}$)
        between a track from an energetic   
        particle in the magnetic spectrometer and the back projection of
the  
        position on the MCP detector versus the time of flight (TOF)  
        of the atomic shell particle for a muonium measurement (left)  
        and for all data recorded within 1290 h of   
        data taking while searching for   
        anti-muonium (right) (From reference \cite{Willmann:1998gd}).   
        One single event falls within 3 standard deviations region  
        of the expected TOF and $R_{dca}$ which is indicated by the
ellipse.  
         The events concentrated at early times   
       and low $R_{dca}$correspond   
        to a background signal from the allowed decay  
        $\mu \to 3e2\nu$. In a new experiment such background could be
        suppressed significantly through the characteristically different
        time evolution of a potential anti-muonium signal and the background.
        Taken with kind permission of Physical Review Letters from figure 4 in
        reference \cite{Willmann:1998gd}.  
        Copyrighted by the American Physical Society.
          }\label{Fmmbar_res}  
   
\end{figure}  
  
With a new and intense, pulsed beam the characteristic time dependence
of the conversion process could be exploited only if the decay of
atoms  that have survived several muon lifetimes, $\tau_{\mu}$, can
be observed. 
Whereas all beam-muon-related background decays exponentially, the
anti-atom population increases quadratically with time, giving the
signal an advantage over background which, for a 3-fold coincidence
signature as in the PSI experiment, can be expected to decay with a
time constant of $\tau_{\mu}/3$ (compare equation (\ref{t_oscill})).
Some two orders of magnitude improvement can be envisaged 
\cite{Jungmann:2000qp} with no significant background arising from the
$\mu \to 3e2\nu$ process or internal Bhabha scattering 
in which the positron from $\mu^+$ decay would transfer its
energy to the electron in the atomic shell and mimic a signal
event (figure \ref{Fmmbar_res}).
The requirements for radiation hardness and rate capability of the
set-up are similar to those of a $\mu \to 3e$ experiment. As before, a common
approach to these two measurements may be found.

%% file: 07-Muon-Physics/07-05-normal-decay-WF/07-05-normal-decay-WF.tex
\subsection{Normal muon decay}\label{sec:2.1}

\subsubsection{Theoretical background}\label{subsec:2.1.1}

All measurements of normal muon decay, $\mu^{-} \to e^{-}
\overline{\nu}_{e} \nu_{\mu}$, and its inverse, $\nu_{\mu} e^{-} \to
\mu^{-} \nu_{e}$, are successfully described by the `V--A'
interaction, which is a particular case of the local, derivative-free,
lepton-number-conserving, four-fermion interaction \cite{Michel:1949qe}.
The `V--A' form and the nature of the neutrinos ($\overline{\nu}_{e}$
and $\nu_{e}$) have been determined by experiment
\cite{Fetscher:1986uj,Langacker:1988fp}.

The observables in muon decay (energy spectra, polarisations and
angular distributions) and in inverse muon decay (the reaction cross
section) at energies well below $m_{W}c^{2}$ may be parameterised in
terms of the dimensionless coupling constants $g_{\varepsilon
\mu}^{\gamma}$ and the Fermi coupling constant $G_{\mathrm{F}}$.  The
matrix element is:
\begin{equation}
        {\mathcal M} = \frac{4 G_{\mathrm{F}}}{\sqrt{2}}\,
        \sum_{\substack{\gamma = \mathrm{S, V, T} \\ \varepsilon, \mu =
        \mathrm{R, L} }}g^{\gamma}_{\varepsilon \mu} \langle
        \overline{e}_{\varepsilon} |\Gamma^{\gamma}|(\nu_{e})_{n}\rangle
        \langle (\overline{\nu}_{\mu})_{m}|
        \Gamma_{\gamma}|\mu_{\mu}\rangle \;.\label{eq:wfetscher:1}
\end{equation}
We use here the notation of Fetscher et al., \cite{Fetscher:1986uj,Fetscher:2000th}
who in turn use the sign conventions and definitions of Scheck
\cite{Scheck:1996ur}.  Here $\gamma = \mathrm{S,V,T}$ indicates a (Lorentz)
scalar, vector, or tensor interaction, and the chirality of the
electron or muon (right- or left-handed) is labelled by $\varepsilon,
\mu = \mathrm{R, L}$.  The chiralities $n$ and $m$ of the $\nu_{e}$
and the $\overline{\nu}_{\mu}$ are determined by given values of
$\gamma$, $\varepsilon$ and $\mu$.  The 10 complex amplitudes,
$g_{\varepsilon \mu}^{\gamma}$, and $G_{\mathrm{F}}$ constitute 19
independent parameters to be determined by experiment.  The `V--A'
interaction corresponds to $g_{\mathrm{LL}}^{\mathrm{V}} = 1$, with
all other amplitudes being 0.

With the deduction from experiments that the interaction is
predominantly of the vector type and left-handed
($g_{\mathrm{LL}}^{\mathrm{V}} > 0.96~(90~\%\mathrm{CL}$)), there
remain several separate routes of investigation of normal muon decay,
which will be discussed in the following.

\subsubsection{Muon-lifetime measurements}\label{subsec:2.1.3}

The measurement of the muon lifetime yields the most precise
determination of the Fermi coupling constant $G_{\mathrm{F}}$, which
until recently was known with a relative precision 
of $9 \times 10^{-6}$~\cite{Yao:2006px}.
Improving this measurement is certainly an interesting
goal \cite{Marciano:1999ih} since $G_{\mathrm{F}}$ is one of the
fundamental parameters of the Standard Model. Until recently,
the ability to extract $G_{\mathrm{F}}$ from $\tau_\mu$ was limited by
theory, the recent the radiative corrections calculated
by van Ritbergen and Stuart
\cite{vanRitbergen:1999fi,vanRitbergen:1998hn,vanRitbergen:1998yd}
have removed this uncertainty.

A clean beam pulse
structure with very good suppression of particles between pulses is
indispensable.  Presently three experiments are in progress, two of
which are at PSI \cite{muLan,FAST} and one is located at RAL
\cite{mu_lifetime_RAL}. The MuLan experiment at PSI has recently released
an 11~ppm measurement of $\tau_\mu$ obtained
from their 2004 data set \cite{Chitwood:2007pa}, which gives a new world average
$\tau_\mu = 2.197\,019(21)~\mu{\rm s}$ and determines the Fermi 
constant to be $G_F =
1.166\,371(6) \times 10^{-5}\ {\rm GeV}^{-2}$ ($\pm5$ ppm). 
The 2006 MuLan data set has
$10^{12}$ $\mu^+$ decays on tape which, in principle, will give a 1~ppm
measurement.  A final data run in 2007 should accumulate a data set of
equal size.
These new data should result in 
an improvement in the precision of $\tau_{\mu}$ by about a factor of
20 over the previous world average \cite{Yao:2006px}.
An additional order of magnitude could be gained
at a Neutrino Factory primarily from increased muon flux, with the
major systematics being pile-up and detector timing stability.

There are two caveats, however:
                reducing the error on $G_{\mathrm{F}}$ by precise measurements
                of the muon lifetime would not improve the
                electroweak fits, because the error on the dimensionless input
                $G_{\mathrm{F}} M_{Z}^{2}$ is dominated by the uncertainty on
                $M_{Z}^{2}$, which is now $23 \times 10^{-6}$ (or 23 ppm), 
                where $(M_{Z} =
                (91\,187.6 \pm 2.1)$~MeV \cite{Yao:2006px}).  Also
                $G_{\mathrm{F}}$ is commonly determined assuming exclusively
                V-A interactions.  A somewhat more general formula has been
                given by Greub et al. \cite{Greub:1994kg}:
                \begin{align}
                        G_{\mathrm{F}}^{2} = & \frac{192 \pi^{3} \hbar}{
                        \tau_{\mu} m^{5}_{\mu}} \left[ 1 + \frac{\alpha}{ 2 \pi}
                        \left(\pi^{2} -\frac{25}{4}\right)\right] \left[ 1 -
                        \frac{3}{5} \left(\frac{m_{\mu}}{m_{W}}\right)^{2}\right]
                        \nonumber \\[5mm] & \times \left[1 - 4 \eta
                        \frac{m_{e}}{m_{\mu}} - 4 \lambda
                        \frac{m_{\nu_{\mu}}}{m_{\mu}} + 8
                        \left(\frac{m_{e}}{m_{\mu}}\right)^{2} + 8
                        \left(\frac{m_{\nu_{\mu}}}{m_{\mu}}\right)^{2}\right] ,
                        \label{eq:wfetscher:2}
                \end{align}
                \begin{equation}
                        \eta = \frac{1}{2} {\rm Re} \left[
                        g^{\mathrm{V}}_{\mathrm{LL}} 
                        g^{\mathrm{S}*}_{\mathrm{RR}} + 
                        g^{\mathrm{V}}_{\mathrm{RR}}
                        g^{\mathrm{S}*}_{\mathrm{LL}} +
                        g^{\mathrm{V}}_{\mathrm{LR}} (
                        g^{\mathrm{S}*}_{\mathrm{RL}} +
                        g^{\mathrm{T}*}_{\mathrm{RL}} ) +
                        g^{\mathrm{V}}_{\mathrm{RL}} (
                        g^{\mathrm{S}*}_{\mathrm{LR}} +
                        g^{\mathrm{T}*}_{\mathrm{LR}} ) \right] ,
                        \label{eq:wfetscher:3}
                \end{equation}
                \begin{equation}
                        \lambda = \frac{1}{2} {\rm Re} \left[
                        g^{\mathrm{S}}_{\mathrm{LL}} 
                        g^{\mathrm{S}*}_{\mathrm{LR}} + 
                        g^{\mathrm{S}}_{\mathrm{RR}}
                        g^{\mathrm{S}*}_{\mathrm{RL}} -2
                        g^{\mathrm{V}}_{\mathrm{RR}} 
                        g^{\mathrm{V}*}_{\mathrm{RL}} -2
                        g^{\mathrm{V}}_{\mathrm{LL}} 
                        g^{\mathrm{V}*}_{\mathrm{LR}}  \right] .
                \end{equation}
                Here, besides the muon lifetime and the muon mass, radiative
                corrections to first order and mass terms are included.  Most
                important is the muon-decay parameter $\eta$ which is 0 in the
                SM. If we assume that only one additional interaction
                contributes to muon decay, then $\eta \simeq \frac{1}{2}{\rm
                Re} g^{\mathrm{S}}_{\mathrm{RR}}$, where
                $g^{\mathrm{S}}_{\mathrm{RR}}$ corresponds to a scalar
                coupling with right-handed charged leptons.  Including the
                experimental value of $\eta = (-7 \pm 13) \times 10^{-3}$
                \cite{Burkard:1985wn} the error on $G_{\mathrm{F}}$ increases by a
                factor of 20.

\subsubsection{Precision measurement of the Michel parameters}\label{subsec:2.1.4}

The measurement of individual decay parameters alone generally does
not give conclusive information about the decay interaction owing to
the many different couplings and interference terms.  An example is
the spectrum Michel parameter, $\varrho$.  A precise measurement
yielding the V--A value of 3/4 by no means establishes the V--A
interaction.  In fact, any interaction consisting of an
arbitrary combination of $g_{\mathrm{LL}}^{\mathrm{S}}$,
$g_{\mathrm{LR}}^{\mathrm{S}}$, $g_{\mathrm{RL}}^{\mathrm{S}}$,
$g_{\mathrm{RR}}^{\mathrm{S}}$, $g_{\mathrm{RR}}^{\mathrm{V}}$, and
$g_{\mathrm{LL}}^{\mathrm{V}}$ will yield exactly $\varrho = 3/4$
\cite{Fetscher:1990su}.  This can be seen if we write $\varrho$ in the
form \cite{Gerber:1987cr}:
\begin{equation}
        \varrho - \textstyle{\frac{3}{4}} = - \textstyle{\frac{3}{4}} 
        \displaystyle \left\{|g_{\mathrm{LR}}^{\mathrm{V}}|^{2} +
        |g_{\mathrm{RL}}^{\mathrm{V}}|^{2}\right\} + 
        2 \left(|g_{\mathrm{LR}}^{\mathrm{T}}|^{2} +
        |g_{\mathrm{RL}}^{\mathrm{T}}|^{2} \right) + Re \left( 
        g_{\mathrm{LR}}^{\mathrm{S}}g_{\mathrm{LR}}^{\mathrm{T*}} + 
        g_{\mathrm{RL}}^{\mathrm{S}}g_{\mathrm{RL}}^{\mathrm{T*}}
        \right)\;.
        \label{eq:wfetscher:5}
\end{equation}
For $\varrho = \textstyle{\frac{3}{4}}$ and
$g_{\mathrm{RL}}^{\mathrm{T}} = g_{\mathrm{LR}}^{\mathrm{T}} = 0$ (no
tensor interaction) one finds $g_{\mathrm{RL}}^{\mathrm{V}} =
g_{\mathrm{LR}}^{\mathrm{V}} = 0$, with all of the remaining six
couplings being arbitrary.  On the other hand, any deviation from the
canonical value certainly would signify new physics.  Tree-level new
physics contributions to the Michel parameters occur in supersymmetric
theories with $R$-parity violation or theories with left-right
symmetric gauge groups.  For instance, the $R$-parity violating
interactions $\lambda_{311} L_{\mathrm{L}}^{(3)} L_{\mathrm{L}}^{(1)}
{\bar E}_{\mathrm{R}}^{(1)} +\lambda_{322} L_{\mathrm{L}}^{(3)}
L_{\mathrm{L}}^{(2)} {\bar E}_{\mathrm{R}}^{(2)}$ (where the index
denotes the lepton generation) give the following
contributions \cite{Hall:1983id,Barger:1989rk,Dreiner:1997uz}:
\begin{equation}
        \Delta \rho = \frac{3 \epsilon^{2}}{16},~~ \Delta \eta
        =\frac{\epsilon}{2},~~ \Delta \xi = -\frac{\epsilon^{2}}{4},~~
        \Delta \delta =0,~~~~ \epsilon \equiv \frac{\lambda_{311}
        \lambda_{322}}{4\sqrt{2} G_{\mathrm{F}}{\tilde
       m}^{2}_{e_{\mathrm{L}}^{(3)}}}\;.
        \label{eq:wfetscher:6}
\end{equation}
For a left-right model, one finds:
\begin{equation}
        \Delta \rho =-\frac{3}{2} \vartheta_{W_{\mathrm{R}}}^{2},~~\Delta
        \xi = -2 \vartheta_{W_{\mathrm{R}}}^{2} - 2 \left(
        \frac{M_{W_{1}}}{M_{W_{2}}}\right)^{4},
        \label{eq:wfetscher:7}
\end{equation}
where $\vartheta_{W_{\mathrm{R}}}$ is the
$W_{\mathrm{L}}$-$W_{\mathrm{R}}$ mixing angle, and $M_{W_{1}}$
($M_{W_{2}}$) is the mass of the mainly left (right) charged gauge
boson.  Measurements of $\varrho$ and $\xi$ with a precision of
$10^{-4}$ can probe $W_{\mathrm{R}}$ masses of about 1~TeV (in the
most unfavourable case $\vartheta_{W_{\mathrm{R}}} = 0$) and values of
the $R$-parity violating couplings $\lambda_{311} \approx
\lambda_{322} \approx 0.2$ (for a slepton mass of 200~GeV).  These
tests are competitive with direct searches at high-energy colliders.

There exist also observables which yield valuable information even if
they assume their canonical values, all of which are related to the
spin variables of the muon and the electron:
\begin{itemize}
        \item
                A measurement of the decay asymmetry yields the parameters
                $\delta$ and $P_{\mu}\xi$.  Especially interesting is the
                combination $P_{\mu}\xi\delta/\varrho$, which has been
                measured at TRIUMF \cite{Jodidio:1986mz} with a precision of
                $\approx 3 \times 10^{-3}$.  A new, ambitious experiment of
                the TWIST collaboration at TRIUMF measuring $\varrho$,
                $\delta$ and $P_{\mu}\xi$ has published improved results on
                the former two decay parameters ($\rho = 750.80 \pm
                0.44_{\mathrm{stat.}} \pm 0.93_{\mathrm{syst.}} \pm
                0.23_{\eta}) \times 10^{-3}$, where the last uncertainty is
                due to the correlation of $\rho$ with $\eta$ \cite{Musser:2004zw},
                $\delta = (749.64 \pm 0.66_{\mathrm{stat.}} \pm
                1.12_{\mathrm{syst.}}) \times 10^{-3}$ \cite{Gaponenko:2004mi} and
                $P_{\mu} \xi = (1000.3 \pm 0.6_{\mathrm{stat.}} \pm
                3.8_{\mathrm{syst.}}) \times 10^{-3}$ \cite{Jamieson:2006cf});
                
        \item
                A measurement of the longitudinal polarisation of the decay
                electrons $P_{\mathrm{L}}$ consistent with 1 yields limits for
                all five couplings where the electrons are right-handed.  This
                is a difficult experiment due to the lack of highly polarised
                electron targets used as analysers.  The present precision is
                $\Delta P_{\mathrm{L}} = 45 \times 10^{-3}$;
        \item
                The angular dependence of the longitudinal polarisation of
                decay positrons at the endpoint energy is currently being
                measured at PSI by the Louvain-la-Neuve-PSI-ETH Z\"urich
                collaboration \cite{Prieels_97}.  This yields the parameter
                $\xi''$ which is sensitive to the right-handed vector and the
                tensor currents; and
        \item
                A measurement of the transverse polarisation of the decay
                positrons requires a highly polarised, pulsed muon beam.  From
                the energy dependence of the component $P_{\mathrm{T}_{1}}$
                one can deduce the low-energy decay parameter $\eta$ which is
                needed for a model-independent value of the Fermi coupling
                constant.  The second component $P_{\mathrm{T}_{2}}$, which is
                transverse to the positron momentum and the muon polarisation,
                is non-invariant under time reversal.  A second generation
                experiment has been performed at PSI by the ETH
                Z\"urich-Cracow-PSI collaboration \cite{Barnett:2000ew}.  They
                obtained, among several other results, for the energy averaged
                transverse polarisation components $\langle P_{\mathrm{T}_{1}}
                \rangle = (6.3 \pm 7.7_{\mathrm{stat.}} \pm
                3.4_{\mathrm{syst.}}) \times 10^{-3}$, $\langle
                P_{\mathrm{T}_{2}} \rangle = (-3.7 \pm 7.7_{\mathrm{stat.}}
                \pm 3.4_{\mathrm{syst.}}) \times 10^{-3}$ and for the decay
                parameter $\eta = -2.1 \pm 7.0_{\mathrm{stat.}} \pm
                1.0_{\mathrm{syst.}}) \times 10^{-3}$ \cite{Danneberg:2005xv}.
                This last value has been obtained considering only terms
                interfering with the dominant $V-A$ interaction.
\end{itemize}

\subsubsection{Experimental prospects}\label{subsec:2.1.5}

As mentioned above, the precision on the muon lifetime can presumably
be increased over the ongoing measurements by one order of magnitude.
Improvement in measurements of the decay parameters seems more
difficult.  Most ambitious is the TRIUMF project which has published
first results on the parameters $\varrho$ (positron energy spectrum),
$P_{\mu} \xi$ and $\delta$ (decay 
asymmetry) \cite{Musser:2004zw,Gaponenko:2004mi,Jamieson:2006cf}.
  Their final goal is an
improvement by more than one order of magnitude.  The limits on most
other observables are not given by the muon rates which usually are
high enough already ($\approx 3 \times 10^{8}$~s$^{-1}$ at the $\mu$E1
beam at PSI, for example), but rather by effects such as positron
depolarisation in matter or by the small available polarisation ($< 7
\%$) of the electron targets used as analysers.  The measurement of
the transverse positron polarisation might be improved with a smaller
phase space (lateral beam dimension of a few millimetres or better).
This experiment needs a pulsed beam with high polarisation.

%% file: 07-Muon-Physics/07-06-conclusions/07-06-conclusions.tex
\subsection{Muon-Physics Conclusions}

The main conclusion of this study is that the physics potential of
a new slow muon facility,
such as the one that will become available as a necessary step on the
way to building a muon storage ring (Neutrino Factory),
is very rich and compelling, with a large variety of applications in many 
fields of basic research.  Indeed, 
muon physics, that has already played an important
role in establishing the Standard Model,
may provide us with crucial information regarding the theory that lies beyond, 
proving itself to be still far from having exhausted its potential.

This new low-energy muon source will have unprecedented intensity, three to
four orders of magnitude larger than presently available. It can have the large
degree of flexibility necessary to satisfy the requirements of very different 
experiments, providing muon beams with a wide variety of momenta and time 
structures. Both continuous and pulsed beams are possible. In addition, 
it is capable of producing physics results at the very early stages of the 
muon complex, well before the completion of muon cooling, 
acceleration, and storage sections.

Only preliminary ideas on the design of this facility are introduced
here, suggesting ways by which the muon flux could be boosted orders of
magnitude above present or foreseen facilities. The tasks of detailed
conceptual design of target and capture systems and of quantitative 
estimates of beam
performances are still entirely ahead of us.  The possibility of
using pions/muons produced in the backward direction is actively being
studied, which if feasible would permit the Neutrino Factory to 
take forward muons simultaneously and operate simultaneously with the
muon facility.

A major interest in muon physics lies in the searches for
rare processes that violate muon number conservation, or for a permanent
electric-dipole moment of the muon.
In many extensions of
the Standard Model, such as supersymmetry, lepton flavour violation
 may occur at rates close to the current experimental bounds.
Their discovery would have far-reaching consequences.
The most interesting processes are $\mu^+\to e^+\gamma$, $\mu^+\to
e^+e^-e^+$, and $\mu^-$--$e^-$ conversion in nuclei. 
We emphasise that all the different processes should be pursued,
along with a search for a muon EDM. Indeed,
the relative rates of the different modes provide a powerful 
tool for discriminating
different manifestations of new physics. 

 The muon
facility discussed here has enough flexibility to allow
the study of different muon processes, and 
promises to be more sensitive 
by at least a few orders of magnitude, when compared with 
current experiments.  In closing, we should 
mention that if such a facility existed, a number of 
fundamental studies with muonium and other muonic atoms
would also be possible.  Such studies  would permit increased
precision of the measurement of fundamental constants, and would 
serve to attract an additional community to such a facility.

%% file: 08-Appendix/08-Appendix.tex
\appendix
\section{Origin of the ISS and its Committees}

Further information on the activities that took place during the
course of the International Scoping Study of a future Neutrino Factory
and super-beam facility (the ISS) and links to the working groups can
be found at: {\tt http://www.hep.ph.ic.ac.uk/iss/}.

\subsection{Origin}

The international scoping study of a future Neutrino Factory and
super-beam facility (the ISS) was carried by the international
community between NuFact05, (the $7^{\rm th}$ International Workshop
on Neutrino Factories and Superbeams, Laboratori Nazionali di
Frascati, Rome, June 21--26, 2005) and NuFact06 (Irvine, California,
24--30 August 2006).  

The physics case for the facility was evaluated and options for the
accelerator complex and the neutrino detection systems were studied.
The principal objective of the study was to lay the foundations for a
full conceptual-design study of the facility. 
The plan for the scoping study was prepared in collaboration by the
international community that wished to carry it out; the ECFA/BENE
network in Europe, the Japanese NuFact-J collaboration, the US
Neutrino Factory and Muon Collider collaboration, and the UK Neutrino
Factory collaboration. 
STFC's Rutherford Appleton Laboratory was the host laboratory for the
study.

The study was directed by a Programme Committee advised by a
Stakeholders Board. 
The work of the study was carried out by three working groups: the
Physics Group; the Accelerator Group; and the Detector Group.  
Four plenary meetings at CERN, KEK, RAL, and Irvine were held during
the study period; workshops on specific topics were organised by the
individual working groups in between the plenary meetings. 
The conclusions of the study were presented at NuFact06.
This document, which presents the Physics Group's conclusions, was
prepared as the physics section of the ISS study group.

\subsection{Committee}

\noindent
\textbf{Programme Committee}\\
\begin{tabular}{lll}
Chairman: &Peter Dornan, &Imperial College London \\
Physics convener: &Yorikiyo Nagashima, &Osaka University \\
Accelerator convener: &Michael S. Zisman, &Lawrence Berkeley\\
\multicolumn{2}{l}{}&National Laboratory\\
Detector convener: &Alain Blondel, &University of Geneva 
\end{tabular}\\
\bigskip
\vfil\eject

\noindent
\textbf{Physics Working Subgroup Conveners}\\
\begin{tabular}{lll}
Theoretical: &Steve King, &University of Southampton\\
Phenomenological:&Osamu Yasuda, &Tokyo Metropolitan University\\
Experimental: &Ken Long, &Imperial College London\\
Muon: &Lee Roberts, &Boston University
\end{tabular}\\
\bigskip

\noindent
\textbf{Council Members of the Physics Working Group}\\
\begin{tabular}{ll}
Deborah A. Harris, &Fermilab\\
Pilar Hernandez, &University of Valencia\\ 
Steve King, &University of Southampton\\ 
Manfred Lindner, &Max-Planck-Institut f\"{u}r Kernphysik\\
Ken Long,  &Imperial College London\\ 
William Marciano, &Brookhaven National Laboratory\\
Mauro Mezzetto, &INFN Padova\\ 
Hitoshi Murayama, &Lawrence Berkeley National Laboratory\\
Yorikiyo Nagashima, &Osaka University \\
Kenzo Nakamura, &KEK\\
Lee Roberts, &Boston University\\
Osamu Yasuda, &Tokyo Metropolitan University\\

\end{tabular}\\